\documentclass[12pt,onecolumn]{IEEEtran}
\usepackage{color,graphicx,psfrag,epsfig,epsf,latexsym,theorem,hhline,amsmath,amssymb,array}
\renewcommand {\marginpar}[1]{}
\newcommand {\mymarginpar}[1]{\marginpar{#1}}

\def\_{\rule{.3em}{.15ex}}      

\newtheorem {definition}{Definition}
\newtheorem {axiom}[definition]{Axiom}
\newtheorem {lemma}[definition]{Lemma}
\newtheorem {theorem}[definition]{Theorem}
\newtheorem {corollary}[definition]{Corollary}
\newtheorem {property}[definition]{Property}
\newtheorem {proposition}[definition]{Proposition}
\newtheorem {example}[definition]{Example}
\newtheorem {remark}[definition]{Remark}
\newtheorem {algorithm}[definition]{Algorithm}
\newtheorem {conjecture}[definition]{Conjecture}
\newcommand {\bdefinition}[1]{\begin{definition}
                              \mymarginpar{definition:#1}
                              \label{definition:#1}}
\newcommand {\edefinition}   {\end{definition}}
\newcommand {\rdefinition}[1]{Definition~\ref{definition:#1}}

\newcommand {\baxiom}[1]{\begin{axiom}
                         \mymarginpar{axiom:#1}
                         \label{axiom:#1} }
\newcommand {\eaxiom}   {\end{axiom}}

\newcommand {\blemma}[1]{\begin{lemma}
                         \mymarginpar{lemma:#1}
                         \label{lemma:#1} }
\newcommand {\elemma}   {\end{lemma}}
\newcommand {\rlemma}[1]{Lemma~\ref{lemma:#1}}

\newcommand {\btheorem}[1]{\begin{theorem}
                           \mymarginpar{theorem:#1}
                           \label{theorem:#1}}
\newcommand {\etheorem}   {\end{theorem}}
\newcommand {\rtheorem}[1]{Theorem~\ref{theorem:#1}}

\newcommand {\bcorollary}[1]{\begin{corollary}
                             \mymarginpar{corollary:#1}
                             \label{corollary:#1} }
\newcommand {\ecorollary}   {\end{corollary}}
\newcommand {\rcorollary}[1]{Corollary~\ref{corollary:#1}}

\newcommand {\bproperty}[1]{\begin{property}
                            \mymarginpar{property:#1}
                            \label{property:#1}}
\newcommand {\eproperty}   {\end{property}}

\newcommand {\bproposition}[1]{\begin{proposition}
                               \mymarginpar{proposition:#1}
                               \label{proposition:#1}}
\newcommand {\eproposition}   {\end{proposition}}

\newcommand {\bproof}{\noindent {\bf Proof.} \ }
\newcommand {\eproof} {\hspace*{\fill}~\mbox{\rule[0pt]{1.3ex}{1.3ex}}}

\newcommand {\bexample}[1]{\begin{example}
                           \mymarginpar{example:#1}
                           \label{example:#1}}
\newcommand {\eexample}   {\end{example}}

\newcommand {\bremark}[1]{\begin{remark}
                          \mymarginpar{remark:#1}
                          \label{remark:#1}}
\newcommand {\eremark}   {\end{remark}}

\newcommand {\balgorithm}[1]{\begin{algorithm}
                         \mymarginpar{algorithm:#1}
                         \label{algorithm:#1} }
\newcommand {\ealgorithm}   {\end{algorithm}}
\newcommand {\ralgorithm}[1]{Algorithm~\ref{algorithm:#1}}

\newcommand {\bconjecture}[1]{\begin{conjecture}
                         \mymarginpar{conjecture:#1}
                         \label{conjecture:#1} }
\newcommand {\econjecture}   {\end{conjecture}}
\newcommand {\rconjecture}[1]{Conjecture~\ref{conjecture:#1}}

\newcommand {\bexercise}[1]{\begin{exercise}
                         \mymarginpar{exercise:#1}
                         \label{exercise:#1} }
\newcommand {\eexercise}   {\end{exercise}}

\newcommand {\barray}{\begin{array}}
\newcommand {\earray}{\end{array}}

\newcommand {\beqn}[1]{\begin{equation}
                       \mymarginpar{eqn:#1}
                       \label{eqn:#1}}
\newcommand {\eeqn}   {\end{equation}}

\newcommand {\beqnarray}[1]{\begin{eqnarray}
                            \mymarginpar{eqn:#1}
                            \label{eqn:#1} }
\newcommand {\eeqnarray}   {\end{eqnarray}}

\newcommand {\beqnarraynn}[1]{\begin{eqnarray*}
                              \mymarginpar{eqn:#1}}
\newcommand {\eeqnarraynn}   {\end{eqnarray*}}

\newcommand {\reqnarray}[1]{(\ref{eqn:#1})}

\newcommand {\bcases}{\begin{cases}}
\newcommand {\ecases}{\end{cases}}

\newcommand {\bselection}{\left\{ \begin{array}{ll}}
\newcommand {\eselection}{\end{array} \right.}

\newcommand {\bsection}[2]{\section{#1}
                           \mymarginpar{section:#2}
                           \label{section:#2} }
\newcommand {\rsection}[1]{Section~\ref{section:#1}}

\newcommand {\bappendix}[2]{\section{#1}
                           \mymarginpar{appendix:#2}
                           \label{appendix:#2} }
\newcommand {\rappendix}[1]{Appendix~\ref{appendix:#1}}

\newcommand {\bfigure}[2]{\begin{figure}[htbp]
                          \centerline {
                          \epsfig{figure={#1},clip=,width={#2}}}}

\newcommand {\efigure}[2]{\caption{#2}
                          \label{figure:#1}
                          \end{figure}
                          \mymarginpar{figure:#1}}

\newcommand {\bpdffigure}[2]{\begin{figure}[htbp]
                          \centering
                          \includegraphics[width={#2}]{#1}}

\newcommand {\epdffigure}[2]{\caption{#2}
                          \label{figure:#1}
                          \end{figure}
                          \mymarginpar{figure:#1}}

\newcommand {\brotatefigure}[2]{\begin{figure}[htbp]
                                \centerline {
                                \epsfig{figure={#1},clip=,angle=-90,width={#2}}}}

\newcommand {\erotatefigure}[2]{\caption{#2}
                                \label{figure:#1}
                                \end{figure}
                                \mymarginpar{figure:#1}}

\newcommand {\bfigurefirst}[2] {\begin{figure}[h]
                                \centerline {
                                \setlength{\epsfxsize}{#2}
                                \epsffile{#1}}}

\newcommand {\rfigure}[1]{Figure~\ref{figure:#1}}

\newcommand {\btable}[2]{\begin{table}[#1]
                         \begin{center}
                         \begin{tabular}{#2}}
\newcommand {\etable}[2]{\end{tabular}
                         \end{center}
                         \caption{#2}
                         \label{table:#1}
                         \end{table}
                         \mymarginpar{table:#1}
                         \vspace{.1in}}
\newcommand {\rtable}[1]{Table~\ref{table:#1}}

\newcommand {\btabular}[1]{\begin{center}
                           \begin{tabular}{#1}}
\newcommand {\etabular}{\end{tabular}
                        \end{center}}

\newcommand {\btinytable}[2]{\begin{table}[#1]
                         \tiny
                         \begin{center}
                         \begin{tabular}{#2}}
\newcommand {\etinytable}[2]{\end{tabular}
                         \end{center}
                         \caption{#2}
                         \label{table:#1}
                         \end{table}
                         \mymarginpar{table:#1}
                         \vspace{.1in}}

\newcommand {\bscriptsizetable}[2]{\begin{table}[#1]
                         \scriptsize
                         \begin{center}
                         \begin{tabular}{#2}}
\newcommand {\escriptsizetable}[2]{\end{tabular}
                         \end{center}
                         \caption{#2}
                         \label{table:#1}
                         \end{table}
                         \mymarginpar{table:#1}
                         \vspace{.1in}}

\newcommand {\blargetable}[2]{\begin{table}[#1]
                         \LARGE
                         \begin{center}
                         \begin{tabular}{#2}}
\newcommand {\elargetable}[2]{\end{tabular}
                         \end{center}
                         \caption{#2}
                         \label{table:#1}
                         \end{table}
                         \mymarginpar{table:#1}
                         \vspace{.1in}}

\newcommand{\bitemize}{\begin{itemize}}
\newcommand{\eitemize}{\end{itemize}}

\newcommand{\benumerate}{\begin{enumerate}}
\newcommand{\eenumerate}{\end{enumerate}}

\newcommand {\bdescription} {\begin{description}}
\newcommand {\edescription} {\end{description}}

\newcommand{\indep}{\mathop{\lower .12em\hbox{\underbar{\raise .3ex\hbox{$\|$}}}\kern .5pt}}
\newcommand{\nn}{\nonumber}
\newcommand{\aligneq}{\hspace*{-0.1in}&=&\hspace*{-0.1in}}
\newcommand{\alignleq}{\hspace*{-0.1in}&\leq&\hspace*{-0.1in}}
\newcommand{\alignless}{\hspace*{-0.1in}&<&\hspace*{-0.1in}}
\newcommand{\aligngeq}{\hspace*{-0.1in}&\geq&\hspace*{-0.1in}}

\newcommand{\alignspace}{\hspace*{-0.1in}& &\hspace*{-0.1in}}

\def\0bf{{\bf 0}}
\def\1bf{{\bf 1}}
\def\2bf{{\bf 2}}
\def\3bf{{\bf 3}}
\def\4bf{{\bf 4}}
\def\5bf{{\bf 5}}
\def\6bf{{\bf 6}}
\def\7bf{{\bf 7}}
\def\8bf{{\bf 8}}
\def\9bf{{\bf 9}}

\def\dbf{{\bf d}}
\def\ebf{{\bf e}}

\def\lbf{{\bf l}}
\def\mbf{{\bf m}}
\def\nbf{{\bf n}}

\def\Zbf{{\bf Z}}

\def\Bcal{{\cal B}}

\def\Gcal{{\cal G}}

\def\Ncal{{\cal N}}

\begin{document}

\baselineskip16pt

\title{Constructions of Optical Queues With a Limited Number of
Recirculations--Part~II: Optimal Constructions}

\author{Xuan-Chao Huang and Jay Cheng \\
\thanks{This work was supported in part by the National Science Council, Taiwan, R.O.C.,
        under Contract NSC 96-2221-E-007-076, Contract NSC 97-2221-E-007-105-MY3,
        and the Program for Promoting Academic Excellence of Universities NSC 94-2752-E-007-002-PAE.}
\thanks{The authors are with the Department of Electrical Engineering and
        the Institute of Communications Engineering,
        National Tsing Hua University, Hsinchu 30013, Taiwan, R.O.C.
        (e-mails: d9761812@oz.nthu.edu.tw; jcheng@ee.nthu.edu.tw).}
}

\maketitle
\thispagestyle{empty}
\begin{abstract}

\baselineskip12pt

One of the main problems in all-optical packet-switched networks
is the lack of optical buffers,
and one feasible technology for the constructions of optical buffers
is to use optical crossbar Switches and fiber Delay Lines (SDL).
In this two-part paper, we consider SDL constructions of optical queues
with a limited number of recirculations through
the optical switches and the fiber delay lines.
Such a problem arises from practical feasibility considerations.
In Part~I, we have proposed a class of greedy constructions
for certain types of optical queues, including linear compressors,
linear decompressors, and 2-to-1 FIFO multiplexers,
and have shown that every optimal construction among our previous constructions
of these types of optical queues under the constraint of
a limited number of recirculations must be a greedy construction.
Specifically, given $M\geq 2$ and $1\leq k\leq M-1$,
we have shown that to find an optimal construction,
it suffices to find an optimal sequence ${\dbf^*}_1^M\in \Gcal_{M,k}$ such that
$B({\dbf^*}_1^M;k)=\max_{\dbf_1^M\in \Gcal_{M,k}}B(\dbf_1^M;k)$,
where $B(\dbf_1^M;k)$ is the maximum representable integer
with respect to $\dbf_1^M$ and $k$ (defined in Part~I)
and $\Gcal_{M,k}$ is the set of sequences $\dbf_1^M=(d_1,d_2,\ldots,d_M)$
given by $d_{s_i+j}=B(\dbf_1^{s_i+j-1};i+1)+1$
for $i=0,1,\ldots,k-1$ and $j=1,2,\ldots,n_{i+1}$
for some sequence $\nbf_1^k=(n_1,n_2,\ldots,n_k)$
with $n_1\geq 2$, $n_2,n_3,\ldots,n_k\geq 1$, and $\sum_{i=1}^{k}n_i=M$,
in which $s_0=0$ and $s_i=\sum_{\ell=1}^{i}n_{\ell}$ for $i=1,2,\ldots,k$.

In Part~II, the present paper,
we further show that there are at most two optimal constructions
and give a simple algorithm to obtain the optimal construction(s).
The main idea in Part~II is to use \emph{pairwise comparison}
to remove a sequence $\dbf_1^M\in \Gcal_{M,k}$ such that
$B(\dbf_1^M;k)<B({\dbf'}_1^M;k)$ for some ${\dbf'}_1^M\in \Gcal_{M,k}$.
To our surprise, the simple algorithm for obtaining the optimal construction(s)
is related to the well-known \emph{Euclid's algorithm}
for finding the greatest common divisor (gcd) of two integers.
In particular, we show that if $\gcd(M,k)=1$, then there is only one optimal construction;
if $\gcd(M,k)=2$, then there are two optimal constructions;
and if $\gcd(M,k)\geq 3$, then there are at most two optimal constructions.
\end{abstract}


\pagestyle{empty}
\pagestyle{headings}
\pagenumbering{arabic}

\newpage
\bsection{Introduction}{introduction}

One of the bottlenecks toward all-optical packet-switched networks
is the O-E-O (optical-electrical-optical) conversion
due to the lack of optical buffers.
Currently, the only known way to ``store'' optical packets
without converting them into other media
is to use optical Switches and fiber Delay Lines (SDL)
to direct optical packets to the right place at the right time.
Although the optical buffers constructed by the SDL approach
can only be used as sequential buffers with fixed storage times
so that they do not have the random access capability,
research results in the SDL literature (see the references in Part~I \cite{CCCLL10} of this paper)
show that they can still be used to construct many types of optical queues
commonly encountered in practice:
including output-buffered switches, FIFO multiplexers, FIFO queues, LIFO queues, priority queues,
time slot interchanges, linear compressors, linear decompressors,
non-overtaking delay lines, and flexible delay lines.

\bpdffigure{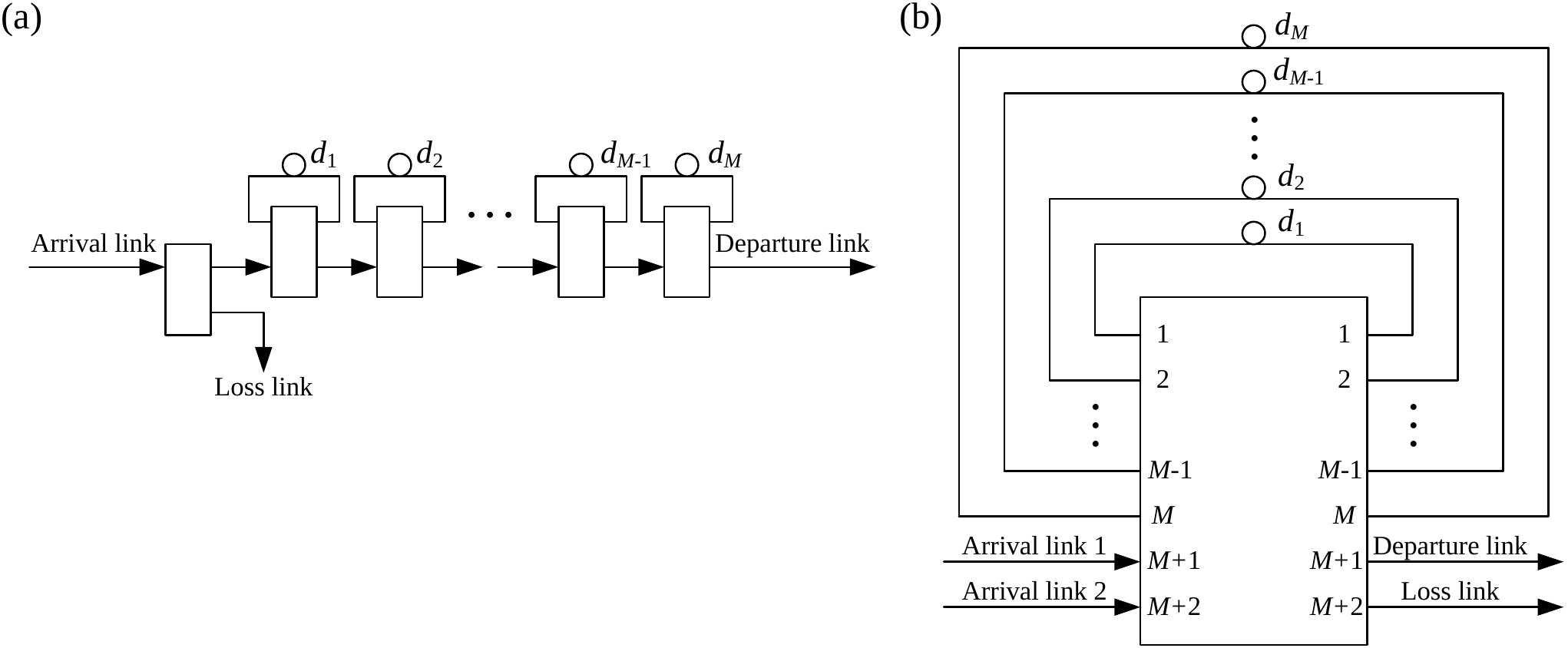}{6.0in}
\epdffigure{LC-2-to-1-FIFO-MUX}
{(a) A construction of a linear compressor.
(b) A construction of a 2-to-1 FIFO multiplexer.}

In this two-part paper, we address an important practical feasibility issue
that is of great concern in the SDL constructions of optical queues:
the constructions of optical queues with a limited number of
recirculations through the optical switches and the fiber delay lines.
We recall that it was shown in \cite{CCCL09} (resp., \cite{CCLC06}) that the construction
in \rfigure{LC-2-to-1-FIFO-MUX}(a)/mirror image of \rfigure{LC-2-to-1-FIFO-MUX}(a)
(resp., \rfigure{LC-2-to-1-FIFO-MUX}(b))
can be operated as a linear compressor/decompressor
(resp., 2-to-1 FIFO multiplexer) under a simple packet routing scheme.
Suppose that there is a limitation on the number, say $k$,
of recirculations through the $M$ fibers in \rfigure{LC-2-to-1-FIFO-MUX}
due to practical feasibility considerations.
For the nontrivial case that $M\geq 2$ and $1\leq k\leq M-1$,
we have proposed in Part~I a class of greedy constructions
by specifying a class $\Gcal_{M,k}$ of sequences
of the delays of the $M$ fiber delay lines in \rfigure{LC-2-to-1-FIFO-MUX}
such that every sequence $\dbf_1^M=(d_1,d_2,\ldots,d_M)$ in $\Gcal_{M,k}$ is given by
\beqnarray{OQ-LR-delays-greedy-1}
d_{s_i+j}=B(\dbf_1^{s_i+j-1};i+1)+1,
\emph{ for } i=0,1,\ldots,k-1 \textrm{ and } j=1,2,\ldots,n_{i+1},
\eeqnarray
for some sequence $\nbf_1^k=(n_1,n_2,\ldots,n_k)$
with $n_1\geq 2$, $n_2,n_3,\ldots,n_k\geq 1$, and $\sum_{i=1}^{k}n_i=M$,
where $s_0=0$, $s_i=\sum_{\ell=1}^{i}n_{\ell}$ for $i=1,2,\ldots,k$,
and $B(\dbf_1^{s_i+j-1};i+1)$ is the maximum representable integer
with respect to $\dbf_1^{s_i+j-1}$ and $i+1$
(see \cite{CCCLL10} for a definition of the maximum representable integer).
Furthermore, we have shown that every optimal construction among our previous constructions
of linear compressors/decompressors in \cite{CCCL09}
and 2-to-1 FIFO multiplexers in \cite{CCLC06}
under the constraint of a limited number of recirculations
must be a greedy construction.
Specifically, let
\beqnarray{G-M-k}
\Gcal_{M,k}=
\left\{\dbf_1^M\in (\Zbf^+)^M: \dbf_1^M \textrm{ is given by } \reqnarray{OQ-LR-delays-greedy-1}
\textrm{ for some } \nbf_1^k\in \Ncal_{M,k}\right\},
\eeqnarray
where
\beqnarray{N-M-k}
\Ncal_{M,k}=
\left\{\nbf_1^k\in (\Zbf^+)^k: n_1\geq 2 \textrm{ and } \sum_{i=1}^{k}n_i=M\right\},
\eeqnarray
then to find an optimal construction,
it suffices to find an optimal sequence over $\Gcal_{M,k}$,
i.e., to find a sequence ${\dbf^*}_1^M\in \Gcal_{M,k}$ such that
$B({\dbf^*}_1^M;k)=\max_{\dbf_1^M\in \Gcal_{M,k}}B(\dbf_1^M;k)$.
We call a sequence ${\nbf^*}_1^M\in \Ncal_{M,k}$
an \emph{optimal} sequence over $\Ncal_{M,k}$
if the sequence ${\dbf^*}_1^M\in \Gcal_{M,k}$
obtained by using ${\nbf^*}_1^M$ in \reqnarray{OQ-LR-delays-greedy-1}
is an optimal sequence over $\Gcal_{M,k}$.
Therefore, to find an optimal construction,
it suffices to find an optimal sequence over $\Ncal_{M,k}$.

Our contribution in Part~II, the present paper,
is to show in \rsection{optimal constructions} and \rsection{proof of the main result}
that there are at most two optimal sequences over $\Ncal_{M,k}$
and give a simple algorithm to obtain the optimal sequence(s).
The main idea in \rsection{optimal constructions} and \rsection{proof of the main result}
is to use \emph{pairwise comparison} to remove a sequence $\nbf_1^M\in \Ncal_{M,k}$
such that $B(\dbf_1^M;k)<B({\dbf'}_1^M;k)$ for some ${\nbf'}_1^M\in \Ncal_{M,k}$,
where $\dbf_1^M$ and ${\dbf'}_1^M$ are obtained by using $\nbf_1^M$ and ${\nbf'}_1^M$,
respectively, in \reqnarray{OQ-LR-delays-greedy-1}.
To our surprise, the simple algorithm for obtaining the optimal sequence(s)
is related to the well-known \emph{Euclid's algorithm}
for finding the greatest common divisor (gcd) of two integers.
In particular, we show that if $\gcd(M,k)=1$, then there is only one optimal sequence;
if $\gcd(M,k)=2$, then there are two optimal sequences;
and if $\gcd(M,k)\geq 3$, then there are at most two optimal sequences.
We conclude this paper in \rsection{conclusion}.

\bsection{The Optimal Constructions}{optimal constructions}

To simplify the presentation in this paper,
in the following we first define left-imbedded sequences, left pre-sequences,
right-imbedded sequences, and right pre-sequences.

\bpdffigure{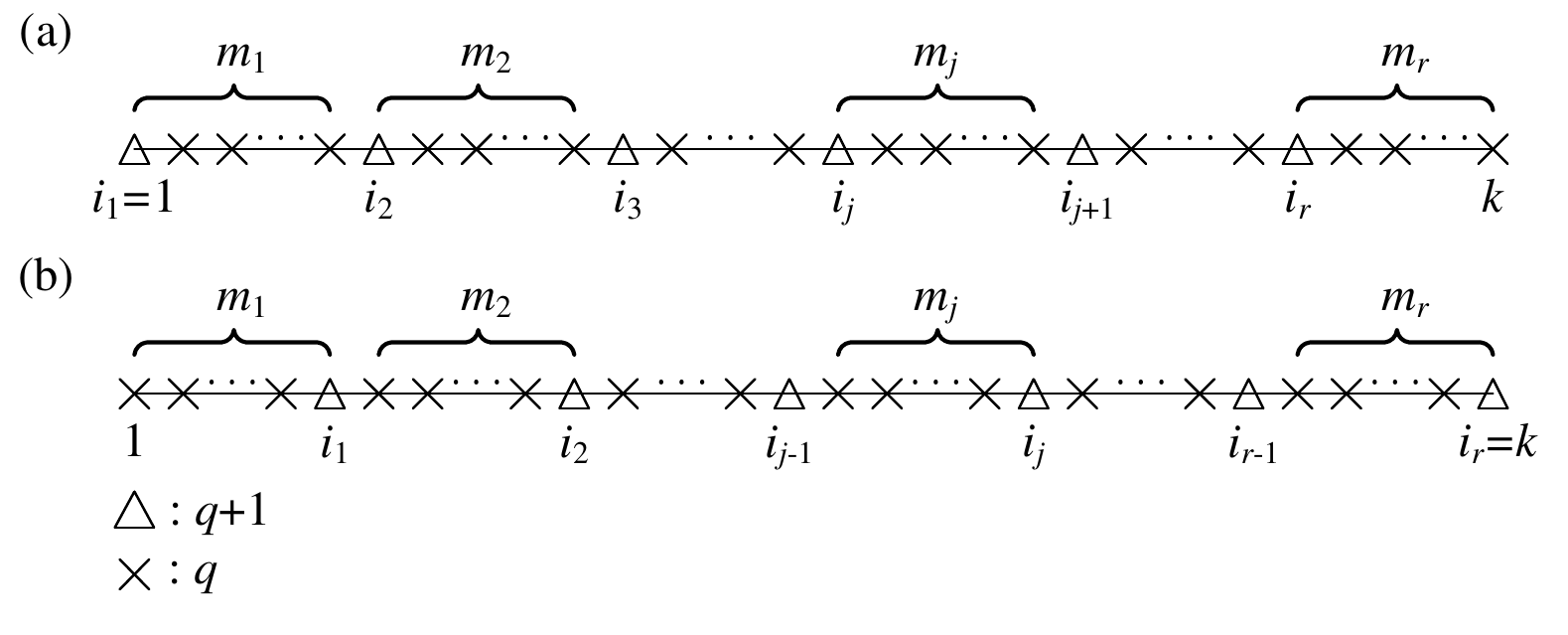}{4.5in}
\epdffigure{left-right-imbedded-sequences}
{(a) An illustration of \reqnarray{left-imbedded sequences-2}
in the definition of left-imbedded sequences in \rdefinition{left-imbedded sequences}
and \reqnarray{left pre-sequences-2}
in the definition of left pre-sequences in \rdefinition{left pre-sequences}.
(b) An illustration of \reqnarray{right-imbedded sequences-2}
in the definition of right-imbedded sequences in \rdefinition{right-imbedded sequences}
and \reqnarray{right pre-sequences-2}
in the definition of right pre-sequences in \rdefinition{right pre-sequences}.}

\bdefinition{left-imbedded sequences}\textbf{(Left-imbedded sequences)}
Suppose that $M\geq 2$ and $1\leq k\leq M-1$.
Let $M=qk+r$, where $q$ and $r$ are the quotient and the remainder,
respectively, of $M$ divided by $k$.
Suppose that $r\neq 0$ and $\nbf_1^k=(n_1,n_2,\ldots,n_k)$ is a sequence of positive integers
such that
\beqnarray{left-imbedded sequences-1}
n_i=
\bselection
q+1, &\textrm{if } i=i_1,i_2,\ldots,i_r, \\
q, &\textrm{otherwise},
\eselection
\eeqnarray
for some $1=i_1<i_2<\cdots <i_r\leq k$ (note that $\sum_{i=1}^{k}n_i=qk+r=M$).
The \emph{left-imbedded sequence} $\mbf_1^r=(m_1,m_2,\ldots,m_r)$ of the sequence $\nbf_1^k$
with respect to $M$ and $k$, denoted $\mbf_1^r=L_{M,k}^I(\nbf_1^k)$,
is a sequence of positive integers given by
(see \rfigure{left-right-imbedded-sequences}(a) for an illustration)
\beqnarray{left-imbedded sequences-2}
m_j=
\bselection
|\{i_j,i_j+1,\ldots,i_{j+1}-1\}|=i_{j+1}-i_j, &\textrm{if } j=1,2,\ldots,r-1, \\
|\{i_r,i_r+1,\ldots,k\}|=k-i_r+1, &\textrm{if } j=r.
\eselection
\eeqnarray
Note that $\sum_{j=1}^{r}m_j=k-i_1+1=k$.
\edefinition

\bdefinition{left pre-sequences}\textbf{(Left pre-sequences)}
Suppose that $M\geq 2$ and $1\leq k\leq M-1$.
Let $M=qk+r$, where $q$ and $r$ are the quotient and the remainder,
respectively, of $M$ divided by $k$.
Suppose that $r\neq 0$ and $\mbf_1^r=(m_1,m_2,\ldots,m_r)$ is a sequence of positive integers
such that $\sum_{j=1}^{r}m_j=k$.
The \emph{left pre-sequence} $\nbf_1^k=(n_1,n_2,\ldots,n_k)$ of the sequence $\mbf_1^r$
with respect to $M$ and $k$, denoted $\nbf_1^k=L_{M,k}(\mbf_1^r)$,
is a sequence of positive integers given by
\beqnarray{left pre-sequences-1}
n_i=
\bselection
q+1, &\textrm{if } i=i_1,i_2,\ldots,i_r, \\
q, &\textrm{otherwise},
\eselection
\eeqnarray
where
\beqnarray{left pre-sequences-2}
i_j=\sum_{\ell=1}^{j-1}m_{\ell}+1, \textrm{ for } j=1,2,\ldots,r,
\eeqnarray
(see \rfigure{left-right-imbedded-sequences}(a) for an illustration).
Note that $i_1=\sum_{\ell=1}^{0}m_{\ell}+1=1$ and $\sum_{i=1}^{k}n_i=qk+r=M$.
\edefinition

\bexample{left-imbedded sequences and left pre-sequences}
Suppose that $M=13$ and $k=5$.
Then the quotient and the remainder of $M$ divided by $k$
are $q=2$ and $r=3$, respectively.

(i) Suppose that $\nbf_1^k=(3,3,2,3,2)$, then the left-imbedded sequence $\mbf_1^r$
of the sequence $\nbf_1^k$ with respect to $M$ and $k$ is given by
\beqnarray{}
\mbf_1^r=L_{M,k}^I(\nbf_1^k)=L_{13,5}^I((3,3,2,3,2))=(1,2,2). \nn
\eeqnarray

(ii) Suppose that $\mbf_1^r=(1,2,2)$, then the left pre-sequence $\nbf_1^k$
of the sequence $\mbf_1^r$ with respect to $M$ and $k$ is given by
\beqnarray{}
\nbf_1^k=L_{M,k}(\mbf_1^r)=L_{13,5}((1,2,2))=(3,3,2,3,2). \nn
\eeqnarray
\eexample

Clearly, left-imbedded sequences and left pre-sequences are closely related
as can be seen in the following lemma.

\blemma{left-imbedded sequences-left pre-sequences}
Suppose that $M\geq 2$ and $1\leq k\leq M-1$.
Let $M=qk+r$, where $q$ and $r$ are the quotient and the remainder,
respectively, of $M$ divided by $k$.
Suppose that $r\neq 0$.

(i) If $\mbf_1^r=L_{M,k}^I(\nbf_1^k)$,
where $\nbf_1^k$ is given by \reqnarray{left-imbedded sequences-1}
for some $1=i_1<i_2<\cdots <i_r\leq k$,
then we have $\nbf_1^k=L_{M,k}(\mbf_1^r)$.

(ii) Conversely, if $\nbf_1^k=L_{M,k}(\mbf_1^r)$,
where $\mbf_1^r$ is a sequence of positive integers
such that $\sum_{j=1}^{r}m_j=k$,
then we have $\mbf_1^r=L_{M,k}^I(\nbf_1^k)$.
\elemma

\bproof
(i) If $\mbf_1^r=L_{M,k}^I(\nbf_1^k)$,
where $\nbf_1^k$ is given by \reqnarray{left-imbedded sequences-1}
for some $1=i_1<i_2<\cdots <i_r\leq k$,
then we see from \reqnarray{left-imbedded sequences-2} that
\beqnarray{}
\alignspace
\sum_{\ell=1}^{j-1}m_{\ell}+1=
\bselection
\sum_{\ell=1}^{0}m_{\ell}+1=1=i_1=i_j, &\textrm{if } j=1, \\
\sum_{\ell=1}^{j-1}(i_{\ell+1}-i_{\ell})+1=i_j-i_1+1=i_j, &\textrm{if } j=2,3,\ldots,r,
\eselection
\label{eqn:proof-left-imbedded sequences-left pre-sequences-111} \\
\alignspace
\sum_{j=1}^{r}m_j=\sum_{\ell=1}^{r-1}(i_{\ell+1}-i_{\ell})+k-i_r+1=k-i_1+1=k.
\label{eqn:proof-left-imbedded sequences-left pre-sequences-222}
\eeqnarray

As we have $\sum_{j=1}^{r}m_j=k$ in
\reqnarray{proof-left-imbedded sequences-left pre-sequences-222},
the left pre-sequence of $\mbf_1^r$ with respect to $M$ and $k$ is well defined,
say ${\nbf'}_1^k=L_{M,k}(\mbf_1^r)$.
From the definition of left pre-sequences
in \rdefinition{left pre-sequences}
and \reqnarray{proof-left-imbedded sequences-left pre-sequences-111},
we have
\beqnarray{proof-left-imbedded sequences-left pre-sequences-333}
n_i'=
\bselection
q+1, &\textrm{if } i=i_1',i_2',\ldots,i_r', \\
q, &\textrm{otherwise},
\eselection
\eeqnarray
where
\beqnarray{proof-left-imbedded sequences-left pre-sequences-444}
i_j'=\sum_{\ell=1}^{j-1}m_{\ell}+1=i_j, \textrm{ for } j=1,2,\ldots,r,
\eeqnarray
As such, it follows from \reqnarray{left-imbedded sequences-1},
\reqnarray{proof-left-imbedded sequences-left pre-sequences-333},
\reqnarray{proof-left-imbedded sequences-left pre-sequences-444},
and ${\nbf'}_1^k=L_{M,k}(\mbf_1^r)$ that $\nbf_1^k={\nbf'}_1^k=L_{M,k}(\mbf_1^r)$.

(ii) Conversely, if $\nbf_1^k=L_{M,k}(\mbf_1^r)$,
where $\mbf_1^r$ is a sequence of positive integers
such that $\sum_{j=1}^{r}m_j=k$,
then we see from \reqnarray{left pre-sequences-2} that
\beqnarray{}
\alignspace
i_1=\sum_{\ell=1}^{0}m_{\ell}+1=1,
\label{eqn:proof-left-imbedded sequences-left pre-sequences-555} \\
\alignspace
i_{j+1}-i_j=\left(\sum_{\ell=1}^{j}m_{\ell}+1\right)
-\left(\sum_{\ell=1}^{j-1}m_{\ell}+1\right)=m_j,
\textrm{ for } j=1,2,\ldots,r-1,
\label{eqn:proof-left-imbedded sequences-left pre-sequences-666}
\eeqnarray
\beqnarray{}
\hspace*{-1.36in}
k-i_r+1=\sum_{j=1}^{r}m_j-\left(\sum_{\ell=1}^{r-1}m_{\ell}+1\right)+1=m_r.
\label{eqn:proof-left-imbedded sequences-left pre-sequences-777}
\eeqnarray

From $i_1=1$ in \reqnarray{proof-left-imbedded sequences-left pre-sequences-555}
and \reqnarray{left pre-sequences-1}
we see that $\nbf_1^k$ satisfies \reqnarray{left-imbedded sequences-1}
and hence the left-imbedded sequence of $\nbf_1^k$
with respect to $M$ and $k$ is well defined,
say ${\mbf'}_1^r=L_{M,k}^I(\nbf_1^k)$.
From the definition of left-imbedded sequences
in \rdefinition{left-imbedded sequences},
we have
\beqnarray{proof-left-imbedded sequences-left pre-sequences-888}
m_j'=
\bselection
i_{j+1}-i_j, &\textrm{if } j=1,2,\ldots,r-1, \\
k-i_r+1, &\textrm{if } j=r.
\eselection
\eeqnarray
As such, it follows from \reqnarray{proof-left-imbedded sequences-left pre-sequences-666},
\reqnarray{proof-left-imbedded sequences-left pre-sequences-777},
\reqnarray{proof-left-imbedded sequences-left pre-sequences-888},
and ${\mbf'}_1^r=L_{M,k}^I(\nbf_1^k)$ that $\mbf_1^r={\mbf'}_1^r=L_{M,k}^I(\nbf_1^k)$.
\eproof

\bdefinition{right-imbedded sequences}\textbf{(Right-imbedded sequences)}
Suppose that $M\geq 2$ and $1\leq k\leq M-1$.
Let $M=qk+r$, where $q$ and $r$ are the quotient and the remainder,
respectively, of $M$ divided by $k$.
Suppose that $r\neq 0$ and $\nbf_1^k=(n_1,n_2,\ldots,n_k)$ is a sequence of positive integers
such that
\beqnarray{right-imbedded sequences-1}
n_i=
\bselection
q+1, &\textrm{if } i=i_1,i_2,\ldots,i_r, \\
q, &\textrm{otherwise},
\eselection
\eeqnarray
for some $1\leq i_1<i_2<\cdots <i_r=k$ (note that $\sum_{i=1}^{k}n_i=qk+r=M$).
The \emph{right-imbedded sequence} $\mbf_1^r=(m_1,m_2,\ldots,m_r)$ of the sequence $\nbf_1^k$
with respect to $M$ and $k$, denoted $\mbf_1^r=R_{M,k}^I(\nbf_1^k)$,
is a sequence of positive integers given by
(see \rfigure{left-right-imbedded-sequences}(b) for an illustration)
\beqnarray{right-imbedded sequences-2}
m_j=
\bselection
|\{1,2,\ldots,i_1\}|=i_1, &\textrm{if } j=1, \\
|\{i_{j-1}+1,i_{j-1}+2,\ldots,i_j\}|=i_j-i_{j-1}, &\textrm{if } j=2,3,\ldots,r.
\eselection
\eeqnarray
Note that $\sum_{j=1}^{r}m_j=i_r=k$.
\edefinition

\bdefinition{right pre-sequences}\textbf{(Right pre-sequences)}
Suppose that $M\geq 2$ and $1\leq k\leq M-1$.
Let $M=qk+r$, where $q$ and $r$ are the quotient and the remainder,
respectively, of $M$ divided by $k$.
Suppose that $r\neq 0$ and $\mbf_1^r=(m_1,m_2,\ldots,m_r)$ is a sequence of positive integers
such that $\sum_{j=1}^{r}m_j=k$.
The \emph{right pre-sequence} $\nbf_1^k=(n_1,n_2,\ldots,n_k)$ of the sequence $\mbf_1^r$
with respect to $M$ and $k$, denoted $\nbf_1^k=R_{M,k}(\mbf_1^r)$,
is a sequence of positive integers given by
\beqnarray{right pre-sequences-1}
n_i=
\bselection
q+1, &\textrm{if } i=i_1,i_2,\ldots,i_r, \\
q, &\textrm{otherwise},
\eselection
\eeqnarray
where
\beqnarray{right pre-sequences-2}
i_j=\sum_{\ell=1}^{j}m_{\ell}, &\textrm{for } j=1,2,\ldots,r,
\eeqnarray
(see \rfigure{left-right-imbedded-sequences}(b) for an illustration).
Note that $i_r=\sum_{\ell=1}^{r}m_{\ell}=k$ and $\sum_{i=1}^{k}n_i=qk+r=M$.
\edefinition

\bexample{right-imbedded sequences and right pre-sequences}
Suppose that $M=13$ and $k=5$.
Then the quotient and the remainder of $M$ divided by $k$
are $q=2$ and $r=3$, respectively.

(i) Suppose that $\nbf_1^k=(2,3,2,3,3)$, then the right-imbedded sequence $\mbf_1^r$
of the sequence $\nbf_1^k$ with respect to $M$ and $k$ is given by
\beqnarray{}
\mbf_1^r=R_{M,k}^I(\nbf_1^k)=R_{13,5}^I((2,3,2,3,3))=(2,2,1). \nn
\eeqnarray

(ii) Suppose that $\mbf_1^r=(2,2,1)$, then the right pre-sequence $\nbf_1^k$
of the sequence $\mbf_1^r$ with respect to $M$ and $k$ is given by
\beqnarray{}
\nbf_1^k=R_{M,k}(\mbf_1^r)=R_{13,5}((2,2,1))=(2,3,2,3,3). \nn
\eeqnarray
\eexample

Right-imbedded sequences and right pre-sequences are also closely related
as can be seen in the following lemma.

\blemma{right-imbedded sequences-right pre-sequences}
Suppose that $M\geq 2$ and $1\leq k\leq M-1$.
Let $M=qk+r$, where $q$ and $r$ are the quotient and the remainder,
respectively, of $M$ divided by $k$.
Suppose that $r\neq 0$.

(i) If $\mbf_1^r=R_{M,k}^I(\nbf_1^k)$,
where $\nbf_1^k$ is given by \reqnarray{right-imbedded sequences-1}
for some $1\leq i_1<i_2<\cdots <i_r=k$,
then we have $\nbf_1^k=R_{M,k}(\mbf_1^r)$.

(ii) Conversely, if $\nbf_1^k=R_{M,k}(\mbf_1^r)$,
where $\mbf_1^r$ is a sequence of positive integers
such that $\sum_{j=1}^{r}m_j=k$,
then we have $\mbf_1^r=R_{M,k}^I(\nbf_1^k)$.
\elemma

\bproof
(i) If $\mbf_1^r=R_{M,k}^I(\nbf_1^k)$,
where $\nbf_1^k$ is given by \reqnarray{right-imbedded sequences-1}
for some $1\leq i_1<i_2<\cdots <i_r=k$,
then we see from \reqnarray{right-imbedded sequences-2} that
\beqnarray{proof-right-imbedded sequences-right pre-sequences-111}
\alignspace
\sum_{\ell=1}^{j}m_{\ell}=i_j, \textrm{ for } j=1,2,\ldots,r.
\eeqnarray
From $i_r=k$ and \reqnarray{proof-right-imbedded sequences-right pre-sequences-111},
we can see that $\sum_{j=1}^{r}m_j=i_r=k$
and hence the right pre-sequence of $\mbf_1^r$ with respect to $M$ and $k$ is well defined,
say ${\nbf'}_1^k=R_{M,k}(\mbf_1^r)$.
From the definition of right pre-sequences
in \rdefinition{right pre-sequences}
and \reqnarray{proof-right-imbedded sequences-right pre-sequences-111},
we have
\beqnarray{proof-right-imbedded sequences-right pre-sequences-222}
n_i'=
\bselection
q+1, &\textrm{if } i=i_1',i_2',\ldots,i_r', \\
q, &\textrm{otherwise},
\eselection
\eeqnarray
where
\beqnarray{proof-right-imbedded sequences-right pre-sequences-333}
i_j'=\sum_{\ell=1}^{j}m_{\ell}=i_j, \textrm{ for } j=1,2,\ldots,r,
\eeqnarray
As such, it follows from \reqnarray{right-imbedded sequences-1},
\reqnarray{proof-right-imbedded sequences-right pre-sequences-222},
\reqnarray{proof-right-imbedded sequences-right pre-sequences-333},
and ${\nbf'}_1^k=R_{M,k}(\mbf_1^r)$ that $\nbf_1^k={\nbf'}_1^k=R_{M,k}(\mbf_1^r)$.

(ii) Conversely, if $\nbf_1^k=R_{M,k}(\mbf_1^r)$,
where $\mbf_1^r$ is a sequence of positive integers
such that $\sum_{j=1}^{r}m_j=k$,
then we see from \reqnarray{right pre-sequences-2} that
\beqnarray{}
\alignspace
i_1=m_1,
\label{eqn:proof-right-imbedded sequences-right pre-sequences-444} \\
\alignspace
i_j-i_{j-1}=\sum_{\ell=1}^{j}m_{\ell}-\sum_{\ell=1}^{j-1}m_{\ell}=m_j,
\textrm{ for } j=2,3,\ldots,r,
\label{eqn:proof-right-imbedded sequences-right pre-sequences-555} \\
\alignspace
i_r=\sum_{j=1}^{r}m_j=k.
\label{eqn:proof-right-imbedded sequences-right pre-sequences-666}
\eeqnarray

From \reqnarray{right pre-sequences-1}
and $i_r=k$ in \reqnarray{proof-right-imbedded sequences-right pre-sequences-666},
we see that $\nbf_1^k$ satisfies \reqnarray{right-imbedded sequences-1}
and hence the right-imbedded sequence of $\nbf_1^k$
with respect to $M$ and $k$ is well defined,
say ${\mbf'}_1^r=R_{M,k}^I(\nbf_1^k)$.
From the definition of right-imbedded sequences
in \rdefinition{right-imbedded sequences},
we have
\beqnarray{proof-right-imbedded sequences-right pre-sequences-777}
m_j'=
\bselection
i_1, &\textrm{if } j=1, \\
i_j-i_{j-1}, &\textrm{if } j=2,3,\ldots,r.
\eselection
\eeqnarray
As such, it follows from \reqnarray{proof-right-imbedded sequences-right pre-sequences-444},
\reqnarray{proof-right-imbedded sequences-right pre-sequences-555},
\reqnarray{proof-right-imbedded sequences-right pre-sequences-777},
and ${\mbf'}_1^r=R_{M,k}^I(\nbf_1^k)$ that $\mbf_1^r={\mbf'}_1^r=R_{M,k}^I(\nbf_1^k)$.
\eproof

In the following theorem,
we state the main result in this paper on optimal sequences over $\Ncal_{M,k}$.
The proof of the theorem will be given in \rsection{proof of the main result}.

\btheorem{main result}
Let $M\geq 2$ and $1\leq k\leq M-1$.

(i) Suppose that $\gcd(M,k)=1$.
Then there is only one optimal sequence over $\Ncal_{M,k}$,
and the optimal sequence is given by the sequence $\nbf_1^k(1)$
obtained in Step 2 or Step 3 of \ralgorithm{main result} below
(depending on which of the two steps is executed in \ralgorithm{main result}).

(ii) Suppose that $\gcd(M,k)=2$.
Then there are two optimal sequences over $\Ncal_{M,k}$,
and the two optimal sequences are given by the two sequences
$\nbf_1^k(1)$ and $\mbf_1^k(1)$
obtained in Step 2 or Step 3 of \ralgorithm{main result}
(depending on which of the two steps is executed in \ralgorithm{main result}).

(iii) Suppose that $\gcd(M,k)\geq 3$.
Then there are at most two optimal sequences over $\Ncal_{M,k}$,
and the two possible optimal sequences are given by the two sequences
$\nbf_1^k(1)$ and $\mbf_1^k(1)$
obtained in Step 2 or Step 3 of \ralgorithm{main result}
(depending on which of the two steps is executed in \ralgorithm{main result}).
\etheorem

\balgorithm{main result}
Given $M\geq 2$ and $1\leq k\leq M-1$.

Step 1. (Euclid's algorithm) Let $r_{-1}=M$, $r_0=k$.
If $r_{i-1}\neq 0$ for $i\geq 1$, then define $q_i\geq 1$ and $0\leq r_i<r_{i-1}$ recursively
as the quotient and the remainder, respectively, of $r_{i-2}$ divided by $r_{i-1}$ so that
\beqnarray{Euclid's algorithm}
r_{i-2}=q_i\cdot r_{i-1}+r_i.
\eeqnarray
It is well known that such a recursive process will stop after a finite number of recursions
when the remainder is zero, say $r_N=0$ for some $N\geq 1$,
at which point we obtain the greatest common divisor of $M$ and $k$ as $\gcd(M,k)=r_{N-1}$.
If $N$ is an odd integer, then go to Step 2;
otherwise, if $N$ is an even integer, then go to Step 3.

Step 2. (i) Let $\nbf_1^{r_{N-1}}(N)$ be given by $n_j(N)=q_N$ for $1\leq j\leq r_{N-1}$.
For $i=N-2,N-4,\ldots,1$ (in that order), recursively compute
\beqnarray{}
\nbf_1^{r_{i}}(i+1)\aligneq R_{r_{i-1},r_{i}}(\nbf_1^{r_{i+1}}(i+2)),\label{eqn:main result-111}\\
\nbf_1^{r_{i-1}}(i)\aligneq L_{r_{i-2},r_{i-1}}(\nbf_1^{r_{i}}(i+1)).\label{eqn:main result-222}
\eeqnarray

(ii) If $r_{N-1}\geq 2$, let $\mbf_1^{r_{N-1}}(N)$ be given by
$m_1(N)=q_N+1$, $m_{r_{N-1}}(N)=q_N-1$
(note that $q_N\geq 2$ as $r_{N-2}=q_N\cdot r_{N-1}+r_N=q_N\cdot r_{N-1}$ and $r_{N-2}>r_{N-1}$),
and $m_j(N)=q_N$ for $2\leq j\leq r_{N-1}-1$.
For $i=N-2,N-4,\ldots,1$ (in that order), recursively compute
\beqnarray{}
\mbf_1^{r_{i}}(i+1)\aligneq R_{r_{i-1},r_{i}}(\mbf_1^{r_{i+1}}(i+2)),\label{eqn:main result-333}\\
\mbf_1^{r_{i-1}}(i)\aligneq L_{r_{i-2},r_{i-1}}(\mbf_1^{r_{i}}(i+1)).\label{eqn:main result-444}
\eeqnarray

Step 3. (i) Let $\nbf_1^{r_{N-1}}(N)$ be given by $n_j(N)=q_N$ for $1\leq j\leq r_{N-1}$.
First compute
\beqnarray{main result-555}
\nbf_1^{r_{N-2}}(N-1)=L_{r_{N-3},r_{N-2}}(\nbf_1^{r_{N-1}}(N)).
\eeqnarray
Then for $i=N-3,N-5,\ldots,1$ (in that order), recursively compute
\beqnarray{}
\nbf_1^{r_{i}}(i+1)\aligneq R_{r_{i-1},r_{i}}(\nbf_1^{r_{i+1}}(i+2)),\label{eqn:main result-666}\\
\nbf_1^{r_{i-1}}(i)\aligneq L_{r_{i-2},r_{i-1}}(\nbf_1^{r_{i}}(i+1)).\label{eqn:main result-777}
\eeqnarray

(ii) If $r_{N-1}\geq 2$, let $\mbf_1^{r_{N-1}}(N)$ be given by
$m_1(N)=q_N-1$ (note that $q_N\geq 2$), $m_{r_{N-1}}(N)=q_N+1$,
and $m_j(N)=q_N$ for $2\leq j\leq r_{N-1}-1$.
First compute
\beqnarray{main result-888}
\mbf_1^{r_{N-2}}(N-1)=L_{r_{N-3},r_{N-2}}(\mbf_1^{r_{N-1}}(N)).
\eeqnarray
Then for $i=N-3,N-5,\ldots,1$ (in that order), recursively compute
\beqnarray{}
\mbf_1^{r_{i}}(i+1)\aligneq R_{r_{i-1},r_{i}}(\mbf_1^{r_{i+1}}(i+2)),\label{eqn:main result-999}\\
\mbf_1^{r_{i-1}}(i)\aligneq L_{r_{i-2},r_{i-1}}(\mbf_1^{r_{i}}(i+1)).\label{eqn:main result-aaa}
\eeqnarray
\ealgorithm

Note that in Step~2(i) (for the case that $N$ is an odd integer),
we begin with $\nbf_1^{r_{N-1}}(N)=(q_N,q_N,\ldots,q_N)$ and then compute
\beqnarray{}
\nbf_1^{r_{N-2}}(N-1) \aligneq R_{r_{N-3},r_{N-2}}(\nbf_1^{r_{N-1}}(N)), \nn\\
\nbf_1^{r_{N-3}}(N-2) \aligneq L_{r_{N-4},r_{N-3}}(\nbf_1^{r_{N-2}}(N-1)), \nn\\
\nbf_1^{r_{N-4}}(N-3) \aligneq R_{r_{N-5},r_{N-4}}(\nbf_1^{r_{N-3}}(N-2)), \nn\\
\nbf_1^{r_{N-5}}(N-4) \aligneq L_{r_{N-6},r_{N-5}}(\nbf_1^{r_{N-4}}(N-3)), \nn\\
&\vdots& \nn\\
\nbf_1^{r_1}(2) \aligneq R_{r_0,r_1}(\nbf_1^{r_2}(3)), \nn\\
\nbf_1^{r_0}(1) \aligneq L_{r_{-1},r_0}(\nbf_1^{r_1}(2)). \nn
\eeqnarray
As $r_N=0$, we have $\sum_{i=1}^{r_{N-1}}n_i(N)=q_N\cdot r_{N-1}=q_N\cdot r_{N-1}+r_N=r_{N-2}$,
and hence the right pre-sequence $\nbf_1^{r_{N-2}}(N-1)=R_{r_{N-3},r_{N-2}}(\nbf_1^{r_{N-1}}(N))$
of the sequence $\nbf_1^{r_{N-1}}(N)$ with respect to $r_{N-3}$ and $r_{N-2}$ is well defined.
From the definition of right pre-sequences in \rdefinition{right pre-sequences},
we can see that $\sum_{i=1}^{r_{N-2}}n_i(N-1)=r_{N-3}$,
and thus the left pre-sequence $\nbf_1^{r_{N-3}}(N-2)=L_{r_{N-4},r_{N-3}}(\nbf_1^{r_{N-2}}(N-1))$
of the sequence $\nbf_1^{r_{N-2}}(N-1)$ with respect to $r_{N-4}$ and $r_{N-3}$ is well defined.
We can repeat the above argument and see that
the right pre-sequence $\nbf_1^{r_{N-4}}(N-3)=R_{r_{N-5},r_{N-4}}(\nbf_1^{r_{N-3}}(N-2))$
of the sequence $\nbf_1^{r_{N-3}}(N-2)$ with respect to $r_{N-5}$ and $r_{N-4}$ is well defined,
the left pre-sequence $\nbf_1^{r_{N-5}}(N-4)=L_{r_{N-6},r_{N-5}}(\nbf_1^{r_{N-4}}(N-3))$
of the sequence $\nbf_1^{r_{N-4}}(N-3)$ with respect to $r_{N-6}$ and $r_{N-5}$ is well defined, $\ldots$,
the right pre-sequence $\nbf_1^{r_1}(2)=R_{r_0,r_1}(\nbf_1^{r_2}(3))$
of the sequence $\nbf_1^{r_2}(3)$ with respect to $r_0$ and $r_1$ is well defined,
and the left pre-sequence $\nbf_1^{r_0}(1)=L_{r_{-1},r_0}(\nbf_1^{r_1}(2))$
of the sequence $\nbf_1^{r_1}(2)$ with respect to $r_{-1}$ and $r_0$ is well defined.
Similarly, in Step~2(ii) (for the case that $N$ is an odd integer and $r_{N-1}\geq 2$),
we begin with $\mbf_1^{r_{N-1}}(N)=(q_N+1,q_N,\ldots,q_N,q_N-1)$
and we can see that the left pre-sequences and the right pre-sequences in Step~2(ii) are all well defined.

Also note that in Step~3(i) (for the case that $N$ is an even integer),
we begin with $\nbf_1^{r_{N-1}}(N)=(q_N,q_N,\ldots,q_N)$ and then compute
\beqnarray{}
\nbf_1^{r_{N-2}}(N-1) \aligneq L_{r_{N-3},r_{N-2}}(\nbf_1^{r_{N-1}}(N)), \nn\\
\nbf_1^{r_{N-3}}(N-2) \aligneq R_{r_{N-4},r_{N-3}}(\nbf_1^{r_{N-2}}(N-1)), \nn\\
\nbf_1^{r_{N-4}}(N-3) \aligneq L_{r_{N-5},r_{N-4}}(\nbf_1^{r_{N-3}}(N-2)), \nn\\
&\vdots& \nn\\
\nbf_1^{r_1}(2) \aligneq R_{r_0,r_1}(\nbf_1^{r_2}(3)), \nn\\
\nbf_1^{r_0}(1) \aligneq L_{r_{-1},r_0}(\nbf_1^{r_1}(2)). \nn
\eeqnarray
As in Step~2(i), we can argue that the left pre-sequences and the right pre-sequences in Step~3(i)
are all well defined.
Similarly, in Step~3(ii) (for the case that $N$ is an even integer and $r_{N-1}\geq 2$),
we begin with $\mbf_1^{r_{N-1}}(N)=(q_N-1,q_N,\ldots,q_N,q_N+1)$
and we can see that the left pre-sequences and the right pre-sequences in Step~3(ii) are all well defined.

In the following, we give a few examples to illustrate how
\rtheorem{main result} and \ralgorithm{main result} work.

\bexample{main result-1}
Suppose that $M=11$ and $k=3$.
In Step 1 of \ralgorithm{main result},
we obtain $r_{-1}=11$, $r_0=3$, $q_1=3$, $r_1=2$, $q_2=1$, $r_2=1$, $q_3=2$, and $r_3=0$.
As a result, we have $N=3$ and $\gcd(M,k)=r_{N-1}=r_2=1$.
Since $N=3$ is an odd integer and $r_{N-1}=1$,
we proceed to Step 2(i) of \ralgorithm{main result} and obtain
\beqnarray{}
\alignspace \nbf_1^{r_2}(3)=(q_3)=(2), \nn\\
\alignspace \nbf_1^{r_1}(2)=R_{r_0,r_1}(\nbf_1^{r_2}(3))=R_{3,2}((2))=(1,2), \nn\\
\alignspace \nbf_1^k(1)=\nbf_1^{r_0}(1)=L_{r_{-1},r_0}(\nbf_1^{r_1}(2))=L_{11,3}((1,2))=(4,4,3). \nn
\eeqnarray
It follows from \rtheorem{main result}(i) that
there is only one optimal sequence over $\Ncal_{M,k}$,
and the optimal sequence is given by $\nbf_1^k(1)=(4,4,3)$.
\eexample

\bexample{main result-2}
Suppose that $M=13$ and $k=5$.
In Step 1 of \ralgorithm{main result},
we obtain $r_{-1}=13$, $r_0=5$, $q_1=2$, $r_1=3$, $q_2=1$, $r_2=2$,
$q_3=1$, $r_3=1$, $q_4=2$, and $r_4=0$.
Consequently, we have $N=4$ and $\gcd(M,k)=r_{N-1}=r_3=1$.
Since $N=4$ is an even integer and $r_{N-1}=1$,
we proceed to Step 3(i) of \ralgorithm{main result} and obtain
\beqnarray{}
\alignspace \nbf_1^{r_3}(4)=(q_4)=(2), \nn\\
\alignspace \nbf_1^{r_2}(3)=L_{r_1,r_2}(\nbf_1^{r_3}(4))=L_{3,2}((2))=(2,1), \nn\\
\alignspace \nbf_1^{r_1}(2)=R_{r_0,r_1}(\nbf_1^{r_2}(3))=R_{5,3}((2,1))=(1,2,2), \nn\\
\alignspace \nbf_1^k(1)=\nbf_1^{r_0}(1)=L_{r_{-1},r_0}(\nbf_1^{r_1}(2))=L_{13,5}((1,2,2))=(3,3,2,3,2). \nn
\eeqnarray
It follows from \rtheorem{main result}(i) that
there is only one optimal sequence over $\Ncal_{M,k}$,
and the optimal sequence is given by $\nbf_1^k(1)=(3,3,2,3,2)$.
\eexample

\bexample{main result-3}
Suppose that $M=16$ and $k=6$.
In Step 1 of \ralgorithm{main result},
we obtain $r_{-1}=16$, $r_0=6$, $q_1=2$, $r_1=4$, $q_2=1$, $r_2=2$,
$q_3=2$, and $r_3=0$.
As a result, we have $N=3$ and $\gcd(M,k)=r_{N-1}=r_2=2$.
Since $N=3$ is an odd integer and $r_{N-1}=2$,
we proceed to Step 2(i) and Step 2(ii) of \ralgorithm{main result} and obtain
\beqnarray{}
\alignspace \nbf_1^{r_2}(3)=(q_3,q_3)=(2,2), \nn\\
\alignspace \nbf_1^{r_1}(2)=R_{r_0,r_1}(\nbf_1^{r_2}(3))=R_{6,4}((2,2))=(1,2,1,2), \nn\\
\alignspace \nbf_1^k(1)=\nbf_1^{r_0}(1)=L_{r_{-1},r_0}(\nbf_1^{r_1}(2))=L_{16,6}((1,2,1,2))=(3,3,2,3,3,2); \nn\\
\alignspace \mbf_1^{r_2}(3)=(q_3+1,q_3-1)=(3,1), \nn\\
\alignspace \mbf_1^{r_1}(2)=R_{r_0,r_1}(\mbf_1^{r_2}(3))=R_{6,4}((3,1))=(1,1,2,2), \nn\\
\alignspace \mbf_1^k(1)=\mbf_1^{r_0}(1)=L_{r_{-1},r_0}(\mbf_1^{r_1}(2))=L_{16,6}((1,1,2,2))=(3,3,3,2,3,2). \nn
\eeqnarray
It follows from \rtheorem{main result}(ii) that
there are two optimal sequences over $\Ncal_{M,k}$,
and the two optimal sequences are given by
$\nbf_1^k(1)=(3,3,2,3,3,2)$ and $\mbf_1^k(1)=(3,3,3,2,3,2)$.
\eexample

\bexample{main result-4}
Suppose that $M=26$ and $k=10$.
In Step 1 of \ralgorithm{main result},
we obtain $r_{-1}=26$, $r_0=10$, $q_1=2$, $r_1=6$, $q_2=1$, $r_2=4$,
$q_3=1$, $r_3=2$, $q_4=2$, and $r_4=0$.
Consequently, we have $N=4$ and $\gcd(M,k)=r_{N-1}=r_3=2$.
Since $N=4$ is an even integer and $r_{N-1}=2$,
we proceed to Step 3(i) and Step 3(ii) of \ralgorithm{main result} and obtain
\beqnarray{}
\alignspace \nbf_1^{r_3}(4)=(q_4,q_4)=(2,2), \nn\\
\alignspace \nbf_1^{r_2}(3)=L_{r_1,r_2}(\nbf_1^{r_3}(4))=L_{6,4}((2,2))=(2,1,2,1), \nn\\
\alignspace \nbf_1^{r_1}(2)=R_{r_0,r_1}(\nbf_1^{r_2}(3))=R_{10,6}((2,1,2,1))=(1,2,2,1,2,2), \nn\\
\alignspace \nbf_1^k(1)=\nbf_1^{r_0}(1)=L_{r_{-1},r_0}(\nbf_1^{r_1}(2))=L_{26,10}((1,2,2,1,2,2))=(3,3,2,3,2,3,3,2,3,2); \nn\\
\alignspace \mbf_1^{r_3}(4)=(q_4-1,q_4+1)=(1,3), \nn\\
\alignspace \mbf_1^{r_2}(3)=L_{r_1,r_2}(\mbf_1^{r_3}(4))=L_{6,4}((1,3))=(2,2,1,1), \nn\\
\alignspace \mbf_1^{r_1}(2)=R_{r_0,r_1}(\mbf_1^{r_2}(3))=R_{10,6}((2,2,1,1))=(1,2,1,2,2,2), \nn\\
\alignspace \mbf_1^k(1)=\mbf_1^{r_0}(1)=L_{r_{-1},r_0}(\mbf_1^{r_1}(2))=L_{26,10}((1,2,1,2,2,2))=(3,3,2,3,3,2,3,2,3,2). \nn
\eeqnarray
It follows from \rtheorem{main result}(ii) that
there are two optimal sequences over $\Ncal_{M,k}$,
and the two optimal sequences are given by
$\nbf_1^k(1)=(3,3,2,3,2,3,3,2,3,2)$ and $\mbf_1^k(1)=(3,3,2,3,3,2,3,2,3,2)$.
\eexample

\bexample{main result-5}
Suppose that $M=24$ and $k=9$.
In Step 1 of \ralgorithm{main result},
we obtain $r_{-1}=24$, $r_0=9$, $q_1=2$, $r_1=6$, $q_2=1$, $r_2=3$,
$q_3=2$, and $r_3=0$.
As a result, we have $N=3$ and $\gcd(M,k)=r_{N-1}=r_2=3$.
Since $N=3$ is an odd integer and $r_{N-1}=3$,
we proceed to Step 2(i) and Step 2(ii) of \ralgorithm{main result} and obtain
\beqnarray{}
\alignspace \nbf_1^{r_2}(3)=(q_3,q_3,q_3)=(2,2,2), \nn\\
\alignspace \nbf_1^{r_1}(2)=R_{r_0,r_1}(\nbf_1^{r_2}(3))=R_{9,6}((2,2,2))=(1,2,1,2,1,2), \nn\\
\alignspace \nbf_1^k(1)=\nbf_1^{r_0}(1)=L_{r_{-1},r_0}(\nbf_1^{r_1}(2))=L_{24,9}((1,2,1,2,1,2))=(3,3,2,3,3,2,3,3,2); \nn\\
\alignspace \mbf_1^{r_2}(3)=(q_3+1,q_3,q_3-1)=(3,2,1), \nn\\
\alignspace \mbf_1^{r_1}(2)=R_{r_0,r_1}(\mbf_1^{r_2}(3))=R_{9,6}((3,2,1))=(1,1,2,1,2,2), \nn\\
\alignspace \mbf_1^k(1)=\mbf_1^{r_0}(1)=L_{r_{-1},r_0}(\mbf_1^{r_1}(2))=L_{24,9}((1,1,2,1,2,2))=(3,3,3,2,3,3,2,3,2). \nn
\eeqnarray
It follows from \rtheorem{main result}(iii) that
there are at most two optimal sequences over $\Ncal_{M,k}$,
and the two possible optimal sequences are given by
$\nbf_1^k(1)=(3,3,2,3,3,2,3,3,2)$ and $\mbf_1^k(1)=(3,3,3,2,3,3,2,3,2)$.
\eexample

\bexample{main result-6}
Suppose that $M=39$ and $k=15$.
In Step 1 of \ralgorithm{main result},
we obtain $r_{-1}=39$, $r_0=15$, $q_1=2$, $r_1=9$, $q_2=1$, $r_2=6$,
$q_3=1$, $r_3=3$, $q_4=2$, and $r_4=0$.
Consequently, we have $N=4$ and $\gcd(M,k)=r_{N-1}=r_3=3$.
Since $N=4$ is an even integer and $r_{N-1}=3$,
we proceed to Step 3(i) and Step 3(ii) of \ralgorithm{main result} and obtain
\beqnarray{}
\alignspace \nbf_1^{r_3}(4)=(q_4,q_4,q_4)=(2,2,2), \nn\\
\alignspace \nbf_1^{r_2}(3)=L_{r_1,r_2}(\nbf_1^{r_3}(4))=L_{9,6}((2,2,2))=(2,1,2,1,2,1), \nn\\
\alignspace \nbf_1^{r_1}(2)=R_{r_0,r_1}(\nbf_1^{r_2}(3))=R_{15,9}((2,1,2,1,2,1))=(1,2,2,1,2,2,1,2,2), \nn\\
\alignspace \nbf_1^k(1)=\nbf_1^{r_0}(1)=L_{r_{-1},r_0}(\nbf_1^{r_1}(2))=L_{39,15}((1,2,2,1,2,2,1,2,2)) \nn\\
\alignspace \hspace*{0.43in} =(3,3,2,3,2,3,3,2,3,2,3,3,2,3,2); \nn\\
\alignspace \mbf_1^{r_3}(4)=(q_4-1,q_4,q_4+1)=(1,2,3), \nn\\
\alignspace \mbf_1^{r_2}(3)=L_{r_1,r_2}(\mbf_1^{r_3}(4))=L_{9,6}((1,2,3))=(2,2,1,2,1,1), \nn\\
\alignspace \mbf_1^{r_1}(2)=R_{r_0,r_1}(\mbf_1^{r_2}(3))=R_{15,9}((2,2,1,2,1,1))=(1,2,1,2,2,1,2,2,2), \nn\\
\alignspace \mbf_1^k(1)=\mbf_1^{r_0}(1)=L_{r_{-1},r_0}(\mbf_1^{r_1}(2))=L_{39,15}((1,2,1,2,2,1,2,2,2))\nn\\
\alignspace \hspace*{0.48in} =(3,3,2,3,3,2,3,2,3,3,2,3,2,3,2). \nn
\eeqnarray
It follows from \rtheorem{main result}(iii) that
there are at most two optimal sequences over $\Ncal_{M,k}$,
and the two possible optimal sequences are given by
$\nbf_1^k(1)=(3,3,2,3,2,3,3,2,3,2,3,3,2,3,2)$ and $\mbf_1^k(1)=(3,3,2,3,3,2,3,2,3,3,2,3,2,3,2)$.
\eexample

\bsection{Proof of \rtheorem{main result}}{proof of the main result}

The main idea in our proof of \rtheorem{main result}
is to use \emph{pairwise comparison} to remove a sequence $\nbf_1^M\in \Ncal_{M,k}$
such that $B(\dbf_1^M;k)<B({\dbf'}_1^M;k)$ for some ${\nbf'}_1^M\in \Ncal_{M,k}$,
where $\dbf_1^M$ and ${\dbf'}_1^M$ are obtained by using $\nbf_1^M$ and ${\nbf'}_1^M$,
respectively, in \reqnarray{OQ-LR-delays-greedy-1}.

To simplify the presentation of the proof,
we first introduce a few notations that will be used in the proof of \rtheorem{main result}.
Suppose that $M\geq 2$ and $1\leq k\leq M-1$. Let $r_{-1}=M$, $r_0=k$, and let $q_i$ and $r_i$, $i=1,2,\ldots,N$,
be recursively obtained as in Step 1 of \ralgorithm{main result}.
For $1\leq h\leq N$, let $\Ncal_{M,k}(h)$ be the set of sequences of positive integers
$\nbf_1^{r_{h-1}}(h)=(n_1(h),n_2(h),\ldots,n_{r_{h-1}}(h))$
such that: (i) $\sum_{i=1}^{r_{h-1}}n_i(h)=r_{h-2}$, and (ii) $n_1(h)\geq 2$ in the case that $h=1$,
i.e.,
\beqnarray{N-M-k-h}
\Ncal_{M,k}(h)=
\bselection
\{\nbf_1^{r_{h-1}}(h)\in (\Zbf^+)^{r_{h-1}}:
\sum_{i=1}^{r_{h-1}}n_i(h)=r_{h-2} \textrm{ and } n_1(h)\geq 2\}, &\textrm{if } h=1, \\
\{\nbf_1^{r_{h-1}}(h)\in (\Zbf^+)^{r_{h-1}}:
\sum_{i=1}^{r_{h-1}}n_i(h)=r_{h-2}\}, &\textrm{if } 2\leq h\leq N.
\eselection
\eeqnarray
Note that as $r_{-1}=M$ and $r_0=k$,
it is clear from \reqnarray{N-M-k-h} and \reqnarray{N-M-k} that $\Ncal_{M,k}(1)=\Ncal_{M,k}$.

Let $\nbf_1^{r_{h-1}}(h)\in \Ncal_{M,k}(h)$ and ${\nbf'}_1^{r_{h-1}}(h)\in \Ncal_{M,k}(h)$,
where $1\leq h\leq N$.
If $2\leq h\leq N$ and $h$ is an odd integer,
then we recursively compute for $i=h-2,h-4,\ldots,1$ (in that order)
\beqnarray{}
\nbf_1^{r_i}(i+1)\aligneq R_{r_{i-1},r_i}(\nbf_1^{r_{i+1}}(i+2)), \label{eqn:order relation-111}\\
\nbf_1^{r_{i-1}}(i)\aligneq L_{r_{i-2},r_{i-1}}(\nbf_1^{r_i}(i+1)). \label{eqn:order relation-222}
\eeqnarray
If $2\leq h\leq N$ and $h$ is an even integer,
then we first compute
\beqnarray{order relation-333}
\nbf_1^{r_{h-2}}(h-1)=L_{r_{h-3},r_{h-2}}(\nbf_1^{r_{h-1}}(h)),
\eeqnarray
and then we recursively compute for $i=h-3,h-5,\ldots,1$ (in that order)
\beqnarray{}
\nbf_1^{r_i}(i+1)\aligneq R_{r_{i-1},r_i}(\nbf_1^{r_{i+1}}(i+2)), \label{eqn:order relation-444}\\
\nbf_1^{r_{i-1}}(i)\aligneq L_{r_{i-2},r_{i-1}}(\nbf_1^{r_i}(i+1)). \label{eqn:order relation-555}
\eeqnarray
Therefore, by using $\nbf_1^{r_{h-1}}(h)$,
we obtain sequences of positive integers $\nbf_1^{r_{i-1}}(i)$, $i=1,2,\ldots,h-1$,
such that: (i) $\sum_{j=1}^{r_{i-1}}n_j(i)=r_{i-2}$
(according to the definitions of left pre-sequences in \rdefinition{left pre-sequences}
and right pre-sequences in \rdefinition{right pre-sequences}),
and (ii) $n_1(i)=q_i+1\geq 2$ in the case that $i$ is an odd integer
(according to the definition of left pre-sequences in \rdefinition{left pre-sequences}).
As such, we see from \reqnarray{N-M-k-h} that
$\nbf_1^{r_{i-1}}(i)\in \Ncal_{M,k}(i)$ for $i=1,2,\ldots,h-1$.
Similarly, by using ${\nbf'}_1^{r_{h-1}}(h)$,
we can also obtain ${\nbf'}_1^{r_{i-1}}(i)\in \Ncal_{M,k}(i)$ for $i=1,2,\ldots,h-1$.
Let $\dbf_1^M$ and ${\dbf'}_1^M$ be obtained by using $\nbf_1^k(1)=\nbf_1^{r_0}(1)$
and ${\nbf'}_1^k(1)={\nbf'}_1^{r_0}(1)$, respectively, in \reqnarray{OQ-LR-delays-greedy-1}.
We define the binary relation $\prec$ (resp., $\equiv$, $\succ$, $\preceq$, $\succeq$)
on $\Ncal_{M,k}(h)$ as follows:
\beqnarray{order relation-666}
\nbf_1^{r_{h-1}}(h)\prec (\textrm{resp.}, \equiv, \succ, \preceq, \succeq)\
{\nbf'}_1^{r_{h-1}}(h)
\textrm{ if }
B(\dbf_1^M;k)< (\textrm{resp.}, =, >, \leq, \geq)\ B({\dbf'}_1^M;k).
\eeqnarray
We call a sequence of positive integers $\nbf_1^{r_{h-1}}(h)$
an \emph{optimal} sequence over $\Ncal_{M,k}(h)$
if $\nbf_1^{r_{h-1}}(h)\in \Ncal_{M,k}(h)$
and $\nbf_1^{r_{h-1}}(h)\succeq{\nbf'}_1^{r_{h-1}}(h)$
for all ${\nbf'}_1^{r_{h-1}}(h)\in \Ncal_{M,k}(h)$.

It is clear that if $\nbf_1^{r_{i-1}}(i)$ (resp., ${\nbf'}_1^{r_{i-1}}(i)$),
where $1\leq i\leq h-1$,
is given by \reqnarray{order relation-111}--\reqnarray{order relation-222}
or \reqnarray{order relation-333}--\reqnarray{order relation-555}
(depending on whether $h$ is an odd or an even integer),
then we have from the definition of the binary relation
$\prec$ (resp., $\equiv$, $\succ$, $\preceq$, $\succeq$)
in \reqnarray{order relation-666} that
\beqnarray{order relation-777}
\nbf_1^{r_{h-1}}(h)\prec
(\textrm{resp.}, \equiv, \succ, \preceq, \succeq)\ {\nbf'}_1^{r_{h-1}}(h)
\textrm{ iff } \nbf_1^{r_{i-1}}(i)\prec
(\textrm{resp.}, \equiv, \succ, \preceq, \succeq)\ {\nbf'}_1^{r_{i-1}}(i).
\eeqnarray

We need the following theorem from Part~I to prove \rtheorem{main result}.

\btheorem{OQ-LR-delays-greedy}  \cite{CCCLL10}
Suppose that $M\geq 2$ and $1\leq k\leq M-1$.
Let $\dbf_1^M\in \Gcal_{M,k}$ so that there exists a sequence $\nbf_1^k\in \Ncal_{M,k}$
such that $d_{s_i+j}$ is given by \reqnarray{OQ-LR-delays-greedy-1},
i.e., $d_{s_i+j}=B(\dbf_1^{s_i+j-1};i+1)+1$,
for $i=0,1,\ldots,k-1$ and $j=1,2,\ldots,n_{i+1}$,
where $s_0=0$ and $s_i=\sum_{\ell=1}^{i}n_{\ell}$ for $i=1,2,\ldots,k$.
Then $d_1,d_2,\ldots,d_M$ can be recursively expressed as
\beqnarray{}
\alignspace
d_j=j, \textrm{ for } j=1,2,\ldots,s_1,
\label{eqn:OQ-LR-delays-greedy-2}\\
\alignspace
d_{s_i+j}=2d_{s_i}+(j-1)(d_{s_1}+d_{s_2}+\cdots+d_{s_i}+1), \nn\\
\alignspace \hspace*{0.2in}
\textrm{ for } i=1,2,\ldots,k-1 \textrm{ and } j=1,2,\ldots,n_{i+1},
\label{eqn:OQ-LR-delays-greedy-3}
\eeqnarray
and we have
\beqnarray{OQ-LR-delays-greedy-4}
\dbf_1^{s_i+j}\in \Bcal_{s_i+j},
\textrm{ for } i=0,1,\ldots,k-1 \textrm{ and } j=1,2,\ldots,n_{i+1}.
\eeqnarray
Furthermore, we have
\beqnarray{}
\alignspace
B(\dbf_1^j;1)=j, \textrm{ for } j=1,2,\ldots,s_1,
\label{eqn:OQ-LR-delays-greedy-5}\\
\alignspace
B(\dbf_1^{s_i+j};i+1)=d_{s_i+j}+d_{s_1}+d_{s_2}+\cdots+d_{s_i}, \nn\\
\alignspace \hspace*{0.1in}
\textrm{ for } i=1,2,\ldots,k-1 \textrm{ and } j=1,2,\ldots,n_{i+1}.
\label{eqn:OQ-LR-delays-greedy-6}
\eeqnarray
In particular, we have
\beqnarray{OQ-LR-delays-greedy-7}
B(\dbf_1^{s_i};i)=d_{s_1}+d_{s_2}+\cdots+d_{s_i}, \textrm{ for } i=1,2,\ldots,k.
\eeqnarray
\etheorem

We also need the following eight lemmas
(i.e., \rlemma{adjacent distance larger than one}, \rlemma{comparison rule A},
\rlemma{nonadjacent distance larger than one}, \rlemma{main lemma},
\rlemma{adjacent distance larger than one II}, \rlemma{comparison rule B},
\rlemma{nonadjacent distance larger than one II}, and \rlemma{main lemma II})
to prove \rtheorem{main result}.
The first four lemmas are for the case that $1\leq h\leq N$ is an odd integer
and the last four lemmas are for the case that $1\leq h\leq N$ is an even integer.
The proofs of these lemmas are given in
\rappendix{proof of adjacent distance larger than one with h=1}--\rappendix{proof of main lemma II}.
Note that if $r_{h-1}=1$,
then it is clear from \reqnarray{N-M-k-h} that $\Ncal_{M,k}(h)=\{(r_{h-2})\}$.
As $(r_{h-2})$ is the only sequence in $\Ncal_{M,k}(h)$,
it is the only optimal sequence over $\Ncal_{M,k}(h)$.
Therefore, we only consider the nontrivial case that $r_{h-1}\geq 2$
in the following eight lemmas.

In the following lemma, we show some pairwise comparison results
for a sequence $\nbf_1^{r_{h-1}}(h)\in \Ncal_{M,k}(h)$,
where $1\leq h\leq N$ is an odd integer and $r_{h-1}\geq 2$,
such that the absolute value of the difference of two ``adjacent'' entries
of $\nbf_1^{r_{h-1}}(h)$ is greater than or equal to two.

\blemma{adjacent distance larger than one}
Suppose that $M\geq 2$ and $1\leq k\leq M-1$. Let $r_{-1}=M$, $r_0=k$, and let $q_i$ and $r_i$, $i=1,2,\ldots,N$, be recursively obtained
as in Step 1 of \ralgorithm{main result}.
Assume that $1\leq h\leq N$ is an odd integer and $r_{h-1}\geq 2$.
Let $\nbf_1^{r_{h-1}}(h)=(n_1(h),n_2(h),\ldots,n_{r_{h-1}}(h))\in \Ncal_{M,k}(h)$.

(i) Suppose that $n_a(h)-n_{a+1}(h)\leq -2$ for some $1\leq a\leq r_{h-1}-1$.
Let ${\nbf'}_1^{r_{h-1}}(h)=(n'_1(h),n'_2(h),\ldots,n'_{r_{h-1}}(h))$ be a sequence of positive integers such that
$n'_a(h)=n_a(h)+1$, $n'_{a+1}(h)=n_{a+1}(h)-1$, and $n'_i(h)=n_i(h)$ for $i\neq a$ and $a+1$.
Then we have ${\nbf'}_1^{r_{h-1}}(h)\in \Ncal_{M,k}(h)$ and
\beqnarray{adjacent distance larger than one-1}
\nbf_1^{r_{h-1}}(h)\prec{\nbf'}_1^{r_{h-1}}(h).
\eeqnarray

(ii) Suppose that $n_a(h)-n_{a+1}(h)\geq 2$ for some $1\leq a\leq r_{h-1}-1$.
Let ${\nbf'}_1^{r_{h-1}}(h)=(n'_1(h),n'_2(h),\ldots,n'_{r_{h-1}}(h))$ be a sequence of positive integers such that
$n'_a(h)=n_a(h)-1$, $n'_{a+1}(h)=n_{a+1}(h)+1$, and $n'_i(h)=n_i(h)$ for $i\neq a$ and $a+1$.
Then we have ${\nbf'}_1^{r_{h-1}}(h)\in \Ncal_{M,k}(h)$ and
\beqnarray{adjacent distance larger than one-2}
\nbf_1^{r_{h-1}}(h)\preceq{\nbf'}_1^{r_{h-1}}(h),
\eeqnarray
where $\nbf_1^{r_{h-1}}(h)\equiv{\nbf'}_1^{r_{h-1}}(h)$ if and only if $r_{h-1}=2$ and $n_1(h)=n_2(h)+2$.
\elemma

\bexample{adjacent distance larger than one}
Suppose that $M=16$ and $k=6$.
In Step 1 of \ralgorithm{main result},
we obtain $r_{-1}=16$, $r_0=6$, $q_1=2$, $r_1=4$, $q_2=1$, $r_2=2$,
$q_3=2$, and $r_3=0$.

(i) Assume that $h=1$ and hence $r_{h-1}=r_0=6\geq 2$.
Let
\beqnarray{}
{\nbf''}_1^{r_0}(1) \aligneq (3,3,2,1,5,2), \nn\\
{\nbf'}_1^{r_0}(1) \aligneq (3,3,2,2,4,2), \nn\\
\nbf_1^{r_0}(1) \aligneq (3,3,2,3,3,2), \nn\\
{\mbf'}_1^{r_0}(1) \aligneq (3,3,2,4,2,2), \nn\\
{\mbf''}_1^{r_0}(1) \aligneq (3,3,2,5,1,2). \nn
\eeqnarray
Then it follows from \reqnarray{adjacent distance larger than one-1}
in \rlemma{adjacent distance larger than one}(i) (with $h=1$ and $a=4$) that
\beqnarray{}
{\nbf''}_1^{r_0}(1)\prec{\nbf'}_1^{r_0}(1)\prec\nbf_1^{r_0}(1). \nn
\eeqnarray
We also have from \reqnarray{adjacent distance larger than one-2}
in \rlemma{adjacent distance larger than one}(ii) (with $h=1$ and $a=4$) that
\beqnarray{}
\nbf_1^{r_0}(1)\succ{\mbf'}_1^{r_0}(1)\succ{\mbf''}_1^{r_0}(1). \nn
\eeqnarray
These results can be verified by the numerical results in \rtable{adjacent distance larger than one},
where we compute the maximum representable integers $B(\dbf_1^M;k)$ with $\dbf_1^M$ obtained by using
${\nbf''}_1^{r_0}(1)$, ${\nbf'}_1^{r_0}(1)$, $\nbf_1^{r_0}(1)$, ${\mbf'}_1^{r_0}(1)$,
and ${\mbf''}_1^{r_0}(1)$, respectively, in \reqnarray{OQ-LR-delays-greedy-1}.

\btable{htbp}{|r|c|}
\hline  &  $B(d_1^M;k)$  \\
\hline ${\nbf''}_1^{r_0}(1)=(3,3,2,1,5,2)$ & $3543$   \\
\hline ${\nbf'}_1^{r_0}(1)=(3,3,2,2,4,2)$  & $4327$   \\
\hline $\nbf_1^{r_0}(1)=(3,3,2,3,3,2)$   & $4599$   \\
\hline ${\mbf'}_1^{r_0}(1)=(3,3,2,4,2,2)$  & $4359$   \\
\hline ${\mbf''}_1^{r_0}(1)=(3,3,2,5,1,2)$ & $3607$   \\
\hline
\etable{adjacent distance larger than one}
{The maximum representable integers $B(\dbf_1^M;k)$ with $\dbf_1^M$ obtained by using
${\nbf''}_1^{r_0}(1)$, ${\nbf'}_1^{r_0}(1)$, $\nbf_1^{r_0}(1)$, ${\mbf'}_1^{r_0}(1)$,
and ${\mbf''}_1^{r_0}(1)$, respectively, in \reqnarray{OQ-LR-delays-greedy-1},
where $M=16$ and $k=6$.}

(ii) Assume that $h=3$ and hence $r_{h-1}=r_2=2$.
Let $\nbf_1^{r_2}(3)=(3,1)$ and ${\nbf'}_1^{r_2}(3)=(2,2)$.
As $r_2=2$ and $n_1(3)=n_2(3)+2$,
it follows from \reqnarray{adjacent distance larger than one-2}
in \rlemma{adjacent distance larger than one}(ii) (with $h=3$ and $a=1$) that
\beqnarray{}
\nbf_1^{r_2}(3)\equiv{\nbf'}_1^{r_2}(3). \nn
\eeqnarray
This result can also be verified numerically.
To see this, note that from \reqnarray{order relation-111}
and \reqnarray{order relation-222} with $h=3$, we have
\beqnarray{}
\alignspace \nbf_1^{r_1}(2)=R_{r_0,r_1}(\nbf_1^{r_2}(3))=R_{6,4}((3,1))=(1,1,2,2), \nn\\
\alignspace \nbf_1^{r_0}(1)=L_{r_{-1},r_0}(\nbf_1^{r_1}(2))=L_{16,6}((1,1,2,2))=(3,3,3,2,3,2); \nn\\
\alignspace {\nbf'}_1^{r_1}(2)=R_{r_0,r_1}({\nbf'}_1^{r_2}(3))=R_{6,4}((2,2))=(1,2,1,2), \nn\\
\alignspace {\nbf'}_1^{r_0}(1)=L_{r_{-1},r_0}({\nbf'}_1^{r_1}(2))=L_{16,6}((1,2,1,2))=(3,3,2,3,3,2). \nn
\eeqnarray
Let $\dbf_1^M$ and ${\dbf'}_1^M$ be obtained by using $\nbf_1^{r_0}(1)=(3,3,3,2,3,2)$
and ${\nbf'}_1^{r_0}(1)=(3,3,2,3,3,2)$, respectively, in \reqnarray{OQ-LR-delays-greedy-1}.
We then compute that $B(\dbf_1^M;k)=B({\dbf'}_1^M;k)=4599$
and this verifies that $\nbf_1^{r_2}(3)\equiv{\nbf'}_1^{r_2}(3)$.
\eexample

We have the following corollary to \rlemma{adjacent distance larger than one}.

\bcorollary{adjacent distance larger than one}
Suppose that $M\geq 2$ and $1\leq k\leq M-1$. Let $r_{-1}=M$, $r_0=k$, and let $q_i$ and $r_i$, $i=1,2,\ldots,N$, be recursively obtained
as in Step 1 of \ralgorithm{main result}.
Assume that $1\leq h\leq N$ is an odd integer and $r_{h-1}\geq 2$.

(i) Suppose that $r_{h-1}\neq 2$ or $r_h\neq 0$.
Then an optimal sequence $\nbf_1^{r_{h-1}}(h)$ over $\Ncal_{M,k}(h)$
must satisfy the condition that the absolute value of the difference
of any two adjacent entries of $\nbf_1^{r_{h-1}}(h)$ is less than or equal to one, i.e.,
\beqnarray{adjacent distance larger than one-3}
|n_i(h)-n_{i+1}(h)|\leq 1, \textrm{ for } i=1,2,\ldots,r_{h-1}-1.
\eeqnarray

(ii) Suppose that $r_{h-1}=2$ and $r_h=0$.
Then there are two optimal sequences over $\Ncal_{M,k}(h)$,
and the two optimal sequences, say $\nbf_1^{r_{h-1}}(h)$ and $\mbf_1^{r_{h-1}}(h)$,
are given by
\beqnarray{adjacent distance larger than one-4}
\nbf_1^{r_{h-1}}(h)=(q_h,q_h) \textrm{ and } \mbf_1^{r_{h-1}}(h)=(q_h+1,q_h-1)
\eeqnarray
(note that $q_h\geq 2$ as $r_{h-2}=q_h\cdot r_{h-1}+r_h=q_h\cdot r_{h-1}$ and $r_{h-2}>r_{h-1}$).
\ecorollary

\bproof
(i) Let $\nbf_1^{r_{h-1}}(h)$ be an optimal sequence over $\Ncal_{M,k}(h)$.
We show that $|n_i(h)-n_{i+1}(h)|\leq 1$ for $i=1,2,\ldots,r_{h-1}-1$ by contradiction.
Assume on the contrary that $|n_a(h)-n_{a+1}(h)|\geq 2$ for some $1\leq a\leq r_{h-1}-1$.
Note that in \rcorollary{adjacent distance larger than one}(i),
we have $r_{h-1}\neq 2$ or $r_h\neq 0$.
As such, if $r_{h-1}=2$, then we have $r_h\neq 0$
and it must be the case that $n_1(h)\neq n_2(h)+2$.
Otherwise, if $n_1(h)=n_2(h)+2$,
then we see from $\nbf_1^{r_{h-1}}(h)\in \Ncal_{M,k}(h)$ and \reqnarray{N-M-k-h} that
\beqnarray{}
r_{h-2}=\sum_{i=1}^{r_{h-1}}n_i(h)=n_1(h)+n_2(h)=2n_2(h)+2=(n_2(h)+1)\cdot r_{h-1}. \nn
\eeqnarray
Thus, the remainder $r_h$ of $r_{h-2}$ divided by $r_{h-1}$ is equal to zero,
contradicting to $r_h\neq 0$.
Since it cannot be the case that $r_{h-1}=2$ and $n_1(h)=n_2(h)+2$,
we see from \rlemma{adjacent distance larger than one}
that there exists ${\nbf'}_1^{r_{h-1}}(h)\in \Ncal_{M,k}(h)$
such that $\nbf_1^{r_{h-1}}(h)\prec{\nbf'}_1^{r_{h-1}}(h)$,
contradicting to the optimality of $\nbf_1^{r_{h-1}}(h)$.

(ii) Let $\nbf_1^{r_{h-1}}(h)$ be an optimal sequence over $\Ncal_{M,k}(h)$.
As we have from $r_{h-1}=2$ and $r_h=0$
that $r_{h-2}=q_h\cdot r_{h-1}+r_h=2q_h$
and we have from $\nbf_1^{r_{h-1}}(h)\in \Ncal_{M,k}(h)$, \reqnarray{N-M-k-h}, and $r_{h-1}=2$
that $r_{h-2}=\sum_{i=1}^{r_{h-1}}n_i(h)=n_1(h)+n_2(h)$,
we see that $n_1(h)+n_2(h)=2q_h$ is an even integer.
As such, it follows that $n_1(h)-n_2(h)$ is also an even integer.

If $n_1(h)-n_2(h)\leq -2$ or $n_1(h)-n_2(h)\geq 4$,
then we see from \rlemma{adjacent distance larger than one}
that there exists ${\nbf'}_1^{r_{h-1}}(h)\in \Ncal_{M,k}(h)$
such that $\nbf_1^{r_{h-1}}(h)\prec{\nbf'}_1^{r_{h-1}}(h)$,
contradicting to the optimality of $\nbf_1^{r_{h-1}}(h)$.
Therefore, we must have $n_1(h)-n_2(h)=0$ or 2,
i.e., $\nbf_1^{r_{h-1}}(h)=(q_h,q_h)$ or $\nbf_1^{r_{h-1}}(h)=(q_h+1,q_h-1)$.
It is easy to see from \rlemma{adjacent distance larger than one}(ii)
that $(q_h,q_h)\equiv (q_h+1,q_h-1)$,
and hence both $(q_h,q_h)$ and $(q_h+1,q_h-1)$ are optimal sequences over $\Ncal_{M,k}(h)$.
\eproof

In the following lemma, we show some pairwise comparison results
for a sequence $\nbf_1^{r_{h-1}}(h)\in \Ncal_{M,k}(h)$,
where $1\leq h\leq N$ is an odd integer and $r_{h-1}\geq 2$,
such that the absolute value of the difference of two ``adjacent'' entries
of $\nbf_1^{r_{h-1}}(h)$ is equal to one.

\blemma{comparison rule A} (\textbf{Comparison rule A})
Suppose that $M\geq 2$ and $1\leq k\leq M-1$. Let $r_{-1}=M$, $r_0=k$, and let $q_i$ and $r_i$, $i=1,2,\ldots,N$, be recursively obtained
as in Step 1 of \ralgorithm{main result}.
Assume that $1\leq h\leq N$ is an odd integer and $r_{h-1}\geq 2$.
Let $\nbf_1^{r_{h-1}}(h)=(n_1(h),n_2(h),\ldots,n_{r_{h-1}}(h))\in \Ncal_{M,k}(h)$,
$n_a(h)-n_{a+1}(h)=1$ for some $1\leq a\leq r_{h-1}-1$,
and $n_1(h)\geq 3$ in the case that $h=1$ and $a=1$.
Let ${\nbf'}_1^{r_{h-1}}(h)=(n'_1(h),n'_2(h),\ldots,n'_{r_{h-1}}(h))$
be a sequence of positive integers such that $n'_a(h)=n_a(h)-1$, $n'_{a+1}(h)=n_{a+1}(h)+1$,
and $n'_i(h)=n_i(h)$ for $i\neq a$ and $a+1$.
Then we have ${\nbf'}_1^{r_{h-1}}(h)\in \Ncal_{M,k}(h)$.
Furthermore, we have the following pairwise comparison results.

(i) Suppose that $a=1$ or $a=r_{h-1}-1$.
Then we have
\beqnarray{comparison rule A-1}
\nbf_1^{r_{h-1}}(h)\succ {\nbf'}_1^{r_{h-1}}(h).
\eeqnarray

(ii) Suppose that $2\leq a\leq r_{h-1}-2$ and there exists a positive integer $j$ such that
$1\leq j\leq \min\{a-1,r_{h-1}-a-1\}$, $n_{a-j'}(h)=n_{a+1+j'}(h)$ for $j'=1,2,\ldots,j-1$,
and $n_{a-j}(h)\neq n_{a+1+j}(h)$.
If $n_{a-j}(h)<n_{a+1+j}(h)$, then we have
\beqnarray{comparison rule A-2}
\nbf_1^{r_{h-1}}(h)\succ{\nbf'}_1^{r_{h-1}}(h).
\eeqnarray
On the other hand, if $n_{a-j}(h)>n_{a+1+j}(h)$, then we have
\beqnarray{comparison rule A-3}
\nbf_1^{r_{h-1}}(h)\preceq{\nbf'}_1^{r_{h-1}}(h),
\eeqnarray
where $\nbf_1^{r_{h-1}}(h)\equiv {\nbf'}_1^{r_{h-1}}(h)$
if and only if $a-j=1$, $a+1+j=r_{h-1}$, and $n_1(h)=n_{r_{h-1}}(h)+1$.

(iii) Suppose that $2\leq a\leq r_{h-1}-2$ and $n_{a-j'}(h)=n_{a+1+j'}(h)$
for $j'=1,2,\ldots,\min\{a-1,r_{h-1}-a-1\}$.
Then we have
\beqnarray{comparison rule A-4}
\nbf_1^{r_{h-1}}(h)\succ{\nbf'}_1^{r_{h-1}}(h).
\eeqnarray
\elemma

\bexample{comparison rule A}
Suppose that $M=16$ and $k=6$.
In Step 1 of \ralgorithm{main result},
we obtain $r_{-1}=16$, $r_0=6$, $q_1=2$, $r_1=4$, $q_2=1$, $r_2=2$,
$q_3=2$, and $r_3=0$.
Assume that $h=1$ and hence $r_{h-1}=r_0=6\geq 2$.
Let
\beqnarray{}
{\nbf'''''}_1^{r_0}(1) \aligneq (2,3,2,3,3,3), \nn\\
{\nbf''''}_1^{r_0}(1) \aligneq (3,2,2,3,3,3), \nn\\
{\nbf'''}_1^{r_0}(1) \aligneq (3,2,3,2,3,3), \nn\\
{\nbf''}_1^{r_0}(1) \aligneq (3,2,3,3,2,3), \nn\\
{\nbf'}_1^{r_0}(1) \aligneq (3,2,3,3,3,2), \nn\\
\nbf_1^{r_0}(1) \aligneq (3,3,2,3,3,2), \nn\\
\mbf_1^{r_0}(1) \aligneq (3,3,3,2,3,2). \nn
\eeqnarray
Then we have
\beqnarray{}
{\nbf'''''}_1^{r_0}(1)\prec{\nbf''''}_1^{r_0}(1)\prec{\nbf'''}_1^{r_0}(1)\prec{\nbf''}_1^{r_0}(1)
\prec{\nbf'}_1^{r_0}(1)\prec\nbf_1^{r_0}(1)\equiv\mbf_1^{r_0}(1). \nn
\eeqnarray
This can be proved by using Comparison rule A in \rlemma{comparison rule A}.
First, it is easy to see from $n_1''''(1)-n_2''''(1)=1$ and \reqnarray{comparison rule A-1}
in \rlemma{comparison rule A}(i) (with $h=1$ and $a=1$) that
\beqnarray{}
{\nbf'''''}_1^{r_0}(1)\prec{\nbf''''}_1^{r_0}(1). \nn
\eeqnarray
From $n_3'''(1)-n_4'''(1)=1$, $n_2'''(1)<n_5'''(1)$,
and \reqnarray{comparison rule A-2} in \rlemma{comparison rule A}(ii) (with $h=1$, $a=3$, and $j=1$)
we see that
\beqnarray{}
{\nbf''''}_1^{r_0}(1)\prec{\nbf'''}_1^{r_0}(1). \nn
\eeqnarray
From $n_4''(1)-n_5''(1)=1$, $n_3''(1)=n_6''(1)$,
and \reqnarray{comparison rule A-4} in \rlemma{comparison rule A}(iii) (with $h=1$ and $a=4$),
we have
\beqnarray{}
{\nbf'''}_1^{r_0}(1)\prec{\nbf''}_1^{r_0}(1). \nn
\eeqnarray
From $n_5'(1)-n_6'(1)=1$ and \reqnarray{comparison rule A-1}
in \rlemma{comparison rule A}(i) (with $h=1$ and $a=r_0-1=5$),
we have
\beqnarray{}
{\nbf''}_1^{r_0}(1)\prec{\nbf'}_1^{r_0}(1). \nn
\eeqnarray
Since $n_2(1)-n_3(1)=1$ and $n_1(1)=n_4(1)$,
it follows from \reqnarray{comparison rule A-4}
in \rlemma{comparison rule A}(iii) (with $h=1$ and $a=2$)
that
\beqnarray{}
{\nbf'}_1^{r_0}(1)\prec\nbf_1^{r_0}(1). \nn
\eeqnarray
Finally, as $m_3(1)-m_4(1)=1$, $m_2(1)=m_5(1)$, and $m_1(1)=m_6(1)+1$,
it follows from \reqnarray{comparison rule A-3} in \rlemma{comparison rule A}(ii) (with $h=1$, $a=3$, and $j=2$)
that
\beqnarray{}
\nbf_1^{r_0}(1)\equiv\mbf_1^{r_0}(1). \nn
\eeqnarray
These results can also be verified by the numerical results in \rtable{comparison rule A},
where we compute the maximum representable integers $B(\dbf_1^M;k)$ with $\dbf_1^M$ obtained by using
${\nbf'''''}_1^{r_0}(1)$, ${\nbf''''}_1^{r_0}(1)$, ${\nbf'''}_1^{r_0}(1)$, ${\nbf''}_1^{r_0}(1)$,
${\nbf'}_1^{r_0}(1)$, $\nbf_1^{r_0}(1)$, and $\mbf_1^{r_0}(1)$, respectively, in \reqnarray{OQ-LR-delays-greedy-1}.

\btable{htbp}{|r|c|}
\hline  & $B(d_1^M;k)$  \\
\hline ${\nbf'''''}_1^{r_0}(1)=(2,3,2,3,3,3)$  & $4231$   \\
\hline ${\nbf''''}_1^{r_0}(1)=(3,2,2,3,3,3)$   & $4395$   \\
\hline ${\nbf'''}_1^{r_0}(1)=(3,2,3,2,3,3)$    & $4439$    \\
\hline ${\nbf''}_1^{r_0}(1)=(3,2,3,3,2,3)$     & $4455$   \\
\hline ${\nbf'}_1^{r_0}(1)=(3,2,3,3,3,2)$      & $4579$   \\
\hline $\nbf_1^{r_0}(1)=(3,3,2,3,3,2)$       & $4599$   \\
\hline $\mbf_1^{r_0}(1)=(3,3,3,2,3,2)$       & $4599$   \\
\hline
\etable{comparison rule A}
{The maximum representable integers $B(\dbf_1^M;k)$ with $\dbf_1^M$ obtained by using
${\nbf'''''}_1^{r_0}(1)$, ${\nbf''''}_1^{r_0}(1)$, ${\nbf'''}_1^{r_0}(1)$, ${\nbf''}_1^{r_0}(1)$,
${\nbf'}_1^{r_0}(1)$, $\nbf_1^{r_0}(1)$, and $\mbf_1^{r_0}(1)$, respectively,
in \reqnarray{OQ-LR-delays-greedy-1}, where $M=16$ and $k=6$.}
\eexample

From \rcorollary{adjacent distance larger than one}(i),
we know that if $1\leq h\leq N$ is an odd integer and $r_{h-1}\geq 3$,
then an optimal sequence $\nbf_1^{r_{h-1}}(h)$ over $\Ncal_{M,k}(h)$
must satisfy the condition that the absolute value of the difference
of any two adjacent entries of $\nbf_1^{r_{h-1}}(h)$ is less than or equal to one.
In the following lemma, we show some pairwise comparison results
for a sequence $\nbf_1^{r_{h-1}}(h)\in \Ncal_{M,k}(h)$,
where $1\leq h\leq N$ is an odd integer and $r_{h-1}\geq 3$,
such that the absolute value of the difference of any two adjacent entries
of $\nbf_1^{r_{h-1}}(h)$ is less than or equal to one
and the absolute value of the difference of two ``nonadjacent'' entries
of $\nbf_1^{r_{h-1}}(h)$ is greater than or equal to two.

\blemma{nonadjacent distance larger than one}
Suppose that $M\geq 2$ and $1\leq k\leq M-1$. Let $r_{-1}=M$, $r_0=k$, and let $q_i$ and $r_i$, $i=1,2,\ldots,N$, be recursively obtained
as in Step 1 of \ralgorithm{main result}.
Assume that $1\leq h\leq N$ is an odd integer and $r_{h-1}\geq 3$.
Let $\nbf_1^{r_{h-1}}(h)=(n_1(h),n_2(h),\ldots,n_{r_{h-1}}(h))\in \Ncal_{M,k}(h)$
and $|n_i(h)-n_{i+1}(h)|\leq 1$ for $i=1,2,\ldots,r_{h-1}-1$.

(i) Suppose that $n_a(h)-n_b(h)\geq 2$ for some $1\leq a<b\leq r_{h-1}$ and $b\geq a+2$.
If $n_1(h)\neq n_{r_{h-1}}(h)+2$ or $n_i(h)\neq n_{r_{h-1}}(h)+1$ for some $2\leq i\leq r_{h-1}-1$,
then there exists a sequence of positive integers
${\nbf'}_1^{r_{h-1}}(h)=(n'_1(h),n'_2(h),\ldots,n'_{r_{h-1}}(h))\in \Ncal_{M,k}(h)$
such that
\beqnarray{nonadjacent distance larger than one-1}
{\nbf'}_1^{r_{h-1}}(h)\succ\nbf_1^{r_{h-1}}(h).
\eeqnarray

(ii) Suppose that $n_a(h)-n_b(h)\leq -2$ for some $1\leq a<b\leq r_{h-1}$ and $b\geq a+2$.
Then there exists a sequence of positive integers
${\nbf'}_1^{r_{h-1}}(h)=(n'_1(h),n'_2(h),\ldots,n'_{r_{h-1}}(h))\in \Ncal_{M,k}(h)$
such that
\beqnarray{nonadjacent distance larger than one-2}
{\nbf'}_1^{r_{h-1}}(h)\succ\nbf_1^{r_{h-1}}(h).
\eeqnarray
\elemma

We have the following corollary to \rlemma{nonadjacent distance larger than one}.

\bcorollary{nonadjacent distance larger than one}
Suppose that $M\geq 2$ and $1\leq k\leq M-1$. Let $r_{-1}=M$, $r_0=k$, and let $q_i$ and $r_i$, $i=1,2,\ldots,N$, be recursively obtained
as in Step 1 of \ralgorithm{main result}.
Assume that $1\leq h\leq N$ is an odd integer.

(i) Suppose that $r_{h-1}\geq 3$ and $r_h\neq 0$.
Then an optimal sequence $\nbf_1^{r_{h-1}}(h)$ over $\Ncal_{M,k}(h)$
must satisfy the condition that the absolute value of the difference
of any two entries (adjacent or nonadjacent) of $\nbf_1^{r_{h-1}}(h)$
is less than or equal to one, i.e.,
\beqnarray{nonadjacent distance larger than one-3}
|n_a(h)-n_b(h)|\leq 1, \textrm{ for all } 1\leq a<b\leq r_{h-1}.
\eeqnarray

(ii) Suppose that $r_{h-1}\geq 3$ and $r_h=0$.
Then there are at most two optimal sequences over $\Ncal_{M,k}(h)$,
and the two possible optimal sequences, say $\nbf_1^{r_{h-1}}(h)$ and $\mbf_1^{r_{h-1}}(h)$,
are given by
\beqnarray{nonadjacent distance larger than one-4}
\nbf_1^{r_{h-1}}(h)=(q_h,q_h,\ldots,q_h)
\textrm{ and } \mbf_1^{r_{h-1}}(h)=(q_h+1,q_h,\ldots,q_h,q_h-1)
\eeqnarray
(note that $q_h\geq 2$ as $r_{h-2}=q_h\cdot r_{h-1}+r_h=q_h\cdot r_{h-1}$ and $r_{h-2}>r_{h-1}$).
\ecorollary

\bproof
(i) Let $\nbf_1^{r_{h-1}}(h)$ be an optimal sequence over $\Ncal_{M,k}(h)$.
As we have $r_{h-1}\geq 3$ and hence $r_{h-1}\neq 2$,
it follows from \rcorollary{adjacent distance larger than one}(i)
that $\nbf_1^{r_{h-1}}(h)$ must satisfy the condition that
\beqnarray{proof-nonadjacent distance larger than one-corollary-111}
|n_i(h)-n_{i+1}(h)|\leq 1, \textrm{ for all } i=1,2,\ldots,r_{h-1}-1.
\eeqnarray
Furthermore, as we have $r_h\neq 0$, it must be the case that $n_1(h)\neq n_{r_{h-1}}(h)+2$
or $n_i(h)\neq n_{r_{h-1}}(h)+1$ for some $2\leq i\leq r_{h-1}-1$.
Otherwise, if $n_1(h)=n_{r_{h-1}}(h)+2$ and $n_i(h)=n_{r_{h-1}}(h)+1$
for all $2\leq i\leq r_{h-1}-1$,
then we see from $\nbf_1^{r_{h-1}}(h)\in \Ncal_{M,k}(h)$ and \reqnarray{N-M-k-h} that
\beqnarray{}
r_{h-2}=\sum_{i=1}^{r_{h-1}}n_i(h)=(n_{r_{h-1}}(h)+1)\cdot r_{h-1}. \nn
\eeqnarray
Thus, the remainder $r_h$ of $r_{h-2}$ divided by $r_{h-1}$ is equal to zero,
contradicting to $r_h\neq 0$.

To show \reqnarray{nonadjacent distance larger than one-3},
it is clear from \reqnarray{proof-nonadjacent distance larger than one-corollary-111}
that it suffices to show that $|n_a(h)-n_b(h)|\leq 1$
for all $1\leq a<b\leq r_{h-1}$ and $b\geq a+2$ by contradiction.
Assume on the contrary that $|n_a(h)-n_b(h)|\geq 2$
for some $1\leq a<b\leq r_{h-1}$ and $b\geq a+2$.
Since $n_1(h)\neq n_{r_{h-1}}(h)+2$
or $n_i(h)\neq n_{r_{h-1}}(h)+1$ for some $2\leq i\leq r_{h-1}-1$,
we see from \rlemma{nonadjacent distance larger than one}
that there exists ${\nbf'}_1^{r_{h-1}}(h)\in \Ncal_{M,k}(h)$
such that ${\nbf'}_1^{r_{h-1}}(h)\succ \nbf_1^{r_{h-1}}(h)$,
contradicting to the optimality of $\nbf_1^{r_{h-1}}(h)$.

(ii) Let $\nbf_1^{r_{h-1}}(h)$ be an optimal sequence over $\Ncal_{M,k}(h)$.
As we have $r_{h-1}\geq 3$,
\reqnarray{proof-nonadjacent distance larger than one-corollary-111} still holds.
Furthermore, as we have $r_h=0$,
it is clear that $r_{h-2}=q_h\cdot r_{h-1}+r_h=q_h\cdot r_{h-1}$.
It then follows from $\nbf_1^{r_{h-1}}(h)\in \Ncal_{M,k}(h)$ and \reqnarray{N-M-k-h} that
\beqnarray{proof-nonadjacent distance larger than one-corollary-222}
\sum_{i=1}^{r_{h-1}}n_i(h)=r_{h-2}=q_h\cdot r_{h-1}.
\eeqnarray

We need to consider the following two cases.

\emph{Case 1: $|n_a(h)-n_b(h)|\leq 1$ for all $1\leq a<b\leq r_{h-1}$.}
In this case, it is easy to see from
\reqnarray{proof-nonadjacent distance larger than one-corollary-222} that
\beqnarray{proof-nonadjacent distance larger than one-corollary-333}
n_i(h)=q_h, \textrm{ for } i=1,2,\ldots,r_{h-1}.
\eeqnarray

\emph{Case 2: $|n_a(h)-n_b(h)|\geq 2$ for some $1\leq a<b\leq r_{h-1}$.}
In this case, it is clear from
\reqnarray{proof-nonadjacent distance larger than one-corollary-111} that $b\geq a+2$.
If $n_a(h)-n_b(h)\geq 2$ and $n_1(h)\neq n_{r_{h-1}}(h)+2$
or $n_i(h)\neq n_{r_{h-1}}(h)+1$ for some $2\leq i\leq r_{h-1}-1$,
then it follows from \rlemma{nonadjacent distance larger than one}(i)
that there exists ${\nbf'}_1^{r_{h-1}}(h)\in \Ncal_{M,k}(h)$
such that ${\nbf'}_1^{r_{h-1}}(h)\succ \nbf_1^{r_{h-1}}(h)$,
contradicting to the optimality of $\nbf_1^{r_{h-1}}(h)$.
Also, if $n_a(h)-n_b(h)\leq -2$,
then it follows from \rlemma{nonadjacent distance larger than one}(ii)
that there exists ${\nbf'}_1^{r_{h-1}}(h)\in \Ncal_{M,k}(h)$
such that ${\nbf'}_1^{r_{h-1}}(h)\succ \nbf_1^{r_{h-1}}(h)$,
contradicting to the optimality of $\nbf_1^{r_{h-1}}(h)$.
As such, it must be the case that
\beqnarray{proof-nonadjacent distance larger than one-corollary-444}
n_a(h)-n_b(h)\geq 2,\ n_1(h)=n_{r_{h-1}}(h)+2,
\textrm{ and } n_i(h)=n_{r_{h-1}}(h)+1 \textrm{ for } 2\leq i\leq r_{h-1}-1.
\eeqnarray
It follows from \reqnarray{proof-nonadjacent distance larger than one-corollary-222}
and \reqnarray{proof-nonadjacent distance larger than one-corollary-444} that
\beqnarray{proof-nonadjacent distance larger than one-corollary-555}
n_1(h)=q_h+1,\ n_{r_{h-1}}(h)=q_h-1,
\textrm{ and } n_i(h)=q_h \textrm{ for } 2\leq i\leq r_{h-1}-1.
\eeqnarray

By combining \reqnarray{proof-nonadjacent distance larger than one-corollary-333}
and \reqnarray{proof-nonadjacent distance larger than one-corollary-555},
we see that $(q_h,q_h,\ldots,q_h)$ and $(q_h+1,q_h,\ldots,q_h,q_h-1)$
are the two possible optimal sequences over $\Ncal_{M,k}(h)$,
and the proof is completed.
\eproof

Suppose that $1\leq h\leq N$ is an odd integer, $r_{h-1}\geq 2$, $r_h\neq 0$,
and $\nbf_1^{r_{h-1}}(h)$ is an optimal sequence over $\Ncal_{M,k}(h)$.
If $r_{h-1}=2$, then it follows from \rcorollary{adjacent distance larger than one}(i)
that $|n_1(h)-n_2(h)|\leq 1$, i.e., \reqnarray{nonadjacent distance larger than one-3} holds.
On the other hand, if $r_{h-1}\geq 3$,
then it follows from \rcorollary{nonadjacent distance larger than one}(i)
that \reqnarray{nonadjacent distance larger than one-3} also holds.
As such, we see from \reqnarray{nonadjacent distance larger than one-3},
$\sum_{i=1}^{r_{h-1}}n_i(h)=r_{h-2}$ in \reqnarray{N-M-k-h},
and $r_{h-2}=q_h\cdot r_{h-1}+r_h$ that
\beqnarray{}
n_i(h)=
\bselection
q_h+1, &\textrm{if } i=i_1,i_2,\ldots,i_{r_h}, \\
q_h, &\textrm{otherwise},
\eselection \nn
\eeqnarray
for some $1\leq i_1<i_2<\cdots <i_{r_h}\leq r_{h-1}$.
In \rlemma{main lemma} below, we further show that $i_1$ must be equal to 1.

\blemma{main lemma}
Suppose that $M\geq 2$ and $1\leq k\leq M-1$. Let $r_{-1}=M$, $r_0=k$, and let $q_i$ and $r_i$, $i=1,2,\ldots,N$, be recursively obtained
as in Step 1 of \ralgorithm{main result}.
Assume that $1\leq h\leq N$ is an odd integer, $r_{h-1}\geq 2$, and $r_h\neq 0$.
Then an optimal sequence $\nbf_1^{r_{h-1}}(h)$ over $\Ncal_{M,k}(h)$
must satisfy the condition that
\beqnarray{main lemma-1}
n_i(h)=
\bselection
q_h+1, &\textrm{if } i=i_1,i_2,\ldots,i_{r_h}, \\
q_h, &\textrm{otherwise},
\eselection
\eeqnarray
for some $1=i_1<i_2<\cdots <i_{r_h}\leq r_{h-1}$.
\elemma

The following four lemmas (for the case that $1\leq h\leq N$ is an even integer)
are the counterparts of the above four lemmas (for the case that $1\leq h\leq N$ is an odd integer).

In the following lemma, we show some pairwise comparison results
for a sequence $\nbf_1^{r_{h-1}}(h)\in \Ncal_{M,k}(h)$,
where $1\leq h\leq N$ is an even integer and $r_{h-1}\geq 2$,
such that the absolute value of the difference of two ``adjacent'' entries
of $\nbf_1^{r_{h-1}}(h)$ is greater than or equal to two.

\blemma{adjacent distance larger than one II}
Suppose that $M\geq 2$ and $1\leq k\leq M-1$. Let $r_{-1}=M$, $r_0=k$, and let $q_i$ and $r_i$, $i=1,2,\ldots,N$, be recursively obtained
as in Step 1 of \ralgorithm{main result}.
Assume that $1\leq h\leq N$ is an even integer and $r_{h-1}\geq 2$.
Let $\nbf_1^{r_{h-1}}(h)=(n_1(h),n_2(h),\ldots,n_{r_{h-1}}(h))\in \Ncal_{M,k}(h)$.

(i) Suppose that $n_a(h)-n_{a+1}(h)\geq 2$ for some $1\leq a\leq r_{h-1}-1$.
Let ${\nbf'}_1^{r_{h-1}}(h)=(n'_1(h),n'_2(h),\ldots,n'_{r_{h-1}}(h))$
be a sequence of positive integers such that
$n'_a(h)=n_a(h)-1$, $n'_{a+1}(h)=n_{a+1}(h)+1$, and $n'_i(h)=n_i(h)$ for $i\neq a$ and $a+1$.
Then we have ${\nbf'}_1^{r_{h-1}}(h)\in \Ncal_{M,k}(h)$ and
\beqnarray{adjacent distance larger than one II-1}
\nbf_1^{r_{h-1}}(h)\prec{\nbf'}_1^{r_{h-1}}(h).
\eeqnarray

(ii) Suppose that $n_a(h)-n_{a+1}(h)\leq -2$ for some $1\leq a\leq r_{h-1}-1$.
Let ${\nbf'}_1^{r_{h-1}}(h)=(n'_1(h),n'_2(h),\ldots,n'_{r_{h-1}}(h))$
be a sequence of positive integers such that
$n'_a(h)=n_a(h)+1$, $n'_{a+1}(h)=n_{a+1}(h)-1$, and $n'_i(h)=n_i(h)$ for $i\neq a$ and $a+1$.
Then we have ${\nbf'}_1^{r_{h-1}}(h)\in \Ncal_{M,k}(h)$ and
\beqnarray{adjacent distance larger than one II-2}
\nbf_1^{r_{h-1}}(h)\preceq{\nbf'}_1^{r_{h-1}}(h),
\eeqnarray
where $\nbf_1^{r_{h-1}}(h)\equiv {\nbf'}_1^{r_{h-1}}(h)$ if and only if $r_{h-1}=2$ and $n_1(h)=n_2(h)-2$.
\elemma

\bexample{adjacent distance larger than one II}
Suppose that $M=26$ and $k=10$.
In Step 1 of \ralgorithm{main result},
we obtain $r_{-1}=26$, $r_0=10$, $q_1=2$, $r_1=6$, $q_2=1$, $r_2=4$,
$q_3=1$, $r_3=2$, $q_4=2$, and $r_4=0$.

(i) Assume that $h=2$ and hence $r_{h-1}=r_1=6\geq 2$.
Let
\beqnarray{}
{\nbf''}_1^{r_1}(2) \aligneq (1,1,5,1,1,1), \nn\\
{\nbf'}_1^{r_1}(2) \aligneq (1,1,4,2,1,1), \nn\\
\nbf_1^{r_1}(2) \aligneq (1,1,3,3,1,1), \nn\\
{\mbf'}_1^{r_1}(2) \aligneq (1,1,2,4,1,1), \nn\\
{\mbf''}_1^{r_1}(2) \aligneq (1,1,1,5,1,1). \nn
\eeqnarray
Then it follows from \reqnarray{adjacent distance larger than one II-1}
in \rlemma{adjacent distance larger than one II}(i) (with $h=2$ and $a=3$) that
\beqnarray{}
{\nbf''}_1^{r_1}(2)\prec{\nbf'}_1^{r_1}(2)\prec\nbf_1^{r_1}(2). \nn
\eeqnarray
From \reqnarray{adjacent distance larger than one II-2}
in \rlemma{adjacent distance larger than one II}(ii) (with $h=2$ and $a=3$),
we also have
\beqnarray{}
\nbf_1^{r_1}(2)\succ{\mbf'}_1^{r_1}(2)\succ{\mbf''}_1^{r_1}(2). \nn
\eeqnarray
These results can be verified by the numerical results in \rtable{adjacent distance larger than one II},
where we compute the maximum representable integers $B(\dbf_1^M;k)$ with $\dbf_1^M$ obtained by using
\beqnarray{}
{\nbf''}_1^{r_0}(1)=L_{r_{-1},r_0}({\nbf''}_1^{r_1}(2))
=L_{26,10}((1,1,5,1,1,1)) \aligneq (3,3,3,2,2,2,2,3,3,3), \nn\\
{\nbf'}_1^{r_0}(1)=L_{r_{-1},r_0}({\nbf'}_1^{r_1}(2))
=L_{26,10}((1,1,4,2,1,1)) \aligneq (3,3,3,2,2,2,3,2,3,3), \nn\\
\nbf_1^{r_0}(1)=L_{r_{-1},r_0}(\nbf_1^{r_1}(2))
=L_{26,10}((1,1,3,3,1,1)) \aligneq (3,3,3,2,2,3,2,2,3,3), \nn\\
{\mbf'}_1^{r_0}(1)=L_{r_{-1},r_0}({\mbf'}_1^{r_1}(2))
=L_{26,10}((1,1,2,4,1,1)) \aligneq (3,3,3,2,3,2,2,2,3,3), \nn\\
{\mbf''}_1^{r_0}(1)=L_{r_{-1},r_0}({\mbf''}_1^{r_1}(2))
=L_{26,10}((1,1,1,5,1,1)) \aligneq (3,3,3,3,2,2,2,2,3,3), \nn
\eeqnarray
respectively, in \reqnarray{OQ-LR-delays-greedy-1}.

\btable{htbp}{|r|r|c|}
\hline  &  & $B(d_1^M;k)$  \\
\hline ${\nbf''}_1^{r_1}(2)=(1,1,5,1,1,1)$ & ${\nbf''}_1^{r_0}(1)=(3,3,3,2,2,2,2,3,3,3)$ & $1072727$   \\
\hline ${\nbf'}_1^{r_1}(2)=(1,1,4,2,1,1)$  & ${\nbf'}_1^{r_0}(1)=(3,3,3,2,2,2,3,2,3,3)$  & $1084591$   \\
\hline $\nbf_1^{r_1}(2)=(1,1,3,3,1,1)$   & $\nbf_1^{r_0}(1)=(3,3,3,2,2,3,2,2,3,3)$   & $1086295$   \\
\hline ${\mbf'}_1^{r_1}(2)=(1,1,2,4,1,1)$  & ${\mbf'}_1^{r_0}(1)=(3,3,3,2,3,2,2,2,3,3)$  & $1084655$   \\
\hline ${\mbf''}_1^{r_1}(2)=(1,1,1,5,1,1)$ & ${\mbf''}_1^{r_0}(1)=(3,3,3,3,2,2,2,2,3,3)$ & $1073111$   \\
\hline
\etable{adjacent distance larger than one II}
{The maximum representable integers $B(\dbf_1^M;k)$ with $\dbf_1^M$ obtained by using
${\nbf''}_1^{r_0}(1)=L_{r_{-1},r_0}({\nbf''}_1^{r_1}(2))$,
${\nbf'}_1^{r_0}(1)=L_{r_{-1},r_0}({\nbf'}_1^{r_1}(2))$,
$\nbf_1^{r_0}(1)=L_{r_{-1},r_0}(\nbf_1^{r_1}(2))$,
${\mbf'}_1^{r_0}(1)=L_{r_{-1},r_0}({\mbf'}_1^{r_1}(2))$,
and ${\mbf''}_1^{r_0}(1)=L_{r_{-1},r_0}({\mbf''}_1^{r_1}(2))$,
respectively, in \reqnarray{OQ-LR-delays-greedy-1},
where $M=26$ and $k=10$.}

(ii)  Assume that $h=4$ and hence $r_{h-1}=r_3=2$.
Let $\nbf_1^{r_3}(4)=(1,3)$ and ${\nbf'}_1^{r_3}(4)=(2,2)$.
As $r_3=2$ and $n_1(4)=n_2(4)-2$,
it follows from \reqnarray{adjacent distance larger than one II-2}
in \rlemma{adjacent distance larger than one II}(ii) (with $h=4$ and $a=1$) that
\beqnarray{}
\nbf_1^{r_3}(4)\equiv{\nbf'}_1^{r_3}(4). \nn
\eeqnarray
This result can also be verified numerically.
To see this, note that from \reqnarray{order relation-333}--\reqnarray{order relation-555}
with $h=4$, we have
\beqnarray{}
\alignspace \nbf_1^{r_2}(3)=L_{r_1,r_2}(\nbf_1^{r_3}(4))=L_{6,4}((1,3))=(2,2,1,1), \nn\\
\alignspace \nbf_1^{r_1}(2)=R_{r_0,r_1}(\nbf_1^{r_2}(3))=R_{10,6}((2,2,1,1))=(1,2,1,2,2,2), \nn\\
\alignspace \nbf_1^{r_0}(1)=L_{r_{-1},r_0}(\nbf_1^{r_1}(2))=L_{26,10}((1,2,1,2,2,2))=(3,3,2,3,3,2,3,2,3,2); \nn\\
\alignspace {\nbf'}_1^{r_2}(3)=L_{r_1,r_2}({\nbf'}_1^{r_3}(4))=L_{6,4}((2,2))=(2,1,2,1), \nn\\
\alignspace {\nbf'}_1^{r_1}(2)=R_{r_0,r_1}({\nbf'}_1^{r_2}(3))=R_{10,6}((2,1,2,1))=(1,2,2,1,2,2), \nn\\
\alignspace {\nbf'}_1^{r_0}(1)=L_{r_{-1},r_0}({\nbf'}_1^{r_1}(2))=L_{26,10}((1,2,2,1,2,2))=(3,3,2,3,2,3,3,2,3,2). \nn
\eeqnarray
Let $\dbf_1^M$ and ${\dbf'}_1^M$ be obtained by using $\nbf_1^{r_0}(1)=(3,3,2,3,3,2,3,2,3,2)$
and ${\nbf'}_1^{r_0}(1)=(3,3,2,3$, $2,3,3,2,3,2)$, respectively, in \reqnarray{OQ-LR-delays-greedy-1}.
We then compute that $B(\dbf_1^M;k)=B({\dbf'}_1^M;k)=1141023$
and this verifies that $\nbf_1^{r_2}(3)\equiv{\nbf'}_1^{r_2}(3)$.
\eexample

We have the following corollary to \rlemma{adjacent distance larger than one II}.

\bcorollary{adjacent distance larger than one II}
Suppose that $M\geq 2$ and $1\leq k\leq M-1$. Let $r_{-1}=M$, $r_0=k$, and let $q_i$ and $r_i$, $i=1,2,\ldots,N$, be recursively obtained
as in Step 1 of \ralgorithm{main result}.
Assume that $1\leq h\leq N$ is an even integer and $r_{h-1}\geq 2$.

(i) Suppose that $r_{h-1}\neq 2$ or $r_h\neq 0$.
Then an optimal sequence $\nbf_1^{r_{h-1}}(h)$ over $\Ncal_{M,k}(h)$
must satisfy the condition that the absolute value of the difference
of any two adjacent entries of $\nbf_1^{r_{h-1}}(h)$ is less than or equal to one, i.e.,
\beqnarray{adjacent distance larger than one II-3}
|n_i(h)-n_{i+1}(h)|\leq 1, \textrm{ for } i=1,2,\ldots,r_{h-1}-1.
\eeqnarray

(ii) Suppose that $r_{h-1}=2$ and $r_h=0$.
Then there are two optimal sequences over $\Ncal_{M,k}(h)$,
and the two optimal sequences, say $\nbf_1^{r_{h-1}}(h)$ and $\mbf_1^{r_{h-1}}(h)$,
are given by
\beqnarray{adjacent distance larger than one II-4}
\nbf_1^{r_{h-1}}(h)=(q_h,q_h) \textrm{ and } \mbf_1^{r_{h-1}}(h)=(q_h-1,q_h+1)
\eeqnarray
(note that $q_h\geq 2$ as $r_{h-2}=q_h\cdot r_{h-1}+r_h=q_h\cdot r_{h-1}$ and $r_{h-2}>r_{h-1}$).
\ecorollary

\bproof
(i) Let $\nbf_1^{r_{h-1}}(h)$ be an optimal sequence over $\Ncal_{M,k}(h)$.
We show that $|n_i(h)-n_{i+1}(h)|\leq 1$ for $i=1,2,\ldots,r_{h-1}-1$ by contradiction.
Assume on the contrary that $|n_a(h)-n_{a+1}(h)|\geq 2$ for some $1\leq a\leq r_{h-1}-1$.
Note that in \rcorollary{adjacent distance larger than one II}(i),
we have $r_{h-1}\neq 2$ or $r_h\neq 0$.
As such, if $r_{h-1}=2$, then we have $r_h\neq 0$
and it must be the case that $n_1(h)\neq n_2(h)-2$.
Otherwise, if $n_1(h)=n_2(h)-2$,
then we see from $\nbf_1^{r_{h-1}}(h)\in \Ncal_{M,k}(h)$ and \reqnarray{N-M-k-h} that
\beqnarray{}
r_{h-2}=\sum_{i=1}^{r_{h-1}}n_i(h)=n_1(h)+n_2(h)=2n_2(h)-2=(n_2(h)-1)\cdot r_{h-1}. \nn
\eeqnarray
Thus, the remainder $r_h$ of $r_{h-2}$ divided by $r_{h-1}$ is equal to zero,
contradicting to $r_h\neq 0$.
Since it cannot be the case that $r_{h-1}=2$ and $n_1(h)=n_2(h)-2$,
we see from \rlemma{adjacent distance larger than one II}
that there exists ${\nbf'}_1^{r_{h-1}}(h)\in \Ncal_{M,k}(h)$
such that $\nbf_1^{r_{h-1}}(h)\prec{\nbf'}_1^{r_{h-1}}(h)$,
contradicting to the optimality of $\nbf_1^{r_{h-1}}(h)$.

(ii) Let $\nbf_1^{r_{h-1}}(h)$ be an optimal sequence over $\Ncal_{M,k}(h)$.
As we have from $r_{h-1}=2$ and $r_h=0$
that $r_{h-2}=q_h\cdot r_{h-1}+r_h=2q_h$
and we have from $\nbf_1^{r_{h-1}}(h)\in \Ncal_{M,k}(h)$, \reqnarray{N-M-k-h}, and $r_{h-1}=2$
that $r_{h-2}=\sum_{i=1}^{r_{h-1}}n_i(h)=n_1(h)+n_2(h)$,
we see that $n_1(h)+n_2(h)=2q_h$ is an even integer.
As such, it follows that $n_1(h)-n_2(h)$ is also an even integer.

If $n_1(h)-n_2(h)\geq 2$ or $n_1(h)-n_2(h)\leq -4$,
then we see from \rlemma{adjacent distance larger than one II}
that there exists ${\nbf'}_1^{r_{h-1}}(h)\in \Ncal_{M,k}(h)$
such that $\nbf_1^{r_{h-1}}(h)\prec{\nbf'}_1^{r_{h-1}}(h)$,
contradicting to the optimality of $\nbf_1^{r_{h-1}}(h)$.
Therefore, we must have $n_1(h)-n_2(h)=0$ or $-2$,
i.e., $\nbf_1^{r_{h-1}}(h)=(q_h,q_h)$ or $\nbf_1^{r_{h-1}}(h)=(q_h-1,q_h+1)$.
It is easy to see from \rlemma{adjacent distance larger than one II}(ii)
that $(q_h,q_h)\equiv (q_h-1,q_h+1)$,
and hence both $(q_h,q_h)$ and $(q_h-1,q_h+1)$ are optimal sequences over $\Ncal_{M,k}(h)$.
\eproof

In the following lemma, we show some pairwise comparison results
for a sequence $\nbf_1^{r_{h-1}}(h)\in \Ncal_{M,k}(h)$,
where $1\leq h\leq N$ is an even integer and $r_{h-1}\geq 2$,
such that the absolute value of the difference of two ``adjacent'' entries
of $\nbf_1^{r_{h-1}}(h)$ is equal to one.

\blemma{comparison rule B} (\textbf{Comparison rule B}).
Suppose that $M\geq 2$ and $1\leq k\leq M-1$. Let $r_{-1}=M$, $r_0=k$, and let $q_i$ and $r_i$, $i=1,2,\ldots,N$, be recursively obtained
as in Step 1 of \ralgorithm{main result}.
Assume that $1\leq h\leq N$ is an even integer and $r_{h-1}\geq 2$.
Let $\nbf_1^{r_{h-1}}(h)=(n_1(h),n_2(h),\ldots,n_{r_{h-1}}(h))\in \Ncal_{M,k}(h)$
and $n_a(h)-n_{a+1}(h)=1$ for some $1\leq a\leq r_{h-1}-1$.
Let ${\nbf'}_1^{r_{h-1}}(h)=(n'_1(h),n'_2(h),$ $\ldots,n'_{r_{h-1}}(h))$
be a sequence of positive integers such that
$n'_a(h)=n_a(h)-1$, $n'_{a+1}(h)=n_{a+1}(h)+1$, and $n'_i(h)=n_i(h)$ for $i\neq a$ and $a+1$.
Then we have ${\nbf'}_1^{r_{h-1}}(h)\in \Ncal_{M,k}(h)$.
Furthermore, we have the following pairwise comparison results.

(i) Suppose that $a=1$ or $a=r_{h-1}-1$.
Then we have
\beqnarray{comparison rule B-1}
\nbf_1^{r_{h-1}}(h)\prec{\nbf'}_1^{r_{h-1}}(h).
\eeqnarray

(ii) Suppose that $2\leq a\leq r_{h-1}-2$ and there exists a positive integer $j$ such that
$1\leq j\leq \min\{a-1,r_{h-1}-a-1\}$, $n_{a-j'}(h)=n_{a+1+j'}(h)$ for $j'=1,2,\ldots,j-1$,
and $n_{a-j}(h)\neq n_{a+1+j}(h)$.
If $n_{a-j}(h)>n_{a+1+j}(h)$, then we have
\beqnarray{comparison rule B-2}
\nbf_1^{r_{h-1}}(h)\prec{\nbf'}_1^{r_{h-1}}(h).
\eeqnarray
On the other hand, if $n_{a-j}(h)<n_{a+1+j}(h)$, then we have
\beqnarray{comparison rule B-3}
\nbf_1^{r_{h-1}}(h)\succeq{\nbf'}_1^{r_{h-1}}(h),
\eeqnarray
where $\nbf_1^{r_{h-1}}(h)\equiv {\nbf'}_1^{r_{h-1}}(h)$
if and only if $a-j=1$, $a+1+j=r_{h-1}$, and $n_1(h)=n_{r_{h-1}}(h)-1$.

(iii) Suppose that $2\leq a\leq r_{h-1}-2$ and $n_{a-j'}(h)=n_{a+1+j'}(h)$
for $j'=1,2,\ldots,\min\{a-1,r_{h-1}-a-1\}$.
Then we have
\beqnarray{comparison rule B-4}
\nbf_1^{r_{h-1}}(h)\prec{\nbf'}_1^{r_{h-1}}(h).
\eeqnarray
\elemma

\bexample{comparison rule B}
Suppose that $M=26$ and $k=10$.
In Step 1 of \ralgorithm{main result},
we obtain $r_{-1}=26$, $r_0=10$, $q_1=2$, $r_1=6$, $q_2=1$, $r_2=4$,
$q_3=1$, $r_3=2$, $q_4=2$, and $r_4=0$.
Assume that $h=2$ and hence $r_{h-1}=r_1=6\geq 2$.
Let
\beqnarray{}
{\nbf'''''}_1^{r_1}(2) \aligneq (2,2,2,1,2,1), \nn\\
{\nbf''''}_1^{r_1}(2) \aligneq (2,2,1,2,2,1), \nn\\
{\nbf'''}_1^{r_1}(2) \aligneq (2,2,1,2,1,2), \nn\\
{\nbf''}_1^{r_1}(2) \aligneq (2,1,2,2,1,2), \nn\\
{\nbf'}_1^{r_1}(2) \aligneq (1,2,2,2,1,2), \nn\\
\nbf_1^{r_1}(2) \aligneq (1,2,2,1,2,2), \nn\\
\mbf_1^{r_1}(2) \aligneq (1,2,1,2,2,2). \nn
\eeqnarray
Then we have
\beqnarray{}
{\nbf'''''}_1^{r_1}(2)\prec{\nbf''''}_1^{r_1}(2)\prec{\nbf'''}_1^{r_1}(2)
\prec{\nbf''}_1^{r_1}(2)\prec{\nbf'}_1^{r_1}(2)\prec\nbf_1^{r_1}(2)\equiv\mbf_1^{r_1}(2). \nn
\eeqnarray
This can be proved by using Comparison rule B in \rlemma{comparison rule B}.
First, it is easy to see from $n'''''_3(2)-n'''''_4(2)=1$, $n'''''_2(2)=n'''''_5(2)$,
$n'''''_1(2)>n'''''_6(2)$, and \reqnarray{comparison rule B-2}
in \rlemma{comparison rule B}(ii) (with $h=2$, $a=3$, and $j=2$) that
\beqnarray{}
{\nbf'''''}_1^{r_1}(2)\prec{\nbf''''}_1^{r_1}(2). \nn
\eeqnarray
From $n''''_5(2)-n''''_6(2)=1$ and \reqnarray{comparison rule B-1} in
\rlemma{comparison rule B}(i) (with $h=2$ and $a=r_1-1=5$), we have
\beqnarray{}
{\nbf''''}_1^{r_1}(2)\prec{\nbf'''}_1^{r_1}(2). \nn
\eeqnarray
From $n'''_2(2)-n'''_3(2)=1$, $n'''_1(2)=n'''_4(2)$,
and \reqnarray{comparison rule B-4} in \rlemma{comparison rule B}(iii)
(with $h=2$ and $a=2$), we have
\beqnarray{}
{\nbf'''}_1^{r_1}(2)\prec{\nbf''}_1^{r_1}(2). \nn
\eeqnarray
From $n''_1(2)-n''_2(2)=1$ and \reqnarray{comparison rule B-1} in
\rlemma{comparison rule B}(i) (with $h=2$ and $a=1$), we have
\beqnarray{}
{\nbf''}_1^{r_1}(2)\prec{\nbf'}_1^{r_1}(2). \nn
\eeqnarray
Since $n'_4(2)-n'_5(2)=1$ and $n'_3(2)=n'_6(2)$,
it follows from \reqnarray{comparison rule B-4}
in \rlemma{comparison rule B}(iii) (with $h=2$ and $a=4$) that
\beqnarray{}
{\nbf'}_1^{r_1}(2)\prec\nbf_1^{r_1}(2). \nn
\eeqnarray
Finally, as $n_3(2)-n_4(2)=1$, $n_2(2)=n_5(2)$, and $n_1(2)=n_6(2)-1$,
it follows from \reqnarray{comparison rule B-3}
in \rlemma{comparison rule B}(ii) (with $h=2$, $a=3$, and $j=2$) that
\beqnarray{}
\nbf_1^{r_1}(2)\equiv\mbf_1^{r_1}(2). \nn
\eeqnarray
These results can also be verified by the numerical results in \rtable{comparison rule B},
where we compute the maximum representable integers $B(\dbf_1^M;k)$ with $\dbf_1^M$ obtained by using
\beqnarray{}
{\nbf'''''}_1^{r_0}(1)=L_{r_{-1},r_0}({\nbf'''''}_1^{r_1}(2))
=L_{26,10}((2,2,2,1,2,1)) \aligneq (3,2,3,2,3,2,3,3,2,3), \nn\\
{\nbf''''}_1^{r_0}(1)=L_{r_{-1},r_0}({\nbf''''}_1^{r_1}(2))
=L_{26,10}((2,2,1,2,2,1)) \aligneq (3,2,3,2,3,3,2,3,2,3), \nn\\
{\nbf'''}_1^{r_0}(1)=L_{r_{-1},r_0}({\nbf'''}_1^{r_1}(2))
=L_{26,10}((2,2,1,2,1,2)) \aligneq (3,2,3,2,3,3,2,3,3,2), \nn\\
{\nbf''}_1^{r_0}(1)=L_{r_{-1},r_0}({\nbf''}_1^{r_1}(2))
=L_{26,10}((2,1,2,2,1,2)) \aligneq (3,2,3,3,2,3,2,3,3,2), \nn\\
{\nbf'}_1^{r_0}(1)=L_{r_{-1},r_0}({\nbf'}_1^{r_1}(2))
=L_{26,10}((1,2,2,2,1,2)) \aligneq (3,3,2,3,2,3,2,3,3,2), \nn\\
\nbf_1^{r_0}(1)=L_{r_{-1},r_0}(\nbf_1^{r_1}(2))
=L_{26,10}((1,2,2,1,2,2)) \aligneq (3,3,2,3,2,3,3,2,3,2), \nn\\
\mbf_1^{r_0}(1)=L_{r_{-1},r_0}(\mbf_1^{r_1}(2))
=L_{26,10}((1,2,1,2,2,2)) \aligneq (3,3,2,3,3,2,3,2,3,2), \nn
\eeqnarray
respectively, in \reqnarray{OQ-LR-delays-greedy-1}.

\btable{htbp}{|r|r|c|}
\hline                                      &                                              & $B(d_1^M;k)$ \\
\hline ${\nbf'''''}_1^{r_1}(2)=(2,2,2,1,2,1)$ & ${\nbf'''''}_1^{r_0}(1)=(3,2,3,2,3,2,3,3,2,3)$ & $1104735$   \\
\hline ${\nbf''''}_1^{r_1}(2)=(2,2,1,2,2,1)$  & ${\nbf''''}_1^{r_0}(1)=(3,2,3,2,3,3,2,3,2,3)$  & $1104799$   \\
\hline ${\nbf'''}_1^{r_1}(2)=(2,2,1,2,1,2)$   & ${\nbf'''}_1^{r_0}(1)=(3,2,3,2,3,3,2,3,3,2)$   & $1136415$   \\
\hline ${\nbf''}_1^{r_1}(2)=(2,1,2,2,1,2)$    & ${\nbf''}_1^{r_0}(1)=(3,2,3,3,2,3,2,3,3,2)$    & $1136495$   \\
\hline ${\nbf'}_1^{r_1}(2)=(1,2,2,2,1,2)$     & ${\nbf'}_1^{r_0}(1)=(3,3,2,3,2,3,2,3,3,2)$     & $1140511$   \\
\hline $\nbf_1^{r_1}(2)=(1,2,2,1,2,2)$      & $\nbf_1^{r_0}(1)=(3,3,2,3,2,3,3,2,3,2)$      & $1141023$   \\
\hline $\mbf_1^{r_1}(2)=(1,2,1,2,2,2)$      & $\mbf_1^{r_0}(1)=(3,3,2,3,3,2,3,2,3,2)$      & $1141023$   \\
\hline
\etable{comparison rule B}
{The maximum representable integers $B(\dbf_1^M;k)$ with $\dbf_1^M$ obtained by using
${\nbf'''''}_1^{r_0}(1)=L_{r_{-1},r_0}({\nbf'''''}_1^{r_1}(2))$,
${\nbf''''}_1^{r_0}(1)=L_{r_{-1},r_0}({\nbf''''}_1^{r_1}(2))$,
${\nbf'''}_1^{r_0}(1)=L_{r_{-1},r_0}({\nbf'''}_1^{r_1}(2))$,
${\nbf''}_1^{r_0}(1)=L_{r_{-1},r_0}({\nbf''}_1^{r_1}(2))$,
${\nbf'}_1^{r_0}(1)=L_{r_{-1},r_0}({\nbf'}_1^{r_1}(2))$,
$\nbf_1^{r_0}(1)=L_{r_{-1},r_0}(\nbf_1^{r_1}(2))$,
and $\mbf_1^{r_0}(1)=L_{r_{-1},r_0}(\mbf_1^{r_1}(2))$, respectively,
in \reqnarray{OQ-LR-delays-greedy-1}, where $M=26$ and $k=10$.}
\eexample

From \rcorollary{adjacent distance larger than one II}(i),
we know that if $1\leq h\leq N$ is an even integer and $r_{h-1}\geq 3$,
then an optimal sequence $\nbf_1^{r_{h-1}}(h)$ over $\Ncal_{M,k}(h)$
must satisfy the condition that the absolute value of the difference
of any two adjacent entries of $\nbf_1^{r_{h-1}}(h)$ is less than or equal to one.
In the following lemma, we show some pairwise comparison results
for a sequence $\nbf_1^{r_{h-1}}(h)\in \Ncal_{M,k}(h)$,
where $1\leq h\leq N$ is an even integer and $r_{h-1}\geq 3$,
such that the absolute value of the difference of any two adjacent entries
of $\nbf_1^{r_{h-1}}(h)$ is less than or equal to one
and the absolute value of the difference of two ``nonadjacent'' entries
of $\nbf_1^{r_{h-1}}(h)$ is greater than or equal to two.

\blemma{nonadjacent distance larger than one II}
Suppose that $M\geq 2$ and $1\leq k\leq M-1$. Let $r_{-1}=M$, $r_0=k$, and let $q_i$ and $r_i$, $i=1,2,\ldots,N$, be recursively obtained
as in Step 1 of \ralgorithm{main result}.
Assume that $1\leq h\leq N$ is an even integer and $r_{h-1}\geq 3$.
Let $\nbf_1^{r_{h-1}}(h)=(n_1(h),n_2(h),\ldots,n_{r_{h-1}}(h))\in \Ncal_{M,k}(h)$
and $|n_i(h)-n_{i+1}(h)|\leq 1$ for $i=1,2,\ldots,r_{h-1}-1$.

(i) Suppose that $n_a(h)-n_b(h)\leq -2$ for some $1\leq a<b\leq r_{h-1}$ and $b\geq a+2$.
If $n_{r_{h-1}}(h)\neq n_1(h)+2$ or $n_i(h)\neq n_1(h)+1$ for some $2\leq i\leq r_{h-1}-1$,
then there exists a sequence of positive integers
${\nbf'}_1^{r_{h-1}}(h)=(n'_1(h),n'_2(h),\ldots,n'_{r_{h-1}}(h))\in \Ncal_{M,k}(h)$
such that
\beqnarray{nonadjacent distance larger than one II-1}
{\nbf'}_1^{r_{h-1}}(h)\succ\nbf_1^{r_{h-1}}(h).
\eeqnarray

(ii) Suppose that $n_a(h)-n_b(h)\geq 2$ for some $1\leq a<b\leq r_{h-1}$ and $b\geq a+2$.
Then there exists a sequence of positive integers
${\nbf'}_1^{r_{h-1}}(h)=(n'_1(h),n'_2(h),\ldots,n'_{r_{h-1}}(h))\in \Ncal_{M,k}(h)$
such that
\beqnarray{nonadjacent distance larger than one II-2}
{\nbf'}_1^{r_{h-1}}(h)\succ\nbf_1^{r_{h-1}}(h).
\eeqnarray
\elemma

We have the following corollary to \rlemma{nonadjacent distance larger than one}.

\bcorollary{nonadjacent distance larger than one II}
Suppose that $M\geq 2$ and $1\leq k\leq M-1$. Let $r_{-1}=M$, $r_0=k$, and let $q_i$ and $r_i$, $i=1,2,\ldots,N$, be recursively obtained
as in Step 1 of \ralgorithm{main result}.
Assume that $1\leq h\leq N$ is an even integer.

(i) Suppose that $r_{h-1}\geq 3$ and $r_h\neq 0$.
Then an optimal sequence $\nbf_1^{r_{h-1}}(h)$ over $\Ncal_{M,k}(h)$
must satisfy the condition that the absolute value of the difference
of any two entries (adjacent or nonadjacent) of $\nbf_1^{r_{h-1}}(h)$
is less than or equal to one, i.e.,
\beqnarray{nonadjacent distance larger than one II-3}
|n_a(h)-n_b(h)|\leq 1, \textrm{ for all } 1\leq a<b\leq r_{h-1}.
\eeqnarray

(ii) Suppose that $r_{h-1}\geq 3$ and $r_h=0$.
Then there are at most two optimal sequences over $\Ncal_{M,k}(h)$,
and the two possible optimal sequences, say $\nbf_1^{r_{h-1}}(h)$ and $\mbf_1^{r_{h-1}}(h)$,
are given by
\beqnarray{nonadjacent distance larger than one II-4}
\nbf_1^{r_{h-1}}(h)=(q_h,q_h,\ldots,q_h)
\textrm{ and } \mbf_1^{r_{h-1}}(h)=(q_h-1,q_h,\ldots,q_h,q_h+1)
\eeqnarray
(note that $q_h\geq 2$ as $r_{h-2}=q_h\cdot r_{h-1}+r_h=q_h\cdot r_{h-1}$ and $r_{h-2}>r_{h-1}$).
\ecorollary

\bproof
(i) Let $\nbf_1^{r_{h-1}}(h)$ be an optimal sequence over $\Ncal_{M,k}(h)$.
As we have $r_{h-1}\geq 3$ and hence $r_{h-1}\neq 2$,
it follows from \rcorollary{adjacent distance larger than one II}(i)
that $\nbf_1^{r_{h-1}}(h)$ must satisfy the condition that
\beqnarray{proof-nonadjacent distance larger than one II-corollary-111}
|n_i(h)-n_{i+1}(h)|\leq 1, \textrm{ for all } i=1,2,\ldots,r_{h-1}-1.
\eeqnarray
Furthermore, as we have $r_h\neq 0$, it must be the case that $n_{r_{h-1}}(h)\neq n_1(h)+2$
or $n_i(h)\neq n_1(h)+1$ for some $2\leq i\leq r_{h-1}-1$.
Otherwise, if $n_{r_{h-1}}(h)=n_1(h)+2$ and $n_i(h)=n_1(h)+1$
for all $2\leq i\leq r_{h-1}-1$,
then we see from $\nbf_1^{r_{h-1}}(h)\in \Ncal_{M,k}(h)$ and \reqnarray{N-M-k-h} that
\beqnarray{}
r_{h-2}=\sum_{i=1}^{r_{h-1}}n_i(h)=(n_1(h)+1)\cdot r_{h-1}. \nn
\eeqnarray
Thus, the remainder $r_h$ of $r_{h-2}$ divided by $r_{h-1}$ is equal to zero,
contradicting to $r_h\neq 0$.

To show \reqnarray{nonadjacent distance larger than one II-3},
it is clear from \reqnarray{proof-nonadjacent distance larger than one II-corollary-111}
that it suffices to show that $|n_a(h)-n_b(h)|\leq 1$
for all $1\leq a<b\leq r_{h-1}$ and $b\geq a+2$ by contradiction.
Assume on the contrary that $|n_a(h)-n_b(h)|\geq 2$
for some $1\leq a<b\leq r_{h-1}$ and $b\geq a+2$.
Since $n_{r_{h-1}}(h)\neq n_1(h)+2$
or $n_i(h)\neq n_1(h)+1$ for some $2\leq i\leq r_{h-1}-1$,
we see from \rlemma{nonadjacent distance larger than one II}
that there exists ${\nbf'}_1^{r_{h-1}}(h)\in \Ncal_{M,k}(h)$
such that ${\nbf'}_1^{r_{h-1}}(h)\succ \nbf_1^{r_{h-1}}(h)$,
contradicting to the optimality of $\nbf_1^{r_{h-1}}(h)$.

(ii) Let $\nbf_1^{r_{h-1}}(h)$ be an optimal sequence over $\Ncal_{M,k}(h)$.
As we have $r_{h-1}\geq 3$,
\reqnarray{proof-nonadjacent distance larger than one II-corollary-111} still holds.
Furthermore, as we have $r_h=0$,
it is clear that $r_{h-2}=q_h\cdot r_{h-1}+r_h=q_h\cdot r_{h-1}$.
It then follows from $\nbf_1^{r_{h-1}}(h)\in \Ncal_{M,k}(h)$ and \reqnarray{N-M-k-h} that
\beqnarray{proof-nonadjacent distance larger than one II-corollary-222}
\sum_{i=1}^{r_{h-1}}n_i(h)=r_{h-2}=q_h\cdot r_{h-1}.
\eeqnarray

We need to consider the following two cases.

\emph{Case 1: $|n_a(h)-n_b(h)|\leq 1$ for all $1\leq a<b\leq r_{h-1}$.}
In this case, it is easy to see from
\reqnarray{proof-nonadjacent distance larger than one II-corollary-222} that
\beqnarray{proof-nonadjacent distance larger than one II-corollary-333}
n_i(h)=q_h, \textrm{ for } i=1,2,\ldots,r_{h-1}.
\eeqnarray

\emph{Case 2: $|n_a(h)-n_b(h)|\geq 2$ for some $1\leq a<b\leq r_{h-1}$.}
In this case, it is clear from
\reqnarray{proof-nonadjacent distance larger than one II-corollary-111} that $b\geq a+2$.
If $n_a(h)-n_b(h)\leq -2$ and $n_{r_{h-1}}(h)\neq n_1(h)+2$
or $n_i(h)\neq n_1(h)+1$ for some $2\leq i\leq r_{h-1}-1$,
then it follows from \rlemma{nonadjacent distance larger than one II}(i)
that there exists ${\nbf'}_1^{r_{h-1}}(h)\in \Ncal_{M,k}(h)$
such that ${\nbf'}_1^{r_{h-1}}(h)\succ \nbf_1^{r_{h-1}}(h)$,
contradicting to the optimality of $\nbf_1^{r_{h-1}}(h)$.
Also, if $n_a(h)-n_b(h)\geq 2$,
then it follows from \rlemma{nonadjacent distance larger than one II}(ii)
that there exists ${\nbf'}_1^{r_{h-1}}(h)\in \Ncal_{M,k}(h)$
such that ${\nbf'}_1^{r_{h-1}}(h)\succ \nbf_1^{r_{h-1}}(h)$,
contradicting to the optimality of $\nbf_1^{r_{h-1}}(h)$.
As such, it must be the case that
\beqnarray{proof-nonadjacent distance larger than one II-corollary-444}
n_a(h)-n_b(h)\leq -2,\ n_{r_{h-1}}(h)=n_1(h)+2,
\textrm{ and } n_i(h)=n_1(h)+1 \textrm{ for } 2\leq i\leq r_{h-1}-1.
\eeqnarray
It follows from \reqnarray{proof-nonadjacent distance larger than one II-corollary-222}
and \reqnarray{proof-nonadjacent distance larger than one II-corollary-444} that
\beqnarray{proof-nonadjacent distance larger than one II-corollary-555}
n_1(h)=q_h-1,\ n_{r_{h-1}}(h)=q_h+1,
\textrm{ and } n_i(h)=q_h \textrm{ for } 2\leq i\leq r_{h-1}-1.
\eeqnarray

By combining \reqnarray{proof-nonadjacent distance larger than one II-corollary-333}
and \reqnarray{proof-nonadjacent distance larger than one II-corollary-555},
we see that $(q_h,q_h,\ldots,q_h)$ and $(q_h-1,q_h,\ldots,q_h,q_h+1)$
are the two possible optimal sequences over $\Ncal_{M,k}(h)$,
and the proof is completed.
\eproof

Suppose that $1\leq h\leq N$ is an even integer, $r_{h-1}\geq 2$, $r_h\neq 0$,
and $\nbf_1^{r_{h-1}}(h)$ is an optimal sequence over $\Ncal_{M,k}(h)$.
If $r_{h-1}=2$, then it follows from \rcorollary{adjacent distance larger than one II}(i)
that $|n_1(h)-n_2(h)|\leq 1$, i.e., \reqnarray{nonadjacent distance larger than one II-3} holds.
On the other hand, if $r_{h-1}\geq 3$,
then it follows from \rcorollary{nonadjacent distance larger than one II}(i)
that \reqnarray{nonadjacent distance larger than one II-3} also holds.
As such, we see from \reqnarray{nonadjacent distance larger than one II-3},
$\sum_{i=1}^{r_{h-1}}n_i(h)=r_{h-2}$ in \reqnarray{N-M-k-h},
and $r_{h-2}=q_h\cdot r_{h-1}+r_h$ that
\beqnarray{}
n_i(h)=
\bselection
q_h+1, &\textrm{if } i=i_1,i_2,\ldots,i_{r_h}, \\
q_h, &\textrm{otherwise},
\eselection \nn
\eeqnarray
for some $1\leq i_1<i_2<\cdots <i_{r_h}\leq r_{h-1}$.
In \rlemma{main lemma II} below, we further show that $i_{r_h}$ must be equal to $r_{h-1}$.

\blemma{main lemma II}
Suppose that $M\geq 2$ and $1\leq k\leq M-1$. Let $r_{-1}=M$, $r_0=k$, and let $q_i$ and $r_i$, $i=1,2,\ldots,N$, be recursively obtained
as in Step 1 of \ralgorithm{main result}.
Assume that $1\leq h\leq N$ is an even integer, $r_{h-1}\geq 2$, and $r_h\neq 0$.
Then an optimal sequence $\nbf_1^{r_{h-1}}(h)$ over $\Ncal_{M,k}(h)$
must satisfy the condition that
\beqnarray{main lemma II-1}
n_i(h)=
\bselection
q_h+1, &\textrm{if } i=i_1,i_2,\ldots,i_{r_h}, \\
q_h, &\textrm{otherwise},
\eselection
\eeqnarray
for some $1\leq i_1<i_2<\cdots <i_{r_h}=r_{h-1}$.
\elemma

In the appendices,
we first show that \rlemma{adjacent distance larger than one}
and Comparison rule A in \rlemma{comparison rule A} hold for $h=1$
in \rappendix{proof of adjacent distance larger than one with h=1}
and \rappendix{proof of comparison rule A with h=1}, respectively.
We also show that if Comparison rule A in \rlemma{comparison rule A}
holds for some odd integer $h-1$, where $1\leq h-1\leq N-1$,
then \rlemma{adjacent distance larger than one II}
and Comparison rule B in \rlemma{comparison rule B} hold for the even integer $h$
in \rappendix{proof of adjacent distance larger than one II for an even integer h
by using comparison rule A for the odd integer h-1}
and \rappendix{proof of comparison rule B for an even integer h
by using comparison rule A for the odd integer h-1}, respectively.
Similarly, we show that if Comparison rule B in \rlemma{comparison rule B}
holds for some even integer $h-1$, where $2\leq h-1\leq N-1$,
then \rlemma{adjacent distance larger than one}
and Comparison rule A in \rlemma{comparison rule A} hold for the odd integer $h$
in \rappendix{proof of adjacent distance larger than one for an odd integer h
by using comparison rule B for the even integer h-1}
and \rappendix{proof of comparison rule A for an odd integer h
by using comparison rule B for the even integer h-1}, respectively.
Therefore, \rlemma{adjacent distance larger than one},
Comparison rule A in \rlemma{comparison rule A},
\rlemma{adjacent distance larger than one II},
and Comparison rule B in \rlemma{comparison rule B}
are proved by induction on $h$
(see \rfigure{proof-of-lemmas-by-induction} for an illustration).

\bpdffigure{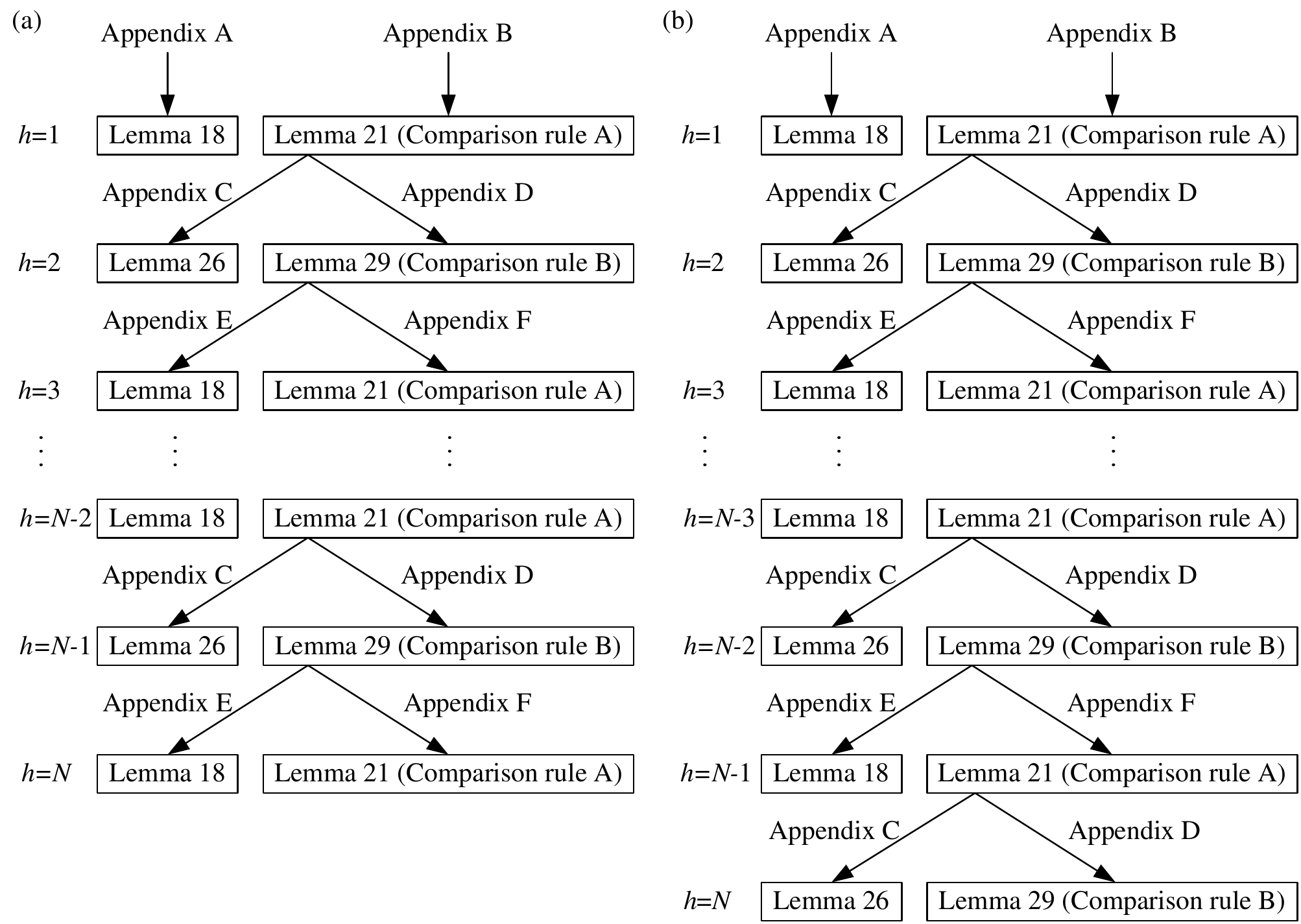}{5.5in}
\epdffigure{proof-of-lemmas-by-induction}
{Proof of \rlemma{adjacent distance larger than one},
Comparison rule A in \rlemma{comparison rule A},
\rlemma{adjacent distance larger than one II},
and Comparison rule B in \rlemma{comparison rule B} by induction on $h$:
(a) $N$ is an odd integer. (b) $N$ is an even integer.}

Then we use \rlemma{adjacent distance larger than one} and Comparison rule A in \rlemma{comparison rule A}
to prove \rlemma{nonadjacent distance larger than one} in \rappendix{proof of nonadjacent distance larger than one},
and use \rcorollary{adjacent distance larger than one}(i) (corollary to \rlemma{adjacent distance larger than one}),
\rcorollary{nonadjacent distance larger than one}(i) (corollary to \rlemma{nonadjacent distance larger than one}),
and Comparison rule A in \rlemma{comparison rule A}
to prove \rlemma{main lemma} in \rappendix{proof of main lemma}
(see \rfigure{proof-of-the-main-lemmas}(a) for an illustration).
Finally, we use \rlemma{adjacent distance larger than one II} and Comparison rule B in \rlemma{comparison rule B}
to prove \rlemma{nonadjacent distance larger than one II} in \rappendix{proof of nonadjacent distance larger than one II},
and use \rcorollary{adjacent distance larger than one II}(i) (corollary to \rlemma{adjacent distance larger than one II}),
\rcorollary{nonadjacent distance larger than one II}(i) (corollary to \rlemma{nonadjacent distance larger than one II}),
and Comparison rule B in \rlemma{comparison rule B}
to prove \rlemma{main lemma II} in \rappendix{proof of main lemma II}
(see \rfigure{proof-of-the-main-lemmas}(b) for an illustration).

\bpdffigure{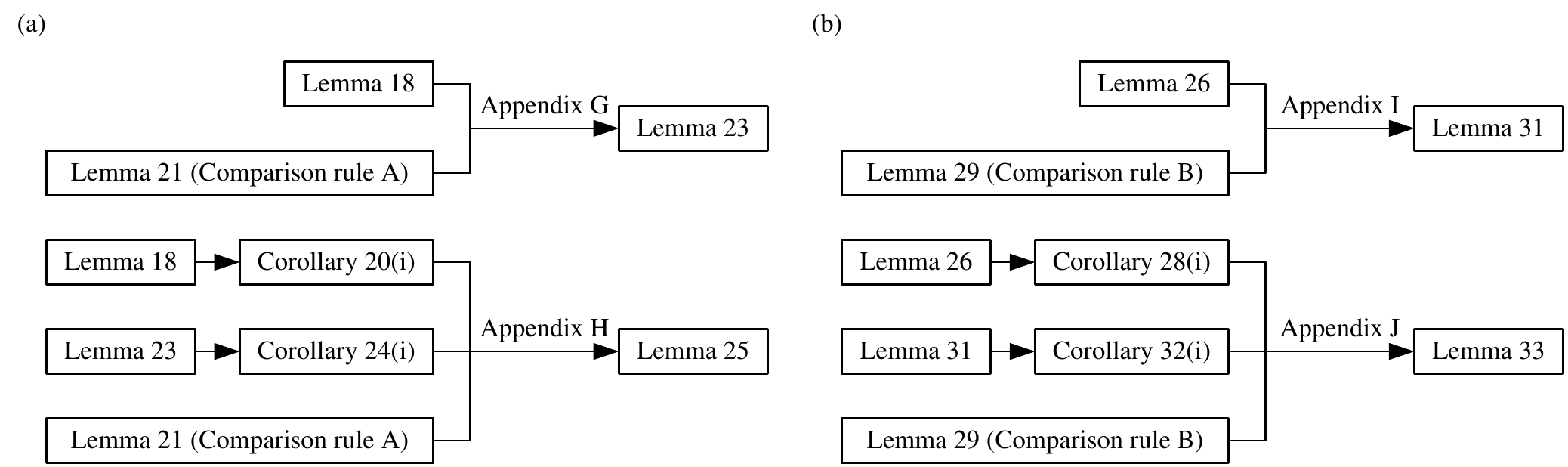}{6.0in}
\epdffigure{proof-of-the-main-lemmas}
{(a) Proof of \rlemma{nonadjacent distance larger than one} and \rlemma{main lemma}.
(b) Proof of \rlemma{nonadjacent distance larger than one II} and \rlemma{main lemma II}.}

We now use \rcorollary{adjacent distance larger than one}(ii),
\rcorollary{nonadjacent distance larger than one}(ii), \rlemma{main lemma},
\rcorollary{adjacent distance larger than one II}(ii),
\rcorollary{nonadjacent distance larger than one II}(ii), and \rlemma{main lemma II}
to prove \rtheorem{main result}.

\bproof \textbf{(Proof of \rtheorem{main result})}
Let $r_{-1}=M$, $r_0=k$, and let $q_i$ and $r_i$, $i=1,2,\ldots,N$, be recursively obtained
as in Step 1 of \ralgorithm{main result}.
Then we have $\gcd(M,k)=r_{N-1}$.
Note that as $r_{-1}>r_0>r_1>\cdots>r_{N-1}>r_N$ and $r_N=0$,
it is easy to see that
\beqnarray{proof-main result-111}
r_{h-1}\geq 2 \textrm{ and } r_h\neq 0, \textrm{ for } h=0,1,\ldots,N-1.
\eeqnarray

(i) Note that in \rtheorem{main result}(i), we have $\gcd(M,k)=r_{N-1}=1$.
Thus, it is easy to see from \reqnarray{N-M-k-h} that $\Ncal_{M,k}(N)=\{(r_{N-2})\}$.
Since $r_{N-1}=1$ and $r_N=0$,
we have $r_{N-2}=q_N\cdot r_{N-1}+r_N=q_N$
and it follows that $\Ncal_{M,k}(N)=\{(q_N)\}$.
We need to consider the following three cases.

\emph{Case 1: $N=1$.}
In this case, we have $\Ncal_{M,k}(1)=\{(q_1)\}$.
Let $\nbf_1^{r_0}(1)=(q_1)$.
As $\nbf_1^{r_0}(1)=(q_1)$ is the only sequence in $\Ncal_{M,k}(1)$,
it follows that there is only one optimal sequence over $\Ncal_{M,k}(1)$
and the optimal sequence is given by $\nbf_1^{r_0}(1)=(q_1)$.
It is also clear that $\nbf_1^{r_0}(1)=(q_1)$ is the sequence obtained
in Step 2(i) of \ralgorithm{main result} (note that $N=1$ is an odd integer).

\emph{Case 2: $N\geq 2$ and $N$ is an odd integer.}
Let $\nbf_1^{r_0}(1)\in \Ncal_{M,k}(1)$ be an optimal sequence over $\Ncal_{M,k}(1)$.
From \reqnarray{proof-main result-111}, we have $r_0\geq 2$ and $r_1\neq 0$.
It then follows from \rlemma{main lemma} (for the odd integer $h=1$)
that $\nbf_1^{r_0}(1)$ must satisfy the condition that
\beqnarray{}
n_i(1)=
\bselection
q_1+1, &\textrm{if } i=i_1,i_2,\ldots,i_{r_1}, \\
q_1, &\textrm{otherwise},
\eselection \nn
\eeqnarray
for some $1=i_1<i_2<\cdots <i_{r_1}\leq r_0$.
Thus, the left-imbedded sequence of $\nbf_1^{r_0}(1)$
with respect to $r_{-1}$ and $r_0$ is well defined, say
\beqnarray{proof-main result-(i)-case-2-111}
\nbf_1^{r_1}(2)=L_{r_{-1},r_0}^I(\nbf_1^{r_0}(1)).
\eeqnarray
From the definition of left-imbedded sequences in \rdefinition{left-imbedded sequences},
we know that $\nbf_1^{r_1}(2)$ is a sequence of positive integers such that $\sum_{i=1}^{r_1}n_i(2)=r_0$,
i.e., $\nbf_1^{r_1}(2)\in \Ncal_{M,k}(2)$.

We prove by contradiction that $\nbf_1^{r_1}(2)$ is an optimal sequence over $\Ncal_{M,k}(2)$.
Assume that there exists a sequence ${\nbf'}_1^{r_1}(2)\in \Ncal_{M,k}(2)$
such that ${\nbf'}_1^{r_1}(2)\succ\nbf_1^{r_1}(2)$.
As ${\nbf'}_1^{r_1}(2)\in \Ncal_{M,k}(2)$, we have $\sum_{i=1}^{r_1}n_i'(2)=r_0$.
Thus, the left pre-sequence of ${\nbf'}_1^{r_1}(2)$
with respect to $r_{-1}$ and $r_0$ is well defined, say
\beqnarray{proof-main result-(i)-case-2-222}
{\nbf'}_1^{r_0}(1)=L_{r_{-1},r_0}({\nbf'}_1^{r_1}(2)).
\eeqnarray
From \reqnarray{proof-main result-(i)-case-2-111} and \rlemma{left-imbedded sequences-left pre-sequences}(i),
we have
\beqnarray{proof-main result-(i)-case-2-333}
\nbf_1^{r_0}(1)=L_{r_{-1},r_0}(\nbf_1^{r_1}(2)).
\eeqnarray
As such, we have from
${\nbf'}_1^{r_1}(2)\succ\nbf_1^{r_1}(2)$,
\reqnarray{proof-main result-(i)-case-2-222},
\reqnarray{proof-main result-(i)-case-2-333},
and \reqnarray{order relation-777} that ${\nbf'}_1^{r_0}(1)\succ\nbf_1^{r_0}(1)$,
and we have reached a contradiction to the optimality of $\nbf_1^{r_0}(1)$.

As $N\geq 2$ and $N$ is an odd integer, we have $N\geq 3$.
Thus, it is clear from \reqnarray{proof-main result-111} that $r_1\geq 2$ and $r_2\neq 0$.
As $\nbf_1^{r_1}(2)$ is an optimal sequence over $\Ncal_{M,k}(2)$,
it then follows from \rlemma{main lemma II} (for the even integer $h=2$)
that $\nbf_1^{r_1}(2)$ must satisfy the condition that
\beqnarray{}
n_i(2)=
\bselection
q_2+1, &\textrm{if } i=i_1,i_2,\ldots,i_{r_2}, \\
q_2, &\textrm{otherwise},
\eselection \nn
\eeqnarray
for some $1\leq i_1<i_2<\cdots <i_{r_2}=r_1$.
Thus, the right-imbedded sequence of $\nbf_1^{r_1}(2)$
with respect to $r_0$ and $r_1$ is well defined, say
\beqnarray{proof-main result-(i)-case-2-444}
\nbf_1^{r_2}(3)=R_{r_0,r_1}^I(\nbf_1^{r_1}(2)).
\eeqnarray
From the definition of right-imbedded sequences in \rdefinition{right-imbedded sequences},
we know that $\nbf_1^{r_2}(3)$ is a sequence of positive integers such that
$\sum_{i=1}^{r_2}n_i(3)=r_1$, i.e., $\nbf_1^{r_2}(3)\in \Ncal_{M,k}(3)$.

We prove by contradiction that $\nbf_1^{r_2}(3)$ is an optimal sequence over $\Ncal_{M,k}(3)$.
Assume that there exists a sequence ${\nbf'}_1^{r_2}(3)\in \Ncal_{M,k}(3)$
such that ${\nbf'}_1^{r_2}(3)\succ\nbf_1^{r_2}(3)$.
As ${\nbf'}_1^{r_2}(3)\in \Ncal_{M,k}(3)$, we have $\sum_{i=1}^{r_2}n_i'(3)=r_1$.
Thus, the right pre-sequence of ${\nbf'}_1^{r_2}(3)$
with respect to $r_0$ and $r_1$ is well defined, say
\beqnarray{proof-main result-(i)-case-2-555}
{\nbf'}_1^{r_1}(2)=R_{r_0,r_1}({\nbf'}_1^{r_2}(3)).
\eeqnarray
From \reqnarray{proof-main result-(i)-case-2-444} and \rlemma{right-imbedded sequences-right pre-sequences}(i),
we have
\beqnarray{proof-main result-(i)-case-2-666}
\nbf_1^{r_1}(2)=R_{r_0,r_1}(\nbf_1^{r_2}(3)).
\eeqnarray
As such, we have from
${\nbf'}_1^{r_2}(3)\succ\nbf_1^{r_2}(3)$,
\reqnarray{proof-main result-(i)-case-2-555},
\reqnarray{proof-main result-(i)-case-2-666},
and \reqnarray{order relation-777} that ${\nbf'}_1^{r_1}(2)\succ\nbf_1^{r_1}(2)$,
and we have reached a contradiction to the optimality of $\nbf_1^{r_1}(2)$.

Clearly, for the optimal sequence $\nbf_1^{r_0}(1)$ over $\Ncal_{M,k}(1)$,
we can repeat the above argument and obtain a corresponding optimal sequence
$\nbf_1^{r_{h-1}}(h)$ over $\Ncal_{M,k}(h)$ for $h=2,3,\ldots,N$, where
\beqnarray{}
\nbf_1^{r_1}(2) \aligneq L_{r_{-1},r_0}^I(\nbf_1^{r_0}(1)),
\label{eqn:proof-main result-(i)-case-2-777} \\
\nbf_1^{r_2}(3) \aligneq R_{r_0,r_1}^I(\nbf_1^{r_1}(2)),
\label{eqn:proof-main result-(i)-case-2-888} \\
\nbf_1^{r_3}(4) \aligneq L_{r_1,r_2}^I(\nbf_1^{r_2}(3)),
\label{eqn:proof-main result-(i)-case-2-999} \\
\nbf_1^{r_4}(5) \aligneq R_{r_2,r_3}^I(\nbf_1^{r_3}(4)),
\label{eqn:proof-main result-(i)-case-2-aaa} \\
&\vdots& \nn\\
\nbf_1^{r_{N-2}}(N-1) \aligneq L_{r_{N-4},r_{N-3}}^I(\nbf_1^{r_{N-3}}(N-2)),
\label{eqn:proof-main result-(i)-case-2-bbb} \\
\nbf_1^{r_{N-1}}(N) \aligneq R_{r_{N-3},r_{N-2}}^I(\nbf_1^{r_{N-2}}(N-1)).
\label{eqn:proof-main result-(i)-case-2-ccc}
\eeqnarray
As $\Ncal_{M,k}(N)=\{(q_N)\}$, we see that $\nbf_1^{r_{N-1}}(N)=(q_N)$.
It then follows from \reqnarray{proof-main result-(i)-case-2-777}--\reqnarray{proof-main result-(i)-case-2-ccc},
\rlemma{left-imbedded sequences-left pre-sequences}(i),
and \rlemma{right-imbedded sequences-right pre-sequences}(i)
that $\nbf_1^{r_0}(1)$ can be obtained from $\nbf_1^{r_{N-1}}(N)=(q_N)$ as follows:
\beqnarray{}
\nbf_1^{r_{N-2}}(N-1) \aligneq R_{r_{N-3},r_{N-2}}(\nbf_1^{r_{N-1}}(N)),
\label{eqn:proof-main result-(i)-case-2-ddd} \\
\nbf_1^{r_{N-3}}(N-2) \aligneq L_{r_{N-4},r_{N-3}}(\nbf_1^{r_{N-2}}(N-1)),
\label{eqn:proof-main result-(i)-case-2-eee} \\
&\vdots& \nn\\
\nbf_1^{r_3}(4) \aligneq R_{r_2,r_3}(\nbf_1^{r_4}(5)),
\label{eqn:proof-main result-(i)-case-2-fff} \\
\nbf_1^{r_2}(3) \aligneq L_{r_1,r_2}(\nbf_1^{r_3}(4)),
\label{eqn:proof-main result-(i)-case-2-ggg} \\
\nbf_1^{r_1}(2) \aligneq R_{r_0,r_1}(\nbf_1^{r_2}(3)),
\label{eqn:proof-main result-(i)-case-2-hhh} \\
\nbf_1^{r_0}(1) \aligneq L_{r_{-1},r_0}(\nbf_1^{r_1}(2)).
\label{eqn:proof-main result-(i)-case-2-iii}
\eeqnarray

Suppose that there exists an optimal sequence ${\nbf'}_1^{r_0}(1)$
over $\Ncal_{M,k}(1)$ such that ${\nbf'}_1^{r_0}(1)\neq \nbf_1^{r_0}(1)$.
For the optimal sequence ${\nbf'}_1^{r_0}(1)$,
we can obtain a corresponding optimal sequence ${\nbf'}_1^{r_{N-1}}(N)$ over $\Ncal_{M,k}(N)$
in a similar way that the optimal sequence $\nbf_1^{r_{N-1}}(N)$ is obtained from $\nbf_1^{r_0}(1)$
by using \reqnarray{proof-main result-(i)-case-2-777}--\reqnarray{proof-main result-(i)-case-2-ccc}.
As $\Ncal_{M,k}(N)=\{(q_N)\}$, we have ${\nbf'}_1^{r_{N-1}}(N)=(q_N)=\nbf_1^{r_{N-1}}(N)$.
Since ${\nbf'}_1^{r_0}(1)$ can be obtained from ${\nbf'}_1^{r_{N-1}}(N)$
in a similar way that $\nbf_1^{r_0}(1)$ is obtained from $\nbf_1^{r_{N-1}}(N)$
by using \reqnarray{proof-main result-(i)-case-2-ddd}--\reqnarray{proof-main result-(i)-case-2-iii},
we see from $\nbf_1^{r_{N-1}}(N)={\nbf'}_1^{r_{N-1}}(N)=(q_N)$ that
$\nbf_1^{r_0}(1)={\nbf'}_1^{r_0}(1)$, contradicting to $\nbf_1^{r_0}(1)\neq {\nbf'}_1^{r_0}(1)$.
Therefore, we conclude that $\nbf_1^{r_0}(1)$
is the only optimal sequence over $\Ncal_{M,k}(1)$.
As $\nbf_1^{r_{N-1}}(N)=(q_N)$ and \reqnarray{proof-main result-(i)-case-2-ddd}--\reqnarray{proof-main result-(i)-case-2-iii}
are the same as \reqnarray{main result-111}--\reqnarray{main result-222}
in Step 2(i) of \ralgorithm{main result} (note that $N$ is an odd integer),
it then follows that the optimal sequence $\nbf_1^{r_0}(1)$ obtained from $\nbf_1^{r_{N-1}}(N)=(q_N)$
by using \reqnarray{proof-main result-(i)-case-2-ddd}--\reqnarray{proof-main result-(i)-case-2-iii}
is the sequence obtained in Step 2(i) of \ralgorithm{main result}.

\emph{Case 3: $N\geq 2$ and $N$ is an even integer.}
Let $\nbf_1^{r_0}(1)\in \Ncal_{M,k}(1)$ be an optimal sequence over $\Ncal_{M,k}(1)$.
By a similar argument as in Case~2 above,
for the optimal sequence $\nbf_1^{r_0}(1)$ over $\Ncal_{M,k}(1)$,
we can obtain a corresponding optimal sequence
$\nbf_1^{r_{h-1}}(h)$ over $\Ncal_{M,k}(h)$ for $h=2,3,\ldots,N$, where
\beqnarray{}
\nbf_1^{r_1}(2) \aligneq L_{r_{-1},r_0}^I(\nbf_1^{r_0}(1)),
\label{eqn:proof-main result-(i)-case-3-111} \\
\nbf_1^{r_2}(3) \aligneq R_{r_0,r_1}^I(\nbf_1^{r_1}(2)),
\label{eqn:proof-main result-(i)-case-3-222} \\
\nbf_1^{r_3}(4) \aligneq L_{r_1,r_2}^I(\nbf_1^{r_2}(3)),
\label{eqn:proof-main result-(i)-case-3-333} \\
\nbf_1^{r_4}(5) \aligneq R_{r_2,r_3}^I(\nbf_1^{r_3}(4)),
\label{eqn:proof-main result-(i)-case-3-444} \\
&\vdots& \nn\\
\nbf_1^{r_{N-3}}(N-2) \aligneq L_{r_{N-5},r_{N-4}}^I(\nbf_1^{r_{N-4}}(N-3)),
\label{eqn:proof-main result-(i)-case-3-555} \\
\nbf_1^{r_{N-2}}(N-1) \aligneq R_{r_{N-4},r_{N-3}}^I(\nbf_1^{r_{N-3}}(N-2)),
\label{eqn:proof-main result-(i)-case-3-666} \\
\nbf_1^{r_{N-1}}(N) \aligneq L_{r_{N-3},r_{N-2}}^I(\nbf_1^{r_{N-2}}(N-1)).
\label{eqn:proof-main result-(i)-case-3-777}
\eeqnarray
As $\Ncal_{M,k}(N)=\{(q_N)\}$, we see that $\nbf_1^{r_{N-1}}(N)=(q_N)$.
It then follows from \reqnarray{proof-main result-(i)-case-3-111}--\reqnarray{proof-main result-(i)-case-3-777},
\rlemma{left-imbedded sequences-left pre-sequences}(i),
and \rlemma{right-imbedded sequences-right pre-sequences}(i)
that $\nbf_1^{r_0}(1)$ can be obtained from $\nbf_1^{r_{N-1}}(N)=(q_N)$ as follows:
\beqnarray{}
\nbf_1^{r_{N-2}}(N-1) \aligneq L_{r_{N-3},r_{N-2}}(\nbf_1^{r_{N-1}}(N)),
\label{eqn:proof-main result-(i)-case-3-888} \\
\nbf_1^{r_{N-3}}(N-2) \aligneq R_{r_{N-4},r_{N-3}}(\nbf_1^{r_{N-2}}(N-1)),
\label{eqn:proof-main result-(i)-case-3-999} \\
\nbf_1^{r_{N-4}}(N-3) \aligneq L_{r_{N-5},r_{N-4}}(\nbf_1^{r_{N-3}}(N-2)),
\label{eqn:proof-main result-(i)-case-3-aaa} \\
&\vdots& \nn\\
\nbf_1^{r_3}(4) \aligneq R_{r_2,r_3}(\nbf_1^{r_4}(5)),
\label{eqn:proof-main result-(i)-case-3-bbb} \\
\nbf_1^{r_2}(3) \aligneq L_{r_1,r_2}(\nbf_1^{r_3}(4)),
\label{eqn:proof-main result-(i)-case-3-ccc} \\
\nbf_1^{r_1}(2) \aligneq R_{r_0,r_1}(\nbf_1^{r_2}(3)),
\label{eqn:proof-main result-(i)-case-3-ddd} \\
\nbf_1^{r_0}(1) \aligneq L_{r_{-1},r_0}(\nbf_1^{r_1}(2)).
\label{eqn:proof-main result-(i)-case-3-eee}
\eeqnarray

Furthermore, we can argue as in Case~2 above that
$\nbf_1^{r_0}(1)$ is the only optimal sequence over $\Ncal_{M,k}(1)$.
As $\nbf_1^{r_{N-1}}(N)=(q_N)$ and \reqnarray{proof-main result-(i)-case-3-888}--\reqnarray{proof-main result-(i)-case-3-eee}
are the same as \reqnarray{main result-555}--\reqnarray{main result-777}
in Step 3(i) of \ralgorithm{main result} (note that $N$ is an even integer),
it then follows that the optimal sequence $\nbf_1^{r_0}(1)$ obtained from $\nbf_1^{r_{N-1}}(N)=(q_N)$
by using \reqnarray{proof-main result-(i)-case-3-888}--\reqnarray{proof-main result-(i)-case-3-eee}
is the sequence obtained in Step 3(i) of \ralgorithm{main result}.

(ii) Note that in \rtheorem{main result}(ii), we have $\gcd(M,k)=r_{N-1}=2$.
We consider the following three cases.

\emph{Case 1: $N=1$.}
As $r_{N-1}=2$ and $r_N=0$,
it follows from \rcorollary{adjacent distance larger than one}(ii) (for the odd integer $h=N=1$)
that there are two optimal sequences over $\Ncal_{M,k}(1)$,
and the two optimal sequences, say $\nbf_1^{r_0}(1)$ and $\mbf_1^{r_0}(1)$,
are given by $\nbf_1^{r_0}(1)=(q_1,q_1)$ and $\mbf_1^{r_0}(1)=(q_1+1,q_1-1)$.
Also, it is clear that $\nbf_1^{r_0}(1)=(q_1,q_1)$
and $\mbf_1^{r_0}=(q_1+1,q_1-1)$
are the sequences obtained in Step 2(i) and Step 2(ii), respectively,
of \ralgorithm{main result} (note that $N=1$ is an odd integer).

\emph{Case 2: $N\geq 2$ and $N$ is an odd integer.}
As $r_{N-1}=2$ and $r_N=0$,
it follows from \rcorollary{adjacent distance larger than one}(ii) (for the odd integer $h=N$)
that there are two optimal sequences over $\Ncal_{M,k}(N)$,
and the two optimal sequences, say $\nbf_1^{r_{N-1}}(N)$ and $\mbf_1^{r_{N-1}}(N)$,
are given by $\nbf_1^{r_{N-1}}(N)=(q_N,q_N)$
and $\mbf_1^{r_{N-1}}(N)=(q_N+1,q_N-1)$.

Let $\nbf_1^{r_0}(1)\in \Ncal_{M,k}(1)$ be an optimal sequence over $\Ncal_{M,k}(1)$.
As in Case~2 of (i) above,
for the optimal sequence $\nbf_1^{r_0}(1)$,
we can obtain a corresponding optimal sequence over $\Ncal_{M,k}(N)$,
say $\nbf_1^{r_{N-1}}(N)=(q_N,q_N)$,
and $\nbf_1^{r_0}(1)$ can be uniquely obtained from $\nbf_1^{r_{N-1}}(N)=(q_N,q_N)$
by using \reqnarray{proof-main result-(i)-case-2-ddd}--\reqnarray{proof-main result-(i)-case-2-iii}.
Furthermore, let
\beqnarray{}
\mbf_1^{r_{N-2}}(N-1) \aligneq R_{r_{N-3},r_{N-2}}(\mbf_1^{r_{N-1}}(N)),
\label{eqn:proof-main result-(ii)-case-2-111} \\
\mbf_1^{r_{N-3}}(N-2) \aligneq L_{r_{N-4},r_{N-3}}(\mbf_1^{r_{N-2}}(N-1)),
\label{eqn:proof-main result-(ii)-case-2-222} \\
&\vdots& \nn\\
\mbf_1^{r_3}(4) \aligneq R_{r_2,r_3}(\mbf_1^{r_4}(5)),
\label{eqn:proof-main result-(ii)-case-2-333} \\
\mbf_1^{r_2}(3) \aligneq L_{r_1,r_2}(\mbf_1^{r_3}(4)),
\label{eqn:proof-main result-(ii)-case-2-444} \\
\mbf_1^{r_1}(2) \aligneq R_{r_0,r_1}(\mbf_1^{r_2}(3)),
\label{eqn:proof-main result-(ii)-case-2-555} \\
\mbf_1^{r_0}(1) \aligneq L_{r_{-1},r_0}(\mbf_1^{r_1}(2)).
\label{eqn:proof-main result-(ii)-case-2-666}
\eeqnarray
As $\nbf_1^{r_{N-1}}(N)=(q_N,q_N)$ and $\mbf_1^{r_{N-1}}(N)=(q_N+1,q_N-1)$
are optimal sequences over $\Ncal_{M,k}(N)$,
we have $\nbf_1^{r_{N-1}}(N)\equiv \mbf_1^{r_{N-1}}(N)$.
It then follows from $\nbf_1^{r_{N-1}}(N)\equiv \mbf_1^{r_{N-1}}(N)$,
\reqnarray{proof-main result-(i)-case-2-ddd}--\reqnarray{proof-main result-(i)-case-2-iii},
\reqnarray{proof-main result-(ii)-case-2-111}--\reqnarray{proof-main result-(ii)-case-2-666},
and \reqnarray{order relation-777} that $\nbf_1^{r_0}(1)\equiv \mbf_1^{r_0}(1)$,
i.e., $\mbf_1^{r_0}(1)$ is also an optimal sequence over $\Ncal_{M,k}(1)$.

Since there are two optimal sequences over $\Ncal_{M,k}(N)$,
we see that there are two optimal sequences over $\Ncal_{M,k}(1)$,
and the two optimal sequences are $\nbf_1^{r_0}(1)$
obtained from $\nbf_1^{r_{N-1}}(N)=(q_N,q_N)$
by using \reqnarray{proof-main result-(i)-case-2-ddd}--\reqnarray{proof-main result-(i)-case-2-iii}
and $\mbf_1^{r_0}(1)$ obtained from $\mbf_1^{r_{N-1}}(N)=(q_N+1,q_N-1)$
by using \reqnarray{proof-main result-(ii)-case-2-111}--\reqnarray{proof-main result-(ii)-case-2-666}.
It is clear that the two optimal sequences $\nbf_1^{r_0}(1)$ and $\mbf_1^{r_0}(1)$
are the sequences obtained in Step 2(i) and Step 2(ii), respectively,
of \ralgorithm{main result} (note that $N$ is an odd integer).

\emph{Case 3: $N\geq 2$ and $N$ is an even integer.}
As $r_{N-1}=2$ and $r_N=0$,
it follows from \rcorollary{adjacent distance larger than one II}(ii) (for the even integer $h=N$)
that there are two optimal sequences over $\Ncal_{M,k}(N)$,
and the two optimal sequences, say $\nbf_1^{r_{N-1}}(N)$ and $\mbf_1^{r_{N-1}}(N)$,
are given by $\nbf_1^{r_{N-1}}(N)=(q_N,q_N)$
and $\mbf_1^{r_{N-1}}(N)=(q_N-1,q_N+1)$.

Let $\nbf_1^{r_0}(1)\in \Ncal_{M,k}(1)$ be an optimal sequence over $\Ncal_{M,k}(1)$.
As in Case~3 of (i) above,
for the optimal sequence $\nbf_1^{r_0}(1)$,
we can obtain a corresponding optimal sequence over $\Ncal_{M,k}(N)$,
say $\nbf_1^{r_{N-1}}(N)=(q_N,q_N)$,
and $\nbf_1^{r_0}(1)$ can be uniquely obtained from $\nbf_1^{r_{N-1}}(N)=(q_N,q_N)$
by using \reqnarray{proof-main result-(i)-case-3-888}--\reqnarray{proof-main result-(i)-case-3-eee}.
Furthermore, let
\beqnarray{}
\mbf_1^{r_{N-2}}(N-1) \aligneq L_{r_{N-3},r_{N-2}}(\mbf_1^{r_{N-1}}(N)),
\label{eqn:proof-main result-(ii)-case-3-111} \\
\mbf_1^{r_{N-3}}(N-2) \aligneq R_{r_{N-4},r_{N-3}}(\mbf_1^{r_{N-2}}(N-1)),
\label{eqn:proof-main result-(ii)-case-3-222} \\
\mbf_1^{r_{N-4}}(N-3) \aligneq L_{r_{N-5},r_{N-4}}(\mbf_1^{r_{N-3}}(N-2)),
\label{eqn:proof-main result-(ii)-case-3-333} \\
&\vdots& \nn\\
\mbf_1^{r_3}(4) \aligneq R_{r_2,r_3}(\mbf_1^{r_4}(5)),
\label{eqn:proof-main result-(ii)-case-3-444} \\
\mbf_1^{r_2}(3) \aligneq L_{r_1,r_2}(\mbf_1^{r_3}(4)),
\label{eqn:proof-main result-(ii)-case-3-555} \\
\mbf_1^{r_1}(2) \aligneq R_{r_0,r_1}(\mbf_1^{r_2}(3)),
\label{eqn:proof-main result-(ii)-case-3-666} \\
\mbf_1^{r_0}(1) \aligneq L_{r_{-1},r_0}(\mbf_1^{r_1}(2)).
\label{eqn:proof-main result-(ii)-case-3-777}
\eeqnarray
As $\nbf_1^{r_{N-1}}(N)=(q_N,q_N)$ and $\mbf_1^{r_{N-1}}(N)=(q_N-1,q_N+1)$
are optimal sequences over $\Ncal_{M,k}(N)$,
we have $\nbf_1^{r_{N-1}}(N)\equiv \mbf_1^{r_{N-1}}(N)$.
It then follows from $\nbf_1^{r_{N-1}}(N)\equiv \mbf_1^{r_{N-1}}(N)$,
\reqnarray{proof-main result-(i)-case-3-888}--\reqnarray{proof-main result-(i)-case-3-eee},
\reqnarray{proof-main result-(ii)-case-3-111}--\reqnarray{proof-main result-(ii)-case-3-777},
and \reqnarray{order relation-777} that $\nbf_1^{r_0}(1)\equiv \mbf_1^{r_0}(1)$,
i.e., $\mbf_1^{r_0}(1)$ is also an optimal sequence over $\Ncal_{M,k}(1)$.

Since there are two optimal sequences over $\Ncal_{M,k}(N)$,
we see that there are two optimal sequences over $\Ncal_{M,k}(1)$,
and the two optimal sequences are $\nbf_1^{r_0}(1)$
obtained from $\nbf_1^{r_{N-1}}(N)=(q_N,q_N)$
by using \reqnarray{proof-main result-(i)-case-3-888}--\reqnarray{proof-main result-(i)-case-3-eee}
and $\mbf_1^{r_0}(1)$ obtained from $\mbf_1^{r_{N-1}}(N)=(q_N-1,q_N+1)$
by using \reqnarray{proof-main result-(ii)-case-3-111}--\reqnarray{proof-main result-(ii)-case-3-777}.
It is clear that the two optimal sequences $\nbf_1^{r_0}(1)$ and $\mbf_1^{r_0}(1)$
are the sequences obtained in Step 3(i) and Step 3(ii), respectively,
of \ralgorithm{main result} (note that $N$ is an even integer).

(iii) Note that we assume that $\gcd(M,k)=r_{N-1}\geq 3$.
The proof is similar to that of (ii)
and we consider the following three cases.

\emph{Case 1: $N=1$.}
As $r_{N-1}\geq 3$ and $r_N=0$,
we have from \rcorollary{nonadjacent distance larger than one}(ii) (for the odd integer $h=N=1$)
that there are at most two optimal sequences over $\Ncal_{M,k}(1)$,
and the two possible optimal sequences, say $\nbf_1^{r_0}(1)$ and $\mbf_1^{r_0}(1)$,
are given by $\nbf_1^{r_0}(1)=(q_1,q_1,\ldots,q_1)$ and $\mbf_1^{r_0}(1)=(q_1+1,q_1,\ldots,q_1,q_1-1)$.
Clearly, $\nbf_1^{r_0}(1)=(q_1,q_1,\ldots,q_1)$ and $\mbf_1^{r_0}=(q_1+1,q_1,\ldots,q_1,q_1-1)$
are the sequences obtained in Step 2(i) and Step 2(ii), respectively,
of \ralgorithm{main result} (note that $N=1$ is an odd integer).

\emph{Case 2: $N\geq 2$ and $N$ is an odd integer.}
As $r_{N-1}\geq 3$ and $r_N=0$,
we have from \rcorollary{nonadjacent distance larger than one}(ii) (for the odd integer $h=N$)
that there are at most two optimal sequences over $\Ncal_{M,k}(N)$,
and the two possible optimal sequences, say $\nbf_1^{r_{N-1}}(N)$ and $\mbf_1^{r_{N-1}}(N)$,
are given by $\nbf_1^{r_{N-1}}(N)=(q_N,q_N,\ldots,q_N)$
and $\mbf_1^{r_{N-1}}(N)=(q_N+1,q_N,\ldots,q_N,q_N-1)$.
Therefore, there are at most two optimal sequences over $\Ncal_{M,k}(1)$,
and the two possible optimal sequences are $\nbf_1^{r_0}(1)$
obtained from $\nbf_1^{r_{N-1}}(N)=(q_N,q_N,\ldots,q_N)$
by using \reqnarray{proof-main result-(i)-case-2-ddd}--\reqnarray{proof-main result-(i)-case-2-iii}
and $\mbf_1^{r_0}(1)$ obtained from $\mbf_1^{r_{N-1}}(N)=(q_N+1,q_N,\ldots,q_N,q_N-1)$
by using \reqnarray{proof-main result-(ii)-case-2-111}--\reqnarray{proof-main result-(ii)-case-2-666}.
It is clear that the two optimal sequences $\nbf_1^{r_0}(1)$ and $\mbf_1^{r_0}(1)$
are the sequences obtained in Step 2(i) and Step 2(ii), respectively,
of \ralgorithm{main result} (note that $N$ is an odd integer).

\emph{Case 3: $N\geq 2$ and $N$ is an even integer.}
As $r_{N-1}\geq 3$ and $r_N=0$,
we have from \rcorollary{nonadjacent distance larger than one II}(ii) (for the even integer $h=N$)
that there are at most two optimal sequences over $\Ncal_{M,k}(N)$,
and the two possible optimal sequences, say $\nbf_1^{r_{N-1}}(N)$ and $\mbf_1^{r_{N-1}}(N)$,
are given by $\nbf_1^{r_{N-1}}(N)=(q_N,q_N,\ldots,q_N)$
and $\mbf_1^{r_{N-1}}(N)=(q_N-1,q_N,\ldots,q_N,q_N+1)$.
Therefore, there are at most two optimal sequences over $\Ncal_{M,k}(1)$,
and the two possible optimal sequences are $\nbf_1^{r_0}(1)$
obtained from $\nbf_1^{r_{N-1}}(N)=(q_N,q_N,\ldots,q_N)$
by using \reqnarray{proof-main result-(i)-case-3-888}--\reqnarray{proof-main result-(i)-case-3-eee}
and $\mbf_1^{r_0}(1)$ obtained from $\mbf_1^{r_{N-1}}(N)=(q_N-1,q_N,\ldots,q_N,q_N+1)$
by using \reqnarray{proof-main result-(ii)-case-3-111}--\reqnarray{proof-main result-(ii)-case-3-777}.
It is clear that the two optimal sequences $\nbf_1^{r_0}(1)$ and $\mbf_1^{r_0}(1)$
are the sequences obtained in Step 3(i) and Step 3(ii), respectively,
of \ralgorithm{main result} (note that $N$ is an even integer).
\eproof

\bsection{Conclusion}{conclusion}

In this two-part paper, we address an important practical feasibility issue
that is of great concern in the SDL constructions of optical queues:
the constructions of optical queues with a limited number of
recirculations through the optical switches and the fiber delay lines.
In Part~I, we have proposed a class of greedy constructions
for certain types of optical queues, including linear compressors,
linear decompressors, and 2-to-1 FIFO multiplexers,
and have shown that every optimal construction among our previous constructions
of these types of optical queues under the constraint of
a limited number of recirculations must be a greedy construction.
In Part~II, we have further shown that there are at most two optimal constructions
and give a simple algorithm to obtain the optimal construction(s).

\appendices

\bappendix{Proof of \rlemma{adjacent distance larger than one} with $h=1$}
{proof of adjacent distance larger than one with h=1}

In this appendix, we prove \rlemma{adjacent distance larger than one} for the case that $h=1$.
We need the following lemma for the proof of \rlemma{adjacent distance larger than one} with $h=1$.

\blemma{d>B+1}
Suppose that $M\geq 2$ and $1\leq k\leq M-1$.
Let $\nbf_1^k\in \Ncal_{M,k}$,
and let $s_0=0$ and $s_i=\sum_{\ell=1}^{i}n_{\ell}$ for $i=1,2,\ldots,k$.
Let $\dbf_1^M$ be obtained by using $\nbf_1^k$ in \reqnarray{OQ-LR-delays-greedy-1}.
Then
\beqnarray{d>B+1}
d_{s_{i+1}}>B(\dbf_1^{s_i};i)+1,
\eeqnarray
for $i=0,1,\ldots,k-1$.
\elemma

\bproof
We prove this lemma by induction on $i$.
As $\nbf_1^k\in \Ncal_{M,k}$,
we have $n_1\geq 2$, $n_2,n_3,\ldots,n_k\geq 1$, and $\sum_{i=1}^{k}n_{i}=M$.
From \reqnarray{OQ-LR-delays-greedy-2}, $n_1\geq 2$, and $B(\dbf_1^{s_0};0)=0$,
we have
\beqnarray{proof-d>B+1-111}
d_{s_1}=s_1=n_1\geq 2>1=B(\dbf_1^{s_0};0)+1.
\eeqnarray
It follows from \reqnarray{proof-d>B+1-111} that \reqnarray{d>B+1} holds for $i=0$.

Suppose as the induction hypothesis that \reqnarray{d>B+1} holds for some $0\leq i\leq k-2$.
From \reqnarray{OQ-LR-delays-greedy-3} and \reqnarray{OQ-LR-delays-greedy-7},
we have
\beqnarray{proof-d>B+1-222}
\alignspace d_{s_{i+2}}-(B(\dbf_1^{s_{i+1}};i+1)+1) \nn \\
\alignspace =2d_{s_{i+1}}+(n_{i+2}-1)(B(\dbf_1^{s_{i+1}};i+1)+1)-(B(\dbf_1^{s_{i+1}};i+1)+1) \nn \\
\alignspace =2d_{s_{i+1}}+(n_{i+2}-2)(B(\dbf_1^{s_{i+1}};i+1)+1).
\eeqnarray
If $n_{i+2}\geq 2$, then it follows from \reqnarray{proof-d>B+1-222} that
\beqnarray{proof-d>B+1-333}
d_{s_{i+2}}-(B(\dbf_1^{s_{i+1}};i+1)+1) \geq 2d_{s_{i+1}}>0.
\eeqnarray
On the other hand, if $n_{i+2}=1$, then we have from
\reqnarray{proof-d>B+1-222}, \reqnarray{OQ-LR-delays-greedy-7}, and the induction hypothesis that
\beqnarray{proof-d>B+1-444}
d_{s_{i+2}}-(B(\dbf_1^{s_{i+1}};i+1)+1)
\aligneq 2d_{s_{i+1}}-(B(\dbf_1^{s_{i+1}};i+1)+1) \nn\\
\aligneq 2d_{s_{i+1}}-(B(\dbf_1^{s_{i}};i)+d_{s_{i+1}}+1)\nn\\
\aligneq d_{s_{i+1}}-(B(\dbf_1^{s_{i}};i)+1)>0.
\eeqnarray
The induction is completed by combining
\reqnarray{proof-d>B+1-333} and \reqnarray{proof-d>B+1-444}.
\eproof

Now we use \rlemma{d>B+1} to prove \rlemma{adjacent distance larger than one}
for the case that $h=1$.
Suppose that $h=1$ in \rlemma{adjacent distance larger than one}.
For simplicity, let $\nbf_1^k=\nbf_1^{r_0}(1)$ and ${\nbf'}_1^k={\nbf'}_1^{r_0}(1)$
(note that $r_0=k$), i.e., $n_i=n_i(1)$ and $n'_i=n'_i(1)$ for $i=1,2,\ldots,k$.

Let $\dbf_1^M$ and ${\dbf'}_1^M$ be obtained by using
$\nbf_1^k$ and ${\nbf'}_1^k$, respectively,
in \reqnarray{OQ-LR-delays-greedy-1}.
Let $s_0=0$ and $s_i=\sum_{\ell=1}^{i}n_{\ell}$ for $i=1,2,\ldots,k$,
and let $s'_0=0$ and $s'_i=\sum_{\ell=1}^{i}n'_{\ell}$ for $i=1,2,\ldots,k$.
Let
\beqnarray{}
\alpha_i \aligneq d_{s_i}-d'_{s'_i}, \textrm{ for } i=1,2,\ldots,k,
\label{eqn:proof-adjacent distance larger than one-111}\\
\beta_i \aligneq B(\dbf_1^{s_i};i)-B({\dbf'}_1^{s'_i};i), \textrm{ for } i=0,1,2,\ldots,k.
\label{eqn:proof-adjacent distance larger than one-222}
\eeqnarray
It follows from \reqnarray{proof-adjacent distance larger than one-222},
\reqnarray{OQ-LR-delays-greedy-7},
and \reqnarray{proof-adjacent distance larger than one-111} that
\beqnarray{proof-adjacent distance larger than one-333}
\beta_i
\aligneq B(\dbf_1^{s_i};i)-B({\dbf'}_1^{s'_i};i)\nn \\
\aligneq B(\dbf_1^{s_{i-1}};i-1)+d_{s_i}-B({\dbf'}_1^{s'_{i-1}};i-1)-d'_{s'_i}\nn \\
\aligneq \alpha_i+\beta_{i-1},
\eeqnarray
for $i=1,2,\ldots,k$.

Note that in both \rlemma{adjacent distance larger than one}(i)
and \rlemma{adjacent distance larger than one}(ii),
we have $n'_i=n_i$ for $i=a+2,a+3,\ldots,k$,
and hence it follows from \reqnarray{proof-adjacent distance larger than one-111},
\reqnarray{OQ-LR-delays-greedy-3},
and \reqnarray{proof-adjacent distance larger than one-222} that
\beqnarray{proof-adjacent distance larger than one-444}
\alpha_i
\aligneq d_{s_i}-d'_{s'_i} \nn \\
\aligneq 2d_{s_{i-1}}+(n_i-1)(B(\dbf_1^{s_{i-1}};i-1)+1)
         -2d_{s'_{i-1}}-(n'_i-1)(B({\dbf'}_1^{s'_{i-1}};i-1)+1) \nn \\
\aligneq 2\alpha_{i-1}+(n_i-1)\beta_{i-1},
\eeqnarray
for $i=a+2,a+3,\ldots,k$.

(i) Note that in \rlemma{adjacent distance larger than one}(i),
we have $\nbf_1^k \in \Ncal_{M,k}(1)$,
$n_a-n_{a+1}\leq -2$ for some $1\leq a\leq k-1$,
$n'_a=n_a+1$, $n'_{a+1}=n_{a+1}-1$, and $n'_i=n_i$ for $i\neq a$ and $a+1$.
It is clear that
\beqnarray{proof-adjacent distance larger than one-(i)-111}
n'_a=n_a+1\geq 2,\ n'_{a+1}=n_{a+1}-1\geq n_a+1\geq 2,
\textrm{ and } n'_i=n_i\geq 1 \textrm{ for } i\neq a, a+1.
\eeqnarray
Also, we have from $n'_a=n_a+1$, $n'_{a+1}=n_{a+1}-1$, $n'_i=n_i$ for $i\neq a$ and $a+1$,
$\nbf_1^k \in \Ncal_{M,k}(1)$, and \reqnarray{N-M-k-h} that
\beqnarray{proof-adjacent distance larger than one-(i)-222}
\sum_{i=1}^{k}n'_i
\aligneq \sum_{i\neq a,a+1}n'_i+n'_a+n'_{a+1} \nn\\
\aligneq \sum_{i\neq a,a+1}n_i+(n_a+1)+(n_{a+1}-1) \nn\\
\aligneq \sum_{i=1}^{k}n_i=M=r_{-1}.
\eeqnarray
As such, it follows from \reqnarray{proof-adjacent distance larger than one-(i)-111},
\reqnarray{proof-adjacent distance larger than one-(i)-222},
and \reqnarray{N-M-k-h} that ${\nbf'}_1^k\in \Ncal_{M,k}(1)$.

To show \reqnarray{adjacent distance larger than one-1} with $h=1$,
i.e., $\nbf_1^k\prec {\nbf'}_1^k$,
we see from the definition of the binary relation
$\prec$ in \reqnarray{order relation-666}
that we need to show that $B(\dbf_1^M;k)<B({\dbf'}_1^M;k)$.
It follows from \reqnarray{proof-adjacent distance larger than one-222}
that we need to show that
\beqnarray{proof-adjacent distance larger than one-(i)-333}
\beta_k=B(\dbf_1^{s_k};k)-B({\dbf'}_1^{s_k};k)=B(\dbf_1^M;k)-B({\dbf'}_1^M;k)<0.
\eeqnarray
We discuss the two cases $a=1$ and $2\leq a \leq k-1$ separately.

\emph{Case 1: $a=1$.}
In this case, we have from \reqnarray{proof-adjacent distance larger than one-111},
\reqnarray{proof-adjacent distance larger than one-222},
\reqnarray{OQ-LR-delays-greedy-7}, \reqnarray{OQ-LR-delays-greedy-2}, and $n'_1=n_1+1$ that
\beqnarray{proof-adjacent distance larger than one-(i)-case-1-111}
\alpha_1=\beta_1=d_{s_1}-d'_{s'_1}=s_1-s'_1=n_1-n'_1=-1.
\eeqnarray
From \reqnarray{proof-adjacent distance larger than one-111},
\reqnarray{OQ-LR-delays-greedy-3}, $n'_2=n_2-1$,
\reqnarray{proof-adjacent distance larger than one-222},
\reqnarray{proof-adjacent distance larger than one-(i)-case-1-111},
\reqnarray{OQ-LR-delays-greedy-7}, \reqnarray{OQ-LR-delays-greedy-2},
and $n_1-n_2\leq -2$, we have
\beqnarray{proof-adjacent distance larger than one-(i)-case-1-222}
\alpha_2
\aligneq d_{s_2}-d'_{s'_2} \nn\\
\aligneq 2d_{s_1}+(n_2-1)(B(\dbf_1^{s_{1}};1)+1)-2d'_{s'_1}-(n'_2-1)(B({\dbf'}_1^{s'_{1}};1)+1) \nn\\
\aligneq 2\alpha_1+(n_2-2)\beta_1+(B(\dbf_1^{s_{1}};1)+1) \nn\\
\aligneq -2-(n_2-2)+(d_{s_1}+1) \nn\\
\aligneq n_1-n_2+1 \nn\\
\alignleq -1.
\eeqnarray

As a result of $\beta_1<0$ in \reqnarray{proof-adjacent distance larger than one-(i)-case-1-111},
$\alpha_2<0$ in \reqnarray{proof-adjacent distance larger than one-(i)-case-1-222},
and $n_i\geq 1$ for $i=3,4,\ldots,k$,
we can use \reqnarray{proof-adjacent distance larger than one-333}
and \reqnarray{proof-adjacent distance larger than one-444}
(note that \reqnarray{proof-adjacent distance larger than one-444}
holds for $i=3,4,\ldots,k$ as we have $a=1$ in this case)
repeatedly to show that
$\beta_2<0,\ \alpha_3<0,\ \beta_3<0,\ \alpha_4<0, \ldots,\ \beta_{k-1}<0,\ \alpha_k<0$,
and $\beta_k<0$.

\emph{Case 2: $2\leq a\leq k-1$.}
In this case, we have from $n_i=n'_i$ for $i=1,2,\ldots,a-1$
that $s_i=s'_i$ for $i=1,2,\ldots,a-1$.
Thus, it is easy to see from \reqnarray{OQ-LR-delays-greedy-2}
and \reqnarray{OQ-LR-delays-greedy-3} that
\beqnarray{proof-adjacent distance larger than one-(i)-case-2-111}
d_i=d'_i, \textrm{ for } i=1,2,\ldots,s_{a-1}.
\eeqnarray
It follows from \reqnarray{proof-adjacent distance larger than one-111},
\reqnarray{proof-adjacent distance larger than one-222},
and \reqnarray{proof-adjacent distance larger than one-(i)-case-2-111} that
\beqnarray{proof-adjacent distance larger than one-(i)-case-2-222}
\alpha_i=\beta_i=0, \textrm{ for } i=1,2,\ldots,a-1.
\eeqnarray
From \reqnarray{proof-adjacent distance larger than one-111},
\reqnarray{OQ-LR-delays-greedy-3}, $n'_a=n_a+1$,
\reqnarray{proof-adjacent distance larger than one-222},
and $\alpha_{a-1}=\beta_{a-1}=0$ in \reqnarray{proof-adjacent distance larger than one-(i)-case-2-222},
we have
\beqnarray{proof-adjacent distance larger than one-(i)-case-2-333}
\alpha_a
\aligneq d_{s_a}-d'_{s'_a} \nn \\
\aligneq 2d_{s_{a-1}}+(n_a-1)(B(\dbf_1^{s_{a-1}};a-1)+1)-2d'_{s'_{a-1}}-(n'_a-1)(B({\dbf'}_1^{s'_{a-1}};a-1)+1) \nn\\
\aligneq 2\alpha_{a-1}+n_a\beta_{a-1}-(B(\dbf_1^{s_{a-1}};a-1)+1) \nn\\
\aligneq -(B(\dbf_1^{s_{a-1}};a-1)+1).
\eeqnarray
It then follows from \reqnarray{proof-adjacent distance larger than one-333},
$\beta_{a-1}=0$ in \reqnarray{proof-adjacent distance larger than one-(i)-case-2-222},
and \reqnarray{proof-adjacent distance larger than one-(i)-case-2-333} that
\beqnarray{proof-adjacent distance larger than one-(i)-case-2-444}
\beta_a=\alpha_a+\beta_{a-1}=-(B(\dbf_1^{s_{a-1}};a-1)+1).
\eeqnarray
From \reqnarray{proof-adjacent distance larger than one-111},
\reqnarray{OQ-LR-delays-greedy-3}, $n'_{a+1}=n_{a+1}-1$,
\reqnarray{proof-adjacent distance larger than one-222},
\reqnarray{OQ-LR-delays-greedy-7},
\reqnarray{proof-adjacent distance larger than one-(i)-case-2-333},
\reqnarray{proof-adjacent distance larger than one-(i)-case-2-444},
\reqnarray{OQ-LR-delays-greedy-3}, \reqnarray{OQ-LR-delays-greedy-7},
and $n_a-n_{a+1}\leq -2$, we have
\beqnarray{}
\alpha_{a+1}
\aligneq d_{s_{a+1}}-d'_{s'_{a+1}} \nn \\
\aligneq 2d_{s_a}+(n_{a+1}-1)(B(\dbf_1^{s_{a}};a)+1)-2d'_{s'_a}-(n'_{a+1}-1)(B({\dbf'}_1^{s'_{a}};a)+1) \nn\\
\aligneq 2\alpha_a+(n_{a+1}-2)\beta_a+(B(\dbf_1^{s_{a}};a)+1) \nn\\
\aligneq 2\alpha_a+(n_{a+1}-2)\beta_a+(B(\dbf_1^{s_{a-1}};a-1)+d_{s_a}+1) \nn\\
\aligneq -(n_{a+1}-1)(B(\dbf_1^{s_{a-1}};a-1)+1)+d_{s_a} \nn \\
\aligneq -(n_{a+1}-1)(B(\dbf_1^{s_{a-1}};a-1)+1)+2d_{s_{a-1}}+(n_a-1)(B(\dbf_1^{s_{a-1}};a-1)+1) \nn\\
\aligneq (n_a-n_{a+1})(B(\dbf_1^{s_{a-1}};a-1)+1)+2d_{s_{a-1}} \nn
\eeqnarray
\beqnarray{proof-adjacent distance larger than one-(i)-case-2-555}
\phantom{\alpha_{a+1}}
\alignspace \phantom{-(n_{a+1}-1)(B(\dbf_1^{s_{a-1}};a-1)+1)+2d_{s_{a-1}}+(n_a-1)(B(\dbf_1^{s_{a-1}};a-1)+1)} \nn\\
\aligneq (n_a-n_{a+1})(B(\dbf_1^{s_{a-2}};a-2)+d_{s_{a-1}}+1)+2d_{s_{a-1}}\nn\\
\aligneq (n_a-n_{a+1})(B(\dbf_1^{s_{a-2}};a-2)+1)+(n_a-n_{a+1}+2)d_{s_{a-1}}\nn\\
\alignleq -2(B(\dbf_1^{s_{a-2}};a-2)+1).
\eeqnarray

As a result of $\beta_a<0$ in \reqnarray{proof-adjacent distance larger than one-(i)-case-2-444},
$\alpha_{a+1}<0$ in \reqnarray{proof-adjacent distance larger than one-(i)-case-2-555},
and $n_i\geq 1$ for $i=a+2,a+3,\ldots,k$,
we can use \reqnarray{proof-adjacent distance larger than one-333}
and \reqnarray{proof-adjacent distance larger than one-444}
(note that \reqnarray{proof-adjacent distance larger than one-444} holds for $i=a+2,a+3,\ldots,k$)
repeatedly to show that
$\beta_{a+1}<0,\ \alpha_{a+2}<0,\ \beta_{a+2}<0,\ \alpha_{a+3}<0, \ldots,\ \beta_{k-1}<0,\ \alpha_k<0$,
and $\beta_k<0$.

(ii) Note that in \rlemma{adjacent distance larger than one}(ii),
we have $\nbf_1^k \in \Ncal_{M,k}(1)$,
$n_a-n_{a+1}\geq 2$ for some $1\leq a\leq k-1$,
$n'_a=n_a-1$, $n'_{a+1}=n_{a+1}+1$, and $n'_i=n_i$ for $i\neq a$ and $a+1$.
It is clear that
\beqnarray{proof-adjacent distance larger than one-(ii)-111}
n'_a=n_a-1\geq n_{a+1}+1\geq 2,\ n'_{a+1}=n_{a+1}+1\geq 2,
\textrm{ and } n'_i=n_i\geq 1 \textrm{ for } i\neq a,a+1.
\eeqnarray
Also, we have from $n'_a=n_a-1$, $n'_{a+1}=n_{a+1}+1$, $n'_i=n_i$ for $i\neq a$ and $a+1$,
$\nbf_1^k \in \Ncal_{M,k}(1)$, and \reqnarray{N-M-k-h} that
\beqnarray{proof-adjacent distance larger than one-(ii)-222}
\sum_{i=1}^{k}n'_i
\aligneq \sum_{i\neq a,a+1}n'_i+n'_a+n'_{a+1} \nn\\
\aligneq \sum_{i\neq a,a+1}n_i+(n_a-1)+(n_{a+1}+1) \nn\\
\aligneq \sum_{i=1}^{k}n_i=M=r_{-1}.
\eeqnarray
As such, it follows from \reqnarray{proof-adjacent distance larger than one-(ii)-111},
\reqnarray{proof-adjacent distance larger than one-(ii)-222},
and \reqnarray{N-M-k-h} that ${\nbf'}_1^k\in \Ncal_{M,k}(1)$.

To show \reqnarray{adjacent distance larger than one-2} with $h=1$,
i.e., $\nbf_1^k\preceq {\nbf'}_1^k$,
where $\nbf_1^k\equiv {\nbf'}_1^k$ if and only if $k=2$ and $n_1=n_2+2$,
we see from \reqnarray{order relation-666}
and \reqnarray{proof-adjacent distance larger than one-222}
that we need to show that
\beqnarray{proof-adjacent distance larger than one-(ii)-333}
\beta_k=B(\dbf_1^{s_k};k)-B({\dbf'}_1^{s_k};k)=B(\dbf_1^M;k)-B({\dbf'}_1^M;k) \leq 0,
\eeqnarray
where $\beta_k=0$ if and only if $k=2$ and $n_1=n_2+2$.
We discuss the two cases $a=1$ and $2\leq a \leq k-1$ separately.

\emph{Case 1: $a=1$.}
In this case, we have from \reqnarray{proof-adjacent distance larger than one-111},
\reqnarray{proof-adjacent distance larger than one-222},
\reqnarray{OQ-LR-delays-greedy-7}, \reqnarray{OQ-LR-delays-greedy-2}, and $n'_1=n_1-1$ that
\beqnarray{proof-adjacent distance larger than one-(ii)-case-1-111}
\alpha_1=\beta_1=d_{s_1}-d'_{s'_1}=s_1-s'_1=n_1-n'_1=1.
\eeqnarray
From \reqnarray{proof-adjacent distance larger than one-111},
\reqnarray{OQ-LR-delays-greedy-3}, $n'_2=n_2+1$,
\reqnarray{proof-adjacent distance larger than one-222},
\reqnarray{proof-adjacent distance larger than one-(ii)-case-1-111},
\reqnarray{OQ-LR-delays-greedy-7}, \reqnarray{OQ-LR-delays-greedy-2},
and $n_1-n_2\geq 2$, we have
\beqnarray{}
\alpha_2
\aligneq d_{s_2}-d'_{s'_2} \nn\\
\aligneq 2d_{s_1}+(n_2-1)(B(\dbf_1^{s_{1}};1)+1)-2d'_{s'_1}-(n'_2-1)(B({\dbf'}_1^{s'_{1}};1)+1) \nn\\
\aligneq 2\alpha_1+n_2\beta_1-(B(\dbf_1^{s_{1}};1)+1) \nn\\
\aligneq 2+n_2-(d_{s_1}+1)\nn\\
\aligneq n_2-n_1+1
\label{eqn:proof-adjacent distance larger than one-(ii)-case-1-222}\\
\alignleq -1,
\label{eqn:proof-adjacent distance larger than one-(ii)-case-1-333}
\eeqnarray
where the inequality holds with equality if and only if $n_1=n_2+2$.
It follows from \reqnarray{proof-adjacent distance larger than one-333},
\reqnarray{proof-adjacent distance larger than one-(ii)-case-1-333},
and \reqnarray{proof-adjacent distance larger than one-(ii)-case-1-111} that
\beqnarray{proof-adjacent distance larger than one-(ii)-case-1-444}
\beta_2=\alpha_2+\beta_1\leq -1+1=0,
\eeqnarray
where the inequality holds with equality if and only if $n_1=n_2+2$.
If $k=2$, then we have from \reqnarray{proof-adjacent distance larger than one-(ii)-case-1-444}
that $\beta_k=\beta_2\leq 0$,
where $\beta_k=\beta_2=0$ if and only if $n_1=n_2+2$.
On the other hand, if $k>2$,
then as a result of $\alpha_2<0$ in \reqnarray{proof-adjacent distance larger than one-(ii)-case-1-333},
$\beta_2\leq 0$ in \reqnarray{proof-adjacent distance larger than one-(ii)-case-1-444},
and $n_i\geq 1$ for $i=3,4,\ldots,k$,
we can use \reqnarray{proof-adjacent distance larger than one-444}
(note that \reqnarray{proof-adjacent distance larger than one-444}
holds for $i=3,4,\ldots,k$ as we have $a=1$ in this case)
and \reqnarray{proof-adjacent distance larger than one-333}
repeatedly to show that $\alpha_3<0,\ \beta_3<0,\ \alpha_4<0,\ \beta_4<0, \ldots,\ \alpha_k<0,\ \beta_k<0$.

\emph{Case 2: $2\leq a\leq k-1$.}
As in Case~2 of (i) above,
\reqnarray{proof-adjacent distance larger than one-(i)-case-2-222} also holds in this case.
From \reqnarray{proof-adjacent distance larger than one-111},
\reqnarray{OQ-LR-delays-greedy-3}, $n'_a=n_a-1$,
\reqnarray{proof-adjacent distance larger than one-222},
and $\alpha_{a-1}=\beta_{a-1}=0$ in \reqnarray{proof-adjacent distance larger than one-(i)-case-2-222},
we have
\beqnarray{proof-adjacent distance larger than one-(ii)-case-2-111}
\alpha_a
\aligneq d_{s_a}-d'_{s'_a} \nn \\
\aligneq 2d_{s_{a-1}}+(n_a-1)(B(\dbf_1^{s_{a-1}};a-1)+1)-2d'_{s'_{a-1}}-(n'_a-1)(B({\dbf'}_1^{s'_{a-1}};a-1)+1) \nn\\
\aligneq 2\alpha_{a-1}+(n_a-2)\beta_{a-1}+(B(\dbf_1^{s_{a-1}};a-1)+1) \nn\\
\aligneq B(\dbf_1^{s_{a-1}};a-1)+1.
\eeqnarray
It then follows from \reqnarray{proof-adjacent distance larger than one-333},
$\beta_{a-1}=0$ in \reqnarray{proof-adjacent distance larger than one-(i)-case-2-222},
and \reqnarray{proof-adjacent distance larger than one-(ii)-case-2-111} that
\beqnarray{proof-adjacent distance larger than one-(ii)-case-2-222}
\beta_a=\alpha_a+\beta_{a-1}=B(\dbf_1^{s_{a-1}};a-1)+1.
\eeqnarray

From \reqnarray{proof-adjacent distance larger than one-111},
\reqnarray{OQ-LR-delays-greedy-3}, $n'_{a+1}=n_{a+1}+1$,
\reqnarray{proof-adjacent distance larger than one-222},
\reqnarray{OQ-LR-delays-greedy-7},
\reqnarray{proof-adjacent distance larger than one-(ii)-case-2-111},
\reqnarray{proof-adjacent distance larger than one-(ii)-case-2-222},
\reqnarray{OQ-LR-delays-greedy-3}, \reqnarray{OQ-LR-delays-greedy-7},
and $n_a-n_{a+1}\geq 2$, we have
\beqnarray{}
\alpha_{a+1}
\aligneq d_{s_{a+1}}-d'_{s'_{a+1}} \nn \\
\aligneq 2d_{s_a}+(n_{a+1}-1)(B(\dbf_1^{s_{a}};a)+1)-2d'_{s'_a}-(n'_{a+1}-1)(B({\dbf'}_1^{s'_{a}};a)+1) \nn\\
\aligneq 2\alpha_a+n_{a+1}\beta_a-(B(\dbf_1^{s_{a}};a)+1) \nn\\
\aligneq 2\alpha_a+n_{a+1}\beta_a-(B(\dbf_1^{s_{a-1}};a-1)+d_{s_a}+1) \nn\\
\aligneq (n_{a+1}+1)(B(\dbf_1^{s_{a-1}};a-1)+1)-d_{s_a}\nn\\
\aligneq (n_{a+1}+1)(B(\dbf_1^{s_{a-1}};a-1)+1)-2d_{s_{a-1}}-(n_a-1)(B(\dbf_1^{s_{a-1}};a-1)+1) \nn\\
\aligneq (n_{a+1}-n_a+2)(B(\dbf_1^{s_{a-1}};a-1)+1)-2d_{s_{a-1}} \nn\\
\aligneq (n_{a+1}-n_a+2)(B(\dbf_1^{s_{a-2}};a-2)+d_{s_{a-1}}+1)-2d_{s_{a-1}}\nn\\
\aligneq (n_{a+1}-n_a+2)(B(\dbf_1^{s_{a-2}};a-2)+1)+(n_{a+1}-n_a)d_{s_{a-1}}
\label{eqn:proof-adjacent distance larger than one-(ii)-case-2-333}\\
\alignleq -2d_{s_{a-1}}.
\label{eqn:proof-adjacent distance larger than one-(ii)-case-2-444}
\eeqnarray
Also, from \reqnarray{proof-adjacent distance larger than one-333},
\reqnarray{proof-adjacent distance larger than one-(ii)-case-2-333},
\reqnarray{proof-adjacent distance larger than one-(ii)-case-2-222},
\reqnarray{OQ-LR-delays-greedy-7},
$n_a-n_{a+1}\geq 2$, and \rlemma{d>B+1}, we have
\beqnarray{}
\beta_{a+1}
\aligneq \alpha_{a+1}+\beta_a \nn\\
\aligneq (n_{a+1}-n_a+2)(B(\dbf_1^{s_{a-2}};a-2)+1)+(n_{a+1}-n_a)d_{s_{a-1}}+B(\dbf_1^{s_{a-1}};a-1)+1 \nn\\
\aligneq (n_{a+1}-n_a+2)(B(\dbf_1^{s_{a-2}};a-2)+1)+(n_{a+1}-n_a)d_{s_{a-1}}+B(\dbf_1^{s_{a-2}};a-2)+d_{s_{a-1}}+1 \nn
\eeqnarray
\beqnarray{}
\phantom{\beta_{a+1}}
\alignspace \phantom{(n_{a+1}-n_a+2)(B(\dbf_1^{s_{a-2}};a-2)+1)+(n_{a+1}-n_a)d_{s_{a-1}}+B(\dbf_1^{s_{a-2}};a-2)+d_{s_{a-1}}+1} \nn\\
\aligneq (n_{a+1}-n_a+3)(B(\dbf_1^{s_{a-2}};a-2)+1)+(n_{a+1}-n_a+1)d_{s_{a-1}}
\label{eqn:proof-adjacent distance larger than one-(ii)-case-2-555}\\
\alignleq B(\dbf_1^{s_{a-2}};a-2)+1-d_{s_{a-1}} \nn\\
\alignless 0.
\label{eqn:proof-adjacent distance larger than one-(ii)-case-2-666}
\eeqnarray

As a result of $\alpha_{a+1}<0$ in \reqnarray{proof-adjacent distance larger than one-(ii)-case-2-444},
$\beta_{a+1}<0$ in \reqnarray{proof-adjacent distance larger than one-(ii)-case-2-666},
and $n_i\geq 1$ for $i=a+2,a+3,\ldots,k$,
we can use \reqnarray{proof-adjacent distance larger than one-444}
(note that \reqnarray{proof-adjacent distance larger than one-444} holds for $i=a+2,a+3,\ldots,k$)
and \reqnarray{proof-adjacent distance larger than one-333} repeatedly to show that
$\alpha_{a+2}<0,\ \beta_{a+2}<0,\ \alpha_{a+3}<0,\ \beta_{a+3}<0,\ldots,\ \alpha_k<0,\ \beta_k<0$.
\eproof

\bappendix{Proof of Comparison rule A in \rlemma{comparison rule A} with $h=1$}
{proof of comparison rule A with h=1}

In this appendix, we prove Comparison rule A in \rlemma{comparison rule A} for the case that $h=1$.

Suppose that $h=1$ in Comparison rule A in \rlemma{comparison rule A}.
For simplicity, let $\nbf_1^k=\nbf_1^{r_0}(1)$ and ${\nbf'}_1^k={\nbf'}_1^{r_0}(1)$
(note that $r_0=k$).
Let $\dbf_1^M$ and ${\dbf'}_1^M$ be obtained by using
$\nbf_1^k$ and ${\nbf'}_1^k$, respectively, in \reqnarray{OQ-LR-delays-greedy-1}.
Let $s_0=0$ and $s_i=\sum_{\ell=1}^{i}n_{\ell}$ for $i=1,2,\ldots,k$,
and let $s'_0=0$ and $s'_i=\sum_{\ell=1}^{i}n'_{\ell}$ for $i=1,2,\ldots,k$.
Also let $\alpha_i=d_{s_i}-d'_{s'_i}$ for $i=1,2,\ldots,k$
as in \reqnarray{proof-adjacent distance larger than one-111},
and $\beta_i=B(\dbf_1^{s_i};i)-B({\dbf'}_1^{s'_i};i)$ for $i=0,1,2,\ldots,k$
as in \reqnarray{proof-adjacent distance larger than one-222}.

Note that in \rlemma{comparison rule A},
we have $\nbf_1^k \in \Ncal_{M,k}(1)$,
$n_a-n_{a+1}=1$ for some $1\leq a\leq k-1$,
$n_1\geq 3$ in the case that $a=1$,
$n'_a=n_a-1$, $n'_{a+1}=n_{a+1}+1$, and $n'_i=n_i$ for $i\neq a$ and $a+1$.
As ${\nbf'}_1^k$ is obtained from $\nbf_1^k$ in exactly the same way as that
in \rlemma{adjacent distance larger than one}(ii),
it is clear that \reqnarray{proof-adjacent distance larger than one-333}--\reqnarray{proof-adjacent distance larger than one-444},
\reqnarray{proof-adjacent distance larger than one-(ii)-222}--\reqnarray{proof-adjacent distance larger than one-(ii)-case-1-222},
\reqnarray{proof-adjacent distance larger than one-(ii)-case-2-111}--\reqnarray{proof-adjacent distance larger than one-(ii)-case-2-333},
and \reqnarray{proof-adjacent distance larger than one-(ii)-case-2-555}
in the proof of \rlemma{adjacent distance larger than one}(ii) still hold.
It is also clear that
\beqnarray{proof-comparison rule A-111}
n'_a=n_a-1=n_{a+1}\geq 1,\ n'_{a+1}=n_{a+1}+1\geq 2,
\textrm{ and } n'_i=n_i\geq 1,\ i\neq a, a+1.
\eeqnarray
In the case that $a=1$, we have $n_1\geq 3$ and it follows that
\beqnarray{proof-comparison rule A-222}
n'_a=n_a-1=n_1-1\geq 3-1=2.
\eeqnarray
As such, it follows from \reqnarray{proof-comparison rule A-111},
\reqnarray{proof-comparison rule A-222},
\reqnarray{proof-adjacent distance larger than one-(ii)-222},
and \reqnarray{N-M-k-h} that ${\nbf'}_1^k\in \Ncal_{M,k}(1)$.

(i) Note that in \rlemma{comparison rule A}(i), we have $a=1$ or $a=k-1$.
To show \reqnarray{comparison rule A-1} with $h=1$,
i.e., $\nbf_1^k\succ {\nbf'}_1^k$,
we see from \reqnarray{order relation-666}
and \reqnarray{proof-adjacent distance larger than one-222}
that we need to show that $\beta_k>0$.

For the case that $a=1$,
we have from \reqnarray{proof-adjacent distance larger than one-(ii)-case-1-222} and $n_1-n_2=1$ that
\beqnarray{proof-comparison rule A-(i)-111}
\alpha_2=n_2-n_1+1=0.
\eeqnarray
It then follows from \reqnarray{proof-adjacent distance larger than one-333},
\reqnarray{proof-comparison rule A-(i)-111},
and \reqnarray{proof-adjacent distance larger than one-(ii)-case-1-111} that
\beqnarray{proof-comparison rule A-(i)-222}
\beta_2=\alpha_2+\beta_1=0+1=1.
\eeqnarray
As a result of $\alpha_2=0$ in \reqnarray{proof-comparison rule A-(i)-111},
$\beta_2>0$ in \reqnarray{proof-comparison rule A-(i)-222},
and $n_i\geq 1$ for $i=3,4,\ldots,k$,
we can use \reqnarray{proof-adjacent distance larger than one-444}
(note that \reqnarray{proof-adjacent distance larger than one-444}
holds for $i=3,4,\ldots,k$ as we have $a=1$ in this case)
and \reqnarray{proof-adjacent distance larger than one-333} repeatedly to show that
$\alpha_3\geq 0,\ \beta_3>0,\ \alpha_4\geq 0,\ \beta_4>0, \ldots,\ \alpha_k\geq 0,\ \beta_k>0$.

For the case that $a=k-1\geq 2$ (note that the case $a=1$ has just been discussed),
we have from \reqnarray{proof-adjacent distance larger than one-(ii)-case-2-555} and $n_{k-1}-n_k=1$ that
\beqnarray{}
\beta_k=2(B(\dbf_1^{s_{k-3}};k-3)+1)>0. \nn
\eeqnarray

(ii) Note that in \rlemma{comparison rule A}(ii),
we have $2\leq a\leq r_{h-1}-2$ and there exists a positive integer $j$ such that
$1\leq j\leq \min\{a-1,r_{h-1}-a-1\}$,
$n_{a-j'}=n_{a+1+j'}$ for $j'=1,2,\ldots,j-1$, and $n_{a-j}\neq n_{a+1+j}$.

First we use $n_{a-j'}=n_{a+1+j'}$ for $j'=1,2,\ldots,j-1$ to show that
\beqnarray{}
\alpha_{a+j'}\aligneq 2^{j'-1}(B(\dbf_1^{s_{a-j'-1}};a-j'-1)+1-d_{s_{a-j'}}),
\label{eqn:proof-comparison rule A-(ii)-111}\\
\beta_{a+j'}\aligneq 2^{j'}(B(\dbf_1^{s_{a-j'-1}};a-j'-1)+1),
\label{eqn:proof-comparison rule A-(ii)-222}
\eeqnarray
for $j'=1,2,\ldots,j$.
We prove \reqnarray{proof-comparison rule A-(ii)-111} and \reqnarray{proof-comparison rule A-(ii)-222}
by induction on $j'$.
From \reqnarray{proof-adjacent distance larger than one-(ii)-case-2-333},
\reqnarray{proof-adjacent distance larger than one-(ii)-case-2-555},
and $n_a-n_{a+1}=1$, we have
\beqnarray{}
\alpha_{a+1}\aligneq B(\dbf_1^{s_{a-2}};a-2)+1-d_{s_{a-1}},
\label{eqn:proof-comparison rule A-(ii)-333}\\
\beta_{a+1}\aligneq 2(B(\dbf_1^{s_{a-2}};a-2)+1).
\label{eqn:proof-comparison rule A-(ii)-444}
\eeqnarray
It follows from
\reqnarray{proof-comparison rule A-(ii)-333} and \reqnarray{proof-comparison rule A-(ii)-444}
that \reqnarray{proof-comparison rule A-(ii)-111} and \reqnarray{proof-comparison rule A-(ii)-222}
hold for $j'=1$.

Assume as the induction hypothesis that \reqnarray{proof-comparison rule A-(ii)-111}
and \reqnarray{proof-comparison rule A-(ii)-222} hold for some $1\leq j'\leq j-1$.
From \reqnarray{proof-adjacent distance larger than one-444},
the induction hypothesis, \reqnarray{OQ-LR-delays-greedy-3} (note that $a-j'-1\geq a-j\geq 1$),
$n_{a-j'}=n_{a+1+j'}$ (as $1\leq j'\leq j-1$), and \reqnarray{OQ-LR-delays-greedy-7}, we have
\beqnarray{proof-comparison rule A-(ii)-555}
\alignspace \hspace*{-0.3in} \alpha_{a+(j'+1)} \nn\\
\aligneq 2\alpha_{a+j'}+(n_{a+1+j'}-1)\beta_{a+j'} \nn \\
\aligneq 2\cdot 2^{j'-1}(B(\dbf_1^{s_{a-j'-1}};a-j'-1)+1-d_{s_{a-j'}}) \nn\\
\alignspace +(n_{a+1+j'}-1)\cdot 2^{j'}(B(\dbf_1^{s_{a-j'-1}};a-j'-1)+1) \nn \\
\aligneq 2^{j'}\left(B(\dbf_1^{s_{a-j'-1}};a-j'-1)+1-2d_{s_{a-j'-1}}
         -(n_{a-j'}-1)(B(\dbf_1^{s_{a-j'-1}};a-j'-1)+1)\right)\nn\\
\alignspace +2^{j'}(n_{a+1+j'}-1)(B(\dbf_1^{s_{a-j'-1}};a-j'-1)+1) \nn \\
\aligneq 2^{j'}(B(\dbf_1^{s_{a-j'-1}};a-j'-1)+1-2d_{s_{a-j'-1}})\nn \\
\aligneq 2^{j'}(B(\dbf_1^{s_{a-j'-2}};a-j'-2)+d_{s_{a-j'-1}}+1-2d_{s_{a-j'-1}})\nn \\
\aligneq 2^{j'}(B(\dbf_1^{s_{a-j'-2}};a-j'-2)+1-d_{s_{a-j'-1}}).
\eeqnarray
From \reqnarray{proof-adjacent distance larger than one-333},
\reqnarray{proof-comparison rule A-(ii)-555}, the induction hypothesis,
and \reqnarray{OQ-LR-delays-greedy-7}, we have
\beqnarray{proof-comparison rule A-(ii)-666}
\alignspace \hspace*{-0.3in} \beta_{a+(j'+1)} \nn\\
\aligneq \alpha_{a+j'+1}+\beta_{a+j'} \nn\\
\aligneq 2^{j'}(B(\dbf_1^{s_{a-j'-2}};a-j'-2)+1-d_{s_{a-j'-1}})+2^{j'}(B(\dbf_1^{s_{a-j'-1}};a-j'-1)+1) \nn\\
\aligneq 2^{j'}(B(\dbf_1^{s_{a-j'-2}};a-j'-2)+1-d_{s_{a-j'-1}})+2^{j'}(B(\dbf_1^{s_{a-j'-2}};a-j'-2)+d_{s_{a-j'-1}}+1) \nn\\
\aligneq 2^{j'+1}(B(\dbf_1^{s_{a-j'-2}};a-j'-2)+1).
\eeqnarray
The induction is completed by combining \reqnarray{proof-comparison rule A-(ii)-555}
and \reqnarray{proof-comparison rule A-(ii)-666}.

Now we show that if $n_{a-j}<n_{a+1+j}$,
then \reqnarray{comparison rule A-2} holds with $h=1$,
i.e., $\beta_k>0$;
on the other hand, if $n_{a-j}>n_{a+1+j}$,
then \reqnarray{comparison rule A-3} holds with $h=1$,
i.e., $\beta_k\leq 0$,
where $\beta_k=0$ if and only if $a-j=1$, $a+1+j=k$, and $n_1=n_k+1$.
Note that as $j\leq \min\{a-1,k-a-1\}$, we have $j+1\leq a\leq k-j-1$.
We consider the two cases $a=j+1$ and $j+2\leq a\leq k-j-1$ separately.

\emph{Case 1: $a=j+1$.}
In this case, we have from \reqnarray{proof-comparison rule A-(ii)-222} (with $j'=j$),
$a=j+1$, and $B(\dbf_1^{s_0};0)=0$ that
\beqnarray{proof-comparison rule A-(ii)-case-1-111}
\beta_{a+j}=2^j(B(\dbf_1^{s_{a-j-1}};a-j-1)+1)=2^j(B(\dbf_1^{s_0};0)+1)=2^j.
\eeqnarray
We also have from \reqnarray{proof-adjacent distance larger than one-444},
\reqnarray{proof-comparison rule A-(ii)-111} (with $j'=j$),
\reqnarray{proof-comparison rule A-(ii)-case-1-111}, $a=j+1$,
$B(\dbf_1^{s_0};0)=0$, and \reqnarray{OQ-LR-delays-greedy-2} that
\beqnarray{proof-comparison rule A-(ii)-case-1-222}
\alpha_{a+j+1}
\aligneq 2\alpha_{a+j}+(n_{a+1+j}-1)\beta_{a+j} \nn \\
\aligneq 2\cdot 2^{j-1}(B(\dbf_1^{s_{a-j-1}};a-j-1)+1-d_{s_{a-j}})+(n_{a+1+j}-1)\cdot 2^j \nn \\
\aligneq 2^j(B(\dbf_1^{s_0};0)+1-s_1)+2^j(n_{a+1+j}-1) \nn\\
\aligneq 2^j(1-n_1)+2^j(n_{a+1+j}-1) \nn\\
\aligneq 2^j(n_{a+1+j}-n_{a-j}).
\eeqnarray

If $n_{a-j}<n_{a+1+j}$,
then it follows from \reqnarray{proof-comparison rule A-(ii)-case-1-222}
that $\alpha_{a+j+1}>0$.
As a result of $\beta_{a+j}>0$ in \reqnarray{proof-comparison rule A-(ii)-case-1-111},
$\alpha_{a+j+1}>0$, and $n_i\geq 1$ for $i=a+j+2,a+j+3,\ldots,k$,
we can use \reqnarray{proof-adjacent distance larger than one-333}
and \reqnarray{proof-adjacent distance larger than one-444}
(note that \reqnarray{proof-adjacent distance larger than one-444} holds for $i=a+2,a+3,\ldots,k$)
repeatedly to show that $\beta_{a+j+1}>0,\ \alpha_{a+j+2}>0,\ \beta_{a+j+2}>0,\ \alpha_{a+j+3}>0,
\ldots,\ \beta_{k-1}>0,\ \alpha_k>0$, and $\beta_k>0$.

On the other hand, if $n_{a-j}>n_{a+1+j}$, then it follows from \reqnarray{proof-comparison rule A-(ii)-case-1-222}
that $\alpha_{a+j+1}<0$.
From \reqnarray{proof-adjacent distance larger than one-333},
\reqnarray{proof-comparison rule A-(ii)-case-1-222}, \reqnarray{proof-comparison rule A-(ii)-case-1-111},
and $n_{a-j}>n_{a+1+j}$, we have
\beqnarray{}
\beta_{a+j+1}
\aligneq \alpha_{a+j+1}+\beta_{a+j} \nn\\
\aligneq 2^j(n_{a+1+j}-n_{a-j}+1)
\label{eqn:proof-comparison rule A-(ii)-case-1-333}\\
\alignleq 0,
\label{eqn:proof-comparison rule A-(ii)-case-1-444}
\eeqnarray
where the inequality holds with equality if and only if $n_{a-j}=n_{a+1+j}+1$.
For the case that $a=k-j-1$,
we see from \reqnarray{proof-comparison rule A-(ii)-case-1-444} that $\beta_k=\beta_{a+j+1}\leq 0$,
where $\beta_k=\beta_{a+j+1}=0$ if and only if $n_{a-j}=n_{a+1+j}+1$,
i.e., $n_1=n_k+1$ (as we have $a=j+1$ and $a=k-j-1$ in this case).
For the case that $a<k-j-1$,
we see that as a result of $\alpha_{a+j+1}<0$,
$\beta_{a+j+1}\leq 0$ in \reqnarray{proof-comparison rule A-(ii)-case-1-444},
and $n_i\geq 1$ for $i=a+j+2,a+j+3,\ldots,k$,
we can use \reqnarray{proof-adjacent distance larger than one-444}
(note that \reqnarray{proof-adjacent distance larger than one-444} holds for $i=a+2,a+3,\ldots,k$)
and \reqnarray{proof-adjacent distance larger than one-333} repeatedly
to show that $\alpha_{a+j+2}<0,\ \beta_{a+j+2}<0,\ \alpha_{a+j+3}<0,\ \beta_{a+j+3}<0,
\ldots,\ \alpha_k<0,\ \beta_k<0$.

\emph{Case 2: $j+2\leq a\leq k-j-1$.}
In this case, we have from \reqnarray{proof-comparison rule A-(ii)-222} (with $j'=j$) that
\beqnarray{proof-comparison rule A-(ii)-case-2-111}
\beta_{a+j}=2^j(B(\dbf_1^{s_{a-j-1}};a-j-1)+1).
\eeqnarray
We also have from \reqnarray{proof-adjacent distance larger than one-444},
\reqnarray{proof-comparison rule A-(ii)-111} (with $j'=j$),
\reqnarray{proof-comparison rule A-(ii)-case-2-111},
\reqnarray{OQ-LR-delays-greedy-3}, and \reqnarray{OQ-LR-delays-greedy-7} that
\beqnarray{proof-comparison rule A-(ii)-case-2-222}
\alignspace \hspace*{-0.3in} \alpha_{a+j+1} \nn \\
\aligneq 2\alpha_{a+j}+(n_{a+1+j}-1)\beta_{a+j} \nn \\
\aligneq 2\cdot 2^{j-1}(B(\dbf_1^{s_{a-j-1}};a-j-1)+1-d_{s_{a-j}})
         +(n_{a+1+j}-1)\cdot 2^j(B(\dbf_1^{s_{a-j-1}};a-j-1)+1) \nn \\
\aligneq 2^j\Bigl(B(\dbf_1^{s_{a-j-1}};a-j-1)+1-2d_{s_{a-j-1}}-(n_{a-j}-1)(B(\dbf_1^{s_{a-j-1}};a-j-1)+1)\Bigr)\nn\\
\alignspace +2^j(n_{a+1+j}-1)(B(\dbf_1^{s_{a-j-1}};a-j-1)+1) \nn \\
\aligneq 2^j\Bigl((n_{a+1+j}-n_{a-j}+1)(B(\dbf_1^{s_{a-j-1}};a-j-1)+1)-2d_{s_{a-j-1}}\Bigr)\nn\\
\aligneq 2^j\Bigl((n_{a+1+j}-n_{a-j}+1)(B(\dbf_1^{s_{a-j-2}};a-j-2)+d_{s_{a-j-1}}+1)-2d_{s_{a-j-1}}\Bigr)\nn\\
\aligneq 2^j\Bigl((n_{a+1+j}-n_{a-j}+1)(B(\dbf_1^{s_{a-j-2}};a-j-2)+1)+(n_{a+1+j}-n_{a-j}-1)d_{s_{a-j-1}}\Bigr).
\eeqnarray

If $n_{a-j}<n_{a+1+j}$, then it follows from \reqnarray{proof-comparison rule A-(ii)-case-2-222} that
\beqnarray{proof-comparison rule A-(ii)-case-2-333}
\alpha_{a+j+1} \geq 2^j\cdot 2(B(\dbf_1^{s_{a-j-2}};a-j-2)+1)>0.
\eeqnarray
As we have $\beta_{a+j}>0$ in \reqnarray{proof-comparison rule A-(ii)-case-2-111},
$\alpha_{a+j+1}>0$ in \reqnarray{proof-comparison rule A-(ii)-case-2-333},
and $n_i\geq 1$ for $i=a+j+2,a+j+3,\ldots,k$,
we can use \reqnarray{proof-adjacent distance larger than one-333}
and \reqnarray{proof-adjacent distance larger than one-444}
(note that \reqnarray{proof-adjacent distance larger than one-444} holds for $i=a+2,a+3,\ldots,k$)
repeatedly to show that $\beta_{a+j+1}>0,\ \alpha_{a+j+2}>0,\ \beta_{a+j+2}>0,\ \alpha_{a+j+3}>0,
\ldots,\ \beta_{k-1}>0,\ \alpha_k>0$, and $\beta_k>0$.

On the other hand, if $n_{a-j}>n_{a+1+j}$, then it follows from \reqnarray{proof-comparison rule A-(ii)-case-2-222} that
\beqnarray{proof-comparison rule A-(ii)-case-2-444}
\alpha_{a+j+1} \leq 2^j(-2d_{s_{a-j-1}})< 0.
\eeqnarray
From \reqnarray{proof-adjacent distance larger than one-333},
\reqnarray{proof-comparison rule A-(ii)-case-2-222}, \reqnarray{proof-comparison rule A-(ii)-case-2-111},
\reqnarray{OQ-LR-delays-greedy-7}, $n_{a-j}>n_{a+1+j}$,
and \rlemma{d>B+1} (note that $0\leq a-j-2\leq k-2j-3\leq k-1$),
we have
\beqnarray{proof-comparison rule A-(ii)-case-2-555}
\alignspace \hspace*{-0.3in} \beta_{a+j+1} \nn\\
\aligneq \alpha_{a+j+1}+\beta_{a+j} \nn\\
\aligneq 2^j\Bigl((n_{a+1+j}-n_{a-j}+1)(B(\dbf_1^{s_{a-j-2}};a-j-2)+1)+(n_{a+1+j}-n_{a-j}-1)d_{s_{a-j-1}}\Bigr) \nn\\
\alignspace + 2^j(B(\dbf_1^{s_{a-j-1}};a-j-1)+1) \nn\\
\aligneq 2^j\Bigl((n_{a+1+j}-n_{a-j}+1)(B(\dbf_1^{s_{a-j-2}};a-j-2)+1)+(n_{a+1+j}-n_{a-j}-1)d_{s_{a-j-1}}\Bigr) \nn\\
\alignspace + 2^j(B(\dbf_1^{s_{a-j-2}};a-j-2)+d_{s_{a-j-1}}+1) \nn\\
\aligneq 2^j\Bigl((n_{a+1+j}-n_{a-j}+2)(B(\dbf_1^{s_{a-j-2}};a-j-2)+1)+(n_{a+1+j}-n_{a-j})d_{s_{a-j-1}}\Bigr) \nn\\
\alignleq 2^j(B(\dbf_1^{s_{a-j-2}};a-j-2)+1-d_{s_{a-j-1}}) \nn\\
\alignless 0.
\eeqnarray
As a result of $\alpha_{a+j+1}<0$ in \reqnarray{proof-comparison rule A-(ii)-case-2-444},
$\beta_{a+j+1}< 0$ in \reqnarray{proof-comparison rule A-(ii)-case-2-555},
and $n_i\geq 1$ for $i=a+j+2,a+j+3,\ldots,k$,
we can use \reqnarray{proof-adjacent distance larger than one-444}
(note that \reqnarray{proof-adjacent distance larger than one-444} holds for $i=a+2,a+3,\ldots,k$)
and \reqnarray{proof-adjacent distance larger than one-333} repeatedly
to show that $\alpha_{a+j+2}<0,\ \beta_{a+j+2}<0,\ \alpha_{a+j+3}<0,\ \beta_{a+j+3}<0,
\ldots,\ \alpha_k<0,\ \beta_k<0$.

(iii) Note that in \rlemma{comparison rule A}(iii),
we have $2\leq a\leq k-2$ and
$n_{a-j'}=n_{a+1+j'}$ for $j'=1,2,\ldots,\min\{a-1,k-a-1\}$.
To show \reqnarray{comparison rule A-4} with $h=1$,
i.e., $\nbf_1^k\succ {\nbf'}_1^k$,
we see from \reqnarray{order relation-666}
and \reqnarray{proof-adjacent distance larger than one-222}
that we need to show that $\beta_k>0$.

If $a-1\leq k-a-1$, then $\min\{a-1,k-a-1\}=a-1$
and we have $n_{a-j'}=n_{a+1+j'}$ for $j'=1,2,\ldots,a-1$.
From $n_{a-j'}=n_{a+1+j'}$, $j'=1,2,\ldots,a-1$,
we can show as in (ii) above that
\reqnarray{proof-comparison rule A-(ii)-111} and \reqnarray{proof-comparison rule A-(ii)-222}
hold for $j'=1,2,\ldots,a$,
and \reqnarray{proof-comparison rule A-(ii)-case-1-222} and \reqnarray{proof-comparison rule A-(ii)-case-1-333}
hold for $j=a-1$.
As $n_{a-j'}=n_{a+1+j'}$ for $j'=a-1$, we have $n_1=n_{2a}$.
It then follows from \reqnarray{proof-comparison rule A-(ii)-case-1-222} (with $j=a-1$),
\reqnarray{proof-comparison rule A-(ii)-case-1-333} (with $j=a-1$),
and $n_1=n_{2a}$ that
\beqnarray{proof-comparison rule A-(iii)-111}
\alpha_{2a}=0 \textrm{ and } \beta_{2a}=2^{a-1}>0.
\eeqnarray
As a result of \reqnarray{proof-comparison rule A-(iii)-111}
and $n_i\geq 1$ for $i=2a+1,2a+2,\ldots,k$,
we can use \reqnarray{proof-adjacent distance larger than one-444}
(note that \reqnarray{proof-adjacent distance larger than one-444}
holds for $i=a+2,a+3,\ldots,k$ and $2a+1\geq a+2$)
\reqnarray{proof-adjacent distance larger than one-333} repeatedly
to show that $\alpha_{2a+1}\geq 0,\ \beta_{2a+1}>0,\ \alpha_{2a+2}\geq 0,\ \beta_{2a+2}>0,
\ldots,\ \alpha_k\geq 0,\ \beta_k>0$.

On the other hand, if $a-1>k-a-1$,
then $\min\{a-1,k-a-1\}=k-a-1$
and we have $n_{a-j'}=n_{a+1+j'}$ for $j'=1,2,\ldots,k-a-1$.
We can show as in (ii) above that
\reqnarray{proof-comparison rule A-(ii)-111} and \reqnarray{proof-comparison rule A-(ii)-222}
hold for $j'=1,2,\ldots,k-a$.
It then follows from \reqnarray{proof-comparison rule A-(ii)-222} (with $j'=k-a$)
and $2a-k-1\geq 0$ (as $a-1>k-a-1$) that
\beqnarray{}
\beta_k=2^{k-a}(B(\dbf_1^{s_{2a-k-1}};2a-k-1)+1)>0. \nn
\eeqnarray

\bappendix{Proof of \rlemma{adjacent distance larger than one II} for an even integer $2\leq h\leq N$
by using Comparison rule A in \rlemma{comparison rule A} for the odd integer $h-1$}
{proof of adjacent distance larger than one II for an even integer h
by using comparison rule A for the odd integer h-1}

In this appendix, we assume that Comparison rule A in \rlemma{comparison rule A} holds
for some odd integer $h-1$, where $1\leq h-1\leq N-1$,
and show that \rlemma{adjacent distance larger than one II} holds for the even integer $h$.

Let
\beqnarray{}
\nbf_1^{r_{h-2}}(h-1)\aligneq L_{r_{h-3},r_{h-2}}(\nbf_1^{r_{h-1}}(h)),
\label{eqn:proof of adjacent distance larger than one II-111} \\
{\nbf'}_1^{r_{h-2}}(h-1)\aligneq L_{r_{h-3},r_{h-2}}({\nbf'}_1^{r_{h-1}}(h)).
\label{eqn:proof of adjacent distance larger than one II-222}
\eeqnarray
For simplicity, let $\mbf_1^{r_{h-1}}=\nbf_1^{r_{h-1}}(h)$,
${\mbf'}_1^{r_{h-1}}={\nbf'}_1^{r_{h-1}}(h)$,
$\nbf_1^{r_{h-2}}=\nbf_1^{r_{h-2}}(h-1)$,
and ${\nbf'}_1^{r_{h-2}}={\nbf'}_1^{r_{h-2}}(h-1)$.
Then we have from \reqnarray{proof of adjacent distance larger than one II-111},
\reqnarray{proof of adjacent distance larger than one II-222},
and \reqnarray{order relation-777} that
\beqnarray{proof of adjacent distance larger than one II-333}
\mbf_1^{r_{h-1}}\prec
(\textrm{resp.}, \equiv, \succ, \preceq, \succeq)\ {\mbf'}_1^{r_{h-1}}
\textrm{ iff } \nbf_1^{r_{h-2}}\prec
(\textrm{resp.}, \equiv, \succ, \preceq, \succeq)\ {\nbf'}_1^{r_{h-2}}.
\eeqnarray
Furthermore, from \reqnarray{proof of adjacent distance larger than one II-111},
\reqnarray{proof of adjacent distance larger than one II-222},
and the definition of left pre-sequences in \rdefinition{left pre-sequences},
we have
\beqnarray{proof of adjacent distance larger than one II-444}
n_i=
\bselection
q_{h-1}+1, &\textrm{if } i=i_1,i_2,\ldots,i_{r_{h-1}}, \\
q_{h-1}, &\textrm{otherwise},
\eselection
\eeqnarray
where
\beqnarray{proof of adjacent distance larger than one II-555}
i_j=\sum_{\ell=1}^{j-1}m_{\ell}+1, \textrm{ for } j=1,2,\ldots,r_{h-1},
\eeqnarray
and
\beqnarray{proof of adjacent distance larger than one II-666}
n'_i=
\bselection
q_{h-1}+1, &\textrm{if } i=i'_1,i'_2,\ldots,i'_{r_{h-1}}, \\
q_{h-1}, &\textrm{otherwise},
\eselection
\eeqnarray
where
\beqnarray{proof of adjacent distance larger than one II-777}
i'_j=\sum_{\ell=1}^{j-1}m'_{\ell}+1, \textrm{ for } j=1,2,\ldots,r_{h-1}.
\eeqnarray
Note that in \rlemma{adjacent distance larger than one II}, we have $r_{h-1}\geq 2$.
As such, it follows from $r_{h-2}>r_{h-1}$ that
\beqnarray{proof of adjacent distance larger than one II-888}
r_{h-2}\geq 2.
\eeqnarray

(i) Note that in \rlemma{adjacent distance larger than one II}(i),
we have $\mbf_1^{r_{h-1}}\in \Ncal_{M,k}(h)$,
$m_a-m_{a+1}\geq 2$ for some $1\leq a\leq r_{h-1}-1$,
$m'_a=m_a-1$, $m'_{a+1}=m_{a+1}+1$, and $m'_i=m_i$ for $i\neq a$ and $a+1$.
It is easy to see that
\beqnarray{proof of adjacent distance larger than one II-(i)-111}
m'_a=m_a-1\geq m_{a+1}+1\geq 2,\ m'_{a+1}=m_{a+1}+1\geq 2,
\textrm{ and } m'_i=m_i \textrm{ for } i\neq a, a+1.
\eeqnarray
Also, we have from $m'_a=m_a-1$, $m'_{a+1}=m_{a+1}+1$, and $m'_i=m_i$ for $i\neq a$ and $a+1$,
$\mbf_1^{r_{h-1}}\in \Ncal_{M,k}(h)$, and \reqnarray{N-M-k-h} that
\beqnarray{proof of adjacent distance larger than one II-(i)-222}
\sum_{i=1}^{r_{h-1}}m'_i=\sum_{i=1}^{r_{h-1}}m_i=r_{h-2}.
\eeqnarray
As such, it follows from \reqnarray{proof of adjacent distance larger than one II-(i)-111},
\reqnarray{proof of adjacent distance larger than one II-(i)-222},
and \reqnarray{N-M-k-h} that ${\mbf'}_1^{r_{h-1}}\in \Ncal_{M,k}(h)$.

As $\mbf_1^{r_{h-1}}\in \Ncal_{M,k}(h)$ and ${\mbf'}_1^{r_{h-1}}\in \Ncal_{M,k}(h)$,
we see from \reqnarray{proof of adjacent distance larger than one II-111},
\reqnarray{proof of adjacent distance larger than one II-222},
and the argument in the paragraph after \reqnarray{N-M-k-h} that
\beqnarray{proof of adjacent distance larger than one II-(i)-333}
\nbf_1^{r_{h-2}}\in \Ncal_{M,k}(h-1) \textrm{ and } {\nbf'}_1^{r_{h-2}}\in \Ncal_{M,k}(h-1).
\eeqnarray
To show \reqnarray{adjacent distance larger than one II-1},
i.e., $\mbf_1^{r_{h-1}}\prec{\mbf'}_1^{r_{h-1}}$,
we see from \reqnarray{proof of adjacent distance larger than one II-333}
that it suffices to show that
\beqnarray{proof of adjacent distance larger than one II-(i)-444}
\nbf_1^{r_{h-2}}\prec{\nbf'}_1^{r_{h-2}}.
\eeqnarray

\bpdffigure{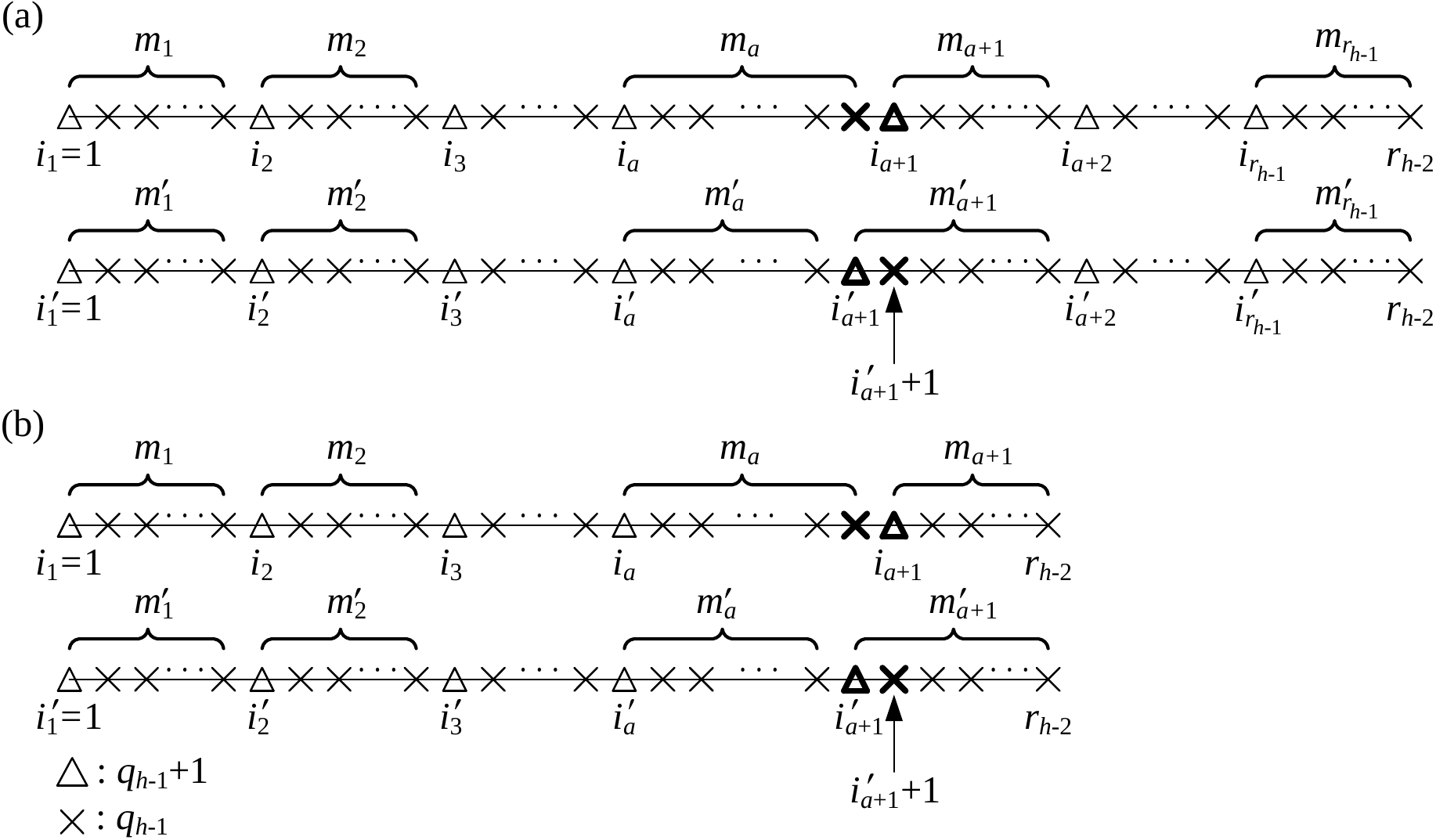}{5.5in}
\epdffigure{appendix-C-(i)}
{An illustration of
\reqnarray{proof of adjacent distance larger than one II-(i)-666}--\reqnarray{proof of adjacent distance larger than one II-(i)-bbb}:
(a) $1\leq a\leq r_{h-1}-2$
(note that in this case we have $i_a<i'_{a+1}<i_{a+1}$
in \reqnarray{proof of adjacent distance larger than one II-(i)-ccc}
and $i'_{a+1}<i'_{a+1}+1<i'_{a+2}$
in \reqnarray{proof of adjacent distance larger than one II-(i)-ddd});
(b) $a=r_{h-1}-1$
(note that in this case we have $i_a<i'_{a+1}<i_{a+1}$
in \reqnarray{proof of adjacent distance larger than one II-(i)-ccc}
and $i'_{r_{h-1}}<i'_{a+1}+1\leq r_{h-2}$
in \reqnarray{proof of adjacent distance larger than one II-(i)-eee}
and \reqnarray{proof of adjacent distance larger than one II-(i)-fff}).}

Note that from $m'_a=m_a-1$, $m'_{a+1}=m_{a+1}+1$, $m'_i=m_i$ for $i\neq a$ and $a+1$,
\reqnarray{proof of adjacent distance larger than one II-555},
and \reqnarray{proof of adjacent distance larger than one II-777},
it is easy to see that
\beqnarray{proof of adjacent distance larger than one II-(i)-555}
i'_j=
\bselection
i_j-1, &\textrm{if } j=a+1, \\
i_j, &\textrm{otherwise}.
\eselection
\eeqnarray
In the following, we show that
\beqnarray{}
\alignspace n_{i'_{a+1}}=q_{h-1},
\label{eqn:proof of adjacent distance larger than one II-(i)-666}\\
\alignspace n'_{i'_{a+1}+1}=q_{h-1}.
\label{eqn:proof of adjacent distance larger than one II-(i)-777}
\eeqnarray
It then follows from \reqnarray{proof of adjacent distance larger than one II-444},
\reqnarray{proof of adjacent distance larger than one II-666},
\reqnarray{proof of adjacent distance larger than one II-(i)-555},
\reqnarray{proof of adjacent distance larger than one II-(i)-666},
and \reqnarray{proof of adjacent distance larger than one II-(i)-777} that
\beqnarray{}
\alignspace n'_{i'_{a+1}}-n'_{i'_{a+1}+1}=(q_{h-1}+1)-q_{h-1}=1,
\label{eqn:proof of adjacent distance larger than one II-(i)-888}\\
\alignspace n_{i'_{a+1}}=q_{h-1}=n'_{i'_{a+1}}-1,
\label{eqn:proof of adjacent distance larger than one II-(i)-999}\\
\alignspace n_{i'_{a+1}+1}=n_{i_{a+1}}=q_{h-1}+1=n'_{i'_{a+1}+1}+1,
\label{eqn:proof of adjacent distance larger than one II-(i)-aaa}\\
\alignspace n_i=n'_i, \textrm{ for } i\neq i'_{a+1} \textrm{ and } i'_{a+1}+1.
\label{eqn:proof of adjacent distance larger than one II-(i)-bbb}
\eeqnarray
An illustration of \reqnarray{proof of adjacent distance larger than one II-(i)-666}--\reqnarray{proof of adjacent distance larger than one II-(i)-bbb}
is given in \rfigure{appendix-C-(i)}.

To prove \reqnarray{proof of adjacent distance larger than one II-(i)-666},
note from \reqnarray{proof of adjacent distance larger than one II-(i)-555} that
\beqnarray{proof of adjacent distance larger than one II-(i)-ccc}
i'_{a+1}>i'_a=i_a \textrm{ and } i'_{a+1}=i_{a+1}-1<i_{a+1}.
\eeqnarray
Thus, \reqnarray{proof of adjacent distance larger than one II-(i)-666}
follows from \reqnarray{proof of adjacent distance larger than one II-444}
and $i_a<i'_{a+1}<i_{a+1}$ in \reqnarray{proof of adjacent distance larger than one II-(i)-ccc}.

To prove \reqnarray{proof of adjacent distance larger than one II-(i)-777},
note that if $1\leq a\leq r_{h-1}-2$,
then we have from \reqnarray{proof of adjacent distance larger than one II-(i)-555} that
\beqnarray{proof of adjacent distance larger than one II-(i)-ddd}
i'_{a+1}+1=i_{a+1}<i_{a+2}=i'_{a+2}.
\eeqnarray
Thus, \reqnarray{proof of adjacent distance larger than one II-(i)-777}
follows from \reqnarray{proof of adjacent distance larger than one II-666}
and $i'_{a+1}<i'_{a+1}+1<i'_{a+2}$ in \reqnarray{proof of adjacent distance larger than one II-(i)-ddd}.
On the other hand, if $a=r_{h-1}-1$, then we have
\beqnarray{proof of adjacent distance larger than one II-(i)-eee}
i'_{a+1}+1>i'_{a+1}=i'_{r_{h-1}},
\eeqnarray
and we have from \reqnarray{proof of adjacent distance larger than one II-(i)-555} that
\beqnarray{proof of adjacent distance larger than one II-(i)-fff}
i'_{a+1}+1=i_{a+1}=i_{r_{h-1}}\leq r_{h-2}.
\eeqnarray
Thus, \reqnarray{proof of adjacent distance larger than one II-(i)-777}
also follows from \reqnarray{proof of adjacent distance larger than one II-666}
and $i'_{r_{h-1}}<i'_{a+1}+1\leq r_{h-2}$
in \reqnarray{proof of adjacent distance larger than one II-(i)-eee}
and \reqnarray{proof of adjacent distance larger than one II-(i)-fff}.

Note that from \reqnarray{proof of adjacent distance larger than one II-777}, we have
\beqnarray{proof of adjacent distance larger than one II-(i)-ggg}
i'_{a+1}=\sum_{\ell=1}^{a}m'_{\ell}+1\geq m'_a+1 \geq 2,
\eeqnarray
and
\beqnarray{proof of adjacent distance larger than one II-(i)-hhh}
i'_{j+1}=\sum_{\ell=1}^{j}m'_{\ell}+1
=\left(\sum_{\ell=1}^{j-1}m'_{\ell}+1\right)+m'_j
=i'_j+m'_j, \textrm{ for } j=1,2,\ldots,r_{h-1}-1.
\eeqnarray
We then consider the two cases $1\leq a\leq r_{h-1}-2$ and $a=r_{h-1}-1$ separately.

\emph{Case 1: $1\leq a\leq r_{h-1}-2$.}
From \reqnarray{proof of adjacent distance larger than one II-777},
\reqnarray{proof of adjacent distance larger than one II-(i)-222},
$a+1\leq r_{h-1}-1$,
and $m'_{a+1}\geq 2$ in \reqnarray{proof of adjacent distance larger than one II-(i)-111},
we have
\beqnarray{proof of adjacent distance larger than one II-(i)-case-1-111}
i'_{a+1}
\aligneq \sum_{\ell=1}^{a}m'_{\ell}+1=\sum_{\ell=1}^{r_{h-1}}m'_{\ell}-\sum_{\ell=a+1}^{r_{h-1}}m'_{\ell}+1 \nn\\
\alignleq r_{h-2}-(m'_{a+1}+m'_{r_{h-1}})+1\nn\\
\alignleq r_{h-2}-(2+1)+1 \nn\\
\aligneq r_{h-2}-2.
\eeqnarray
As such, we see from \reqnarray{proof of adjacent distance larger than one II-(i)-ggg}
and \reqnarray{proof of adjacent distance larger than one II-(i)-case-1-111} that
\beqnarray{proof of adjacent distance larger than one II-(i)-case-1-222}
2\leq i'_{a+1}\leq r_{h-2}-2.
\eeqnarray

\bpdffigure{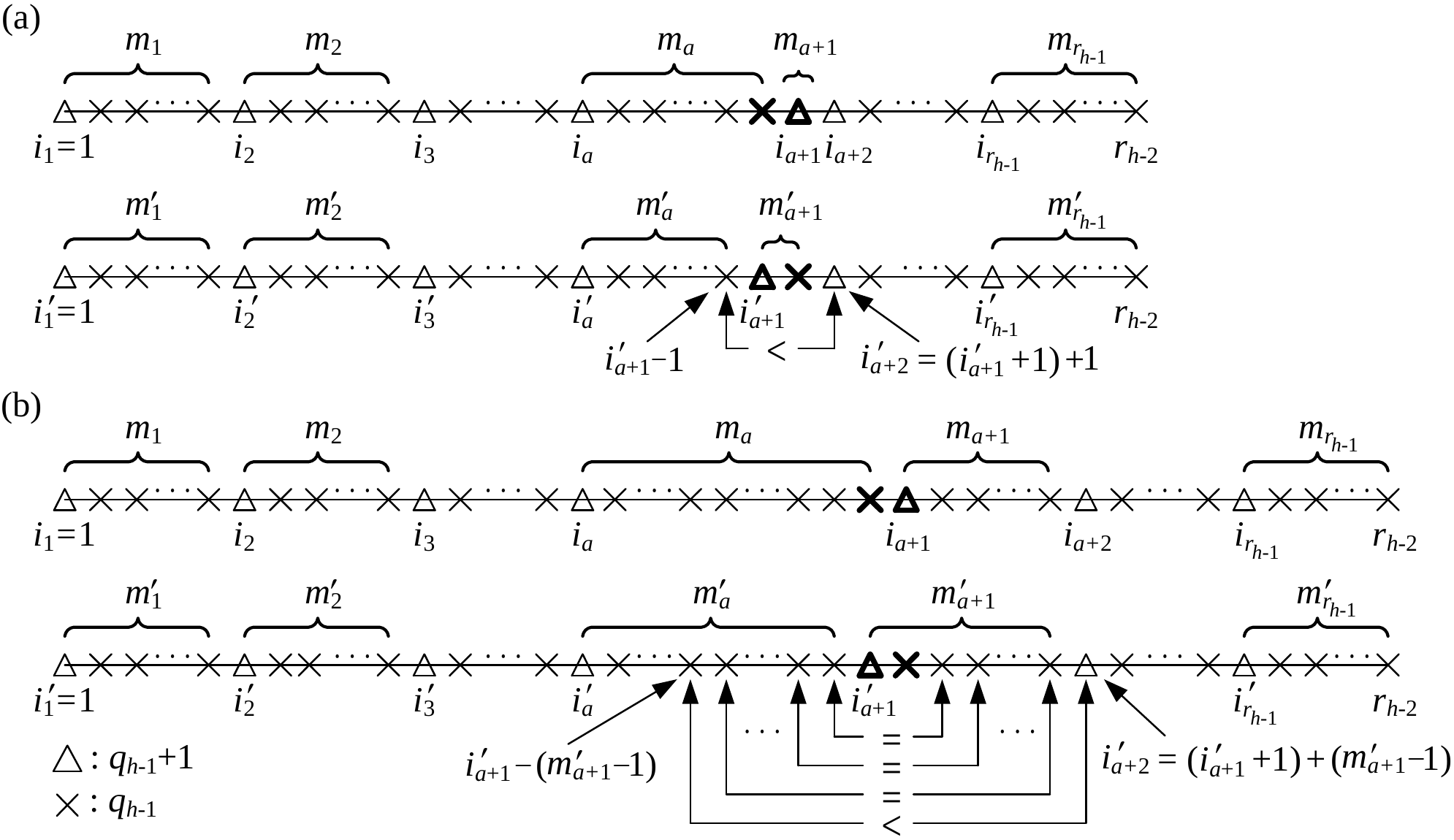}{5.5in}
\epdffigure{appendix-C-(i)-case-1}
{(a) An illustration of \reqnarray{proof of adjacent distance larger than one II-(i)-case-1-333}
in the case that $1\leq a\leq r_{h-1}-2$ and $m_{a+1}=1$
(note that in this case we have $i'_a<i'_{a+1}-1<i'_{a+1}$
in \reqnarray{proof of adjacent distance larger than one II-(i)-case-1-444}
and $(i'_{a+1}+1)+1=i'_{a+2}$
in \reqnarray{proof of adjacent distance larger than one II-(i)-case-1-666});
(b) An illustration of \reqnarray{proof of adjacent distance larger than one II-(i)-case-1-999}
and \reqnarray{proof of adjacent distance larger than one II-(i)-case-1-aaa}
in the case that $1\leq a\leq r_{h-1}-2$ and $m_{a+1}\geq 2$
(note that in this case we have $i'_a<i'_{a+1}-(m'_{a+1}-1)<i'_{a+1}$
in \reqnarray{proof of adjacent distance larger than one II-(i)-case-1-ddd}
and $(i'_{a+1}+1)+(m'_{a+1}-1)=i'_{a+2}$
in \reqnarray{proof of adjacent distance larger than one II-(i)-case-1-hhh}).}

If $m_{a+1}=1$, then we show that
\beqnarray{proof of adjacent distance larger than one II-(i)-case-1-333}
n'_{i'_{a+1}-1}=q_{h-1}<n'_{(i'_{a+1}+1)+1}=q_{h-1}+1.
\eeqnarray
An illustration of \reqnarray{proof of adjacent distance larger than one II-(i)-case-1-333}
is given in \rfigure{appendix-C-(i)-case-1}(a).
Therefore, it follows from
\reqnarray{proof of adjacent distance larger than one II-888},
${\nbf'}_1^{r_{h-2}}\in \Ncal_{M,k}(h-1)$ in
\reqnarray{proof of adjacent distance larger than one II-(i)-333},
\reqnarray{proof of adjacent distance larger than one II-(i)-888}--\reqnarray{proof of adjacent distance larger than one II-(i)-bbb},
\reqnarray{proof of adjacent distance larger than one II-(i)-case-1-222},
\reqnarray{proof of adjacent distance larger than one II-(i)-case-1-333},
and \reqnarray{comparison rule A-2} in \rlemma{comparison rule A}(ii)
(for the odd integer $h-1$) that ${\nbf'}_1^{r_{h-2}}\succ\nbf_1^{r_{h-2}}$,
i.e., \reqnarray{proof of adjacent distance larger than one II-(i)-444} holds.

To prove \reqnarray{proof of adjacent distance larger than one II-(i)-case-1-333},
note that from \reqnarray{proof of adjacent distance larger than one II-(i)-hhh}, $1\leq a\leq r_{h-1}-2$,
and $m'_a\geq 2$ in \reqnarray{proof of adjacent distance larger than one II-(i)-111}, we shave
\beqnarray{proof of adjacent distance larger than one II-(i)-case-1-444}
i'_{a+1}-1=(i'_a+m'_a)-1\geq i'_a+1.
\eeqnarray
It follows from \reqnarray{proof of adjacent distance larger than one II-666}
and $i'_a<i'_{a+1}-1<i'_{a+1}$ in
\reqnarray{proof of adjacent distance larger than one II-(i)-case-1-444} that
\beqnarray{proof of adjacent distance larger than one II-(i)-case-1-555}
n'_{i'_{a+1}-1}=q_{h-1}.
\eeqnarray
Also, from $m'_{a+1}=m_{a+1}+1=1+1=2$,
\reqnarray{proof of adjacent distance larger than one II-(i)-hhh},
and $1\leq a\leq r_{h-1}-2$, we have
\beqnarray{proof of adjacent distance larger than one II-(i)-case-1-666}
(i'_{a+1}+1)+1=i'_{a+1}+2=i'_{a+1}+m'_{a+1}=i'_{a+2}.
\eeqnarray
It follows from \reqnarray{proof of adjacent distance larger than one II-(i)-case-1-666},
\reqnarray{proof of adjacent distance larger than one II-666},
and $1\leq a\leq r_{h-1}-2$ that
\beqnarray{proof of adjacent distance larger than one II-(i)-case-1-777}
n'_{(i'_{a+1}+1)+1}=n'_{i'_{a+2}}=q_{h-1}+1.
\eeqnarray
Thus, \reqnarray{proof of adjacent distance larger than one II-(i)-case-1-333}
follows from \reqnarray{proof of adjacent distance larger than one II-(i)-case-1-555}
and \reqnarray{proof of adjacent distance larger than one II-(i)-case-1-777}.

On the other hand, if $m_{a+1}\geq 2$,
then we show that
\beqnarray{}
\alignspace 1\leq m'_{a+1}-1\leq \min\{i'_{a+1}-1,r_{h-2}-i'_{a+1}-1\},
\label{eqn:proof of adjacent distance larger than one II-(i)-case-1-888}\\
\alignspace n'_{i'_{a+1}-j'}=n'_{(i'_{a+1}+1)+j'}=q_{h-1}, \textrm{ for } j'=1,2,\ldots,m'_{a+1}-2,
\label{eqn:proof of adjacent distance larger than one II-(i)-case-1-999}\\
\alignspace n'_{i'_{a+1}-(m'_{a+1}-1)}=q_{h-1}<n'_{(i'_{a+1}+1)+(m'_{a+1}-1)}=q_{h-1}+1,
\label{eqn:proof of adjacent distance larger than one II-(i)-case-1-aaa}
\eeqnarray
where we note that $m'_{a+1}-2=(m_{a+1}+1)-2\geq 1$ (as $m_{a+1}\geq 2$).
An illustration of \reqnarray{proof of adjacent distance larger than one II-(i)-case-1-999}
and \reqnarray{proof of adjacent distance larger than one II-(i)-case-1-aaa}
is given in \rfigure{appendix-C-(i)-case-1}(b).
Therefore, it follows from
\reqnarray{proof of adjacent distance larger than one II-888},
${\nbf'}_1^{r_{h-2}}\in \Ncal_{M,k}(h-1)$ in
\reqnarray{proof of adjacent distance larger than one II-(i)-333},
\reqnarray{proof of adjacent distance larger than one II-(i)-888}--\reqnarray{proof of adjacent distance larger than one II-(i)-bbb},
\reqnarray{proof of adjacent distance larger than one II-(i)-case-1-222},
\reqnarray{proof of adjacent distance larger than one II-(i)-case-1-888}--\reqnarray{proof of adjacent distance larger than one II-(i)-case-1-aaa},
and \reqnarray{comparison rule A-2} in \rlemma{comparison rule A}(ii)
(for the odd integer $h-1$) that ${\nbf'}_1^{r_{h-2}}\succ\nbf_1^{r_{h-2}}$,
i.e., \reqnarray{proof of adjacent distance larger than one II-(i)-444} holds.

To prove \reqnarray{proof of adjacent distance larger than one II-(i)-case-1-888}--\reqnarray{proof of adjacent distance larger than one II-(i)-case-1-aaa},
observe from $m'_{a+1}=m_{a+1}+1$ that
\beqnarray{proof of adjacent distance larger than one II-(i)-case-1-bbb}
m'_{a+1}-1=m_{a+1}\geq 1.
\eeqnarray
From \reqnarray{proof of adjacent distance larger than one II-(i)-hhh},
$1\leq a\leq r_{h-1}-2$, $m'_a=m_a-1$, $m'_{a+1}=m_{a+1}+1$,
and $m_a-m_{a+1}\geq 2$, we have
\beqnarray{}
i'_{a+1}-(m'_{a+1}-1)
\aligneq i'_a+m'_a-m'_{a+1}+1 \nn\\
\aligneq i'_a+(m_a-1)-(m_{a+1}+1)+1 \nn\\
\aligneq i'_a+m_a-m_{a+1}-1
\label{eqn:proof of adjacent distance larger than one II-(i)-case-1-ccc} \\
\aligngeq i'_a+1.
\label{eqn:proof of adjacent distance larger than one II-(i)-case-1-ddd}
\eeqnarray
We immediately see from \reqnarray{proof of adjacent distance larger than one II-(i)-case-1-ddd} that
\beqnarray{}
\alignspace m'_{a+1}-1\leq i'_{a+1}-i'_a-1<i'_{a+1}-1,
\label{eqn:proof of adjacent distance larger than one II-(i)-case-1-eee}\\
\alignspace
i'_a<i'_{a+1}-j'<i'_{a+1}, \textrm{ for } j'=1,2,\ldots,m'_{a+1}-1.
\label{eqn:proof of adjacent distance larger than one II-(i)-case-1-fff}
\eeqnarray
It is clear from \reqnarray{proof of adjacent distance larger than one II-666}
and \reqnarray{proof of adjacent distance larger than one II-(i)-case-1-fff} that
\beqnarray{proof of adjacent distance larger than one II-(i)-case-1-ggg}
n'_{i'_{a+1}-j'}=q_{h-1}, \textrm{ for } j'=1,2,\ldots,m'_{a+1}-1.
\eeqnarray
Furthermore, we have from \reqnarray{proof of adjacent distance larger than one II-(i)-hhh}
and $1\leq a\leq r_{h-1}-2$ that
\beqnarray{proof of adjacent distance larger than one II-(i)-case-1-hhh}
(i'_{a+1}+1)+(m'_{a+1}-1)=i'_{a+1}+m'_{a+1}=i'_{a+2}.
\eeqnarray
We immediately see from \reqnarray{proof of adjacent distance larger than one II-(i)-case-1-hhh}
and $i'_{a+2}\leq r_{h-2}$ that
\beqnarray{proof of adjacent distance larger than one II-(i)-case-1-iii}
m'_{a+1}-1=i'_{a+2}-i'_{a+1}-1\leq r_{h-2}-i'_{a+1}-1.
\eeqnarray
Thus, \reqnarray{proof of adjacent distance larger than one II-(i)-case-1-888}
follows from \reqnarray{proof of adjacent distance larger than one II-(i)-case-1-bbb},
\reqnarray{proof of adjacent distance larger than one II-(i)-case-1-eee},
and \reqnarray{proof of adjacent distance larger than one II-(i)-case-1-iii}.
Also, it is clear from \reqnarray{proof of adjacent distance larger than one II-666}
and \reqnarray{proof of adjacent distance larger than one II-(i)-case-1-hhh} that
\beqnarray{}
\alignspace n'_{(i'_{a+1}+1)+j'}=q_{h-1}, \textrm{ for } j'=1,2,\ldots,m'_{a+1}-2,
\label{eqn:proof of adjacent distance larger than one II-(i)-case-1-jjj} \\
\alignspace n'_{(i'_{a+1}+1)+(m'_{a+1}-1)}=n'_{i'_{a+2}}=q_{h-1}+1.
\label{eqn:proof of adjacent distance larger than one II-(i)-case-1-kkk}
\eeqnarray
Thus, \reqnarray{proof of adjacent distance larger than one II-(i)-case-1-999}
and \reqnarray{proof of adjacent distance larger than one II-(i)-case-1-aaa}
follow from \reqnarray{proof of adjacent distance larger than one II-(i)-case-1-ggg},
\reqnarray{proof of adjacent distance larger than one II-(i)-case-1-jjj},
and \reqnarray{proof of adjacent distance larger than one II-(i)-case-1-kkk}.

\emph{Case 2: $a=r_{h-1}-1$.}

\bpdffigure{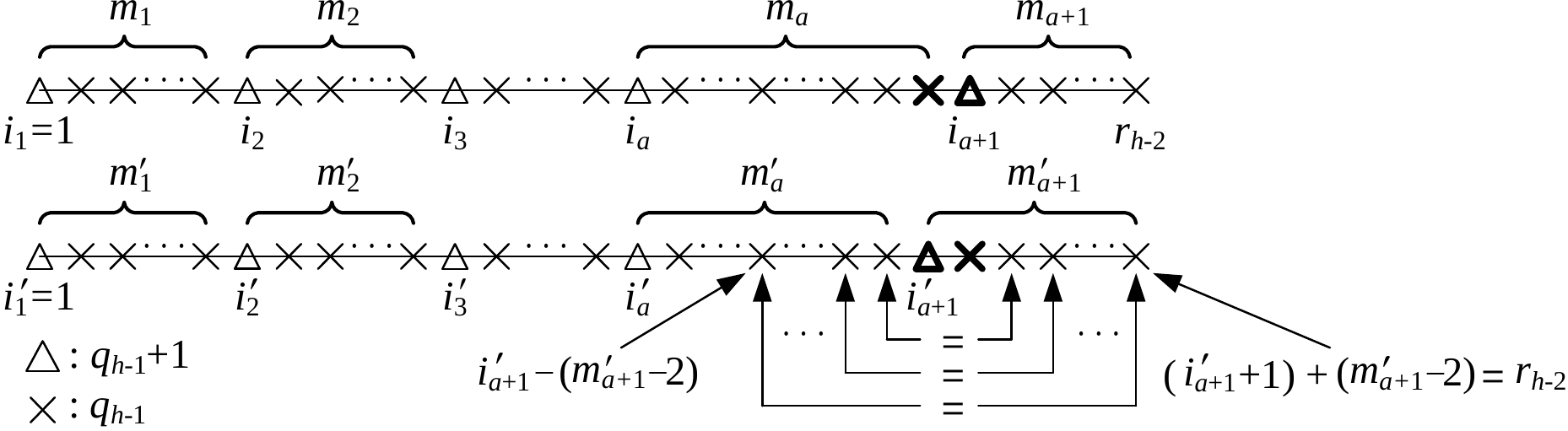}{5.5in}
\epdffigure{appendix-C-(i)-case-2}
{An illustration of \reqnarray{proof of adjacent distance larger than one II-(i)-case-2-444}
in the case that $a=r_{h-1}-1$ and $m_{a+1}\geq 2$
(note that in this case we have
$\min\{i'_{a+1}-1,r_{h-2}-i'_{a+1}-1\}=m'_{a+1}-2$
in \reqnarray{proof of adjacent distance larger than one II-(i)-case-2-777},
$i'_a<i'_{a+1}-(m'_{a+1}-2)<i'_{a+1}$
in \reqnarray{proof of adjacent distance larger than one II-(i)-case-2-999},
and $(i'_{a+1}+1)+(m'_{a+1}-2)=r_{h-2}$
in \reqnarray{proof of adjacent distance larger than one II-(i)-case-2-ccc}).}

If $m_{a+1}=1$, then we have from $a=r_{h-1}-1$,
\reqnarray{proof of adjacent distance larger than one II-777},
\reqnarray{proof of adjacent distance larger than one II-(i)-222},
and $m'_{a+1}=m_{a+1}+1=1+1=2$ that
\beqnarray{}
i'_{a+1}
\aligneq i'_{r_{h-1}}=\sum_{\ell=1}^{r_{h-1}-1}m'_{\ell}+1 \nn\\
\aligneq \sum_{\ell=1}^{r_{h-1}}m'_{\ell}-m'_{r_{h-1}}+1 \nn\\
\aligneq r_{h-2}-m'_{a+1}+1
\label{eqn:proof of adjacent distance larger than one II-(i)-case-2-111} \\
\aligneq r_{h-2}-1.
\label{eqn:proof of adjacent distance larger than one II-(i)-case-2-222}
\eeqnarray
Therefore, it follows from
\reqnarray{proof of adjacent distance larger than one II-888},
${\nbf'}_1^{r_{h-2}}\in \Ncal_{M,k}(h-1)$ in
\reqnarray{proof of adjacent distance larger than one II-(i)-333},
\reqnarray{proof of adjacent distance larger than one II-(i)-888}--\reqnarray{proof of adjacent distance larger than one II-(i)-bbb},
\reqnarray{proof of adjacent distance larger than one II-(i)-case-2-222},
and \reqnarray{comparison rule A-1} in \rlemma{comparison rule A}(i)
(for the odd integer $h-1$) that ${\nbf'}_1^{r_{h-2}}\succ\nbf_1^{r_{h-2}}$,
i.e., \reqnarray{proof of adjacent distance larger than one II-(i)-444} holds.

On the other hand, if $m_{a+1}\geq 2$,
then we show that
\beqnarray{}
\alignspace 2\leq i'_{a+1}\leq r_{h-2}-2,
\label{eqn:proof of adjacent distance larger than one II-(i)-case-2-333}\\
\alignspace n'_{i'_{a+1}-j'}=n'_{(i'_{a+1}+1)+j'}=q_{h-1},
\textrm{ for } j'=1,2,\ldots,\min\{i'_{a+1}-1,r_{h-2}-i'_{a+1}-1\}.
\label{eqn:proof of adjacent distance larger than one II-(i)-case-2-444}
\eeqnarray
An illustration of \reqnarray{proof of adjacent distance larger than one II-(i)-case-2-444}
is given in \rfigure{appendix-C-(i)-case-2}.
Therefore, it follows from
\reqnarray{proof of adjacent distance larger than one II-888},
${\nbf'}_1^{r_{h-2}}\in \Ncal_{M,k}(h-1)$ in
\reqnarray{proof of adjacent distance larger than one II-(i)-333},
\reqnarray{proof of adjacent distance larger than one II-(i)-888}--\reqnarray{proof of adjacent distance larger than one II-(i)-bbb},
\reqnarray{proof of adjacent distance larger than one II-(i)-case-2-333},
\reqnarray{proof of adjacent distance larger than one II-(i)-case-2-444},
and \reqnarray{comparison rule A-4} in \rlemma{comparison rule A}(iii)
(for the odd integer $h-1$) that ${\nbf'}_1^{r_{h-2}}\succ\nbf_1^{r_{h-2}}$,
i.e., \reqnarray{proof of adjacent distance larger than one II-(i)-444} holds.

To prove \reqnarray{proof of adjacent distance larger than one II-(i)-case-2-333},
note that from \reqnarray{proof of adjacent distance larger than one II-(i)-case-2-111}
and $m'_{a+1}=m_{a+1}+1\geq 2+1=3$, we have
\beqnarray{proof of adjacent distance larger than one II-(i)-case-2-555}
i'_{a+1}=r_{h-2}-m'_{a+1}+1\leq r_{h-2}-2.
\eeqnarray
Thus, \reqnarray{proof of adjacent distance larger than one II-(i)-case-2-333}
follows from \reqnarray{proof of adjacent distance larger than one II-(i)-ggg}
and \reqnarray{proof of adjacent distance larger than one II-(i)-case-2-555}.

To prove \reqnarray{proof of adjacent distance larger than one II-(i)-case-2-444},
note that from \reqnarray{proof of adjacent distance larger than one II-(i)-case-2-111},
\reqnarray{proof of adjacent distance larger than one II-777},
$m'_a=m_a-1$, $m'_{a+1}=m_{a+1}+1$, and $m_a-m_{a+1}\geq 2$, we have
\beqnarray{proof of adjacent distance larger than one II-(i)-case-2-666}
(i'_{a+1}-1)-(r_{h-2}-i'_{a+1}-1)
\aligneq (i'_{a+1}-1)-(m'_{a+1}-2)=\sum_{\ell=1}^{a}m'_{\ell}-m'_{a+1}+2\nn\\
\aligngeq m'_a-m'_{a+1}+2=(m_a-1)-(m_{a+1}+1)+2 \nn\\
\aligneq m_a-m_{a+1}>0.
\eeqnarray
It then follows from \reqnarray{proof of adjacent distance larger than one II-(i)-case-2-666}
and \reqnarray{proof of adjacent distance larger than one II-(i)-case-2-111} that
\beqnarray{proof of adjacent distance larger than one II-(i)-case-2-777}
\min\{i'_{a+1}-1,r_{h-2}-i'_{a+1}-1\}=r_{h-2}-i'_{a+1}-1=m'_{a+1}-2.
\eeqnarray
From \reqnarray{proof of adjacent distance larger than one II-(i)-case-1-ccc}
and $m_a-m_{a+1}\geq 2$, we have
\beqnarray{proof of adjacent distance larger than one II-(i)-case-2-888}
i'_{a+1}-(m'_{a+1}-2)=i'_a+m_a-m_{a+1}>i'_a.
\eeqnarray
From \reqnarray{proof of adjacent distance larger than one II-(i)-case-2-888},
we immediately see that
\beqnarray{proof of adjacent distance larger than one II-(i)-case-2-999}
i'_a<i'_{a+1}-j'<i'_{a+1}, \textrm{ for } j'=1,2,\ldots,m'_{a+1}-2.
\eeqnarray
It then follows from \reqnarray{proof of adjacent distance larger than one II-666}
and \reqnarray{proof of adjacent distance larger than one II-(i)-case-2-999} that
\beqnarray{proof of adjacent distance larger than one II-(i)-case-2-aaa}
n'_{i'_{a+1}-j'}=q_{h-1}, \textrm{ for } j'=1,2,\ldots,m'_{a+1}-2.
\eeqnarray
Also, we have from $a=r_{h-1}-1$ that
\beqnarray{proof of adjacent distance larger than one II-(i)-case-2-bbb}
i'_{a+1}+1=i'_{r_{h-1}}+1>i'_{r_{h-1}},
\eeqnarray
and we have from \reqnarray{proof of adjacent distance larger than one II-(i)-case-2-111} that
\beqnarray{proof of adjacent distance larger than one II-(i)-case-2-ccc}
(i'_{a+1}+1)+(m'_{a+1}-2)=r_{h-2}.
\eeqnarray
It then follows from \reqnarray{proof of adjacent distance larger than one II-666},
\reqnarray{proof of adjacent distance larger than one II-(i)-case-2-bbb},
and \reqnarray{proof of adjacent distance larger than one II-(i)-case-2-ccc} that
\beqnarray{proof of adjacent distance larger than one II-(i)-case-2-ddd}
\alignspace n'_{(i'_{a+1}+1)+j'}=q_{h-1}, \textrm{ for } j=1,2,\ldots,m'_{a+1}-2.
\eeqnarray
Thus, \reqnarray{proof of adjacent distance larger than one II-(i)-case-2-444}
follows from \reqnarray{proof of adjacent distance larger than one II-(i)-case-2-777},
\reqnarray{proof of adjacent distance larger than one II-(i)-case-2-aaa},
and \reqnarray{proof of adjacent distance larger than one II-(i)-case-2-ddd}.

(ii) Note that in \rlemma{adjacent distance larger than one II}(ii),
we have $\mbf_1^{r_{h-1}}\in \Ncal_{M,k}(h)$,
$m_a-m_{a+1}\leq -2$ for some $1\leq a\leq r_{h-1}-1$,
$m'_a=m_a+1$, $m'_{a+1}=m_{a+1}-1$, and $m'_i=m_i$ for $i\neq a$ and $a+1$.
It is easy to see that
\beqnarray{proof of adjacent distance larger than one II-(ii)-111}
m'_a=m_a+1\geq 2,\ m'_{a+1}=m_{a+1}-1\geq m_a+1\geq 2,
\textrm{ and } m'_i=m_i \textrm{ for } i\neq a, a+1.
\eeqnarray
Also, we have from $m'_a=m_a+1$, $m'_{a+1}=m_{a+1}-1$, and $m'_i=m_i$ for $i\neq a$ and $a+1$,
$\mbf_1^{r_{h-1}}\in \Ncal_{M,k}(h)$, and \reqnarray{N-M-k-h} that
\beqnarray{proof of adjacent distance larger than one II-(ii)-222}
\sum_{i=1}^{r_{h-1}}m'_i=\sum_{i=1}^{r_{h-1}}m_i=r_{h-2}.
\eeqnarray
As such, it follows from \reqnarray{proof of adjacent distance larger than one II-(ii)-111},
\reqnarray{proof of adjacent distance larger than one II-(ii)-222},
and \reqnarray{N-M-k-h} that ${\mbf'}_1^{r_{h-1}}\in \Ncal_{M,k}(h)$.

As $\mbf_1^{r_{h-1}}\in \Ncal_{M,k}(h)$ and ${\mbf'}_1^{r_{h-1}}\in \Ncal_{M,k}(h)$,
we see from \reqnarray{proof of adjacent distance larger than one II-111},
\reqnarray{proof of adjacent distance larger than one II-222},
and the argument in the paragraph after \reqnarray{N-M-k-h} that
\beqnarray{proof of adjacent distance larger than one II-(ii)-333}
\nbf_1^{r_{h-2}}\in \Ncal_{M,k}(h-1) \textrm{ and } {\nbf'}_1^{r_{h-2}}\in \Ncal_{M,k}(h-1).
\eeqnarray
To show \reqnarray{adjacent distance larger than one II-2},
i.e., $\mbf_1^{r_{h-1}}\preceq {\mbf'}_1^{r_{h-1}}$,
where $\mbf_1^{r_{h-1}}\equiv {\mbf'}_1^{r_{h-1}}$ if and only if
$r_{h-1}=2$ and $m_1=m_2-2$,
we see from \reqnarray{proof of adjacent distance larger than one II-333}
that it suffices to show that
\beqnarray{proof of adjacent distance larger than one II-(ii)-444}
\nbf_1^{r_{h-2}}\preceq {\nbf'}_1^{r_{h-2}},
\eeqnarray
where $\nbf_1^{r_{h-2}}\equiv {\nbf'}_1^{r_{h-2}}$ if and only if
\beqnarray{proof of adjacent distance larger than one II-(ii)-555}
r_{h-1}=2 \textrm{ and } m_1=m_2-2.
\eeqnarray

Note that from $m_a=m'_a-1$, $m_{a+1}=m'_{a+1}+1$, $m_i=m'_i$ for $i\neq a$ and $a+1$,
\reqnarray{proof of adjacent distance larger than one II-444}--\reqnarray{proof of adjacent distance larger than one II-888},
we can show as in the proof of
\reqnarray{proof of adjacent distance larger than one II-(i)-555}--\reqnarray{proof of adjacent distance larger than one II-(i)-bbb}
in (i) above
(with the roles of $\mbf_1^{r_{h-1}}$ and ${\mbf'}_1^{r_{h-1}}$ interchanged
and the roles of $\nbf_1^{r_{h-2}}$ and ${\nbf'}_1^{r_{h-2}}$ interchanged) that
\beqnarray{}
\alignspace n_{i_{a+1}+1}=q_{h-1},
\label{eqn:proof of adjacent distance larger than one II-(ii)-666} \\
\alignspace n'_{i_{a+1}}=q_{h-1},
\label{eqn:proof of adjacent distance larger than one II-(ii)-777}\\
\alignspace n_{i_{a+1}}-n_{i_{a+1}+1}=(q_{h-1}+1)-q_{h-1}=1,
\label{eqn:proof of adjacent distance larger than one II-(ii)-888}\\
\alignspace n'_{i_{a+1}}=q_{h-1}=n_{i_{a+1}}-1,
\label{eqn:proof of adjacent distance larger than one II-(ii)-999}\\
\alignspace n'_{i_{a+1}+1}=n'_{i'_{a+1}}=q_{h-1}+1=n_{i_{a+1}+1}+1,
\label{eqn:proof of adjacent distance larger than one II-(ii)-aaa}\\
\alignspace n'_i=n_i, \textrm{ for } i\neq i_{a+1} \textrm{ and } i_{a+1}+1.
\label{eqn:proof of adjacent distance larger than one II-(ii)-bbb}
\eeqnarray

\bpdffigure{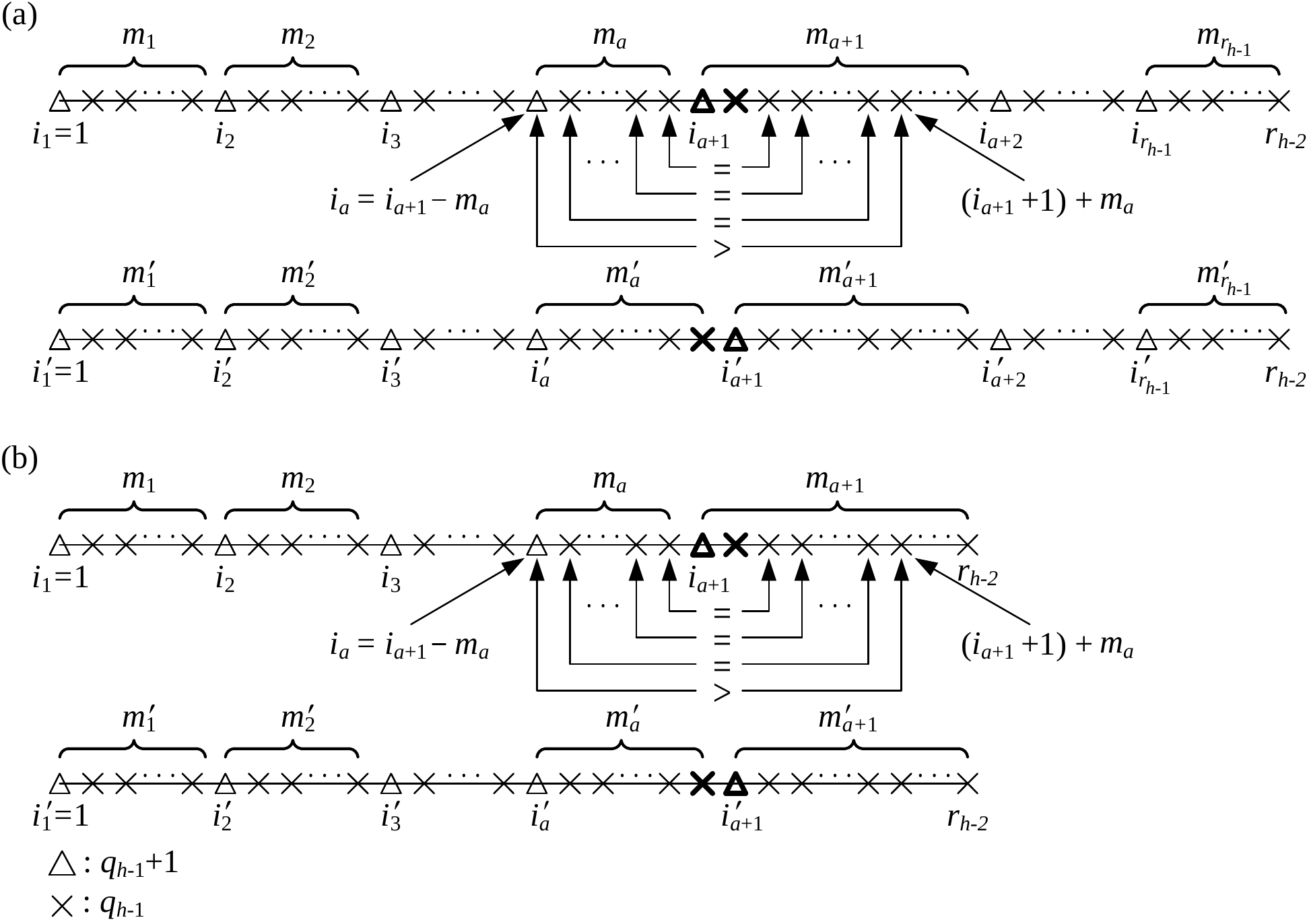}{5.5in}
\epdffigure{appendix-C-(ii)}
{An illustration of \reqnarray{proof of adjacent distance larger than one II-(ii)-eee}
and \reqnarray{proof of adjacent distance larger than one II-(ii)-fff}:
(a) $1\leq a\leq r_{h-1}-2$
(note that in this case we have $i_{a+1}-m_a=i_a$
in \reqnarray{proof of adjacent distance larger than one II-(ii)-2222}
and $i_{a+1}<(i_{a+1}+1)+m_a<i_{a+2}$
in \reqnarray{proof of adjacent distance larger than one II-(ii)-8888});
(b) $a=r_{h-1}-1$
(note that in this case we have $i_{a+1}-m_a=i_a$
in \reqnarray{proof of adjacent distance larger than one II-(ii)-2222}
and $i_{r_{h-1}}=i_{a+1}<(i_{a+1}+1)+m_a\leq r_{h-2}$
in \reqnarray{proof of adjacent distance larger than one II-(ii)-cccc}).}

In the following, we show that
\beqnarray{}
\alignspace 2\leq i_{a+1}\leq r_{h-2}-2,
\label{eqn:proof of adjacent distance larger than one II-(ii)-ccc}\\
\alignspace 1\leq m_a\leq \min\{i_{a+1}-1,r_{h-2}-i_{a+1}-1\},
\label{eqn:proof of adjacent distance larger than one II-(ii)-ddd}\\
\alignspace n_{i_{a+1}-j'}=n_{(i_{a+1}+1)+j'}=q_{h-1}, \textrm{ for } j'=1,2,\ldots,m_a-1,
\label{eqn:proof of adjacent distance larger than one II-(ii)-eee}\\
\alignspace n_{i_{a+1}-m_a}=q_{h-1}+1>n_{(i_{a+1}+1)+m_a}=q_{h-1}.
\label{eqn:proof of adjacent distance larger than one II-(ii)-fff}
\eeqnarray
An illustration of \reqnarray{proof of adjacent distance larger than one II-(ii)-eee}
and \reqnarray{proof of adjacent distance larger than one II-(ii)-fff}
is given in \rfigure{appendix-C-(ii)}.
Therefore, it follows from
\reqnarray{proof of adjacent distance larger than one II-888},
$\nbf_1^{r_{h-2}}\in \Ncal_{M,k}(h-1)$ in
\reqnarray{proof of adjacent distance larger than one II-(ii)-333},
\reqnarray{proof of adjacent distance larger than one II-(ii)-888}--\reqnarray{proof of adjacent distance larger than one II-(ii)-fff},
and \reqnarray{comparison rule A-3} in \rlemma{comparison rule A}(ii)
(for the odd integer $h-1$) that $\nbf_1^{r_{h-2}}\preceq{\nbf'}_1^{r_{h-2}}$,
where $\nbf_1^{r_{h-2}}\equiv{\nbf'}_1^{r_{h-2}}$ if and only if
\beqnarray{proof of adjacent distance larger than one II-(ii)-ggg}
i_{a+1}-m_a=1,\ (i_{a+1}+1)+m_a=r_{h-2}, \textrm{ and } n_1=n_{r_{h-2}}+1.
\eeqnarray

To prove \reqnarray{proof of adjacent distance larger than one II-(ii)-ccc}--\reqnarray{proof of adjacent distance larger than one II-(ii)-fff},
note that from \reqnarray{proof of adjacent distance larger than one II-555}
and $1\leq a\leq r_{h-1}-1$, we have
\beqnarray{proof of adjacent distance larger than one II-(ii)-1111}
i_{a+1}=\sum_{\ell=1}^{a}m_{\ell}+1\geq m_a+1,
\eeqnarray
and
\beqnarray{proof of adjacent distance larger than one II-(ii)-2222}
i_{j+1}=\sum_{\ell=1}^{j}m_{\ell}+1
=\left(\sum_{\ell=1}^{j-1}m_{\ell}+1\right)+m_j
=i_j+m_j, \textrm{ for } j=1,2,\ldots,r_{h-1}-1.
\eeqnarray
From $i_{a+1}-m_a=i_a$ in \reqnarray{proof of adjacent distance larger than one II-(ii)-2222},
we immediately see that
\beqnarray{proof of adjacent distance larger than one II-(ii)-3333}
i_a<i_{a+1}-j'<i_{a+1}, \textrm{ for } j'=1,2,\ldots,m_a-1.
\eeqnarray
It then follows from \reqnarray{proof of adjacent distance larger than one II-444},
\reqnarray{proof of adjacent distance larger than one II-(ii)-3333},
and $i_{a+1}-m_a=i_a$ in \reqnarray{proof of adjacent distance larger than one II-(ii)-2222} that
\beqnarray{}
\alignspace n_{i_{a+1}-j'}=q_{h-1}, \textrm{ for } j'=1,2,\ldots,m_a-1,
\label{eqn:proof of adjacent distance larger than one II-(ii)-4444}\\
\alignspace n_{i_{a+1}-m_a}=n_{i_a}=q_{h-1}+1.
\label{eqn:proof of adjacent distance larger than one II-(ii)-5555}
\eeqnarray

If $1\leq a\leq r_{h-1}-2$, then we have from
\reqnarray{proof of adjacent distance larger than one II-(ii)-2222},
$i_{a+2}\leq r_{h-2}$, and $m_a-m_{a+1}\leq -2$ that
\beqnarray{proof of adjacent distance larger than one II-(ii)-6666}
i_{a+1}=i_{a+2}-m_{a+1}\leq r_{h-2}-m_a-2.
\eeqnarray
Thus, \reqnarray{proof of adjacent distance larger than one II-(ii)-ccc}
follows from $i_{a+1}\geq m_a+1\geq 2$
in \reqnarray{proof of adjacent distance larger than one II-(ii)-1111}
and $i_{a+1}\leq r_{h-2}-m_a-2\leq r_{h-2}-2$
in \reqnarray{proof of adjacent distance larger than one II-(ii)-6666},
and \reqnarray{proof of adjacent distance larger than one II-(ii)-ddd}
follows from $m_a\leq i_{a+1}-1$
in \reqnarray{proof of adjacent distance larger than one II-(ii)-1111}
and $m_a\leq r_{h-2}-i_{a+1}-2\leq r_{h-2}-i_{a+1}-1$
in \reqnarray{proof of adjacent distance larger than one II-(ii)-6666}.
Also, we have from $m_a-m_{a+1}\leq -2$,
\reqnarray{proof of adjacent distance larger than one II-(ii)-2222},
and $1\leq a\leq r_{h-1}-2$ that
\beqnarray{proof of adjacent distance larger than one II-(ii)-7777}
(i_{a+1}+1)+m_a\leq i_{a+1}+m_{a+1}-1=i_{a+2}-1.
\eeqnarray
From \reqnarray{proof of adjacent distance larger than one II-(ii)-7777},
we immediately see that
\beqnarray{proof of adjacent distance larger than one II-(ii)-8888}
i_{a+1}<(i_{a+1}+1)+j'<i_{a+2}, \textrm{ for } j'=1,2,\ldots,m_a.
\eeqnarray
It then follows from \reqnarray{proof of adjacent distance larger than one II-444}
and \reqnarray{proof of adjacent distance larger than one II-(ii)-8888} that
\beqnarray{proof of adjacent distance larger than one II-(ii)-9999}
n_{(i_{a+1}+1)+j'}=q_{h-1}, \textrm{ for } j'=1,2,\ldots,m_a.
\eeqnarray
Thus, \reqnarray{proof of adjacent distance larger than one II-(ii)-eee}
and \reqnarray{proof of adjacent distance larger than one II-(ii)-fff}
follow from \reqnarray{proof of adjacent distance larger than one II-(ii)-4444},
\reqnarray{proof of adjacent distance larger than one II-(ii)-5555},
and \reqnarray{proof of adjacent distance larger than one II-(ii)-9999}.

On the other hand, if $a=r_{h-1}-1$,
then we have from
\reqnarray{proof of adjacent distance larger than one II-555},
\reqnarray{proof of adjacent distance larger than one II-(ii)-222},
and $m_a-m_{a+1}\leq -2$ that
\beqnarray{proof of adjacent distance larger than one II-(ii)-aaaa}
i_{a+1}
\aligneq i_{r_{h-1}}=\sum_{\ell=1}^{r_{h-1}-1}m_{\ell}+1
=\sum_{\ell=1}^{r_{h-1}}m_{\ell}-m_{r_{h-1}}+1 \nn\\
\aligneq r_{h-2}-m_{a+1}+1 \leq r_{h-2}-m_a-1,
\eeqnarray
where the equality holds if and only if $m_a-m_{a+1}=-2$.
Thus, \reqnarray{proof of adjacent distance larger than one II-(ii)-ccc}
follows from $i_{a+1}\geq m_a+1\geq 2$
in \reqnarray{proof of adjacent distance larger than one II-(ii)-1111}
and $i_{a+1}\leq r_{h-2}-m_a-1\leq r_{h-2}-2$
in \reqnarray{proof of adjacent distance larger than one II-(ii)-aaaa},
and \reqnarray{proof of adjacent distance larger than one II-(ii)-ddd}
follows from $m_a\leq i_{a+1}-1$
in \reqnarray{proof of adjacent distance larger than one II-(ii)-1111}
and $m_a\leq r_{h-2}-i_{a+1}-1$
in \reqnarray{proof of adjacent distance larger than one II-(ii)-aaaa}.
Also, we have from \reqnarray{proof of adjacent distance larger than one II-(ii)-aaaa} that
\beqnarray{proof of adjacent distance larger than one II-(ii)-bbbb}
(i_{a+1}+1)+m_a\leq r_{h-2},
\eeqnarray
where the equality holds if and only if $m_a-m_{a+1}=-2$.
From $a=r_{h-1}-1$ and \reqnarray{proof of adjacent distance larger than one II-(ii)-bbbb},
we immediately see that
\beqnarray{proof of adjacent distance larger than one II-(ii)-cccc}
i_{r_{h-1}}=i_{a+1}<(i_{a+1}+1)+j'\leq r_{h-2}, \textrm{ for } j'=1,2,\ldots,m_a.
\eeqnarray
It then follows from \reqnarray{proof of adjacent distance larger than one II-444}
and \reqnarray{proof of adjacent distance larger than one II-(ii)-cccc} that
\beqnarray{proof of adjacent distance larger than one II-(ii)-dddd}
n_{(i_{a+1}+1)+j'}=q_{h-1}, \textrm{ for } j'=1,2,\ldots,m_a.
\eeqnarray
As such, \reqnarray{proof of adjacent distance larger than one II-(ii)-eee}
and \reqnarray{proof of adjacent distance larger than one II-(ii)-fff}
follow from \reqnarray{proof of adjacent distance larger than one II-(ii)-4444},
\reqnarray{proof of adjacent distance larger than one II-(ii)-5555},
and \reqnarray{proof of adjacent distance larger than one II-(ii)-dddd}.

To complete the proof,
we need to show that the condition in \reqnarray{proof of adjacent distance larger than one II-(ii)-ggg}
is equivalent to the condition in \reqnarray{proof of adjacent distance larger than one II-(ii)-555}.
Note that if $i_{a+1}-m_a=1$ and $(i_{a+1}+1)+m_a=r_{h-2}$,
then we have from $n_1=n_{i_{a+1}-m_a}=q_{h-1}+1$ and $n_{r_{h-2}}=n_{(i_{a+1}+1)+m_a}=q_{h-1}$
in \reqnarray{proof of adjacent distance larger than one II-(ii)-fff} that
\beqnarray{}
n_1=n_{r_{h-2}}+1.\nn
\eeqnarray
As such, we see that the condition in \reqnarray{proof of adjacent distance larger than one II-(ii)-ggg}
is equivalent to the following condition:
\beqnarray{proof of adjacent distance larger than one II-(ii)-eeee}
i_{a+1}-m_a=1 \textrm{ and } (i_{a+1}+1)+m_a=r_{h-2}.
\eeqnarray
As we have $i_{a+1}-m_a=i_a$ in \reqnarray{proof of adjacent distance larger than one II-(ii)-2222}
and it is clear from \reqnarray{proof of adjacent distance larger than one II-555}
that $i_a=1$ if and only if $a=1$,
it follows that
\beqnarray{proof of adjacent distance larger than one II-(ii)-ffff}
i_{a+1}-m_a=1 \textrm{ iff } a=1.
\eeqnarray
Furthermore, if $1\leq a\leq r_{h-1}-2$,
then we see from \reqnarray{proof of adjacent distance larger than one II-(ii)-7777}
and $i_{a+2}\leq r_{h-2}$ that
\beqnarray{}
(i_{a+1}+1)+m_a\leq i_{a+2}-1 \leq r_{h-2}-1. \nn
\eeqnarray
On the other hand, if $a=r_{h-1}-1$, then we see from
\reqnarray{proof of adjacent distance larger than one II-(ii)-bbbb} that
\beqnarray{}
(i_{a+1}+1)+m_a\leq r_{h-2}, \nn
\eeqnarray
where the equality holds if and only if $m_a=m_{a+1}-2$.
As such, it is easy to see that
\beqnarray{proof of adjacent distance larger than one II-(ii)-gggg}
(i_{a+1}+1)+m_a=r_{h-2} \textrm{ iff } a=r_{h-1}-1 \textrm{ and } m_a=m_{a+1}-2.
\eeqnarray
From \reqnarray{proof of adjacent distance larger than one II-(ii)-ffff}
and \reqnarray{proof of adjacent distance larger than one II-(ii)-gggg},
we deduce that the condition in \reqnarray{proof of adjacent distance larger than one II-(ii)-eeee}
is equivalent to the following condition:
\beqnarray{proof of adjacent distance larger than one II-(ii)-hhhh}
a=1,\ a=r_{h-1}-1, \textrm{ and } m_a=m_{a+1}-2.
\eeqnarray
It is clear that if $a=1$, $a=r_{h-1}-1$, and $m_a=m_{a+1}-2$,
then we have $r_{h-1}=2$ and $m_1=m_2-2$.
Conversely, if $r_{h-1}=2$ and $m_1=m_2-2$,
then it follows from $1\leq a\leq r_{h-1}-1$ that $a=1$
and hence we have $a=1=r_{h-1}-1$ and $m_a=m_{a+1}-2$.
Therefore, the condition in \reqnarray{proof of adjacent distance larger than one II-(ii)-hhhh}
is equivalent to the condition that $r_{h-1}=2$ and $m_1=m_2-2$
in \reqnarray{proof of adjacent distance larger than one II-(ii)-555},
and the proof is completed.

\bappendix{Proof of Comparison rule B in \rlemma{comparison rule B} for an even integer $2\leq h\leq N$
by using Comparison rule A in \rlemma{comparison rule A} for the odd integer $h-1$}
{proof of comparison rule B for an even integer h by using comparison rule A for the odd integer h-1}

In this appendix, we assume that Comparison rule A in \rlemma{comparison rule A} holds
for some odd integer $h-1$, where $1\leq h-1\leq N-1$,
and show that Comparison rule B in \rlemma{comparison rule B} holds for the even integer $h$.

Let
\beqnarray{}
\nbf_1^{r_{h-2}}(h-1)\aligneq L_{r_{h-3},r_{h-2}}(\nbf_1^{r_{h-1}}(h)),
\label{eqn:proof of comparison rule B-111} \\
{\nbf'}_1^{r_{h-2}}(h-1)\aligneq L_{r_{h-3},r_{h-2}}({\nbf'}_1^{r_{h-1}}(h)).
\label{eqn:proof of comparison rule B-222}
\eeqnarray
For simplicity, let $\mbf_1^{r_{h-1}}=\nbf_1^{r_{h-1}}(h)$,
${\mbf'}_1^{r_{h-1}}={\nbf'}_1^{r_{h-1}}(h)$,
$\nbf_1^{r_{h-2}}=\nbf_1^{r_{h-2}}(h-1)$,
and ${\nbf'}_1^{r_{h-2}}={\nbf'}_1^{r_{h-2}}(h-1)$.
Then \reqnarray{proof of adjacent distance larger than one II-333}--\reqnarray{proof of adjacent distance larger than one II-777}
in \rappendix{proof of adjacent distance larger than one II for an even integer h
by using comparison rule A for the odd integer h-1} still hold.
Note that in \rlemma{comparison rule B},
we have $r_{h-1}\geq 2$, $\mbf_1^{r_{h-1}}\in \Ncal_{M,k}(h)$,
$m_a-m_{a+1}=1$ for some $1\leq a\leq r_{h-1}-1$,
$m'_a=m_a-1$, $m'_{a+1}=m_{a+1}+1$, and $m'_i=m_i$ for $i\neq a$ and $a+1$.
As $r_{h-1}\geq 2$,
we see that \reqnarray{proof of adjacent distance larger than one II-888}
in \rappendix{proof of adjacent distance larger than one II for an even integer h
by using comparison rule A for the odd integer h-1} also holds.
It is easy to see that
\beqnarray{proof of comparison rule B-333}
m'_a=m_a-1=m_{a+1}\geq 1,\ m'_{a+1}=m_{a+1}+1\geq 2,
\textrm{ and } m'_i=m_i \textrm{ for } i\neq a, a+1.
\eeqnarray
From $m'_a=m_a-1$, $m'_{a+1}=m_{a+1}+1$, and $m'_i=m_i$ for $i\neq a$ and $a+1$,
$\mbf_1^{r_{h-1}}\in \Ncal_{M,k}(h)$, and \reqnarray{N-M-k-h},
we can see that \reqnarray{proof of adjacent distance larger than one II-(i)-222}
in \rappendix{proof of adjacent distance larger than one II for an even integer h
by using comparison rule A for the odd integer h-1} also holds.
As such, it follows from \reqnarray{proof of comparison rule B-333},
\reqnarray{proof of adjacent distance larger than one II-(i)-222},
$2\leq h\leq N$, and \reqnarray{N-M-k-h} that ${\mbf'}_1^{r_{h-1}}\in \Ncal_{M,k}(h)$.

From $\mbf_1^{r_{h-1}}\in \Ncal_{M,k}(h)$, ${\mbf'}_1^{r_{h-1}}\in \Ncal_{M,k}(h)$,
\reqnarray{proof of comparison rule B-111}, and \reqnarray{proof of comparison rule B-222},
we can see that \reqnarray{proof of adjacent distance larger than one II-(i)-333}
in \rappendix{proof of adjacent distance larger than one II for an even integer h
by using comparison rule A for the odd integer h-1} also holds.
Furthermore, since ${\mbf'}_1^{r_{h-1}}$ is obtained from $\mbf_1^{r_{h-1}}$
in exactly the same way as that in \rlemma{adjacent distance larger than one II}(i),
it is clear that \reqnarray{proof of adjacent distance larger than one II-(i)-555}--\reqnarray{proof of adjacent distance larger than one II-(i)-hhh}
in \rappendix{proof of adjacent distance larger than one II for an even integer h
by using comparison rule A for the odd integer h-1} also hold.
We also note that from $m'_a=m_a-1$, $m'_{a+1}=m_{a+1}+1$, and $m_a-m_{a+1}=1$, we have
\beqnarray{proof of comparison rule B-444}
m'_a-m'_{a+1}=(m_a-1)-(m_{a+1}+1)=m_a-m_{a+1}-2=-1.
\eeqnarray

(i) Note that in \rlemma{comparison rule B}(i), we have $a=1$ or $a=r_{h-1}-1$.
To show \reqnarray{comparison rule B-1},
i.e., $\mbf_1^{r_{h-1}}\prec {\mbf'}_1^{r_{h-1}}$,
we see from \reqnarray{proof of adjacent distance larger than one II-333}
that it suffices to show that
\beqnarray{proof of comparison rule B-(i)-111}
\nbf_1^{r_{h-2}}\prec{\nbf'}_1^{r_{h-2}}.
\eeqnarray
We consider the two cases $a=1\neq r_{h-1}-1$ and $a=r_{h-1}-1$ separately.

\emph{Case 1: $a=1\neq r_{h-1}-1$.}
In this case, we have $a=1$ and $a\leq r_{h-1}-2$.
It follows that
\reqnarray{proof of adjacent distance larger than one II-(i)-case-1-111}
and \reqnarray{proof of adjacent distance larger than one II-(i)-case-1-222}
in \rappendix{proof of adjacent distance larger than one II for an even integer h
by using comparison rule A for the odd integer h-1} also hold.

\bpdffigure{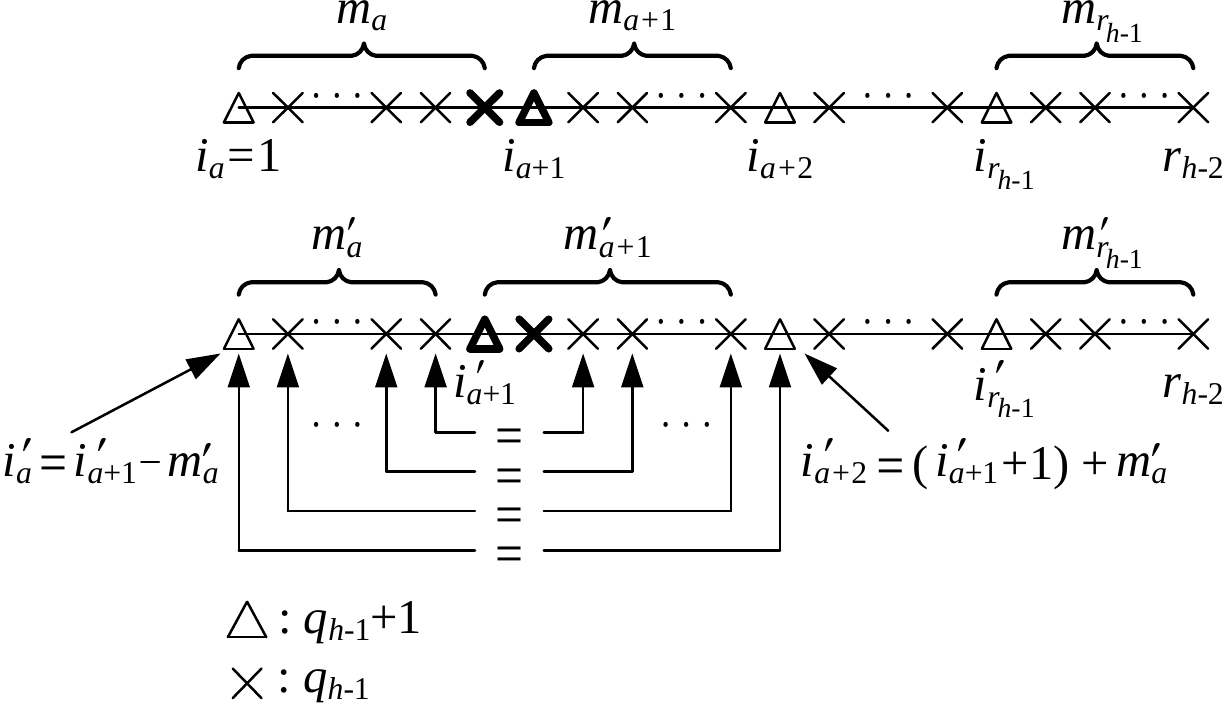}{3.5in}
\epdffigure{appendix-D-(i)-case-1}
{An illustration of \reqnarray{proof of comparison rule B-(i)-case-1-111}
in the case that $a=1\neq r_{h-1}-1$
(note that in this case we have
$\min\{i'_{a+1}-1,r_{h-2}-i'_{a+1}-1\}=m'_a$
in \reqnarray{proof of comparison rule B-(i)-case-1-555},
$i'_{a+1}-m'_a=i'_a$ in \reqnarray{proof of comparison rule B-(i)-case-1-333},
and $(i'_{a+1}+1)+m'_a=i'_{a+2}$ in \reqnarray{proof of comparison rule B-(i)-case-1-444}).}

In the following, we will show that
\beqnarray{proof of comparison rule B-(i)-case-1-111}
n'_{i'_{a+1}-j'}=n'_{(i'_{a+1}+1)+j'}, \textrm{ for } j'=1,2,\ldots,\min\{i'_{a+1}-1,r_{h-2}-i'_{a+1}-1\}.
\eeqnarray
An illustration of \reqnarray{proof of comparison rule B-(i)-case-1-111}
is given in \rfigure{appendix-D-(i)-case-1}.
Therefore, it follows from
\reqnarray{proof of adjacent distance larger than one II-888},
${\nbf'}_1^{r_{h-2}}\in \Ncal_{M,k}(h-1)$ in
\reqnarray{proof of adjacent distance larger than one II-(i)-333},
\reqnarray{proof of adjacent distance larger than one II-(i)-888}--\reqnarray{proof of adjacent distance larger than one II-(i)-bbb},
\reqnarray{proof of adjacent distance larger than one II-(i)-case-1-222},
\reqnarray{proof of comparison rule B-(i)-case-1-111},
and \reqnarray{comparison rule A-4} in \rlemma{comparison rule A}(iii)
(for the odd integer $h-1$) that ${\nbf'}_1^{r_{h-2}}\succ\nbf_1^{r_{h-2}}$,
i.e., \reqnarray{proof of comparison rule B-(i)-111} holds.

To prove \reqnarray{proof of comparison rule B-(i)-case-1-111},
note that from \reqnarray{proof of adjacent distance larger than one II-(i)-hhh}
and $a\leq r_{h-1}-2$, we have
\beqnarray{proof of comparison rule B-(i)-case-1-333}
i'_{a+1}-m'_a=i'_a.
\eeqnarray
From \reqnarray{proof of comparison rule B-444},
\reqnarray{proof of adjacent distance larger than one II-(i)-hhh},
and $a\leq r_{h-1}-2$, we have
\beqnarray{proof of comparison rule B-(i)-case-1-444}
(i'_{a+1}+1)+m'_a=i'_{a+1}+m'_{a+1}=i'_{a+2}.
\eeqnarray
As we have from $i'_1=1$ in \reqnarray{proof of adjacent distance larger than one II-777},
$a=1$, \reqnarray{proof of comparison rule B-(i)-case-1-333}, \reqnarray{proof of comparison rule B-(i)-case-1-444},
and $i'_{a+2}\leq r_{h-2}$ that
\beqnarray{}
i'_{a+1}-1=i'_{a+1}-i'_1=i'_{a+1}-i'_a=m'_a=i'_{a+2}-(i'_{a+1}+1)\leq r_{h-2}-i'_{a+1}-1, \nn
\eeqnarray
it is clear that
\beqnarray{proof of comparison rule B-(i)-case-1-555}
\min\{i'_{a+1}-1,r_{h-2}-i'_{a+1}-1\}=i'_{a+1}-1=m'_a.
\eeqnarray
It is easy to see from \reqnarray{proof of adjacent distance larger than one II-666}
and \reqnarray{proof of comparison rule B-(i)-case-1-333} that
\beqnarray{}
\alignspace n'_{i'_{a+1}-j'}=q_{h-1}, \textrm{ for } j'=1,2,\ldots,m'_a-1,
\label{eqn:proof of comparison rule B-(i)-case-1-666} \\
\alignspace n'_{i'_{a+1}-m'_a}=n'_{i'_a}=q_{h-1}+1,
\label{eqn:proof of comparison rule B-(i)-case-1-777}
\eeqnarray
and it is also easy to see from \reqnarray{proof of adjacent distance larger than one II-666}
and \reqnarray{proof of comparison rule B-(i)-case-1-444} that
\beqnarray{}
\alignspace n'_{(i'_{a+1}+1)+j'}=q_{h-1}, \textrm{ for } j'=1,2,\ldots,m'_a-1,
\label{eqn:proof of comparison rule B-(i)-case-1-888} \\
\alignspace n'_{(i'_{a+1}+1)+m'_a}=n'_{i'_{a+2}}=q_{h-1}+1.
\label{eqn:proof of comparison rule B-(i)-case-1-999}
\eeqnarray
Thus, \reqnarray{proof of comparison rule B-(i)-case-1-111}
follows from \reqnarray{proof of comparison rule B-(i)-case-1-555}--\reqnarray{proof of comparison rule B-(i)-case-1-999}.

\emph{Case 2: $a=r_{h-1}-1$.}
If $m_{a+1}=1$, then \reqnarray{proof of adjacent distance larger than one II-(i)-case-2-222} in
\rappendix{proof of adjacent distance larger than one II for an even integer h
by using comparison rule A for the odd integer h-1} also holds,
and hence it follows from
\reqnarray{proof of adjacent distance larger than one II-888},
${\nbf'}_1^{r_{h-2}}\in \Ncal_{M,k}(h-1)$ in
\reqnarray{proof of adjacent distance larger than one II-(i)-333},
\reqnarray{proof of adjacent distance larger than one II-(i)-888}--\reqnarray{proof of adjacent distance larger than one II-(i)-bbb},
\reqnarray{proof of adjacent distance larger than one II-(i)-case-2-222},
and \reqnarray{comparison rule A-1} in \rlemma{comparison rule A}(i)
(for the odd integer $h-1$) that ${\nbf'}_1^{r_{h-2}}\succ\nbf_1^{r_{h-2}}$,
i.e., \reqnarray{proof of comparison rule B-(i)-111} holds.
On the other hand, if $m_{a+1}\geq 2$,
then it is easy to see that
\reqnarray{proof of adjacent distance larger than one II-(i)-case-2-111}--\reqnarray{proof of adjacent distance larger than one II-(i)-case-2-ddd}
still hold (as we only need $m_a-m_{a+1}>0$ to prove
\reqnarray{proof of adjacent distance larger than one II-(i)-case-2-666}
and \reqnarray{proof of adjacent distance larger than one II-(i)-case-2-888}),
and hence we can show that $\nbf_1^{r_{h-2}}\prec{\nbf'}_1^{r_{h-2}}$,
i.e., \reqnarray{proof of comparison rule B-(i)-111} holds,
as in the proof of Case~2 of \rlemma{adjacent distance larger than one II}(i)
in \rappendix{proof of adjacent distance larger than one II for an even integer h
by using comparison rule A for the odd integer h-1}.

(ii) Note that in \rlemma{comparison rule B}(ii),
we have $2\leq a\leq r_{h-1}-2$ and there exists a positive integer $j$ such that
$1\leq j\leq \min\{a-1,r_{h-1}-a-1\}$, $m_{a-\ell}=m_{a+1+\ell}$ for $\ell=1,2,\ldots,j-1$,
and $m_{a-j}\neq m_{a+1+j}$.
As $2\leq a\leq r_{h-1}-2$,
we see that \reqnarray{proof of adjacent distance larger than one II-(i)-case-1-222}
in \rappendix{proof of adjacent distance larger than one II for an even integer h
by using comparison rule A for the odd integer h-1} also holds.
Furthermore, from $m'_i=m_i$ for $i\neq a$ and $a+1$
and $m_{a-\ell}=m_{a+1+\ell}$ for $\ell=1,2,\ldots,j-1$,
it is clear that
\beqnarray{proof of comparison rule B-(ii)-111}
m'_{a-\ell}=m'_{a+1+\ell}, \textrm{ for } \ell=1,2,\ldots,j-1.
\eeqnarray

\bpdffigure{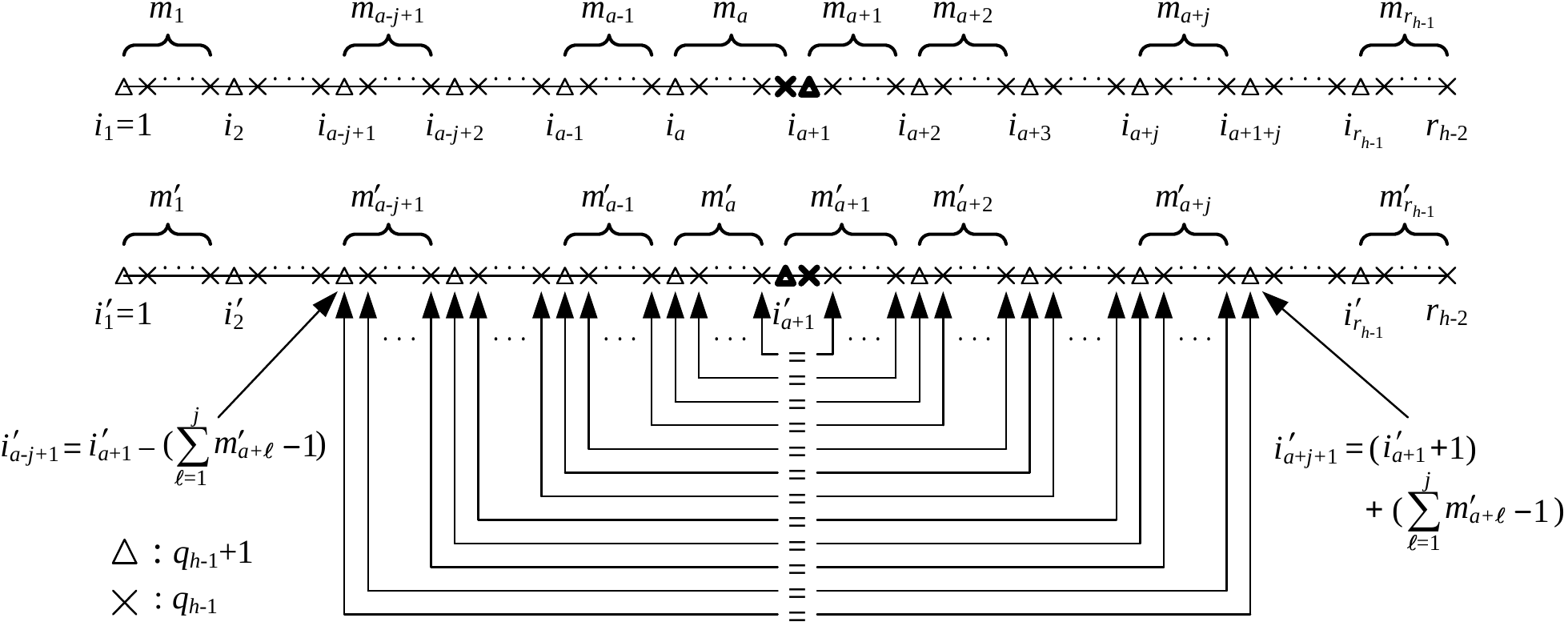}{6.0in}
\epdffigure{appendix-D-(ii)}
{An illustration of \reqnarray{proof of comparison rule B-(ii)-222}
(note that we have
$i'_{a+1}-(\sum_{\ell=1}^{j''}m'_{a+\ell}-1)=i'_{a-j''+1}$
for $1\leq j''\leq j$ in \reqnarray{proof of comparison rule B-(ii)-444},
$(i'_{a+1}+1)+(\sum_{\ell=1}^{j''}m'_{a+\ell}-1)=i'_{a+j''+1}$
for $1\leq j''\leq r_{h-1}-a-1$ in \reqnarray{proof of comparison rule B-(ii)-666},
and $j\leq \min\{a-1,r_{h-1}-a-1\}\leq r_{h-1}-a-1$).}

By using \reqnarray{proof of comparison rule B-(ii)-111},
we can show that
\beqnarray{proof of comparison rule B-(ii)-222}
n'_{i'_{a+1}-j'}=n'_{(i'_{a+1}+1)+j'},
\textrm{ for } j'=1,2,\ldots,\sum_{\ell=1}^{j}m'_{a+\ell}-1.
\eeqnarray
An illustration of \reqnarray{proof of comparison rule B-(ii)-222}
is given in \rfigure{appendix-D-(ii)}.
To prove \reqnarray{proof of comparison rule B-(ii)-222},
observe that for $1\leq j''\leq j$,
we have from \reqnarray{proof of comparison rule B-444}
and $m'_{a-\ell}=m'_{a+1+\ell}$ for $\ell=1,2,\ldots,j''-1$
in \reqnarray{proof of comparison rule B-(ii)-111} that
\beqnarray{proof of comparison rule B-(ii)-333}
\sum_{\ell=1}^{j''}m'_{a+\ell}-1
\aligneq m'_{a+1}+\sum_{\ell=2}^{j''}m'_{a+\ell}-1=m'_a+\sum_{\ell=1}^{j''-1}m'_{a+1+\ell} \nn\\
\aligneq m'_a+\sum_{\ell=1}^{j''-1}m'_{a-\ell}=\sum_{\ell=0}^{j''-1}m'_{a-\ell}
         =\sum_{\ell=a-j''+1}^{a}m'_{\ell}.
\eeqnarray
As such, for $1\leq j''\leq j$,
we have from \reqnarray{proof of adjacent distance larger than one II-777}
and \reqnarray{proof of comparison rule B-(ii)-333} that
\beqnarray{proof of comparison rule B-(ii)-444}
i'_{a+1}-\left(\sum_{\ell=1}^{j''}m'_{a+\ell}-1\right)
=\left(\sum_{\ell=1}^{a}m'_{\ell}+1\right)-\sum_{\ell=a-j''+1}^{a}m'_{\ell}
=\sum_{\ell=1}^{a-j''}m'_{\ell}+1=i'_{a-j''+1}.
\eeqnarray
It follows from \reqnarray{proof of adjacent distance larger than one II-666}
and \reqnarray{proof of comparison rule B-(ii)-444} that
\beqnarray{proof of comparison rule B-(ii)-555}
n'_{i'_{a+1}-j'}=
\bselection
q_{h-1}+1, &\textrm{for } j'=m'_{a+1}-1, \sum_{\ell=1}^{2}m'_{a+\ell}-1,\ldots,\sum_{\ell=1}^{j}m'_{a+\ell}-1, \\
q_{h-1}, &\textrm{for } 1\leq j'\leq \sum_{\ell=1}^{j}m'_{a+\ell}-1 \\
&\textrm{and } j'\neq m'_{a+1}-1, \sum_{\ell=1}^{2}m'_{a+\ell}-1,\ldots,\sum_{\ell=1}^{j}m'_{a+\ell}-1.
\eselection
\eeqnarray
Furthermore, for $1\leq j''\leq r_{h-1}-a-1$,
we have from \reqnarray{proof of adjacent distance larger than one II-777} that
\beqnarray{proof of comparison rule B-(ii)-666}
(i'_{a+1}+1)+\left(\sum_{\ell=1}^{j''}m'_{a+\ell}-1\right)
=\left(\sum_{\ell=1}^{a}m'_{\ell}+1\right)+\sum_{\ell=a+1}^{a+j''}m'_{\ell}
=\sum_{\ell=1}^{a+j''}m'_{\ell}+1=i'_{a+j''+1}.
\eeqnarray
It then follows from \reqnarray{proof of adjacent distance larger than one II-666}
and \reqnarray{proof of comparison rule B-(ii)-666} that
\beqnarray{proof of comparison rule B-(ii)-777}
n'_{(i'_{a+1}+1)+j'}=
\bselection
q_{h-1}+1, &\textrm{for } j'=m'_{a+1}-1, \sum_{\ell=1}^{2}m'_{a+\ell}-1,\ldots,\sum_{\ell=1}^{r_{h-1}-a-1}m'_{a+\ell}-1, \\
q_{h-1}, &\textrm{for } 1\leq j'\leq \sum_{\ell=1}^{r_{h-1}-a-1}m'_{a+\ell}-1 \\
&\textrm{and } j'\neq m'_{a+1}-1, \sum_{\ell=1}^{2}m'_{a+\ell}-1,\ldots,\sum_{\ell=1}^{r_{h-1}-a-1}m'_{a+\ell}-1.
\eselection
\eeqnarray
As we have $j\leq \min\{a-1,r_{h-1}-a-1\}\leq r_{h-1}-a-1$,
it is clear that \reqnarray{proof of comparison rule B-(ii)-222}
follows from \reqnarray{proof of comparison rule B-(ii)-555}
and \reqnarray{proof of comparison rule B-(ii)-777}.

From \reqnarray{proof of adjacent distance larger than one II-333},
we see that:
(a) If $m_{a-j}>m_{a+1+j}$,
then to show \reqnarray{comparison rule B-2},
i.e., $\mbf_1^{r_{h-1}}\prec{\mbf'}_1^{r_{h-1}}$,
it suffices to show that $\nbf_1^{r_{h-2}}\prec{\nbf'}_1^{r_{h-2}}$;
(b) If $m_{a-j}<m_{a+1+j}$,
then to show \reqnarray{comparison rule B-3},
i.e., $\mbf_1^{r_{h-1}}\succeq {\mbf'}_1^{r_{h-1}}$,
where $\mbf_1^{r_{h-1}}\equiv {\mbf'}_1^{r_{h-1}}$
if and only if $a-j=1$, $a+1+j=r_{h-1}$, and $m_1=m_{r_{h-1}}-1$,
it suffices to show that $\nbf_1^{r_{h-2}}\succeq{\nbf'}_1^{r_{h-2}}$,
where $\nbf_1^{r_{h-2}}\equiv{\nbf'}_1^{r_{h-2}}$
if and only if $a-j=1$, $a+1+j=r_{h-1}$, and $m_1=m_{r_{h-1}}-1$.

(a) First we assume that $m_{a-j}>m_{a+1+j}$ and show that
\beqnarray{proof of comparison rule B-(ii)-(a)-111}
\nbf_1^{r_{h-2}}\prec{\nbf'}_1^{r_{h-2}}.
\eeqnarray
As $j\leq \min\{a-1,r_{h-1}-a-1\}$, we have $j+1\leq a\leq r_{h-1}-j-1$.
We consider the two cases $j+1\leq a\leq r_{h-1}-j-2$ and $a=r_{h-1}-j-1$ separately.

\bpdffigure{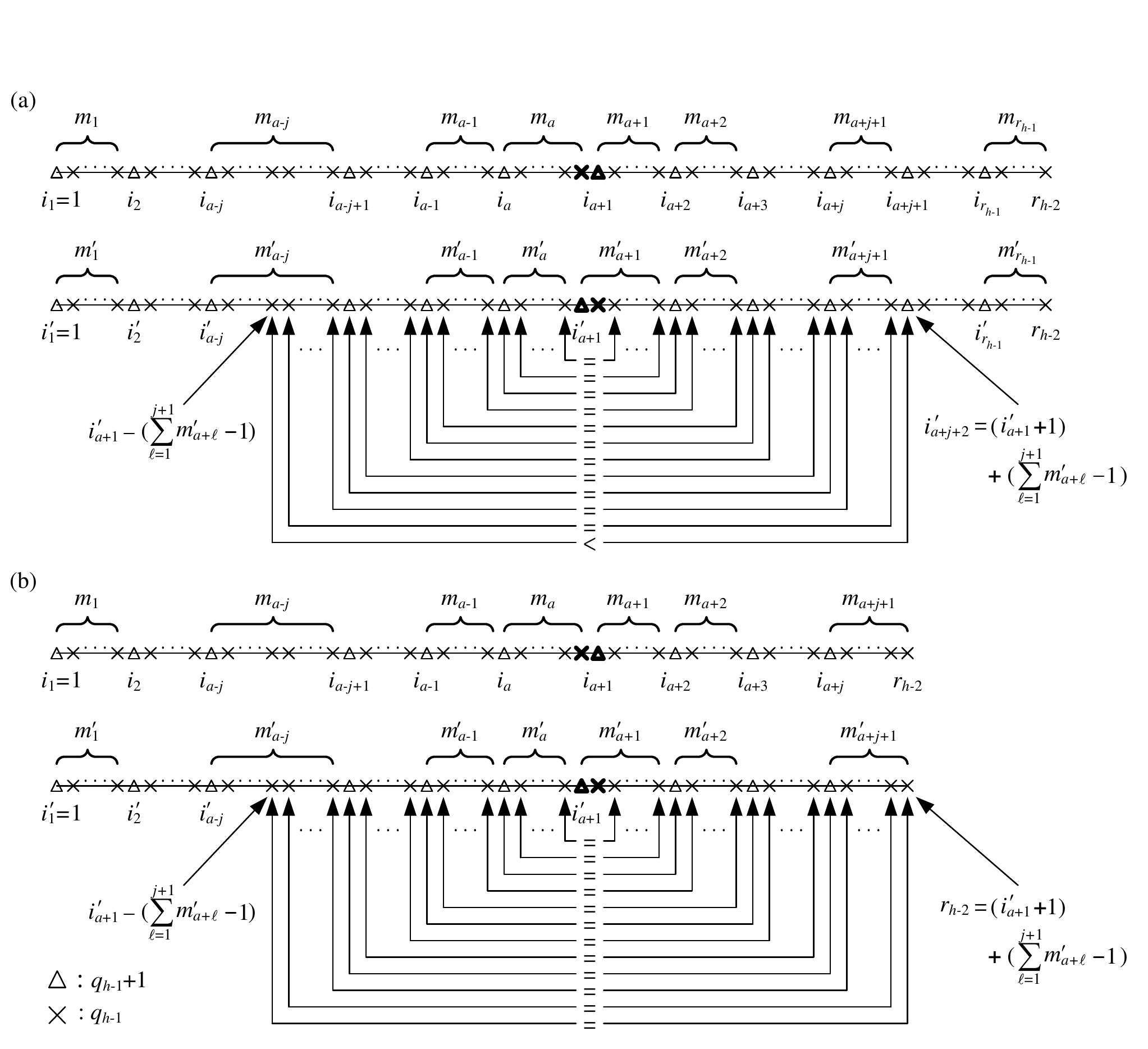}{6.0in}
\epdffigure{appendix-D-(ii)-(a)}
{(a) An illustration of \reqnarray{proof of comparison rule B-(ii)-(a)-case-1-222}
and \reqnarray{proof of comparison rule B-(ii)-(a)-case-1-333}
in the case that $m_{a-j}>m_{a+1+j}$ and $j+1\leq a\leq r_{h-1}-j-2$
(note that in this case we have
$i'_{a+1}-(\sum_{\ell=1}^{j''}m'_{a+\ell}-1)=i'_{a-j''+1}$
for $1\leq j''\leq j$ in \reqnarray{proof of comparison rule B-(ii)-444},
$i'_{a+1}-(\sum_{\ell=1}^{j+1}m'_{a+\ell}-1)>i'_{a-j}$
in \reqnarray{proof of comparison rule B-(ii)-(a)-case-1-999},
$(i'_{a+1}+1)+(\sum_{\ell=1}^{j''}m'_{a+\ell}-1)=i'_{a+j''+1}$
for $1\leq j''\leq r_{h-1}-a-1$ in \reqnarray{proof of comparison rule B-(ii)-666},
and $j+1\leq r_{h-1}-a-1$).
(b) An illustration of \reqnarray{proof of comparison rule B-(ii)-(a)-case-2-111}
in the case that $m_{a-j}>m_{a+1+j}$ and $a=r_{h-1}-j-1$
(note that in this case we have
$i'_{a+1}-(\sum_{\ell=1}^{j''}m'_{a+\ell}-1)=i'_{a-j''+1}$
for $1\leq j''\leq j$ in \reqnarray{proof of comparison rule B-(ii)-444},
$i'_{a+1}-(\sum_{\ell=1}^{j+1}m'_{a+\ell}-2)>i'_{a-j}$
in \reqnarray{proof of comparison rule B-(ii)-(a)-case-1-999},
$(i'_{a+1}+1)+(\sum_{\ell=1}^{j''}m'_{a+\ell}-1)=i'_{a+j''+1}$
for $1\leq j''\leq r_{h-1}-a-1=j$ in \reqnarray{proof of comparison rule B-(ii)-666},
and $(i'_{a+1}+1)+(\sum_{\ell=1}^{j+1}m'_{a+\ell}-2)=r_{h-2}$
in \reqnarray{proof of comparison rule B-(ii)-(a)-case-2-666}).}

\emph{Case 1: $j+1\leq a\leq r_{h-1}-j-2$.}
In this case, we show that
\beqnarray{}
\alignspace
1\leq \sum_{\ell=1}^{j+1}m'_{a+\ell}-1\leq \min\{i'_{a+1}-1,r_{h-2}-i'_{a+1}-1\},
\label{eqn:proof of comparison rule B-(ii)-(a)-case-1-111}\\
\alignspace
n'_{i'_{a+1}-j'}=n'_{(i'_{a+1}+1)+j'},
\textrm{ for } j'=1,2,\ldots,\sum_{\ell=1}^{j+1}m'_{a+\ell}-2,
\label{eqn:proof of comparison rule B-(ii)-(a)-case-1-222}\\
\alignspace
n'_{i'_{a+1}-(\sum_{\ell=1}^{j+1}m'_{a+\ell}-1)}=q_{h-1}
<n'_{(i'_{a+1}+1)+(\sum_{\ell=1}^{j+1}m'_{a+\ell}-1)}=q_{h-1}+1.
\label{eqn:proof of comparison rule B-(ii)-(a)-case-1-333}
\eeqnarray
An illustration of \reqnarray{proof of comparison rule B-(ii)-(a)-case-1-222}
and \reqnarray{proof of comparison rule B-(ii)-(a)-case-1-333}
is given in \rfigure{appendix-D-(ii)-(a)}(a).
Therefore, it follows from \reqnarray{proof of adjacent distance larger than one II-888},
${\nbf'}_1^{r_{h-2}}\in \Ncal_{M,k}(h-1)$ in
\reqnarray{proof of adjacent distance larger than one II-(i)-333},
\reqnarray{proof of adjacent distance larger than one II-(i)-888}--\reqnarray{proof of adjacent distance larger than one II-(i)-bbb},
\reqnarray{proof of adjacent distance larger than one II-(i)-case-1-222},
\reqnarray{proof of comparison rule B-(ii)-(a)-case-1-111}--\reqnarray{proof of comparison rule B-(ii)-(a)-case-1-333},
and \reqnarray{comparison rule A-2} in \rlemma{comparison rule A}(ii)
(for the odd integer $h-1$) that ${\nbf'}_1^{r_{h-2}}\succ\nbf_1^{r_{h-2}}$,
i.e., \reqnarray{proof of comparison rule B-(ii)-(a)-111} holds.

To prove \reqnarray{proof of comparison rule B-(ii)-(a)-case-1-111}, note that
\beqnarray{proof of comparison rule B-(ii)-(a)-case-1-444}
\sum_{\ell=1}^{j+1}m'_{a+\ell}-1\geq m'_{a+1}+m'_{a+j+1}-1\geq 1.
\eeqnarray
From \reqnarray{proof of comparison rule B-(ii)-444} (with $j''=j$),
$m'_{a-j}=m_{a-j}>m_{a+1+j}=m'_{a+1+j}$,
\reqnarray{proof of adjacent distance larger than one II-(i)-hhh},
and $i'_{a-j}\geq 1$, we have
\beqnarray{}
\sum_{\ell=1}^{j+1}m'_{a+\ell}-1
\aligneq \left(\sum_{\ell=1}^{j}m'_{a+\ell}-1\right)+m'_{a+1+j}=(i'_{a+1}-i'_{a-j+1})+m'_{a+1+j} \nn\\
\alignless i'_{a+1}-i'_{a-j+1}+m'_{a-j}=i'_{a+1}-i'_{a-j}
\label{eqn:proof of comparison rule B-(ii)-(a)-case-1-555}\\
\alignleq i'_{a+1}-1.
\label{eqn:proof of comparison rule B-(ii)-(a)-case-1-666}
\eeqnarray
As in this case we have $j+1\leq r_{h-1}-a-1$,
it follows from \reqnarray{proof of comparison rule B-(ii)-666} (with $j''=j+1$)
and $i'_{a+j+2}\leq r_{h-2}$ that
\beqnarray{proof of comparison rule B-(ii)-(a)-case-1-777}
\sum_{\ell=1}^{j+1}m'_{a+\ell}-1
=i'_{a+j+2}-(i'_{a+1}+1)
\leq r_{h-2}-i'_{a+1}-1.
\eeqnarray
Thus, \reqnarray{proof of comparison rule B-(ii)-(a)-case-1-111} follows from
\reqnarray{proof of comparison rule B-(ii)-(a)-case-1-444},
\reqnarray{proof of comparison rule B-(ii)-(a)-case-1-666},
and \reqnarray{proof of comparison rule B-(ii)-(a)-case-1-777}.

To prove \reqnarray{proof of comparison rule B-(ii)-(a)-case-1-222}
and \reqnarray{proof of comparison rule B-(ii)-(a)-case-1-333},
note that from \reqnarray{proof of comparison rule B-(ii)-444} (with $j''=j$), we have
\beqnarray{proof of comparison rule B-(ii)-(a)-case-1-888}
i'_{a+1}-\sum_{\ell=1}^{j}m'_{a+\ell}=i'_{a-j+1}-1<i'_{a-j+1},
\eeqnarray
and from \reqnarray{proof of comparison rule B-(ii)-(a)-case-1-555}, we have
\beqnarray{proof of comparison rule B-(ii)-(a)-case-1-999}
i'_{a+1}-\left(\sum_{\ell=1}^{j+1}m'_{a+\ell}-1\right)>i'_{a-j}.
\eeqnarray
Thus, we see from \reqnarray{proof of adjacent distance larger than one II-666},
\reqnarray{proof of comparison rule B-(ii)-(a)-case-1-888},
and \reqnarray{proof of comparison rule B-(ii)-(a)-case-1-999} that
\beqnarray{proof of comparison rule B-(ii)-(a)-case-1-aaa}
n'_{i'_{a+1}-j'}=q_{h-1},
\textrm{ for } \sum_{\ell=1}^{j}m'_{a+\ell}\leq j'\leq \sum_{\ell=1}^{j+1}m'_{a+\ell}-1.
\eeqnarray
Since in this case we have $j+1\leq r_{h-1}-a-1$,
it follows from \reqnarray{proof of comparison rule B-(ii)-777} that
\beqnarray{proof of comparison rule B-(ii)-(a)-case-1-bbb}
n'_{(i'_{a+1}+1)+j'}=
\bselection
q_{h-1}, &\textrm{ for } \sum_{\ell=1}^{j}m'_{a+\ell}\leq j'\leq \sum_{\ell=1}^{j+1}m'_{a+\ell}-2,\\
q_{h-1}+1, &\textrm{ for } j'=\sum_{\ell=1}^{j+1}m'_{a+\ell}-1.
\eselection
\eeqnarray
By combining \reqnarray{proof of comparison rule B-(ii)-222},
\reqnarray{proof of comparison rule B-(ii)-(a)-case-1-aaa},
and \reqnarray{proof of comparison rule B-(ii)-(a)-case-1-bbb},
we obtain \reqnarray{proof of comparison rule B-(ii)-(a)-case-1-222}
and \reqnarray{proof of comparison rule B-(ii)-(a)-case-1-333}.

\emph{Case 2: $a=r_{h-1}-j-1$.}
In this case, we show that
\beqnarray{proof of comparison rule B-(ii)-(a)-case-2-111}
n'_{i'_{a+1}-j'}=n'_{(i'_{a+1}+1)+j'},
\textrm{ for } j'=1,2,\ldots,\min\{i'_{a+1}-1,r_{h-2}-i'_{a+1}-1\}.
\eeqnarray
An illustration of \reqnarray{proof of comparison rule B-(ii)-(a)-case-2-111}
is given in \rfigure{appendix-D-(ii)-(a)}(b).
Therefore, it follows from \reqnarray{proof of adjacent distance larger than one II-888},
${\nbf'}_1^{r_{h-2}}\in \Ncal_{M,k}(h-1)$ in
\reqnarray{proof of adjacent distance larger than one II-(i)-333},
\reqnarray{proof of adjacent distance larger than one II-(i)-888}--\reqnarray{proof of adjacent distance larger than one II-(i)-bbb},
\reqnarray{proof of adjacent distance larger than one II-(i)-case-1-222},
\reqnarray{proof of comparison rule B-(ii)-(a)-case-2-111},
and \reqnarray{comparison rule A-4} in \rlemma{comparison rule A}(iii)
(for the odd integer $h-1$) that ${\nbf'}_1^{r_{h-2}}\succ\nbf_1^{r_{h-2}}$,
i.e., \reqnarray{proof of comparison rule B-(ii)-(a)-111} holds.

To prove \reqnarray{proof of comparison rule B-(ii)-(a)-case-2-111}, observe that
\reqnarray{proof of comparison rule B-(ii)-(a)-case-1-555}--\reqnarray{proof of comparison rule B-(ii)-(a)-case-1-666}
and \reqnarray{proof of comparison rule B-(ii)-(a)-case-1-888}--\reqnarray{proof of comparison rule B-(ii)-(a)-case-1-aaa}
still hold in this case.
From \reqnarray{proof of adjacent distance larger than one II-(i)-222},
$a+1+j=r_{h-1}$, and \reqnarray{proof of adjacent distance larger than one II-777},
we obtain
\beqnarray{proof of comparison rule B-(ii)-(a)-case-2-222}
r_{h-2}\aligneq \sum_{\ell=1}^{r_{h-1}}m'_{\ell}=\sum_{\ell=1}^{a+1+j}m'_{\ell}
=\sum_{\ell=1}^{a}m'_{\ell}+\sum_{\ell=a+1}^{a+1+j}m'_{\ell}
= i'_{a+1}-1+\sum_{\ell=1}^{j+1}m'_{a+\ell}.
\eeqnarray
From \reqnarray{proof of comparison rule B-(ii)-(a)-case-2-222},
\reqnarray{proof of comparison rule B-(ii)-(a)-case-1-555},
and $i'_{a-j}\geq i'_1=1$, we have
\beqnarray{proof of comparison rule B-(ii)-(a)-case-2-333}
r_{h-2}-i'_{a+1}-1=\sum_{\ell=1}^{j+1}m'_{a+\ell}-2
<i'_{a+1}-i'_{a-j}\leq i'_{a+1}-1.
\eeqnarray
It then follows from \reqnarray{proof of comparison rule B-(ii)-(a)-case-2-333} that
\beqnarray{proof of comparison rule B-(ii)-(a)-case-2-444}
\min\{i'_{a+1}-1,r_{h-2}-i'_{a+1}-1\}=r_{h-2}-i'_{a+1}-1=\sum_{\ell=1}^{j+1}m'_{a+\ell}-2.
\eeqnarray
From \reqnarray{proof of comparison rule B-(ii)-666} (with $j''=j$)
and $a=r_{h-1}-j-1$, we have
\beqnarray{proof of comparison rule B-(ii)-(a)-case-2-555}
(i'_{a+1}+1)+\left(\sum_{\ell=1}^{j}m'_{a+\ell}-1\right)=i'_{a+j+1}=i'_{r_{h-1}},
\eeqnarray
and from \reqnarray{proof of comparison rule B-(ii)-(a)-case-2-222}, we have
\beqnarray{proof of comparison rule B-(ii)-(a)-case-2-666}
(i'_{a+1}+1)+\left(\sum_{\ell=1}^{j+1}m'_{a+\ell}-2\right)=r_{h-2}.
\eeqnarray
Thus, we see from \reqnarray{proof of adjacent distance larger than one II-666},
\reqnarray{proof of comparison rule B-(ii)-(a)-case-2-555},
and \reqnarray{proof of comparison rule B-(ii)-(a)-case-2-666} that
\beqnarray{proof of comparison rule B-(ii)-(a)-case-2-777}
n'_{(i'_{a+1}+1)+j'}=q_{h-1},
\textrm{ for } \sum_{\ell=1}^{j}m'_{a+\ell}\leq j'\leq \sum_{\ell=1}^{j+1}m'_{a+\ell}-2.
\eeqnarray
By combining \reqnarray{proof of comparison rule B-(ii)-222},
\reqnarray{proof of comparison rule B-(ii)-(a)-case-1-aaa},
\reqnarray{proof of comparison rule B-(ii)-(a)-case-2-777},
and \reqnarray{proof of comparison rule B-(ii)-(a)-case-2-444},
we obtain \reqnarray{proof of comparison rule B-(ii)-(a)-case-2-111}.

(b) Now we assume that $m_{a-j}<m_{a+1+j}$ and show that
\beqnarray{proof of comparison rule B-(ii)-(b)-111}
\nbf_1^{r_{h-2}}\succeq {\nbf'}_1^{r_{h-2}},
\eeqnarray
where $\nbf_1^{r_{h-2}}\equiv{\nbf'}_1^{r_{h-2}}$ if and only if
\beqnarray{proof of comparison rule B-(ii)-(b)-222}
a-j=1,\ a+1+j=r_{h-1}, \textrm{ and } m_1=m_{r_{h-1}}-1.
\eeqnarray

\bpdffigure{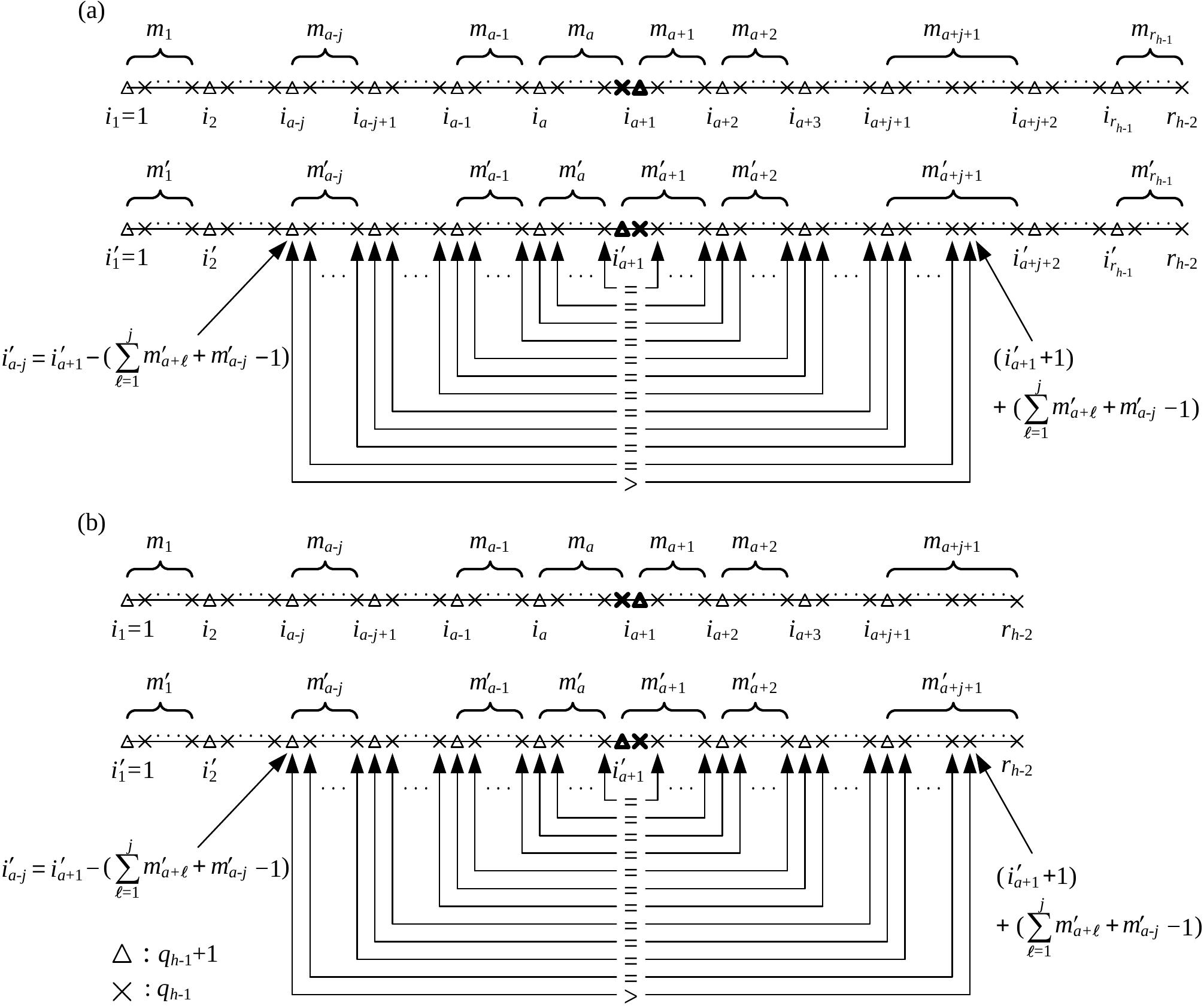}{6.0in}
\epdffigure{appendix-D-(ii)-(b)}
{An illustration of \reqnarray{proof of comparison rule B-(ii)-(b)-333}
and \reqnarray{proof of comparison rule B-(ii)-(b)-444}
in the case that $m_{a-j}<m_{a+1+j}$
(note that in this case we have
$i'_{a+1}-(\sum_{\ell=1}^{j''}m'_{a+\ell}-1)=i'_{a-j''+1}$
for $1\leq j''\leq j$ in \reqnarray{proof of comparison rule B-(ii)-444},
$i'_{a+1}-(\sum_{\ell=1}^{j}m'_{a+\ell}+m'_{a-j}-1)=i'_{a-j}$
in \reqnarray{proof of comparison rule B-(ii)-(b)-999},
and $(i'_{a+1}+1)+(\sum_{\ell=1}^{j''}m'_{a+\ell}-1)=i'_{a+j''+1}$
for $1\leq j''\leq r_{h-1}-a-1$ in \reqnarray{proof of comparison rule B-(ii)-666}):
(a) $j+1\leq a\leq r_{h-1}-j-2$
(note that in this case we have
$(i'_{a+1}+1)+(\sum_{\ell=1}^{j}m'_{a+\ell}+m'_{a-j}-1)<i'_{a+j+2}$
in \reqnarray{proof of comparison rule B-(ii)-(b)-ccc});
(b) $a=r_{h-1}-j-1$
(note that in this case we have
$(i'_{a+1}+1)+(\sum_{\ell=1}^{j}m'_{a+\ell}+m'_{a-j}-1)\leq r_{h-2}$
in \reqnarray{proof of comparison rule B-(ii)-(b)-fff}).}

In the following, we show that
\beqnarray{}
\alignspace
n'_{i'_{a+1}-j'}=n'_{(i'_{a+1}+1)+j'},
\textrm{ for } j'=1,2,\ldots,\sum_{\ell=1}^{j}m'_{a+\ell}+m'_{a-j}-2,
\label{eqn:proof of comparison rule B-(ii)-(b)-333} \\
\alignspace
n'_{i'_{a+1}-(\sum_{\ell=1}^{j}m'_{a+\ell}+m'_{a-j}-1)}=q_{h-1}+1
>n'_{(i'_{a+1}+1)+(\sum_{\ell=1}^{j}m'_{a+\ell}+m'_{a-j}-1)}=q_{h-1}.
\label{eqn:proof of comparison rule B-(ii)-(b)-444}
\eeqnarray
An illustration of \reqnarray{proof of comparison rule B-(ii)-(b)-333}
and \reqnarray{proof of comparison rule B-(ii)-(b)-444}
is given in \rfigure{appendix-D-(ii)-(b)}.
Therefore, it follows from \reqnarray{proof of adjacent distance larger than one II-888},
${\nbf'}_1^{r_{h-2}}\in \Ncal_{M,k}(h-1)$ in
\reqnarray{proof of adjacent distance larger than one II-(i)-333},
\reqnarray{proof of adjacent distance larger than one II-(i)-888}--\reqnarray{proof of adjacent distance larger than one II-(i)-bbb},
\reqnarray{proof of adjacent distance larger than one II-(i)-case-1-222},
\reqnarray{proof of comparison rule B-(ii)-(b)-333}--\reqnarray{proof of comparison rule B-(ii)-(b)-444},
and \reqnarray{comparison rule A-3} in \rlemma{comparison rule A}(ii)
(for the odd integer $h-1$) that ${\nbf'}_1^{r_{h-2}}\preceq\nbf_1^{r_{h-2}}$,
where ${\nbf'}_1^{r_{h-2}}\equiv\nbf_1^{r_{h-2}}$ if and only if
\beqnarray{}
\alignspace i'_{a+1}-\left(\sum_{\ell=1}^{j}m'_{a+\ell}+m'_{a-j}-1\right)=1,
\label{eqn:proof of comparison rule B-(ii)-(b)-555}\\
\alignspace (i'_{a+1}+1)+\left(\sum_{\ell=1}^{j}m'_{a+\ell}+m'_{a-j}-1\right)=r_{h-2},
\label{eqn:proof of comparison rule B-(ii)-(b)-666}\\
\alignspace n'_1=n'_{r_{h-2}}+1.
\label{eqn:proof of comparison rule B-(ii)-(b)-777}
\eeqnarray

To prove \reqnarray{proof of comparison rule B-(ii)-(b)-333}
and \reqnarray{proof of comparison rule B-(ii)-(b)-444},
note that from \reqnarray{proof of comparison rule B-(ii)-444} (with $j''=j$)
and \reqnarray{proof of adjacent distance larger than one II-(i)-hhh}, we have
\beqnarray{}
\alignspace
i'_{a+1}-\sum_{\ell=1}^{j}m'_{a+\ell}=i'_{a-j+1}-1<i'_{a-j+1},
\label{eqn:proof of comparison rule B-(ii)-(b)-888}\\
\alignspace
i'_{a+1}-\left(\sum_{\ell=1}^{j}m'_{a+\ell}+m'_{a-j}-1\right)=i'_{a-j+1}-m'_{a-j}=i'_{a-j}.
\label{eqn:proof of comparison rule B-(ii)-(b)-999}
\eeqnarray
Thus, we see from \reqnarray{proof of adjacent distance larger than one II-666},
\reqnarray{proof of comparison rule B-(ii)-(b)-888},
and \reqnarray{proof of comparison rule B-(ii)-(b)-999} that
\beqnarray{proof of comparison rule B-(ii)-(b)-aaa}
n'_{i'_{a+1}-j'}=
\bselection
q_{h-1}, &\textrm{ for } \sum_{\ell=1}^{j}m'_{a+\ell}\leq j'\leq \sum_{\ell=1}^{j}m'_{a+\ell}+m'_{a-j}-2,\\
q_{h-1}+1, &\textrm{ for } j'=\sum_{\ell=1}^{j}m'_{a+\ell}+m'_{a-j}-1.
\eselection
\eeqnarray
If $j+1\leq a\leq r_{h-1}-j-2$,
then we have from \reqnarray{proof of comparison rule B-(ii)-666}
(with $j''=j$), $m'_{a-j}=m_{a-j}<m_{a+1+j}=m'_{a+1+j}$,
and \reqnarray{proof of adjacent distance larger than one II-(i)-hhh} that
\beqnarray{}
\alignspace \hspace*{-0.2in}
(i'_{a+1}+1)+\sum_{\ell=1}^{j}m'_{a+\ell}=i'_{a+j+1}+1>i'_{a+j+1},
\label{eqn:proof of comparison rule B-(ii)-(b)-bbb}\\
\alignspace \hspace*{-0.2in}
(i'_{a+1}+1)+\left(\sum_{\ell=1}^{j}m'_{a+\ell}+m'_{a-j}-1\right)
=i'_{a+j+1}+m'_{a-j}< i'_{a+j+1}+m'_{a+1+j}=i'_{a+j+2}.
\label{eqn:proof of comparison rule B-(ii)-(b)-ccc}
\eeqnarray
Thus, we see from \reqnarray{proof of adjacent distance larger than one II-666},
\reqnarray{proof of comparison rule B-(ii)-(b)-bbb},
and \reqnarray{proof of comparison rule B-(ii)-(b)-ccc} that
\beqnarray{proof of comparison rule B-(ii)-(b)-ddd}
n'_{(i'_{a+1}+1)+j'}=q_{h-1},
\textrm{ for } \sum_{\ell=1}^{j}m'_{a+\ell}\leq j'\leq \sum_{\ell=1}^{j}m'_{a+\ell}+m'_{a-j}-1.
\eeqnarray
By combining \reqnarray{proof of comparison rule B-(ii)-222},
\reqnarray{proof of comparison rule B-(ii)-(b)-aaa},
and \reqnarray{proof of comparison rule B-(ii)-(b)-ddd},
we obtain \reqnarray{proof of comparison rule B-(ii)-(b)-333}
and \reqnarray{proof of comparison rule B-(ii)-(b)-444}.
On the other hand, if $a=r_{h-1}-j-1$,
then we have from \reqnarray{proof of comparison rule B-(ii)-666} (with $j''=j$) that
\beqnarray{proof of comparison rule B-(ii)-(b)-eee}
(i'_{a+1}+1)+\sum_{\ell=1}^{j}m'_{a+\ell}=i'_{a+j+1}+1=i'_{r_{h-1}}+1>i'_{r_{h-1}},
\eeqnarray
and we have from $m'_{a-j}=m_{a-j}<m_{a+1+j}=m'_{a+1+j}$
and \reqnarray{proof of comparison rule B-(ii)-(a)-case-2-222} that
\beqnarray{proof of comparison rule B-(ii)-(b)-fff}
(i'_{a+1}+1)+\left(\sum_{\ell=1}^{j}m'_{a+\ell}+m'_{a-j}-1\right)
\alignleq i'_{a+1}+\sum_{\ell=1}^{j}m'_{a+\ell}+(m'_{a+1+j}-1) \nn\\
\aligneq i'_{a+1}+\sum_{\ell=1}^{j+1}m'_{a+\ell}-1=r_{h-2},
\eeqnarray
where the equality holds if and only if $m_{a-j}=m_{a+1+j}-1$.
Thus, we see from \reqnarray{proof of adjacent distance larger than one II-666},
\reqnarray{proof of comparison rule B-(ii)-(b)-eee},
and \reqnarray{proof of comparison rule B-(ii)-(b)-fff} that
\beqnarray{proof of comparison rule B-(ii)-(b)-ggg}
n'_{(i'_{a+1}+1)+j'}=q_{h-1},
\textrm{ for }\sum_{\ell=1}^{j}m'_{a+\ell}\leq j'\leq \sum_{\ell=1}^{j}m'_{a+\ell}+m'_{a-j}-1.
\eeqnarray
By combining \reqnarray{proof of comparison rule B-(ii)-222},
\reqnarray{proof of comparison rule B-(ii)-(b)-aaa},
and \reqnarray{proof of comparison rule B-(ii)-(b)-ggg},
we obtain \reqnarray{proof of comparison rule B-(ii)-(b)-333}
and \reqnarray{proof of comparison rule B-(ii)-(b)-444}.

To complete the proof,
we need to show that the condition in
\reqnarray{proof of comparison rule B-(ii)-(b)-555}--\reqnarray{proof of comparison rule B-(ii)-(b)-777}
is equivalent to the condition in \reqnarray{proof of comparison rule B-(ii)-(b)-222}.
Note that if $i'_{a+1}-(\sum_{\ell=1}^{j}m'_{a+\ell}+m'_{a-j}-1)=1$
and $(i'_{a+1}+1)+(\sum_{\ell=1}^{j}m'_{a+\ell}+m'_{a-j}-1)=r_{h-2}$,
then we have from $n'_1=n'_{i'_{a+1}-(\sum_{\ell=1}^{j}m'_{a+\ell}+m'_{a-j}-1)}=q_{h-1}+1$
and $n'_{r_{h-2}}=n'_{(i'_{a+1}+1)+(\sum_{\ell=1}^{j}m'_{a+\ell}+m'_{a-j}-1)}=q_{h-1}$
in \reqnarray{proof of comparison rule B-(ii)-(b)-444} that
\beqnarray{}
n'_1=n'_{r_{h-2}}+1.\nn
\eeqnarray
As such, we see that the condition in
\reqnarray{proof of comparison rule B-(ii)-(b)-555}--\reqnarray{proof of comparison rule B-(ii)-(b)-777}
is equivalent to the following condition:
\beqnarray{proof of comparison rule B-(ii)-(b)-hhh}
i'_{a+1}-\left(\sum_{\ell=1}^{j}m'_{a+\ell}+m'_{a-j}-1\right)=1
\textrm{ and } (i'_{a+1}+1)+\left(\sum_{\ell=1}^{j}m'_{a+\ell}+m'_{a-j}-1\right)=r_{h-2}.
\eeqnarray
As we have $i'_{a+1}-\left(\sum_{\ell=1}^{j}m'_{a+\ell}+m'_{a-j}-1\right)=i'_{a-j}$
in \reqnarray{proof of comparison rule B-(ii)-(b)-999}
and it is clear from \reqnarray{proof of adjacent distance larger than one II-777}
that $i'_{a-j}=1$ if and only if $a-j=1$,
it follows that
\beqnarray{proof of comparison rule B-(ii)-(b)-iii}
i'_{a+1}-\left(\sum_{\ell=1}^{j}m'_{a+\ell}+m'_{a-j}-1\right)=1 \textrm{ iff } a-j=1.
\eeqnarray
Furthermore, if $j+1\leq a\leq r_{h-1}-j-2$,
then we see from \reqnarray{proof of comparison rule B-(ii)-(b)-ccc}
and $i'_{a+j+2}\leq r_{h-2}$ that
\beqnarray{}
(i'_{a+1}+1)+\left(\sum_{\ell=1}^{j}m'_{a+\ell}+m'_{a-j}-1\right)<i'_{a+j+2}\leq r_{h-2}. \nn
\eeqnarray
On the other hand, if $a=r_{h-1}-j-1$, then we see from
\reqnarray{proof of comparison rule B-(ii)-(b)-fff} that
\beqnarray{}
(i'_{a+1}+1)+\left(\sum_{\ell=1}^{j}m'_{a+\ell}+m'_{a-j}-1\right)\leq r_{h-2}, \nn
\eeqnarray
where the equality holds if and only if $m_{a-j}=m_{a+1+j}-1$.
As such, it is easy to see that
\beqnarray{proof of comparison rule B-(ii)-(b)-jjj}
(i'_{a+1}+1)+\left(\sum_{\ell=1}^{j}m'_{a+\ell}+m'_{a-j}-1\right)=r_{h-2}
\textrm{ iff } a=r_{h-1}-j-1 \textrm{ and } m_{a-j}=m_{a+1+j}-1.
\eeqnarray
Therefore, we deduce from \reqnarray{proof of comparison rule B-(ii)-(b)-iii}
and \reqnarray{proof of comparison rule B-(ii)-(b)-jjj}
that the condition in \reqnarray{proof of comparison rule B-(ii)-(b)-hhh}
is equivalent to the following condition:
\beqnarray{}
a-j=1,\ a=r_{h-1}-j-1, \textrm{ and } m_{a-j}=m_{a+1+j}-1, \nn
\eeqnarray
which is clearly equivalent to the condition
that $a-j=1$, $a+1+j=r_{h-1}$, and $m_1=m_{r_{h-1}}-1$
in \reqnarray{proof of comparison rule B-(ii)-(b)-222},
and the proof is completed.

(iii) Note that in \rlemma{comparison rule B}(iii),
we have $2\leq a\leq r_{h-1}-2$
and $m_{a-\ell}=m_{a+1+\ell}$ for $\ell=1,2,\ldots,\min\{a-1,r_{h-1}-a-1\}$.
To show \reqnarray{comparison rule B-4},
i.e., $\mbf_1^{r_{h-1}}\prec{\mbf'}_1^{r_{h-1}}$,
we see from \reqnarray{proof of adjacent distance larger than one II-333}
that it suffices to show that
\beqnarray{proof of comparison rule B-(iii)-111}
\nbf_1^{r_{h-2}}\prec{\nbf'}_1^{r_{h-2}}.
\eeqnarray
Note that as we have $a\leq r_{h-1}-2$,
it follows that \reqnarray{proof of adjacent distance larger than one II-(i)-case-1-222}
in \rappendix{proof of adjacent distance larger than one II for an even integer h
by using comparison rule A for the odd integer h-1} also holds.

\bpdffigure{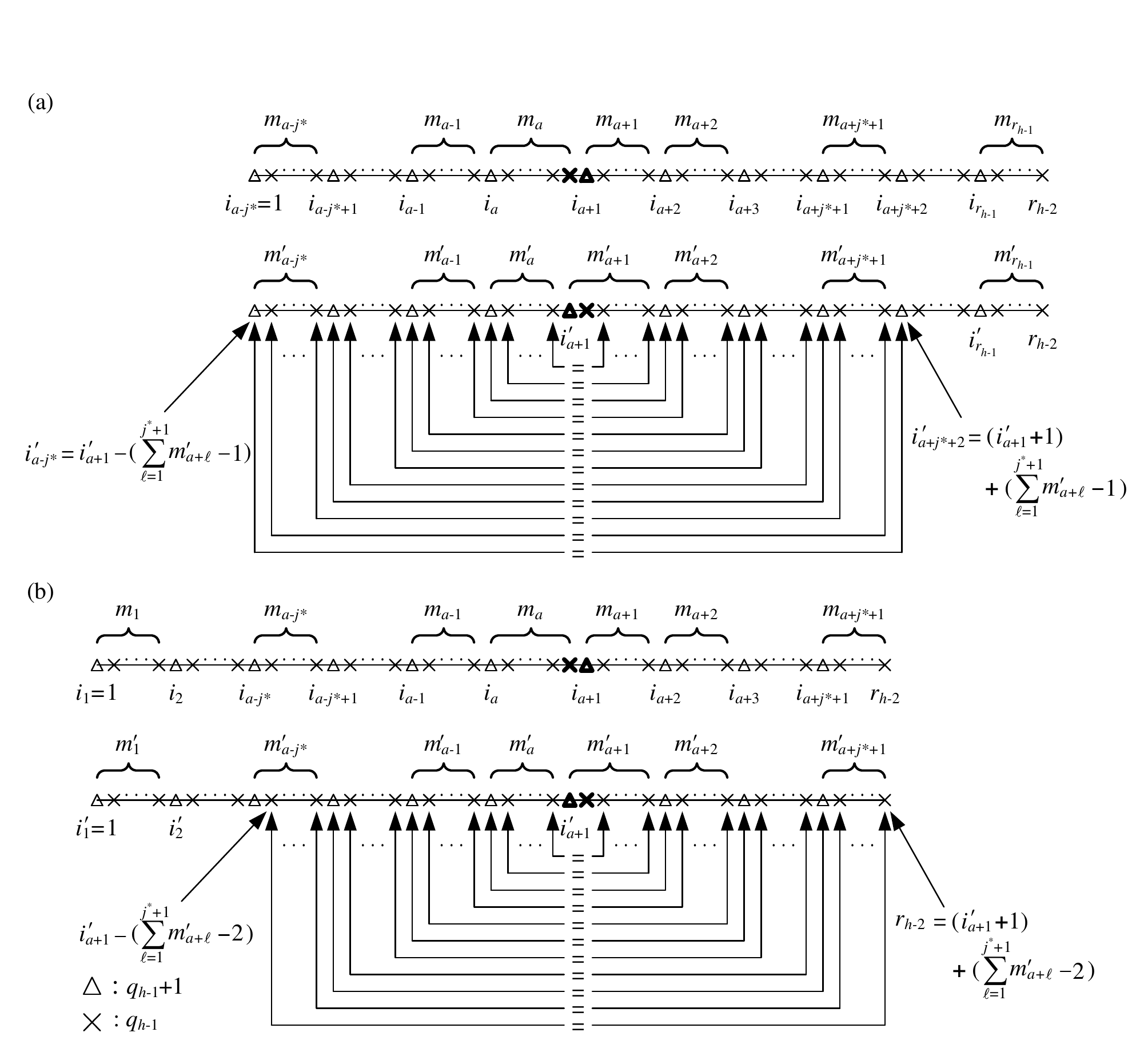}{6.0in}
\epdffigure{appendix-D-(iii)}
{An illustration of \reqnarray{proof of comparison rule B-(iii)-222}
(note that we have
$i'_{a+1}-(\sum_{\ell=1}^{j''}m'_{a+\ell}-1)=i'_{a-j''+1}$
for $1\leq j''\leq j^*+1$ in \reqnarray{proof of comparison rule B-(ii)-444}
and $(i'_{a+1}+1)+(\sum_{\ell=1}^{j''}m'_{a+\ell}-1)=i'_{a+j''+1}$
for $1\leq j''\leq r_{h-1}-a-1$ in \reqnarray{proof of comparison rule B-(ii)-666}):
(a) $a-1<r_{h-1}-a-1$
(note that in this case we have $\min\{i'_{a+1}-1,r_{h-2}-i'_{a+1}-1\}=\sum_{\ell=1}^{j^*+1}m'_{a+\ell}-1$
in \reqnarray{proof of comparison rule B-(iii)-case-1-666},
and $j^*=a-1$ and $j^*+1\leq r_{h-1}-a-1$ in \reqnarray{proof of comparison rule B-(iii)-case-1-111});
(b) $a-1\geq r_{h-1}-a-1$
(note that in this case we have
$\min\{i'_{a+1}-1,r_{h-2}-i'_{a+1}-1\}=\sum_{\ell=1}^{j^*+1}m'_{a+\ell}-2$
in \reqnarray{proof of comparison rule B-(iii)-case-2-333},
$j^*=r_{h-1}-a-1$ in \reqnarray{proof of comparison rule B-(iii)-case-2-111},
and $(i'_{a+1}+1)+(\sum_{\ell=1}^{j^*+1}m'_{a+\ell}-2)=r_{h-2}$
in \reqnarray{proof of comparison rule B-(ii)-(a)-case-2-666}).}

In the following, we show that
\beqnarray{proof of comparison rule B-(iii)-222}
n'_{i'_{a+1}-j'}=n'_{(i'_{a+1}+1)+j'},
\textrm{ for } j'=1,2,\ldots,\min\{i'_{a+1}-1,r_{h-2}-i'_{a+1}-1\}.
\eeqnarray
An illustration of \reqnarray{proof of comparison rule B-(iii)-222}
is given in \rfigure{appendix-D-(iii)}.
Therefore, it follows from
\reqnarray{proof of adjacent distance larger than one II-888},
${\nbf'}_1^{r_{h-2}}\in \Ncal_{M,k}(h-1)$ in
\reqnarray{proof of adjacent distance larger than one II-(i)-333},
\reqnarray{proof of adjacent distance larger than one II-(i)-888}--\reqnarray{proof of adjacent distance larger than one II-(i)-bbb},
\reqnarray{proof of adjacent distance larger than one II-(i)-case-1-222},
\reqnarray{proof of comparison rule B-(iii)-222},
and \reqnarray{comparison rule A-4} in \rlemma{comparison rule A}(iii)
(for the odd integer $h-1$) that ${\nbf'}_1^{r_{h-2}}\succ\nbf_1^{r_{h-2}}$,
i.e., \reqnarray{proof of comparison rule B-(iii)-111} holds.

To prove \reqnarray{proof of comparison rule B-(iii)-222},
let $j^*=\min\{a-1,r_{h-1}-a-1\}$.
Since $m'_{a-\ell}=m_{a-\ell}=m_{a+1+\ell}=m'_{a+1+\ell}$ for $\ell=1,2,\ldots,j^*$,
we see from the same argument as in (ii) above that
\reqnarray{proof of comparison rule B-(ii)-444} holds for $1\leq j''\leq j^*+1$
and \reqnarray{proof of comparison rule B-(ii)-555} holds for $j=j^*+1$.
It is clear that \reqnarray{proof of comparison rule B-(ii)-777} also holds.
We then consider the two cases
$a-1<r_{h-1}-a-1$ and $a-1\geq r_{h-1}-a-1$ separately.

\emph{Case 1: $a-1<r_{h-1}-a-1$.}
In this case, we have
\beqnarray{proof of comparison rule B-(iii)-case-1-111}
j^*=a-1 \textrm{ and } j^*<r_{h-1}-a-1.
\eeqnarray
From \reqnarray{proof of adjacent distance larger than one II-(i)-222},
$j^*\leq r_{h-1}-a-2$ in \reqnarray{proof of comparison rule B-(iii)-case-1-111},
and \reqnarray{proof of adjacent distance larger than one II-777},
we see that
\beqnarray{proof of comparison rule B-(iii)-case-1-222}
r_{h-2}=\sum_{\ell=1}^{r_{h-1}}m'_{\ell}
=\sum_{\ell=1}^{a}m'_{\ell}+\sum_{\ell=a+1}^{r_{h-1}}m'_{\ell}
=(i'_{a+1}-1)+\sum_{\ell=1}^{r_{h-1}-a}m'_{a+\ell}
\geq i'_{a+1}-1+\sum_{\ell=1}^{j^*+2}m'_{a+\ell}.
\eeqnarray
From \reqnarray{proof of comparison rule B-(ii)-444} (with $j''=j^*+1$),
$a-j^*=1$ in \reqnarray{proof of comparison rule B-(iii)-case-1-111},
and $i'_1=1$ in \reqnarray{proof of adjacent distance larger than one II-777}, we have
\beqnarray{}
i'_{a+1}
\aligneq \left(\sum_{\ell=1}^{j^*+1}m'_{a+\ell}-1\right)+i'_{a-j^*}
\label{eqn:proof of comparison rule B-(iii)-case-1-333}\\
\aligneq \sum_{\ell=1}^{j^*+1}m'_{a+\ell}-1+i'_1=\sum_{\ell=1}^{j^*+1}m'_{a+\ell}.
\label{eqn:proof of comparison rule B-(iii)-case-1-444}
\eeqnarray
It follows from \reqnarray{proof of comparison rule B-(iii)-case-1-222}
and \reqnarray{proof of comparison rule B-(iii)-case-1-444} that
\beqnarray{proof of comparison rule B-(iii)-case-1-555}
r_{h-2}-i'_{a+1}-1
\aligngeq \sum_{\ell=1}^{j^*+2}m'_{a+\ell}-2
=\sum_{\ell=1}^{j^*+1}m'_{a+\ell}+m'_{a+j^*+2}-2 \nn\\
\aligneq i'_{a+1}+m'_{a+j^*+2}-2 \nn\\
\aligngeq i'_{a+1}-1.
\eeqnarray
Thus, we see from \reqnarray{proof of comparison rule B-(iii)-case-1-555}
and \reqnarray{proof of comparison rule B-(iii)-case-1-444} that
\beqnarray{proof of comparison rule B-(iii)-case-1-666}
\min\{i'_{a+1}-1,r_{h-2}-i'_{a+1}-1\}=i'_{a+1}-1=\sum_{\ell=1}^{j^*+1}m'_{a+\ell}-1.
\eeqnarray

From \reqnarray{proof of comparison rule B-(ii)-555} (with $j=j^*+1$),
\reqnarray{proof of comparison rule B-(ii)-777},
and $j^*+1\leq r_{h-1}-a-1$ in \reqnarray{proof of comparison rule B-(iii)-case-1-111},
we see that \reqnarray{proof of comparison rule B-(ii)-222}
holds for $j=j^*+1$, i.e.,
\beqnarray{proof of comparison rule B-(iii)-case-1-777}
n'_{i'_{a+1}-j'}=n'_{(i'_{a+1}+1)+j'},
\textrm{ for } j'=1,2,\ldots,\sum_{\ell=1}^{j^*+1}m'_{a+\ell}-1.
\eeqnarray
As such, \reqnarray{proof of comparison rule B-(iii)-222}
follows from \reqnarray{proof of comparison rule B-(iii)-case-1-666}
and \reqnarray{proof of comparison rule B-(iii)-case-1-777}.

\emph{Case 2: $a-1\geq r_{h-1}-a-1$.}
In this case, we have
\beqnarray{proof of comparison rule B-(iii)-case-2-111}
j^*=r_{h-1}-a-1 \textrm{ and } j^*\leq a-1.
\eeqnarray
As \reqnarray{proof of comparison rule B-(ii)-666} holds for $j''=j^*$
and we have $a=r_{h-1}-j^*-1$ in \reqnarray{proof of comparison rule B-(iii)-case-2-111},
it is easy to see that \reqnarray{proof of comparison rule B-(ii)-(a)-case-2-222}
and \reqnarray{proof of comparison rule B-(ii)-(a)-case-2-555}--\reqnarray{proof of comparison rule B-(ii)-(a)-case-2-777}
hold with $j=j^*$.
From \reqnarray{proof of comparison rule B-(ii)-(a)-case-2-222} (with $j=j^*$)
and \reqnarray{proof of comparison rule B-(ii)-444} (with $j=j^*+1$), we have
\beqnarray{proof of comparison rule B-(iii)-case-2-222}
r_{h-2}-i'_{a+1}-1=\sum_{\ell=1}^{j^*+1}m'_{a+\ell}-2
=i'_{a+1}-i'_{a-j^*}-1<i'_{a+1}-1.
\eeqnarray
Thus, we see from \reqnarray{proof of comparison rule B-(iii)-case-2-222} that
\beqnarray{proof of comparison rule B-(iii)-case-2-333}
\min\{i'_{a+1}-1,r_{h-2}-i'_{a+1}-1\}=r_{h-2}-i'_{a+1}-1=\sum_{\ell=1}^{j^*+1}m'_{a+\ell}-2.
\eeqnarray

From \reqnarray{proof of comparison rule B-(ii)-555} (with $j=j^*+1$),
\reqnarray{proof of comparison rule B-(ii)-777} (note that $r_{h-1}-a-1=j^*$),
and \reqnarray{proof of comparison rule B-(ii)-(a)-case-2-777} (with $j=j^*$),
we can see that
\beqnarray{proof of comparison rule B-(iii)-case-2-444}
n'_{i'_{a+1}-j'}=n'_{(i'_{a+1}+1)+j'},
\textrm{ for } j'=1,2,\ldots,\sum_{\ell=1}^{j^*+1}m'_{a+\ell}-2.
\eeqnarray
As such, \reqnarray{proof of comparison rule B-(iii)-222}
follows from \reqnarray{proof of comparison rule B-(iii)-case-2-333}
and \reqnarray{proof of comparison rule B-(iii)-case-2-444}.

\bappendix{Proof of \rlemma{adjacent distance larger than one} for an odd integer $3\leq h\leq N$
by using Comparison rule B in \rlemma{comparison rule B} for the even integer $h-1$}
{proof of adjacent distance larger than one for an odd integer h by using comparison rule B for the even integer h-1}

In this appendix, we assume that Comparison rule B in \rlemma{comparison rule B} holds
for some even integer $h-1$, where $2\leq h-1\leq N-1$,
and show that \rlemma{adjacent distance larger than one} holds for the odd integer $h$.

Let
\beqnarray{}
\nbf_1^{r_{h-2}}(h-1)\aligneq R_{r_{h-3},r_{h-2}}(\nbf_1^{r_{h-1}}(h)),
\label{eqn:proof of adjacent distance larger than one-111} \\
{\nbf'}_1^{r_{h-2}}(h-1)\aligneq R_{r_{h-3},r_{h-2}}({\nbf'}_1^{r_{h-1}}(h)).
\label{eqn:proof of adjacent distance larger than one-222}
\eeqnarray
For simplicity, we let $\mbf_1^{r_{h-1}}=\nbf_1^{r_{h-1}}(h)$,
${\mbf'}_1^{r_{h-1}}={\nbf'}_1^{r_{h-1}}(h)$,
$\nbf_1^{r_{h-2}}=\nbf_1^{r_{h-2}}(h-1)$,
and ${\nbf'}_1^{r_{h-2}}={\nbf'}_1^{r_{h-2}}(h-1)$.
Then we have from \reqnarray{proof of adjacent distance larger than one-111},
\reqnarray{proof of adjacent distance larger than one-222}, and \reqnarray{order relation-777} that
\beqnarray{proof of adjacent distance larger than one-333}
\mbf_1^{r_{h-1}}\prec
(\textrm{resp.}, \equiv, \succ, \preceq, \succeq)\ {\mbf'}_1^{r_{h-1}}
\textrm{ iff } \nbf_1^{r_{h-2}}\prec
(\textrm{resp.}, \equiv, \succ, \preceq, \succeq)\ {\nbf'}_1^{r_{h-2}}.
\eeqnarray
Furthermore, from \reqnarray{proof of adjacent distance larger than one-111},
\reqnarray{proof of adjacent distance larger than one-222},
and the definition of right pre-sequences in \rdefinition{right pre-sequences},
we have
\beqnarray{proof of adjacent distance larger than one-444}
n_i=
\bselection
q_{h-1}+1, &\textrm{if } i=i_1,i_2,\ldots,i_{r_{h-1}}, \\
q_{h-1}, &\textrm{otherwise},
\eselection
\eeqnarray
where
\beqnarray{proof of adjacent distance larger than one-555}
i_j=\sum_{\ell=1}^{j}m_{\ell}, \textrm{ for } j=1,2,\ldots,r_{h-1},
\eeqnarray
and
\beqnarray{proof of adjacent distance larger than one-666}
n'_i=
\bselection
q_{h-1}+1, &\textrm{if } i=i'_1,i'_2,\ldots,i'_{r_{h-1}}, \\
q_{h-1}, &\textrm{otherwise},
\eselection
\eeqnarray
where
\beqnarray{proof of adjacent distance larger than one-777}
i'_j=\sum_{\ell=1}^{j}m'_{\ell}, \textrm{ for } j=1,2,\ldots,r_{h-1}.
\eeqnarray
Note that in \rlemma{adjacent distance larger than one}, we have $r_{h-1}\geq 2$.
As such, it follows from $r_{h-2}>r_{h-1}$ that
\beqnarray{proof of adjacent distance larger than one-888}
r_{h-2}\geq 2.
\eeqnarray

(i) Note that in \rlemma{adjacent distance larger than one}(i),
we have $\mbf_1^{r_{h-1}}\in \Ncal_{M,k}(h)$,
$m_a-m_{a+1}\leq -2$ for some $1\leq a\leq r_{h-1}-1$,
$m'_a=m_a+1$, $m'_{a+1}=m_{a+1}-1$, and $m'_i=m_i$ for $i\neq a$ and $a+1$.
It is easy to see that
\beqnarray{proof of adjacent distance larger than one-(i)-111}
m'_a=m_a+1\geq 2,\ m'_{a+1}=m_{a+1}-1\geq m_{a}+1\geq 2,
\textrm{ and } m'_i=m_i \textrm{ for } i\neq a, a+1.
\eeqnarray
Also, we have from $m'_a=m_a+1$, $m'_{a+1}=m_{a+1}-1$, and $m'_i=m_i$ for $i\neq a$ and $a+1$,
$\mbf_1^{r_{h-1}}\in \Ncal_{M,k}(h)$, and \reqnarray{N-M-k-h} that
\beqnarray{proof of adjacent distance larger than one-(i)-222}
\sum_{i=1}^{r_{h-1}}m'_i=\sum_{i=1}^{r_{h-1}}m_i=r_{h-2}.
\eeqnarray
As such, it follows from \reqnarray{proof of adjacent distance larger than one-(i)-111},
\reqnarray{proof of adjacent distance larger than one-(i)-222},
and \reqnarray{N-M-k-h} that ${\mbf'}_1^{r_{h-1}}\in \Ncal_{M,k}(h)$.

As $\mbf_1^{r_{h-1}}\in \Ncal_{M,k}(h)$ and ${\mbf'}_1^{r_{h-1}}\in \Ncal_{M,k}(h)$,
we see from \reqnarray{proof of adjacent distance larger than one-111},
\reqnarray{proof of adjacent distance larger than one-222},
and the argument in the paragraph after \reqnarray{N-M-k-h} that
\beqnarray{proof of adjacent distance larger than one-(i)-333}
\nbf_1^{r_{h-2}}\in \Ncal_{M,k}(h-1) \textrm{ and } {\nbf'}_1^{r_{h-2}}\in \Ncal_{M,k}(h-1).
\eeqnarray
To show \reqnarray{adjacent distance larger than one-1},
i.e., $\mbf_1^{r_{h-1}}\prec{\mbf'}_1^{r_{h-1}}$,
we see from \reqnarray{proof of adjacent distance larger than one-333}
that it suffices to show that
\beqnarray{proof of adjacent distance larger than one-(i)-444}
\nbf_1^{r_{h-2}}\prec{\nbf'}_1^{r_{h-2}}.
\eeqnarray

\bpdffigure{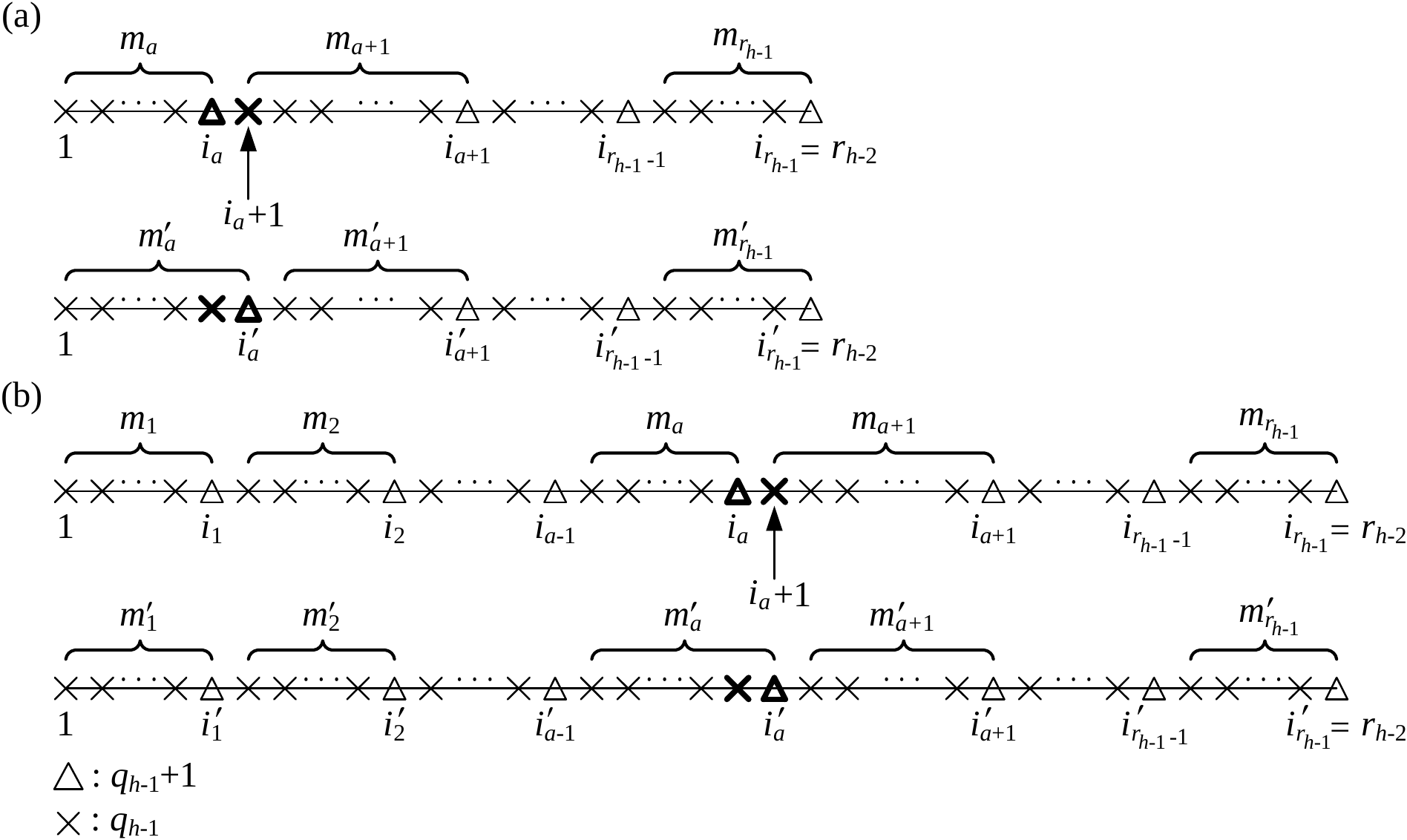}{5.5in}
\epdffigure{appendix-E-(i)}
{An illustration of
\reqnarray{proof of adjacent distance larger than one-(i)-666}--\reqnarray{proof of adjacent distance larger than one-(i)-bbb}:
(a) $a=1$
(note that in this case we have $1\leq i_a<i'_1$
in \reqnarray{proof of adjacent distance larger than one-(i)-ddd}
and $i_a<i_a+1<i_{a+1}$
in \reqnarray{proof of adjacent distance larger than one-(i)-ccc});
(b) $2\leq a\leq r_{h-1}-1$
(note that in this case we have $i'_{a-1}<i_a<i'_a$
in \reqnarray{proof of adjacent distance larger than one-(i)-eee}
and $i_a<i_a+1<i_{a+1}$
in \reqnarray{proof of adjacent distance larger than one-(i)-ccc}).}

Note that from $m_a=m'_a-1$, $m_{a+1}=m'_{a+1}+1$, $m_i=m'_i$ for $i\neq a$ and $a+1$,
\reqnarray{proof of adjacent distance larger than one-555},
and \reqnarray{proof of adjacent distance larger than one-777},
it is easy to see that
\beqnarray{proof of adjacent distance larger than one-(i)-555}
i_j=
\bselection
i'_j-1, &\textrm{if } j=a, \\
i'_j, &\textrm{otherwise}.
\eselection
\eeqnarray
In the following, we show that
\beqnarray{}
\alignspace n'_{i_a}=q_{h-1},
\label{eqn:proof of adjacent distance larger than one-(i)-666}\\
\alignspace n_{i_a+1}=q_{h-1}.
\label{eqn:proof of adjacent distance larger than one-(i)-777}
\eeqnarray
It then follows from \reqnarray{proof of adjacent distance larger than one-444},
\reqnarray{proof of adjacent distance larger than one-666},
and \reqnarray{proof of adjacent distance larger than one-(i)-555}--\reqnarray{proof of adjacent distance larger than one-(i)-777} that
\beqnarray{}
\alignspace n_{i_a}-n_{i_a+1}=(q_{h-1}+1)-q_{h-1}=1,
\label{eqn:proof of adjacent distance larger than one-(i)-888}\\
\alignspace n'_{i_a}=q_{h-1}=n_{i_a}-1,
\label{eqn:proof of adjacent distance larger than one-(i)-999}\\
\alignspace n'_{i_a+1}=n'_{i'_a}=q_{h-1}+1=n_{i_a+1}+1,
\label{eqn:proof of adjacent distance larger than one-(i)-aaa}\\
\alignspace n'_i=n_i, \textrm{ for } i\neq i_a \textrm{ and } i_a+1.
\label{eqn:proof of adjacent distance larger than one-(i)-bbb}
\eeqnarray
An illustration of \reqnarray{proof of adjacent distance larger than one-(i)-666}--\reqnarray{proof of adjacent distance larger than one-(i)-bbb}
is given in \rfigure{appendix-E-(i)}.

To prove \reqnarray{proof of adjacent distance larger than one-(i)-666},
note that if $a=1$, then we have from
\reqnarray{proof of adjacent distance larger than one-(i)-555} that
\beqnarray{proof of adjacent distance larger than one-(i)-ddd}
1\leq i_a=i'_a-1<i'_a=i'_1.
\eeqnarray
Thus, \reqnarray{proof of adjacent distance larger than one-(i)-666}
follows from \reqnarray{proof of adjacent distance larger than one-666}
and $1\leq i_a<i'_1$ in \reqnarray{proof of adjacent distance larger than one-(i)-ddd}.
On the other hand, if $2\leq a\leq r_{h-1}-1$,
then we have from \reqnarray{proof of adjacent distance larger than one-(i)-555} that
\beqnarray{proof of adjacent distance larger than one-(i)-eee}
i_a>i_{a-1}=i'_{a-1} \textrm{ and } i_a=i'_a-1<i'_a.
\eeqnarray
Thus, \reqnarray{proof of adjacent distance larger than one-(i)-666}
follows from \reqnarray{proof of adjacent distance larger than one-666} and $i'_{a-1}<i_a<i'_a$
in \reqnarray{proof of adjacent distance larger than one-(i)-eee}.

To prove \reqnarray{proof of adjacent distance larger than one-(i)-777},
note that from \reqnarray{proof of adjacent distance larger than one-(i)-555} we have
\beqnarray{proof of adjacent distance larger than one-(i)-ccc}
i_a<i_a+1=i'_a<i'_{a+1}=i_{a+1}.
\eeqnarray
Thus, \reqnarray{proof of adjacent distance larger than one-(i)-777}
follows from \reqnarray{proof of adjacent distance larger than one-444}
and $i_a<i_a+1<i_{a+1}$ in \reqnarray{proof of adjacent distance larger than one-(i)-ccc}.

Note that from \reqnarray{proof of adjacent distance larger than one-555},
$a\leq r_{h-1}-1$,
$\sum_{\ell=1}^{r_{h-1}}m_{\ell}=r_{h-2}$ in \reqnarray{proof of adjacent distance larger than one-(i)-222},
and $m_a-m_{a+1}\leq -2$, we have
\beqnarray{proof of adjacent distance larger than one-(i)-fff}
i_a=\sum_{\ell=1}^{a}m_{\ell}=\sum_{\ell=1}^{r_{h-1}}m_{\ell}-\sum_{\ell=a+1}^{r_{h-1}}m_{\ell}
\leq r_{h-2}-m_{a+1} \leq r_{h-2}-m_a-2\leq r_{h-2}-2,
\eeqnarray
and
\beqnarray{proof of adjacent distance larger than one-(i)-ggg}
i_j=\sum_{\ell=1}^{j}m_{\ell}=\sum_{\ell=1}^{j-1}m_{\ell}+m_j=i_{j-1}+m_j,
\textrm{ for } j=2,3,\ldots,r_{h-1}.
\eeqnarray
We then consider the two cases $a=1$ and $2\leq a\leq r_{h-1}-1$ separately.

\emph{Case 1: $a=1$}.
\bpdffigure{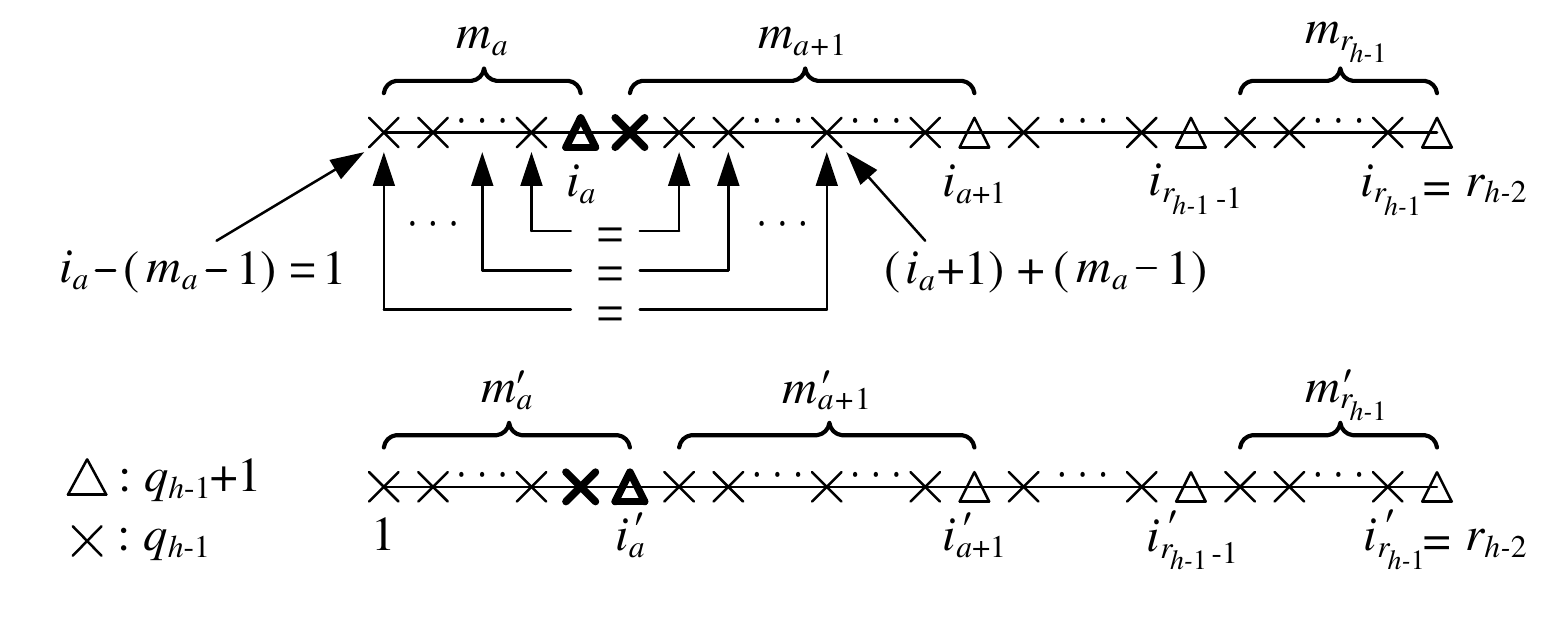}{4.0in}
\epdffigure{appendix-E-(i)-case-1}
{An illustration of \reqnarray{proof of adjacent distance larger than one-(i)-case-1-333}
for the case that $a=1$ and $m_a\geq 2$
(note that in this case we have
$\min\{i_a-1,r_{h-2}-i_a-1\}=m_a-1$
in \reqnarray{proof of adjacent distance larger than one-(i)-case-1-555},
$i_a-(m_a-1)=1$
in \reqnarray{proof of adjacent distance larger than one-(i)-case-1-666},
and $(i_a+1)-(m_a-1)<i_{a+1}$
in \reqnarray{proof of adjacent distance larger than one-(i)-case-1-999}).}

In this case, we have from
\reqnarray{proof of adjacent distance larger than one-555} and $a=1$ that
\beqnarray{proof of adjacent distance larger than one-(i)-case-1-111}
i_a=\sum_{\ell=1}^{a}m_{\ell}=m_1=m_a.
\eeqnarray
If $m_a=1$, then we have from \reqnarray{proof of adjacent distance larger than one-(i)-case-1-111} that $i_a=1$.
Therefore, it follows from
\reqnarray{proof of adjacent distance larger than one-888},
$\nbf_1^{r_{h-2}}\in \Ncal_{M,k}(h-1)$ in
\reqnarray{proof of adjacent distance larger than one-(i)-333},
\reqnarray{proof of adjacent distance larger than one-(i)-888}--\reqnarray{proof of adjacent distance larger than one-(i)-bbb},
$i_a=1$, and \reqnarray{comparison rule B-1} in \rlemma{comparison rule B}(i)
(for the even integer $h-1$) that $\nbf_1^{r_{h-2}}\prec{\nbf'}_1^{r_{h-2}}$,
i.e., \reqnarray{proof of adjacent distance larger than one-(i)-444} holds.

On the other hand, if $m_a\geq 2$,
then we show that
\beqnarray{}
\alignspace
2\leq i_a\leq r_{h-2}-2,
\label{eqn:proof of adjacent distance larger than one-(i)-case-1-222}\\
\alignspace
n_{i_a-j}=n_{(i_a+1)+j}=q_{h-1}, \textrm{ for } j=1,2,\ldots,\min\{i_a-1,r_{h-2}-i_a-1\}.
\label{eqn:proof of adjacent distance larger than one-(i)-case-1-333}
\eeqnarray
An illustration of \reqnarray{proof of adjacent distance larger than one-(i)-case-1-333}
is given in \rfigure{appendix-E-(i)-case-1}.
Therefore, it follows from
\reqnarray{proof of adjacent distance larger than one-888},
$\nbf_1^{r_{h-2}}\in \Ncal_{M,k}(h-1)$ in
\reqnarray{proof of adjacent distance larger than one-(i)-333},
\reqnarray{proof of adjacent distance larger than one-(i)-888}--\reqnarray{proof of adjacent distance larger than one-(i)-bbb},
\reqnarray{proof of adjacent distance larger than one-(i)-case-1-222},
\reqnarray{proof of adjacent distance larger than one-(i)-case-1-333},
and \reqnarray{comparison rule B-4} in \rlemma{comparison rule B}(iii)
(for the even integer $h-1$) that $\nbf_1^{r_{h-2}}\prec{\nbf'}_1^{r_{h-2}}$,
i.e., \reqnarray{proof of adjacent distance larger than one-(i)-444} holds.

To prove \reqnarray{proof of adjacent distance larger than one-(i)-case-1-222},
note that from \reqnarray{proof of adjacent distance larger than one-(i)-case-1-111} and $m_a\geq 2$,
we have $i_a\geq 2$.
Thus, \reqnarray{proof of adjacent distance larger than one-(i)-case-1-222}
follows from $i_a\geq 2$ and \reqnarray{proof of adjacent distance larger than one-(i)-fff}.

To prove \reqnarray{proof of adjacent distance larger than one-(i)-case-1-333},
note that from \reqnarray{proof of adjacent distance larger than one-(i)-case-1-111},
$\sum_{\ell=1}^{r_{h-1}}m_{\ell}=r_{h-2}$ in \reqnarray{proof of adjacent distance larger than one-(i)-222},
and $m_a-m_{a+1}\leq -2$, we have
\beqnarray{proof of adjacent distance larger than one-(i)-case-1-444}
(i_a-1)-(r_{h-2}-i_a-1)
\aligneq 2i_a-r_{h-2}=2m_a-\sum_{\ell=1}^{r_{h-1}}m_{\ell} \nn\\
\alignleq 2m_a-(m_a+m_{a+1})=m_a-m_{a+1}<0.
\eeqnarray
It then follows from \reqnarray{proof of adjacent distance larger than one-(i)-case-1-444}
and \reqnarray{proof of adjacent distance larger than one-(i)-case-1-111} that
\beqnarray{proof of adjacent distance larger than one-(i)-case-1-555}
\min\{i_a-1,r_{h-2}-i_a-1\}=i_a-1=m_a-1.
\eeqnarray
From \reqnarray{proof of adjacent distance larger than one-(i)-case-1-111},
we have
\beqnarray{proof of adjacent distance larger than one-(i)-case-1-666}
i_a-(m_a-1)=1.
\eeqnarray
From \reqnarray{proof of adjacent distance larger than one-(i)-case-1-666}
and $a=1$, we immediately see that
\beqnarray{proof of adjacent distance larger than one-(i)-case-1-777}
1\leq i_a-j<i_a=i_1, \textrm{ for } j=1,2,\ldots,m_a-1.
\eeqnarray
It then follows from \reqnarray{proof of adjacent distance larger than one-444}
and \reqnarray{proof of adjacent distance larger than one-(i)-case-1-777} that
\beqnarray{proof of adjacent distance larger than one-(i)-case-1-888}
n_{i_a-j}=q_{h-1}, \textrm{ for } j=1,2,\ldots,m_a-1.
\eeqnarray
Also, we have from $m_a-m_{a+1}\leq -2$ and
$i_a+m_{a+1}=i_{a+1}$ in \reqnarray{proof of adjacent distance larger than one-(i)-ggg} that
\beqnarray{proof of adjacent distance larger than one-(i)-case-1-999}
(i_a+1)+(m_a-1)\leq i_a+m_{a+1}-2=i_{a+1}-2<i_{a+1}.
\eeqnarray
It then follows from \reqnarray{proof of adjacent distance larger than one-444}
and $i_a<(i_a+1)+(m_a-1)<i_{a+1}$ in
\reqnarray{proof of adjacent distance larger than one-(i)-case-1-999} that
\beqnarray{proof of adjacent distance larger than one-(i)-case-1-aaa}
\alignspace n_{(i_a+1)+j}=q_{h-1}, \textrm{ for } j=1,2,\ldots,m_a-1.
\eeqnarray
Thus, \reqnarray{proof of adjacent distance larger than one-(i)-case-1-333}
follows from \reqnarray{proof of adjacent distance larger than one-(i)-case-1-555},
\reqnarray{proof of adjacent distance larger than one-(i)-case-1-888},
and \reqnarray{proof of adjacent distance larger than one-(i)-case-1-aaa}.

\textbf{Remark:} It is to be noted that if $m_a-m_{a+1}=-1$
(instead of $m_a-m_{a+1}\leq -2$ as in \rlemma{adjacent distance larger than one}(i)),
then \reqnarray{proof of adjacent distance larger than one-(i)-fff}--\reqnarray{proof of adjacent distance larger than one-(i)-case-1-aaa}
still hold (as we only need $m_a-m_{a+1}\leq -1$ to prove
\reqnarray{proof of adjacent distance larger than one-(i)-fff},
\reqnarray{proof of adjacent distance larger than one-(i)-case-1-444},
and \reqnarray{proof of adjacent distance larger than one-(i)-case-1-999}),
and hence $\nbf_1^{r_{h-2}}\prec{\nbf'}_1^{r_{h-2}}$ in
\reqnarray{proof of adjacent distance larger than one-(i)-444} also holds in such a case.
As such, we conclude that if $3\leq h\leq N$ is an odd integer,
$\mbf_1^{r_{h-1}}\in \Ncal_{M,k}(h)$, $m_a-m_{a+1}=-1$,
$m'_a=m_a+1$, $m'_{a+1}=m_{a+1}-1$, and $m'_i=m_i$ for $i\neq a$ and $a+1$,
where $a=1$, then we have $\mbf_1^{r_{h-1}}\prec{\mbf'}_1^{r_{h-1}}$.
This result will be used later in the proof of Case 1 of \rlemma{comparison rule A}(i) in
\rappendix{proof of comparison rule A for an odd integer h by using comparison rule B for the even integer h-1}.

\emph{Case 2: $2\leq a\leq r_{h-1}-1$}.
First note that from $a\geq 2$ and \reqnarray{proof of adjacent distance larger than one-555},
we have
\beqnarray{proof of adjacent distance larger than one-(i)-case-2-111}
i_a=\sum_{\ell=1}^{a}m_a\geq m_1+m_a\geq 2.
\eeqnarray
As such, we see from \reqnarray{proof of adjacent distance larger than one-(i)-fff}
and \reqnarray{proof of adjacent distance larger than one-(i)-case-2-111} that
\beqnarray{proof of adjacent distance larger than one-(i)-case-2-222}
2\leq i_a\leq r_{h-2}-2.
\eeqnarray

\bpdffigure{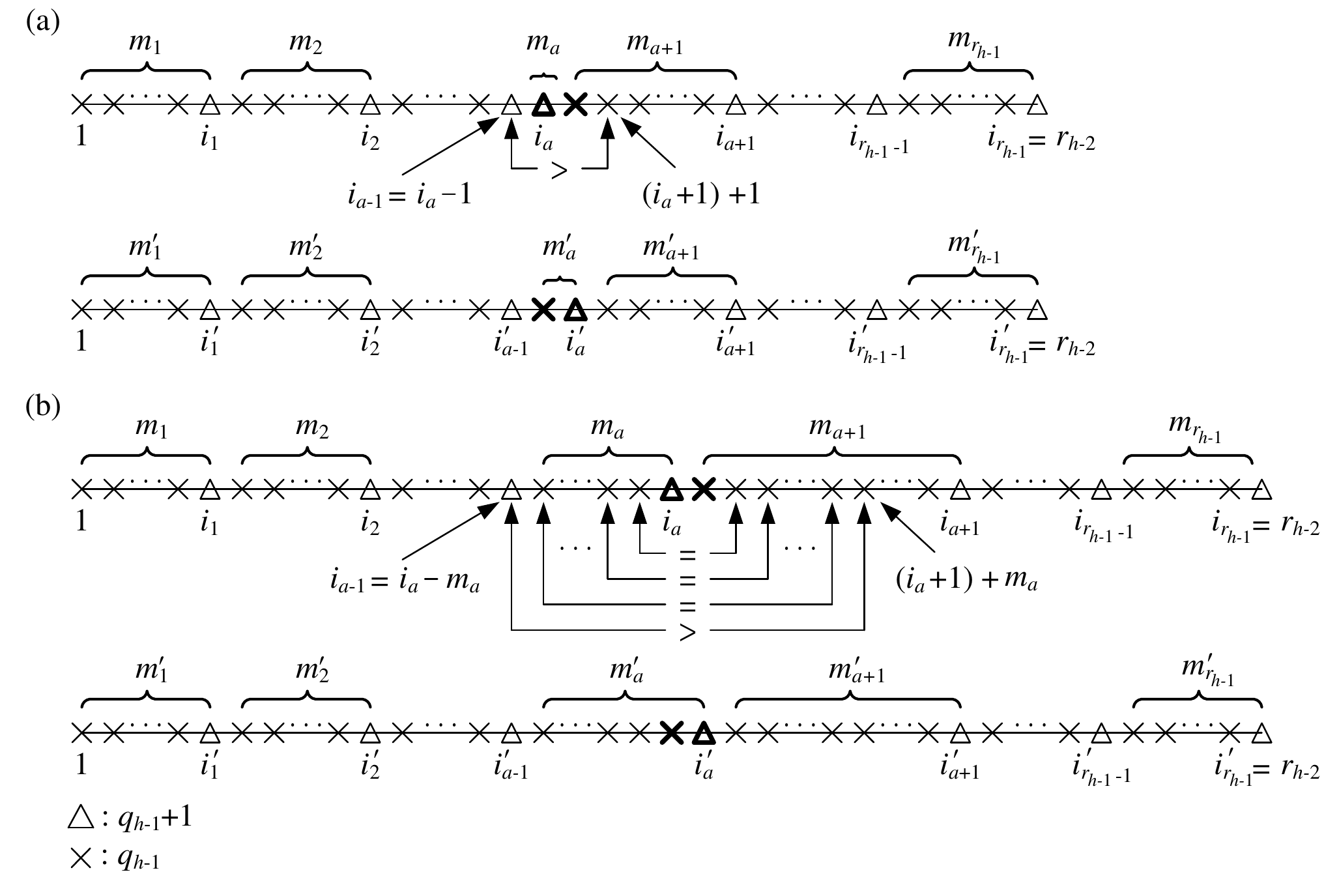}{5.5in}
\epdffigure{appendix-E-(i)-case-2}
{(a) An illustration of \reqnarray{proof of adjacent distance larger than one-(i)-case-2-333}
for the case that $2\leq a\leq r_{h-1}-1$ and $m_a=1$
(note that in this case we have $i_{a-1}=i_a-1$
in \reqnarray{proof of adjacent distance larger than one-(i)-case-2-444}
and $i_a<(i_a+1)+1<i_{a+1}$
in \reqnarray{proof of adjacent distance larger than one-(i)-case-2-666});
(b) An illustration of \reqnarray{proof of adjacent distance larger than one-(i)-case-2-888}
and \reqnarray{proof of adjacent distance larger than one-(i)-case-2-999}
for the case that $2\leq a\leq r_{h-1}-1$ and $m_a\geq 2$
(note that in this case we have $i_{a-1}=i_a-m_a$
in \reqnarray{proof of adjacent distance larger than one-(i)-case-2-aaa}
and $i_a<(i_a+1)+m_a<i_{a+1}$
in \reqnarray{proof of adjacent distance larger than one-(i)-case-2-ddd}).}

If $m_a=1$, then we show that
\beqnarray{proof of adjacent distance larger than one-(i)-case-2-333}
n_{i_a-1}=q_{h-1}+1>n_{(i_a+1)+1}=q_{h-1}.
\eeqnarray
An illustration of \reqnarray{proof of adjacent distance larger than one-(i)-case-2-333}
is given in \rfigure{appendix-E-(i)-case-2}(a).
Therefore, it follows from
\reqnarray{proof of adjacent distance larger than one-888},
$\nbf_1^{r_{h-2}}\in \Ncal_{M,k}(h-1)$ in
\reqnarray{proof of adjacent distance larger than one-(i)-333},
\reqnarray{proof of adjacent distance larger than one-(i)-888}--\reqnarray{proof of adjacent distance larger than one-(i)-bbb},
\reqnarray{proof of adjacent distance larger than one-(i)-case-2-222},
\reqnarray{proof of adjacent distance larger than one-(i)-case-2-333},
and \reqnarray{comparison rule B-2} in \rlemma{comparison rule B}(ii)
(for the even integer $h-1$) that $\nbf_1^{r_{h-2}}\prec{\nbf'}_1^{r_{h-2}}$,
i.e., \reqnarray{proof of adjacent distance larger than one-(i)-444} holds.

To prove \reqnarray{proof of adjacent distance larger than one-(i)-case-2-333},
note that from $m_a=1$, \reqnarray{proof of adjacent distance larger than one-(i)-ggg},
and $2\leq a\leq r_{h-1}-1$, we have
\beqnarray{proof of adjacent distance larger than one-(i)-case-2-444}
i_a-1=i_a-m_a=i_{a-1}.
\eeqnarray
It follows from \reqnarray{proof of adjacent distance larger than one-444}
and $i_a-1=i_{a-1}$ in
\reqnarray{proof of adjacent distance larger than one-(i)-case-2-444} that
\beqnarray{proof of adjacent distance larger than one-(i)-case-2-555}
n_{i_a-1}=n_{i_{a-1}}=q_{h-1}+1.
\eeqnarray
Also, from $m_{a+1}\geq 3$ (as $m_a-m_{a+1}\leq -2$),
\reqnarray{proof of adjacent distance larger than one-(i)-ggg},
and $2\leq a\leq r_{h-1}-1$,
we have
\beqnarray{proof of adjacent distance larger than one-(i)-case-2-666}
i_a<(i_a+1)+1<i_a+m_{a+1}=i_{a+1}.
\eeqnarray
It follows from \reqnarray{proof of adjacent distance larger than one-444}
and $i_a<(i_a+1)+1<i_{a+1}$ in \reqnarray{proof of adjacent distance larger than one-(i)-case-2-666} that
\beqnarray{proof of adjacent distance larger than one-(i)-case-2-777}
n_{(i_a+1)+1}=q_{h-1}.
\eeqnarray
Thus, \reqnarray{proof of adjacent distance larger than one-(i)-case-2-333}
follows from \reqnarray{proof of adjacent distance larger than one-(i)-case-2-555}
and \reqnarray{proof of adjacent distance larger than one-(i)-case-2-777}.

On the other hand, if $m_a\geq 2$,
then we show that
\beqnarray{}
\alignspace n_{i_a-j}=n_{(i_a+1)+j}=q_{h-1}, \textrm{ for } j=1,2,\ldots,m_a-1,
\label{eqn:proof of adjacent distance larger than one-(i)-case-2-888} \\
\alignspace n_{i_a-m_a}=q_{h-1}+1>n_{(i_a+1)+m_a}=q_{h-1},
\label{eqn:proof of adjacent distance larger than one-(i)-case-2-999}
\eeqnarray
An illustration of \reqnarray{proof of adjacent distance larger than one-(i)-case-2-888}
and \reqnarray{proof of adjacent distance larger than one-(i)-case-2-999}
is given in \rfigure{appendix-E-(i)-case-2}(b).
Therefore, it follows from
\reqnarray{proof of adjacent distance larger than one-888},
$\nbf_1^{r_{h-2}}\in \Ncal_{M,k}(h-1)$ in
\reqnarray{proof of adjacent distance larger than one-(i)-333},
\reqnarray{proof of adjacent distance larger than one-(i)-888}--\reqnarray{proof of adjacent distance larger than one-(i)-bbb},
\reqnarray{proof of adjacent distance larger than one-(i)-case-2-222},
\reqnarray{proof of adjacent distance larger than one-(i)-case-2-888},
\reqnarray{proof of adjacent distance larger than one-(i)-case-2-999},
and \reqnarray{comparison rule B-2} in \rlemma{comparison rule B}(ii)
(for the even integer $h-1$) that $\nbf_1^{r_{h-2}}\prec{\nbf'}_1^{r_{h-2}}$,
i.e., \reqnarray{proof of adjacent distance larger than one-(i)-444} holds.

To prove \reqnarray{proof of adjacent distance larger than one-(i)-case-2-888}
and \reqnarray{proof of adjacent distance larger than one-(i)-case-2-999},
observe from \reqnarray{proof of adjacent distance larger than one-(i)-ggg}
and $2\leq a\leq r_{h-1}-1$ that
\beqnarray{proof of adjacent distance larger than one-(i)-case-2-aaa}
i_a-m_a=i_{a-1}.
\eeqnarray
It then follows from \reqnarray{proof of adjacent distance larger than one-444}
and \reqnarray{proof of adjacent distance larger than one-(i)-case-2-aaa} that
\beqnarray{}
\alignspace n_{i_a-j}=q_{h-1}, \textrm{ for } j=1,2,\ldots,m_a-1,
\label{eqn:proof of adjacent distance larger than one-(i)-case-2-bbb} \\
\alignspace n_{i_a-m_a}=n_{i_{a-1}}=q_{h-1}+1.
\label{eqn:proof of adjacent distance larger than one-(i)-case-2-ccc}
\eeqnarray
Also, from $m_a-m_{a+1}\leq -2$,
\reqnarray{proof of adjacent distance larger than one-(i)-ggg},
and $2\leq a\leq r_{h-1}-1$, we see that
\beqnarray{proof of adjacent distance larger than one-(i)-case-2-ddd}
(i_a+1)+m_a\leq i_a+m_{a+1}-1=i_{a+1}-1<i_{a+1}.
\eeqnarray
From \reqnarray{proof of adjacent distance larger than one-(i)-case-2-ddd},
we immediately see that
\beqnarray{proof of adjacent distance larger than one-(i)-case-2-eee}
i_a<(i_a+1)+j<i_{a+1}, \textrm{ for } j=1,2,\ldots,m_a.
\eeqnarray
It then follows from \reqnarray{proof of adjacent distance larger than one-444}
and \reqnarray{proof of adjacent distance larger than one-(i)-case-2-eee} that
\beqnarray{proof of adjacent distance larger than one-(i)-case-2-fff}
n_{(i_a+1)+j}=q_{h-1}, \textrm{ for } j=1,2,\ldots,m_a.
\eeqnarray
Thus, \reqnarray{proof of adjacent distance larger than one-(i)-case-2-888}
and \reqnarray{proof of adjacent distance larger than one-(i)-case-2-999}
follow from \reqnarray{proof of adjacent distance larger than one-(i)-case-2-bbb},
\reqnarray{proof of adjacent distance larger than one-(i)-case-2-ccc},
and \reqnarray{proof of adjacent distance larger than one-(i)-case-2-fff}.

(ii) Note that in \rlemma{adjacent distance larger than one}(ii),
we have $\mbf_1^{r_{h-1}}\in \Ncal_{M,k}(h)$,
$m_a-m_{a+1}\geq 2$ for some $1\leq a\leq r_{h-1}-1$,
$m'_a=m_a-1$, $m'_{a+1}=m_{a+1}+1$, and $m'_i=m_i$ for $i\neq a$ and $a+1$.
It is easy to see that
\beqnarray{proof of adjacent distance larger than one-(ii)-111}
m'_a=m_a-1\geq m_{a+1}+1\geq 2,\ m'_{a+1}=m_{a+1}+1\geq 2,
\textrm{ and } m'_i=m_i \textrm{ for } i\neq a, a+1.
\eeqnarray
Also, we have from $m'_a=m_a-1$, $m'_{a+1}=m_{a+1}+1$, and $m'_i=m_i$ for $i\neq a$ and $a+1$,
$\mbf_1^{r_{h-1}}\in \Ncal_{M,k}(h)$, and \reqnarray{N-M-k-h} that
\beqnarray{proof of adjacent distance larger than one-(ii)-222}
\sum_{i=1}^{r_{h-1}}m'_i=\sum_{i=1}^{r_{h-1}}m_i=r_{h-2}.
\eeqnarray
As such, it follows from \reqnarray{proof of adjacent distance larger than one-(ii)-111},
\reqnarray{proof of adjacent distance larger than one-(ii)-222},
and \reqnarray{N-M-k-h} that ${\mbf'}_1^{r_{h-1}}\in \Ncal_{M,k}(h)$.

As $\mbf_1^{r_{h-1}}\in \Ncal_{M,k}(h)$ and ${\mbf'}_1^{r_{h-1}}\in \Ncal_{M,k}(h)$,
we see from \reqnarray{proof of adjacent distance larger than one-111},
\reqnarray{proof of adjacent distance larger than one-222},
and the argument in the paragraph after \reqnarray{N-M-k-h} that
\beqnarray{proof of adjacent distance larger than one-(ii)-333}
\nbf_1^{r_{h-2}}\in \Ncal_{M,k}(h-1) \textrm{ and } {\nbf'}_1^{r_{h-2}}\in \Ncal_{M,k}(h-1).
\eeqnarray
To show \reqnarray{adjacent distance larger than one-2},
i.e., $\mbf_1^{r_{h-1}}\preceq {\mbf'}_1^{r_{h-1}}$,
where $\mbf_1^{r_{h-1}}\equiv {\mbf'}_1^{r_{h-1}}$ if and only if
$r_{h-1}=2$ and $m_1=m_2+2$,
we see from \reqnarray{proof of adjacent distance larger than one-333}
that it suffices to show that
\beqnarray{proof of adjacent distance larger than one-(ii)-444}
\nbf_1^{r_{h-2}}\preceq {\nbf'}_1^{r_{h-2}},
\eeqnarray
where $\nbf_1^{r_{h-2}}\equiv {\nbf'}_1^{r_{h-2}}$ if and only if
\beqnarray{proof of adjacent distance larger than one-(ii)-555}
r_{h-1}=2 \textrm{ and } m_1=m_2+2.
\eeqnarray

Note that from $m_a=m'_a+1$, $m_{a+1}=m'_{a+1}-1$, $m_i=m'_i$ for $i\neq a$ and $a+1$,
\reqnarray{proof of adjacent distance larger than one-444}--\reqnarray{proof of adjacent distance larger than one-888},
we can show as in the proof of
\reqnarray{proof of adjacent distance larger than one-(i)-666}--\reqnarray{proof of adjacent distance larger than one-(i)-bbb}
in (i) above (with the roles of $\mbf_1^{r_{h-1}}$ and ${\mbf'}_1^{r_{h-1}}$ interchanged
and the roles of $\nbf_1^{r_{h-2}}$ and ${\nbf'}_1^{r_{h-2}}$ interchanged) that
\beqnarray{}
\alignspace n_{i'_a}=q_{h-1},
\label{eqn:proof of adjacent distance larger than one-(ii)-666} \\
\alignspace n'_{i'_a+1}=q_{h-1},
\label{eqn:proof of adjacent distance larger than one-(ii)-777}\\
\alignspace n'_{i'_a}-n'_{i'_a+1}=(q_{h-1}+1)-q_{h-1}=1,
\label{eqn:proof of adjacent distance larger than one-(ii)-888}\\
\alignspace n_{i'_a}=q_{h-1}=n'_{i'_a}-1,
\label{eqn:proof of adjacent distance larger than one-(ii)-999}\\
\alignspace n_{i'_a+1}=n_{i_a}=q_{h-1}+1=n'_{i'_a+1}+1,
\label{eqn:proof of adjacent distance larger than one-(ii)-aaa}\\
\alignspace n_i=n'_i, \textrm{ for } i\neq i'_a \textrm{ and } i'_a+1.
\label{eqn:proof of adjacent distance larger than one-(ii)-bbb}
\eeqnarray

\bpdffigure{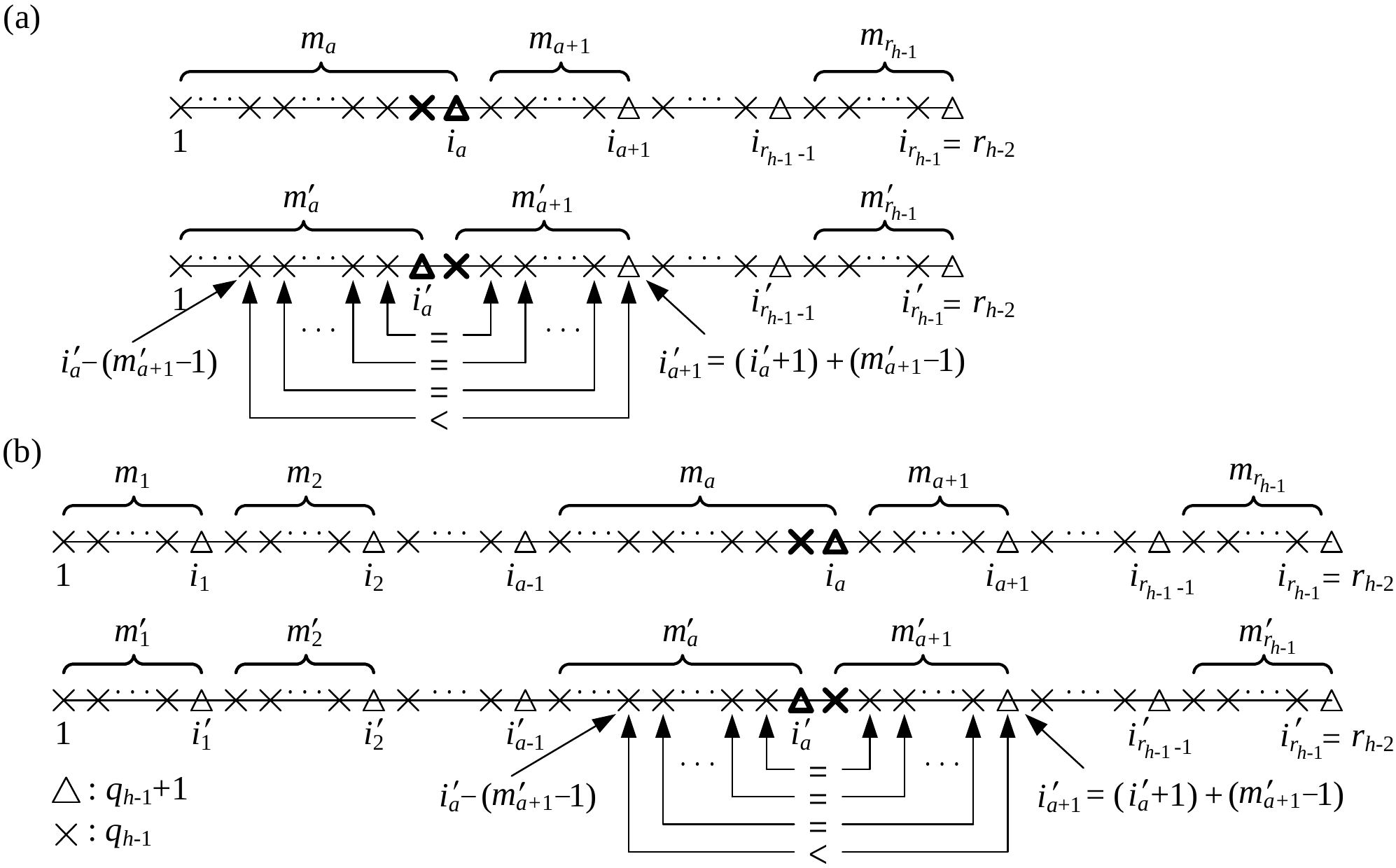}{5.5in}
\epdffigure{appendix-E-(ii)}
{An illustration of \reqnarray{proof of adjacent distance larger than one-(ii)-eee}
and \reqnarray{proof of adjacent distance larger than one-(ii)-fff}:
(a) $a=1$
(note that in this case we have $1\leq i'_a-(m'_{a+1}-1)<i'_1$
in \reqnarray{proof of adjacent distance larger than one-(ii)-8888}
and $(i'_a+1)+(m'_{a+1}-1)=i'_{a+1}$
in \reqnarray{proof of adjacent distance larger than one-(ii)-3333});
(b) $2\leq a\leq r_{h-1}-1$
(note that in this case we have $i'_{a-1}<i'_a-(m'_{a+1}-1)<i'_a$
in \reqnarray{proof of adjacent distance larger than one-(ii)-bbbb}
and $(i'_a+1)+(m'_{a+1}-1)=i'_{a+1}$
in \reqnarray{proof of adjacent distance larger than one-(ii)-3333}).}

In the following, we show that
\beqnarray{}
\alignspace 2\leq i'_a\leq r_{h-2}-2,
\label{eqn:proof of adjacent distance larger than one-(ii)-ccc}\\
\alignspace 1\leq m'_{a+1}-1\leq \min\{i'_a-1,r_{h-2}-i'_a-1\},
\label{eqn:proof of adjacent distance larger than one-(ii)-ddd}\\
\alignspace n'_{i'_a-j'}=n'_{(i'_a+1)+j'}=q_{h-1}, \textrm{ for } j'=1,2,\ldots,m'_{a+1}-2,
\label{eqn:proof of adjacent distance larger than one-(ii)-eee}\\
\alignspace n'_{i'_a-(m'_{a+1}-1)}=q_{h-1}<n'_{(i'_a+1)+(m'_{a+1}-1)}=q_{h-1}+1.
\label{eqn:proof of adjacent distance larger than one-(ii)-fff}
\eeqnarray
An illustration of \reqnarray{proof of adjacent distance larger than one-(ii)-eee}
and \reqnarray{proof of adjacent distance larger than one-(ii)-fff}
is given in \rfigure{appendix-E-(ii)}.
Therefore, it follows from
\reqnarray{proof of adjacent distance larger than one-888},
${\nbf'}_1^{r_{h-2}}\in \Ncal_{M,k}(h-1)$ in
\reqnarray{proof of adjacent distance larger than one-(ii)-333},
\reqnarray{proof of adjacent distance larger than one-(ii)-888}--\reqnarray{proof of adjacent distance larger than one-(ii)-fff},
and \reqnarray{comparison rule B-3} in \rlemma{comparison rule B}(ii)
(for the even integer $h-1$) that ${\nbf'}_1^{r_{h-2}}\succeq\nbf_1^{r_{h-2}}$,
where ${\nbf'}_1^{r_{h-2}}\equiv\nbf_1^{r_{h-2}}$ if and only if
\beqnarray{proof of adjacent distance larger than one-(ii)-ggg}
i'_a-(m'_{a+1}-1)=1,\ (i'_a+1)+(m'_{a+1}-1)=r_{h-2}, \textrm{ and } n'_1=n'_{r_{h-2}}-1.
\eeqnarray

To prove \reqnarray{proof of adjacent distance larger than one-(ii)-ccc}--\reqnarray{proof of adjacent distance larger than one-(ii)-fff},
note that from \reqnarray{proof of adjacent distance larger than one-777},
$m'_a=m_a-1$,
$m_a-m_{a+1}\geq 2$,
and $m'_{a+1}=m_{a+1}+1$,
we have
\beqnarray{proof of adjacent distance larger than one-(ii)-1111}
i'_a
=\sum_{\ell=1}^{a}m'_{\ell}\geq m'_a
=m_a-1 \geq m_{a+1}+1=m'_{a+1},
\eeqnarray
where the first inequality holds with equality if and only if $a=1$
and the second inequality holds with equality if and only if $m_a-m_{a+1}=2$.
Also, we have from \reqnarray{proof of adjacent distance larger than one-777},
$1\leq a\leq r_{h-1}-1$,
and \reqnarray{proof of adjacent distance larger than one-(ii)-222} that
\beqnarray{proof of adjacent distance larger than one-(ii)-2222}
i'_a
=\sum_{\ell=1}^{a}m'_{\ell}
=\sum_{\ell=1}^{r_{h-1}}m'_{\ell}-\sum_{\ell=a+1}^{r_{h-1}}m'_{\ell}
\leq r_{h-2}-m'_{a+1}.
\eeqnarray
Thus, \reqnarray{proof of adjacent distance larger than one-(ii)-ccc}
follows from $i'_a\geq m'_{a+1}$ in \reqnarray{proof of adjacent distance larger than one-(ii)-1111},
$i'_a\leq r_{h-2}-m'_{a+1}$ in \reqnarray{proof of adjacent distance larger than one-(ii)-2222},
and $m'_{a+1}\geq 2$ in \reqnarray{proof of adjacent distance larger than one-(ii)-111},
and \reqnarray{proof of adjacent distance larger than one-(ii)-ddd}
follows from $m'_{a+1}\leq i'_a$ in \reqnarray{proof of adjacent distance larger than one-(ii)-1111}
and $m'_{a+1}\leq r_{h-2}-i_a$ in \reqnarray{proof of adjacent distance larger than one-(ii)-2222}.
From \reqnarray{proof of adjacent distance larger than one-777}  we see that
\beqnarray{proof of adjacent distance larger than one-(ii)-3333}
i'_j=\sum_{\ell=1}^{j}m'_{\ell}
=\sum_{\ell=1}^{j-1}m'_{\ell}+m'_j
=i'_{j-1}+m'_j, \textrm{ for } j=2,3,\ldots,r_{h-1}.
\eeqnarray
From $i'_a+m'_{a+1}=i'_{a+1}$ in \reqnarray{proof of adjacent distance larger than one-(ii)-3333},
we immediately see that
\beqnarray{proof of adjacent distance larger than one-(ii)-4444}
i'_a<(i'_a+1)+j'<i'_{a+1}, \textrm{ for } j'=1,2,\ldots,m'_{a+1}-2.
\eeqnarray
It then follows from \reqnarray{proof of adjacent distance larger than one-666},
\reqnarray{proof of adjacent distance larger than one-(ii)-4444},
and $i'_a+m'_{a+1}=i'_{a+1}$ in \reqnarray{proof of adjacent distance larger than one-(ii)-3333} that
\beqnarray{}
\alignspace n'_{(i'_a+1)+j'}=q_{h-1}, \textrm{ for } j'=1,2,\ldots,m'_{a+1}-2,
\label{eqn:proof of adjacent distance larger than one-(ii)-5555}\\
\alignspace n'_{(i'_a+1)+(m'_{a+1}-1)}=n'_{i'_{a+1}}=q_{h-1}+1.
\label{eqn:proof of adjacent distance larger than one-(ii)-6666}
\eeqnarray

If $a=1$,
then we see from \reqnarray{proof of adjacent distance larger than one-(ii)-1111} that
\beqnarray{proof of adjacent distance larger than one-(ii)-7777}
i'_a-(m'_{a+1}-1)\geq 1,
\eeqnarray
where the inequality holds if and only if $a=1$ and $m_a-m_{a+1}=2$.
It then follows from $a=1$ and \reqnarray{proof of adjacent distance larger than one-(ii)-7777} that
\beqnarray{proof of adjacent distance larger than one-(ii)-8888}
1\leq i'_a-j'<i'_a=i'_1, \textrm{ for } j'=1,2,\ldots,m'_{a+1}-1.
\eeqnarray
From \reqnarray{proof of adjacent distance larger than one-666} and \reqnarray{proof of adjacent distance larger than one-(ii)-8888}
we have
\beqnarray{proof of adjacent distance larger than one-(ii)-9999}
n'_{i'_a-j'}=q_{h-1}, \textrm{ for } j'=1,2,\ldots,m'_{a+1}-1.
\eeqnarray
Thus, \reqnarray{proof of adjacent distance larger than one-(ii)-eee}
and \reqnarray{proof of adjacent distance larger than one-(ii)-fff}
follow from \reqnarray{proof of adjacent distance larger than one-(ii)-5555},
\reqnarray{proof of adjacent distance larger than one-(ii)-6666},
and \reqnarray{proof of adjacent distance larger than one-(ii)-9999}.

On the other hand, if $2\leq a\leq r_{h-1}-1$,
then from $m'_a=m_a-1\geq (m_{a+1}+2)-1=m'_{a+1}$,
and \reqnarray{proof of adjacent distance larger than one-(ii)-3333} we have
\beqnarray{proof of adjacent distance larger than one-(ii)-aaaa}
i'_a-(m'_{a+1}-1)\geq i'_a-(m'_a-1)=i'_{a-1}+1>i'_{a-1}.
\eeqnarray
From \reqnarray{proof of adjacent distance larger than one-(ii)-aaaa},
we immediately see that
\beqnarray{proof of adjacent distance larger than one-(ii)-bbbb}
i'_{a-1}<i'_a-j'<i'_a, \textrm{ for } j'=1,2,\ldots,m'_{a+1}-1.
\eeqnarray
It then follows from \reqnarray{proof of adjacent distance larger than one-666}
and \reqnarray{proof of adjacent distance larger than one-(ii)-bbbb} that
\beqnarray{proof of adjacent distance larger than one-(ii)-cccc}
n'_{i'_a-j'}=q_{h-1}, \textrm{ for } j'=1,2,\ldots,m'_{a+1}-1.
\eeqnarray
As such, \reqnarray{proof of adjacent distance larger than one-(ii)-eee}
and \reqnarray{proof of adjacent distance larger than one-(ii)-fff}
follow from \reqnarray{proof of adjacent distance larger than one-(ii)-5555},
\reqnarray{proof of adjacent distance larger than one-(ii)-6666},
and \reqnarray{proof of adjacent distance larger than one-(ii)-cccc}.

To complete the proof,
we need to show that the condition in \reqnarray{proof of adjacent distance larger than one-(ii)-ggg}
is equivalent to the condition in \reqnarray{proof of adjacent distance larger than one-(ii)-555}.
Note that if $i'_a-(m'_{a+1}-1)=1$ and $(i'_a+1)+(m'_{a+1}-1)=r_{h-2}$,
then we have from $n'_1=n'_{i'_a-(m'_{a+1}-1)}=q_{h-1}$ and $n'_{r_{h-2}}=n'_{(i'_a+1)+(m'_{a+1}-1)}=q_{h-1}+1$
in \reqnarray{proof of adjacent distance larger than one-(ii)-fff} that
\beqnarray{}
n'_1=n'_{r_{h-2}}-1.\nn
\eeqnarray
As such, we see that the condition in \reqnarray{proof of adjacent distance larger than one-(ii)-ggg}
is equivalent to the following condition:
\beqnarray{proof of adjacent distance larger than one-(ii)-dddd}
i'_a-(m'_{a+1}-1)=1 \textrm{ and } (i'_a+1)+(m'_{a+1}-1)=r_{h-2}.
\eeqnarray
From \reqnarray{proof of adjacent distance larger than one-(ii)-7777}, we see that
\beqnarray{proof of adjacent distance larger than one-(ii)-eeee}
i'_a-(m'_{a+1}-1)=1 \textrm{ iff } a=1 \textrm{ and } m_a=m_{a+1}+2.
\eeqnarray
As we have $i'_a+m'_{a+1}=i'_{a+1}$ in \reqnarray{proof of adjacent distance larger than one-(ii)-3333}
and it is clear from \reqnarray{proof of adjacent distance larger than one-777} and \reqnarray{proof of adjacent distance larger than one-(ii)-222}
that $i'_{a+1}=r_{h-2}$ if and only if $a+1=r_{h-1}$,
it follows that
\beqnarray{proof of adjacent distance larger than one-(ii)-ffff}
(i'_a+1)+(m'_{a+1}-1)=r_{h-2} \textrm{ iff } a+1=r_{h-1}.
\eeqnarray
From \reqnarray{proof of adjacent distance larger than one-(ii)-eeee}
and \reqnarray{proof of adjacent distance larger than one-(ii)-ffff},
we deduce that the condition in \reqnarray{proof of adjacent distance larger than one-(ii)-dddd}
is equivalent to the following condition:
\beqnarray{proof of adjacent distance larger than one-(ii)-gggg}
a=1,\ a+1=r_{h-1}, \textrm{ and } m_a=m_{a+1}+2.
\eeqnarray
It is clear that if $a=1$, $a+1=r_{h-1}$, and $m_a=m_{a+1}+2$,
then we have $r_{h-1}=2$ and $m_1=m_2+2$.
Conversely, if $r_{h-1}=2$ and $m_1=m_2+2$,
then it follows from $1\leq a\leq r_{h-1}-1$ that $a=1$
and hence we have $a=1$, $a+1=r_{h-1}$, and $m_a=m_{a+1}+2$.
Therefore, the condition in \reqnarray{proof of adjacent distance larger than one-(ii)-gggg}
is equivalent to the condition that $r_{h-1}=2$ and $m_1=m_2+2$
in \reqnarray{proof of adjacent distance larger than one-(ii)-555},
and the proof is completed.

\bappendix{Proof of Comparison rule A in \rlemma{comparison rule A} for an odd integer $3\leq h\leq N$
by using Comparison rule B in \rlemma{comparison rule B} for the even integer $h-1$}
{proof of comparison rule A for an odd integer h by using comparison rule B for the even integer h-1}

In this appendix, we assume that Comparison rule B in \rlemma{comparison rule B} holds
for some even integer $h-1$, where $2\leq h-1\leq N-1$,
and show that Comparison rule A in \rlemma{comparison rule A} holds for the odd integer $h$.

Let
\beqnarray{}
\nbf_1^{r_{h-2}}(h-1)\aligneq R_{r_{h-3},r_{h-2}}(\nbf_1^{r_{h-1}}(h)),
\label{eqn:proof of comparison rule A-111} \\
{\nbf'}_1^{r_{h-2}}(h-1)\aligneq R_{r_{h-3},r_{h-2}}({\nbf'}_1^{r_{h-1}}(h)).
\label{eqn:proof of comparison rule A-222}
\eeqnarray
For simplicity, let $\mbf_1^{r_{h-1}}=\nbf_1^{r_{h-1}}(h)$,
${\mbf'}_1^{r_{h-1}}={\nbf'}_1^{r_{h-1}}(h)$,
$\nbf_1^{r_{h-2}}=\nbf_1^{r_{h-2}}(h-1)$,
and ${\nbf'}_1^{r_{h-2}}={\nbf'}_1^{r_{h-2}}(h-1)$.
Then \reqnarray{proof of adjacent distance larger than one-333}--\reqnarray{proof of adjacent distance larger than one-777}
in \rappendix{proof of adjacent distance larger than one for an odd integer h by using comparison rule B for the even integer h-1} still hold.
It follows from \reqnarray{proof of adjacent distance larger than one-777} that
\reqnarray{proof of adjacent distance larger than one-(ii)-3333}
in \rappendix{proof of adjacent distance larger than one for an odd integer h by using comparison rule B for the even integer h-1} also holds.
Note that in \rlemma{comparison rule A},
we have $r_{h-1}\geq 2$, $\mbf_1^{r_{h-1}}\in \Ncal_{M,k}(h)$,
$m_a-m_{a+1}=1$ for some $1\leq a\leq r_{h-1}-1$,
$m'_a=m_a-1$, $m'_{a+1}=m_{a+1}+1$, and $m'_i=m_i$ for $i\neq a$ and $a+1$.
As $r_{h-1}\geq 2$,
we see that \reqnarray{proof of adjacent distance larger than one-888}
in \rappendix{proof of adjacent distance larger than one for an odd integer h by using comparison rule B for the even integer h-1} also holds.
It is easy to see that
\beqnarray{proof of comparison rule A-333}
m'_a=m_a-1=m_{a+1}\geq 1,\ m'_{a+1}=m_{a+1}+1\geq 2,
\textrm{ and } m'_i=m_i \textrm{ for } i\neq a, a+1.
\eeqnarray
From $m'_a=m_a-1$, $m'_{a+1}=m_{a+1}+1$, and $m'_i=m_i$ for $i\neq a$ and $a+1$,
$\mbf_1^{r_{h-1}}\in \Ncal_{M,k}(h)$, and \reqnarray{N-M-k-h},
we can see that \reqnarray{proof of adjacent distance larger than one-(ii)-222}
in \rappendix{proof of adjacent distance larger than one for an odd integer h by using comparison rule B for the even integer h-1} also holds.
As such, it follows from \reqnarray{proof of comparison rule A-333},
\reqnarray{proof of adjacent distance larger than one-(ii)-222},
$2\leq h\leq N$,
and \reqnarray{N-M-k-h} that ${\mbf'}_1^{r_{h-1}}\in \Ncal_{M,k}(h)$.

Note that from $\mbf_1^{r_{h-1}}\in \Ncal_{M,k}(h)$, ${\mbf'}_1^{r_{h-1}}\in \Ncal_{M,k}(h)$,
\reqnarray{proof of comparison rule A-111}, and \reqnarray{proof of comparison rule A-222},
we can see that \reqnarray{proof of adjacent distance larger than one-(ii)-333}
in \rappendix{proof of adjacent distance larger than one for an odd integer h by using comparison rule B for the even integer h-1} also holds.
Furthermore, since ${\mbf'}_1^{r_{h-1}}$ is obtained from $\mbf_1^{r_{h-1}}$
in exactly the same way as that in \rlemma{adjacent distance larger than one}(ii),
it is clear that \reqnarray{proof of adjacent distance larger than one-(ii)-666}--\reqnarray{proof of adjacent distance larger than one-(ii)-bbb} also hold.
We also note that from $m'_a=m_a-1$, $m'_{a+1}=m_{a+1}+1$, and $m_a-m_{a+1}=1$, we have
\beqnarray{proof of comparison rule A-444}
m'_a-m'_{a+1}=(m_a-1)-(m_{a+1}+1)=m_a-m_{a+1}-2=-1.
\eeqnarray

(i) Note that in \rlemma{comparison rule A}(i), we have $a=1$ or $a=r_{h-1}-1$.
To show \reqnarray{comparison rule A-1},
i.e., $\mbf_1^{r_{h-1}}\succ {\mbf'}_1^{r_{h-1}}$,
we see from \reqnarray{proof of adjacent distance larger than one-333}
that it suffices to show that
\beqnarray{proof of comparison rule A-(i)-111}
\nbf_1^{r_{h-2}}\succ{\nbf'}_1^{r_{h-2}}.
\eeqnarray
We consider the two cases $a=1$ and $a=r_{h-1}-1\neq 1$ separately.

\emph{Case 1: $a=1$.}
As $3\leq h\leq N$ is an odd integer, ${\mbf'}_1^{r_{h-1}}\in \Ncal_{M,k}(h)$,
$m'_a-m'_{a+1}=-1$ in \reqnarray{proof of comparison rule A-444},
$m_a=m'_a+1$, $m_{a+1}=m'_{a+1}-1$, and $m_i=m'_i$ for $i\neq a$ and $a+1$,
where we have $a=1$ in this case,
it then follows from the remark after \reqnarray{proof of adjacent distance larger than one-(i)-case-1-aaa}
in \rappendix{proof of adjacent distance larger than one for an odd integer h by using comparison rule B for the even integer h-1}
(with the roles of $\mbf_1^{r_{h-1}}$ and ${\mbf'}_1^{r_{h-1}}$ interchanged)
that ${\mbf'}_1^{r_{h-1}}\prec\mbf_1^{r_{h-1}}$.

\emph{Case 2: $a=r_{h-1}-1\neq 1$.}

\bpdffigure{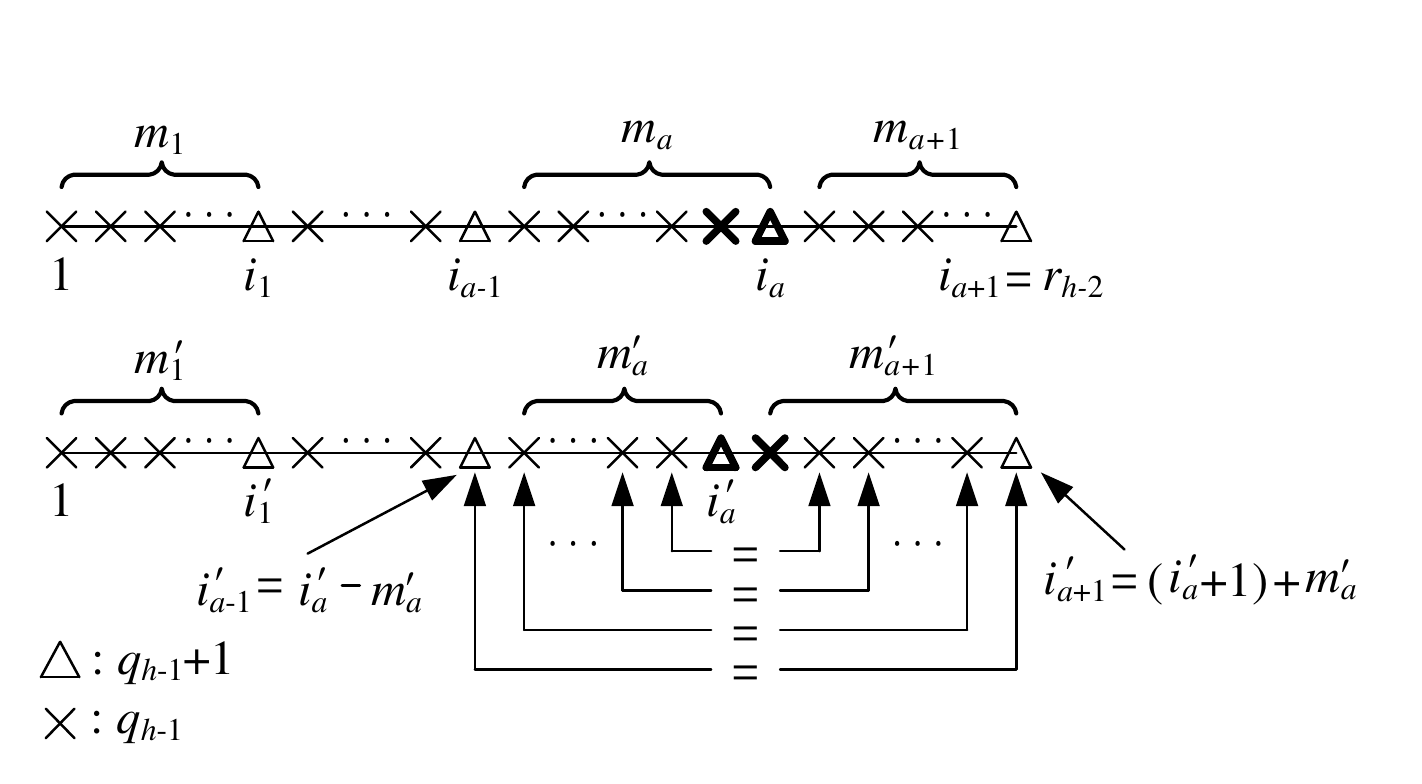}{3.5in}
\epdffigure{appendix-F-(i)-case-2}
{An illustration of \reqnarray{proof of comparison rule A-(i)-case-2-222}
in the case that $a=r_{h-1}-1\neq 1$
(note that in this case we have
$\min\{i'_{a+1}-1,r_{h-2}-i'_{a+1}-1\}=m'_a$
in \reqnarray{proof of comparison rule A-(i)-case-2-666},
$i'_a-m'_a=i'_{a-1}$ in \reqnarray{proof of adjacent distance larger than one-(ii)-3333},
and $(i'_a+1)+m'_a=i'_{a+1}$ in \reqnarray{proof of comparison rule A-(i)-case-2-999}).}

In this case, we have $a=r_{h-1}-1\geq 2$.
In the following, we will show that
\beqnarray{}
\alignspace 2\leq i'_a\leq r_{h-2}-2,
\label{eqn:proof of comparison rule A-(i)-case-2-111} \\
\alignspace n'_{i'_a-j'}=n'_{(i'_a+1)+j'}, \textrm{ for } j'=1,2,\ldots,\min\{i'_a-1,r_{h-2}-i'_a-1\}.
\label{eqn:proof of comparison rule A-(i)-case-2-222}
\eeqnarray
An illustration of \reqnarray{proof of comparison rule A-(i)-case-2-222}
is given in \rfigure{appendix-F-(i)-case-2}.
Therefore, it follows from
\reqnarray{proof of adjacent distance larger than one-888},
${\nbf'}_1^{r_{h-2}}\in \Ncal_{M,k}(h-1)$ in
\reqnarray{proof of adjacent distance larger than one-(ii)-333},
\reqnarray{proof of adjacent distance larger than one-(ii)-888}--\reqnarray{proof of adjacent distance larger than one-(ii)-bbb},
\reqnarray{proof of comparison rule A-(i)-case-2-111},
\reqnarray{proof of comparison rule A-(i)-case-2-222},
and \reqnarray{comparison rule B-4} in \rlemma{comparison rule B}(iii)
(for the even integer $h-1$) that ${\nbf'}_1^{r_{h-2}}\prec\nbf_1^{r_{h-2}}$,
i.e., \reqnarray{proof of comparison rule A-(i)-111} holds.

To prove \reqnarray{proof of comparison rule A-(i)-case-2-111},
note that from \reqnarray{proof of adjacent distance larger than one-777} and $a\geq 2$,
we have
\beqnarray{proof of comparison rule A-(i)-case-2-333}
i'_a=\sum_{\ell=1}^{a}m'_{\ell}\geq m'_1+m'_a.
\eeqnarray
Also, from $a=r_{h-1}-1$, \reqnarray{proof of adjacent distance larger than one-777},
$\sum_{\ell=1}^{r_{h-1}}m'_{\ell}=r_{h-2}$
in \reqnarray{proof of adjacent distance larger than one-(ii)-222},
and $m'_a-m'_{a+1}=-1$ in \reqnarray{proof of comparison rule A-444},
we have
\beqnarray{proof of comparison rule A-(i)-case-2-444}
i'_a=i'_{r_{h-1}-1}=\sum_{\ell=1}^{r_{h-1}-1}m'_{\ell}
=\sum_{\ell=1}^{r_{h-1}}m'_{\ell}-m'_{r_{h-1}}
=r_{h-2}-m'_{a+1}=r_{h-2}-m'_a-1.
\eeqnarray
As such, \reqnarray{proof of comparison rule A-(i)-case-2-111} follows from
$i'_a\geq m'_1+m'_a\geq 2$ in \reqnarray{proof of comparison rule A-(i)-case-2-333}
and $i'_a\leq  r_{h-2}-m'_a-1\leq  r_{h-2}-2$ in \reqnarray{proof of comparison rule A-(i)-case-2-444}.

To prove \reqnarray{proof of comparison rule A-(i)-case-2-222},
note that from \reqnarray{proof of comparison rule A-(i)-case-2-444}
and \reqnarray{proof of comparison rule A-(i)-case-2-333},
we have
\beqnarray{proof of comparison rule A-(i)-case-2-555}
r_{h-2}-i'_a-1=m'_a\leq i'_a-m'_1\leq i'_a-1.
\eeqnarray
It is clear from \reqnarray{proof of comparison rule A-(i)-case-2-555} that
\beqnarray{proof of comparison rule A-(i)-case-2-666}
\min\{i'_a-1,r_{h-2}-i'_a-1\}=r_{h-2}-i'_a-1=m'_a.
\eeqnarray
From \reqnarray{proof of adjacent distance larger than one-666}
and $i'_a-m'_a=i'_{a-1}$ in \reqnarray{proof of adjacent distance larger than one-(ii)-3333}
(note that $a\geq 2$ in this case), we immediately see that
\beqnarray{}
\alignspace n'_{i'_a-j'}=q_{h-1}, \textrm{ for } j'=1,2,\ldots,m'_a-1,
\label{eqn:proof of comparison rule A-(i)-case-2-777} \\
\alignspace n'_{i'_a-m'_a}=n'_{i'_{a-1}}=q_{h-1}+1.
\label{eqn:proof of comparison rule A-(i)-case-2-888}
\eeqnarray
Also, from $m'_a-m'_{a+1}=-1$ in \reqnarray{proof of comparison rule A-444}
and \reqnarray{proof of adjacent distance larger than one-(ii)-3333},
we have
\beqnarray{proof of comparison rule A-(i)-case-2-999}
(i'_a+1)+m'_a=i'_a+m'_{a+1}=i'_{a+1}.
\eeqnarray
It then follows from \reqnarray{proof of adjacent distance larger than one-666}
and \reqnarray{proof of comparison rule A-(i)-case-2-999} that
\beqnarray{}
\alignspace n'_{(i'_a+1)+j'}=q_{h-1}, \textrm{ for } j'=1,2,\ldots,m'_a-1,
\label{eqn:proof of comparison rule A-(i)-case-2-aaa} \\
\alignspace n'_{(i'_a+1)+m'_a}=n'_{i'_{a+1}}=q_{h-1}+1.
\label{eqn:proof of comparison rule A-(i)-case-2-bbb}
\eeqnarray
Thus, \reqnarray{proof of comparison rule A-(i)-case-2-222} follows from
\reqnarray{proof of comparison rule A-(i)-case-2-666}--\reqnarray{proof of comparison rule A-(i)-case-2-888} and
\reqnarray{proof of comparison rule A-(i)-case-2-aaa}--\reqnarray{proof of comparison rule A-(i)-case-2-bbb}.

(ii) Note that in \rlemma{comparison rule A}(ii),
we have $2\leq a\leq r_{h-1}-2$ and there exists a positive integer $j$ such that
$1\leq j\leq \min\{a-1,r_{h-1}-a-1\}$, $m_{a-\ell}=m_{a+1+\ell}$ for $\ell=1,2,\ldots,j-1$, and $m_{a-j}\neq m_{a+1+j}$.
As $a\geq 2$, it is clear that \reqnarray{proof of comparison rule A-(i)-case-2-333} holds.
As $a\leq r_{h-1}-2$, we see from \reqnarray{proof of adjacent distance larger than one-777}
and $\sum_{\ell=1}^{r_{h-1}}m'_{\ell}=r_{h-2}$
in \reqnarray{proof of adjacent distance larger than one-(ii)-222} that
\beqnarray{proof of comparison rule A-(ii)-111}
i'_a=\sum_{\ell=1}^{a}m'_{\ell}
=\sum_{\ell=1}^{r_{h-1}}m'_{\ell}-\sum_{\ell=a+1}^{r_{h-1}}m'_{\ell}
\leq r_{h-2}-m'_{a+1}-m'_{r_{h-1}}.
\eeqnarray
It follows from
$i'_a\geq m'_1+m'_a\geq 2$ in \reqnarray{proof of comparison rule A-(i)-case-2-333}
and $i'_a\leq  r_{h-2}-m'_{a+1}-m'_{r_{h-1}}\leq  r_{h-2}-2$ in \reqnarray{proof of comparison rule A-(i)-case-2-444} that
\beqnarray{proof of comparison rule A-(ii)-222}
2\leq i'_a\leq r_{h-2}-2.
\eeqnarray
Furthermore, from $m'_i=m_i$ for $i\neq a$ and $a+1$
and $m_{a-\ell}=m_{a+1+\ell}$ for $\ell=1,2,\ldots,j-1$,
it is clear that
\beqnarray{proof of comparison rule A-(ii)-333}
m'_{a-\ell}=m'_{a+1+\ell}, \textrm{ for } \ell=1,2,\ldots,j-1.
\eeqnarray

\bpdffigure{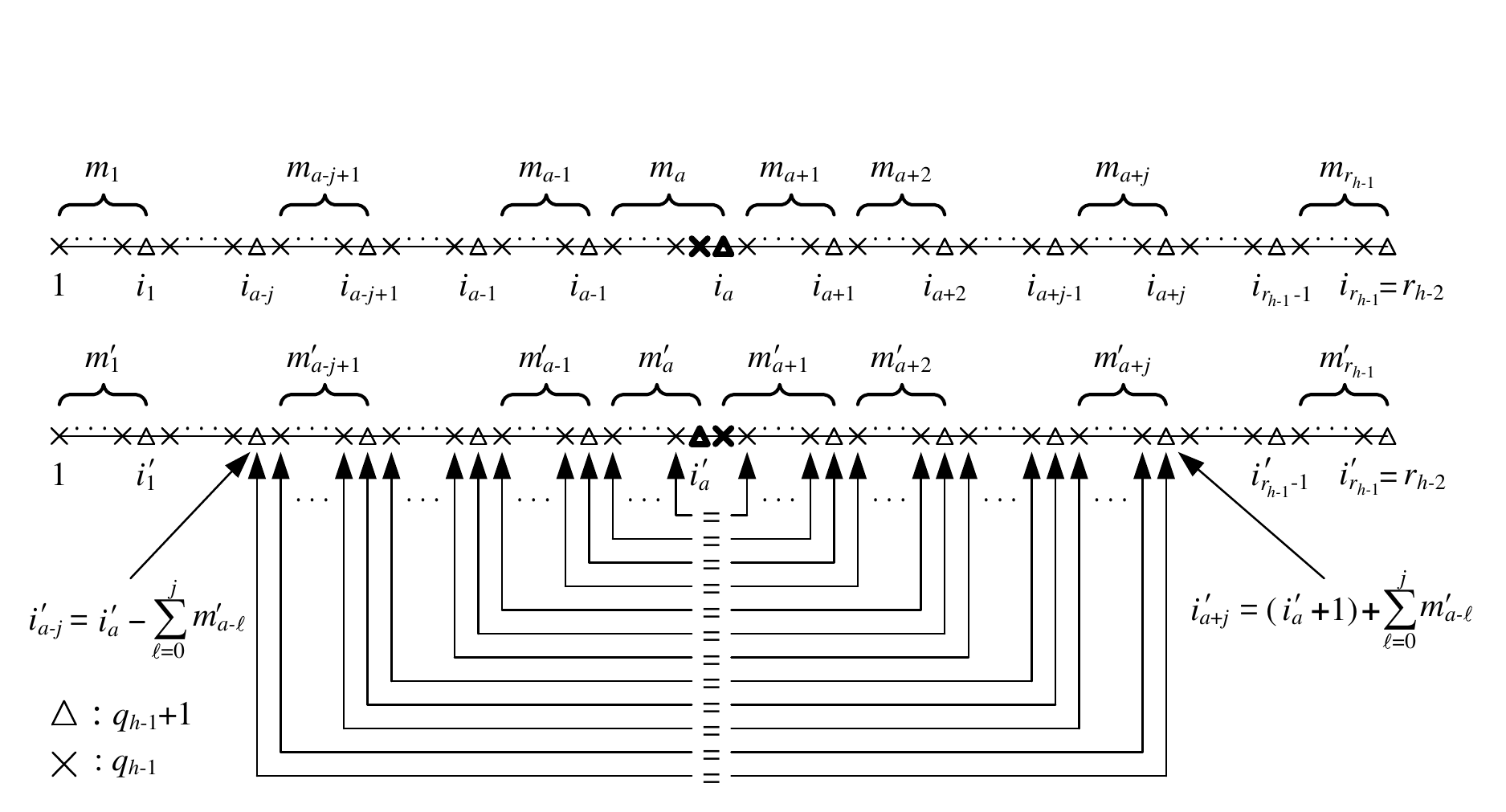}{6.0in}
\epdffigure{appendix-F-(ii)}
{An illustration of \reqnarray{proof of comparison rule A-(ii)-444}
(note that we have
$i'_a-\sum_{\ell=0}^{j''-1}m'_{a-\ell}=i'_{a-j''}$
for $1\leq j''\leq a-1$ in \reqnarray{proof of comparison rule A-(ii)-555},
$(i'_a+1)+\sum_{\ell=0}^{j''-1}m'_{a-\ell}=i'_{a+j''}$
for $1\leq j''\leq j$ in \reqnarray{proof of comparison rule A-(ii)-888},
and $j\leq \min\{a-1,r_{h-1}-a-1\}\leq a-1$).}

By using \reqnarray{proof of comparison rule A-(ii)-333},
we can show that
\beqnarray{proof of comparison rule A-(ii)-444}
n'_{i'_a-j'}=n'_{(i'_a+1)+j'}, \textrm{ for } j'=1,2,\ldots,\sum_{\ell=0}^{j-1}m'_{a-\ell}.
\eeqnarray
An illustration of \reqnarray{proof of comparison rule A-(ii)-444}
is given in \rfigure{appendix-F-(ii)}.
To prove \reqnarray{proof of comparison rule A-(ii)-444},
observe that for $1\leq j''\leq a-1$,
we have from \reqnarray{proof of adjacent distance larger than one-777} that
\beqnarray{proof of comparison rule A-(ii)-555}
i'_a-\sum_{\ell=0}^{j''-1}m'_{a-\ell}
=\sum_{\ell=1}^{a}m'_{\ell}-\sum_{\ell=a-j''+1}^{a}m'_{\ell}
=\sum_{\ell=1}^{a-j''}m'_{\ell}=i'_{a-j''}.
\eeqnarray
It follows from \reqnarray{proof of adjacent distance larger than one-666}
and \reqnarray{proof of comparison rule A-(ii)-555} that
\beqnarray{proof of comparison rule A-(ii)-666}
n'_{i'_a-j'}=
\bselection
q_{h-1}+1, &\textrm{for } j'=m'_a, \sum_{\ell=0}^{1}m'_{a-\ell},\ldots,\sum_{\ell=0}^{a-2}m'_{a-\ell}, \\
q_{h-1}, &\textrm{for } 1\leq j'\leq \sum_{\ell=0}^{a-2}m'_{a-\ell} \\
&\textrm{and } j'\neq m'_a, \sum_{\ell=0}^{1}m'_{a-\ell},\ldots,\sum_{\ell=0}^{a-2}m'_{a-\ell}.
\eselection
\eeqnarray
Furthermore, for $1\leq j''\leq j$,
we have from $m'_a-m'_{a+1}=-1$ in \reqnarray{proof of comparison rule A-444}
and $m'_{a-\ell}=m'_{a+1+\ell}$ for $\ell=1,2,\ldots,j''-1$
in \reqnarray{proof of comparison rule A-(ii)-333} that
\beqnarray{proof of comparison rule A-(ii)-777}
\sum_{\ell=0}^{j''-1}m'_{a-\ell}
=m'_a+\sum_{\ell=1}^{j''-1}m'_{a-\ell}
=m'_{a+1}-1+\sum_{\ell=1}^{j''-1}m'_{a+1+\ell}
=\sum_{\ell=0}^{j''-1}m'_{a+1+\ell}-1.
\eeqnarray
As such, for $1\leq j''\leq j$,
we have from \reqnarray{proof of adjacent distance larger than one-777}
and \reqnarray{proof of comparison rule A-(ii)-777} that
\beqnarray{proof of comparison rule A-(ii)-888}
(i'_a+1)+\sum_{\ell=0}^{j''-1}m'_{a-\ell}
=\sum_{\ell=1}^{a}m'_{\ell}+\sum_{\ell=0}^{j''-1}m'_{a+1+\ell}
=\sum_{\ell=1}^{a+j''}m'_{\ell}
= i'_{a+j''}.
\eeqnarray
It then follows from \reqnarray{proof of adjacent distance larger than one-666}
and \reqnarray{proof of comparison rule A-(ii)-888} that
\beqnarray{proof of comparison rule A-(ii)-999}
n'_{(i'_a+1)+j'}=
\bselection
q_{h-1}+1, &\textrm{for } j'=m'_a, \sum_{\ell=0}^{1}m'_{a-\ell},\ldots,\sum_{\ell=0}^{j-1}m'_{a-\ell}, \\
q_{h-1}, &\textrm{for } 1\leq j'\leq \sum_{\ell=0}^{j-1}m'_{a-\ell} \\
&\textrm{and } j'\neq m'_a, \sum_{\ell=0}^{1}m'_{a-\ell},\ldots,\sum_{\ell=0}^{j-1}m'_{a-\ell}.
\eselection
\eeqnarray
As we have $j\leq \min\{a-1,r_{h-1}-a-1\}\leq a-1$,
it is clear that \reqnarray{proof of comparison rule A-(ii)-444}
follows from \reqnarray{proof of comparison rule A-(ii)-666}
and \reqnarray{proof of comparison rule A-(ii)-999}.

From \reqnarray{proof of adjacent distance larger than one-333},
we see that:
(a) If $m_{a-j}<m_{a+1+j}$,
then to show \reqnarray{comparison rule A-2},
i.e., $\mbf_1^{r_{h-1}}\succ{\mbf'}_1^{r_{h-1}}$,
it suffices to show that $\nbf_1^{r_{h-2}}\succ{\nbf'}_1^{r_{h-2}}$;
(b) If $m_{a-j}>m_{a+1+j}$,
then to show \reqnarray{comparison rule A-3},
i.e., $\mbf_1^{r_{h-1}}\preceq {\mbf'}_1^{r_{h-1}}$,
where $\mbf_1^{r_{h-1}}\equiv {\mbf'}_1^{r_{h-1}}$
if and only if $a-j=1$, $a+1+j=r_{h-1}$, and $m_1=m_{r_{h-1}}+1$,
it suffices to show that $\nbf_1^{r_{h-2}}\preceq{\nbf'}_1^{r_{h-2}}$,
where $\nbf_1^{r_{h-2}}\equiv{\nbf'}_1^{r_{h-2}}$
if and only if $a-j=1$, $a+1+j=r_{h-1}$, and $m_1=m_{r_{h-1}}+1$.

(a) First we assume that $m_{a-j}<m_{a+1+j}$ and show that
\beqnarray{proof of comparison rule A-(ii)-(a)-111}
\nbf_1^{r_{h-2}}\succ{\nbf'}_1^{r_{h-2}}.
\eeqnarray
As $j\leq \min\{a-1,r_{h-1}-a-1\}$, we have $j+1\leq a\leq r_{h-1}-j-1$.
We consider the two cases $a=j+1$ and $j+2\leq a\leq r_{h-1}-j-1$ separately.

\bpdffigure{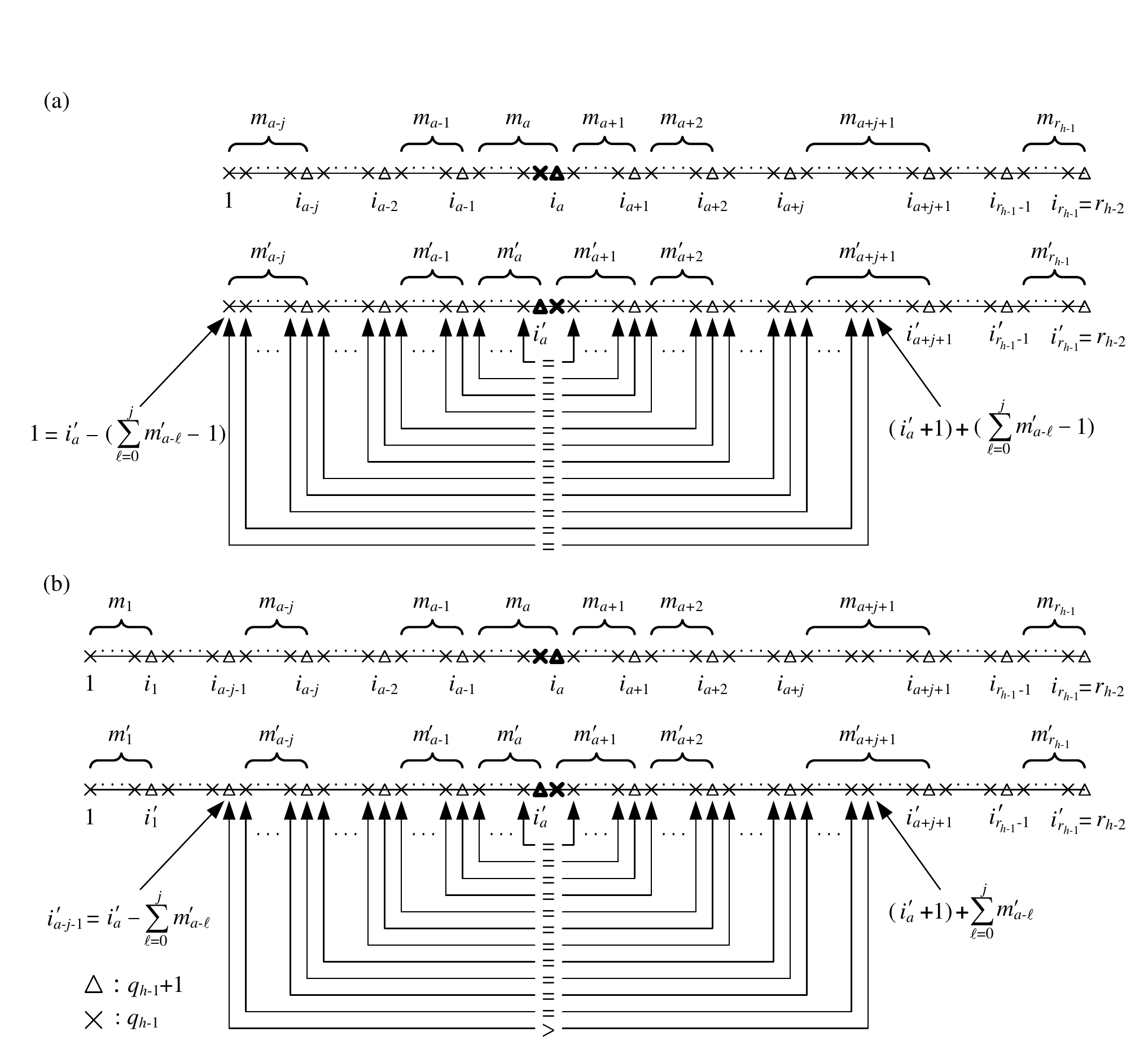}{6.0in}
\epdffigure{appendix-F-(ii)-(a)}
{(a) An illustration of \reqnarray{proof of comparison rule A-(ii)-(a)-case-1-111}
in the case that $m_{a-j}<m_{a+1+j}$ and $a=j+1$
(note that in this case we have
$i'_a-\sum_{\ell=0}^{j''-1}m'_{a-\ell}=i'_{a-j''}$
for $1\leq j''\leq a-1=j$ in \reqnarray{proof of comparison rule A-(ii)-555},
$i'_a-(\sum_{\ell=0}^{j}m'_{a-\ell}-1)=1$
in \reqnarray{proof of comparison rule A-(ii)-(a)-case-1-555},
$(i'_a+1)+\sum_{\ell=0}^{j''-1}m'_{a-\ell}=i'_{a+j''}$
for $1\leq j''\leq j$ in \reqnarray{proof of comparison rule A-(ii)-888},
and $(i'_a+1)+(\sum_{\ell=0}^{j}m'_{a-\ell}-1)<i'_{a+1+j}$
in \reqnarray{proof of comparison rule A-(ii)-(a)-case-1-888}.
(b) An illustration of \reqnarray{proof of comparison rule A-(ii)-(a)-case-2-222}
and \reqnarray{proof of comparison rule A-(ii)-(a)-case-2-333}
in the case that $m_{a-j}<m_{a+1+j}$ and $j+2\leq a\leq r_{h-1}-j-1$
(note that in this case we have $j+1\leq a-1$,
$i'_a-\sum_{\ell=0}^{j''-1}m'_{a-\ell}=i'_{a-j''}$
for $1\leq j''\leq a-1$ in \reqnarray{proof of comparison rule A-(ii)-555},
$(i'_a+1)+\sum_{\ell=0}^{j''-1}m'_{a-\ell}=i'_{a+j''}$
for $1\leq j''\leq j$ in \reqnarray{proof of comparison rule A-(ii)-888},
and $(i'_a+1)+\sum_{\ell=0}^{j}m'_{a-\ell}<i'_{a+1+j}$
in \reqnarray{proof of comparison rule A-(ii)-(a)-case-1-888}.}

\emph{Case 1: $a=j+1$.}
In this case, we show that
\beqnarray{proof of comparison rule A-(ii)-(a)-case-1-111}
n'_{i'_a-j'}=n'_{(i'_a+1)+j'}, \textrm{ for } j'=1,2,\ldots,\min\{i'_a-1,r_{h-2}-i'_a-1\}.
\eeqnarray
An illustration of \reqnarray{proof of comparison rule A-(ii)-(a)-case-1-111}
is given in \rfigure{appendix-F-(ii)-(a)}(a).
Therefore, it follows from \reqnarray{proof of adjacent distance larger than one-888},
${\nbf'}_1^{r_{h-2}}\in \Ncal_{M,k}(h-1)$ in
\reqnarray{proof of adjacent distance larger than one-(ii)-333},
\reqnarray{proof of adjacent distance larger than one-(ii)-888}--\reqnarray{proof of adjacent distance larger than one-(ii)-bbb},
\reqnarray{proof of comparison rule A-(ii)-222},
\reqnarray{proof of comparison rule A-(ii)-(a)-case-1-111},
and \reqnarray{comparison rule B-4} in \rlemma{comparison rule B}(iii)
(for the even integer $h-1$) that ${\nbf'}_1^{r_{h-2}}\prec\nbf_1^{r_{h-2}}$,
i.e., \reqnarray{proof of comparison rule A-(ii)-(a)-111} holds.

To prove \reqnarray{proof of comparison rule A-(ii)-(a)-case-1-111},
note that from \reqnarray{proof of adjacent distance larger than one-777},
$\sum_{\ell=1}^{r_{h-1}}m'_{\ell}=r_{h-2}$ in \reqnarray{proof of adjacent distance larger than one-(ii)-222},
$a\leq r_{h-1}-j-1$, $a=j+1$,
\reqnarray{proof of comparison rule A-444} ,
\reqnarray{proof of comparison rule A-(ii)-333},
and $m'_{a-j}=m_{a-j}<m_{a+1+j}=m'_{a+1+j}$, we have
\beqnarray{proof of comparison rule A-(ii)-(a)-case-1-222}
(i'_a-1)-(r_{h-2}-i'_a-1)
\aligneq 2i'_a-r_{h-2}=2\sum_{\ell=1}^{a}m'_{\ell}-\sum_{\ell=1}^{r_{h-1}}m'_{\ell}
            =\sum_{\ell=1}^{a}m'_{\ell}-\sum_{\ell=a+1}^{r_{h-1}}m'_{\ell} \nn\\
\alignleq \sum_{\ell=1}^{a}m'_{\ell}-\sum_{\ell=a+1}^{a+1+j}m'_{\ell}
            =\sum_{\ell=0}^{a-1}m'_{a-\ell}-\sum_{\ell=0}^{j}m'_{a+1+\ell} \nn\\
\aligneq m'_a+\sum_{\ell=1}^{j}m'_{a-\ell}-m'_{a+1}-\sum_{\ell=1}^{j}m'_{a+1+\ell}\nn\\
\aligneq -1+m'_{a-j}-m'_{a+1+j}<0.
\eeqnarray
It follows from \reqnarray{proof of comparison rule A-(ii)-(a)-case-1-222},
\reqnarray{proof of adjacent distance larger than one-777},
and $a=j+1$ that
\beqnarray{proof of comparison rule A-(ii)-(a)-case-1-333}
\min\{i'_a-1,r_{h-2}-i'_a-1\}
=i'_a-1=\sum_{\ell=1}^{a}m'_{\ell}-1
=\sum_{\ell=0}^{a-1}m'_{a-\ell}-1=\sum_{\ell=0}^{j}m'_{a-\ell}-1.
\eeqnarray
As $a=j+1$, we have from \reqnarray{proof of comparison rule A-(ii)-555} (with $j''=a-1=j$) that
\beqnarray{proof of comparison rule A-(ii)-(a)-case-1-444}
i'_a-\sum_{\ell=0}^{j-1}m'_{a-\ell}=i'_{a-j}=i'_1.
\eeqnarray
From \reqnarray{proof of comparison rule A-(ii)-(a)-case-1-444},
$a=j+1$, and \reqnarray{proof of adjacent distance larger than one-777},
we have
\beqnarray{proof of comparison rule A-(ii)-(a)-case-1-555}
i'_a-\left(\sum_{\ell=0}^{j}m'_{a-\ell}-1\right)=i'_{a-j}-m'_{a-j}+1=i'_1-m'_1+1=1.
\eeqnarray
Thus, we see from \reqnarray{proof of adjacent distance larger than one-666},
\reqnarray{proof of comparison rule A-(ii)-(a)-case-1-444},
and \reqnarray{proof of comparison rule A-(ii)-(a)-case-1-555} that
\beqnarray{proof of comparison rule A-(ii)-(a)-case-1-666}
n'_{i'_a-j'}=q_{h-1},
\textrm{ for } \sum_{\ell=0}^{j-1}m'_{a-\ell}+1\leq j'\leq \sum_{\ell=0}^{j}m'_{a-\ell}-1.
\eeqnarray
Furthermore, from \reqnarray{proof of comparison rule A-(ii)-888} (with $j''=j$),
we have
\beqnarray{proof of comparison rule A-(ii)-(a)-case-1-777}
(i'_a+1)+\sum_{\ell=0}^{j-1}m'_{a-\ell}=i'_{a+j}.
\eeqnarray
From \reqnarray{proof of comparison rule A-(ii)-(a)-case-1-777},
$m'_{a-j}=m_{a-j}<m_{a+1+j}=m'_{a+1+j}$,
and \reqnarray{proof of adjacent distance larger than one-(ii)-3333},
we have
\beqnarray{proof of comparison rule A-(ii)-(a)-case-1-888}
(i'_a+1)+\sum_{\ell=0}^{j}m'_{a-\ell}
= i'_{a+j}+m'_{a-j}
< i'_{a+j}+m'_{a+1+j}=i'_{a+1+j}.
\eeqnarray
Thus, we see from \reqnarray{proof of adjacent distance larger than one-666},
\reqnarray{proof of comparison rule A-(ii)-(a)-case-1-777},
and \reqnarray{proof of comparison rule A-(ii)-(a)-case-1-888} that
\beqnarray{proof of comparison rule A-(ii)-(a)-case-1-999}
n'_{(i'_a+1)+j'}=q_{h-1},
\textrm{ for } \sum_{\ell=0}^{j-1}m'_{a-\ell}+1\leq j'\leq \sum_{\ell=0}^{j}m'_{a-\ell}.
\eeqnarray
By combining \reqnarray{proof of comparison rule A-(ii)-444},
\reqnarray{proof of comparison rule A-(ii)-(a)-case-1-333},
\reqnarray{proof of comparison rule A-(ii)-(a)-case-1-666},
and \reqnarray{proof of comparison rule A-(ii)-(a)-case-1-999},
we obtain \reqnarray{proof of comparison rule A-(ii)-(a)-case-1-111}.

\emph{Case 2: $j+2\leq a\leq r_{h-1}-j-1$.}
In this case, we show that
\beqnarray{}
\alignspace
1\leq \sum_{\ell=0}^{j}m'_{a-\ell}\leq \min\{i'_a-1,r_{h-2}-i'_a-1\},
\label{eqn:proof of comparison rule A-(ii)-(a)-case-2-111}\\
\alignspace
n'_{i'_a-j'}=n'_{(i'_a+1)+j'},\
j'=1,2,\ldots,\sum_{\ell=0}^{j}m'_{a-\ell}-1,
\label{eqn:proof of comparison rule A-(ii)-(a)-case-2-222}\\
\alignspace
n'_{i'_a-\sum_{\ell=0}^{j}m'_{a-\ell}}=q_{h-1}+1>n'_{(i'_a+1)+\sum_{\ell=0}^{j}m'_{a-\ell}}=q_{h-1}.
\label{eqn:proof of comparison rule A-(ii)-(a)-case-2-333}
\eeqnarray
An illustration of \reqnarray{proof of comparison rule A-(ii)-(a)-case-2-222}
and \reqnarray{proof of comparison rule A-(ii)-(a)-case-2-333}
is given in \rfigure{appendix-F-(ii)-(a)}(b).
Therefore, it follows from \reqnarray{proof of adjacent distance larger than one-888},
${\nbf'}_1^{r_{h-2}}\in \Ncal_{M,k}(h-1)$ in
\reqnarray{proof of adjacent distance larger than one-(ii)-333},
\reqnarray{proof of adjacent distance larger than one-(ii)-888}--\reqnarray{proof of adjacent distance larger than one-(ii)-bbb},
\reqnarray{proof of comparison rule A-(ii)-222},
\reqnarray{proof of comparison rule A-(ii)-(a)-case-2-111}--\reqnarray{proof of comparison rule A-(ii)-(a)-case-2-333},
and \reqnarray{comparison rule B-2} in \rlemma{comparison rule B}(ii)
(for the even integer $h-1$) that ${\nbf'}_1^{r_{h-2}}\prec\nbf_1^{r_{h-2}}$,
i.e., \reqnarray{proof of comparison rule A-(ii)-(a)-111} holds.

To prove \reqnarray{proof of comparison rule A-(ii)-(a)-case-2-111},
note that as $j+2\leq a$,
we have from \reqnarray{proof of comparison rule A-(ii)-555} (with $j''=j+1\leq a-1$) that
\beqnarray{proof of comparison rule A-(ii)-(a)-case-2-444}
i'_a-\sum_{\ell=0}^{j}m'_{a-\ell}=i'_{a-j-1}.
\eeqnarray
Also, from \reqnarray{proof of comparison rule A-(ii)-(a)-case-1-888}, we have
\beqnarray{proof of comparison rule A-(ii)-(a)-case-2-555}
(i'_a+1)+\sum_{\ell=0}^{j}m'_{a-\ell}< i'_{a+1+j}\leq r_{h-2}.
\eeqnarray
Thus, \reqnarray{proof of comparison rule A-(ii)-(a)-case-2-111} follows from
$\sum_{\ell=0}^{j}m'_{a-\ell}=i'_a-i'_{a-j-1}\leq i'_a-1$
in \reqnarray{proof of comparison rule A-(ii)-(a)-case-2-444}
and $\sum_{\ell=0}^{j}m'_{a-\ell}\leq r_{h-2}-i'_a-1$
in \reqnarray{proof of comparison rule A-(ii)-(a)-case-2-555}.

To prove \reqnarray{proof of comparison rule A-(ii)-(a)-case-2-222}
and \reqnarray{proof of comparison rule A-(ii)-(a)-case-2-333},
note that from \reqnarray{proof of comparison rule A-(ii)-555} (with $j''=j\leq a-2$),
we have
\beqnarray{proof of comparison rule A-(ii)-(a)-case-2-666}
i'_a-\sum_{\ell=0}^{j-1}m'_{a-\ell}=i'_{a-j}.
\eeqnarray
Thus, we see from
\reqnarray{proof of adjacent distance larger than one-666},
\reqnarray{proof of comparison rule A-(ii)-(a)-case-2-666},
and \reqnarray{proof of comparison rule A-(ii)-(a)-case-2-444} that
\beqnarray{proof of comparison rule A-(ii)-(a)-case-2-777}
n'_{i'_a-j'}=
\bselection
q_{h-1},   &\textrm{for } \sum_{\ell=0}^{j-1}m'_{a-\ell}+1\leq j'\leq \sum_{\ell=0}^{j}m'_{a-\ell}-1. \\
q_{h-1}+1, &\textrm{for } j'=\sum_{\ell=0}^{j}m'_{a-\ell}.
\eselection
\eeqnarray
By combining \reqnarray{proof of comparison rule A-(ii)-444},
\reqnarray{proof of comparison rule A-(ii)-(a)-case-1-999},
and \reqnarray{proof of comparison rule A-(ii)-(a)-case-2-777},
we obtain \reqnarray{proof of comparison rule A-(ii)-(a)-case-2-222}
and \reqnarray{proof of comparison rule A-(ii)-(a)-case-2-333}.

(b) Now we assume that $m_{a-j}>m_{a+1+j}$ and show that
\beqnarray{proof of comparison rule A-(ii)-(b)-111}
\nbf_1^{r_{h-2}}\preceq {\nbf'}_1^{r_{h-2}},
\eeqnarray
where $\nbf_1^{r_{h-2}}\equiv{\nbf'}_1^{r_{h-2}}$ if and only if
\beqnarray{proof of comparison rule A-(ii)-(b)-222}
a-j=1,\ a+1+j=r_{h-1}, \textrm{ and } m_1=m_{r_{h-1}}+1.
\eeqnarray

\bpdffigure{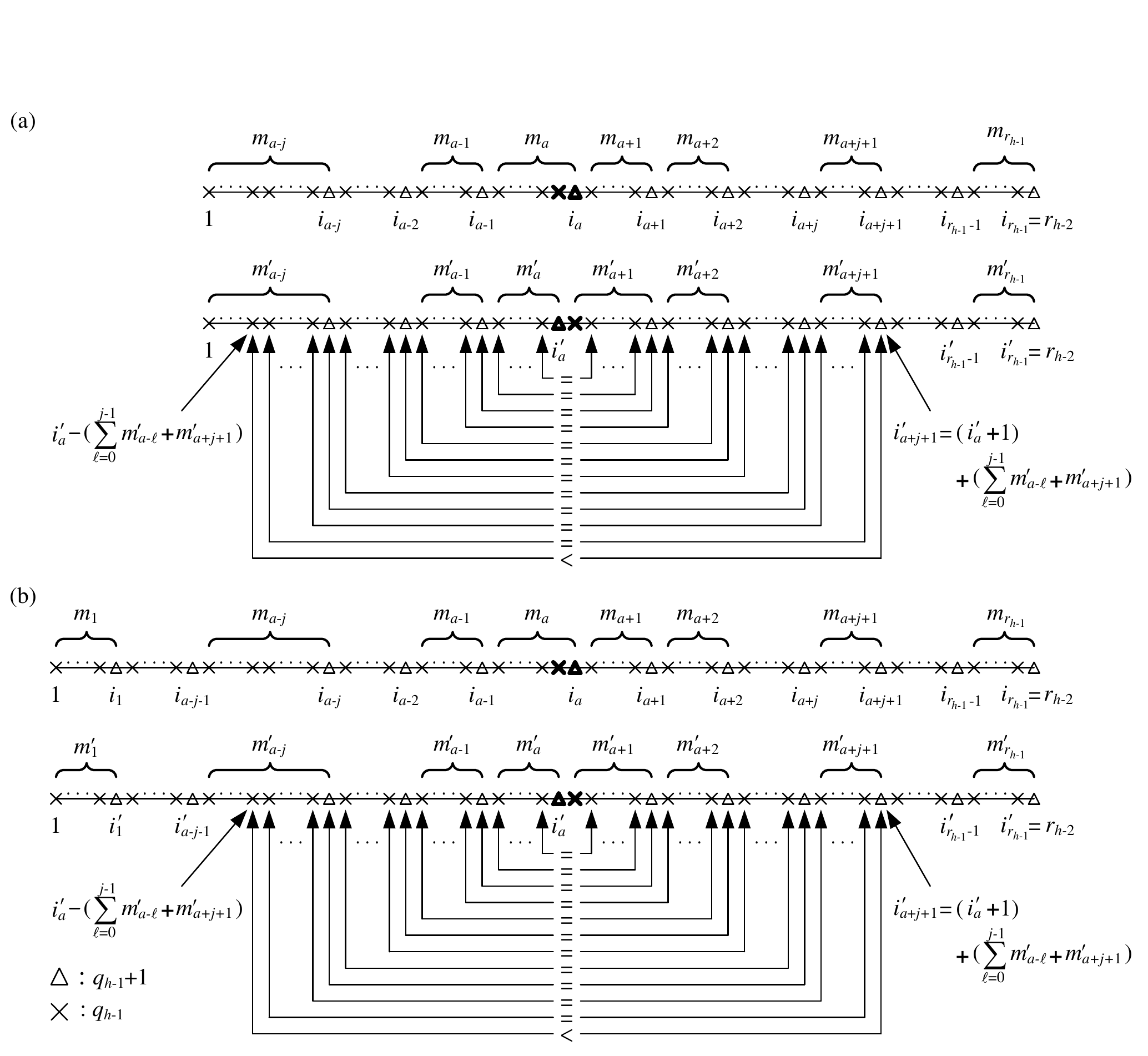}{6.0in}
\epdffigure{appendix-F-(ii)-(b)}
{An illustration of \reqnarray{proof of comparison rule A-(ii)-(b)-333}
and \reqnarray{proof of comparison rule A-(ii)-(b)-444}
in the case that $m_{a-j}>m_{a+1+j}$
(note that in this case we have
$(i'_a+1)+\sum_{\ell=0}^{j''-1}m'_{a-\ell}=i'_{a+j''}$
for $1\leq j''\leq j$ in \reqnarray{proof of comparison rule A-(ii)-888},
$(i'_a+1)+(\sum_{\ell=0}^{j-1}m'_{a-\ell}+m'_{a+1+j})=i'_{a+1+j}$
in \reqnarray{proof of comparison rule A-(ii)-(b)-999},
$i'_a-\sum_{\ell=0}^{j''-1}m'_{a-\ell}=i'_{a-j''}$
for $1\leq j''\leq a-1$ in \reqnarray{proof of comparison rule A-(ii)-555}):
(a) $a=j+1$
(note that in this case we have
$i'_a-(\sum_{\ell=0}^{j-1}m'_{a-\ell}+m'_{a+1+j})\geq 1$
in \reqnarray{proof of comparison rule A-(ii)-(b)-ccc});
(b) $j+2\leq a\leq r_{h-1}-j-1$
(note that in this case we have
$i'_a-(\sum_{\ell=0}^{j-1}m'_{a-\ell}+m'_{a+1+j})>i'_{a-j-1}$
in \reqnarray{proof of comparison rule A-(ii)-(b)-eee}).}

In the following, we show that
\beqnarray{}
\alignspace
n'_{i'_a-j'}=n'_{(i'_a+1)+j'},
\textrm{ for } j'=1,2,\ldots,\sum_{\ell=0}^{j-1}m'_{a-\ell}+m'_{a+1+j}-1,
\label{eqn:proof of comparison rule A-(ii)-(b)-333} \\
\alignspace
n'_{i'_a-(\sum_{\ell=0}^{j-1}m'_{a-\ell}+m'_{a+1+j})}=q_{h-1}
<n'_{(i'_a+1)+(\sum_{\ell=0}^{j-1}m'_{a-\ell}+m'_{a+1+j})}=q_{h-1}+1.
\label{eqn:proof of comparison rule A-(ii)-(b)-444}
\eeqnarray
An illustration of \reqnarray{proof of comparison rule A-(ii)-(b)-333}
and \reqnarray{proof of comparison rule A-(ii)-(b)-444}
is given in \rfigure{appendix-F-(ii)-(b)}.
Therefore, it follows from \reqnarray{proof of adjacent distance larger than one-888},
${\nbf'}_1^{r_{h-2}}\in \Ncal_{M,k}(h-1)$ in
\reqnarray{proof of adjacent distance larger than one-(ii)-333},
\reqnarray{proof of adjacent distance larger than one-(ii)-888}--\reqnarray{proof of adjacent distance larger than one-(ii)-bbb},
\reqnarray{proof of comparison rule A-(ii)-222},
\reqnarray{proof of comparison rule A-(ii)-(b)-333}--\reqnarray{proof of comparison rule A-(ii)-(b)-444},
and \reqnarray{comparison rule B-3} in \rlemma{comparison rule B}(ii)
(for the even integer $h-1$) that ${\nbf'}_1^{r_{h-2}}\succeq\nbf_1^{r_{h-2}}$,
where ${\nbf'}_1^{r_{h-2}}\equiv\nbf_1^{r_{h-2}}$ if and only if
\beqnarray{}
\alignspace i'_a-\left(\sum_{\ell=0}^{j-1}m'_{a-\ell}+m'_{a+1+j}\right)=1,
\label{eqn:proof of comparison rule A-(ii)-(b)-555}\\
\alignspace (i'_a+1)+\left(\sum_{\ell=0}^{j-1}m'_{a-\ell}+m'_{a+1+j}\right)=r_{h-2},
\label{eqn:proof of comparison rule A-(ii)-(b)-666}\\
\alignspace n'_1=n'_{r_{h-2}}-1.
\label{eqn:proof of comparison rule A-(ii)-(b)-777}
\eeqnarray

To prove \reqnarray{proof of comparison rule A-(ii)-(b)-333}
and \reqnarray{proof of comparison rule A-(ii)-(b)-444},
note that from \reqnarray{proof of comparison rule A-(ii)-888} (with $j''=j$)
and \reqnarray{proof of adjacent distance larger than one-(ii)-3333},
we have
\beqnarray{}
\alignspace
(i'_a+1)+\sum_{\ell=0}^{j-1}m'_{a-\ell}=i'_{a+j},
\label{eqn:proof of comparison rule A-(ii)-(b)-888}\\
\alignspace
(i'_a+1)+\left(\sum_{\ell=0}^{j-1}m'_{a-\ell}+m'_{a+1+j}\right)=i'_{a+j}+m'_{a+1+j}=i'_{a+1+j}.
\label{eqn:proof of comparison rule A-(ii)-(b)-999}
\eeqnarray
Thus, we see from \reqnarray{proof of adjacent distance larger than one-666},
\reqnarray{proof of comparison rule A-(ii)-(b)-888},
and \reqnarray{proof of comparison rule A-(ii)-(b)-999} that
\beqnarray{proof of comparison rule A-(ii)-(b)-aaa}
n'_{(i'_a+1)+j'}=
\bselection
q_{h-1}, &\textrm{ for } \sum_{\ell=0}^{j-1}m'_{a-\ell}+1\leq j'\leq \sum_{\ell=0}^{j-1}m'_{a-\ell}+m'_{a+1+j}-1,\\
q_{h-1}+1, &\textrm{ for } j'=\sum_{\ell=0}^{j-1}m'_{a-\ell}+m'_{a+1+j}.
\eselection
\eeqnarray
As $j\leq \min\{a-1,r_{h-1}-a-1\}\leq a-1$,
we have from \reqnarray{proof of comparison rule A-(ii)-555} (with $j''=j\leq a-1$)
and $m'_{a-j}=m_{a-j}>m_{a+1+j}=m'_{a+1+j}$ that
\beqnarray{proof of comparison rule A-(ii)-(b)-bbb}
i'_a-\left(\sum_{\ell=0}^{j-1}m'_{a-\ell}+m'_{a+1+j}\right)=i'_{a-j}-m'_{a+1+j}\geq i'_{a-j}-m'_{a-j}+1,
\eeqnarray
where the equality holds if and only if $m_{a-j}=m_{a+1+j}+1$.
If $a=j+1$, then we have from \reqnarray{proof of comparison rule A-(ii)-(b)-bbb}
and \reqnarray{proof of adjacent distance larger than one-777} that
\beqnarray{proof of comparison rule A-(ii)-(b)-ccc}
i'_a-\left(\sum_{\ell=0}^{j-1}m'_{a-\ell}+m'_{a+1+j}\right)\geq i'_1-m'_1+1=1,
\eeqnarray
where the equality holds if and only if $m_{a-j}=m_{a+1+j}+1$.
Thus, we see from \reqnarray{proof of adjacent distance larger than one-666},
\reqnarray{proof of comparison rule A-(ii)-(a)-case-1-444} (note that $a=j+1$),
and \reqnarray{proof of comparison rule A-(ii)-(b)-ccc} that
\beqnarray{proof of comparison rule A-(ii)-(b)-ddd}
n'_{i'_a-j'}=q_{h-1},
\textrm{ for } \sum_{\ell=0}^{j-1}m'_{a-\ell}+1\leq j'\leq \sum_{\ell=0}^{j-1}m'_{a-\ell}+m'_{a+1+j}.
\eeqnarray
By combining \reqnarray{proof of comparison rule A-(ii)-444},
\reqnarray{proof of comparison rule A-(ii)-(b)-aaa},
and \reqnarray{proof of comparison rule A-(ii)-(b)-ddd},
we obtain \reqnarray{proof of comparison rule A-(ii)-(b)-333}
and \reqnarray{proof of comparison rule A-(ii)-(b)-444}.
On the other hand, if $j+2\leq a\leq r_{h-1}-j-1$,
then we have from \reqnarray{proof of comparison rule A-(ii)-(b)-bbb} and \reqnarray{proof of adjacent distance larger than one-(ii)-3333} that
\beqnarray{proof of comparison rule A-(ii)-(b)-eee}
i'_a-\left(\sum_{\ell=0}^{j-1}m'_{a-\ell}+m'_{a+1+j}\right)\geq i'_{a-j}-m'_{a-j}+1=i'_{a-j-1}+1>i'_{a-j-1}.
\eeqnarray
Thus, we see from \reqnarray{proof of adjacent distance larger than one-666},
\reqnarray{proof of comparison rule A-(ii)-(a)-case-2-666}
(note that $j+2\leq a\leq r_{h-1}-j-1$),
and \reqnarray{proof of comparison rule A-(ii)-(b)-eee} that
\beqnarray{proof of comparison rule A-(ii)-(b)-fff}
n'_{i'_a-j'}=q_{h-1},
\textrm{ for } \sum_{\ell=0}^{j-1}m'_{a-\ell}+1\leq j'\leq \sum_{\ell=0}^{j-1}m'_{a-\ell}+m'_{a+1+j}.
\eeqnarray
By combining \reqnarray{proof of comparison rule A-(ii)-444},
\reqnarray{proof of comparison rule A-(ii)-(b)-aaa},
and \reqnarray{proof of comparison rule A-(ii)-(b)-fff},
we obtain \reqnarray{proof of comparison rule A-(ii)-(b)-333}
and \reqnarray{proof of comparison rule A-(ii)-(b)-444}.

To complete the proof,
we need to show that the condition in
\reqnarray{proof of comparison rule A-(ii)-(b)-555}--\reqnarray{proof of comparison rule A-(ii)-(b)-777}
is equivalent to the condition in \reqnarray{proof of comparison rule A-(ii)-(b)-222}.
Note that if $i'_a-(\sum_{\ell=0}^{j-1}m'_{a-\ell}+m'_{a+1+j})=1$
and $(i'_a+1)+(\sum_{\ell=0}^{j-1}m'_{a-\ell}+m'_{a+1+j})=r_{h-2}$,
then we have from $n'_1=n'_{i'_a-(\sum_{\ell=0}^{j-1}m'_{a-\ell}+m'_{a+1+j})}=q_{h-1}$
and $n'_{r_{h-2}}=n'_{(i'_a+1)+(\sum_{\ell=0}^{j-1}m'_{a-\ell}+m'_{a+1+j})}=q_{h-1}+1$
in \reqnarray{proof of comparison rule A-(ii)-(b)-444} that
\beqnarray{}
n'_1=n'_{r_{h-2}}-1.\nn
\eeqnarray
As such, we see that the condition in
\reqnarray{proof of comparison rule A-(ii)-(b)-555}--\reqnarray{proof of comparison rule A-(ii)-(b)-777}
is equivalent to the following condition:
\beqnarray{proof of comparison rule A-(ii)-(b)-ggg}
i'_a-\left(\sum_{\ell=0}^{j-1}m'_{a-\ell}+m'_{a+1+j}\right)=1
\textrm{ and } (i'_a+1)+\left(\sum_{\ell=0}^{j-1}m'_{a-\ell}+m'_{a+1+j}\right)=r_{h-2}.
\eeqnarray
Note that from \reqnarray{proof of comparison rule A-(ii)-(b)-ccc}
and \reqnarray{proof of comparison rule A-(ii)-(b)-eee},
it is easy to see that
\beqnarray{proof of comparison rule A-(ii)-(b)-hhh}
i'_a-\left(\sum_{\ell=0}^{j-1}m'_{a-\ell}+m'_{a+1+j}\right)=1
\textrm{ iff } a=j+1 \textrm{ and } m_{a-j}=m_{a+1+j}+1.
\eeqnarray
As we have $(i'_a+1)+\left(\sum_{\ell=0}^{j-1}m'_{a-\ell}+m'_{a+1+j}\right)=i'_{a+1+j}$
in \reqnarray{proof of comparison rule A-(ii)-(b)-999}
and it is clear from \reqnarray{proof of adjacent distance larger than one-777}
and \reqnarray{proof of adjacent distance larger than one-(ii)-222}
that $i'_{a+1+j}=r_{h-2}$ if and only if $a+1+j=r_{h-1}$,
it follows that
\beqnarray{proof of comparison rule A-(ii)-(b)-iii}
(i'_a+1)+\left(\sum_{\ell=0}^{j-1}m'_{a-\ell}+m'_{a+1+j}\right)=r_{h-2} \textrm{ iff } a+1+j=r_{h-1}.
\eeqnarray
Therefore, we deduce from \reqnarray{proof of comparison rule A-(ii)-(b)-hhh}
and \reqnarray{proof of comparison rule A-(ii)-(b)-iii}
that the condition in \reqnarray{proof of comparison rule A-(ii)-(b)-ggg}
is equivalent to the following condition:
\beqnarray{}
a=j+1,\ a+1+j=r_{h-1}, \textrm{ and } m_{a-j}=m_{a+1+j}+1, \nn
\eeqnarray
which is clearly equivalent to the condition
that $a-j=1$, $a+1+j=r_{h-1}$, and $m_1=m_{r_{h-1}}+1$
in \reqnarray{proof of comparison rule A-(ii)-(b)-222},
and the proof is completed.

(iii) Note that in \rlemma{comparison rule A}(iii),
we have $2\leq a\leq r_{h-1}-2$
and $m_{a-\ell}=m_{a+1+\ell}$ for $\ell=1,2,\ldots,\min\{a-1,r_{h-1}-a-1\}$.
To show \reqnarray{comparison rule A-4},
i.e., $\mbf_1^{r_{h-1}}\succ{\mbf'}_1^{r_{h-1}}$,
we see from \reqnarray{proof of adjacent distance larger than one-333}
that it suffices to show that
\beqnarray{proof of comparison rule A-(iii)-111}
\nbf_1^{r_{h-2}}\succ{\nbf'}_1^{r_{h-2}}.
\eeqnarray
Note that as we have $2\leq a\leq r_{h-1}-2$,
it follows that \reqnarray{proof of comparison rule A-(ii)-222} also holds.

\bpdffigure{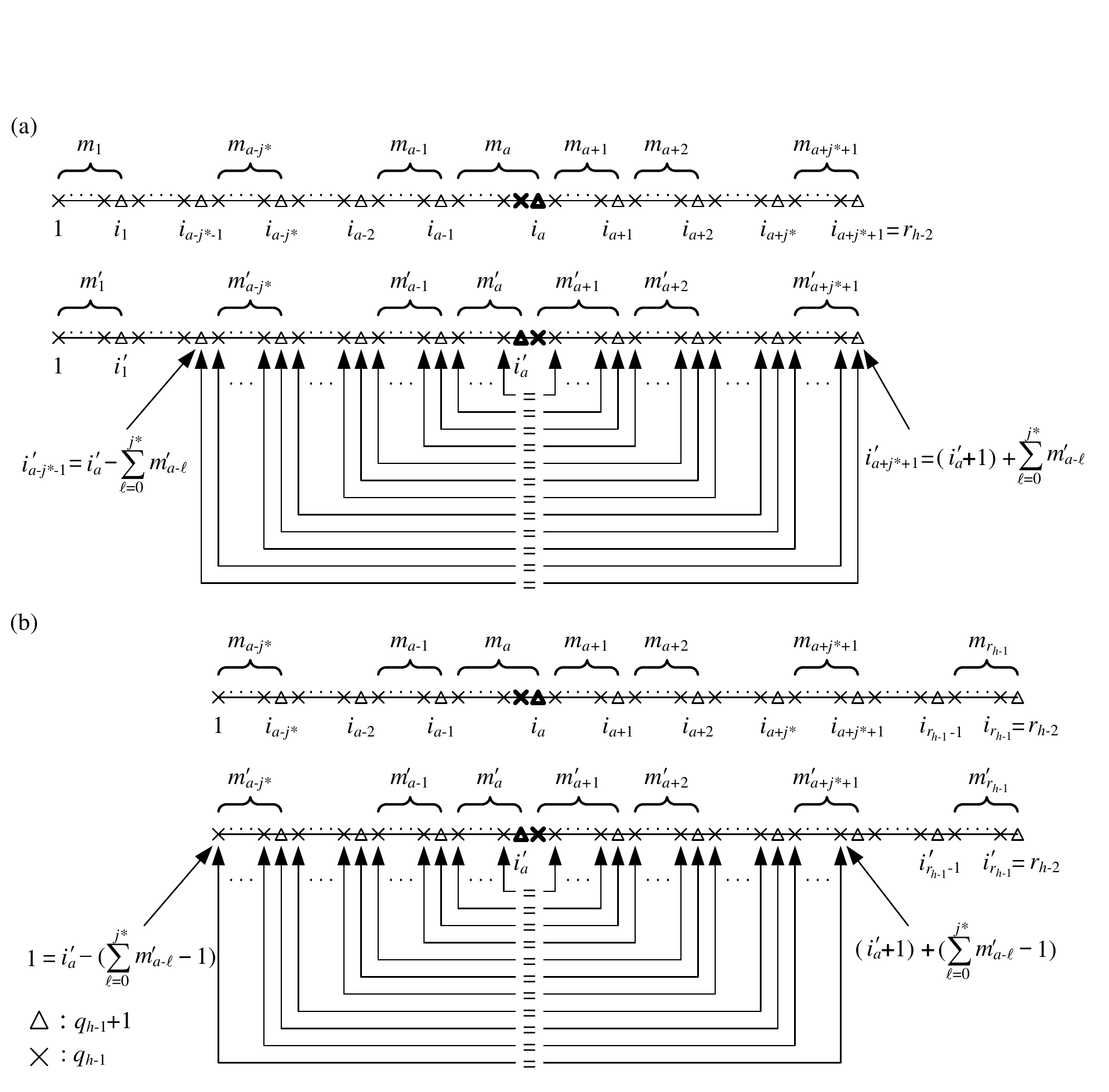}{6.0in}
\epdffigure{appendix-F-(iii)}
{An illustration of \reqnarray{proof of comparison rule A-(iii)-222}
(note that we have
$i'_a-\sum_{\ell=0}^{j''-1}m'_{a-\ell}=i'_{a-j''}$
for $1\leq j''\leq a-1$ in \reqnarray{proof of comparison rule A-(ii)-555}
and $(i'_a+1)+\sum_{\ell=0}^{j''-1}m'_{a-\ell}=i'_{a+j''}$
for $1\leq j''\leq j^*+1$ in \reqnarray{proof of comparison rule A-(ii)-888}):
(a) $a-1>r_{h-1}-a-1$
(note that in this case we have
$j^*=r_{h-1}-a-1$ and $j^*+1\leq a-1$ in \reqnarray{proof of comparison rule A-(iii)-case-1-111}
and $\min\{i'_a-1,r_{h-2}-i'_a-1\}=\sum_{\ell=0}^{j^*}m'_{a-\ell}$
in \reqnarray{proof of comparison rule A-(iii)-case-1-555});
(b) $a-1\leq r_{h-1}-a-1$
(note that in this case we have
$i'_a-(\sum_{\ell=0}^{j*}m'_{a-\ell}-1)=1$
in \reqnarray{proof of comparison rule A-(ii)-(a)-case-1-555},
$j^*=a-1$ in \reqnarray{proof of comparison rule A-(iii)-case-2-111},
and $\min\{i'_a-1,r_{h-2}-i'_a-1\}=\sum_{\ell=1}^{j^*}m'_{a-\ell}-1$
in \reqnarray{proof of comparison rule A-(iii)-case-2-222}).}

In the following, we show that
\beqnarray{proof of comparison rule A-(iii)-222}
n'_{i'_a-j'}=n'_{(i'_a+1)+j'},
\textrm{ for } j'=1,2,\ldots,\min\{i'_a-1,r_{h-2}-i'_a-1\}.
\eeqnarray
An illustration of \reqnarray{proof of comparison rule A-(iii)-222}
is given in \rfigure{appendix-F-(iii)}.
Therefore, it follows from \reqnarray{proof of adjacent distance larger than one-888},
${\nbf'}_1^{r_{h-2}}\in \Ncal_{M,k}(h-1)$ in
\reqnarray{proof of adjacent distance larger than one-(ii)-333},
\reqnarray{proof of adjacent distance larger than one-(ii)-888}--\reqnarray{proof of adjacent distance larger than one-(ii)-bbb},
\reqnarray{proof of comparison rule A-(ii)-222},
\reqnarray{proof of comparison rule A-(iii)-222},
and \reqnarray{comparison rule B-4} in \rlemma{comparison rule B}(iii)
(for the even integer $h-1$) that ${\nbf'}_1^{r_{h-2}}\prec\nbf_1^{r_{h-2}}$,
i.e., \reqnarray{proof of comparison rule A-(iii)-111} holds.

To prove \reqnarray{proof of comparison rule A-(iii)-222},
let $j^*=\min\{a-1,r_{h-1}-a-1\}$.
As $a\geq 2$, it is clear from the same argument as in (ii) above
that \reqnarray{proof of comparison rule A-(ii)-666} still holds.
Furthermore, from $m'_i=m_i$ for $i\neq a$ and $a+1$
and $m_{a-\ell}=m_{a+1+\ell}$ for $\ell=1,2,\ldots,j^*$,
it is clear that
\beqnarray{proof of comparison rule A-(iii)-333}
m'_{a-\ell}=m'_{a+1+\ell}, \textrm{ for } \ell=1,2,\ldots,j^*.
\eeqnarray
From \reqnarray{proof of comparison rule A-(iii)-333},
it is easy to see from the same argument as in (ii) above that
\reqnarray{proof of comparison rule A-(ii)-777} and \reqnarray{proof of comparison rule A-(ii)-888}
hold for $1\leq j''\leq j^*+1$,
and \reqnarray{proof of comparison rule A-(ii)-999} holds for $j=j^*+1$.
We then consider the two cases
$a-1>r_{h-1}-a-1$ and $a-1\leq r_{h-1}-a-1$ separately.

\emph{Case 1: $a-1>r_{h-1}-a-1$.}
In this case, we have
\beqnarray{proof of comparison rule A-(iii)-case-1-111}
j^*=r_{h-1}-a-1 \textrm{ and } j^*<a-1.
\eeqnarray
From $\sum_{\ell=1}^{r_{h-1}}m'_{\ell}=r_{h-2}$ in \reqnarray{proof of adjacent distance larger than one-(ii)-222},
\reqnarray{proof of adjacent distance larger than one-777},
$j^*=r_{h-1}-a-1$ in \reqnarray{proof of comparison rule A-(iii)-case-1-111},
$m'_a-m'_{a+1}=-1$ in \reqnarray{proof of comparison rule A-444},
and \reqnarray{proof of comparison rule A-(iii)-333},
we have
\beqnarray{proof of comparison rule A-(iii)-case-1-222}
r_{h-2}
\aligneq \sum_{\ell=1}^{r_{h-1}}m'_{\ell}=\sum_{\ell=1}^{a}m'_{\ell}+\sum_{\ell=a+1}^{r_{h-1}}m'_{\ell} \nn\\
\aligneq i'_a+\sum_{\ell=0}^{r_{h-1}-a-1}m'_{a+1+\ell}=i'_a+m'_{a+1}+\sum_{\ell=1}^{j^*}m'_{a+1+\ell} \nn\\
\aligneq i'_a+m'_{a}+1+\sum_{\ell=1}^{j^*}m'_{a-\ell}=i'_a+1+\sum_{\ell=0}^{j^*}m'_{a-\ell}.
\eeqnarray{}
As we have $j^*+1\leq a-1$ in \reqnarray{proof of comparison rule A-(iii)-case-1-111},
we see from \reqnarray{proof of comparison rule A-(ii)-555} (with $j''=j^*+1\leq a-1$) that
\beqnarray{proof of comparison rule A-(iii)-case-1-333}
i'_a-\sum_{\ell=0}^{j^*}m'_{a-\ell}=i'_{a-j^*-1}.
\eeqnarray
It follows from \reqnarray{proof of comparison rule A-(iii)-case-1-222}
and \reqnarray{proof of comparison rule A-(iii)-case-1-333} that
\beqnarray{proof of comparison rule A-(iii)-case-1-444}
r_{h-2}-i'_a-1=\sum_{\ell=0}^{j^*}m'_{a-\ell}=i'_a-i'_{a-j^*-1}\leq i'_a-1.
\eeqnarray
Thus, we see from \reqnarray{proof of comparison rule A-(iii)-case-1-444} that
\beqnarray{proof of comparison rule A-(iii)-case-1-555}
\min\{i'_a-1,r_{h-2}-i'_a-1\}=r_{h-2}-i'_a-1=\sum_{\ell=0}^{j^*}m'_{a-\ell}.
\eeqnarray

As such, it follows from \reqnarray{proof of comparison rule A-(ii)-666},
\reqnarray{proof of comparison rule A-(ii)-999} (with $j=j^*+1$),
and $j^*+1\leq a-1$ in \reqnarray{proof of comparison rule A-(iii)-case-1-111}
that \reqnarray{proof of comparison rule A-(ii)-444} holds for $j=j^*+1$, i.e.,
\beqnarray{proof of comparison rule A-(iii)-case-1-777}
n'_{i'_a-j'}=n'_{(i'_a+1)+j'},
\textrm{ for } j'=1,2,\ldots,\sum_{\ell=0}^{j^*}m'_{a-\ell}.
\eeqnarray
Therefore, \reqnarray{proof of comparison rule A-(iii)-222}
follows from \reqnarray{proof of comparison rule A-(iii)-case-1-555}
and \reqnarray{proof of comparison rule A-(iii)-case-1-777}.

\emph{Case 2: $a-1\leq r_{h-1}-a-1$.}
In this case, we have
\beqnarray{proof of comparison rule A-(iii)-case-2-111}
j^*=a-1 \textrm{ and } j^*\leq r_{h-1}-a-1.
\eeqnarray
As we have $a=j^*+1$ and $a\leq r_{h-1}-j^*-1$
in \reqnarray{proof of comparison rule A-(iii)-case-2-111}
and $m'_{a-\ell}=m'_{a+1+\ell}$ for $\ell=1,2,\ldots,j^*$
in \reqnarray{proof of comparison rule A-(iii)-333},
it is easy to see that $(i'_a-1)-(r_{h-2}-i'_a-1)<0$
in \reqnarray{proof of comparison rule A-(ii)-(a)-case-1-222} still holds
and hence \reqnarray{proof of comparison rule A-(ii)-(a)-case-1-333} holds with $j=j^*$,
i.e.,
\beqnarray{proof of comparison rule A-(iii)-case-2-222}
\min\{i'_a-1,r_{h-2}-i'_a-1\}=\sum_{\ell=0}^{j^*}m'_{a-\ell}-1.
\eeqnarray

From \reqnarray{proof of comparison rule A-(ii)-999} (with $j=j^*+1$), we have
\beqnarray{proof of comparison rule A-(iii)-case-2-333}
n'_{(i'_a+1)+j'}=
\bselection
q_{h-1}+1, &\textrm{ for } j'=m'_a, \sum_{\ell=0}^{1}m'_{a-\ell},\ldots,\sum_{\ell=0}^{j^*}m'_{a-\ell}, \\
q_{h-1}, &\textrm{ for } 1\leq j'\leq \sum_{\ell=0}^{j^*}m'_{a-\ell} \\
&\textrm{ and } j'\neq m'_a, \sum_{\ell=0}^{1}m'_{a-\ell},\ldots,\sum_{\ell=0}^{j^*}m'_{a-\ell}.
\eselection
\eeqnarray
As we have $j^*=a-1$ in \reqnarray{proof of comparison rule A-(iii)-case-2-111},
it follows from \reqnarray{proof of comparison rule A-(ii)-666} that
\beqnarray{proof of comparison rule A-(iii)-case-2-444}
n'_{i'_a-j'}=
\bselection
q_{h-1}+1, &\textrm{ for } j'=m'_a, \sum_{\ell=0}^{1}m'_{a-\ell},\ldots,\sum_{\ell=0}^{j^*-1}m'_{a-\ell}, \\
q_{h-1}, &\textrm{ for } 1\leq j'\leq \sum_{\ell=0}^{j^*-1}m'_{a-\ell} \\
&\textrm{ and } j'\neq m'_a, \sum_{\ell=0}^{1}m'_{a-\ell},\ldots,\sum_{\ell=0}^{j^*-1}m'_{a-\ell}.
\eselection
\eeqnarray
Also, it is clear from $j^*=a-1$ in \reqnarray{proof of comparison rule A-(iii)-case-2-111} that
\reqnarray{proof of comparison rule A-(ii)-(a)-case-1-444}--\reqnarray{proof of comparison rule A-(ii)-(a)-case-1-666}
hold for $j=j^*$,
and hence we see from \reqnarray{proof of comparison rule A-(ii)-(a)-case-1-666} (with $j=j^*$) that
\beqnarray{proof of comparison rule A-(iii)-case-2-555}
n'_{i'_a-j'}=q_{h-1},
\textrm{ for } \sum_{\ell=0}^{j^*-1}m'_{a-\ell}+1\leq j'\leq \sum_{\ell=0}^{j^*}m'_{a-\ell}-1.
\eeqnarray
As such, it follows from
\reqnarray{proof of comparison rule A-(iii)-case-2-333}--\reqnarray{proof of comparison rule A-(iii)-case-2-555} that
\beqnarray{proof of comparison rule A-(iii)-case-2-666}
n'_{i'_a-j'}=n'_{(i'_a+1)+j'},
\textrm{ for } j'=1,2,\ldots,\sum_{\ell=0}^{j^*}m'_{a-\ell}-1.
\eeqnarray
Therefore, \reqnarray{proof of comparison rule A-(iii)-222}
follows from \reqnarray{proof of comparison rule A-(iii)-case-2-222}
and \reqnarray{proof of comparison rule A-(iii)-case-2-666}.

\bappendix{Proof of \rlemma{nonadjacent distance larger than one}}
{proof of nonadjacent distance larger than one}

In this appendix, we use \rlemma{adjacent distance larger than one}
and Comparison rule A in \rlemma{comparison rule A}
to prove \rlemma{nonadjacent distance larger than one}.
For simplicity, let $\nbf_1^{r_{h-1}}=\nbf_1^{r_{h-1}}(h)$.
Note that in \rlemma{nonadjacent distance larger than one}, we have
\beqnarray{proof of nonadjacent distance larger than one-111}
r_{h-1}\geq 3,\ \nbf_1^{r_{h-1}}\in \Ncal_{M,k}(h),
\textrm{ and } |n_i-n_{i+1}|\leq 1 \textrm{ for } i=1,2,\ldots,r_{h-1}-1.
\eeqnarray

(i) Note that in \rlemma{nonadjacent distance larger than one}(i),
we have $n_a-n_b\geq 2$ for some $1\leq a<b\leq r_{h-1}$ and $b\geq a+2$.
For ease of presentation, let $n_b=p$.
Then we have from $n_a-n_b\geq 2$ that $n_a\geq p+2$.
Note that the condition $|n_{i+1}-n_i|\leq 1$ for $i=1,2,\ldots,r_{h-1}-1$
in \reqnarray{proof of nonadjacent distance larger than one-111}
says that the absolute value of the difference of any two adjacent entries
of $\nbf_1^{r_{h-1}}$ is at most equal to one.
As such, it is easy to see from $n_a\geq p+2$, $n_b=p$, $b\geq a+2>a$,
and the condition $|n_{i+1}-n_i|\leq 1$ for $i=1,2,\ldots,r_{h-1}-1$
in \reqnarray{proof of nonadjacent distance larger than one-111}
that there must exist a positive integer $c$ such that $a\leq c<b$ and $n_c=p+2$.
Let
\beqnarray{}
a'\aligneq \max\{i:n_i=p+2,\ c\leq i<b\},
\label{eqn:proof of nonadjacent distance larger than one-(i)-111}\\
b'\aligneq \min\{i:n_i=p,\ a'<i\leq b\}.
\label{eqn:proof of nonadjacent distance larger than one-(i)-222}
\eeqnarray
In other words, $a'$ is the largest positive integer $i$ such that $c\leq i<b$ and $n_i=p+2$,
and $b'$ is the smallest positive integer $i$ such that $a'<i\leq b$ and $n_i=p$.
Note that $a'$ and $b'$ are well defined as we have $n_c=p+2$ and $n_b=p$.
Since we have from \reqnarray{proof of nonadjacent distance larger than one-(i)-111}
and \reqnarray{proof of nonadjacent distance larger than one-(i)-222}
that $n_{a'}=p+2$ and $n_{b'}=p$,
it is easy to see from the condition $|n_{i+1}-n_i|\leq 1$ for $i=1,2,\ldots,r_{h-1}-1$
in \reqnarray{proof of nonadjacent distance larger than one-111} that $b'\geq a'+2$.
In summary, we have
\beqnarray{}
\alignspace a\leq c\leq a'<b'\leq b \textrm{ and } b'\geq a'+2,
\label{eqn:proof of nonadjacent distance larger than one-(i)-333}\\
\alignspace n_a\geq p+2,\ n_c=n_{a'}=p+2, \textrm{ and } n_{b'}=n_b=p.
\label{eqn:proof of nonadjacent distance larger than one-(i)-444}
\eeqnarray

\bpdffigure{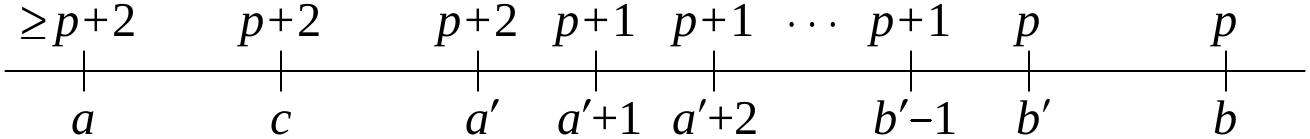}{4.5in}
\epdffigure{appendix-G-(i)}
{An illustration of \reqnarray{proof of nonadjacent distance larger than one-(i)-333}--\reqnarray{proof of nonadjacent distance larger than one-(i)-555}.}

We claim that
\beqnarray{proof of nonadjacent distance larger than one-(i)-555}
n_i=p+1, \textrm{ for } a'<i<b'.
\eeqnarray
An illustration of
\reqnarray{proof of nonadjacent distance larger than one-(i)-333}--\reqnarray{proof of nonadjacent distance larger than one-(i)-555}
is given in \rfigure{appendix-G-(i)}.
We prove \reqnarray{proof of nonadjacent distance larger than one-(i)-555} by contradiction.
First assume that $n_i\geq p+2$ for some $a'<i<b'$.
From $c\leq a'<b'\leq b$ in \reqnarray{proof of nonadjacent distance larger than one-(i)-333} and $a'<i<b'$,
we have $c\leq a'<i<b'\leq b$ and hence it follows from the definition of $a'$
in \reqnarray{proof of nonadjacent distance larger than one-(i)-111} that $n_i\neq p+2$.
Since we assume that $n_i\geq p+2$, it is clear that we must have $n_i>p+2$.
As such, we see from $n_i>p+2$,
$n_{b'}=p$ in \reqnarray{proof of nonadjacent distance larger than one-(i)-444},
$i<b'$, and the condition $|n_{i+1}-n_i|\leq 1$ for $i=1,2,\ldots,r_{h-1}-1$
in \reqnarray{proof of nonadjacent distance larger than one-111}
that there must exist a positive integer $a''$ such that
\beqnarray{proof of nonadjacent distance larger than one-(i)-666}
i<a''<b' \textrm{ and } n_{a''}=p+2.
\eeqnarray
From $c\leq a'<b'\leq b$ in \reqnarray{proof of nonadjacent distance larger than one-(i)-333},
$a'<i<b'$, and $i<a''<b'$ in \reqnarray{proof of nonadjacent distance larger than one-(i)-666},
we have $c\leq a'<i<a''<b'\leq b$ and hence it follows from the definition of $a'$
in \reqnarray{proof of nonadjacent distance larger than one-(i)-111} that $n_{a''}\neq p+2$,
contradicting to $n_{a''}=p+2$ in \reqnarray{proof of nonadjacent distance larger than one-(i)-666}.
Now assume that $n_i\leq p$ for some $a'<i<b'$.
From $a'<b'\leq b$ in \reqnarray{proof of nonadjacent distance larger than one-(i)-333} and $a'<i<b'$,
we have $a'<i<b'\leq b$ and hence it follows from the definition of $b'$
in \reqnarray{proof of nonadjacent distance larger than one-(i)-222} that $n_i\neq p$.
Since we assume that $n_i\leq p$, it is clear that we must have $n_i<p$.
As such, we see from $n_{a'}=p+2$ in \reqnarray{proof of nonadjacent distance larger than one-(i)-444},
$n_i<p$, $a'<i$, and the condition $|n_{i+1}-n_i|\leq 1$ for $i=1,2,\ldots,r_{h-1}-1$
in \reqnarray{proof of nonadjacent distance larger than one-111}
that there must exist a positive integer $b''$ such that
\beqnarray{proof of nonadjacent distance larger than one-(i)-777}
a'<b''<i \textrm{ and } n_{b''}=p.
\eeqnarray
From $a'<b''<i$ in \reqnarray{proof of nonadjacent distance larger than one-(i)-777},
$a'<i<b'$, and $a'<b'\leq b$ in \reqnarray{proof of nonadjacent distance larger than one-(i)-333},
we have $a'<b''<i<b'\leq b$ and hence it follows from the definition of $b'$
in \reqnarray{proof of nonadjacent distance larger than one-(i)-222} that $n_{b''}\neq p$,
contradicting to $n_{b''}=p$ in \reqnarray{proof of nonadjacent distance larger than one-(i)-777}.
The proof of \reqnarray{proof of nonadjacent distance larger than one-(i)-555} is completed.

Note that as we have $b'\geq a'+2$ in \reqnarray{proof of nonadjacent distance larger than one-(i)-333},
it is clear that $a'<a'+1<b'$ and $a'<b'-1<b'$.
It then follows from \reqnarray{proof of nonadjacent distance larger than one-(i)-555} that
\beqnarray{proof of nonadjacent distance larger than one-(i)-888}
n_{a'+1}=p+1 \textrm{ and } n_{b'-1}=p+1.
\eeqnarray

To prove \rlemma{nonadjacent distance larger than one}(i),
we need to show that if $n_1\neq n_{r_{h-1}}+2$
or $n_i\neq n_{r_{h-1}}+1$ for some $2\leq i\leq r_{h-1}-1$,
then there exists a sequence of positive integers
${\nbf'}_1^{r_{h-1}}\in \Ncal_{M,k}(h)$ such that
\beqnarray{proof of nonadjacent distance larger than one-(i)-999}
{\nbf'}_1^{r_{h-1}}\succ\nbf_1^{r_{h-1}}.
\eeqnarray
Note that if $n_1\neq n_{r_{h-1}}+2$
or $n_i\neq n_{r_{h-1}}+1$ for some $2\leq i\leq r_{h-1}-1$,
then we have $a'\geq 2$ or $b'\leq r_{h-1}-1$.
To see this, suppose on the contrary that $a'=1$ and $b'=r_{h-1}$.
Then it follows from $a'=1$, $b'=r_{h-1}$,
$n_{a'}=p+2$ and $n_{b'}=p$
in \reqnarray{proof of nonadjacent distance larger than one-(i)-444},
and $n_i=p+1$ for $a'+1\leq i\leq b'-1$
in \reqnarray{proof of nonadjacent distance larger than one-(i)-555} that
\beqnarray{proof of nonadjacent distance larger than one-(i)-aaa}
n_1=p+2,\ n_{r_{h-1}}=p,
\textrm{ and } n_i=p+1 \textrm{ for } 2\leq i\leq r_{h-1}-1.
\eeqnarray
It is clear from \reqnarray{proof of nonadjacent distance larger than one-(i)-aaa}
that $n_1=n_{r_{h-1}}+2$ and $n_i=n_{r_{h-1}}+1$ for all $2\leq i\leq r_{h-1}-1$,
and a contradiction is reached.

In the following, we show that if $a'\geq 2$ or $b'\leq r_{h-1}-1$,
then there exists a sequence of positive integers ${\nbf'}_1^{r_{h-1}}\in \Ncal_{M,k}(h)$
such that \reqnarray{proof of nonadjacent distance larger than one-(i)-999} holds,
and hence \rlemma{nonadjacent distance larger than one}(i) is proved.

(a) First, we assume that $a'\geq 2$
and show that there exists a sequence of positive integers
${\nbf'}_1^{r_{h-1}}\in \Ncal_{M,k}(h)$ such that
\reqnarray{proof of nonadjacent distance larger than one-(i)-999} holds.
Note that from $b'\geq a'+2$
in \reqnarray{proof of nonadjacent distance larger than one-(i)-333}
and $b'\leq r_{h-1}$, we have
\beqnarray{proof of nonadjacent distance larger than one-(i)-(a)-111}
a'\leq b'-2\leq r_{h-1}-2.
\eeqnarray
As we assume that $a'\geq 2$,
it follows from \reqnarray{proof of nonadjacent distance larger than one-(i)-(a)-111} that
\beqnarray{proof of nonadjacent distance larger than one-(i)-(a)-222}
2\leq a'\leq r_{h-1}-2.
\eeqnarray
Furthermore, we have from $n_{a'}=p+2$
in \reqnarray{proof of nonadjacent distance larger than one-(i)-444}
and $n_{a'+1}=p+1$
in \reqnarray{proof of nonadjacent distance larger than one-(i)-888} that
\beqnarray{proof of nonadjacent distance larger than one-(i)-(a)-333}
n_{a'}-n_{a'+1}=(p+2)-(p+1)=1.
\eeqnarray

We need to consider the following three possible cases.

\emph{Case 1: There exists a positive integer $j$ such that
$1\leq j\leq \min\{a'-1,b'-a'-1\}$,
$n_{a'-j'}=p+1$ for $j'=1,2,\ldots,j-1$, and $n_{a'-j}>p+1$.}
Let $\mbf_1^{r_{h-1}}$ be a sequence of positive integers such that
\beqnarray{proof of nonadjacent distance larger than one-(i)-(a)-case-1-111}
m_{a'}=n_{a'}-1,\ m_{a'+1}=n_{a'+1}+1, \textrm{ and } m_i=n_i \textrm{ for } i\neq a', a'+1.
\eeqnarray
As before, it is easy to show that $\mbf_1^{r_{h-1}}\in \Ncal_{M,k}(h)$.
As we have $j\leq \min\{a'-1,b'-a'-1\}\leq b'-a'-1$,
we consider the two subcases $j<b'-a'-1$ and $j=b'-a'-1$ separately.

\emph{Subcase 1(a): $j<b'-a'-1$.}

\bpdffigure{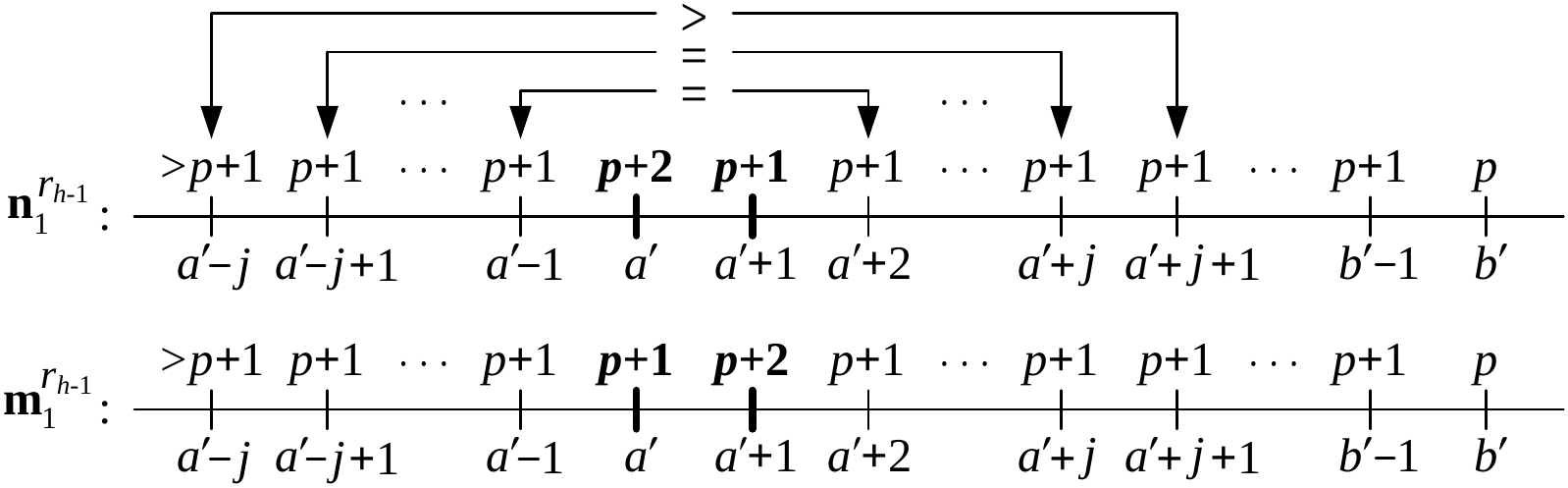}{5.5in}
\epdffigure{appendix-G-(i)-(a)-case-1}
{An illustration of \reqnarray{proof of nonadjacent distance larger than one-(i)-(a)-case-1-333}
and \reqnarray{proof of nonadjacent distance larger than one-(i)-(a)-case-1-444}.}

In this subcase, we show that
\beqnarray{}
\alignspace 1\leq j\leq \min\{a'-1,r_{h-1}-a'-1\}
\label{eqn:proof of nonadjacent distance larger than one-(i)-(a)-case-1-222}\\
\alignspace n_{a'-j'}=n_{(a'+1)+j'}, \textrm{ for } j'=1,2,\ldots,j-1,
\label{eqn:proof of nonadjacent distance larger than one-(i)-(a)-case-1-333}\\
\alignspace n_{a'-j}>n_{(a'+1)+j}.
\label{eqn:proof of nonadjacent distance larger than one-(i)-(a)-case-1-444}
\eeqnarray
An illustration of
\reqnarray{proof of nonadjacent distance larger than one-(i)-(a)-case-1-333}
and \reqnarray{proof of nonadjacent distance larger than one-(i)-(a)-case-1-444}
is given in \rfigure{appendix-G-(i)-(a)-case-1}.
Therefore, it follows from
$\nbf_1^{r_{h-1}}\in \Ncal_{M,k}(h)$
in \reqnarray{proof of nonadjacent distance larger than one-111},
\reqnarray{proof of nonadjacent distance larger than one-(i)-(a)-222}--\reqnarray{proof of nonadjacent distance larger than one-(i)-(a)-case-1-444},
and \reqnarray{comparison rule A-3} in \rlemma{comparison rule A}(ii) that
\beqnarray{proof of nonadjacent distance larger than one-(i)-(a)-case-1-555}
\nbf_1^{r_{h-1}}\preceq \mbf_1^{r_{h-1}},
\eeqnarray
where $\nbf_1^{r_{h-1}}\equiv \mbf_1^{r_{h-1}}$
if and only if $a'-j=1$, $(a'+1)+j=r_{h-1}$,
and $n_1=n_{r_{h-1}}+1$ (i.e., $n_{a'-j}=n_{(a'+1)+j}+1$).
From $j<b'-a'-1$ in this subcase and $b'\leq r_{h-1}$,
we see that $(a'+1)+j<b'\leq r_{h-1}$.
This implies that $(a'+1)+j\neq r_{h-1}$ and hence it cannot be the case that
$\nbf_1^{r_{h-1}}\equiv \mbf_1^{r_{h-1}}$.
As such, we see from \reqnarray{proof of nonadjacent distance larger than one-(i)-(a)-case-1-555}
that $\nbf_1^{r_{h-1}}\prec \mbf_1^{r_{h-1}}$,
i.e., \reqnarray{proof of nonadjacent distance larger than one-(i)-999} holds
with ${\nbf'}_1^{r_{h-1}}=\mbf_1^{r_{h-1}}$.

From $1\leq j\leq \min\{a'-1,b'-a'-1\}$ and $b'\leq r_{h-1}$, we see that
\beqnarray{proof of nonadjacent distance larger than one-(i)-(a)-case-1-666}
1\leq j\leq \min\{a'-1,b'-a'-1\}\leq \min\{a'-1,r_{h-1}-a'-1\}.
\eeqnarray
Thus, \reqnarray{proof of nonadjacent distance larger than one-(i)-(a)-case-1-222}
follows from \reqnarray{proof of nonadjacent distance larger than one-(i)-(a)-case-1-666}.

To prove \reqnarray{proof of nonadjacent distance larger than one-(i)-(a)-case-1-333}
and \reqnarray{proof of nonadjacent distance larger than one-(i)-(a)-case-1-444},
note that in this subcase we have $a'<(a'+1)+j<b'$,
and hence it follows from
\reqnarray{proof of nonadjacent distance larger than one-(i)-555} that
\beqnarray{proof of nonadjacent distance larger than one-(i)-(a)-case-1-777}
n_{(a'+1)+j'}=p+1, \textrm{ for } j'=1,2,\ldots,j.
\eeqnarray
By combining $n_{a'-j'}=p+1$ for $j'=1,2,\ldots,j-1$, $n_{a'-j}>p+1$,
and \reqnarray{proof of nonadjacent distance larger than one-(i)-(a)-case-1-777},
we obtain \reqnarray{proof of nonadjacent distance larger than one-(i)-(a)-case-1-333}
and \reqnarray{proof of nonadjacent distance larger than one-(i)-(a)-case-1-444}.

\emph{Subcase 1(b): $j=b'-a'-1$.}
In this subcase, we have $a'<(a'+1)+j=b'$
and hence it follows from \reqnarray{proof of nonadjacent distance larger than one-(i)-555}
and $n_{b'}=p$ in \reqnarray{proof of nonadjacent distance larger than one-(i)-444} that
\beqnarray{}
\alignspace n_{(a'+1)+j'}=p+1, \textrm{ for } j'=1,2,\ldots,j-1,
\label{eqn:proof of nonadjacent distance larger than one-(i)-(a)-case-1-888}\\
\alignspace n_{(a'+1)+j}=n_{b'}=p.
\label{eqn:proof of nonadjacent distance larger than one-(i)-(a)-case-1-999}
\eeqnarray
By using \reqnarray{proof of nonadjacent distance larger than one-(i)-(a)-case-1-888}
and \reqnarray{proof of nonadjacent distance larger than one-(i)-(a)-case-1-999},
we can argue as in Subcase~1(a) above that
\reqnarray{proof of nonadjacent distance larger than one-(i)-(a)-case-1-222}--\reqnarray{proof of nonadjacent distance larger than one-(i)-(a)-case-1-555}
still hold.
Since it is clear from $n_{a'-j}>p+1$ and $n_{(a'+1)+j}=p$
in \reqnarray{proof of nonadjacent distance larger than one-(i)-(a)-case-1-999}
that $n_{a'-j}\neq n_{(a'+1)+j}+1$,
it cannot be the case that $\nbf_1^{r_{h-1}}\equiv \mbf_1^{r_{h-1}}$.
As such, we see from \reqnarray{proof of nonadjacent distance larger than one-(i)-(a)-case-1-555}
that $\nbf_1^{r_{h-1}}\prec \mbf_1^{r_{h-1}}$,
i.e., \reqnarray{proof of nonadjacent distance larger than one-(i)-999} holds
with ${\nbf'}_1^{r_{h-1}}=\mbf_1^{r_{h-1}}$.

\emph{Case 2: There exists a positive integer $j$ such that
$1\leq j\leq \min\{a'-1,b'-a'-1\}$,
$n_{a'-j'}=p+1$ for $j'=1,2,\ldots,j-1$, and $n_{a'-j}<p+1$.}
In this case, we can show that $j\geq 2$.
To see this, suppose on the contrary that $j=1$,
then we have $n_{a'-1}<p+1$ in this case.
As it is easy to see from $n_{a'}=p+2$
in \reqnarray{proof of nonadjacent distance larger than one-(i)-444}
and the condition $|n_{i+1}-n_i|\leq 1$ for $i=1,2,\ldots,r_{h-1}-1$
in \reqnarray{proof of nonadjacent distance larger than one-111}
that $n_{a'-1}$ must be equal to $p+1$, $p+2$, or $p+3$,
we have reached a contradiction.

Let $\mbf_1^{r_{h-1}}$ be a sequence of positive integers such that
\beqnarray{proof of nonadjacent distance larger than one-(i)-(a)-case-2-111}
m_{a'-1}=n_{a'-1}+1,\ m_{a'}=n_{a'}-1,
\textrm{ and } m_i=n_i \textrm{ for } i\neq a'-1, a'.
\eeqnarray
As before, it is easy to show that $\mbf_1^{r_{h-1}}\in \Ncal_{M,k}(h)$.
As $j\geq 2$, we have $n_{a'-1}=p+1$ in this case.
It then follows from \reqnarray{proof of nonadjacent distance larger than one-(i)-(a)-case-2-111},
$n_{a'-1}=p+1$, and $n_{a'}=p+2$
in \reqnarray{proof of nonadjacent distance larger than one-(i)-444} that
\beqnarray{proof of nonadjacent distance larger than one-(i)-(a)-case-2-222}
m_{a'-1}-m_{a'}=(n_{a'-1}+1)-(n_{a'}-1)=(p+1+1)-(p+2-1)=1.
\eeqnarray

\bpdffigure{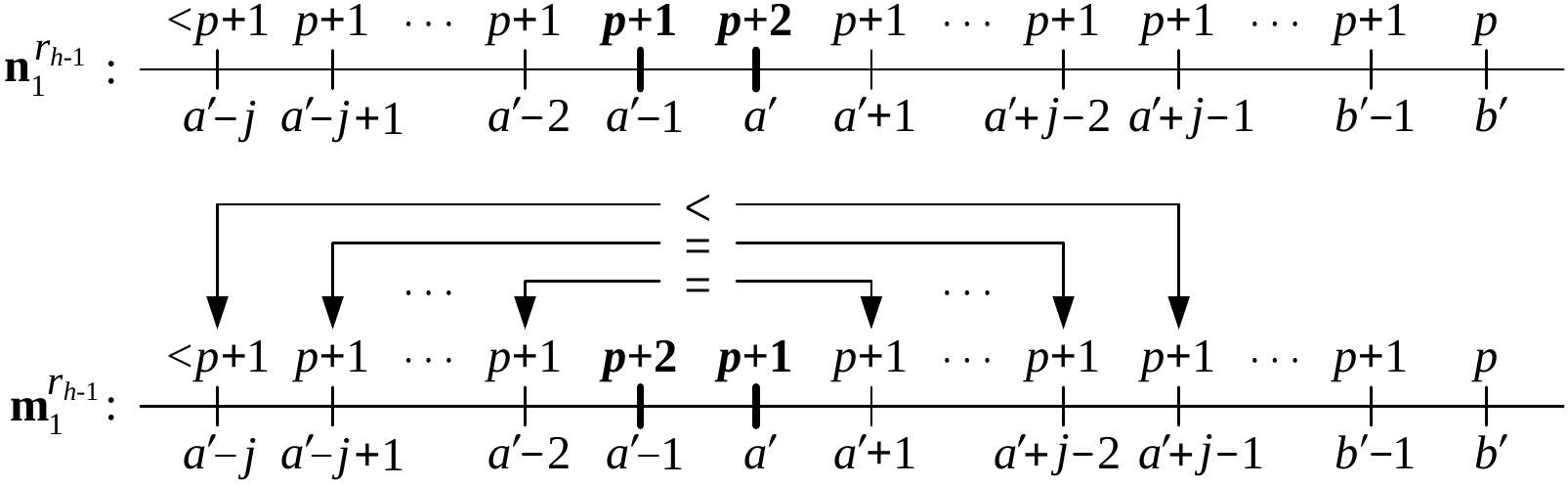}{5.5in}
\epdffigure{appendix-G-(i)-(a)-case-2}
{An illustration of \reqnarray{proof of nonadjacent distance larger than one-(i)-(a)-case-2-555}
and \reqnarray{proof of nonadjacent distance larger than one-(i)-(a)-case-2-666}.}

In the following, we show that
\beqnarray{}
\alignspace
2\leq a'-1\leq r_{h-1}-2,
\label{eqn:proof of nonadjacent distance larger than one-(i)-(a)-case-2-333}\\
\alignspace
1\leq j-1\leq \min\{(a'-1)-1,r_{h-1}-(a'-1)-1\},
\label{eqn:proof of nonadjacent distance larger than one-(i)-(a)-case-2-444}\\
\alignspace m_{(a'-1)-j'}=m_{a'+j'}, \textrm{ for } j'=1,2,\ldots,j-2,
\label{eqn:proof of nonadjacent distance larger than one-(i)-(a)-case-2-555}\\
\alignspace m_{(a'-1)-(j-1)}<m_{a'+(j-1)}.
\label{eqn:proof of nonadjacent distance larger than one-(i)-(a)-case-2-666}
\eeqnarray
An illustration of
\reqnarray{proof of nonadjacent distance larger than one-(i)-(a)-case-2-555}
and \reqnarray{proof of nonadjacent distance larger than one-(i)-(a)-case-2-666}
is given in \rfigure{appendix-G-(i)-(a)-case-2}.
Therefore, it follows from
$\mbf_1^{r_{h-1}}\in \Ncal_{M,k}(h)$,
\reqnarray{proof of nonadjacent distance larger than one-(i)-(a)-case-2-111}--\reqnarray{proof of nonadjacent distance larger than one-(i)-(a)-case-2-666},
and \reqnarray{comparison rule A-2} in \rlemma{comparison rule A}(ii) that
$\mbf_1^{r_{h-1}}\succ \nbf_1^{r_{h-1}}$,
i.e., \reqnarray{proof of nonadjacent distance larger than one-(i)-999} holds
with ${\nbf'}_1^{r_{h-1}}=\mbf_1^{r_{h-1}}$.

From $2\leq j\leq \min\{a'-1,b'-a'-1\}$,
\reqnarray{proof of nonadjacent distance larger than one-(i)-(a)-222},
and $b'\leq r_{h-1}$, we can see that
\beqnarray{}
\alignspace
2\leq \min\{a'-1,b'-a'-1\}\leq a'-1<a'\leq r_{h-1}-2,
\label{eqn:proof of nonadjacent distance larger than one-(i)-(a)-case-2-777}\\
\alignspace
1\leq j-1\leq \min\{a'-2,b'-a'-2\}\leq \min\{(a'-1)-1,r_{h-1}-(a'-1)-1\}.
\label{eqn:proof of nonadjacent distance larger than one-(i)-(a)-case-2-888}
\eeqnarray
Thus, \reqnarray{proof of nonadjacent distance larger than one-(i)-(a)-case-2-333}
follows from \reqnarray{proof of nonadjacent distance larger than one-(i)-(a)-case-2-777},
and \reqnarray{proof of nonadjacent distance larger than one-(i)-(a)-case-2-444}
follows from \reqnarray{proof of nonadjacent distance larger than one-(i)-(a)-case-2-888}.

To prove \reqnarray{proof of nonadjacent distance larger than one-(i)-(a)-case-2-555}
and \reqnarray{proof of nonadjacent distance larger than one-(i)-(a)-case-2-666},
note that as we have $n_{a'-j'}=p+1$ for $j'=1,2,\ldots,j-1$ and $n_{a'-j}<p+1$ in this case,
it is clear from \reqnarray{proof of nonadjacent distance larger than one-(i)-(a)-case-2-111} that
\beqnarray{}
\alignspace m_{(a'-1)-j'}=n_{(a'-1)-j'}=p+1, \textrm{ for } j'=1,2,\ldots,j-2,
\label{eqn:proof of nonadjacent distance larger than one-(i)-(a)-case-2-999}\\
\alignspace m_{(a'-1)-(j-1)}=n_{(a'-1)-(j-1)}=n_{a'-j}<p+1.
\label{eqn:proof of nonadjacent distance larger than one-(i)-(a)-case-2-aaa}
\eeqnarray
Furthermore, we have from $2\leq j\leq \min\{a'-1,b'-a'-1\}$ that
\beqnarray{proof of nonadjacent distance larger than one-(i)-(a)-case-2-bbb}
a'< a'+j-1\leq a'+(b'-a'-1)-1=b'-2<b'.
\eeqnarray
It then follows from \reqnarray{proof of nonadjacent distance larger than one-(i)-(a)-case-2-111},
\reqnarray{proof of nonadjacent distance larger than one-(i)-555},
and \reqnarray{proof of nonadjacent distance larger than one-(i)-(a)-case-2-bbb} that
\beqnarray{proof of nonadjacent distance larger than one-(i)-(a)-case-2-ccc}
m_{a'+j'}=n_{a'+j'}=p+1, \textrm{ for } j'=1,2,\ldots,j-1.
\eeqnarray
By combining \reqnarray{proof of nonadjacent distance larger than one-(i)-(a)-case-2-999},
\reqnarray{proof of nonadjacent distance larger than one-(i)-(a)-case-2-aaa},
and \reqnarray{proof of nonadjacent distance larger than one-(i)-(a)-case-2-ccc},
we obtain \reqnarray{proof of nonadjacent distance larger than one-(i)-(a)-case-2-555}
and \reqnarray{proof of nonadjacent distance larger than one-(i)-(a)-case-2-666}.

\emph{Case 3: $n_{a'-j'}=p+1$ for $j'=1,2,\ldots,\min\{a'-1,b'-a'-1\}$.}
We consider the two subcases $a'-1<b'-a'-1$ and $a'-1\geq b'-a'-1$ separately.

\emph{Subcase 3(a): $a'-1<b'-a'-1$.}
Let $\mbf_1^{r_{h-1}}$ be a sequence of positive integers as given in
\reqnarray{proof of nonadjacent distance larger than one-(i)-(a)-case-2-111}.
As in Case~2 above, we have $\mbf_1^{r_{h-1}}\in \Ncal_{M,k}(h)$.
As it is clear from $a'\geq 2$ and $b'\geq a'+2$
in \reqnarray{proof of nonadjacent distance larger than one-(i)-333}
that $\min\{a'-1,b'-a'-1\}\geq 1$,
we have $n_{a'-1}=p+1$ in this case and hence it is easy to see that
\reqnarray{proof of nonadjacent distance larger than one-(i)-(a)-case-2-222}
still holds in this subcase.

If $a'=2$, then we have $a'-1=1$ and
it follows from $r_{h-1}\geq 3$
in \reqnarray{proof of nonadjacent distance larger than one-111},
$\mbf_1^{r_{h-1}}\in \Ncal_{M,k}(h)$,
\reqnarray{proof of nonadjacent distance larger than one-(i)-(a)-case-2-111},
\reqnarray{proof of nonadjacent distance larger than one-(i)-(a)-case-2-222},
and \reqnarray{comparison rule A-1} in \rlemma{comparison rule A}(i)
that $\mbf_1^{r_{h-1}}\succ \nbf_1^{r_{h-1}}$,
i.e., \reqnarray{proof of nonadjacent distance larger than one-(i)-999} holds
with ${\nbf'}_1^{r_{h-1}}=\mbf_1^{r_{h-1}}$.

\bpdffigure{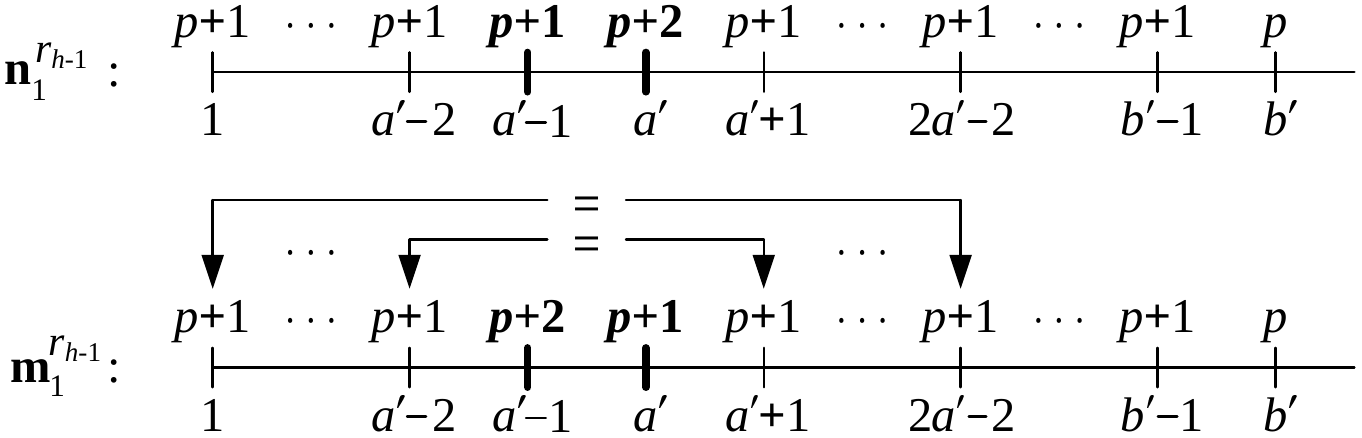}{4.5in}
\epdffigure{appendix-G-(i)-(a)-case-3-1}
{An illustration of \reqnarray{proof of nonadjacent distance larger than one-(i)-(a)-case-3-222}
(note that we have $\min\{(a'-1)-1, r_{h-1}-(a'-1)-1\}=a'-2$
in \reqnarray{proof of nonadjacent distance larger than one-(i)-(a)-case-3-555}).}

On the other hand, if $a'\geq 3$, then we show that
\beqnarray{}
\alignspace \hspace*{-0.2in}
2\leq a'-1\leq r_{h-1}-2,
\label{eqn:proof of nonadjacent distance larger than one-(i)-(a)-case-3-111}\\
\alignspace \hspace*{-0.2in}
m_{(a'-1)-j'}=m_{a'+j'}=p+1, \textrm{ for } j'=1,2,\ldots,\min\{(a'-1)-1, r_{h-1}-(a'-1)-1\}.
\label{eqn:proof of nonadjacent distance larger than one-(i)-(a)-case-3-222}
\eeqnarray
An illustration of \reqnarray{proof of nonadjacent distance larger than one-(i)-(a)-case-3-222}
is given in \rfigure{appendix-G-(i)-(a)-case-3-1}.
Therefore, it follows from
$\mbf_1^{r_{h-1}}\in \Ncal_{M,k}(h)$,
\reqnarray{proof of nonadjacent distance larger than one-(i)-(a)-case-2-111},
\reqnarray{proof of nonadjacent distance larger than one-(i)-(a)-case-2-222},
\reqnarray{proof of nonadjacent distance larger than one-(i)-(a)-case-3-111},
\reqnarray{proof of nonadjacent distance larger than one-(i)-(a)-case-3-222},
and \reqnarray{comparison rule A-4} in \rlemma{comparison rule A}(iii) that
$\mbf_1^{r_{h-1}}\succ \nbf_1^{r_{h-1}}$,
i.e., \reqnarray{proof of nonadjacent distance larger than one-(i)-999} holds
with ${\nbf'}_1^{r_{h-1}}=\mbf_1^{r_{h-1}}$.

From $a'\geq 3$ and \reqnarray{proof of nonadjacent distance larger than one-(i)-(a)-222},
we see that
\beqnarray{proof of nonadjacent distance larger than one-(i)-(a)-case-3-333}
2\leq a'-1<a'\leq r_{h-1}-2.
\eeqnarray
Thus, \reqnarray{proof of nonadjacent distance larger than one-(i)-(a)-case-3-111}
follows from \reqnarray{proof of nonadjacent distance larger than one-(i)-(a)-case-3-333}.

To prove \reqnarray{proof of nonadjacent distance larger than one-(i)-(a)-case-3-222},
note that from $a'-1<b'-a'-1$ in this subcase and $b'\leq r_{h-1}$, we have
\beqnarray{proof of nonadjacent distance larger than one-(i)-(a)-case-3-444}
(a'-1)-1<(b'-a'-1)-1\leq r_{h-1}-a'-2<r_{h-1}-(a'-1)-1.
\eeqnarray
It follows from \reqnarray{proof of nonadjacent distance larger than one-(i)-(a)-case-3-444} that
\beqnarray{proof of nonadjacent distance larger than one-(i)-(a)-case-3-555}
\min\{(a'-1)-1, r_{h-1}-(a'-1)-1\}=(a'-1)-1=a'-2.
\eeqnarray
As in this subcase we have
\beqnarray{proof of nonadjacent distance larger than one-(i)-(a)-case-3-666}
n_{a'-j'}=p+1, \textrm{ for } j'=1,2,\ldots,\min\{a'-1,b'-a'-1\}=a'-1,
\eeqnarray
it is clear from \reqnarray{proof of nonadjacent distance larger than one-(i)-(a)-case-2-111}
and \reqnarray{proof of nonadjacent distance larger than one-(i)-(a)-case-3-666} that
\beqnarray{proof of nonadjacent distance larger than one-(i)-(a)-case-3-777}
m_{(a'-1)-j'}=n_{(a'-1)-j'}=p+1, \textrm{ for } j'=1,2,\ldots,a'-2.
\eeqnarray
Furthermore, we have from $a'\geq 3$ and $a'-1<b'-a'-1$ that
\beqnarray{proof of nonadjacent distance larger than one-(i)-(a)-case-3-888}
a'<a'+(a'-2)<a'+(b'-a'-2)=b'-2<b'.
\eeqnarray
It then follows from \reqnarray{proof of nonadjacent distance larger than one-(i)-(a)-case-2-111},
\reqnarray{proof of nonadjacent distance larger than one-(i)-555},
and \reqnarray{proof of nonadjacent distance larger than one-(i)-(a)-case-3-888} that
\beqnarray{proof of nonadjacent distance larger than one-(i)-(a)-case-3-999}
m_{a'+j'}=n_{a'+j'}=p+1, \textrm{ for } j'=1,2,\ldots,a'-2.
\eeqnarray
By combining
\reqnarray{proof of nonadjacent distance larger than one-(i)-(a)-case-3-555},
\reqnarray{proof of nonadjacent distance larger than one-(i)-(a)-case-3-777},
and \reqnarray{proof of nonadjacent distance larger than one-(i)-(a)-case-3-999},
we obtain \reqnarray{proof of nonadjacent distance larger than one-(i)-(a)-case-3-222}.

\emph{Subcase 3(b): $a'-1\geq b'-a'-1$.}

\bpdffigure{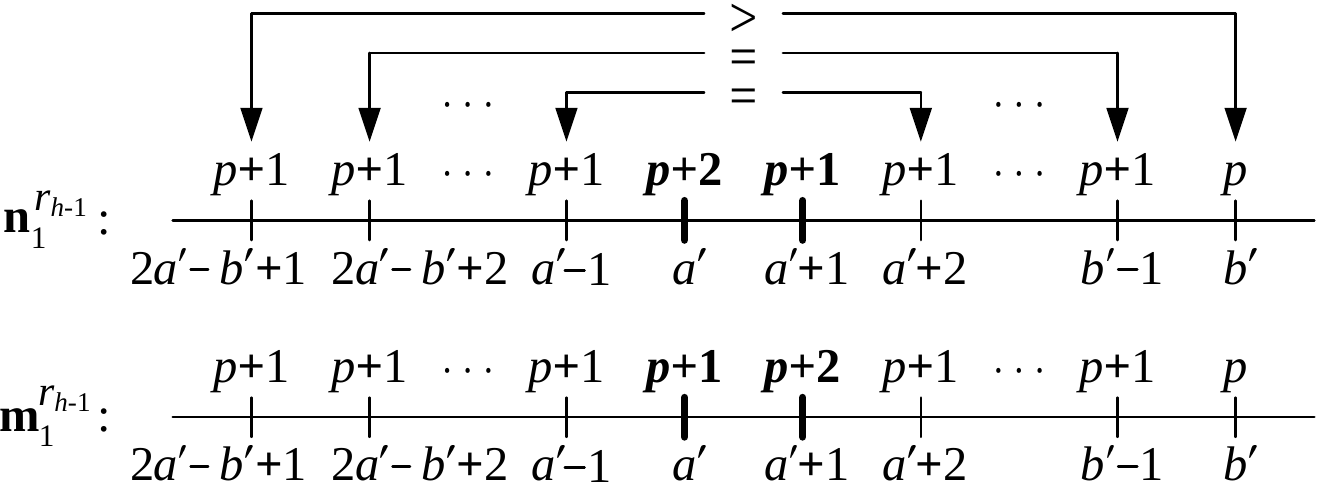}{4.5in}
\epdffigure{appendix-G-(i)-(a)-case-3-2}
{An illustration of \reqnarray{proof of nonadjacent distance larger than one-(i)-(a)-case-3-bbb}
and \reqnarray{proof of nonadjacent distance larger than one-(i)-(a)-case-3-ccc}.}

Let $\mbf_1^{r_{h-1}}$ be a sequence of positive integers as given in
\reqnarray{proof of nonadjacent distance larger than one-(i)-(a)-case-1-111}.
As in Case~1 above, we have $\mbf_1^{r_{h-1}}\in \Ncal_{M,k}(h)$.
In this subcase, we show that
\beqnarray{}
\alignspace 1\leq b'-a'-1\leq \min\{a'-1,r_{h-1}-a'-1\}
\label{eqn:proof of nonadjacent distance larger than one-(i)-(a)-case-3-aaa}\\
\alignspace n_{a'-j'}=n_{(a'+1)+j'}=p+1, \textrm{ for } j'=1,2,\ldots,b'-a'-2,
\label{eqn:proof of nonadjacent distance larger than one-(i)-(a)-case-3-bbb}\\
\alignspace n_{a'-(b'-a'-1)}=p+1>n_{(a'+1)+(b'-a'-1)}=p.
\label{eqn:proof of nonadjacent distance larger than one-(i)-(a)-case-3-ccc}
\eeqnarray
An illustration of
\reqnarray{proof of nonadjacent distance larger than one-(i)-(a)-case-3-bbb}
and \reqnarray{proof of nonadjacent distance larger than one-(i)-(a)-case-3-ccc}
is given in \rfigure{appendix-G-(i)-(a)-case-3-2}.
Therefore, it follows from
$\nbf_1^{r_{h-1}}\in \Ncal_{M,k}(h)$
in \reqnarray{proof of nonadjacent distance larger than one-111},
\reqnarray{proof of nonadjacent distance larger than one-(i)-(a)-222}--\reqnarray{proof of nonadjacent distance larger than one-(i)-(a)-case-1-111},
\reqnarray{proof of nonadjacent distance larger than one-(i)-(a)-case-3-aaa}--\reqnarray{proof of nonadjacent distance larger than one-(i)-(a)-case-3-ccc},
and \reqnarray{comparison rule A-3} in \rlemma{comparison rule A}(ii)
(with $j=b'-a'-1$) that
\beqnarray{proof of nonadjacent distance larger than one-(i)-(a)-case-3-ddd}
\nbf_1^{r_{h-1}}\preceq \mbf_1^{r_{h-1}}.
\eeqnarray

From $b'\geq a'+2$ in \reqnarray{proof of nonadjacent distance larger than one-(i)-333},
$a'-1\geq b'-a'-1$, and $b'\leq r_{h-1}$, we can see that
\beqnarray{proof of nonadjacent distance larger than one-(i)-(a)-case-3-eee}
1\leq b'-a'-1\leq \min\{a'-1,b'-a'-1\}\leq \min\{a'-1,r_{h-1}-a'-1\}.
\eeqnarray
Thus, \reqnarray{proof of nonadjacent distance larger than one-(i)-(a)-case-3-aaa}
follows from \reqnarray{proof of nonadjacent distance larger than one-(i)-(a)-case-3-eee}.

To prove \reqnarray{proof of nonadjacent distance larger than one-(i)-(a)-case-3-bbb}
and \reqnarray{proof of nonadjacent distance larger than one-(i)-(a)-case-3-ccc},
note that in this case we have
\beqnarray{proof of nonadjacent distance larger than one-(i)-(a)-case-3-fff}
n_{a'-j'}=p+1, \textrm{ for } j'=1,2,\ldots,\min\{a'-1,b'-a'-1\}=b'-a'-1.
\eeqnarray
Also, it is clear from
\reqnarray{proof of nonadjacent distance larger than one-(i)-555},
$(a'+1)+(b'-a'-1)=b'$,
and $n_{b'}=p$ in \reqnarray{proof of nonadjacent distance larger than one-(i)-444} that
\beqnarray{}
\alignspace n_{(a'+1)+j'}=p+1, \textrm{ for } j'=1,2,\ldots,b'-a'-2,
\label{eqn:proof of nonadjacent distance larger than one-(i)-(a)-case-3-ggg}\\
\alignspace n_{(a'+1)+(b'-a'-1)}=n_{b'}=p.
\label{eqn:proof of nonadjacent distance larger than one-(i)-(a)-case-3-hhh}
\eeqnarray
By combining
\reqnarray{proof of nonadjacent distance larger than one-(i)-(a)-case-3-fff},
\reqnarray{proof of nonadjacent distance larger than one-(i)-(a)-case-3-ggg},
and \reqnarray{proof of nonadjacent distance larger than one-(i)-(a)-case-3-hhh},
we obtain \reqnarray{proof of nonadjacent distance larger than one-(i)-(a)-case-3-bbb}
and \reqnarray{proof of nonadjacent distance larger than one-(i)-(a)-case-3-ccc}.

Now let ${\mbf'}_1^{r_{h-1}}$ be a sequence of positive integers such that
\beqnarray{proof of nonadjacent distance larger than one-(i)-(a)-case-3-iii}
m'_{a'+1}=m_{a'+1}-1,\ m'_{a'+2}=m_{a'+2}+1, \textrm{ and } m'_i=m_i \textrm{ for } i\neq a'+1, a'+2.
\eeqnarray
Again, it is easy to show that ${\mbf'}_1^{r_{h-1}}\in \Ncal_{M,k}(h)$.

If $b'=a'+2$, then we see from
\reqnarray{proof of nonadjacent distance larger than one-(i)-(a)-case-1-111},
$n_{a'+1}=p+1$ in \reqnarray{proof of nonadjacent distance larger than one-(i)-888},
and $n_{b'}=p$ in \reqnarray{proof of nonadjacent distance larger than one-(i)-444} that
\beqnarray{proof of nonadjacent distance larger than one-(i)-(a)-case-3-jjj}
\alignspace m_{a'+1}-m_{a'+2}=(n_{a'+1}+1)-n_{a'+2}=n_{a'+1}+1-n_{b'}=(p+1)+1-p=2.
\eeqnarray
Therefore, it follows from
$r_{h-1}\geq 3$ in \reqnarray{proof of nonadjacent distance larger than one-111},
$\mbf_1^{r_{h-1}}\in \Ncal_{M,k}(h)$,
\reqnarray{proof of nonadjacent distance larger than one-(i)-(a)-case-3-iii},
\reqnarray{proof of nonadjacent distance larger than one-(i)-(a)-case-3-jjj},
and \reqnarray{adjacent distance larger than one-2} in \rlemma{adjacent distance larger than one}(ii) that
\beqnarray{proof of nonadjacent distance larger than one-(i)-(a)-case-3-kkk}
\mbf_1^{r_{h-1}}\preceq {\mbf'}_1^{r_{h-1}},
\eeqnarray
where $\mbf_1^{r_{h-1}}\equiv {\mbf'}_1^{r_{h-1}}$
if and only if $r_{h-1}=2$ and $m_1=m_2+2$.
Since it is clear from $r_{h-1}\geq 3$ in
\reqnarray{proof of nonadjacent distance larger than one-111}
that $r_{h-1}\neq 2$, it cannot be the case that $\mbf_1^{r_{h-1}}\equiv {\mbf'}_1^{r_{h-1}}$.
As such, we see from \reqnarray{proof of nonadjacent distance larger than one-(i)-(a)-case-3-ddd}
and \reqnarray{proof of nonadjacent distance larger than one-(i)-(a)-case-3-kkk}
that $\nbf_1^{r_{h-1}}\preceq \mbf_1^{r_{h-1}}\prec {\mbf'}_1^{r_{h-1}}$,
i.e., \reqnarray{proof of nonadjacent distance larger than one-(i)-999} holds
with ${\nbf'}_1^{r_{h-1}}={\mbf'}_1^{r_{h-1}}$.

On the other hand, if $b'\geq a'+3$, then we have $a'<a'+2<b'$
and it follows from \reqnarray{proof of nonadjacent distance larger than one-(i)-555} that
\beqnarray{proof of nonadjacent distance larger than one-(i)-(a)-case-3-1111}
n_{a'+1}=n_{a'+2}=p+1.
\eeqnarray
From \reqnarray{proof of nonadjacent distance larger than one-(i)-(a)-case-1-111}
and \reqnarray{proof of nonadjacent distance larger than one-(i)-(a)-case-3-1111},
we see that
\beqnarray{proof of nonadjacent distance larger than one-(i)-(a)-case-3-2222}
\alignspace m_{a'+1}-m_{a'+2}=(n_{a'+1}+1)-n_{a'+2}=1.
\eeqnarray

\bpdffigure{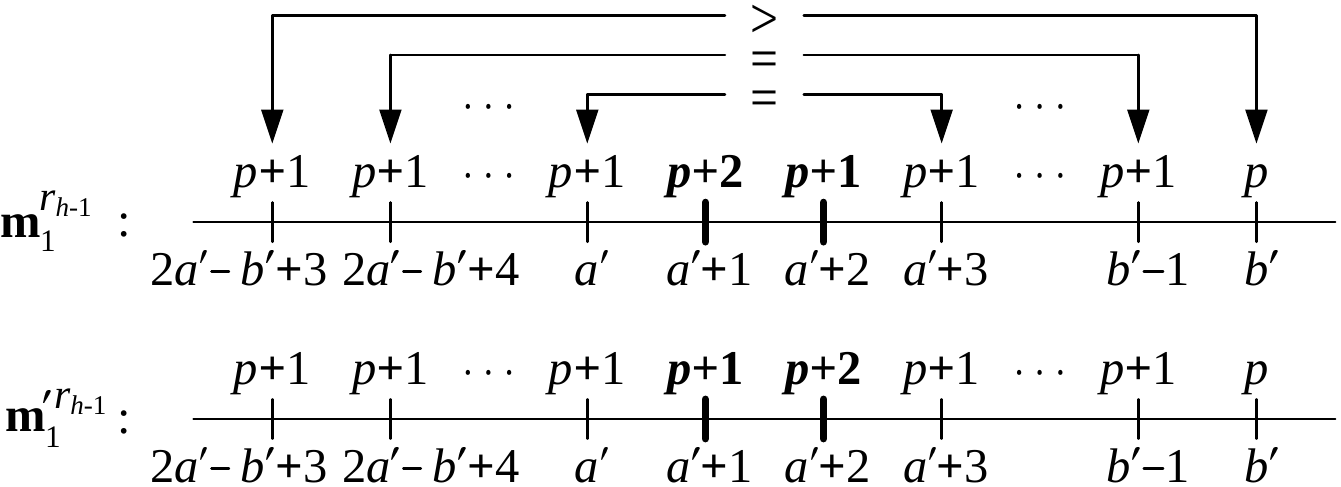}{4.5in}
\epdffigure{appendix-G-(i)-(a)-case-3-3}
{An illustration of \reqnarray{proof of nonadjacent distance larger than one-(i)-(a)-case-3-5555}
and \reqnarray{proof of nonadjacent distance larger than one-(i)-(a)-case-3-6666}.}

We will show that
\beqnarray{}
\alignspace
2\leq a'+1\leq r_{h-1}-2,
\label{eqn:proof of nonadjacent distance larger than one-(i)-(a)-case-3-3333}\\
\alignspace
1\leq b'-a'-2\leq \min\{(a'+1)-1,r_{h-1}-(a'+1)-1\},
\label{eqn:proof of nonadjacent distance larger than one-(i)-(a)-case-3-4444}\\
\alignspace
m_{(a'+1)-j'}=m_{(a'+2)+j'}=p+1, \textrm{ for } j'=1,2,\ldots,b'-a'-3,
\label{eqn:proof of nonadjacent distance larger than one-(i)-(a)-case-3-5555}\\
\alignspace
m_{(a'+1)-(b'-a'-2)}=p+1>m_{(a'+2)+(b'-a'-2)}=p.
\label{eqn:proof of nonadjacent distance larger than one-(i)-(a)-case-3-6666}
\eeqnarray
An illustration of
\reqnarray{proof of nonadjacent distance larger than one-(i)-(a)-case-3-5555}
and \reqnarray{proof of nonadjacent distance larger than one-(i)-(a)-case-3-6666}
is given in \rfigure{appendix-G-(i)-(a)-case-3-3}.
Therefore, it follows from
$\mbf_1^{r_{h-1}}\in \Ncal_{M,k}(h)$,
\reqnarray{proof of nonadjacent distance larger than one-(i)-(a)-case-3-iii},
\reqnarray{proof of nonadjacent distance larger than one-(i)-(a)-case-3-2222}--\reqnarray{proof of nonadjacent distance larger than one-(i)-(a)-case-3-6666},
and \reqnarray{comparison rule A-3} in \rlemma{comparison rule A}(ii)
(with $j=b'-a'-2$) that
\beqnarray{proof of nonadjacent distance larger than one-(i)-(a)-case-3-7777}
\mbf_1^{r_{h-1}}\preceq {\mbf'}_1^{r_{h-1}},
\eeqnarray
where $\mbf_1^{r_{h-1}}\equiv {\mbf'}_1^{r_{h-1}}$
if and only if $(a'+1)-(b'-a'-2)=1$, $(a'+1)+1+(b'-a'-2)=r_{h-1}$, and $m_1=m_{r_{h-1}}+1$.
Since in this subcase we have $a'-1\geq b'-a'-1$,
it is clear that $(a'+1)-(b'-a'-2)\geq (b'-a'+1)-(b'-a'-2)=3$.
This implies that $(a'+1)-(b'-a'-2)\neq 1$
and hence it cannot be the case that $\mbf_1^{r_{h-1}}\equiv {\mbf'}_1^{r_{h-1}}$.
As such, we see from \reqnarray{proof of nonadjacent distance larger than one-(i)-(a)-case-3-ddd}
and \reqnarray{proof of nonadjacent distance larger than one-(i)-(a)-case-3-7777}
that $\nbf_1^{r_{h-1}}\preceq \mbf_1^{r_{h-1}}\prec {\mbf'}_1^{r_{h-1}}$,
i.e., \reqnarray{proof of nonadjacent distance larger than one-(i)-999} holds
with ${\nbf'}_1^{r_{h-1}}={\mbf'}_1^{r_{h-1}}$.

To prove \reqnarray{proof of nonadjacent distance larger than one-(i)-(a)-case-3-3333}
and \reqnarray{proof of nonadjacent distance larger than one-(i)-(a)-case-3-4444},
note from $a'\geq 2$, $b'\geq a'+3$, $b'\leq r_{h-1}$,
and \reqnarray{proof of nonadjacent distance larger than one-(i)-(a)-case-3-aaa} that
\beqnarray{}
\alignspace \hspace*{-0.3in}
2\leq a'<a'+1\leq b'-2\leq r_{h-1}-2,
\label{eqn:proof of nonadjacent distance larger than one-(i)-(a)-case-3-8888}\\
\alignspace \hspace*{-0.3in}
1\leq b'-a'-2\leq \min\{a'-2,r_{h-1}-a'-2\}\leq \min\{(a'+1)-1,r_{h-1}-(a'+1)-1\}.
\label{eqn:proof of nonadjacent distance larger than one-(i)-(a)-case-3-9999}
\eeqnarray
Thus, \reqnarray{proof of nonadjacent distance larger than one-(i)-(a)-case-3-3333}
follows from \reqnarray{proof of nonadjacent distance larger than one-(i)-(a)-case-3-8888},
and \reqnarray{proof of nonadjacent distance larger than one-(i)-(a)-case-3-4444}
follows from \reqnarray{proof of nonadjacent distance larger than one-(i)-(a)-case-3-9999}.

To prove \reqnarray{proof of nonadjacent distance larger than one-(i)-(a)-case-3-5555}
and \reqnarray{proof of nonadjacent distance larger than one-(i)-(a)-case-3-6666},
note that from \reqnarray{proof of nonadjacent distance larger than one-(i)-(a)-case-1-111},
$n_{a'}=p+2$ and $n_{b'}=p$ in \reqnarray{proof of nonadjacent distance larger than one-(i)-444},
and \reqnarray{proof of nonadjacent distance larger than one-(i)-(a)-case-3-fff}--\reqnarray{proof of nonadjacent distance larger than one-(i)-(a)-case-3-hhh},
we have
\beqnarray{}
\alignspace m_{(a'+1)-1}=m_{a'}=n_{a'}-1=(p+2)-1=p+1,
\label{eqn:proof of nonadjacent distance larger than one-(i)-(a)-case-3-aaaa}\\
\alignspace m_{(a'+1)-j'}=n_{(a'+1)-j'}=p+1, \textrm{ for } j'=2,3,\ldots,b'-a',
\label{eqn:proof of nonadjacent distance larger than one-(i)-(a)-case-3-bbbb}\\
\alignspace m_{(a'+2)+j'}=n_{(a'+2)+j'}=p+1, \textrm{ for } j'=1,2,\ldots,b'-a'-3,
\label{eqn:proof of nonadjacent distance larger than one-(i)-(a)-case-3-cccc}\\
\alignspace m_{(a'+2)+(b'-a'-2)}=n_{(a'+2)+(b'-a'-2)}=n_{b'}=p.
\label{eqn:proof of nonadjacent distance larger than one-(i)-(a)-case-3-dddd}
\eeqnarray
Thus, \reqnarray{proof of nonadjacent distance larger than one-(i)-(a)-case-3-5555}
and \reqnarray{proof of nonadjacent distance larger than one-(i)-(a)-case-3-6666}
follow from
\reqnarray{proof of nonadjacent distance larger than one-(i)-(a)-case-3-aaaa}--\reqnarray{proof of nonadjacent distance larger than one-(i)-(a)-case-3-dddd}.

(b) Now we assume that $b'\leq r_{h-1}-1$ and show that there exists a sequence of positive integers
${\nbf'}_1^{r_{h-1}}\in \Ncal_{M,k}(h)$ such that
\reqnarray{proof of nonadjacent distance larger than one-(i)-999} holds.
Note that from $b'\geq a'+2$
in \reqnarray{proof of nonadjacent distance larger than one-(i)-333}
and $a'\geq 1$, we have
\beqnarray{proof of nonadjacent distance larger than one-(i)-(b)-111}
b'-1\geq a'+1\geq 2.
\eeqnarray
As we assume that $b'\leq r_{h-1}-1$,
it follows from \reqnarray{proof of nonadjacent distance larger than one-(i)-(b)-111} that
\beqnarray{proof of nonadjacent distance larger than one-(i)-(b)-222}
2\leq b'-1\leq r_{h-1}-2.
\eeqnarray
Furthermore, we have from $n_{b'-1}=p+1$
in \reqnarray{proof of nonadjacent distance larger than one-(i)-888}
and $n_{b'}=p$
in \reqnarray{proof of nonadjacent distance larger than one-(i)-444} that
\beqnarray{proof of nonadjacent distance larger than one-(i)-(b)-333}
n_{b'-1}-n_{b'}=(p+1)-p=1.
\eeqnarray

We need to consider the following three possible cases.

\emph{Case 1: There exists a positive integer $j$ such that
$1\leq j\leq \min\{b'-a'-1,r_{h-1}-b'\}$,
$n_{b'+j'}=p+1$ for $j'=1,2,\ldots,j-1$, and $n_{b'+j}<p+1$.}
Let $\mbf_1^{r_{h-1}}$ be a sequence of positive integers such that
\beqnarray{proof of nonadjacent distance larger than one-(i)-(b)-case-1-111}
m_{b'-1}=n_{b'-1}-1,\ m_{b'}=n_{b'}+1, \textrm{ and } m_i=n_i \textrm{ for } i\neq b'-1, b'.
\eeqnarray
As before, it is easy to show that $\mbf_1^{r_{h-1}}\in \Ncal_{M,k}(h)$.
As we have $j\leq \min\{b'-a'-1,r_{h-1}-b'\}\leq b'-a'-1$,
we consider the two subcases $j<b'-a'-1$ and $j=b'-a'-1$ separately.

\emph{Subcase 1(a): $j<b'-a'-1$.}

\bpdffigure{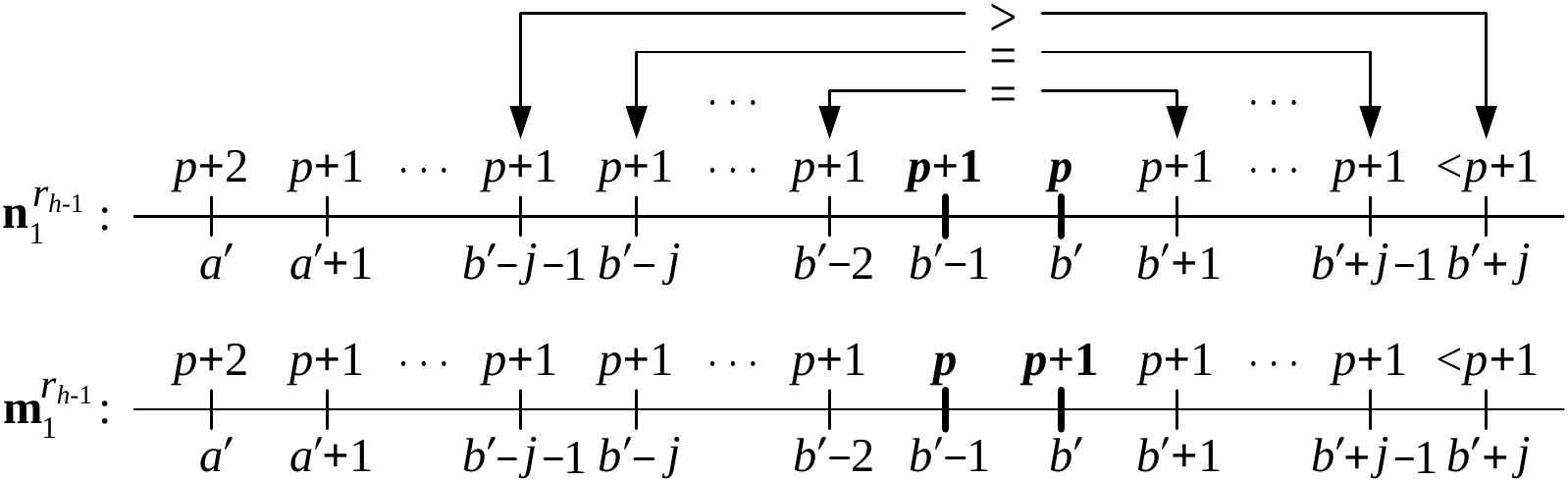}{5.5in}
\epdffigure{appendix-G-(i)-(b)-case-1}
{An illustration of \reqnarray{proof of nonadjacent distance larger than one-(i)-(b)-case-1-333}
and \reqnarray{proof of nonadjacent distance larger than one-(i)-(b)-case-1-444}.}

In this subcase, we show that
\beqnarray{}
\alignspace 1\leq j\leq \min\{(b'-1)-1,r_{h-1}-(b'-1)-1\}
\label{eqn:proof of nonadjacent distance larger than one-(i)-(b)-case-1-222}\\
\alignspace n_{(b'-1)-j'}=n_{b'+j'}, \textrm{ for } j'=1,2,\ldots,j-1,
\label{eqn:proof of nonadjacent distance larger than one-(i)-(b)-case-1-333}\\
\alignspace n_{(b'-1)-j}>n_{b'+j}.
\label{eqn:proof of nonadjacent distance larger than one-(i)-(b)-case-1-444}
\eeqnarray
An illustration of
\reqnarray{proof of nonadjacent distance larger than one-(i)-(b)-case-1-333}
and \reqnarray{proof of nonadjacent distance larger than one-(i)-(b)-case-1-444}
is given in \rfigure{appendix-G-(i)-(b)-case-1}.
Therefore, it follows from
$\nbf_1^{r_{h-1}}\in \Ncal_{M,k}(h)$
in \reqnarray{proof of nonadjacent distance larger than one-111},
\reqnarray{proof of nonadjacent distance larger than one-(i)-(b)-222}--\reqnarray{proof of nonadjacent distance larger than one-(i)-(b)-case-1-444},
and \reqnarray{comparison rule A-3} in \rlemma{comparison rule A}(ii) that
\beqnarray{proof of nonadjacent distance larger than one-(i)-(b)-case-1-555}
\nbf_1^{r_{h-1}}\preceq \mbf_1^{r_{h-1}},
\eeqnarray
where $\nbf_1^{r_{h-1}}\equiv \mbf_1^{r_{h-1}}$
if and only if $(b'-1)-j=1$, $b'+j=r_{h-1}$,
and $n_1=n_{r_{h-1}}+1$ (i.e., $n_{(b'-1)-j}=n_{b'+j}+1$).
As we have $j<b'-a'-1$ in this subcase and $a'\geq 1$,
we immediately see that $(b'-1)-j>a'\geq 1$.
This implies that $(b'-1)-j\neq 1$ and hence it cannot be the case that
$\nbf_1^{r_{h-1}}\equiv \mbf_1^{r_{h-1}}$.
As such, we see from \reqnarray{proof of nonadjacent distance larger than one-(i)-(b)-case-1-555}
that $\nbf_1^{r_{h-1}}\prec \mbf_1^{r_{h-1}}$,
i.e., \reqnarray{proof of nonadjacent distance larger than one-(i)-999} holds
with ${\nbf'}_1^{r_{h-1}}=\mbf_1^{r_{h-1}}$.

From $1\leq j\leq \min\{b'-a'-1,r_{h-1}-b'\}$ and $a'\geq 1$, we see that
\beqnarray{proof of nonadjacent distance larger than one-(i)-(b)-case-1-666}
1\leq j\leq \min\{b'-a'-1,r_{h-1}-b'\}\leq \min\{(b'-1)-1,r_{h-1}-(b'-1)-1\}.
\eeqnarray
Thus, \reqnarray{proof of nonadjacent distance larger than one-(i)-(b)-case-1-222}
follows from \reqnarray{proof of nonadjacent distance larger than one-(i)-(b)-case-1-666}.

To prove \reqnarray{proof of nonadjacent distance larger than one-(i)-(b)-case-1-333}
and \reqnarray{proof of nonadjacent distance larger than one-(i)-(b)-case-1-444},
note that in this subcase we have $a'<(b'-1)-j<b'$,
and hence it follows from
\reqnarray{proof of nonadjacent distance larger than one-(i)-555} that
\beqnarray{proof of nonadjacent distance larger than one-(i)-(b)-case-1-777}
n_{(b'-1)-j'}=p+1, \textrm{ for } j'=1,2,\ldots,j.
\eeqnarray
By combining \reqnarray{proof of nonadjacent distance larger than one-(i)-(b)-case-1-777},
$n_{b'+j'}=p+1$ for $j'=1,2,\ldots,j-1$, and $n_{b'+j}<p+1$,
we obtain \reqnarray{proof of nonadjacent distance larger than one-(i)-(b)-case-1-333}
and \reqnarray{proof of nonadjacent distance larger than one-(i)-(b)-case-1-444}.

\emph{Subcase 1(b): $j=b'-a'-1$.}
In this subcase, we have $(b'-1)-j=a'$
and it follows from \reqnarray{proof of nonadjacent distance larger than one-(i)-555}
and $n_{a'}=p+2$ in \reqnarray{proof of nonadjacent distance larger than one-(i)-444} that
\beqnarray{}
\alignspace n_{(b'-1)-j'}=p+1, \textrm{ for } j'=1,2,\ldots,j-1,
\label{eqn:proof of nonadjacent distance larger than one-(i)-(b)-case-1-888}\\
\alignspace n_{(b'-1)-j}=n_{a'}=p+2.
\label{eqn:proof of nonadjacent distance larger than one-(i)-(b)-case-1-999}
\eeqnarray
By using \reqnarray{proof of nonadjacent distance larger than one-(i)-(b)-case-1-888}
and \reqnarray{proof of nonadjacent distance larger than one-(i)-(b)-case-1-999},
we can argue as in Subcase~1(a) above that
\reqnarray{proof of nonadjacent distance larger than one-(i)-(b)-case-1-222}--\reqnarray{proof of nonadjacent distance larger than one-(i)-(b)-case-1-555}
still hold.
Since it is clear from $n_{(b'-1)-j}=p+2$
in \reqnarray{proof of nonadjacent distance larger than one-(i)-(b)-case-1-999}
and $n_{b'+j}<p+1$ that $n_{(b'-1)-j}\neq n_{b'+j}+1$,
it cannot be the case that $\nbf_1^{r_{h-1}}\equiv \mbf_1^{r_{h-1}}$.
As such, we see from \reqnarray{proof of nonadjacent distance larger than one-(i)-(b)-case-1-555}
that $\nbf_1^{r_{h-1}}\prec \mbf_1^{r_{h-1}}$,
i.e., \reqnarray{proof of nonadjacent distance larger than one-(i)-999} holds
with ${\nbf'}_1^{r_{h-1}}=\mbf_1^{r_{h-1}}$.

\emph{Case 2: There exists a positive integer $j$ such that
$1\leq j\leq \min\{b'-a'-1,r_{h-1}-b'\}$,
$n_{b'+j'}=p+1$ for $j'=1,2,\ldots,j-1$, and $n_{b'+j}>p+1$.}
In this case, we can show that $j\geq 2$.
To see this, suppose on the contrary that $j=1$,
then we have $n_{b'+1}>p+1$ in this case.
As it follows from $n_{b'}=p$ in \reqnarray{proof of nonadjacent distance larger than one-(i)-444}
and the condition $|n_{i+1}-n_i|\leq 1$ for $i=1,2,\ldots,r_{h-1}-1$
in \reqnarray{proof of nonadjacent distance larger than one-111}
that $n_{b'+1}$ must be equal to $p-1$ (provided that $p\geq 2$), $p$, or $p+1$,
we have reached a contradiction.

Let $\mbf_1^{r_{h-1}}$ be a sequence of positive integers such that
\beqnarray{proof of nonadjacent distance larger than one-(i)-(b)-case-2-111}
m_{b'}=n_{b'}+1,\ m_{b'+1}=n_{b'+1}-1,
\textrm{ and } m_i=n_i \textrm{ for } i\neq b', b'+1.
\eeqnarray
As before, it is easy to show that $\mbf_1^{r_{h-1}}\in \Ncal_{M,k}(h)$.
As $j\geq 2$, we have $n_{b'+1}=p+1$ in this case.
It then follows from \reqnarray{proof of nonadjacent distance larger than one-(i)-(b)-case-2-111},
$n_{b'+1}=p+1$, and $n_{b'}=p$
in \reqnarray{proof of nonadjacent distance larger than one-(i)-444} that
\beqnarray{proof of nonadjacent distance larger than one-(i)-(b)-case-2-222}
m_{b'}-m_{b'+1}=(n_{b'}+1)-(n_{b'+1}-1)=(p+1)-(p+1-1)=1.
\eeqnarray

\bpdffigure{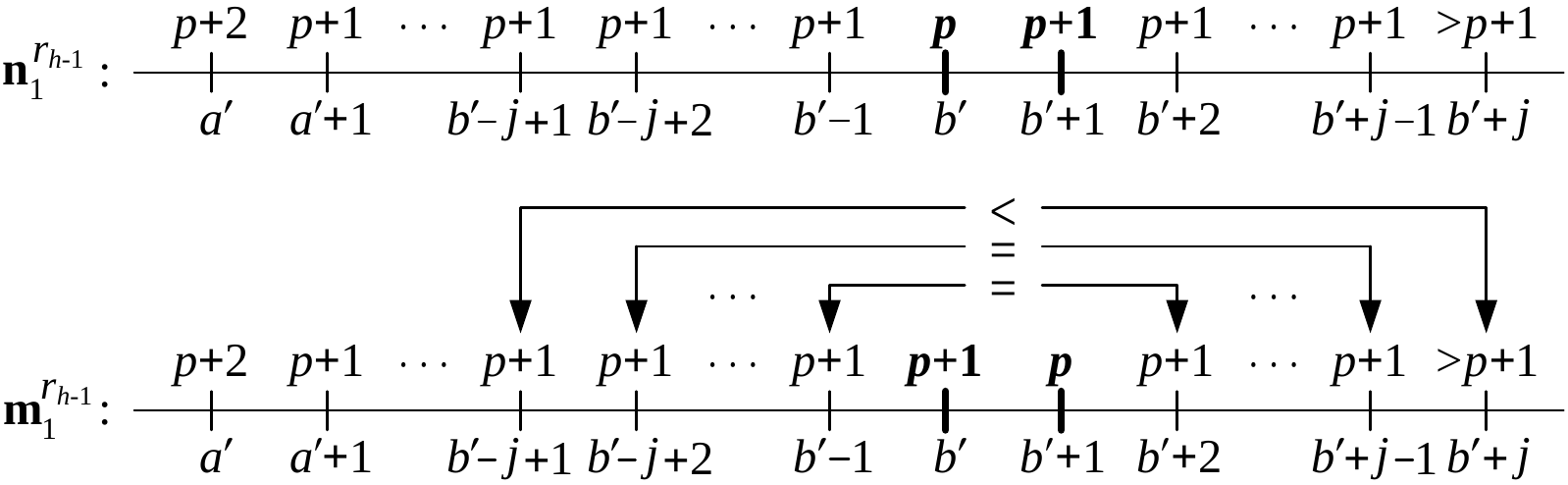}{5.5in}
\epdffigure{appendix-G-(i)-(b)-case-2}
{An illustration of \reqnarray{proof of nonadjacent distance larger than one-(i)-(b)-case-2-555}
and \reqnarray{proof of nonadjacent distance larger than one-(i)-(b)-case-2-666}.}

In the following, we show that
\beqnarray{}
\alignspace
2\leq b'\leq r_{h-1}-2,
\label{eqn:proof of nonadjacent distance larger than one-(i)-(b)-case-2-333}\\
\alignspace
1\leq j-1\leq \min\{b'-1,r_{h-1}-b'-1\},
\label{eqn:proof of nonadjacent distance larger than one-(i)-(b)-case-2-444}\\
\alignspace m_{b'-j'}=m_{(b'+1)+j'}, \textrm{ for } j'=1,2,\ldots,j-2,
\label{eqn:proof of nonadjacent distance larger than one-(i)-(b)-case-2-555}\\
\alignspace m_{b'-(j-1)}<m_{(b'+1)+(j-1)}.
\label{eqn:proof of nonadjacent distance larger than one-(i)-(b)-case-2-666}
\eeqnarray
An illustration of
\reqnarray{proof of nonadjacent distance larger than one-(i)-(b)-case-2-555}
and \reqnarray{proof of nonadjacent distance larger than one-(i)-(b)-case-2-666}
is given in \rfigure{appendix-G-(i)-(b)-case-2}.
Therefore, it follows from
$\mbf_1^{r_{h-1}}\in \Ncal_{M,k}(h)$,
\reqnarray{proof of nonadjacent distance larger than one-(i)-(b)-case-2-111}--\reqnarray{proof of nonadjacent distance larger than one-(i)-(b)-case-2-666},
and \reqnarray{comparison rule A-2} in \rlemma{comparison rule A}(ii) that
$\mbf_1^{r_{h-1}}\succ \nbf_1^{r_{h-1}}$,
i.e., \reqnarray{proof of nonadjacent distance larger than one-(i)-999} holds
with ${\nbf'}_1^{r_{h-1}}=\mbf_1^{r_{h-1}}$.

From $2\leq j\leq \min\{b'-a'-1,r_{h-1}-b'\}$ and $a'\geq 1$,
we can see that
\beqnarray{}
\alignspace
2\leq \min\{b'-a'-1,r_{h-1}-b'\}\leq b'-a'-1\leq b',
\label{eqn:proof of nonadjacent distance larger than one-(i)-(b)-case-2-777}\\
\alignspace
2\leq \min\{b'-a'-1,r_{h-1}-b'\}\leq r_{h-1}-b',
\label{eqn:proof of nonadjacent distance larger than one-(i)-(b)-case-2-888}\\
\alignspace
1\leq j-1\leq \min\{b'-a'-2,r_{h-1}-b'-1\}\leq \min\{b'-1,r_{h-1}-b'-1\}.
\label{eqn:proof of nonadjacent distance larger than one-(i)-(b)-case-2-999}
\eeqnarray
Thus, \reqnarray{proof of nonadjacent distance larger than one-(i)-(b)-case-2-333}
follows from \reqnarray{proof of nonadjacent distance larger than one-(i)-(b)-case-2-777}
and \reqnarray{proof of nonadjacent distance larger than one-(i)-(b)-case-2-888},
and \reqnarray{proof of nonadjacent distance larger than one-(i)-(b)-case-2-444}
follows from \reqnarray{proof of nonadjacent distance larger than one-(i)-(b)-case-2-999}.

To prove \reqnarray{proof of nonadjacent distance larger than one-(i)-(b)-case-2-555}
and \reqnarray{proof of nonadjacent distance larger than one-(i)-(b)-case-2-666},
note that we have from $2\leq j\leq \min\{b'-a'-1,r_{h-1}-b'\}\leq b'-a'-1$ that
\beqnarray{proof of nonadjacent distance larger than one-(i)-(b)-case-2-aaa}
a'< a'+2\leq b'-(j-1)\leq b'-1<b'.
\eeqnarray
It then follows from \reqnarray{proof of nonadjacent distance larger than one-(i)-(b)-case-2-111},
\reqnarray{proof of nonadjacent distance larger than one-(i)-555},
and \reqnarray{proof of nonadjacent distance larger than one-(i)-(b)-case-2-aaa} that
\beqnarray{proof of nonadjacent distance larger than one-(i)-(b)-case-2-bbb}
m_{b'-j'}=n_{b'-j'}=p+1, \textrm{ for } j'=1,2,\ldots,j-1.
\eeqnarray
Furthermore, as we have $n_{b'+j'}=p+1$ for $j'=1,2,\ldots,j-1$ and $n_{b'+j}>p+1$ in this case,
it is clear from \reqnarray{proof of nonadjacent distance larger than one-(i)-(b)-case-2-111} that
\beqnarray{}
\alignspace m_{(b'+1)+j'}=n_{(b'+1)+j'}=p+1, \textrm{ for } j'=1,2,\ldots,j-2,
\label{eqn:proof of nonadjacent distance larger than one-(i)-(b)-case-2-ccc}\\
\alignspace m_{(b'+1)+(j-1)}=n_{(b'+1)+(j-1)}=n_{b'+j}>p+1.
\label{eqn:proof of nonadjacent distance larger than one-(i)-(b)-case-2-ddd}
\eeqnarray
By combining \reqnarray{proof of nonadjacent distance larger than one-(i)-(b)-case-2-bbb},
\reqnarray{proof of nonadjacent distance larger than one-(i)-(b)-case-2-ccc},
and \reqnarray{proof of nonadjacent distance larger than one-(i)-(b)-case-2-ddd},
we obtain \reqnarray{proof of nonadjacent distance larger than one-(i)-(b)-case-2-555}
and \reqnarray{proof of nonadjacent distance larger than one-(i)-(b)-case-2-666}.

\emph{Case 3: $n_{b'+j'}=p+1$  for $j'=1,2,\ldots,\min\{b'-a'-1,r_{h-1}-b'\}$.}
We consider the two subcases $b'-a'-1>r_{h-1}-b'$ and $b'-a'-1\leq r_{h-1}-b'$ separately.

\emph{Subcase 3(a): $b'-a'-1>r_{h-1}-b'$.}
Let $\mbf_1^{r_{h-1}}$ be a sequence of positive integers as given in
\reqnarray{proof of nonadjacent distance larger than one-(i)-(b)-case-2-111}.
As in Case~2 above, we have $\mbf_1^{r_{h-1}}\in \Ncal_{M,k}(h)$.
As it is clear from $b'\geq a'+2$
in \reqnarray{proof of nonadjacent distance larger than one-(i)-333}
and $b'\leq r_{h-1}-1$ that $\min\{b'-a'-1,r_{h-1}-b'\}\geq 1$,
we have $n_{b'+1}=p+1$ in this case and hence it is easy to see that
\reqnarray{proof of nonadjacent distance larger than one-(i)-(b)-case-2-222}
still holds in this subcase.

If $b'=r_{h-1}-1$,
then it follows from $r_{h-1}\geq 3$
in \reqnarray{proof of nonadjacent distance larger than one-111},
$\mbf_1^{r_{h-1}}\in \Ncal_{M,k}(h)$,
\reqnarray{proof of nonadjacent distance larger than one-(i)-(b)-case-2-111},
\reqnarray{proof of nonadjacent distance larger than one-(i)-(b)-case-2-222},
and \reqnarray{comparison rule A-1} in \rlemma{comparison rule A}(i)
that $\mbf_1^{r_{h-1}}\succ \nbf_1^{r_{h-1}}$,
i.e., \reqnarray{proof of nonadjacent distance larger than one-(i)-999} holds
with ${\nbf'}_1^{r_{h-1}}=\mbf_1^{r_{h-1}}$.

\bpdffigure{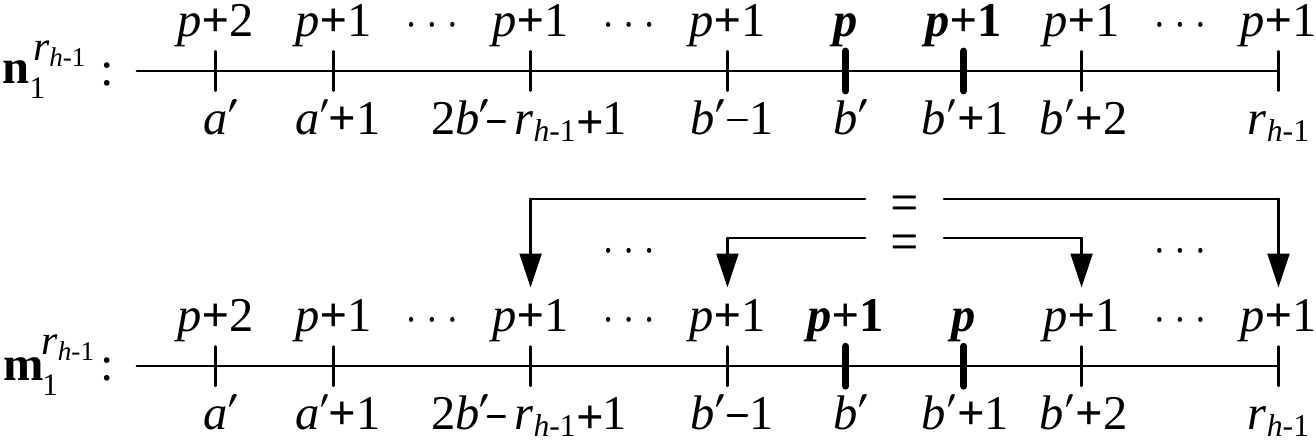}{4.5in}
\epdffigure{appendix-G-(i)-(b)-case-3-1}
{An illustration of \reqnarray{proof of nonadjacent distance larger than one-(i)-(b)-case-3-222}
(note that we have $\min\{b'-1, r_{h-1}-b'-1\}=r_{h-1}-b'-1$
in \reqnarray{proof of nonadjacent distance larger than one-(i)-(b)-case-3-555}).}

On the other hand, if $b'\leq r_{h-1}-2$, then we show that
\beqnarray{}
\alignspace \hspace*{-0.2in}
2\leq b'\leq r_{h-1}-2,
\label{eqn:proof of nonadjacent distance larger than one-(i)-(b)-case-3-111}\\
\alignspace \hspace*{-0.2in}
m_{b'-j'}=m_{(b'+1)+j'}=p+1, \textrm{ for } j'=1,2,\ldots,\min\{b'-1, r_{h-1}-b'-1\}.
\label{eqn:proof of nonadjacent distance larger than one-(i)-(b)-case-3-222}
\eeqnarray
An illustration of \reqnarray{proof of nonadjacent distance larger than one-(i)-(b)-case-3-222}
is given in \rfigure{appendix-G-(i)-(b)-case-3-1}.
Therefore, it follows from
$\mbf_1^{r_{h-1}}\in \Ncal_{M,k}(h)$,
\reqnarray{proof of nonadjacent distance larger than one-(i)-(b)-case-2-111},
\reqnarray{proof of nonadjacent distance larger than one-(i)-(b)-case-2-222},
\reqnarray{proof of nonadjacent distance larger than one-(i)-(b)-case-3-111},
\reqnarray{proof of nonadjacent distance larger than one-(i)-(b)-case-3-222},
and \reqnarray{comparison rule A-4} in \rlemma{comparison rule A}(iii) that
$\mbf_1^{r_{h-1}}\succ \nbf_1^{r_{h-1}}$,
i.e., \reqnarray{proof of nonadjacent distance larger than one-(i)-999} holds
with ${\nbf'}_1^{r_{h-1}}=\mbf_1^{r_{h-1}}$.

From $b'\geq a'+2>2$ and $b'\leq r_{h-1}-2$, we see that
\beqnarray{proof of nonadjacent distance larger than one-(i)-(b)-case-3-333}
2\leq b'\leq r_{h-1}-2.
\eeqnarray
Thus, \reqnarray{proof of nonadjacent distance larger than one-(i)-(b)-case-3-111}
follows from \reqnarray{proof of nonadjacent distance larger than one-(i)-(b)-case-3-333}.

To prove \reqnarray{proof of nonadjacent distance larger than one-(i)-(b)-case-3-222},
note that from $b'-a'-1>r_{h-1}-b'$ in this subcase and $a'\geq 1$, we have
\beqnarray{proof of nonadjacent distance larger than one-(i)-(b)-case-3-444}
r_{h-1}-b'-1<(b'-a'-1)-1<b'-1.
\eeqnarray
It follows from \reqnarray{proof of nonadjacent distance larger than one-(i)-(b)-case-3-444} that
\beqnarray{proof of nonadjacent distance larger than one-(i)-(b)-case-3-555}
\min\{b'-1, r_{h-1}-b'-1\}=r_{h-1}-b'-1.
\eeqnarray
From $b'-a'-1>r_{h-1}-b'$ and $b'\leq r_{h-1}-2$,
we can see that
\beqnarray{}
\alignspace
b'-(r_{h-1}-b'-1)>b'-(b'-a'-2)=a'+2>a',
\label{eqn:proof of nonadjacent distance larger than one-(i)-(b)-case-3-666}\\
\alignspace
b'-(r_{h-1}-b'-1)\leq b'-1<b'.
\label{eqn:proof of nonadjacent distance larger than one-(i)-(b)-case-3-777}
\eeqnarray
It follows from \reqnarray{proof of nonadjacent distance larger than one-(i)-(b)-case-2-111},
\reqnarray{proof of nonadjacent distance larger than one-(i)-555},
\reqnarray{proof of nonadjacent distance larger than one-(i)-(b)-case-3-666},
and \reqnarray{proof of nonadjacent distance larger than one-(i)-(b)-case-3-777} that
\beqnarray{proof of nonadjacent distance larger than one-(i)-(b)-case-3-888}
m_{b'-j'}=n_{b'-j'}=p+1, \textrm{ for } j'=1,2,\ldots,r_{h-1}-b'-1.
\eeqnarray
Furthermore, in this subcase we have
\beqnarray{proof of nonadjacent distance larger than one-(i)-(b)-case-3-999}
n_{b'+j'}=p+1, \textrm{ for } j'=1,2,\ldots,\min\{b'-a'-1,r_{h-1}-b'\}=r_{h-1}-b'.
\eeqnarray
It then follows from \reqnarray{proof of nonadjacent distance larger than one-(i)-(b)-case-2-111}
and \reqnarray{proof of nonadjacent distance larger than one-(i)-(b)-case-3-999} that
\beqnarray{proof of nonadjacent distance larger than one-(i)-(b)-case-3-aaa}
m_{(b'+1)+j'}=n_{(b'+1)+j'}=p+1, \textrm{ for } j'=1,2,\ldots,r_{h-1}-b'-1.
\eeqnarray
By combining
\reqnarray{proof of nonadjacent distance larger than one-(i)-(b)-case-3-555},
\reqnarray{proof of nonadjacent distance larger than one-(i)-(b)-case-3-888},
and \reqnarray{proof of nonadjacent distance larger than one-(i)-(b)-case-3-aaa},
we obtain \reqnarray{proof of nonadjacent distance larger than one-(i)-(b)-case-3-222}.

\emph{Subcase 3(b): $b'-a'-1\leq r_{h-1}-b'$.}

\bpdffigure{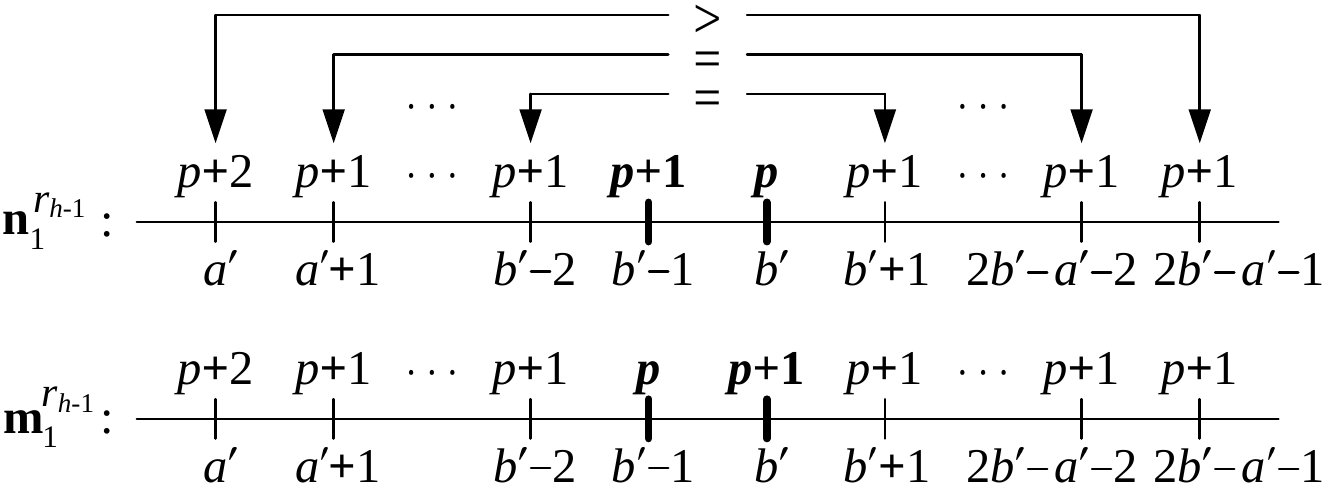}{4.5in}
\epdffigure{appendix-G-(i)-(b)-case-3-2}
{An illustration of \reqnarray{proof of nonadjacent distance larger than one-(i)-(b)-case-3-ccc}
and \reqnarray{proof of nonadjacent distance larger than one-(i)-(b)-case-3-ddd}.}

Let $\mbf_1^{r_{h-1}}$ be a sequence of positive integers as given in
\reqnarray{proof of nonadjacent distance larger than one-(i)-(b)-case-1-111}.
As in Case~1 above, we have $\mbf_1^{r_{h-1}}\in \Ncal_{M,k}(h)$.
In this subcase, we show that
\beqnarray{}
\alignspace 1\leq b'-a'-1\leq \min\{(b'-1)-1,r_{h-1}-(b'-1)-1\}
\label{eqn:proof of nonadjacent distance larger than one-(i)-(b)-case-3-bbb}\\
\alignspace n_{(b'-1)-j'}=n_{b'+j'}=p+1, \textrm{ for } j'=1,2,\ldots,b'-a'-2,
\label{eqn:proof of nonadjacent distance larger than one-(i)-(b)-case-3-ccc}\\
\alignspace n_{(b'-1)-(b'-a'-1)}=p+2>n_{b'+(b'-a'-1)}=p+1.
\label{eqn:proof of nonadjacent distance larger than one-(i)-(b)-case-3-ddd}
\eeqnarray
An illustration of
\reqnarray{proof of nonadjacent distance larger than one-(i)-(b)-case-3-ccc}
and \reqnarray{proof of nonadjacent distance larger than one-(i)-(b)-case-3-ddd}
is given in \rfigure{appendix-G-(i)-(b)-case-3-2}.
Therefore, it follows from
$\nbf_1^{r_{h-1}}\in \Ncal_{M,k}(h)$
in \reqnarray{proof of nonadjacent distance larger than one-111},
\reqnarray{proof of nonadjacent distance larger than one-(i)-(b)-222}--\reqnarray{proof of nonadjacent distance larger than one-(i)-(b)-case-1-111},
\reqnarray{proof of nonadjacent distance larger than one-(i)-(b)-case-3-bbb}--\reqnarray{proof of nonadjacent distance larger than one-(i)-(b)-case-3-ddd},
and \reqnarray{comparison rule A-3} in \rlemma{comparison rule A}(ii)
(with $j=b'-a'-1$) that
\beqnarray{proof of nonadjacent distance larger than one-(i)-(b)-case-3-eee}
\nbf_1^{r_{h-1}}\preceq \mbf_1^{r_{h-1}}.
\eeqnarray

From $b'\geq a'+2$ in \reqnarray{proof of nonadjacent distance larger than one-(i)-333},
$b'-a'-1\leq r_{h-1}-b'$, and $a'\geq 1$, we can see that
\beqnarray{proof of nonadjacent distance larger than one-(i)-(b)-case-3-fff}
1\leq b'-a'-1\leq \min\{b'-a'-1,r_{h-1}-b'\}\leq \min\{(b'-1)-1,r_{h-1}-(b'-1)-1\}.
\eeqnarray
Thus, \reqnarray{proof of nonadjacent distance larger than one-(i)-(b)-case-3-bbb}
follows from \reqnarray{proof of nonadjacent distance larger than one-(i)-(b)-case-3-fff}.

To prove \reqnarray{proof of nonadjacent distance larger than one-(i)-(b)-case-3-ccc}
and \reqnarray{proof of nonadjacent distance larger than one-(i)-(b)-case-3-ddd},
note that it is clear from
\reqnarray{proof of nonadjacent distance larger than one-(i)-555},
$(b'-1)-(b'-a'-1)=a'$,
and $n_{a'}=p+2$ in \reqnarray{proof of nonadjacent distance larger than one-(i)-444} that
\beqnarray{}
\alignspace n_{(b'-1)-j'}=p+1, \textrm{ for } j'=1,2,\ldots,b'-a'-2,
\label{eqn:proof of nonadjacent distance larger than one-(i)-(b)-case-3-ggg}\\
\alignspace n_{(b'-1)-(b'-a'-1)}=n_{a'}=p+2.
\label{eqn:proof of nonadjacent distance larger than one-(i)-(b)-case-3-hhh}
\eeqnarray
Also, in this case we have
\beqnarray{proof of nonadjacent distance larger than one-(i)-(b)-case-3-iii}
n_{b'+j'}=p+1, \textrm{ for } j'=1,2,\ldots,\min\{b'-a'-1,r_{h-1}-b'\}=b'-a'-1.
\eeqnarray
By combining
\reqnarray{proof of nonadjacent distance larger than one-(i)-(b)-case-3-ggg},
\reqnarray{proof of nonadjacent distance larger than one-(i)-(b)-case-3-hhh},
and \reqnarray{proof of nonadjacent distance larger than one-(i)-(b)-case-3-iii},
we obtain \reqnarray{proof of nonadjacent distance larger than one-(i)-(b)-case-3-ccc}
and \reqnarray{proof of nonadjacent distance larger than one-(i)-(b)-case-3-ddd}.

Now let ${\mbf'}_1^{r_{h-1}}$ be a sequence of positive integers such that
\beqnarray{proof of nonadjacent distance larger than one-(i)-(b)-case-3-jjj}
m'_{b'-2}=m_{b'-2}-1,\ m'_{b'-1}=m_{b'-1}+1, \textrm{ and } m'_i=m_i \textrm{ for } i\neq b'-2, b'-1.
\eeqnarray
Again, it is easy to show that ${\mbf'}_1^{r_{h-1}}\in \Ncal_{M,k}(h)$.

If $b'=a'+2$, then we see from
\reqnarray{proof of nonadjacent distance larger than one-(i)-(b)-case-1-111},
$n_{a'}=p+2$ in \reqnarray{proof of nonadjacent distance larger than one-(i)-444},
and $n_{b'-1}=p+1$ in \reqnarray{proof of nonadjacent distance larger than one-(i)-888} that
\beqnarray{proof of nonadjacent distance larger than one-(i)-(b)-case-3-kkk}
\alignspace m_{b'-2}-m_{b'-1}=n_{b'-2}-(n_{b'-1}-1)=n_{a'}-n_{b'-1}+1=(p+2)-(p+1)+1=2.
\eeqnarray
Therefore, it follows from
$r_{h-1}\geq 3$ in \reqnarray{proof of nonadjacent distance larger than one-111},
$\mbf_1^{r_{h-1}}\in \Ncal_{M,k}(h)$,
\reqnarray{proof of nonadjacent distance larger than one-(i)-(b)-case-3-jjj},
\reqnarray{proof of nonadjacent distance larger than one-(i)-(b)-case-3-kkk},
and \reqnarray{adjacent distance larger than one-2} in \rlemma{adjacent distance larger than one}(ii) that
\beqnarray{proof of nonadjacent distance larger than one-(i)-(b)-case-3-1111}
\mbf_1^{r_{h-1}}\preceq {\mbf'}_1^{r_{h-1}},
\eeqnarray
where $\mbf_1^{r_{h-1}}\equiv {\mbf'}_1^{r_{h-1}}$
if and only if $r_{h-1}=2$ and $m_1=m_2+2$.
Since it is clear from $r_{h-1}\geq 3$
in \reqnarray{proof of nonadjacent distance larger than one-111}
that $r_{h-1}\neq 2$, it cannot be the case that $\mbf_1^{r_{h-1}}\equiv {\mbf'}_1^{r_{h-1}}$.
As such, we see from \reqnarray{proof of nonadjacent distance larger than one-(i)-(b)-case-3-eee}
and \reqnarray{proof of nonadjacent distance larger than one-(i)-(b)-case-3-1111}
that $\nbf_1^{r_{h-1}}\preceq \mbf_1^{r_{h-1}}\prec {\mbf'}_1^{r_{h-1}}$,
i.e., \reqnarray{proof of nonadjacent distance larger than one-(i)-999} holds
with ${\nbf'}_1^{r_{h-1}}={\mbf'}_1^{r_{h-1}}$.

On the other hand, if $b'\geq a'+3$, then we have $a'<b'-2<b'$
and it follows from \reqnarray{proof of nonadjacent distance larger than one-(i)-555} that
\beqnarray{proof of nonadjacent distance larger than one-(i)-(b)-case-3-2222}
n_{b'-2}=n_{b'-1}=p+1.
\eeqnarray
From \reqnarray{proof of nonadjacent distance larger than one-(i)-(b)-case-1-111}
and \reqnarray{proof of nonadjacent distance larger than one-(i)-(b)-case-3-2222},
we see that
\beqnarray{proof of nonadjacent distance larger than one-(i)-(b)-case-3-3333}
\alignspace m_{b'-2}-m_{b'-1}=n_{b'-2}-(n_{b'-1}-1)=1.
\eeqnarray

\bpdffigure{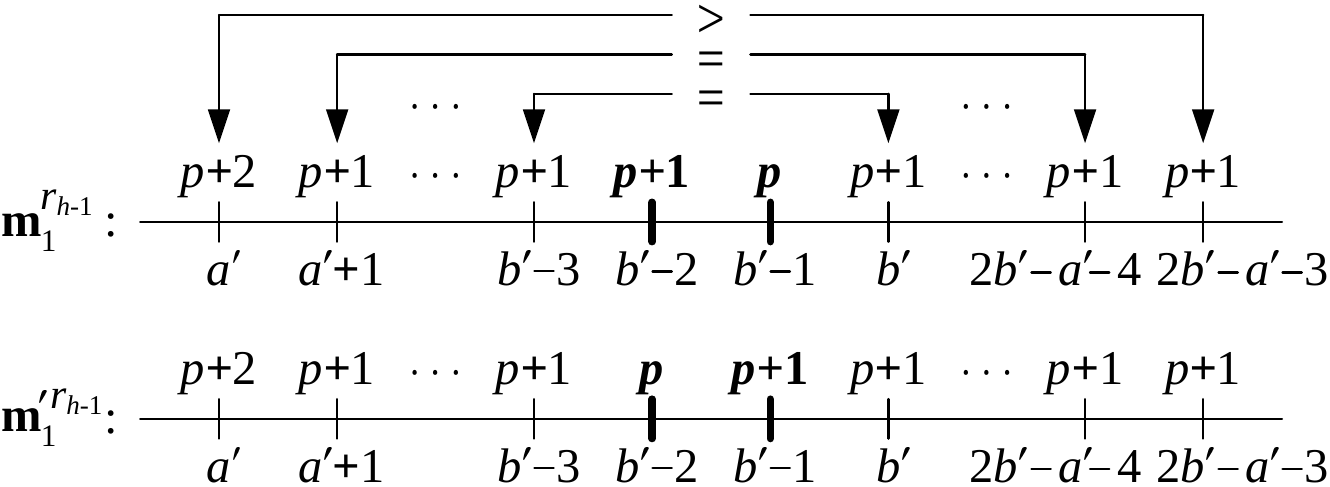}{4.5in}
\epdffigure{appendix-G-(i)-(b)-case-3-3}
{An illustration of \reqnarray{proof of nonadjacent distance larger than one-(i)-(b)-case-3-6666}
and \reqnarray{proof of nonadjacent distance larger than one-(i)-(b)-case-3-7777}.}

We will show that
\beqnarray{}
\alignspace 2\leq b'-2\leq r_{h-1}-2,
\label{eqn:proof of nonadjacent distance larger than one-(i)-(b)-case-3-4444}\\
\alignspace 1\leq b'-a'-2\leq \min\{(b'-2)-1,r_{h-1}-(b'-2)-1\}.
\label{eqn:proof of nonadjacent distance larger than one-(i)-(b)-case-3-5555}\\
\alignspace m_{(b'-2)-j'}=m_{(b'-1)+j'}=p+1, \textrm{ for } j'=1,2,\ldots,b'-a'-3,
\label{eqn:proof of nonadjacent distance larger than one-(i)-(b)-case-3-6666}\\
\alignspace m_{(b'-2)-(b'-a'-2)}=p+2>m_{(b'-1)+(b'-a'-2)}=p+1.
\label{eqn:proof of nonadjacent distance larger than one-(i)-(b)-case-3-7777}
\eeqnarray
An illustration of
\reqnarray{proof of nonadjacent distance larger than one-(i)-(b)-case-3-6666}
and \reqnarray{proof of nonadjacent distance larger than one-(i)-(b)-case-3-7777}
is given in \rfigure{appendix-G-(i)-(b)-case-3-3}.
Therefore, it follows from
$\mbf_1^{r_{h-1}}\in \Ncal_{M,k}(h)$,
\reqnarray{proof of nonadjacent distance larger than one-(i)-(b)-case-3-jjj},
\reqnarray{proof of nonadjacent distance larger than one-(i)-(b)-case-3-3333}--\reqnarray{proof of nonadjacent distance larger than one-(i)-(b)-case-3-7777},
and \reqnarray{comparison rule A-3} in \rlemma{comparison rule A}(ii)
(with $j=b'-a'-2$) that
\beqnarray{proof of nonadjacent distance larger than one-(i)-(b)-case-3-8888}
\mbf_1^{r_{h-1}}\preceq {\mbf'}_1^{r_{h-1}},
\eeqnarray
where $\mbf_1^{r_{h-1}}\equiv {\mbf'}_1^{r_{h-1}}$
if and only if $(b'-2)-(b'-a'-2)=1$, $(b'-1)+(b'-a'-2)=r_{h-1}$, and $m_1=m_{r_{h-1}}+1$.
Since in this subcase we have $b'-a'-1\leq r_{h-1}-b'$,
it is clear that $(b'-1)+(b'-a'-2)\leq (b'-1)+(r_{h-1}-b'-1)=r_{h-1}-2$.
This implies that $(b'-1)+(b'-a'-2)\neq r_{h-1}$
and hence it cannot be the case that $\mbf_1^{r_{h-1}}\equiv {\mbf'}_1^{r_{h-1}}$.
As such, we see from \reqnarray{proof of nonadjacent distance larger than one-(i)-(b)-case-3-eee}
and \reqnarray{proof of nonadjacent distance larger than one-(i)-(b)-case-3-8888}
that $\nbf_1^{r_{h-1}}\preceq \mbf_1^{r_{h-1}}\prec {\mbf'}_1^{r_{h-1}}$,
i.e., \reqnarray{proof of nonadjacent distance larger than one-(i)-999} holds
with ${\nbf'}_1^{r_{h-1}}={\mbf'}_1^{r_{h-1}}$.

To prove \reqnarray{proof of nonadjacent distance larger than one-(i)-(b)-case-3-4444}
and \reqnarray{proof of nonadjacent distance larger than one-(i)-(b)-case-3-5555},
note from $a'\geq 1$, $b'\geq a'+3$, $b'\leq r_{h-1}-1$,
and \reqnarray{proof of nonadjacent distance larger than one-(i)-(b)-case-3-bbb} that
\beqnarray{}
\alignspace \hspace*{-0.3in}
2\leq a'+1\leq b'-2\leq r_{h-1}-2,
\label{eqn:proof of nonadjacent distance larger than one-(i)-(b)-case-3-9999}\\
\alignspace \hspace*{-0.3in}
1\leq b'-a'-2\leq \min\{b'-3,r_{h-1}-b'-1\}\leq \min\{(b'-2)-1,r_{h-1}-(b'-2)-1\}.
\label{eqn:proof of nonadjacent distance larger than one-(i)-(b)-case-3-aaaa}
\eeqnarray
Thus, \reqnarray{proof of nonadjacent distance larger than one-(i)-(b)-case-3-4444}
follows from \reqnarray{proof of nonadjacent distance larger than one-(i)-(b)-case-3-9999},
and \reqnarray{proof of nonadjacent distance larger than one-(i)-(b)-case-3-5555}
follows from \reqnarray{proof of nonadjacent distance larger than one-(i)-(b)-case-3-aaaa}.

To prove \reqnarray{proof of nonadjacent distance larger than one-(i)-(b)-case-3-6666}
and \reqnarray{proof of nonadjacent distance larger than one-(i)-(b)-case-3-7777},
note that from \reqnarray{proof of nonadjacent distance larger than one-(i)-(b)-case-1-111},
$n_{a'}=p+2$ and $n_{b'}=p$ in \reqnarray{proof of nonadjacent distance larger than one-(i)-444},
and \reqnarray{proof of nonadjacent distance larger than one-(i)-(b)-case-3-ggg}--\reqnarray{proof of nonadjacent distance larger than one-(i)-(b)-case-3-iii},
we see that
\beqnarray{}
\alignspace m_{(b'-2)-j'}=n_{(b'-2)-j'}=p+1, \textrm{ for } j'=1,2,\ldots,b'-a'-3,
\label{eqn:proof of nonadjacent distance larger than one-(i)-(b)-case-3-bbbb}\\
\alignspace m_{(b'-2)-(b'-a'-2)}=n_{(b'-2)-(b'-a'-2)}=n_{a'}=p+2.
\label{eqn:proof of nonadjacent distance larger than one-(i)-(b)-case-3-cccc}\\
\alignspace m_{(b'-1)+1}=m_{b'}=n_{b'}+1=p+1,
\label{eqn:proof of nonadjacent distance larger than one-(i)-(b)-case-3-dddd}\\
\alignspace m_{(b'-1)+j'}=n_{(b'-1)+j'}=p+1, \textrm{ for } j'=2,3,\ldots,b'-a',
\label{eqn:proof of nonadjacent distance larger than one-(i)-(b)-case-3-eeee}
\eeqnarray
Thus, \reqnarray{proof of nonadjacent distance larger than one-(i)-(b)-case-3-6666}
and \reqnarray{proof of nonadjacent distance larger than one-(i)-(b)-case-3-7777}
follow from
\reqnarray{proof of nonadjacent distance larger than one-(i)-(b)-case-3-bbbb}--\reqnarray{proof of nonadjacent distance larger than one-(i)-(b)-case-3-eeee}.

(ii) Note that in \rlemma{nonadjacent distance larger than one}(ii),
we have $n_a-n_b\leq -2$ for some $1\leq a<b\leq r_{h-1}$ and $b\geq a+2$.
For ease of presentation, let $n_a=p$.
Then we have from $n_a-n_b\leq -2$ that $n_b\geq p+2$.
It is easy to see from $n_a=p$, $n_b\geq p+2$, $b\geq a+2>a$,
and the condition $|n_{i+1}-n_i|\leq 1$ for $i=1,2,\ldots,r_{h-1}-1$
in \reqnarray{proof of nonadjacent distance larger than one-111}
that there must exist a positive integer $c$ such that $a< c\leq b$ and $n_c=p+2$.
Let
\beqnarray{}
a'\aligneq \max\{i:n_i=p,\ a\leq i<c\},
\label{eqn:proof of nonadjacent distance larger than one-(ii)-111}\\
b'\aligneq \min\{i:n_i=p+2,\ a'<i\leq c\}.
\label{eqn:proof of nonadjacent distance larger than one-(ii)-222}
\eeqnarray
In other words, $a'$ is the largest positive integer $i$ such that $a\leq i<c$ and $n_i=p$,
and $b'$ is the smallest positive integer $i$ such that $a'<i\leq c$ and $n_i=p+2$.
Note that $a'$ and $b'$ are well defined as we have $n_a=p$ and $n_c=p+2$.
Since we have from \reqnarray{proof of nonadjacent distance larger than one-(ii)-111}
and \reqnarray{proof of nonadjacent distance larger than one-(ii)-222}
that $n_{a'}=p$ and $n_{b'}=p+2$, it is easy to see from the condition
$|n_{i+1}-n_i|\leq 1$ for $i=1,2,\ldots,r_{h-1}-1$
in \reqnarray{proof of nonadjacent distance larger than one-111} that $b'\geq a'+2$.
In summary, we have
\beqnarray{}
\alignspace a\leq a'<b'\leq c\leq b \textrm{ and } b'\geq a'+2,
\label{eqn:proof of nonadjacent distance larger than one-(ii)-333}\\
\alignspace n_a=n_{a'}=p,\ n_{b'}=n_c=p+2, \textrm{ and } n_b\geq p+2.
\label{eqn:proof of nonadjacent distance larger than one-(ii)-444}
\eeqnarray

\bpdffigure{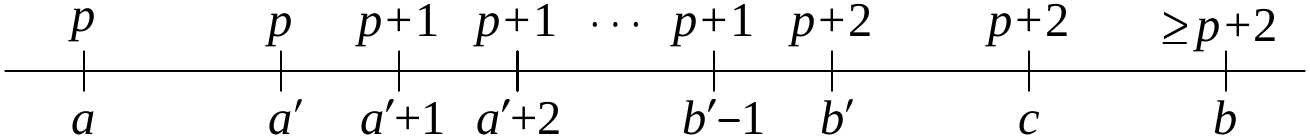}{4.5in}
\epdffigure{appendix-G-(ii)}
{An illustration of \reqnarray{proof of nonadjacent distance larger than one-(ii)-333}--\reqnarray{proof of nonadjacent distance larger than one-(ii)-555}.}

We claim that
\beqnarray{proof of nonadjacent distance larger than one-(ii)-555}
n_i=p+1, \textrm{ for } a'<i<b'.
\eeqnarray
An illustration of
\reqnarray{proof of nonadjacent distance larger than one-(ii)-333}--\reqnarray{proof of nonadjacent distance larger than one-(ii)-555}
is given in \rfigure{appendix-G-(ii)}.
We prove \reqnarray{proof of nonadjacent distance larger than one-(ii)-555} by contradiction.
First assume that $n_i\leq p$ for some $a'<i<b'$.
From $a\leq a'<b'\leq c$ in \reqnarray{proof of nonadjacent distance larger than one-(ii)-333} and $a'<i<b'$,
we have $a\leq a'<i<b'\leq c$ and hence it follows from the definition of $a'$
in \reqnarray{proof of nonadjacent distance larger than one-(ii)-111} that $n_i\neq p$.
Since we assume that $n_i\leq p$, it is clear that we must have $n_i<p$.
As such, we see from $n_i<p$,
$n_{b'}=p+2$ in \reqnarray{proof of nonadjacent distance larger than one-(ii)-444},
$i<b'$, and the condition $|n_{i+1}-n_i|\leq 1$ for $i=1,2,\ldots,r_{h-1}-1$
in \reqnarray{proof of nonadjacent distance larger than one-111}
that there must exist a positive integer $a''$ such that
\beqnarray{proof of nonadjacent distance larger than one-(ii)-666}
i<a''<b' \textrm{ and } n_{a''}=p.
\eeqnarray
From $a\leq a'<b'\leq c$ in \reqnarray{proof of nonadjacent distance larger than one-(ii)-333},
$a'<i<b'$, and $i<a''<b'$ in \reqnarray{proof of nonadjacent distance larger than one-(ii)-666},
we have $a\leq a'<i<a''<b'\leq c$ and hence it follows from the definition of $a'$
in \reqnarray{proof of nonadjacent distance larger than one-(ii)-111} that $n_{a''}\neq p$,
contradicting to $n_{a''}=p$ in \reqnarray{proof of nonadjacent distance larger than one-(ii)-666}.
Now assume that $n_i\geq p+2$ for some $a'<i<b'$.
From $a'<b'\leq c$ in \reqnarray{proof of nonadjacent distance larger than one-(ii)-333} and $a'<i<b'$,
we have $a'<i<b'\leq c$ and hence it follows from the definition of $b'$
in \reqnarray{proof of nonadjacent distance larger than one-(ii)-222} that $n_i\neq p+2$.
Since we assume that $n_i\geq p+2$, it is clear that we must have $n_i>p+2$.
As such, we see from $n_{a'}=p$ in \reqnarray{proof of nonadjacent distance larger than one-(ii)-444},
$n_i>p+2$, $a'<i$, and the condition $|n_{i+1}-n_i|\leq 1$ for $i=1,2,\ldots,r_{h-1}-1$
in \reqnarray{proof of nonadjacent distance larger than one-111}
that there must exist a positive integer $b''$ such that
\beqnarray{proof of nonadjacent distance larger than one-(ii)-777}
a'<b''<i \textrm{ and } n_{b''}=p+2.
\eeqnarray
From $a'<b''<i$ in \reqnarray{proof of nonadjacent distance larger than one-(ii)-777},
$a'<i<b'$, and $a'<b'\leq c$ in \reqnarray{proof of nonadjacent distance larger than one-(ii)-333},
we have $a'<b''<i<b'\leq c$ and hence it follows from the definition of $b'$
in \reqnarray{proof of nonadjacent distance larger than one-(ii)-222} that $n_{b''}\neq p+2$,
contradicting to $n_{b''}=p+2$ in \reqnarray{proof of nonadjacent distance larger than one-(ii)-777}.
The proof of \reqnarray{proof of nonadjacent distance larger than one-(ii)-555} is completed.

To prove \rlemma{nonadjacent distance larger than one}(ii),
we need to show that there exists a sequence of positive integers
${\nbf'}_1^{r_{h-1}}\in \Ncal_{M,k}(h)$ such that
\beqnarray{proof of nonadjacent distance larger than one-(ii)-888}
{\nbf'}_1^{r_{h-1}}\succ\nbf_1^{r_{h-1}}.
\eeqnarray
We consider the following four possible cases.
Note that in Case~2--Case~4 below,
we have $b'\leq r_{h-1}-1$ and hence it follows from $a'\geq 1$
and $b'\geq a'+2$ in \reqnarray{proof of nonadjacent distance larger than one-(ii)-333} that
\beqnarray{proof of nonadjacent distance larger than one-(ii)-999}
2\leq b'-1\leq r_{h-1}-2.
\eeqnarray

\emph{Case 1: $b'=r_{h-1}$.}
Let $\mbf_1^{r_{h-1}}$ be a sequence of positive integers such that
\beqnarray{proof of nonadjacent distance larger than one-(ii)-case-1-111}
m_{b'-1}=n_{b'-1}+1,\ m_{b'}=n_{b'}-1, \textrm{ and } m_i=n_i \textrm{ for } i\neq b'-1, b'.
\eeqnarray
As before, it is easy to show that $\mbf_1^{r_{h-1}}\in \Ncal_{M,k}(h)$.
From \reqnarray{proof of nonadjacent distance larger than one-(ii)-555}
and $b'\geq a'+2$ in \reqnarray{proof of nonadjacent distance larger than one-(ii)-333},
we see that $n_{b'-1}=p+1$.
It then follows from \reqnarray{proof of nonadjacent distance larger than one-(ii)-case-1-111},
$n_{b'-1}=p+1$, and $n_{b'}=p+2$ in \reqnarray{proof of nonadjacent distance larger than one-(ii)-444} that
\beqnarray{proof of nonadjacent distance larger than one-(ii)-case-1-222}
m_{b'-1}-m_{b'}=(n_{b'-1}+1)-(n_{b'}-1)=(p+1+1)-(p+2-1)=1.
\eeqnarray
Therefore, it follows from
$\mbf_1^{r_{h-1}}\in \Ncal_{M,k}(h)$,
\reqnarray{proof of nonadjacent distance larger than one-(ii)-case-1-111},
\reqnarray{proof of nonadjacent distance larger than one-(ii)-case-1-222},
$b'-1=r_{h-1}-1$, and \reqnarray{comparison rule A-1} in \rlemma{comparison rule A}(i)
that $\mbf_1^{r_{h-1}}\succ\nbf_1^{r_{h-1}}$,
i.e., \reqnarray{proof of nonadjacent distance larger than one-(ii)-888} holds
with ${\nbf'}_1^{r_{h-1}}=\mbf_1^{r_{h-1}}$.

\emph{Case 2: $b'\leq r_{h-1}-1$ and there exists a positive integer $j$ such that
$1\leq j\leq \min\{b'-a'-1,r_{h-1}-b'\}$, $n_{b'+j'}=p+1$ for $j'=1,2,\ldots,j-1$, and $n_{b'+j}>p+1$.}
Let $\mbf_1^{r_{h-1}}$ be a sequence of positive integers as given in
\reqnarray{proof of nonadjacent distance larger than one-(ii)-case-1-111}.
As in Case~1 above, we have $\mbf_1^{r_{h-1}}\in \Ncal_{M,k}(h)$
and \reqnarray{proof of nonadjacent distance larger than one-(ii)-case-1-222}
also holds in this case.

\bpdffigure{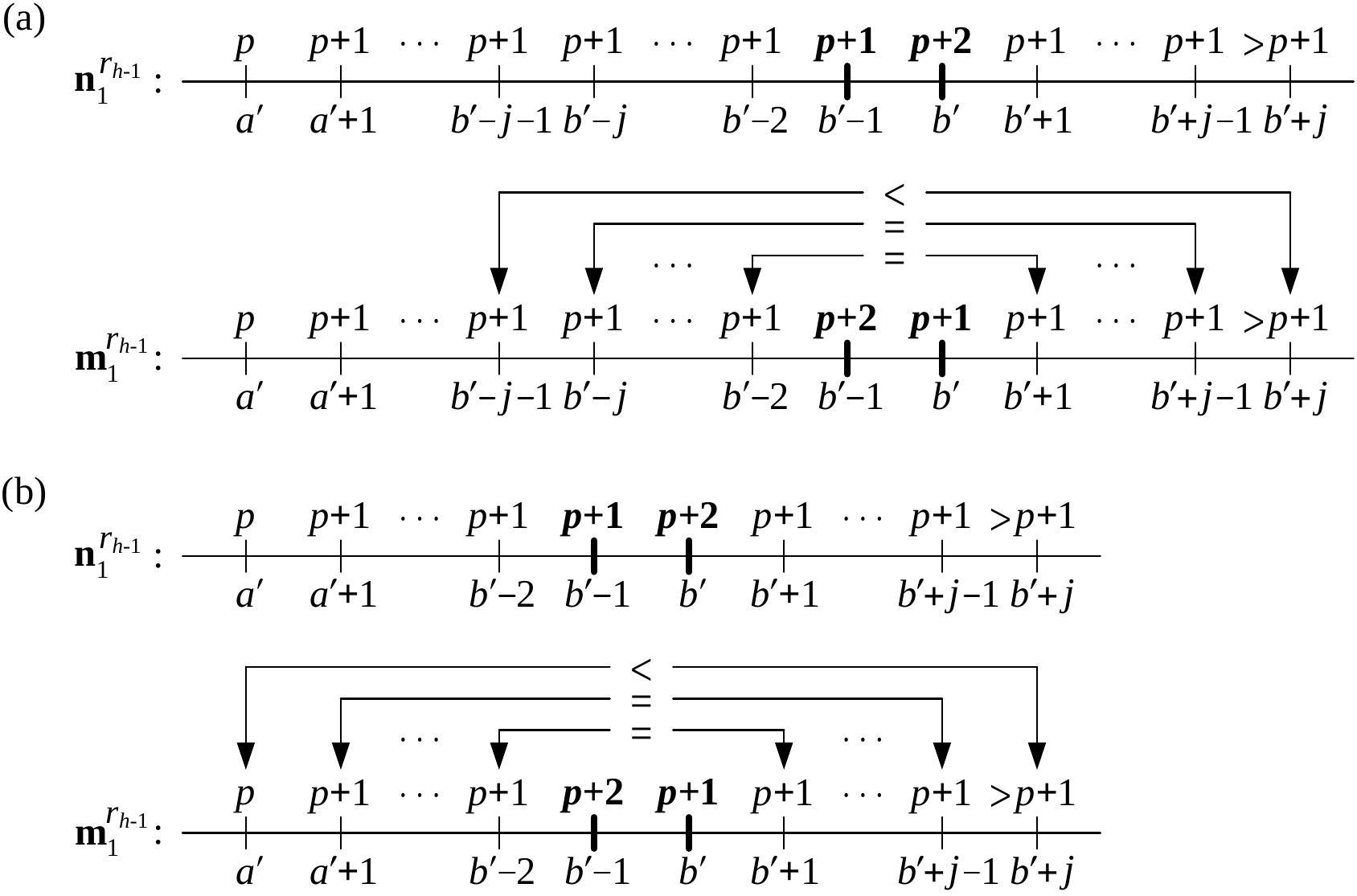}{5.5in}
\epdffigure{appendix-G-(ii)-case-2}
{An illustration of \reqnarray{proof of nonadjacent distance larger than one-(ii)-case-2-333}
and \reqnarray{proof of nonadjacent distance larger than one-(ii)-case-2-444}:
(a) $j<b'-a'-1$; (b) $j=b'-a'-1$.}

In the following, we show that
\beqnarray{}
\alignspace 1\leq j\leq \min\{(b'-1)-1,r_{h-1}-(b'-1)-1\},
\label{eqn:proof of nonadjacent distance larger than one-(ii)-case-2-222}\\
\alignspace m_{(b'-1)-j'}=m_{b'+j'}, \textrm{ for } j'=1,2,\ldots,j-1,
\label{eqn:proof of nonadjacent distance larger than one-(ii)-case-2-333}\\
\alignspace m_{(b'-1)-j}<m_{b'+j}.
\label{eqn:proof of nonadjacent distance larger than one-(ii)-case-2-444}
\eeqnarray
An illustration of
\reqnarray{proof of nonadjacent distance larger than one-(ii)-case-2-333}
and \reqnarray{proof of nonadjacent distance larger than one-(ii)-case-2-444}
is given in \rfigure{appendix-G-(ii)-case-2}.
Therefore, it follows from
$\mbf_1^{r_{h-1}}\in \Ncal_{M,k}(h)$,
\reqnarray{proof of nonadjacent distance larger than one-(ii)-999}--\reqnarray{proof of nonadjacent distance larger than one-(ii)-case-2-444},
and \reqnarray{comparison rule A-2} in \rlemma{comparison rule A}(ii) that
$\mbf_1^{r_{h-1}}\succ \nbf_1^{r_{h-1}}$,
i.e., \reqnarray{proof of nonadjacent distance larger than one-(ii)-888} holds
with ${\nbf'}_1^{r_{h-1}}=\mbf_1^{r_{h-1}}$.

From $1\leq j\leq \min\{b'-a'-1,r_{h-1}-b'\}$ and $a'\geq 1$, we have
\beqnarray{proof of nonadjacent distance larger than one-(ii)-case-2-555}
1\leq j\leq \min\{b'-a'-1,r_{h-1}-b'\}\leq \min\{(b'-1)-1,r_{h-1}-(b'-1)-1\}.
\eeqnarray
Thus, \reqnarray{proof of nonadjacent distance larger than one-(ii)-case-2-222}
follows from \reqnarray{proof of nonadjacent distance larger than one-(ii)-case-2-555}.

To prove \reqnarray{proof of nonadjacent distance larger than one-(ii)-case-2-333}
and \reqnarray{proof of nonadjacent distance larger than one-(ii)-case-2-444},
note that we have $j\leq \min\{b'-a'-1, r_{h-1}-b'\}\leq b'-a'-1$.
If $j<b'-a'-1$, then we have $a'<b'-1-j<b'$ and it follows from
\reqnarray{proof of nonadjacent distance larger than one-(ii)-case-1-111}
and \reqnarray{proof of nonadjacent distance larger than one-(ii)-555} that
\beqnarray{proof of nonadjacent distance larger than one-(ii)-case-2-666}
m_{(b'-1)-j'}=n_{(b'-1)-j'}=p+1, \textrm{ for } j'=1,2,\ldots,j.
\eeqnarray
On the other hand, if $j=b'-a'-1$, then we have $a'=b'-1-j<b'$
and it follows from \reqnarray{proof of nonadjacent distance larger than one-(ii)-case-1-111},
\reqnarray{proof of nonadjacent distance larger than one-(ii)-555},
and $n_{a'}=p$ in \reqnarray{proof of nonadjacent distance larger than one-(ii)-444} that
\beqnarray{}
\alignspace
m_{(b'-1)-j'}=n_{(b'-1)-j'}=p+1, \textrm{ for } j'=1,2,\ldots,j-1,
\label{eqn:proof of nonadjacent distance larger than one-(ii)-case-2-777}\\
\alignspace
m_{(b'-1)-j}=n_{(b'-1)-j}=n_{a'}=p.
\label{eqn:proof of nonadjacent distance larger than one-(ii)-case-2-888}
\eeqnarray
As in this case we have $n_{b'+j'}=p+1$ for $j'=1,2,\ldots,j-1$, and $n_{b'+j}>p+1$,
we immediately see from \reqnarray{proof of nonadjacent distance larger than one-(ii)-case-1-111} that
\beqnarray{}
\alignspace
m_{b'+j'}=n_{b'+j'}=p+1, \textrm{ for } j'=1,2,\ldots,j-1,
\label{eqn:proof of nonadjacent distance larger than one-(ii)-case-2-999}\\
\alignspace
m_{b'+j}=n_{b'+j}>p+1.
\label{eqn:proof of nonadjacent distance larger than one-(ii)-case-2-aaa}
\eeqnarray
By combining
\reqnarray{proof of nonadjacent distance larger than one-(ii)-case-2-666}--\reqnarray{proof of nonadjacent distance larger than one-(ii)-case-2-aaa},
we obtain \reqnarray{proof of nonadjacent distance larger than one-(ii)-case-2-333}
and \reqnarray{proof of nonadjacent distance larger than one-(ii)-case-2-444}.

\emph{Case 3: $b'\leq r_{h-1}-1$ and there exists a positive integer $j$ such that
$1\leq j\leq \min\{b'-a'-1,r_{h-1}-b'\}$, $n_{b'+j'}=p+1$ for $j'=1,2,\ldots,j-1$, and $n_{b'+j}<p+1$.}
In this case, we can show that $j\geq 2$.
To see this, suppose on the contrary that $j=1$,
then we have $n_{b'+1}<p+1$ in this case.
As it follows from $n_{b'}=p+2$ in \reqnarray{proof of nonadjacent distance larger than one-(ii)-444}
and the condition $|n_{i+1}-n_i|\leq 1$ for $i=1,2,\ldots,r_{h-1}-1$
in \reqnarray{proof of nonadjacent distance larger than one-111}
that $n_{b'+1}$ must be equal to $p+1$, $p+2$, or $p+3$,
we have reached a contradiction.
Since $j\geq 2$, we have $n_{b'+j-1}=p+1$ in this case.
It then follows from the condition $|n_{i+1}-n_i|\leq 1$ for $i=1,2,\ldots,r_{h-1}-1$
in \reqnarray{proof of nonadjacent distance larger than one-111}
that $n_{b'+j}$ must be equal to $p$, $p+1$, or $p+2$.
As we also have $n_{b'+j}<p+1$ in this case, we immediately see that $n_{b'+j}=p$.

From $n_{b'}=p+2$, $n_{b'+j'}=p+1$ for $j'=1,2,\ldots,j-1$, $n_{b'+j}=p$,
and $b'\geq 3$ in \reqnarray{proof of nonadjacent distance larger than one-(ii)-999},
we can argue in the same way as in the proof of (i) above
(with the roles of $a'$ and $b'$ in the proof of (i) replaced by $b'$ and $b'+j$, respectively)
that there exists a sequence of positive integers ${\nbf'}_1^{r_{h-1}}\in \Ncal_{M,k}(h)$
such that ${\nbf'}_1^{r_{h-1}}\succ\nbf_1^{r_{h-1}}$.

\emph{Case 4: $b'\leq r_{h-1}-1$ and $n_{b'+j'}=p+1$ for $j'=1,2,\ldots,\min\{b'-a'-1,r_{h-1}-b'\}$.}
Let $\mbf_1^{r_{h-1}}$ be a sequence of positive integers as given in
\reqnarray{proof of nonadjacent distance larger than one-(ii)-case-1-111}.
As in Case~1 above, we have $\mbf_1^{r_{h-1}}\in \Ncal_{M,k}(h)$
and \reqnarray{proof of nonadjacent distance larger than one-(ii)-case-1-222}
also holds in this case.
We then consider the two subcases $b'-a'-1>r_{h-1}-b'$ and $b'-a'-1\leq r_{h-1}-b'$ separately.

\emph{Subcase 4(a): $b'-a'-1>r_{h-1}-b'$.}

\bpdffigure{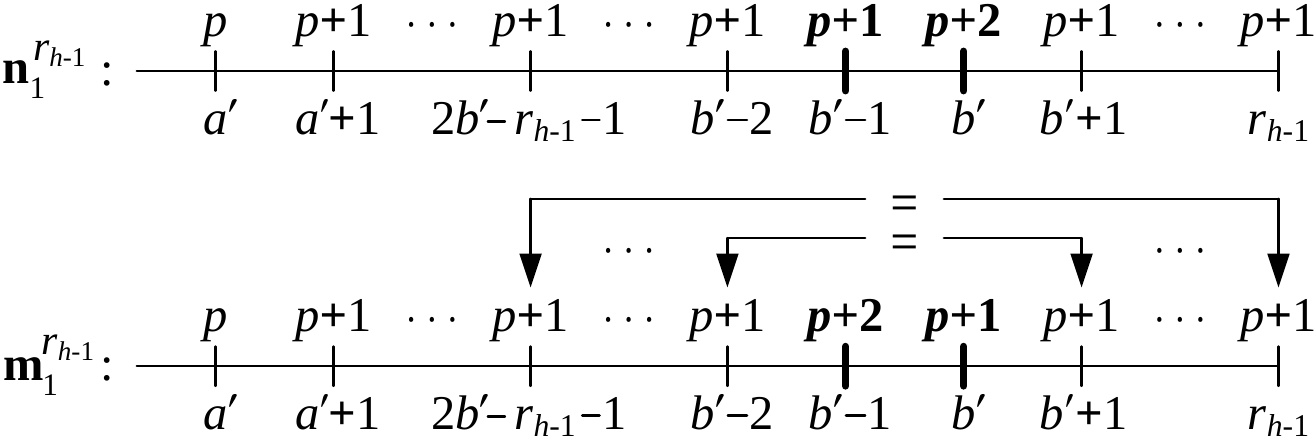}{4.5in}
\epdffigure{appendix-G-(ii)-case-4-1}
{An illustration of \reqnarray{proof of nonadjacent distance larger than one-(ii)-case-4-111}
(note that we have $\min\{(b'-1)-1, r_{h-1}-(b'-1)-1\}=r_{h-1}-b'$
in \reqnarray{proof of nonadjacent distance larger than one-(ii)-case-4-444}).}

In this subcase, we show that
\beqnarray{proof of nonadjacent distance larger than one-(ii)-case-4-111}
m_{(b'-1)-j'}=m_{b'+j'}=p+1, \textrm{ for } j'=1,2,\ldots,\min\{(b'-1)-1, r_{h-1}-(b'-1)-1\}.
\eeqnarray
An illustration of \reqnarray{proof of nonadjacent distance larger than one-(ii)-case-4-111}
is given in \rfigure{appendix-G-(ii)-case-4-1}.
Therefore, it follows from
$\mbf_1^{r_{h-1}}\in \Ncal_{M,k}(h)$,
\reqnarray{proof of nonadjacent distance larger than one-(ii)-999}--\reqnarray{proof of nonadjacent distance larger than one-(ii)-case-1-222},
\reqnarray{proof of nonadjacent distance larger than one-(ii)-case-4-111},
and \reqnarray{comparison rule A-4} in \rlemma{comparison rule A}(iii) that
$\mbf_1^{r_{h-1}}\succ \nbf_1^{r_{h-1}}$,
i.e., \reqnarray{proof of nonadjacent distance larger than one-(ii)-888} holds
with ${\nbf'}_1^{r_{h-1}}=\mbf_1^{r_{h-1}}$.

To prove \reqnarray{proof of nonadjacent distance larger than one-(ii)-case-4-111},
note that from $b'-a'-1>r_{h-1}-b'$, $a'\geq 1$, and $b'\leq r_{h-1}-1$, we have
\beqnarray{}
\alignspace
r_{h-1}-b'<b'-a'-1\leq b'-2,
\label{eqn:proof of nonadjacent distance larger than one-(ii)-case-4-222}\\
\alignspace
a'<(b'-1)-(r_{h-1}-b')\leq (b'-1)-1<b'.
\label{eqn:proof of nonadjacent distance larger than one-(ii)-case-4-333}
\eeqnarray
From \reqnarray{proof of nonadjacent distance larger than one-(ii)-case-4-222},
we see that
\beqnarray{proof of nonadjacent distance larger than one-(ii)-case-4-444}
\min\{(b'-1)-1, r_{h-1}-(b'-1)-1\}=\min\{b'-2, r_{h-1}-b'\}=r_{h-1}-b'.
\eeqnarray
From \reqnarray{proof of nonadjacent distance larger than one-(ii)-case-1-111},
\reqnarray{proof of nonadjacent distance larger than one-(ii)-555},
and \reqnarray{proof of nonadjacent distance larger than one-(ii)-case-4-333},
we have
\beqnarray{proof of nonadjacent distance larger than one-(ii)-case-4-555}
m_{(b'-1)-j'}=n_{(b'-1)-j'}=p+1, \textrm{ for } j'=1,2,\ldots,r_{h-1}-b'.
\eeqnarray
Furthermore, in this subcase we have from
\reqnarray{proof of nonadjacent distance larger than one-(ii)-case-1-111}
and $b'-a'-1>r_{h-1}-b'$ that
\beqnarray{proof of nonadjacent distance larger than one-(ii)-case-4-666}
m_{b'+j'}=n_{b'+j'}=p+1, \textrm{ for } j'=1,2,\ldots,\min\{b'-a'-1,r_{h-1}-b'\}=r_{h-1}-b'.
\eeqnarray
By combining \reqnarray{proof of nonadjacent distance larger than one-(ii)-case-4-444},
\reqnarray{proof of nonadjacent distance larger than one-(ii)-case-4-555},
and \reqnarray{proof of nonadjacent distance larger than one-(ii)-case-4-666},
we obtain \reqnarray{proof of nonadjacent distance larger than one-(ii)-case-4-111}.

\emph{Subcase 4(b): $b'-a'-1\leq r_{h-1}-b'$.}

\bpdffigure{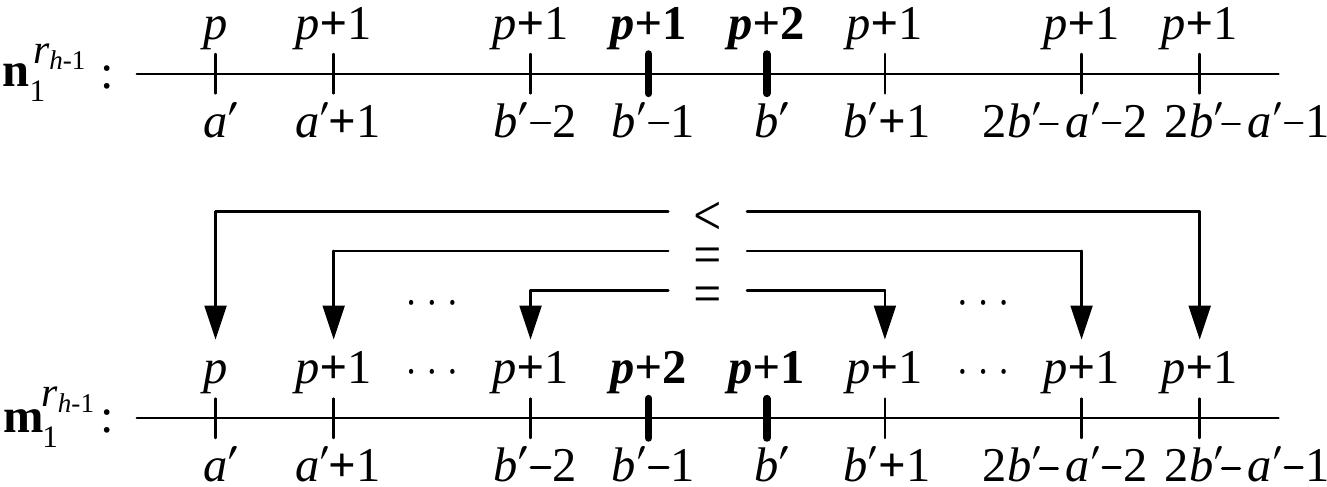}{4.5in}
\epdffigure{appendix-G-(ii)-case-4-2}
{An illustration of \reqnarray{proof of nonadjacent distance larger than one-(ii)-case-4-888}
and \reqnarray{proof of nonadjacent distance larger than one-(ii)-case-4-999}.}

In this subcase, we show that
\beqnarray{}
\alignspace
1\leq b'-a'-1\leq \min\{(b'-1)-1, r_{h-1}-(b'-1)-1\},
\label{eqn:proof of nonadjacent distance larger than one-(ii)-case-4-777}\\
\alignspace
m_{(b'-1)-j'}=m_{b'+j'}=p+1, \textrm{ for } j'=1,2,\ldots,b'-a'-2,
\label{eqn:proof of nonadjacent distance larger than one-(ii)-case-4-888}\\
\alignspace
m_{(b'-1)-(b'-a'-1)}=p<m_{b'+(b'-a'-1)}=p+1.
\label{eqn:proof of nonadjacent distance larger than one-(ii)-case-4-999}
\eeqnarray
An illustration of
\reqnarray{proof of nonadjacent distance larger than one-(ii)-case-4-888}
and \reqnarray{proof of nonadjacent distance larger than one-(ii)-case-4-999}
is given in \rfigure{appendix-G-(ii)-case-4-2}.
Therefore, it follows from
$\mbf_1^{r_{h-1}}\in \Ncal_{M,k}(h)$,
\reqnarray{proof of nonadjacent distance larger than one-(ii)-999}--\reqnarray{proof of nonadjacent distance larger than one-(ii)-case-1-222},
\reqnarray{proof of nonadjacent distance larger than one-(ii)-case-4-777}--\reqnarray{proof of nonadjacent distance larger than one-(ii)-case-4-999},
and \reqnarray{comparison rule A-2} in \rlemma{comparison rule A}(ii) that
$\mbf_1^{r_{h-1}}\succ \nbf_1^{r_{h-1}}$,
i.e., \reqnarray{proof of nonadjacent distance larger than one-(ii)-888} holds
with ${\nbf'}_1^{r_{h-1}}=\mbf_1^{r_{h-1}}$.

From $b'\geq a'+2$ in \reqnarray{proof of nonadjacent distance larger than one-(ii)-333}
and $a'\geq 1$, we see that
\beqnarray{proof of nonadjacent distance larger than one-(ii)-case-4-aaa}
1\leq b'-a'-1\leq b'-2.
\eeqnarray
It then follows from \reqnarray{proof of nonadjacent distance larger than one-(ii)-case-4-aaa}
and $b'-a'-1\leq r_{h-1}-b'$ that
\beqnarray{proof of nonadjacent distance larger than one-(ii)-case-4-bbb}
1\leq b'-a'-1\leq \min\{b'-2, r_{h-1}-b'\}=\min\{(b'-1)-1, r_{h-1}-(b'-1)-1\}.
\eeqnarray
Thus, \reqnarray{proof of nonadjacent distance larger than one-(ii)-case-4-777}
follows from \reqnarray{proof of nonadjacent distance larger than one-(ii)-case-4-bbb}.

To prove \reqnarray{proof of nonadjacent distance larger than one-(ii)-case-4-888}
and \reqnarray{proof of nonadjacent distance larger than one-(ii)-case-4-999},
note that from \reqnarray{proof of nonadjacent distance larger than one-(ii)-case-1-111},
\reqnarray{proof of nonadjacent distance larger than one-(ii)-555},
$a'=(b'-1)-(b'-a'-1)<b'$,
and $n_{a'}=p$ in \reqnarray{proof of nonadjacent distance larger than one-(ii)-444},
we have
\beqnarray{}
\alignspace
m_{(b'-1)-j'}=n_{(b'-1)-j'}=p+1, \textrm{ for } j'=1,2,\ldots,b'-a'-2,
\label{eqn:proof of nonadjacent distance larger than one-(ii)-case-4-ccc}\\
\alignspace
m_{(b'-1)-(b'-a'-1)}=n_{(b'-1)-(b'-a'-1)}=n_{a'}=p.
\label{eqn:proof of nonadjacent distance larger than one-(ii)-case-4-ddd}
\eeqnarray
Furthermore, in this subcase we have from
\reqnarray{proof of nonadjacent distance larger than one-(ii)-case-1-111}
and $b'-a'-1\leq r_{h-1}-b'$ that
\beqnarray{proof of nonadjacent distance larger than one-(ii)-case-4-eee}
m_{b'+j'}=n_{b'+j'}=p+1, \textrm{ for } j'=1,2,\ldots,\min\{b'-a'-1,r_{h-1}-b'\}=b'-a'-1.
\eeqnarray
By combining \reqnarray{proof of nonadjacent distance larger than one-(ii)-case-4-ccc},
\reqnarray{proof of nonadjacent distance larger than one-(ii)-case-4-ddd},
and \reqnarray{proof of nonadjacent distance larger than one-(ii)-case-4-eee},
we obtain \reqnarray{proof of nonadjacent distance larger than one-(ii)-case-4-888}
and \reqnarray{proof of nonadjacent distance larger than one-(ii)-case-4-999}.

\bappendix{Proof of \rlemma{main lemma}}{proof of main lemma}

In this appendix,
we use \rcorollary{adjacent distance larger than one}(i) (corollary to \rlemma{adjacent distance larger than one}),
\rcorollary{nonadjacent distance larger than one}(i) (corollary to \rlemma{nonadjacent distance larger than one}),
and Comparison rule A in \rlemma{comparison rule A} to prove \rlemma{main lemma}.

Let $\nbf_1^{r_{h-1}}(h)$ be an optimal sequence over $\Ncal_{M,k}(h)$.
As commented before the statement of \rlemma{main lemma},
we can use \rcorollary{adjacent distance larger than one}(i)
and \rcorollary{nonadjacent distance larger than one}(i) to show that
\beqnarray{proof of main lemma-111}
n_i(h)=
\bselection
q_h+1, &\textrm{if } i=i_1,i_2,\ldots,i_{r_h}, \\
q_h, &\textrm{otherwise},
\eselection
\eeqnarray
for some $1\leq i_1<i_2<\cdots <i_{r_h}\leq r_{h-1}$.

In the following, we show that $i_1=1$ by contradiction.
Assume on the contrary that $i_1\geq 2$.
If $h=1$ and $q_h=1$,
then we see from $\nbf_1^{r_{h-1}}(1)=\nbf_1^{r_{h-1}}(h)\in \Ncal_{M,k}(h)=\Ncal_{M,k}(1)$
and the definition of $\Ncal_{M,k}(1)$ in \reqnarray{N-M-k-h} that $n_1(1)\geq 2$.
We also see from \reqnarray{proof of main lemma-111} and $i_1\geq 2$ that $n_1(1)=n_1(h)=q_h=1$,
and a contradiction is reached.

On the other hand, if $h\neq 1$ or $q_h\neq 1$,
then we will use Comparison rule A in \rlemma{comparison rule A} to show that
there exists a sequence ${\nbf'}_1^{r_{h-1}}(h)\in \Ncal_{M,k}(h)$
such that ${\nbf'}_1^{r_{h-1}}(h)\succ\nbf_1^{r_{h-1}}(h)$,
contradicting to the optimality of $\nbf_1^{r_{h-1}}(h)$.
For simplicity, let $\nbf_1^{r_{h-1}}=\nbf_1^{r_{h-1}}(h)$.
Let ${\nbf'}_1^{r_{h-1}}$ be a sequence of positive integers such that
\beqnarray{proof of main lemma-222}
n'_{i_1-1}=n_{i_1-1}+1,\ n'_{i_1}=n_{i_1}-1,
\textrm{ and } n'_i=n_i \textrm{ for } i\neq i_1-1, i_1.
\eeqnarray
As $i_1\geq 2$, we have $1\leq i_1-1<i_1$ and it is easy to see from
\reqnarray{proof of main lemma-222}, \reqnarray{proof of main lemma-111},
$\nbf_1^{r_{h-1}}\in \Ncal_{M,k}(h)$, and \reqnarray{N-M-k-h} that
\beqnarray{}
\alignspace
n'_1=
\bselection
n_1+1=q_h+1, &\textrm{if } i_1-1=1, \\
n_1=q_h, &\textrm{otherwise},
\eselection
\label{eqn:proof of main lemma-333}\\
\alignspace
\sum_{i=1}^{r_{h-1}}n'_i=\sum_{i=1}^{r_{h-1}}n_i=r_{h-2}.
\label{eqn:proof of main lemma-444}
\eeqnarray
In the case that $h=1$, we must have $q_h\neq 1$, i.e., $q_h\geq 2$,
and hence it is clear from \reqnarray{proof of main lemma-333}
that $n'_1\geq q_h\geq 2$.
As such, it follows from $\nbf_1^{r_{h-1}}\in \Ncal_{M,k}(h)$,
\reqnarray{proof of main lemma-111}, \reqnarray{proof of main lemma-222},
and \reqnarray{proof of main lemma-444} that ${\nbf'}_1^{r_{h-1}}\in \Ncal_{M,k}(h)$.
Note that from
\reqnarray{proof of main lemma-222} and \reqnarray{proof of main lemma-111},
we have
\beqnarray{proof of main lemma-555}
n'_{i_1-1}-n'_{i_1}=(n_{i_1-1}+1)-(n_{i_1}-1)=(q_h+1)-(q_h+1-1)=1.
\eeqnarray
Furthermore, note that in the case that $h=1$ and $i_1-1=1$,
we have $q_h\geq 2$ and it follows from \reqnarray{proof of main lemma-333}
that $n'_1=q_h+1\geq 3$.

Now we have ${\nbf'}_1^{r_{h-1}}\in \Ncal_{M,k}(h)$,
$n'_{i_1-1}-n'_{i_1}=1$ in \reqnarray{proof of main lemma-555},
$n'_1\geq 3$ in the case that $h=1$ and $i_1-1=1$,
and $n_{i_1-1}=n'_{i_1-1}-1$, $n_{i_1}=n'_{i_1}+1$, and $n_i=n'_i$ for $i\neq i_1-1, i_1$
in \reqnarray{proof of main lemma-222}.
As such, we are in a position to use Comparison rule A in \rlemma{comparison rule A} (with $a=i_1-1$)
to show that ${\nbf'}_1^{r_{h-1}}\succ\nbf_1^{r_{h-1}}$.
We need to consider the two cases $r_h=1$ and $r_h\geq 2$ separately.

\emph{Case 1: $r_h=1$}.
In this case, we have from \reqnarray{proof of main lemma-222},
\reqnarray{proof of main lemma-111}, and $i_1\geq 2$ that
\beqnarray{proof of main lemma-case-1-111}
n'_i=
\bselection
q_h+1, &\textrm{if } i=i_1-1, \\
q_h, &\textrm{otherwise}.
\eselection
\eeqnarray
If $i_1=2$ or $i_1=r_{h-1}$, then we have $i_1-1=1$ or $i_1-1=r_{h-1}-1$,
and it follows from \reqnarray{comparison rule A-1} in \rlemma{comparison rule A}(i)
that ${\nbf'}_1^{r_{h-1}}\succ\nbf_1^{r_{h-1}}$.
On the other hand, if $3\leq i_1\leq r_{h-1}-1$,
then we have
\beqnarray{proof of main lemma-case-1-222}
2\leq i_1-1\leq r_{h-1}-2
\eeqnarray
From \reqnarray{proof of main lemma-case-1-111}, it is easy to see that
\beqnarray{proof of main lemma-case-1-333}
n'_{(i_1-1)-j'}=n'_{i_1+j'}=q_h, \textrm{ for } j'=1,2,\ldots,\min\{(i_1-1)-1,r_{h-1}-(i_1-1)-1\}.
\eeqnarray
Therefore, it follows from
\reqnarray{proof of main lemma-case-1-222}, \reqnarray{proof of main lemma-case-1-333},
and \reqnarray{comparison rule A-4} in \rlemma{comparison rule A}(iii)
that ${\nbf'}_1^{r_{h-1}}\succ\nbf_1^{r_{h-1}}$.

\emph{Case 2: $r_h\geq 2$}.
As $2\leq i_1<i_2<\cdots <i_{r_h}\leq r_{h-1}$ and $r_h\geq 2$,
we have $2\leq i_1\leq i_2-1\leq r_{h-1}-1$ in this case.
If $i_1=2$, then we have $i_1-1=1$, and it follows from \reqnarray{comparison rule A-1}
in \rlemma{comparison rule A}(i) that ${\nbf'}_1^{r_{h-1}}\succ\nbf_1^{r_{h-1}}$.
On the other hand, if $3\leq i_1\leq r_{h-1}-1$,
then \reqnarray{proof of main lemma-case-1-222} holds
and we also have from \reqnarray{proof of main lemma-222}
and \reqnarray{proof of main lemma-111} that
\beqnarray{proof of main lemma-case-2-111}
n'_i(h)=
\bselection
q_h+1, &\textrm{if } i=i_1-1,i_2,\ldots,i_{r_h}, \\
q_h, &\textrm{otherwise},
\eselection
\eeqnarray
We then consider the following two subcases.

\emph{Subcase 2(a): $i_1-2<i_2-i_1$.}
In this subcase, we have $i_1-2<i_2-i_1\leq r_{h-1}-i_1$,
and it follows that
\beqnarray{proof of main lemma-case-2-222}
\min\{(i_1-1)-1,r_{h-1}-(i_1-1)-1\}=\min\{i_1-2,r_{h-1}-i_1\}=i_1-2.
\eeqnarray
As $(i_1-1)-(i_1-2)=1$ and $i_1+(i_1-2)<i_1+(i_2-i_1)=i_2$,
it is easy to see from \reqnarray{proof of main lemma-case-2-111}
and \reqnarray{proof of main lemma-case-2-222} that
\beqnarray{proof of main lemma-case-2-333}
n'_{(i_1-1)-j'}=n'_{i_1+j'}=q_h,\ j'=1,2,\ldots,i_1-2=\min\{(i_1-1)-1,r_{h-1}-(i_1-1)-1\}.
\eeqnarray
Therefore, it follows from
\reqnarray{proof of main lemma-case-1-222}, \reqnarray{proof of main lemma-case-2-333},
and \reqnarray{comparison rule A-4} in \rlemma{comparison rule A}(iii)
that ${\nbf'}_1^{r_{h-1}}\succ\nbf_1^{r_{h-1}}$.

\emph{Subcase 2(b): $i_1-2\geq i_2-i_1$.}
In this subcase, we see from $i_1<i_2$, $i_1-2\geq i_2-i_1$, and $i_2\leq r_{h-1}$ that
\beqnarray{proof of main lemma-case-2-444}
1\leq i_2-i_1\leq \min\{i_1-2,r_{h-1}-i_1\}=\min\{(i_1-1)-1,r_{h-1}-(i_1-1)-1\}.
\eeqnarray
As it is clear that $1\leq (i_1-1)-(i_2-i_1)\leq i_1-2$ and $i_1+(i_2-i_1)=i_2$,
we see from \reqnarray{proof of main lemma-case-2-111} that
\beqnarray{}
\alignspace
n'_{(i_1-1)-j'}=n'_{i_1+j'}=q_h, \textrm{ for } j'=1,2,\ldots,i_2-i_1-1,
\label{eqn:proof of main lemma-case-2-555}\\
\alignspace
n'_{(i_1-1)-(i_2-i_1)}=q_h<n'_{i_1+(i_2-i_1)}=n'_{i_2}=q_h+1.
\label{eqn:proof of main lemma-case-2-666}
\eeqnarray
Therefore, it follows from
\reqnarray{proof of main lemma-case-1-222},
\reqnarray{proof of main lemma-case-2-444}--\reqnarray{proof of main lemma-case-2-666},
and \reqnarray{comparison rule A-2} in \rlemma{comparison rule A}(ii)
that ${\nbf'}_1^{r_{h-1}}\succ\nbf_1^{r_{h-1}}$.

\bappendix{Proof of \rlemma{nonadjacent distance larger than one II}}
{proof of nonadjacent distance larger than one II}

In this appendix, we use \rlemma{adjacent distance larger than one II}
and Comparison rule B in \rlemma{comparison rule B}
to prove \rlemma{nonadjacent distance larger than one II}.
For simplicity, let $\nbf_1^{r_{h-1}}=\nbf_1^{r_{h-1}}(h)$.
Note that in \rlemma{nonadjacent distance larger than one II}, we have
\beqnarray{proof of nonadjacent distance larger than one II-111}
r_{h-1}\geq 3,\ \nbf_1^{r_{h-1}}\in \Ncal_{M,k}(h),
\textrm{ and } |n_i-n_{i+1}|\leq 1 \textrm{ for } i=1,2,\ldots,r_{h-1}-1.
\eeqnarray

(i) Note that in \rlemma{nonadjacent distance larger than one II}(i),
we have $n_a-n_b\leq -2$ for some $1\leq a<b\leq r_{h-1}$ and $b\geq a+2$.
For ease of presentation, let $n_a=p$.
Then we have from $n_a-n_b\leq -2$ that $n_b\geq p+2$.
Note that the condition $|n_{i+1}-n_i|\leq 1$ for $i=1,2,\ldots,r_{h-1}-1$
in \reqnarray{proof of nonadjacent distance larger than one II-111}
says that the absolute value of the difference of any two adjacent entries
of $\nbf_1^{r_{h-1}}$ is at most equal to one.
As such, from $n_a=p$, $n_b\geq p+2$, $b\geq a+2>a$,
and the condition $|n_{i+1}-n_i|\leq 1$ for $i=1,2,\ldots,r_{h-1}-1$
in \reqnarray{proof of nonadjacent distance larger than one II-111},
we can argue as that for \reqnarray{proof of nonadjacent distance larger than one-(ii)-111}--\reqnarray{proof of nonadjacent distance larger than one-(ii)-555}
in the proof of \rlemma{nonadjacent distance larger than one}(ii)
in \rappendix{proof of nonadjacent distance larger than one} that there exist two positive integers $a'$ and $b'$ such that
\beqnarray{}
\alignspace a\leq a'<b'\leq b, \textrm{ and } b'\geq a'+2,
\label{eqn:proof of nonadjacent distance larger than one II-(i)-111} \\
\alignspace n_{a'}=p,\ n_{b'}=p+2, \textrm{ and } n_i=p+1 \textrm{ for } a'<i<b'.
\label{eqn:proof of nonadjacent distance larger than one II-(i)-222}
\eeqnarray
An illustration of
\reqnarray{proof of nonadjacent distance larger than one II-(i)-111}
and \reqnarray{proof of nonadjacent distance larger than one II-(i)-222}
is given in \rfigure{appendix-I-(i)}.
Note that from $b'\geq a'+2$ in \reqnarray{proof of nonadjacent distance larger than one II-(i)-111}
and $n_i=p+1$ for $a'<i<b'$ in \reqnarray{proof of nonadjacent distance larger than one II-(i)-222},
it is clear that
\beqnarray{proof of nonadjacent distance larger than one II-(i)-333}
n_{a'+1}=p+1 \textrm{ and } n_{b'-1}=p+1.
\eeqnarray

\bpdffigure{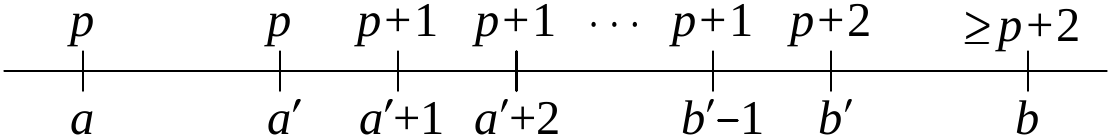}{4.0in}
\epdffigure{appendix-I-(i)}
{An illustration of \reqnarray{proof of nonadjacent distance larger than one II-(i)-111}
and \reqnarray{proof of nonadjacent distance larger than one II-(i)-222}.}

To prove \rlemma{nonadjacent distance larger than one II}(i),
we need to show that if $n_{r_{h-1}}\neq n_1+2$
or $n_i\neq n_1+1$ for some $2\leq i\leq r_{h-1}-1$,
then there exists a sequence of positive integers
${\nbf'}_1^{r_{h-1}}\in \Ncal_{M,k}(h)$ such that
\beqnarray{proof of nonadjacent distance larger than one II-(i)-444}
{\nbf'}_1^{r_{h-1}}\succ\nbf_1^{r_{h-1}}.
\eeqnarray
Note that if $n_{r_{h-1}}\neq n_1+2$
or $n_i\neq n_1+1$ for some $2\leq i\leq r_{h-1}-1$,
then we have $a'\geq 2$ or $b'\leq r_{h-1}-1$.
To see this, suppose on the contrary that $a'=1$ and $b'=r_{h-1}$.
Then it follows from $a'=1$, $b'=r_{h-1}$, and \reqnarray{proof of nonadjacent distance larger than one II-(i)-222} that
\beqnarray{proof of nonadjacent distance larger than one II-(i)-555}
n_1=p,\ n_{r_{h-1}}=p+2,
\textrm{ and } n_i=p+1 \textrm{ for } 2\leq i\leq r_{h-1}-1.
\eeqnarray
It is clear from \reqnarray{proof of nonadjacent distance larger than one II-(i)-555}
that $n_{r_{h-1}}=n_1+2$ and $n_i=n_1+1$ for all $2\leq i\leq r_{h-1}-1$,
and a contradiction is reached.

In the following, we show that if $a'\geq 2$ or $b'\leq r_{h-1}-1$,
then there exists a sequence of positive integers ${\nbf'}_1^{r_{h-1}}\in \Ncal_{M,k}(h)$
such that \reqnarray{proof of nonadjacent distance larger than one II-(i)-444} holds,
and hence \rlemma{nonadjacent distance larger than one II}(i) is proved.

(a) First, we assume that $a'\geq 2$
and show that there exists a sequence of positive integers
${\nbf'}_1^{r_{h-1}}\in \Ncal_{M,k}(h)$ such that
\reqnarray{proof of nonadjacent distance larger than one II-(i)-444} holds.
Note that from $b'\geq a'+2$
in \reqnarray{proof of nonadjacent distance larger than one II-(i)-111}
and $b'\leq r_{h-1}$, we have
\beqnarray{proof of nonadjacent distance larger than one II-(i)-(a)-111}
a'\leq b'-2\leq r_{h-1}-2.
\eeqnarray
As we assume that $a'\geq 2$,
it follows from \reqnarray{proof of nonadjacent distance larger than one II-(i)-(a)-111} that
\beqnarray{proof of nonadjacent distance larger than one II-(i)-(a)-222}
2\leq a'\leq r_{h-1}-2.
\eeqnarray

We need to consider the following three possible cases.

\emph{Case 1: There exists a positive integer $j$ such that
$1\leq j\leq \min\{a'-1,b'-a'-1\}$,
$n_{a'-j'}=p+1$ for $j'=1,2,\ldots,j-1$, and $n_{a'-j}<p+1$.}
Let $\mbf_1^{r_{h-1}}$ be a sequence of positive integers such that
\beqnarray{proof of nonadjacent distance larger than one II-(i)-(a)-case-1-111}
m_{a'}=n_{a'}+1,\ m_{a'+1}=n_{a'+1}-1, \textrm{ and } m_i=n_i \textrm{ for } i\neq a', a'+1.
\eeqnarray
As before, it is easy to show that $\mbf_1^{r_{h-1}}\in \Ncal_{M,k}(h)$.
From \reqnarray{proof of nonadjacent distance larger than one II-(i)-(a)-case-1-111},
$n_{a'}=p$ in \reqnarray{proof of nonadjacent distance larger than one II-(i)-222},
and $n_{a'+1}=p+1$ in \reqnarray{proof of nonadjacent distance larger than one II-(i)-333},
we have
\beqnarray{proof of nonadjacent distance larger than one II-(i)-(a)-case-1-222}
m_{a'}-m_{a'+1}=(n_{a'}+1)-(n_{a'+1}-1)=(p+1)-(p+1-1)=1.
\eeqnarray
As we have $j\leq \min\{a'-1,b'-a'-1\}\leq b'-a'-1$,
we consider the two subcases $j<b'-a'-1$ and $j=b'-a'-1$ separately.

\emph{Subcase 1(a): $j<b'-a'-1$.}

\bpdffigure{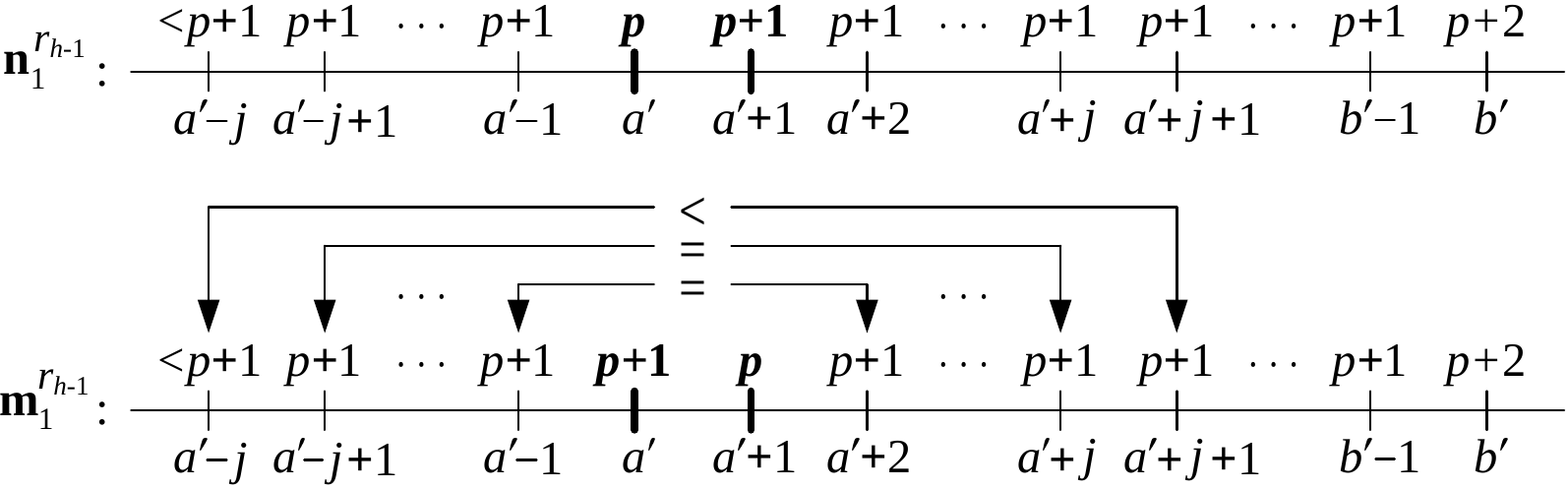}{5.5in}
\epdffigure{appendix-I-(i)-(a)-case-1}
{An illustration of \reqnarray{proof of nonadjacent distance larger than one II-(i)-(a)-case-1-444}
and \reqnarray{proof of nonadjacent distance larger than one II-(i)-(a)-case-1-555}.}

In this subcase, we show that
\beqnarray{}
\alignspace 1\leq j\leq \min\{a'-1,r_{h-1}-a'-1\}
\label{eqn:proof of nonadjacent distance larger than one II-(i)-(a)-case-1-333}\\
\alignspace m_{a'-j'}=m_{(a'+1)+j'}, \textrm{ for } j'=1,2,\ldots,j-1,
\label{eqn:proof of nonadjacent distance larger than one II-(i)-(a)-case-1-444}\\
\alignspace m_{a'-j}<m_{(a'+1)+j}.
\label{eqn:proof of nonadjacent distance larger than one II-(i)-(a)-case-1-555}
\eeqnarray
An illustration of
\reqnarray{proof of nonadjacent distance larger than one II-(i)-(a)-case-1-444}
and \reqnarray{proof of nonadjacent distance larger than one II-(i)-(a)-case-1-555}
is given in \rfigure{appendix-I-(i)-(a)-case-1}.
Therefore, it follows from
$\mbf_1^{r_{h-1}}\in \Ncal_{M,k}(h)$,
\reqnarray{proof of nonadjacent distance larger than one II-(i)-(a)-222}--\reqnarray{proof of nonadjacent distance larger than one II-(i)-(a)-case-1-555},
and \reqnarray{comparison rule B-3} in \rlemma{comparison rule B}(ii) that
\beqnarray{proof of nonadjacent distance larger than one II-(i)-(a)-case-1-666}
\mbf_1^{r_{h-1}}\succeq \nbf_1^{r_{h-1}},
\eeqnarray
where $\mbf_1^{r_{h-1}}\equiv \nbf_1^{r_{h-1}}$
if and only if $a'-j=1$, $(a'+1)+j=r_{h-1}$,
and $m_1=m_{r_{h-1}}-1$ (i.e., $m_{a'-j}=m_{(a'+1)+j}-1$).
From $j<b'-a'-1$ in this subcase and $b'\leq r_{h-1}$,
we see that $(a'+1)+j<b'\leq r_{h-1}$.
This implies that $(a'+1)+j\neq r_{h-1}$ and hence it cannot be the case that
$\mbf_1^{r_{h-1}}\equiv \nbf_1^{r_{h-1}}$.
As such, we see from \reqnarray{proof of nonadjacent distance larger than one II-(i)-(a)-case-1-666}
that $\mbf_1^{r_{h-1}}\succ \nbf_1^{r_{h-1}}$,
i.e., \reqnarray{proof of nonadjacent distance larger than one II-(i)-444} holds
with ${\nbf'}_1^{r_{h-1}}=\mbf_1^{r_{h-1}}$.

From $1\leq j\leq \min\{a'-1,b'-a'-1\}$ and $b'\leq r_{h-1}$, we see that
\beqnarray{proof of nonadjacent distance larger than one II-(i)-(a)-case-1-777}
1\leq j\leq \min\{a'-1,b'-a'-1\}\leq \min\{a'-1,r_{h-1}-a'-1\}.
\eeqnarray
Thus, \reqnarray{proof of nonadjacent distance larger than one II-(i)-(a)-case-1-333}
follows from \reqnarray{proof of nonadjacent distance larger than one II-(i)-(a)-case-1-777}.

To prove \reqnarray{proof of nonadjacent distance larger than one II-(i)-(a)-case-1-444}
and \reqnarray{proof of nonadjacent distance larger than one II-(i)-(a)-case-1-555},
note that as we have $n_{a'-j'}=p+1$ for $j'=1,2,\ldots,j-1$, and $n_{a'-j}<p+1$ in this case,
it is clear from \reqnarray{proof of nonadjacent distance larger than one II-(i)-(a)-case-1-111} that
\beqnarray{}
\alignspace m_{a'-j'}=n_{a'-j'}=p+1, \textrm{ for } j'=1,2,\ldots,j-1,
\label{eqn:proof of nonadjacent distance larger than one II-(i)-(a)-case-1-888}\\
\alignspace m_{a'-j}=n_{a'-j}<p+1.
\label{eqn:proof of nonadjacent distance larger than one II-(i)-(a)-case-1-999}
\eeqnarray
Furthermore, as in this subcase we have $a'<(a'+1)+j<b'$,
it follows from \reqnarray{proof of nonadjacent distance larger than one II-(i)-(a)-case-1-111}
and \reqnarray{proof of nonadjacent distance larger than one II-(i)-222} that
\beqnarray{proof of nonadjacent distance larger than one II-(i)-(a)-case-1-aaa}
m_{(a'+1)+j'}=n_{(a'+1)+j'}=p+1, \textrm{ for } j'=1,2,\ldots,j.
\eeqnarray
By combining
\reqnarray{proof of nonadjacent distance larger than one II-(i)-(a)-case-1-888}--\reqnarray{proof of nonadjacent distance larger than one II-(i)-(a)-case-1-aaa},
we obtain \reqnarray{proof of nonadjacent distance larger than one II-(i)-(a)-case-1-444}
and \reqnarray{proof of nonadjacent distance larger than one II-(i)-(a)-case-1-555}.

\emph{Subcase 1(b): $j=b'-a'-1$.}
In this subcase, we have $a'<(a'+1)+j=b'$
and hence it follows from \reqnarray{proof of nonadjacent distance larger than one II-(i)-(a)-case-1-111}
and \reqnarray{proof of nonadjacent distance larger than one II-(i)-222} that
\beqnarray{}
\alignspace m_{(a'+1)+j'}=n_{(a'+1)+j'}=p+1, \textrm{ for } j'=1,2,\ldots,j-1,
\label{eqn:proof of nonadjacent distance larger than one II-(i)-(a)-case-1-bbb}\\
\alignspace m_{(a'+1)+j'}=n_{(a'+1)+j}=n_{b'}=p+2.
\label{eqn:proof of nonadjacent distance larger than one II-(i)-(a)-case-1-ccc}
\eeqnarray
By using \reqnarray{proof of nonadjacent distance larger than one II-(i)-(a)-case-1-bbb}
and \reqnarray{proof of nonadjacent distance larger than one II-(i)-(a)-case-1-ccc},
we can argue as in Subcase~1(a) above that
\reqnarray{proof of nonadjacent distance larger than one II-(i)-(a)-case-1-333}--\reqnarray{proof of nonadjacent distance larger than one II-(i)-(a)-case-1-666}
still hold.
Since it is clear from $m_{a'-j}<p+1$ in \reqnarray{proof of nonadjacent distance larger than one II-(i)-(a)-case-1-999}
and $m_{(a'+1)+j}=p+2$ in \reqnarray{proof of nonadjacent distance larger than one II-(i)-(a)-case-1-ccc}
that $m_{a'-j}\neq m_{(a'+1)+j}-1$,
it cannot be the case that $\mbf_1^{r_{h-1}}\equiv \nbf_1^{r_{h-1}}$.
As such, we see from \reqnarray{proof of nonadjacent distance larger than one II-(i)-(a)-case-1-666}
that $\mbf_1^{r_{h-1}}\succ \nbf_1^{r_{h-1}}$,
i.e., \reqnarray{proof of nonadjacent distance larger than one II-(i)-444} holds
with ${\nbf'}_1^{r_{h-1}}=\mbf_1^{r_{h-1}}$.

\emph{Case 2: There exists a positive integer $j$ such that
$1\leq j\leq \min\{a'-1,b'-a'-1\}$,
$n_{a'-j'}=p+1$ for $j'=1,2,\ldots,j-1$, and $n_{a'-j}>p+1$.}
In this case, we can show that $j\geq 2$.
To see this, suppose on the contrary that $j=1$,
then we have $n_{a'-1}>p+1$ in this case.
As it is easy to see from $n_{a'}=p$
in \reqnarray{proof of nonadjacent distance larger than one II-(i)-222}
and the condition $|n_{i+1}-n_i|\leq 1$ for $i=1,2,\ldots,r_{h-1}-1$
in \reqnarray{proof of nonadjacent distance larger than one II-111}
that $n_{a'-1}$ must be equal to $p-1$ (provided that $p\geq 2$), $p$, or $p+1$,
we have reached a contradiction.

As $j\geq 2$, we have $n_{a'-1}=p+1$ in this case.
It then follows from
$n_{a'-1}=p+1$ and $n_{a'}=p$
in \reqnarray{proof of nonadjacent distance larger than one II-(i)-222} that
\beqnarray{proof of nonadjacent distance larger than one II-(i)-(a)-case-2-111}
n_{a'-1}-n_{a'}=(p+1)-p=1.
\eeqnarray
Let $\mbf_1^{r_{h-1}}$ be a sequence of positive integers such that
\beqnarray{proof of nonadjacent distance larger than one II-(i)-(a)-case-2-222}
m_{a'-1}=n_{a'-1}-1,\ m_{a'}=n_{a'}+1,
\textrm{ and } m_i=n_i \textrm{ for } i\neq a'-1, a'.
\eeqnarray
As before, it is easy to show that $\mbf_1^{r_{h-1}}\in \Ncal_{M,k}(h)$.

\bpdffigure{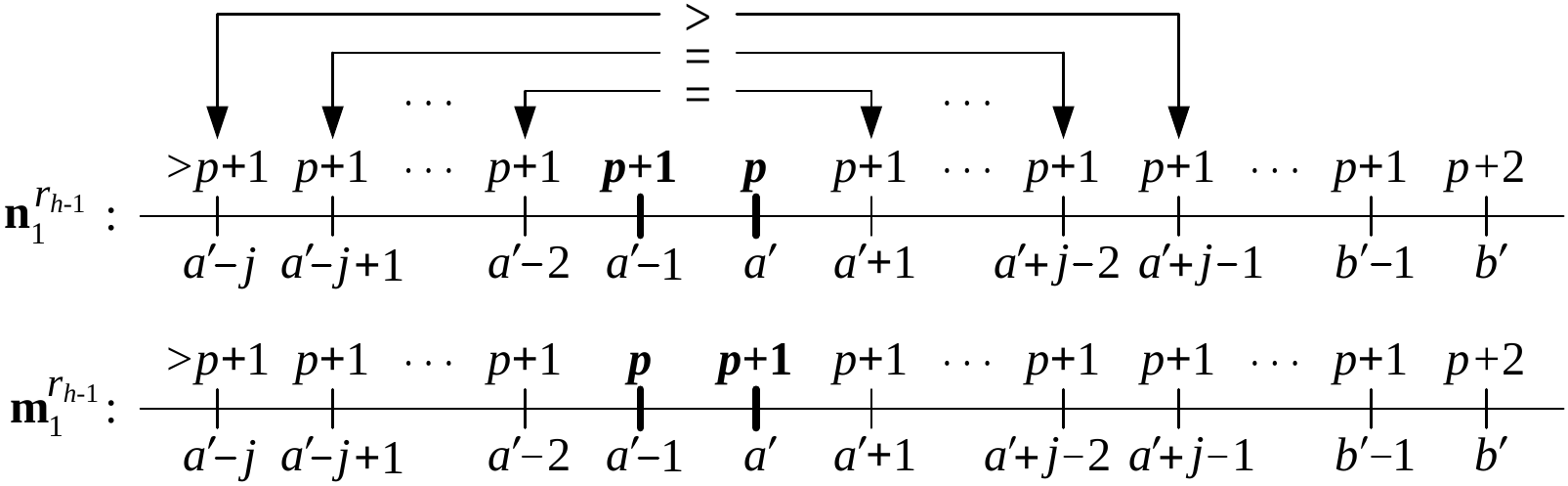}{5.5in}
\epdffigure{appendix-I-(i)-(a)-case-2}
{An illustration of \reqnarray{proof of nonadjacent distance larger than one II-(i)-(a)-case-2-555}
and \reqnarray{proof of nonadjacent distance larger than one II-(i)-(a)-case-2-666}.}

In the following, we show that
\beqnarray{}
\alignspace
2\leq a'-1\leq r_{h-1}-2,
\label{eqn:proof of nonadjacent distance larger than one II-(i)-(a)-case-2-333}\\
\alignspace
1\leq j-1\leq \min\{(a'-1)-1,r_{h-1}-(a'-1)-1\},
\label{eqn:proof of nonadjacent distance larger than one II-(i)-(a)-case-2-444}\\
\alignspace n_{(a'-1)-j'}=n_{a'+j'}, \textrm{ for } j'=1,2,\ldots,j-2,
\label{eqn:proof of nonadjacent distance larger than one II-(i)-(a)-case-2-555}\\
\alignspace n_{(a'-1)-(j-1)}>n_{a'+(j-1)}.
\label{eqn:proof of nonadjacent distance larger than one II-(i)-(a)-case-2-666}
\eeqnarray
An illustration of
\reqnarray{proof of nonadjacent distance larger than one II-(i)-(a)-case-2-555}
and \reqnarray{proof of nonadjacent distance larger than one II-(i)-(a)-case-2-666}
is given in \rfigure{appendix-I-(i)-(a)-case-2}.
Therefore, it follows from
$\nbf_1^{r_{h-1}}\in \Ncal_{M,k}(h)$ in \reqnarray{proof of nonadjacent distance larger than one II-111},
\reqnarray{proof of nonadjacent distance larger than one II-(i)-(a)-case-2-111}--\reqnarray{proof of nonadjacent distance larger than one II-(i)-(a)-case-2-666},
and \reqnarray{comparison rule B-2} in \rlemma{comparison rule B}(ii) that
$\nbf_1^{r_{h-1}}\prec \mbf_1^{r_{h-1}}$,
i.e., \reqnarray{proof of nonadjacent distance larger than one II-(i)-444} holds
with ${\nbf'}_1^{r_{h-1}}=\mbf_1^{r_{h-1}}$.

From $2\leq j\leq \min\{a'-1,b'-a'-1\}$,
\reqnarray{proof of nonadjacent distance larger than one II-(i)-(a)-222},
and $b'\leq r_{h-1}$, we can see that
\beqnarray{}
\alignspace
2\leq \min\{a'-1,b'-a'-1\}\leq a'-1<a'\leq r_{h-1}-2,
\label{eqn:proof of nonadjacent distance larger than one II-(i)-(a)-case-2-777}\\
\alignspace
1\leq j-1\leq \min\{a'-2,b'-a'-2\}\leq \min\{(a'-1)-1,r_{h-1}-(a'-1)-1\}.
\label{eqn:proof of nonadjacent distance larger than one II-(i)-(a)-case-2-888}
\eeqnarray
Thus, \reqnarray{proof of nonadjacent distance larger than one II-(i)-(a)-case-2-333}
follows from \reqnarray{proof of nonadjacent distance larger than one II-(i)-(a)-case-2-777},
and \reqnarray{proof of nonadjacent distance larger than one II-(i)-(a)-case-2-444}
follows from \reqnarray{proof of nonadjacent distance larger than one II-(i)-(a)-case-2-888}.

To prove \reqnarray{proof of nonadjacent distance larger than one II-(i)-(a)-case-2-555}
and \reqnarray{proof of nonadjacent distance larger than one II-(i)-(a)-case-2-666},
note that as we have $n_{a'-j'}=p+1$ for $j'=1,2,\ldots,j-1$ and $n_{a'-j}>p+1$ in this case,
it is clear that
\beqnarray{}
\alignspace n_{(a'-1)-j'}=p+1, \textrm{ for } j'=1,2,\ldots,j-2,
\label{eqn:proof of nonadjacent distance larger than one II-(i)-(a)-case-2-999}\\
\alignspace n_{(a'-1)-(j-1)}=n_{a'-j}>p+1.
\label{eqn:proof of nonadjacent distance larger than one II-(i)-(a)-case-2-aaa}
\eeqnarray
Furthermore, we have from $2\leq j\leq \min\{a'-1,b'-a'-1\}$ that
\beqnarray{proof of nonadjacent distance larger than one II-(i)-(a)-case-2-bbb}
a'< a'+j-1\leq a'+(b'-a'-1)-1=b'-2<b'.
\eeqnarray
It then follows from \reqnarray{proof of nonadjacent distance larger than one II-(i)-222}
and \reqnarray{proof of nonadjacent distance larger than one II-(i)-(a)-case-2-bbb} that
\beqnarray{proof of nonadjacent distance larger than one II-(i)-(a)-case-2-ccc}
n_{a'+j'}=p+1, \textrm{ for } j'=1,2,\ldots,j-1.
\eeqnarray
By combining \reqnarray{proof of nonadjacent distance larger than one II-(i)-(a)-case-2-999},
\reqnarray{proof of nonadjacent distance larger than one II-(i)-(a)-case-2-aaa},
and \reqnarray{proof of nonadjacent distance larger than one II-(i)-(a)-case-2-ccc},
we obtain \reqnarray{proof of nonadjacent distance larger than one II-(i)-(a)-case-2-555}
and \reqnarray{proof of nonadjacent distance larger than one II-(i)-(a)-case-2-666}.

\emph{Case 3: $n_{a'-j'}=p+1$ for $j'=1,2,\ldots,\min\{a'-1,b'-a'-1\}$.}
We consider the two subcases $a'-1<b'-a'-1$ and $a'-1\geq b'-a'-1$ separately.

\emph{Subcase 3(a): $a'-1<b'-a'-1$.}
Let $\mbf_1^{r_{h-1}}$ be a sequence of positive integers as given in
\reqnarray{proof of nonadjacent distance larger than one II-(i)-(a)-case-2-222}.
As in Case~2 above, we have $\mbf_1^{r_{h-1}}\in \Ncal_{M,k}(h)$.
As it is clear from $a'\geq 2$ and $b'\geq a'+2$
in \reqnarray{proof of nonadjacent distance larger than one II-(i)-111}
that $\min\{a'-1,b'-a'-1\}\geq 1$,
we have $n_{a'-1}=p+1$ in this case and hence it is easy to see that
\reqnarray{proof of nonadjacent distance larger than one II-(i)-(a)-case-2-111}
still holds in this subcase.

If $a'=2$, then we have $a'-1=1$ and
it follows from $r_{h-1}\geq 3$
in \reqnarray{proof of nonadjacent distance larger than one II-111},
$\nbf_1^{r_{h-1}}\in \Ncal_{M,k}(h)$ in \reqnarray{proof of nonadjacent distance larger than one II-111},
\reqnarray{proof of nonadjacent distance larger than one II-(i)-(a)-case-2-111},
\reqnarray{proof of nonadjacent distance larger than one II-(i)-(a)-case-2-222},
and \reqnarray{comparison rule B-1} in \rlemma{comparison rule B}(i)
that $\nbf_1^{r_{h-1}}\prec \mbf_1^{r_{h-1}}$,
i.e., \reqnarray{proof of nonadjacent distance larger than one II-(i)-444} holds
with ${\nbf'}_1^{r_{h-1}}=\mbf_1^{r_{h-1}}$.

\bpdffigure{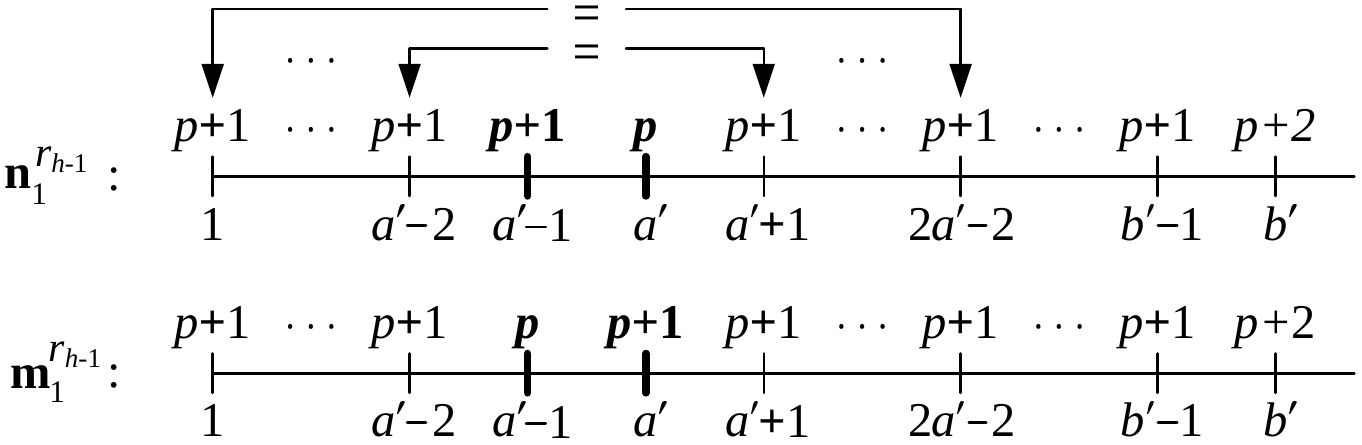}{4.5in}
\epdffigure{appendix-I-(i)-(a)-case-3-1}
{An illustration of \reqnarray{proof of nonadjacent distance larger than one II-(i)-(a)-case-3-222}
(note that we have $\min\{(a'-1)-1, r_{h-1}-(a'-1)-1\}=a'-2$
in \reqnarray{proof of nonadjacent distance larger than one II-(i)-(a)-case-3-555}).}

On the other hand, if $a'\geq 3$, then we show that
\beqnarray{}
\alignspace \hspace*{-0.2in}
2\leq a'-1\leq r_{h-1}-2,
\label{eqn:proof of nonadjacent distance larger than one II-(i)-(a)-case-3-111}\\
\alignspace \hspace*{-0.2in}
n_{(a'-1)-j'}=n_{a'+j'}=p+1, \textrm{ for } j'=1,2,\ldots,\min\{(a'-1)-1, r_{h-1}-(a'-1)-1\}.
\label{eqn:proof of nonadjacent distance larger than one II-(i)-(a)-case-3-222}
\eeqnarray
An illustration of \reqnarray{proof of nonadjacent distance larger than one II-(i)-(a)-case-3-222}
is given in \rfigure{appendix-I-(i)-(a)-case-3-1}.
Therefore, it follows from
$\nbf_1^{r_{h-1}}\in \Ncal_{M,k}(h)$ in \reqnarray{proof of nonadjacent distance larger than one II-111},
\reqnarray{proof of nonadjacent distance larger than one II-(i)-(a)-case-2-111},
\reqnarray{proof of nonadjacent distance larger than one II-(i)-(a)-case-2-222},
\reqnarray{proof of nonadjacent distance larger than one II-(i)-(a)-case-3-111},
\reqnarray{proof of nonadjacent distance larger than one II-(i)-(a)-case-3-222},
and \reqnarray{comparison rule B-4} in \rlemma{comparison rule B}(iii) that
$\nbf_1^{r_{h-1}}\prec \mbf_1^{r_{h-1}}$,
i.e., \reqnarray{proof of nonadjacent distance larger than one II-(i)-444} holds
with ${\nbf'}_1^{r_{h-1}}=\mbf_1^{r_{h-1}}$.

From $a'\geq 3$ and \reqnarray{proof of nonadjacent distance larger than one II-(i)-(a)-222},
we see that
\beqnarray{proof of nonadjacent distance larger than one II-(i)-(a)-case-3-333}
2\leq a'-1<a'\leq r_{h-1}-2.
\eeqnarray
Thus, \reqnarray{proof of nonadjacent distance larger than one II-(i)-(a)-case-3-111}
follows from \reqnarray{proof of nonadjacent distance larger than one II-(i)-(a)-case-3-333}.

To prove \reqnarray{proof of nonadjacent distance larger than one II-(i)-(a)-case-3-222},
note that from $a'-1<b'-a'-1$ in this subcase and $b'\leq r_{h-1}$, we have
\beqnarray{proof of nonadjacent distance larger than one II-(i)-(a)-case-3-444}
(a'-1)-1<(b'-a'-1)-1\leq r_{h-1}-a'-2<r_{h-1}-(a'-1)-1.
\eeqnarray
It follows from \reqnarray{proof of nonadjacent distance larger than one II-(i)-(a)-case-3-444} that
\beqnarray{proof of nonadjacent distance larger than one II-(i)-(a)-case-3-555}
\min\{(a'-1)-1, r_{h-1}-(a'-1)-1\}=(a'-1)-1=a'-2.
\eeqnarray
Note that this subcase we have
\beqnarray{proof of nonadjacent distance larger than one II-(i)-(a)-case-3-666}
n_{a'-j'}=p+1, \textrm{ for } j'=1,2,\ldots,\min\{a'-1,b'-a'-1\}=a'-1.
\eeqnarray
From $a'\geq 3$ and $a'-1<b'-a'-1$, we have that
\beqnarray{proof of nonadjacent distance larger than one II-(i)-(a)-case-3-777}
a'<a'+(a'-2)<a'+(b'-a'-2)=b'-2<b'.
\eeqnarray
It then follows from \reqnarray{proof of nonadjacent distance larger than one II-(i)-222}
and \reqnarray{proof of nonadjacent distance larger than one II-(i)-(a)-case-3-777} that
\beqnarray{proof of nonadjacent distance larger than one II-(i)-(a)-case-3-888}
n_{a'+j'}=p+1, \textrm{ for } j'=1,2,\ldots,a'-2.
\eeqnarray
by combining
\reqnarray{proof of nonadjacent distance larger than one II-(i)-(a)-case-3-555},
\reqnarray{proof of nonadjacent distance larger than one II-(i)-(a)-case-3-666},
and \reqnarray{proof of nonadjacent distance larger than one II-(i)-(a)-case-3-888},
we obtain \reqnarray{proof of nonadjacent distance larger than one II-(i)-(a)-case-3-222}.

\emph{Subcase 3(b): $a'-1\geq b'-a'-1$.}
Let $\mbf_1^{r_{h-1}}$ be a sequence of positive integers as given in
\reqnarray{proof of nonadjacent distance larger than one II-(i)-(a)-case-1-111}.
As in Case~1 above, we have $\mbf_1^{r_{h-1}}\in \Ncal_{M,k}(h)$.
Also note that \reqnarray{proof of nonadjacent distance larger than one II-(i)-(a)-case-1-222} still holds in this subcase.

\bpdffigure{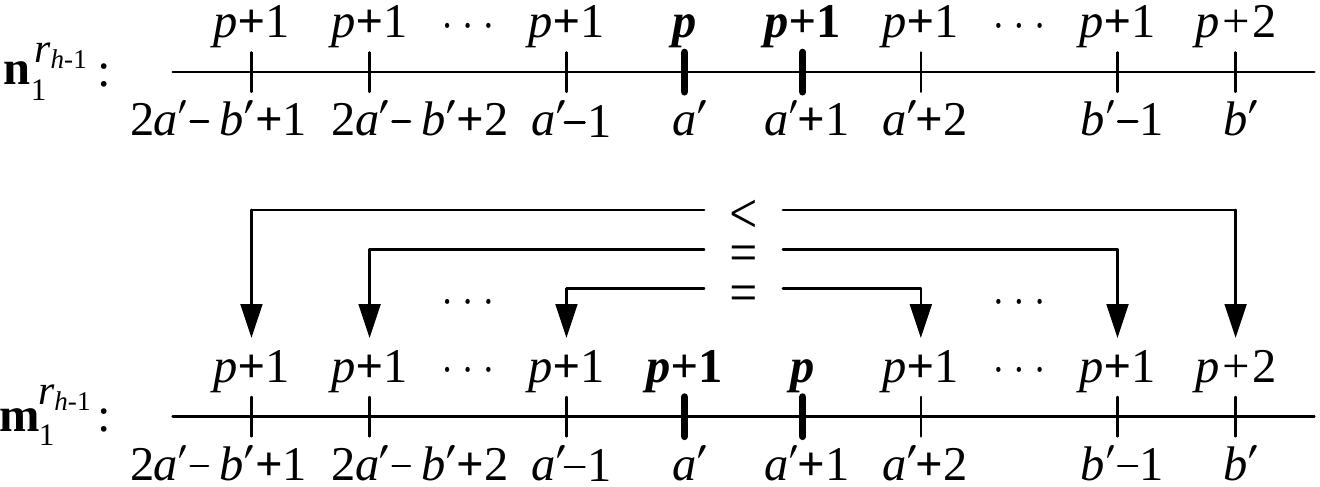}{4.5in}
\epdffigure{appendix-I-(i)-(a)-case-3-2}
{An illustration of \reqnarray{proof of nonadjacent distance larger than one II-(i)-(a)-case-3-aaa}
and \reqnarray{proof of nonadjacent distance larger than one II-(i)-(a)-case-3-bbb}.}

In the following, we show that
\beqnarray{}
\alignspace 1\leq b'-a'-1\leq \min\{a'-1,r_{h-1}-a'-1\}
\label{eqn:proof of nonadjacent distance larger than one II-(i)-(a)-case-3-999}\\
\alignspace m_{a'-j'}=m_{(a'+1)+j'}=p+1, \textrm{ for } j'=1,2,\ldots,b'-a'-2,
\label{eqn:proof of nonadjacent distance larger than one II-(i)-(a)-case-3-aaa}\\
\alignspace m_{a'-(b'-a'-1)}=p+1<m_{(a'+1)+(b'-a'-1)}=p+2.
\label{eqn:proof of nonadjacent distance larger than one II-(i)-(a)-case-3-bbb}
\eeqnarray
An illustration of
\reqnarray{proof of nonadjacent distance larger than one II-(i)-(a)-case-3-aaa}
and \reqnarray{proof of nonadjacent distance larger than one II-(i)-(a)-case-3-bbb}
is given in \rfigure{appendix-I-(i)-(a)-case-3-2}.
Therefore, it follows from
$\mbf_1^{r_{h-1}}\in \Ncal_{M,k}(h)$,
\reqnarray{proof of nonadjacent distance larger than one II-(i)-(a)-222}--\reqnarray{proof of nonadjacent distance larger than one II-(i)-(a)-case-1-222},
\reqnarray{proof of nonadjacent distance larger than one II-(i)-(a)-case-3-999}--\reqnarray{proof of nonadjacent distance larger than one II-(i)-(a)-case-3-bbb},
and \reqnarray{comparison rule B-3} in \rlemma{comparison rule B}(ii)
(with $j=b'-a'-1$) that
\beqnarray{proof of nonadjacent distance larger than one II-(i)-(a)-case-3-ccc}
\mbf_1^{r_{h-1}}\succeq \nbf_1^{r_{h-1}}.
\eeqnarray

From $b'\geq a'+2$ in \reqnarray{proof of nonadjacent distance larger than one II-(i)-111},
$a'-1\geq b'-a'-1$, and $b'\leq r_{h-1}$, we can see that
\beqnarray{proof of nonadjacent distance larger than one II-(i)-(a)-case-3-ddd}
1\leq b'-a'-1\leq \min\{a'-1,b'-a'-1\}\leq \min\{a'-1,r_{h-1}-a'-1\}.
\eeqnarray
Thus, \reqnarray{proof of nonadjacent distance larger than one II-(i)-(a)-case-3-999}
follows from \reqnarray{proof of nonadjacent distance larger than one II-(i)-(a)-case-3-ddd}.

To prove \reqnarray{proof of nonadjacent distance larger than one II-(i)-(a)-case-3-aaa}
and \reqnarray{proof of nonadjacent distance larger than one II-(i)-(a)-case-3-bbb},
note that in this case we have
\beqnarray{proof of nonadjacent distance larger than one II-(i)-(a)-case-3-eee}
n_{a'-j'}=p+1, \textrm{ for } j'=1,2,\ldots,\min\{a'-1,b'-a'-1\}=b'-a'-1.
\eeqnarray
It follows from \reqnarray{proof of nonadjacent distance larger than one II-(i)-(a)-case-1-111}
and \reqnarray{proof of nonadjacent distance larger than one II-(i)-(a)-case-3-eee} that
\beqnarray{proof of nonadjacent distance larger than one II-(i)-(a)-case-3-fff}
m_{a'-j'}=n_{a'-j'}=p+1, \textrm{ for } j'=1,2,\ldots,b'-a'-1.
\eeqnarray
Also, it is clear from \reqnarray{proof of nonadjacent distance larger than one II-(i)-(a)-case-1-111},
\reqnarray{proof of nonadjacent distance larger than one II-(i)-222},
$(a'+1)+(b'-a'-1)=b'$,
and $n_{b'}=p+2$ in \reqnarray{proof of nonadjacent distance larger than one II-(i)-222} that
\beqnarray{}
\alignspace m_{(a'+1)+j'}=n_{(a'+1)+j'}=p+1, \textrm{ for } j'=1,2,\ldots,b'-a'-2,
\label{eqn:proof of nonadjacent distance larger than one II-(i)-(a)-case-3-ggg}\\
\alignspace m_{(a'+1)+(b'-a'-1)}=n_{(a'+1)+(b'-a'-1)}=n_{b'}=p+2.
\label{eqn:proof of nonadjacent distance larger than one II-(i)-(a)-case-3-hhh}
\eeqnarray
By combining
\reqnarray{proof of nonadjacent distance larger than one II-(i)-(a)-case-3-fff},
\reqnarray{proof of nonadjacent distance larger than one II-(i)-(a)-case-3-ggg},
and \reqnarray{proof of nonadjacent distance larger than one II-(i)-(a)-case-3-hhh},
we obtain \reqnarray{proof of nonadjacent distance larger than one II-(i)-(a)-case-3-aaa}
and \reqnarray{proof of nonadjacent distance larger than one II-(i)-(a)-case-3-bbb}.

Now let ${\mbf'}_1^{r_{h-1}}$ be a sequence of positive integers such that
\beqnarray{proof of nonadjacent distance larger than one II-(i)-(a)-case-3-iii}
m'_{a'+1}=m_{a'+1}+1,\ m'_{a'+2}=m_{a'+2}-1, \textrm{ and } m'_i=m_i \textrm{ for } i\neq a'+1, a'+2.
\eeqnarray
Again, it is easy to show that ${\mbf'}_1^{r_{h-1}}\in \Ncal_{M,k}(h)$.

If $b'=a'+2$, then we see from
\reqnarray{proof of nonadjacent distance larger than one II-(i)-(a)-case-1-111},
$n_{a'+1}=p+1$ in \reqnarray{proof of nonadjacent distance larger than one II-(i)-333},
and $n_{b'}=p+2$ in \reqnarray{proof of nonadjacent distance larger than one II-(i)-222} that
\beqnarray{proof of nonadjacent distance larger than one II-(i)-(a)-case-3-jjj}
m_{a'+1}-m_{a'+2}
\aligneq (n_{a'+1}-1)-n_{a'+2}=n_{a'+1}-1-n_{b'} \nn\\
\aligneq (p+1)-1-(p+2)=-2.
\eeqnarray
Therefore, it follows from
$r_{h-1}\geq 3$ in \reqnarray{proof of nonadjacent distance larger than one II-111},
$\mbf_1^{r_{h-1}}\in \Ncal_{M,k}(h)$,
\reqnarray{proof of nonadjacent distance larger than one II-(i)-(a)-case-3-iii},
\reqnarray{proof of nonadjacent distance larger than one II-(i)-(a)-case-3-jjj},
and \reqnarray{adjacent distance larger than one II-2} in \rlemma{adjacent distance larger than one II}(ii) that
\beqnarray{proof of nonadjacent distance larger than one II-(i)-(a)-case-3-kkk}
\mbf_1^{r_{h-1}}\preceq {\mbf'}_1^{r_{h-1}},
\eeqnarray
where $\mbf_1^{r_{h-1}}\equiv {\mbf'}_1^{r_{h-1}}$
if and only if $r_{h-1}=2$ and $m_1=m_2-2$.
Since it is clear from $r_{h-1}\geq 3$ in
\reqnarray{proof of nonadjacent distance larger than one II-111}
that $r_{h-1}\neq 2$, it cannot be the case that $\mbf_1^{r_{h-1}}\equiv {\mbf'}_1^{r_{h-1}}$.
As such, we see from \reqnarray{proof of nonadjacent distance larger than one II-(i)-(a)-case-3-ccc}
and \reqnarray{proof of nonadjacent distance larger than one II-(i)-(a)-case-3-kkk}
that $\nbf_1^{r_{h-1}}\preceq \mbf_1^{r_{h-1}}\prec {\mbf'}_1^{r_{h-1}}$,
i.e., \reqnarray{proof of nonadjacent distance larger than one II-(i)-444} holds
with ${\nbf'}_1^{r_{h-1}}={\mbf'}_1^{r_{h-1}}$.

On the other hand, if $b'\geq a'+3$, then we have $a'<a'+2<b'$
and it follows from \reqnarray{proof of nonadjacent distance larger than one II-(i)-222} that
\beqnarray{proof of nonadjacent distance larger than one II-(i)-(a)-case-3-1111}
n_{a'+1}=n_{a'+2}=p+1.
\eeqnarray
From \reqnarray{proof of nonadjacent distance larger than one II-(i)-(a)-case-3-iii},
\reqnarray{proof of nonadjacent distance larger than one II-(i)-(a)-case-1-111},
and \reqnarray{proof of nonadjacent distance larger than one II-(i)-(a)-case-3-1111},
we see that
\beqnarray{proof of nonadjacent distance larger than one II-(i)-(a)-case-3-2222}
m'_{a'+1}-m'_{a'+2}
\aligneq (m_{a'+1}+1)-(m_{a'+2}-1)=m_{a'+1}-m_{a'+2}+2 \nn\\
\aligneq (n_{a'+1}-1)-n_{a'+2}+2=1.
\eeqnarray

\bpdffigure{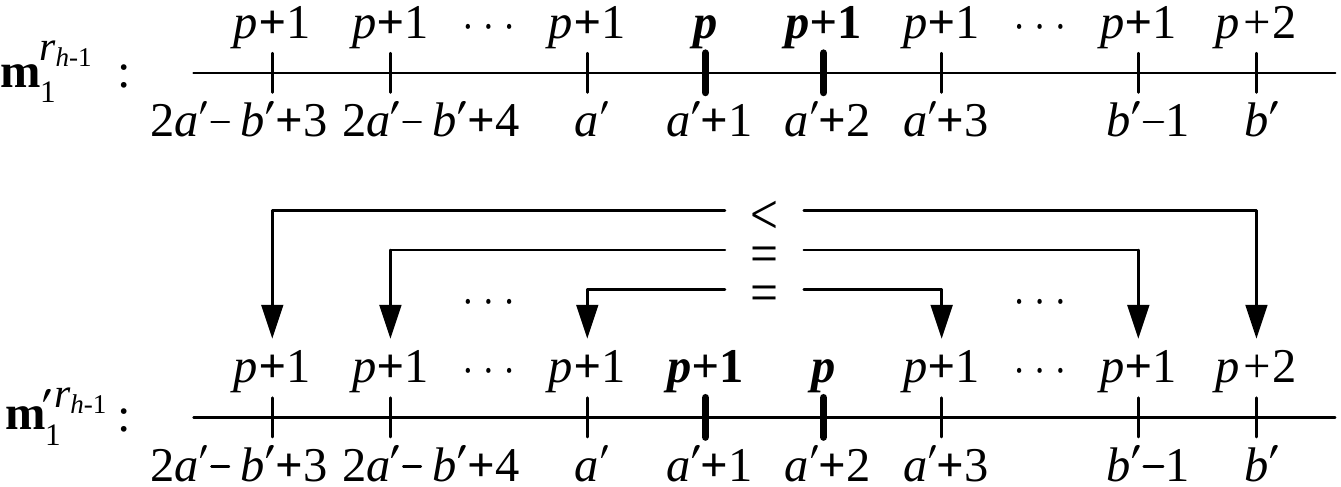}{4.5in}
\epdffigure{appendix-I-(i)-(a)-case-3-3}
{An illustration of \reqnarray{proof of nonadjacent distance larger than one II-(i)-(a)-case-3-5555}
and \reqnarray{proof of nonadjacent distance larger than one II-(i)-(a)-case-3-6666}.}

We will show that
\beqnarray{}
\alignspace
2\leq a'+1\leq r_{h-1}-2,
\label{eqn:proof of nonadjacent distance larger than one II-(i)-(a)-case-3-3333}\\
\alignspace
1\leq b'-a'-2\leq \min\{(a'+1)-1,r_{h-1}-(a'+1)-1\},
\label{eqn:proof of nonadjacent distance larger than one II-(i)-(a)-case-3-4444}\\
\alignspace
m'_{(a'+1)-j'}=m'_{(a'+2)+j'}=p+1, \textrm{ for } j'=1,2,\ldots,b'-a'-3,
\label{eqn:proof of nonadjacent distance larger than one II-(i)-(a)-case-3-5555}\\
\alignspace
m'_{(a'+1)-(b'-a'-2)}=p+1<m'_{(a'+2)+(b'-a'-2)}=p+2.
\label{eqn:proof of nonadjacent distance larger than one II-(i)-(a)-case-3-6666}
\eeqnarray
An illustration of
\reqnarray{proof of nonadjacent distance larger than one II-(i)-(a)-case-3-5555}
and \reqnarray{proof of nonadjacent distance larger than one II-(i)-(a)-case-3-6666}
is given in \rfigure{appendix-I-(i)-(a)-case-3-3}.
Therefore, it follows from
${\mbf'}_1^{r_{h-1}}\in \Ncal_{M,k}(h)$,
\reqnarray{proof of nonadjacent distance larger than one II-(i)-(a)-case-3-iii},
\reqnarray{proof of nonadjacent distance larger than one II-(i)-(a)-case-3-2222}--\reqnarray{proof of nonadjacent distance larger than one II-(i)-(a)-case-3-6666},
and \reqnarray{comparison rule B-3} in \rlemma{comparison rule B}(ii)
(with $j=b'-a'-2$) that
\beqnarray{proof of nonadjacent distance larger than one- II(i)-(a)-case-3-7777}
{\mbf'}_1^{r_{h-1}}\succeq \mbf_1^{r_{h-1}},
\eeqnarray
where ${\mbf'}_1^{r_{h-1}}\equiv \mbf_1^{r_{h-1}}$
if and only if $(a'+1)-(b'-a'-2)=1$, $(a'+1)+1+(b'-a'-2)=r_{h-1}$, and $m'_1=m'_{r_{h-1}}-1$.
Since in this subcase we have $a'-1\geq b'-a'-1$,
it is clear that $(a'+1)-(b'-a'-2)\geq (b'-a'+1)-(b'-a'-2)=3$.
This implies that $(a'+1)-(b'-a'-2)\neq 1$
and hence it cannot be the case that ${\mbf'}_1^{r_{h-1}}\equiv \mbf_1^{r_{h-1}}$.
As such, we see from \reqnarray{proof of nonadjacent distance larger than one II-(i)-(a)-case-3-ccc}
and \reqnarray{proof of nonadjacent distance larger than one- II(i)-(a)-case-3-7777}
that $\nbf_1^{r_{h-1}}\preceq \mbf_1^{r_{h-1}}\prec {\mbf'}_1^{r_{h-1}}$,
i.e., \reqnarray{proof of nonadjacent distance larger than one II-(i)-444} holds
with ${\nbf'}_1^{r_{h-1}}={\mbf'}_1^{r_{h-1}}$.

To prove \reqnarray{proof of nonadjacent distance larger than one II-(i)-(a)-case-3-3333}
and \reqnarray{proof of nonadjacent distance larger than one II-(i)-(a)-case-3-4444},
note from $a'\geq 2$, $b'\geq a'+3$, $b'\leq r_{h-1}$,
and \reqnarray{proof of nonadjacent distance larger than one II-(i)-(a)-case-3-999} that
\beqnarray{}
\alignspace \hspace*{-0.3in}
2\leq a'<a'+1\leq b'-2\leq r_{h-1}-2,
\label{eqn:proof of nonadjacent distance larger than one II-(i)-(a)-case-3-8888}\\
\alignspace \hspace*{-0.3in}
1\leq b'-a'-2\leq \min\{a'-2,r_{h-1}-a'-2\}\leq \min\{(a'+1)-1,r_{h-1}-(a'+1)-1\}.
\label{eqn:proof of nonadjacent distance larger than one II-(i)-(a)-case-3-9999}
\eeqnarray
Thus, \reqnarray{proof of nonadjacent distance larger than one II-(i)-(a)-case-3-3333}
follows from \reqnarray{proof of nonadjacent distance larger than one II-(i)-(a)-case-3-8888},
and \reqnarray{proof of nonadjacent distance larger than one II-(i)-(a)-case-3-4444}
follows from \reqnarray{proof of nonadjacent distance larger than one II-(i)-(a)-case-3-9999}.

To prove \reqnarray{proof of nonadjacent distance larger than one II-(i)-(a)-case-3-5555}
and \reqnarray{proof of nonadjacent distance larger than one II-(i)-(a)-case-3-6666},
note that from \reqnarray{proof of nonadjacent distance larger than one II-(i)-(a)-case-3-iii},
\reqnarray{proof of nonadjacent distance larger than one II-(i)-(a)-case-1-111},
$n_{a'}=p$ in \reqnarray{proof of nonadjacent distance larger than one II-(i)-222},
and \reqnarray{proof of nonadjacent distance larger than one II-(i)-(a)-case-3-fff}--\reqnarray{proof of nonadjacent distance larger than one II-(i)-(a)-case-3-hhh},
we have
\beqnarray{}
\alignspace m'_{(a'+1)-1}=m_{(a'+1)-1}=m_{a'}=n_{a'}+1=p+1,
\label{eqn:proof of nonadjacent distance larger than one II-(i)-(a)-case-3-aaaa}\\
\alignspace m'_{(a'+1)-j'}=m_{(a'+1)-j'}=p+1, \textrm{ for } j'=2,3,\ldots,b'-a',
\label{eqn:proof of nonadjacent distance larger than one II-(i)-(a)-case-3-bbbb}\\
\alignspace m'_{(a'+2)+j'}=m_{(a'+2)+j'}=p+1, \textrm{ for } j'=1,2,\ldots,b'-a'-3,
\label{eqn:proof of nonadjacent distance larger than one II-(i)-(a)-case-3-cccc}\\
\alignspace m'_{(a'+2)+(b'-a'-2)}=m_{(a'+2)+(b'-a'-2)}=p+2.
\label{eqn:proof of nonadjacent distance larger than one II-(i)-(a)-case-3-dddd}
\eeqnarray
Thus, \reqnarray{proof of nonadjacent distance larger than one II-(i)-(a)-case-3-5555}
and \reqnarray{proof of nonadjacent distance larger than one II-(i)-(a)-case-3-6666}
follow from
\reqnarray{proof of nonadjacent distance larger than one II-(i)-(a)-case-3-aaaa}--\reqnarray{proof of nonadjacent distance larger than one II-(i)-(a)-case-3-dddd}.

(b) Now we assume that $b'\leq r_{h-1}-1$ and show that there exists a sequence of positive integers
${\nbf'}_1^{r_{h-1}}\in \Ncal_{M,k}(h)$ such that
\reqnarray{proof of nonadjacent distance larger than one II-(i)-444} holds.
Note that from $b'\geq a'+2$
in \reqnarray{proof of nonadjacent distance larger than one II-(i)-111}
and $a'\geq 1$, we have
\beqnarray{proof of nonadjacent distance larger than one II-(i)-(b)-111}
b'-1\geq a'+1\geq 2.
\eeqnarray
As we assume that $b'\leq r_{h-1}-1$,
it follows from \reqnarray{proof of nonadjacent distance larger than one II-(i)-(b)-111} that
\beqnarray{proof of nonadjacent distance larger than one II-(i)-(b)-222}
2\leq b'-1\leq r_{h-1}-2.
\eeqnarray

We need to consider the following three possible cases.

\emph{Case 1: There exists a positive integer $j$ such that
$1\leq j\leq \min\{b'-a'-1,r_{h-1}-b'\}$,
$n_{b'+j'}=p+1$ for $j'=1,2,\ldots,j-1$, and $n_{b'+j}>p+1$.}
Let $\mbf_1^{r_{h-1}}$ be a sequence of positive integers such that
\beqnarray{proof of nonadjacent distance larger than one II-(i)-(b)-case-1-111}
m_{b'-1}=n_{b'-1}+1,\ m_{b'}=n_{b'}-1, \textrm{ and } m_i=n_i \textrm{ for } i\neq b'-1, b'.
\eeqnarray
As before, it is easy to show that $\mbf_1^{r_{h-1}}\in \Ncal_{M,k}(h)$.
Also, noth that from \reqnarray{proof of nonadjacent distance larger than one II-(i)-(b)-case-1-111},
$n_{b'-1}=p+1$ in \reqnarray{proof of nonadjacent distance larger than one II-(i)-333},
and $n_{b'}=p+2$ in \reqnarray{proof of nonadjacent distance larger than one II-(i)-222},
we have
\beqnarray{proof of nonadjacent distance larger than one II-(i)-(b)-case-1-222}
m_{b'-1}-m_{b'}=(n_{b'-1}+1)-(n_{b'}-1)=(p+1+1)-(p+2-1)=1
\eeqnarray
As we have $j\leq \min\{b'-a'-1,r_{h-1}-b'\}\leq b'-a'-1$,
we consider the two subcases $j<b'-a'-1$ and $j=b'-a'-1$ separately.

\emph{Subcase 1(a): $j<b'-a'-1$.}

\bpdffigure{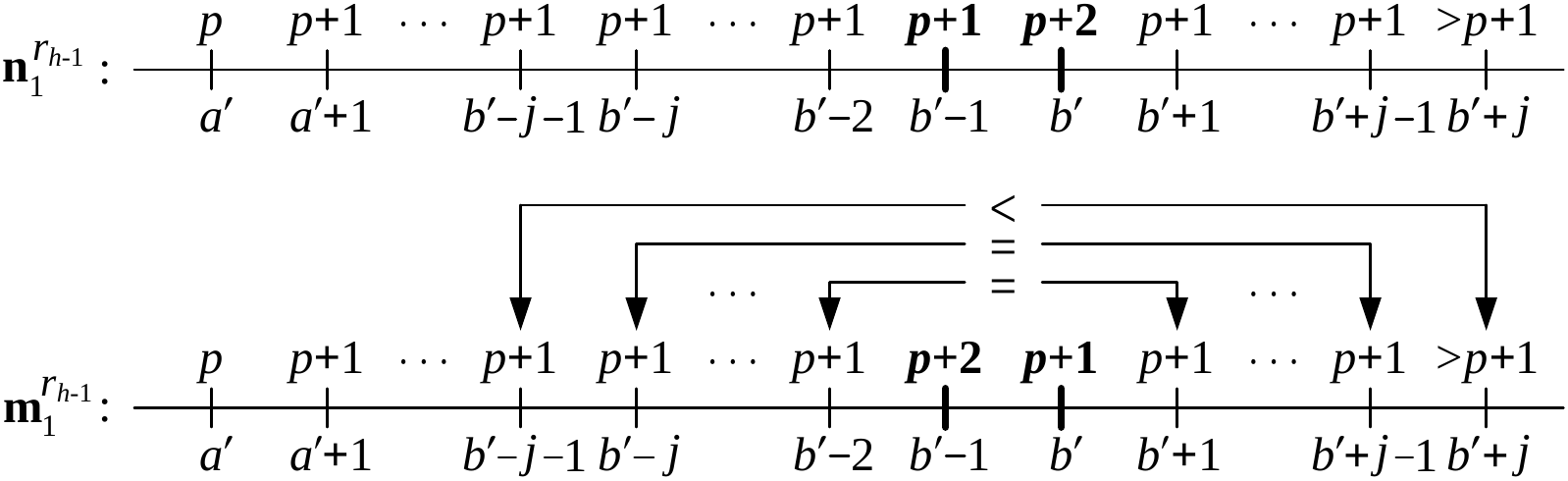}{5.5in}
\epdffigure{appendix-I-(i)-(b)-case-1}
{An illustration of \reqnarray{proof of nonadjacent distance larger than one II-(i)-(b)-case-1-444}
and \reqnarray{proof of nonadjacent distance larger than one II-(i)-(b)-case-1-555}.}

In this subcase, we show that
\beqnarray{}
\alignspace 1\leq j\leq \min\{(b'-1)-1,r_{h-1}-(b'-1)-1\}
\label{eqn:proof of nonadjacent distance larger than one II-(i)-(b)-case-1-333}\\
\alignspace m_{(b'-1)-j'}=m_{b'+j'}, \textrm{ for } j'=1,2,\ldots,j-1,
\label{eqn:proof of nonadjacent distance larger than one II-(i)-(b)-case-1-444}\\
\alignspace m_{(b'-1)-j}<m_{b'+j}.
\label{eqn:proof of nonadjacent distance larger than one II-(i)-(b)-case-1-555}
\eeqnarray
An illustration of
\reqnarray{proof of nonadjacent distance larger than one II-(i)-(b)-case-1-444}
and \reqnarray{proof of nonadjacent distance larger than one II-(i)-(b)-case-1-555}
is given in \rfigure{appendix-I-(i)-(b)-case-1}.
Therefore, it follows from
$\mbf_1^{r_{h-1}}\in \Ncal_{M,k}(h)$,
\reqnarray{proof of nonadjacent distance larger than one II-(i)-(b)-222}--\reqnarray{proof of nonadjacent distance larger than one II-(i)-(b)-case-1-555},
and \reqnarray{comparison rule B-3} in \rlemma{comparison rule B}(ii) that
\beqnarray{proof of nonadjacent distance larger than one II-(i)-(b)-case-1-666}
\mbf_1^{r_{h-1}}\succeq \nbf_1^{r_{h-1}},
\eeqnarray
where $\mbf_1^{r_{h-1}}\equiv \nbf_1^{r_{h-1}}$
if and only if $(b'-1)-j=1$, $b'+j=r_{h-1}$,
and $n_1=n_{r_{h-1}}-1$ (i.e., $n_{(b'-1)-j}=n_{b'+j}-1$).
As we have $j<b'-a'-1$ in this subcase and $a'\geq 1$,
we immediately see that $(b'-1)-j>a'\geq 1$.
This implies that $(b'-1)-j\neq 1$ and hence it cannot be the case that
$\mbf_1^{r_{h-1}}\equiv \nbf_1^{r_{h-1}}$.
As such, we see from \reqnarray{proof of nonadjacent distance larger than one II-(i)-(b)-case-1-666}
that $\mbf_1^{r_{h-1}}\succ \nbf_1^{r_{h-1}}$,
i.e., \reqnarray{proof of nonadjacent distance larger than one II-(i)-444} holds
with ${\nbf'}_1^{r_{h-1}}=\mbf_1^{r_{h-1}}$.

From $1\leq j\leq \min\{b'-a'-1,r_{h-1}-b'\}$ and $a'\geq 1$, we see that
\beqnarray{proof of nonadjacent distance larger than one II-(i)-(b)-case-1-777}
1\leq j\leq \min\{b'-a'-1,r_{h-1}-b'\}\leq \min\{(b'-1)-1,r_{h-1}-(b'-1)-1\}.
\eeqnarray
Thus, \reqnarray{proof of nonadjacent distance larger than one II-(i)-(b)-case-1-333}
follows from \reqnarray{proof of nonadjacent distance larger than one II-(i)-(b)-case-1-777}.

To prove \reqnarray{proof of nonadjacent distance larger than one II-(i)-(b)-case-1-444}
and \reqnarray{proof of nonadjacent distance larger than one II-(i)-(b)-case-1-555},
note that in this subcase we have $a'<(b'-1)-j<b'$,
and hence it follows from \reqnarray{proof of nonadjacent distance larger than one II-(i)-(b)-case-1-111}
and \reqnarray{proof of nonadjacent distance larger than one II-(i)-222} that
\beqnarray{proof of nonadjacent distance larger than one II-(i)-(b)-case-1-888}
m_{(b'-1)-j'}=n_{(b'-1)-j'}=p+1, \textrm{ for } j'=1,2,\ldots,j.
\eeqnarray
Also, as we have $n_{b'+j'}=p+1$ for $j'=1,2,\ldots,j-1$ and $n_{b'+j}>p+1$ in this case,
it follows from \reqnarray{proof of nonadjacent distance larger than one II-(i)-(b)-case-1-111} that
\beqnarray{}
\alignspace m_{b'+j'}=n_{b'+j'}=p+1, \textrm{ for } j'=1,2,\ldots,j-1,
\label{eqn:proof of nonadjacent distance larger than one II-(i)-(b)-case-1-999}\\
\alignspace m_{b'+j}=n_{b'+j}>p+1.
\label{eqn:proof of nonadjacent distance larger than one II-(i)-(b)-case-1-aaa}
\eeqnarray
By combining \reqnarray{proof of nonadjacent distance larger than one II-(i)-(b)-case-1-888},
\reqnarray{proof of nonadjacent distance larger than one II-(i)-(b)-case-1-999},
and \reqnarray{proof of nonadjacent distance larger than one II-(i)-(b)-case-1-aaa},
we obtain \reqnarray{proof of nonadjacent distance larger than one II-(i)-(b)-case-1-444}
and \reqnarray{proof of nonadjacent distance larger than one II-(i)-(b)-case-1-555}.

\emph{Subcase 1(b): $j=b'-a'-1$.}
In this subcase, we have $(b'-1)-j=a'$
and it follows from \reqnarray{proof of nonadjacent distance larger than one II-(i)-(b)-case-1-111}
and \reqnarray{proof of nonadjacent distance larger than one II-(i)-222} that
\beqnarray{}
\alignspace m_{(b'-1)-j'}=n_{(b'-1)-j'}=p+1, \textrm{ for } j'=1,2,\ldots,j-1,
\label{eqn:proof of nonadjacent distance larger than one II-(i)-(b)-case-1-bbb}\\
\alignspace m_{(b'-1)-j'}=n_{(b'-1)-j}=n_{a'}=p.
\label{eqn:proof of nonadjacent distance larger than one II-(i)-(b)-case-1-ccc}
\eeqnarray
By using \reqnarray{proof of nonadjacent distance larger than one II-(i)-(b)-case-1-bbb}
and \reqnarray{proof of nonadjacent distance larger than one II-(i)-(b)-case-1-ccc},
we can argue as in Subcase~1(a) above that
\reqnarray{proof of nonadjacent distance larger than one II-(i)-(b)-case-1-333}--\reqnarray{proof of nonadjacent distance larger than one II-(i)-(b)-case-1-666}
still hold.
Since it is clear from $m_{(b'-1)-j}=p$
in \reqnarray{proof of nonadjacent distance larger than one II-(i)-(b)-case-1-ccc}
and $m_{b'+j}>p+1$ in \reqnarray{proof of nonadjacent distance larger than one II-(i)-(b)-case-1-aaa}
that $m_{(b'-1)-j}\neq m_{b'+j}-1$,
it cannot be the case that $\mbf_1^{r_{h-1}}\equiv \nbf_1^{r_{h-1}}$.
As such, we see from \reqnarray{proof of nonadjacent distance larger than one II-(i)-(b)-case-1-666}
that $\mbf_1^{r_{h-1}}\succ \nbf_1^{r_{h-1}}$,
i.e., \reqnarray{proof of nonadjacent distance larger than one II-(i)-444} holds
with ${\nbf'}_1^{r_{h-1}}=\mbf_1^{r_{h-1}}$.

\emph{Case 2: There exists a positive integer $j$ such that
$1\leq j\leq \min\{b'-a'-1,r_{h-1}-b'\}$,
$n_{b'+j'}=p+1$ for $j'=1,2,\ldots,j-1$, and $n_{b'+j}<p+1$.}
In this case, we can show that $j\geq 2$.
To see this, suppose on the contrary that $j=1$,
then we have $n_{b'+1}<p+1$ in this case.
As it follows from $n_{b'}=p+2$ in \reqnarray{proof of nonadjacent distance larger than one II-(i)-222}
and the condition $|n_{i+1}-n_i|\leq 1$ for $i=1,2,\ldots,r_{h-1}-1$
in \reqnarray{proof of nonadjacent distance larger than one II-111}
that $n_{b'+1}$ must be equal to $p+1$, $p+2$, or $p+3$,
we have reached a contradiction.

Let $\mbf_1^{r_{h-1}}$ be a sequence of positive integers such that
\beqnarray{proof of nonadjacent distance larger than one II-(i)-(b)-case-2-111}
m_{b'}=n_{b'}-1,\ m_{b'+1}=n_{b'+1}+1,
\textrm{ and } m_i=n_i \textrm{ for } i\neq b', b'+1.
\eeqnarray
As before, it is easy to show that $\mbf_1^{r_{h-1}}\in \Ncal_{M,k}(h)$.
As $j\geq 2$, we have $n_{b'+1}=p+1$ in this case.
It then follows from $n_{b'+1}=p+1$ and $n_{b'}=p+2$ in \reqnarray{proof of nonadjacent distance larger than one II-(i)-222} that
\beqnarray{proof of nonadjacent distance larger than one II-(i)-(b)-case-2-222}
n_{b'}-n_{b'+1}=(p+2)-(p+1)=1.
\eeqnarray

\bpdffigure{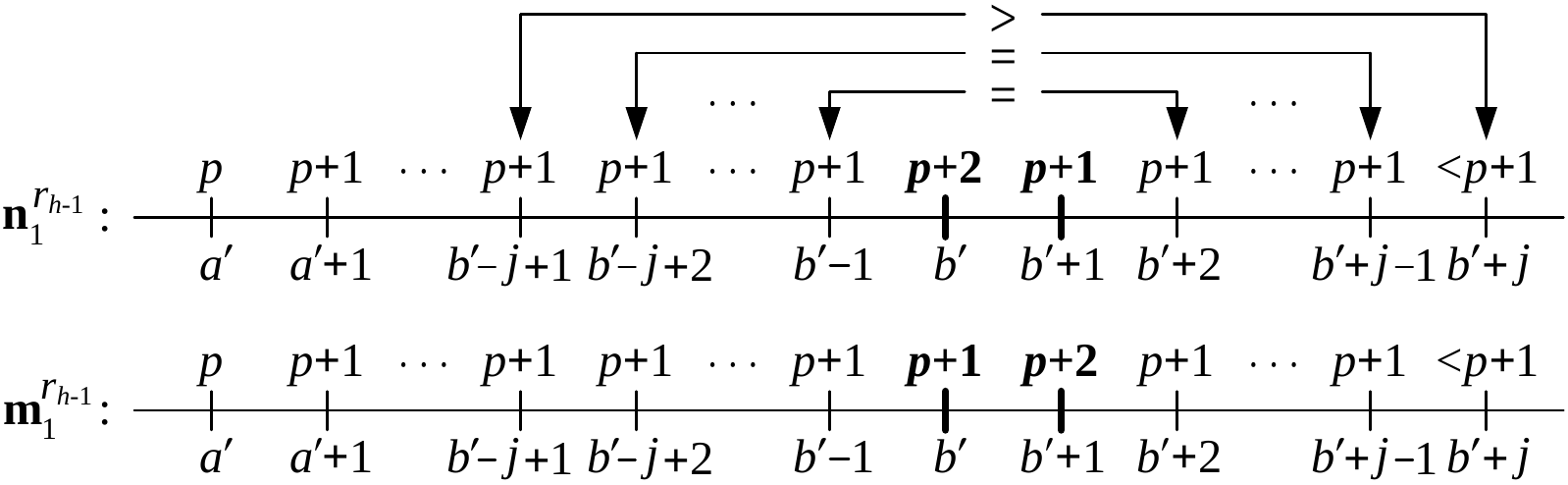}{5.5in}
\epdffigure{appendix-I-(i)-(b)-case-2}
{An illustration of \reqnarray{proof of nonadjacent distance larger than one II-(i)-(b)-case-2-555}
and \reqnarray{proof of nonadjacent distance larger than one II-(i)-(b)-case-2-666}.}

In the following, we show that
\beqnarray{}
\alignspace
2\leq b'\leq r_{h-1}-2,
\label{eqn:proof of nonadjacent distance larger than one II-(i)-(b)-case-2-333}\\
\alignspace
1\leq j-1\leq \min\{b'-1,r_{h-1}-b'-1\},
\label{eqn:proof of nonadjacent distance larger than one II-(i)-(b)-case-2-444}\\
\alignspace n_{b'-j'}=n_{(b'+1)+j'}, \textrm{ for } j'=1,2,\ldots,j-2,
\label{eqn:proof of nonadjacent distance larger than one II-(i)-(b)-case-2-555}\\
\alignspace n_{b'-(j-1)}>n_{(b'+1)+(j-1)}.
\label{eqn:proof of nonadjacent distance larger than one II-(i)-(b)-case-2-666}
\eeqnarray
An illustration of
\reqnarray{proof of nonadjacent distance larger than one II-(i)-(b)-case-2-555}
and \reqnarray{proof of nonadjacent distance larger than one II-(i)-(b)-case-2-666}
is given in \rfigure{appendix-I-(i)-(b)-case-2}.
Therefore, it follows from $\nbf_1^{r_{h-1}}\in \Ncal_{M,k}(h)$ in \reqnarray{proof of nonadjacent distance larger than one II-111},
\reqnarray{proof of nonadjacent distance larger than one II-(i)-(b)-case-2-111}--\reqnarray{proof of nonadjacent distance larger than one II-(i)-(b)-case-2-666},
and \reqnarray{comparison rule B-2} in \rlemma{comparison rule B}(ii) that
$\nbf_1^{r_{h-1}}\prec \mbf_1^{r_{h-1}}$,
i.e., \reqnarray{proof of nonadjacent distance larger than one II-(i)-444} holds
with ${\nbf'}_1^{r_{h-1}}=\mbf_1^{r_{h-1}}$.

From $2\leq j\leq \min\{b'-a'-1,r_{h-1}-b'\}$ and $a'\geq 1$,
we can see that
\beqnarray{}
\alignspace
2\leq \min\{b'-a'-1,r_{h-1}-b'\}\leq b'-a'-1\leq b',
\label{eqn:proof of nonadjacent distance larger than one II-(i)-(b)-case-2-777}\\
\alignspace
2\leq \min\{b'-a'-1,r_{h-1}-b'\}\leq r_{h-1}-b',
\label{eqn:proof of nonadjacent distance larger than one II-(i)-(b)-case-2-888}\\
\alignspace
1\leq j-1\leq \min\{b'-a'-2,r_{h-1}-b'-1\}\leq \min\{b'-1,r_{h-1}-b'-1\}.
\label{eqn:proof of nonadjacent distance larger than one II-(i)-(b)-case-2-999}
\eeqnarray
Thus, \reqnarray{proof of nonadjacent distance larger than one II-(i)-(b)-case-2-333}
follows from \reqnarray{proof of nonadjacent distance larger than one II-(i)-(b)-case-2-777}
and \reqnarray{proof of nonadjacent distance larger than one II-(i)-(b)-case-2-888},
and \reqnarray{proof of nonadjacent distance larger than one II-(i)-(b)-case-2-444}
follows from \reqnarray{proof of nonadjacent distance larger than one II-(i)-(b)-case-2-999}.

To prove \reqnarray{proof of nonadjacent distance larger than one II-(i)-(b)-case-2-555}
and \reqnarray{proof of nonadjacent distance larger than one II-(i)-(b)-case-2-666},
note that we have from $2\leq j\leq \min\{b'-a'-1,r_{h-1}-b'\}\leq b'-a'-1$ that
\beqnarray{proof of nonadjacent distance larger than one II-(i)-(b)-case-2-aaa}
a'< a'+2\leq b'-(j-1)\leq b'-1<b'.
\eeqnarray
It then follows from \reqnarray{proof of nonadjacent distance larger than one II-(i)-222}
and \reqnarray{proof of nonadjacent distance larger than one II-(i)-(b)-case-2-aaa} that
\beqnarray{proof of nonadjacent distance larger than one II-(i)-(b)-case-2-bbb}
n_{b'-j'}=p+1, \textrm{ for } j'=1,2,\ldots,j-1.
\eeqnarray
By combining \reqnarray{proof of nonadjacent distance larger than one II-(i)-(b)-case-2-bbb},
$n_{b'+j'}=p+1$ for $j'=1,2,\ldots,j-1$, and $n_{b'+j}<p+1$,
we obtain \reqnarray{proof of nonadjacent distance larger than one II-(i)-(b)-case-2-555}
and \reqnarray{proof of nonadjacent distance larger than one II-(i)-(b)-case-2-666}.

\emph{Case 3: $n_{b'+j'}=p+1$  for $j'=1,2,\ldots,\min\{b'-a'-1,r_{h-1}-b'\}$.}
We consider the two subcases $b'-a'-1>r_{h-1}-b'$ and $b'-a'-1\leq r_{h-1}-b'$ separately.

\emph{Subcase 3(a): $b'-a'-1>r_{h-1}-b'$.}
Let $\mbf_1^{r_{h-1}}$ be a sequence of positive integers as given in
\reqnarray{proof of nonadjacent distance larger than one II-(i)-(b)-case-2-111}.
As in Case~2 above, we have $\mbf_1^{r_{h-1}}\in \Ncal_{M,k}(h)$.
As it is clear from $b'\geq a'+2$
in \reqnarray{proof of nonadjacent distance larger than one II-(i)-111}
and $b'\leq r_{h-1}-1$ that $\min\{b'-a'-1,r_{h-1}-b'\}\geq 1$,
we have $n_{b'+1}=p+1$ in this case and hence it is easy to see that
\reqnarray{proof of nonadjacent distance larger than one II-(i)-(b)-case-2-222}
still holds in this subcase.

If $b'=r_{h-1}-1$,
then it follows from $r_{h-1}\geq 3$ and $\nbf_1^{r_{h-1}}\in \Ncal_{M,k}(h)$
in \reqnarray{proof of nonadjacent distance larger than one II-111},
\reqnarray{proof of nonadjacent distance larger than one II-(i)-(b)-case-2-111},
\reqnarray{proof of nonadjacent distance larger than one II-(i)-(b)-case-2-222},
and \reqnarray{comparison rule B-1} in \rlemma{comparison rule B}(i)
that $\nbf_1^{r_{h-1}}\prec \mbf_1^{r_{h-1}}$,
i.e., \reqnarray{proof of nonadjacent distance larger than one II-(i)-444} holds
with ${\nbf'}_1^{r_{h-1}}=\mbf_1^{r_{h-1}}$.

\bpdffigure{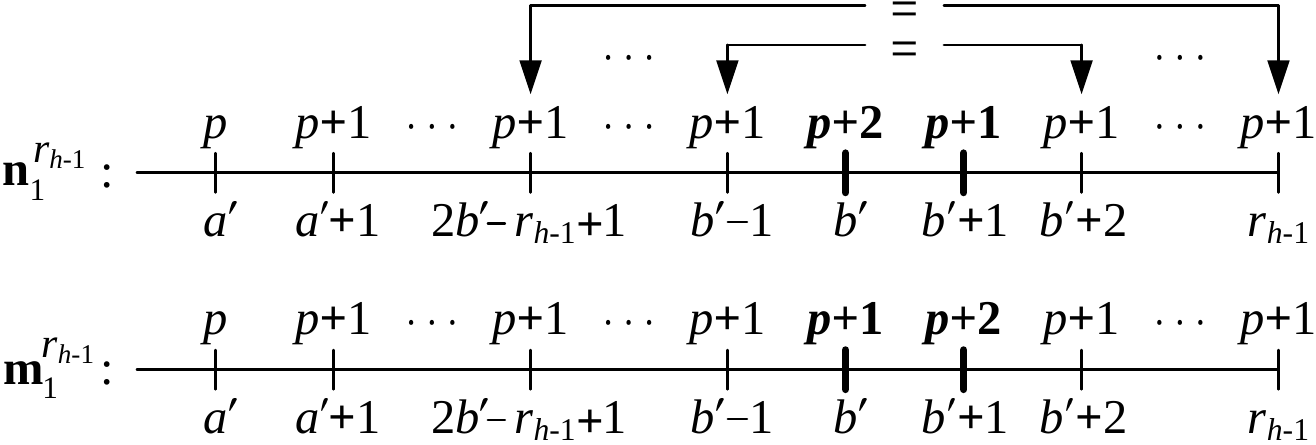}{4.5in}
\epdffigure{appendix-I-(i)-(b)-case-3-1}
{An illustration of \reqnarray{proof of nonadjacent distance larger than one II-(i)-(b)-case-3-222}
(note that we have $\min\{b'-1, r_{h-1}-b'-1\}=r_{h-1}-b'-1$
in \reqnarray{proof of nonadjacent distance larger than one II-(i)-(b)-case-3-555}).}

On the other hand, if $b'\leq r_{h-1}-2$, then we show that
\beqnarray{}
\alignspace \hspace*{-0.2in}
2\leq b'\leq r_{h-1}-2,
\label{eqn:proof of nonadjacent distance larger than one II-(i)-(b)-case-3-111}\\
\alignspace \hspace*{-0.2in}
n_{b'-j'}=n_{(b'+1)+j'}=p+1, \textrm{ for } j'=1,2,\ldots,\min\{b'-1, r_{h-1}-b'-1\}.
\label{eqn:proof of nonadjacent distance larger than one II-(i)-(b)-case-3-222}
\eeqnarray
An illustration of \reqnarray{proof of nonadjacent distance larger than one II-(i)-(b)-case-3-222}
is given in \rfigure{appendix-I-(i)-(b)-case-3-1}.
Therefore, it follows from
$\nbf_1^{r_{h-1}}\in \Ncal_{M,k}(h)$ in \reqnarray{proof of nonadjacent distance larger than one II-111},
\reqnarray{proof of nonadjacent distance larger than one II-(i)-(b)-case-2-111},
\reqnarray{proof of nonadjacent distance larger than one II-(i)-(b)-case-2-222},
\reqnarray{proof of nonadjacent distance larger than one II-(i)-(b)-case-3-111},
\reqnarray{proof of nonadjacent distance larger than one II-(i)-(b)-case-3-222},
and \reqnarray{comparison rule B-4} in \rlemma{comparison rule B}(iii) that
$\nbf_1^{r_{h-1}}\prec \mbf_1^{r_{h-1}}$,
i.e., \reqnarray{proof of nonadjacent distance larger than one II-(i)-444} holds
with ${\nbf'}_1^{r_{h-1}}=\mbf_1^{r_{h-1}}$.

From \reqnarray{proof of nonadjacent distance larger than one II-(i)-(b)-222}
and $b'\leq r_{h-1}-2$, we see that
\beqnarray{proof of nonadjacent distance larger than one II-(i)-(b)-case-3-333}
2\leq b'-1<b'\leq r_{h-1}-2.
\eeqnarray
Thus, \reqnarray{proof of nonadjacent distance larger than one II-(i)-(b)-case-3-111}
follows from \reqnarray{proof of nonadjacent distance larger than one II-(i)-(b)-case-3-333}.

To prove \reqnarray{proof of nonadjacent distance larger than one II-(i)-(b)-case-3-222},
note that from $b'-a'-1>r_{h-1}-b'$ in this subcase and $a'\geq 1$, we have
\beqnarray{proof of nonadjacent distance larger than one II-(i)-(b)-case-3-444}
r_{h-1}-b'-1<(b'-a'-1)-1<b'-1.
\eeqnarray
It follows from \reqnarray{proof of nonadjacent distance larger than one II-(i)-(b)-case-3-444} that
\beqnarray{proof of nonadjacent distance larger than one II-(i)-(b)-case-3-555}
\min\{b'-1, r_{h-1}-b'-1\}=r_{h-1}-b'-1.
\eeqnarray
From $b'-a'-1>r_{h-1}-b'$ and $b'\leq r_{h-1}-2$,
we can see that
\beqnarray{}
\alignspace
b'-(r_{h-1}-b'-1)>b'-(b'-a'-2)=a'+2>a',
\label{eqn:proof of nonadjacent distance larger than one II-(i)-(b)-case-3-666}\\
\alignspace
b'-(r_{h-1}-b'-1)\leq b'-1<b'.
\label{eqn:proof of nonadjacent distance larger than one II-(i)-(b)-case-3-777}
\eeqnarray
It follows from \reqnarray{proof of nonadjacent distance larger than one II-(i)-222},
\reqnarray{proof of nonadjacent distance larger than one II-(i)-(b)-case-3-666},
and \reqnarray{proof of nonadjacent distance larger than one II-(i)-(b)-case-3-777} that
\beqnarray{proof of nonadjacent distance larger than one II-(i)-(b)-case-3-888}
n_{b'-j'}=p+1, \textrm{ for } j'=1,2,\ldots,r_{h-1}-b'-1.
\eeqnarray
Furthermore, in this subcase we have
\beqnarray{proof of nonadjacent distance larger than one II-(i)-(b)-case-3-999}
n_{b'+j'}=p+1, \textrm{ for } j'=1,2,\ldots,\min\{b'-a'-1,r_{h-1}-b'\}=r_{h-1}-b'.
\eeqnarray
It then follows from \reqnarray{proof of nonadjacent distance larger than one II-(i)-(b)-case-3-999} that
\beqnarray{proof of nonadjacent distance larger than one II-(i)-(b)-case-3-aaa}
n_{(b'+1)+j'}=p+1, \textrm{ for } j'=1,2,\ldots,r_{h-1}-b'-1.
\eeqnarray
By combining
\reqnarray{proof of nonadjacent distance larger than one II-(i)-(b)-case-3-555},
\reqnarray{proof of nonadjacent distance larger than one II-(i)-(b)-case-3-888},
and \reqnarray{proof of nonadjacent distance larger than one II-(i)-(b)-case-3-aaa},
we obtain \reqnarray{proof of nonadjacent distance larger than one II-(i)-(b)-case-3-222}.

\emph{Subcase 3(b): $b'-a'-1\leq r_{h-1}-b'$.}
Let $\mbf_1^{r_{h-1}}$ be a sequence of positive integers as given in
\reqnarray{proof of nonadjacent distance larger than one II-(i)-(b)-case-1-111}.
As in Case~1 above, we have $\mbf_1^{r_{h-1}}\in \Ncal_{M,k}(h)$.
Also, note that \reqnarray{proof of nonadjacent distance larger than one II-(i)-(b)-case-1-222} still holds in this subcase.

\bpdffigure{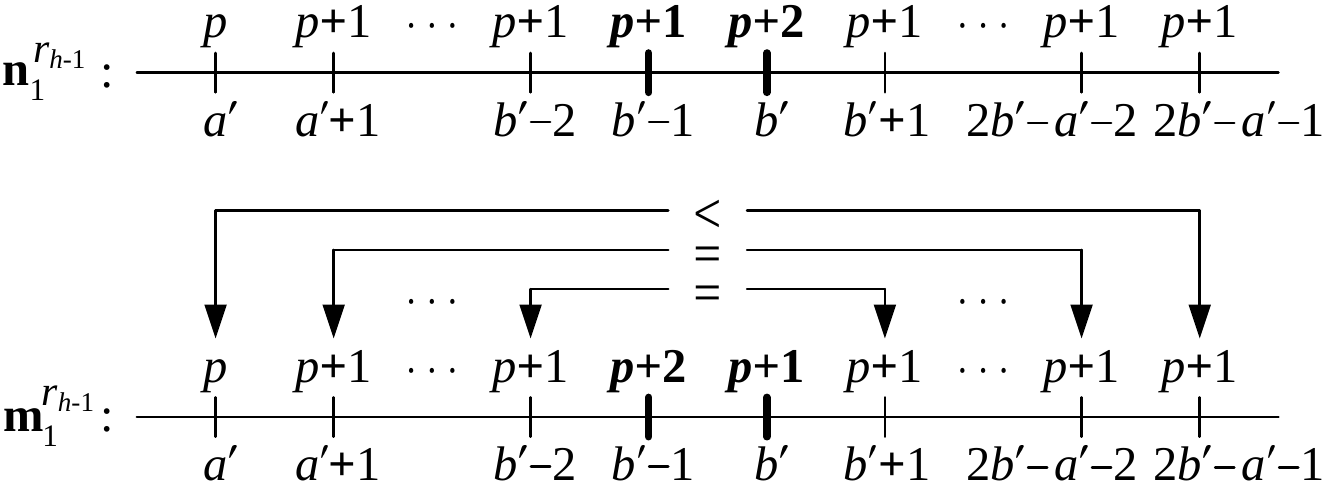}{4.5in}
\epdffigure{appendix-I-(i)-(b)-case-3-2}
{An illustration of \reqnarray{proof of nonadjacent distance larger than one II-(i)-(b)-case-3-ccc}
and \reqnarray{proof of nonadjacent distance larger than one II-(i)-(b)-case-3-ddd}.}

In the following, we show that
\beqnarray{}
\alignspace 1\leq b'-a'-1\leq \min\{(b'-1)-1,r_{h-1}-(b'-1)-1\}
\label{eqn:proof of nonadjacent distance larger than one II-(i)-(b)-case-3-bbb}\\
\alignspace m_{(b'-1)-j'}=m_{b'+j'}=p+1, \textrm{ for } j'=1,2,\ldots,b'-a'-2,
\label{eqn:proof of nonadjacent distance larger than one II-(i)-(b)-case-3-ccc}\\
\alignspace m_{(b'-1)-(b'-a'-1)}=p<m_{b'+(b'-a'-1)}=p+1.
\label{eqn:proof of nonadjacent distance larger than one II-(i)-(b)-case-3-ddd}
\eeqnarray
An illustration of
\reqnarray{proof of nonadjacent distance larger than one II-(i)-(b)-case-3-ccc}
and \reqnarray{proof of nonadjacent distance larger than one II-(i)-(b)-case-3-ddd}
is given in \rfigure{appendix-I-(i)-(b)-case-3-2}.
Therefore, it follows from
$\mbf_1^{r_{h-1}}\in \Ncal_{M,k}(h)$,
\reqnarray{proof of nonadjacent distance larger than one II-(i)-(b)-222}--\reqnarray{proof of nonadjacent distance larger than one II-(i)-(b)-case-1-222},
\reqnarray{proof of nonadjacent distance larger than one II-(i)-(b)-case-3-bbb}--\reqnarray{proof of nonadjacent distance larger than one II-(i)-(b)-case-3-ddd},
and \reqnarray{comparison rule B-3} in \rlemma{comparison rule B}(ii)
(with $j=b'-a'-1$) that
\beqnarray{proof of nonadjacent distance larger than one II-(i)-(b)-case-3-eee}
\mbf_1^{r_{h-1}}\succeq \nbf_1^{r_{h-1}}.
\eeqnarray

From $b'\geq a'+2$ in \reqnarray{proof of nonadjacent distance larger than one II-(i)-111},
$b'-a'-1\leq r_{h-1}-b'$, and $a'\geq 1$, we can see that
\beqnarray{proof of nonadjacent distance larger than one II-(i)-(b)-case-3-fff}
1\leq b'-a'-1\leq \min\{b'-a'-1,r_{h-1}-b'\}\leq \min\{(b'-1)-1,r_{h-1}-(b'-1)-1\}.
\eeqnarray
Thus, \reqnarray{proof of nonadjacent distance larger than one II-(i)-(b)-case-3-bbb}
follows from \reqnarray{proof of nonadjacent distance larger than one II-(i)-(b)-case-3-fff}.

To prove \reqnarray{proof of nonadjacent distance larger than one II-(i)-(b)-case-3-ccc}
and \reqnarray{proof of nonadjacent distance larger than one II-(i)-(b)-case-3-ddd},
note that from
\reqnarray{proof of nonadjacent distance larger than one II-(i)-(b)-case-1-111},
$(b'-1)-(b'-a'-1)=a'$,
and \reqnarray{proof of nonadjacent distance larger than one II-(i)-222},
we have
\beqnarray{}
\alignspace m_{(b'-1)-j'}=n_{(b'-1)-j'}=p+1, \textrm{ for } j'=1,2,\ldots,b'-a'-2,
\label{eqn:proof of nonadjacent distance larger than one II-(i)-(b)-case-3-ggg}\\
\alignspace m_{(b'-1)-(b'-a'-1)}=n_{(b'-1)-(b'-a'-1)}=n_{a'}=p.
\label{eqn:proof of nonadjacent distance larger than one II-(i)-(b)-case-3-hhh}
\eeqnarray
Also, in this case we have
\beqnarray{proof of nonadjacent distance larger than one II-(i)-(b)-case-3-iii}
n_{b'+j'}=p+1, \textrm{ for } j'=1,2,\ldots,\min\{b'-a'-1,r_{h-1}-b'\}=b'-a'-1.
\eeqnarray
It follows from \reqnarray{proof of nonadjacent distance larger than one II-(i)-(b)-case-1-111}
and \reqnarray{proof of nonadjacent distance larger than one II-(i)-(b)-case-3-iii} that
\beqnarray{proof of nonadjacent distance larger than one II-(i)-(b)-case-3-jjj}
m_{b'+j'}=n_{b'+j'}=p+1, \textrm{ for } j'=1,2,\ldots,b'-a'-1.
\eeqnarray
By combining
\reqnarray{proof of nonadjacent distance larger than one II-(i)-(b)-case-3-ggg},
\reqnarray{proof of nonadjacent distance larger than one II-(i)-(b)-case-3-hhh},
and \reqnarray{proof of nonadjacent distance larger than one II-(i)-(b)-case-3-jjj},
we obtain \reqnarray{proof of nonadjacent distance larger than one II-(i)-(b)-case-3-ccc}
and \reqnarray{proof of nonadjacent distance larger than one II-(i)-(b)-case-3-ddd}.

Now let ${\mbf'}_1^{r_{h-1}}$ be a sequence of positive integers such that
\beqnarray{proof of nonadjacent distance larger than one II-(i)-(b)-case-3-kkk}
m'_{b'-2}=m_{b'-2}+1,\ m'_{b'-1}=m_{b'-1}-1, \textrm{ and } m'_i=m_i \textrm{ for } i\neq b'-2, b'-1.
\eeqnarray
Again, it is easy to show that ${\mbf'}_1^{r_{h-1}}\in \Ncal_{M,k}(h)$.

If $b'=a'+2$, then we see from
\reqnarray{proof of nonadjacent distance larger than one II-(i)-(b)-case-1-111},
$n_{a'}=p$ in \reqnarray{proof of nonadjacent distance larger than one II-(i)-222},
and $n_{b'-1}=p+1$ in \reqnarray{proof of nonadjacent distance larger than one II-(i)-333} that
\beqnarray{proof of nonadjacent distance larger than one II-(i)-(b)-case-3-1111}
\alignspace m_{b'-2}-m_{b'-1}=n_{b'-2}-(n_{b'-1}+1)=n_{a'}-n_{b'-1}-1=p-(p+1)-1=-2.
\eeqnarray
Therefore, it follows from
$r_{h-1}\geq 3$ in \reqnarray{proof of nonadjacent distance larger than one II-111},
$\mbf_1^{r_{h-1}}\in \Ncal_{M,k}(h)$,
\reqnarray{proof of nonadjacent distance larger than one II-(i)-(b)-case-3-kkk},
\reqnarray{proof of nonadjacent distance larger than one II-(i)-(b)-case-3-1111},
and \reqnarray{adjacent distance larger than one II-2} in \rlemma{adjacent distance larger than one II}(ii) that
\beqnarray{proof of nonadjacent distance larger than one II-(i)-(b)-case-3-2222}
\mbf_1^{r_{h-1}}\preceq {\mbf'}_1^{r_{h-1}},
\eeqnarray
where $\mbf_1^{r_{h-1}}\equiv {\mbf'}_1^{r_{h-1}}$
if and only if $r_{h-1}=2$ and $m_1=m_2-2$.
Since it is clear from $r_{h-1}\geq 3$
in \reqnarray{proof of nonadjacent distance larger than one II-111}
that $r_{h-1}\neq 2$, it cannot be the case that $\mbf_1^{r_{h-1}}\equiv {\mbf'}_1^{r_{h-1}}$.
As such, we see from \reqnarray{proof of nonadjacent distance larger than one II-(i)-(b)-case-3-eee}
and \reqnarray{proof of nonadjacent distance larger than one II-(i)-(b)-case-3-2222}
that $\nbf_1^{r_{h-1}}\preceq \mbf_1^{r_{h-1}}\prec {\mbf'}_1^{r_{h-1}}$,
i.e., \reqnarray{proof of nonadjacent distance larger than one II-(i)-444} holds
with ${\nbf'}_1^{r_{h-1}}={\mbf'}_1^{r_{h-1}}$.

On the other hand, if $b'\geq a'+3$, then we have $a'<b'-2<b'$
and it follows from \reqnarray{proof of nonadjacent distance larger than one II-(i)-222} that
\beqnarray{proof of nonadjacent distance larger than one II-(i)-(b)-case-3-3333}
n_{b'-2}=n_{b'-1}=p+1.
\eeqnarray
From \reqnarray{proof of nonadjacent distance larger than one II-(i)-(b)-case-3-kkk},
\reqnarray{proof of nonadjacent distance larger than one II-(i)-(b)-case-1-111},
and \reqnarray{proof of nonadjacent distance larger than one II-(i)-(b)-case-3-3333},
we see that
\beqnarray{proof of nonadjacent distance larger than one II-(i)-(b)-case-3-4444}
m'_{b'-2}-m'_{b'-1}
\aligneq (m_{b'-2}+1)-(m_{b'-1}-1)=m_{b'-2}-m_{b'-1}+2 \nn\\
\aligneq n_{b'-2}-(n_{b'-1}+1)+2=(p+1)-(p+1+1)+2=1.
\eeqnarray

\bpdffigure{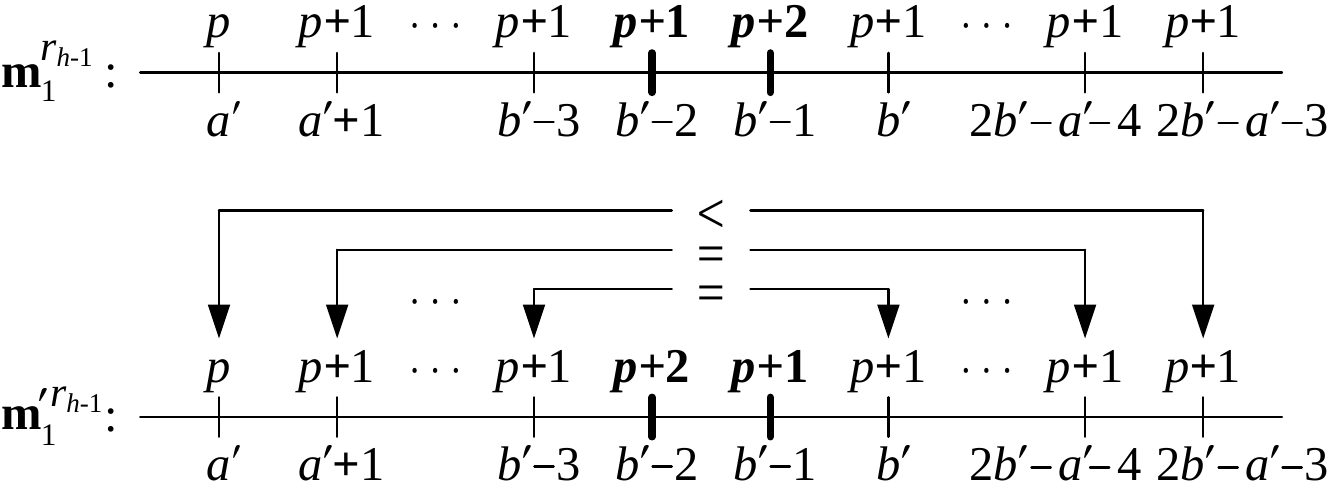}{4.5in}
\epdffigure{appendix-I-(i)-(b)-case-3-3}
{An illustration of \reqnarray{proof of nonadjacent distance larger than one II-(i)-(b)-case-3-7777}
and \reqnarray{proof of nonadjacent distance larger than one II-(i)-(b)-case-3-8888}.}

We will show that
\beqnarray{}
\alignspace 2\leq b'-2\leq r_{h-1}-2,
\label{eqn:proof of nonadjacent distance larger than one II-(i)-(b)-case-3-5555}\\
\alignspace 1\leq b'-a'-2\leq \min\{(b'-2)-1,r_{h-1}-(b'-2)-1\}.
\label{eqn:proof of nonadjacent distance larger than one II-(i)-(b)-case-3-6666}\\
\alignspace m'_{(b'-2)-j'}=m'_{(b'-1)+j'}=p+1, \textrm{ for } j'=1,2,\ldots,b'-a'-3,
\label{eqn:proof of nonadjacent distance larger than one II-(i)-(b)-case-3-7777}\\
\alignspace m'_{(b'-2)-(b'-a'-2)}=p<m'_{(b'-1)+(b'-a'-2)}=p+1.
\label{eqn:proof of nonadjacent distance larger than one II-(i)-(b)-case-3-8888}
\eeqnarray
An illustration of
\reqnarray{proof of nonadjacent distance larger than one II-(i)-(b)-case-3-7777}
and \reqnarray{proof of nonadjacent distance larger than one II-(i)-(b)-case-3-8888}
is given in \rfigure{appendix-I-(i)-(b)-case-3-3}.
Therefore, it follows from
${\mbf'}_1^{r_{h-1}}\in \Ncal_{M,k}(h)$,
\reqnarray{proof of nonadjacent distance larger than one II-(i)-(b)-case-3-kkk},
\reqnarray{proof of nonadjacent distance larger than one II-(i)-(b)-case-3-4444}--\reqnarray{proof of nonadjacent distance larger than one II-(i)-(b)-case-3-8888},
and \reqnarray{comparison rule B-3} in \rlemma{comparison rule B}(ii)
(with $j=b'-a'-2$) that
\beqnarray{proof of nonadjacent distance larger than one II-(i)-(b)-case-3-9999}
{\mbf'}_1^{r_{h-1}}\succeq \mbf_1^{r_{h-1}},
\eeqnarray
where ${\mbf'}_1^{r_{h-1}}\equiv \mbf_1^{r_{h-1}}$
if and only if $(b'-2)-(b'-a'-2)=1$, $(b'-1)+(b'-a'-2)=r_{h-1}$, and $m'_1=m'_{r_{h-1}}-1$.
Since in this subcase we have $b'-a'-1\leq r_{h-1}-b'$,
it is clear that $(b'-1)+(b'-a'-2)\leq (b'-1)+(r_{h-1}-b'-1)=r_{h-1}-2$.
This implies that $(b'-1)+(b'-a'-2)\neq r_{h-1}$
and hence it cannot be the case that ${\mbf'}_1^{r_{h-1}}\equiv \mbf_1^{r_{h-1}}$.
As such, we see from \reqnarray{proof of nonadjacent distance larger than one II-(i)-(b)-case-3-eee}
and \reqnarray{proof of nonadjacent distance larger than one II-(i)-(b)-case-3-9999}
that $\nbf_1^{r_{h-1}}\preceq \mbf_1^{r_{h-1}}\prec {\mbf'}_1^{r_{h-1}}$,
i.e., \reqnarray{proof of nonadjacent distance larger than one II-(i)-444} holds
with ${\nbf'}_1^{r_{h-1}}={\mbf'}_1^{r_{h-1}}$.

To prove \reqnarray{proof of nonadjacent distance larger than one II-(i)-(b)-case-3-5555}
and \reqnarray{proof of nonadjacent distance larger than one II-(i)-(b)-case-3-6666},
note from $a'\geq 1$, $b'\geq a'+3$, $b'\leq r_{h-1}-1$,
and \reqnarray{proof of nonadjacent distance larger than one II-(i)-(b)-case-3-bbb} that
\beqnarray{}
\alignspace \hspace*{-0.3in}
2\leq a'+1\leq b'-2\leq r_{h-1}-2,
\label{eqn:proof of nonadjacent distance larger than one II-(i)-(b)-case-3-aaaa}\\
\alignspace \hspace*{-0.3in}
1\leq b'-a'-2\leq \min\{b'-3,r_{h-1}-b'-1\}\leq \min\{(b'-2)-1,r_{h-1}-(b'-2)-1\}.
\label{eqn:proof of nonadjacent distance larger than one II-(i)-(b)-case-3-bbbb}
\eeqnarray
Thus, \reqnarray{proof of nonadjacent distance larger than one II-(i)-(b)-case-3-5555}
follows from \reqnarray{proof of nonadjacent distance larger than one II-(i)-(b)-case-3-aaaa},
and \reqnarray{proof of nonadjacent distance larger than one II-(i)-(b)-case-3-6666}
follows from \reqnarray{proof of nonadjacent distance larger than one II-(i)-(b)-case-3-bbbb}.

To prove \reqnarray{proof of nonadjacent distance larger than one II-(i)-(b)-case-3-7777}
and \reqnarray{proof of nonadjacent distance larger than one II-(i)-(b)-case-3-8888},
note that from \reqnarray{proof of nonadjacent distance larger than one II-(i)-(b)-case-3-kkk},
\reqnarray{proof of nonadjacent distance larger than one II-(i)-(b)-case-3-ggg}--\reqnarray{proof of nonadjacent distance larger than one II-(i)-(b)-case-3-hhh},
\reqnarray{proof of nonadjacent distance larger than one II-(i)-(b)-case-1-111},
$n_{b'}=p+2$ in \reqnarray{proof of nonadjacent distance larger than one II-(i)-222},
and \reqnarray{proof of nonadjacent distance larger than one II-(i)-(b)-case-3-jjj},
we see that
\beqnarray{}
\alignspace m'_{(b'-2)-j'}=m_{(b'-2)-j'}=p+1, \textrm{ for } j'=1,2,\ldots,b'-a'-3,
\label{eqn:proof of nonadjacent distance larger than one II-(i)-(b)-case-3-cccc}\\
\alignspace m'_{(b'-2)-(b'-a'-2)}=m_{(b'-2)-(b'-a'-2)}=p.
\label{eqn:proof of nonadjacent distance larger than one II-(i)-(b)-case-3-dddd}\\
\alignspace m'_{(b'-1)+1}=m'_{b'}=m_{b'}=n_{b'}-1=p+1,
\label{eqn:proof of nonadjacent distance larger than one II-(i)-(b)-case-3-eeee}\\
\alignspace m'_{(b'-1)+j'}=m_{(b'-1)+j'}=p+1, \textrm{ for } j'=2,3,\ldots,b'-a',
\label{eqn:proof of nonadjacent distance larger than one II-(i)-(b)-case-3-ffff}
\eeqnarray
Thus, \reqnarray{proof of nonadjacent distance larger than one II-(i)-(b)-case-3-7777}
and \reqnarray{proof of nonadjacent distance larger than one II-(i)-(b)-case-3-8888}
follow from
\reqnarray{proof of nonadjacent distance larger than one II-(i)-(b)-case-3-cccc}--\reqnarray{proof of nonadjacent distance larger than one II-(i)-(b)-case-3-ffff}.

(ii) Note that in \rlemma{nonadjacent distance larger than one II}(ii),
we have $n_a-n_b\geq 2$ for some $1\leq a<b\leq r_{h-1}$ and $b\geq a+2$.
For ease of presentation, let $n_b=p$.
Then we have from $n_a-n_b\geq 2$ that $n_a\geq p+2$.
As we have $n_a\geq p+2$, $n_b=p$, $b\geq a+2>a$,
and the condition $|n_{i+1}-n_i|\leq 1$ for $i=1,2,\ldots,r_{h-1}-1$
in \reqnarray{proof of nonadjacent distance larger than one II-111},
we can argue as that for \reqnarray{proof of nonadjacent distance larger than one-(i)-111}--\reqnarray{proof of nonadjacent distance larger than one-(i)-555}
in the proof of \rlemma{nonadjacent distance larger than one}(i)
in \rappendix{proof of nonadjacent distance larger than one} that there exist two positive integers $a'$ and $b'$ such that
\beqnarray{}
\alignspace a\leq a'<b'\leq b \textrm{ and } b'\geq a'+2,
\label{eqn:proof of nonadjacent distance larger than one II-(ii)-111} \\
\alignspace n_{a'}=p+2,\ n_{b'}=p, \textrm{ and } n_i=p+1 \textrm{ for } a'<i<b'.
\label{eqn:proof of nonadjacent distance larger than one II-(ii)-222}
\eeqnarray
An illustration of
\reqnarray{proof of nonadjacent distance larger than one II-(ii)-111} and \reqnarray{proof of nonadjacent distance larger than one II-(ii)-222}
is given in \rfigure{appendix-I-(ii)}.

\bpdffigure{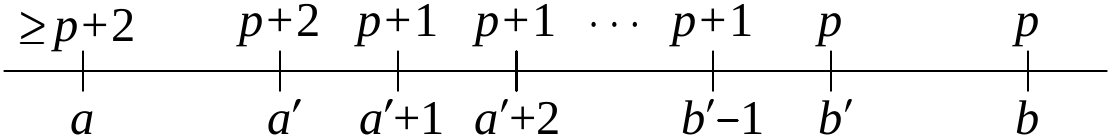}{4.0in}
\epdffigure{appendix-I-(ii)}
{An illustration of \reqnarray{proof of nonadjacent distance larger than one II-(ii)-111} and \reqnarray{proof of nonadjacent distance larger than one II-(ii)-222}.}

To prove \rlemma{nonadjacent distance larger than one II}(ii),
we need to show that there exists a sequence of positive integers
${\nbf'}_1^{r_{h-1}}\in \Ncal_{M,k}(h)$ such that
\beqnarray{proof of nonadjacent distance larger than one II-(ii)-333}
{\nbf'}_1^{r_{h-1}}\succ\nbf_1^{r_{h-1}}.
\eeqnarray
Note that from \reqnarray{proof of nonadjacent distance larger than one II-(ii)-222}
and $b'\geq a'+2$ in \reqnarray{proof of nonadjacent distance larger than one II-(ii)-111},
we see that $n_{b'-1}=p+1$.
It then follows from $n_{b'-1}=p+1$ and $n_{b'}=p$
in \reqnarray{proof of nonadjacent distance larger than one II-(ii)-222} that
\beqnarray{proof of nonadjacent distance larger than one II-(ii)-444}
n_{b'-1}-n_{b'}=(p+1)-p=1.
\eeqnarray
We consider the following four possible cases.
Note that in Case~2--Case~4 below,
we have $b'\leq r_{h-1}-1$ and hence it follows from $a'\geq 1$
and $b'\geq a'+2$ in \reqnarray{proof of nonadjacent distance larger than one II-(ii)-111} that
\beqnarray{proof of nonadjacent distance larger than one II-(ii)-555}
2\leq b'-1\leq r_{h-1}-2.
\eeqnarray

\emph{Case 1: $b'=r_{h-1}$.}
Let $\mbf_1^{r_{h-1}}$ be a sequence of positive integers such that
\beqnarray{proof of nonadjacent distance larger than one II-(ii)-case-1-111}
m_{b'-1}=n_{b'-1}-1,\ m_{b'}=n_{b'}+1, \textrm{ and } m_i=n_i \textrm{ for } i\neq b'-1, b'.
\eeqnarray
As before, it is easy to show that $\mbf_1^{r_{h-1}}\in \Ncal_{M,k}(h)$.
It follows from
$\nbf_1^{r_{h-1}}\in \Ncal_{M,k}(h)$ in \reqnarray{proof of nonadjacent distance larger than one II-111},
\reqnarray{proof of nonadjacent distance larger than one II-(ii)-444},
\reqnarray{proof of nonadjacent distance larger than one II-(ii)-case-1-111},
$b'-1=r_{h-1}-1$, and \reqnarray{comparison rule B-1} in \rlemma{comparison rule B}(i)
that $\nbf_1^{r_{h-1}}\prec\mbf_1^{r_{h-1}}$,
i.e., \reqnarray{proof of nonadjacent distance larger than one II-(ii)-333} holds
with ${\nbf'}_1^{r_{h-1}}=\mbf_1^{r_{h-1}}$.

\emph{Case 2: $b'\leq r_{h-1}-1$ and there exists a positive integer $j$ such that
$1\leq j\leq \min\{b'-a'-1,r_{h-1}-b'\}$, $n_{b'+j'}=p+1$ for $j'=1,2,\ldots,j-1$, and $n_{b'+j}<p+1$.}
Let $\mbf_1^{r_{h-1}}$ be a sequence of positive integers as given in
\reqnarray{proof of nonadjacent distance larger than one II-(ii)-case-1-111}.
As in Case~1 above, we have $\mbf_1^{r_{h-1}}\in \Ncal_{M,k}(h)$.

\bpdffigure{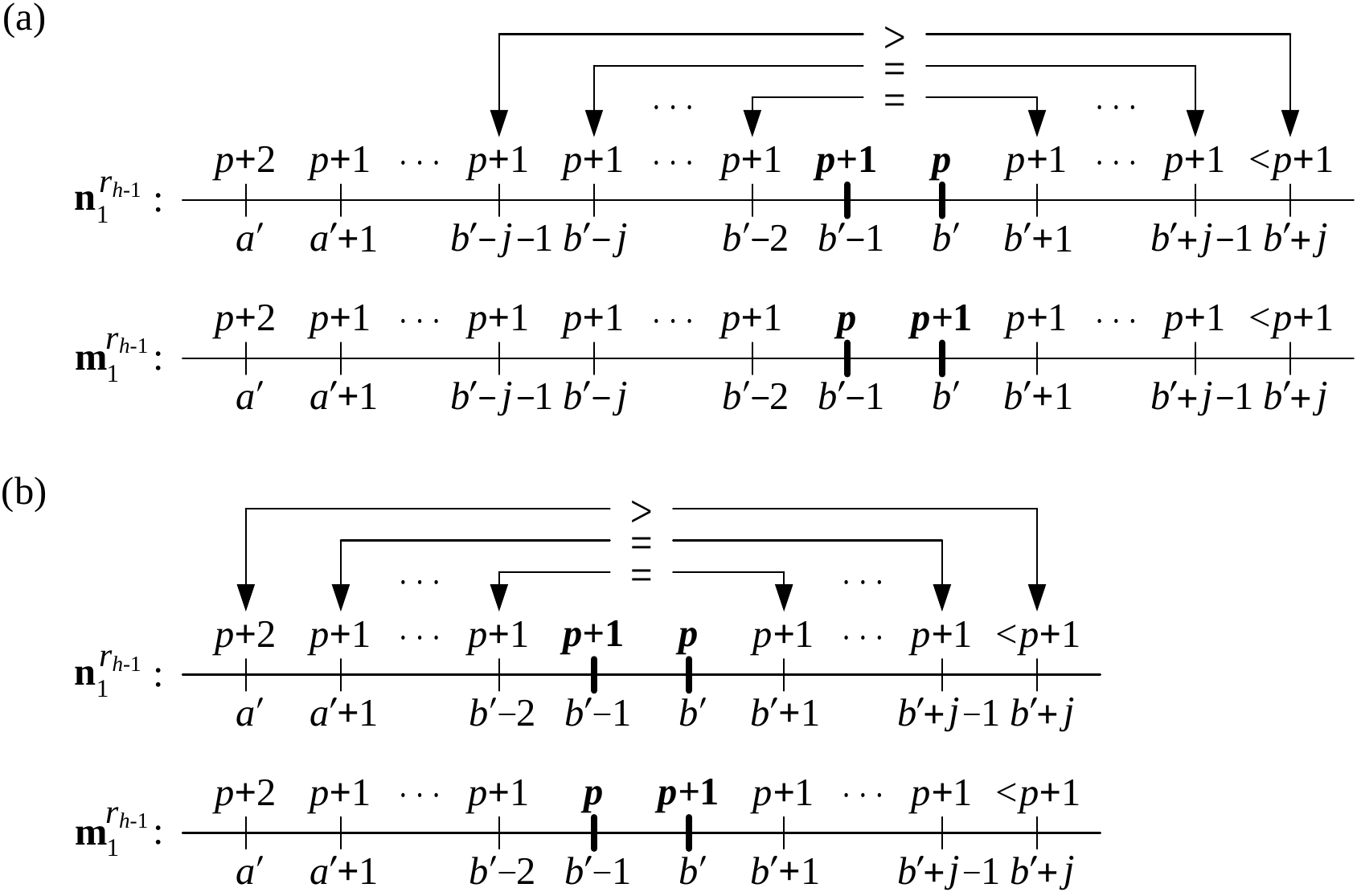}{5.5in}
\epdffigure{appendix-I-(ii)-case-2}
{An illustration of \reqnarray{proof of nonadjacent distance larger than one II-(ii)-case-2-333}
and \reqnarray{proof of nonadjacent distance larger than one II-(ii)-case-2-444}:
(a) $j<b'-a'-1$; (b) $j=b'-a'-1$.}

In the following, we show that
\beqnarray{}
\alignspace 1\leq j\leq \min\{(b'-1)-1,r_{h-1}-(b'-1)-1\},
\label{eqn:proof of nonadjacent distance larger than one II-(ii)-case-2-222}\\
\alignspace n_{(b'-1)-j'}=n_{b'+j'}, \textrm{ for } j'=1,2,\ldots,j-1,
\label{eqn:proof of nonadjacent distance larger than one II-(ii)-case-2-333}\\
\alignspace n_{(b'-1)-j}>n_{b'+j}.
\label{eqn:proof of nonadjacent distance larger than one II-(ii)-case-2-444}
\eeqnarray
An illustration of
\reqnarray{proof of nonadjacent distance larger than one II-(ii)-case-2-333}
and \reqnarray{proof of nonadjacent distance larger than one II-(ii)-case-2-444}
is given in \rfigure{appendix-I-(ii)-case-2}.
Therefore, it follows from
$\nbf_1^{r_{h-1}}\in \Ncal_{M,k}(h)$ in \reqnarray{proof of nonadjacent distance larger than one II-111},
\reqnarray{proof of nonadjacent distance larger than one II-(ii)-444}--\reqnarray{proof of nonadjacent distance larger than one II-(ii)-case-2-444},
and \reqnarray{comparison rule B-2} in \rlemma{comparison rule B}(ii) that
$\nbf_1^{r_{h-1}}\prec \mbf_1^{r_{h-1}}$,
i.e., \reqnarray{proof of nonadjacent distance larger than one II-(ii)-333} holds
with ${\nbf'}_1^{r_{h-1}}=\mbf_1^{r_{h-1}}$.

From $1\leq j\leq \min\{b'-a'-1,r_{h-1}-b'\}$ and $a'\geq 1$, we have
\beqnarray{proof of nonadjacent distance larger than one II-(ii)-case-2-555}
1\leq j\leq \min\{b'-a'-1,r_{h-1}-b'\}\leq \min\{(b'-1)-1,r_{h-1}-(b'-1)-1\}.
\eeqnarray
Thus, \reqnarray{proof of nonadjacent distance larger than one II-(ii)-case-2-222}
follows from \reqnarray{proof of nonadjacent distance larger than one II-(ii)-case-2-555}.

To prove \reqnarray{proof of nonadjacent distance larger than one II-(ii)-case-2-333}
and \reqnarray{proof of nonadjacent distance larger than one II-(ii)-case-2-444},
note that we have $j\leq \min\{b'-a'-1, r_{h-1}-b'\}\leq b'-a'-1$.
If $j<b'-a'-1$, then we have $a'<b'-1-j<b'$ and it follows from
\reqnarray{proof of nonadjacent distance larger than one II-(ii)-222} that
\beqnarray{proof of nonadjacent distance larger than one II-(ii)-case-2-666}
n_{(b'-1)-j'}=p+1, \textrm{ for } j'=1,2,\ldots,j.
\eeqnarray
On the other hand, if $j=b'-a'-1$, then we have $a'=b'-1-j<b'$
and it follows from
\reqnarray{proof of nonadjacent distance larger than one II-(ii)-222} that
\beqnarray{}
\alignspace
n_{(b'-1)-j'}=p+1, \textrm{ for } j'=1,2,\ldots,j-1,
\label{eqn:proof of nonadjacent distance larger than one II-(ii)-case-2-777}\\
\alignspace
n_{(b'-1)-j}=n_{a'}=p+2.
\label{eqn:proof of nonadjacent distance larger than one II-(ii)-case-2-888}
\eeqnarray
By combining \reqnarray{proof of nonadjacent distance larger than one II-(ii)-case-2-666}--\reqnarray{proof of nonadjacent distance larger than one II-(ii)-case-2-888},
$n_{b'+j'}=p+1$ for $j'=1,2,\ldots,j-1$, and $n_{b'+j}<p+1$,
we obtain \reqnarray{proof of nonadjacent distance larger than one II-(ii)-case-2-333}
and \reqnarray{proof of nonadjacent distance larger than one II-(ii)-case-2-444}.

\emph{Case 3: $b'\leq r_{h-1}-1$ and there exists a positive integer $j$ such that
$1\leq j\leq \min\{b'-a'-1,r_{h-1}-b'\}$, $n_{b'+j'}=p+1$ for $j'=1,2,\ldots,j-1$, and $n_{b'+j}>p+1$.}
In this case, we can show that $j\geq 2$.
To see this, suppose on the contrary that $j=1$,
then we have $n_{b'+1}>p+1$ in this case.
As it follows from $n_{b'}=p$ in \reqnarray{proof of nonadjacent distance larger than one II-(ii)-222}
and the condition $|n_{i+1}-n_i|\leq 1$ for $i=1,2,\ldots,r_{h-1}-1$
in \reqnarray{proof of nonadjacent distance larger than one II-111}
that $n_{b'+1}$ must be equal to $p-1$ (provided that $p\geq 2$), $p$, or $p+1$,
we have reached a contradiction.
Since $j\geq 2$, we have $n_{b'+j-1}=p+1$ in this case.
It then follows from the condition $|n_{i+1}-n_i|\leq 1$ for $i=1,2,\ldots,r_{h-1}-1$
in \reqnarray{proof of nonadjacent distance larger than one II-111}
that $n_{b'+j}$ must be equal to $p$, $p+1$, or $p+2$.
As we also have $n_{b'+j}>p+1$ in this case, we immediately see that $n_{b'+j}=p+2$.

From $n_{b'}=p$, $n_{b'+j'}=p+1$ for $j'=1,2,\ldots,j-1$, $n_{b'+j}=p+2$,
and $b'\geq 3$ in \reqnarray{proof of nonadjacent distance larger than one II-(ii)-555},
we can argue in the same way as in the proof of (i) above
(with the roles of $a'$ and $b'$ in the proof of (i) replaced by $b'$ and $b'+j$, respectively)
that there exists a sequence of positive integers ${\nbf'}_1^{r_{h-1}}\in \Ncal_{M,k}(h)$
such that ${\nbf'}_1^{r_{h-1}}\succ\nbf_1^{r_{h-1}}$.

\emph{Case 4: $b'\leq r_{h-1}-1$ and $n_{b'+j'}=p+1$ for $j'=1,2,\ldots,\min\{b'-a'-1,r_{h-1}-b'\}$.}
Let $\mbf_1^{r_{h-1}}$ be a sequence of positive integers as given in
\reqnarray{proof of nonadjacent distance larger than one II-(ii)-case-1-111}.
As in Case~1 above, we have $\mbf_1^{r_{h-1}}\in \Ncal_{M,k}(h)$.
We then consider the two subcases $b'-a'-1>r_{h-1}-b'$ and $b'-a'-1\leq r_{h-1}-b'$ separately.

\emph{Subcase 4(a): $b'-a'-1>r_{h-1}-b'$.}

\bpdffigure{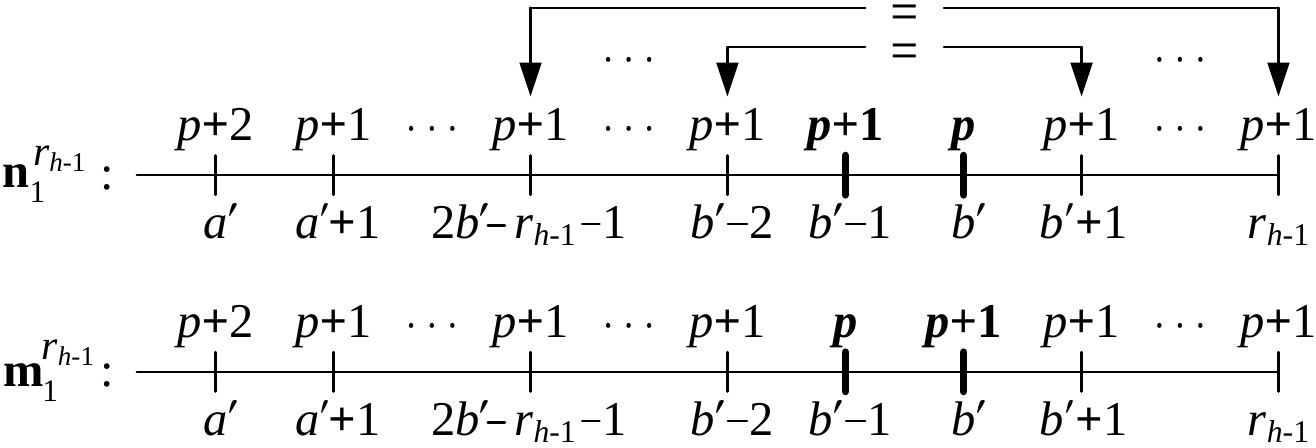}{4.5in}
\epdffigure{appendix-I-(ii)-case-4-1}
{An illustration of \reqnarray{proof of nonadjacent distance larger than one II-(ii)-case-4-111}
(note that we have $\min\{(b'-1)-1, r_{h-1}-(b'-1)-1\}=r_{h-1}-b'$
in \reqnarray{proof of nonadjacent distance larger than one II-(ii)-case-4-444}).}

In this subcase, we show that
\beqnarray{proof of nonadjacent distance larger than one II-(ii)-case-4-111}
n_{(b'-1)-j'}=n_{b'+j'}=p+1, \textrm{ for } j'=1,2,\ldots,\min\{(b'-1)-1, r_{h-1}-(b'-1)-1\}.
\eeqnarray
An illustration of \reqnarray{proof of nonadjacent distance larger than one II-(ii)-case-4-111}
is given in \rfigure{appendix-I-(ii)-case-4-1}.
Therefore, it follows from
$\nbf_1^{r_{h-1}}\in \Ncal_{M,k}(h)$ in \reqnarray{proof of nonadjacent distance larger than one II-111},
\reqnarray{proof of nonadjacent distance larger than one II-(ii)-444}--\reqnarray{proof of nonadjacent distance larger than one II-(ii)-case-1-111},
\reqnarray{proof of nonadjacent distance larger than one II-(ii)-case-4-111},
and \reqnarray{comparison rule B-4} in \rlemma{comparison rule B}(iii) that
$\nbf_1^{r_{h-1}}\prec \mbf_1^{r_{h-1}}$,
i.e., \reqnarray{proof of nonadjacent distance larger than one II-(ii)-333} holds
with ${\nbf'}_1^{r_{h-1}}=\mbf_1^{r_{h-1}}$.

To prove \reqnarray{proof of nonadjacent distance larger than one II-(ii)-case-4-111},
note that from $b'-a'-1>r_{h-1}-b'$, $a'\geq 1$, and $b'\leq r_{h-1}-1$, we have
\beqnarray{}
\alignspace
r_{h-1}-b'<b'-a'-1\leq b'-2,
\label{eqn:proof of nonadjacent distance larger than one II-(ii)-case-4-222}\\
\alignspace
a'<(b'-1)-(r_{h-1}-b')\leq (b'-1)-1<b'.
\label{eqn:proof of nonadjacent distance larger than one II-(ii)-case-4-333}
\eeqnarray
From \reqnarray{proof of nonadjacent distance larger than one II-(ii)-case-4-222},
we see that
\beqnarray{proof of nonadjacent distance larger than one II-(ii)-case-4-444}
\min\{(b'-1)-1, r_{h-1}-(b'-1)-1\}=\min\{b'-2, r_{h-1}-b'\}=r_{h-1}-b'.
\eeqnarray
From \reqnarray{proof of nonadjacent distance larger than one II-(ii)-222}
and \reqnarray{proof of nonadjacent distance larger than one II-(ii)-case-4-333},
we have
\beqnarray{proof of nonadjacent distance larger than one II-(ii)-case-4-555}
n_{(b'-1)-j'}=p+1, \textrm{ for } j'=1,2,\ldots,r_{h-1}-b'.
\eeqnarray
Furthermore, in this subcase we have from $b'-a'-1>r_{h-1}-b'$ that
\beqnarray{proof of nonadjacent distance larger than one II-(ii)-case-4-666}
n_{b'+j'}=p+1, \textrm{ for } j'=1,2,\ldots,\min\{b'-a'-1,r_{h-1}-b'\}=r_{h-1}-b'.
\eeqnarray
By combining \reqnarray{proof of nonadjacent distance larger than one II-(ii)-case-4-444},
\reqnarray{proof of nonadjacent distance larger than one II-(ii)-case-4-555},
and \reqnarray{proof of nonadjacent distance larger than one II-(ii)-case-4-666},
we obtain \reqnarray{proof of nonadjacent distance larger than one II-(ii)-case-4-111}.

\emph{Subcase 4(b): $b'-a'-1\leq r_{h-1}-b'$.}

\bpdffigure{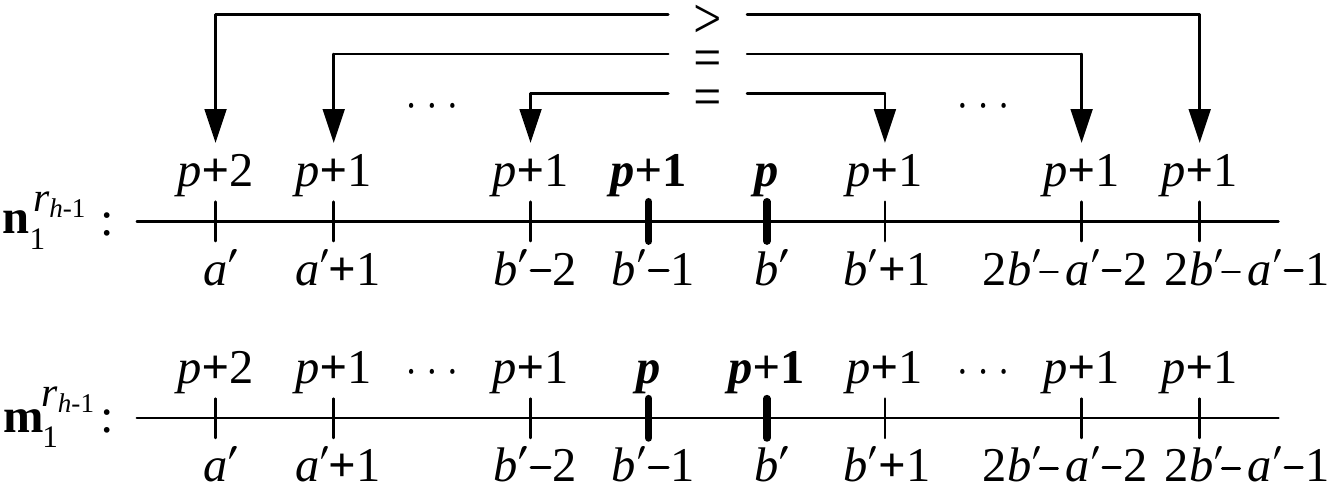}{4.5in}
\epdffigure{appendix-I-(ii)-case-4-2}
{An illustration of \reqnarray{proof of nonadjacent distance larger than one II-(ii)-case-4-888}
and \reqnarray{proof of nonadjacent distance larger than one II-(ii)-case-4-999}.}

In this subcase, we show that
\beqnarray{}
\alignspace
1\leq b'-a'-1\leq \min\{(b'-1)-1, r_{h-1}-(b'-1)-1\},
\label{eqn:proof of nonadjacent distance larger than one II-(ii)-case-4-777}\\
\alignspace
n_{(b'-1)-j'}=n_{b'+j'}=p+1, \textrm{ for } j'=1,2,\ldots,b'-a'-2,
\label{eqn:proof of nonadjacent distance larger than one II-(ii)-case-4-888}\\
\alignspace
n_{(b'-1)-(b'-a'-1)}=p+2>n_{b'+(b'-a'-1)}=p+1.
\label{eqn:proof of nonadjacent distance larger than one II-(ii)-case-4-999}
\eeqnarray
An illustration of
\reqnarray{proof of nonadjacent distance larger than one II-(ii)-case-4-888}
and \reqnarray{proof of nonadjacent distance larger than one II-(ii)-case-4-999}
is given in \rfigure{appendix-I-(ii)-case-4-2}.
Therefore, it follows from
$\nbf_1^{r_{h-1}}\in \Ncal_{M,k}(h)$ in \reqnarray{proof of nonadjacent distance larger than one II-111},
\reqnarray{proof of nonadjacent distance larger than one II-(ii)-444}--\reqnarray{proof of nonadjacent distance larger than one II-(ii)-case-1-111},
\reqnarray{proof of nonadjacent distance larger than one II-(ii)-case-4-777}--\reqnarray{proof of nonadjacent distance larger than one II-(ii)-case-4-999},
and \reqnarray{comparison rule B-2} in \rlemma{comparison rule B}(ii) that
$\nbf_1^{r_{h-1}}\prec \mbf_1^{r_{h-1}}$,
i.e., \reqnarray{proof of nonadjacent distance larger than one II-(ii)-333} holds
with ${\nbf'}_1^{r_{h-1}}=\mbf_1^{r_{h-1}}$.

From $b'\geq a'+2$ in \reqnarray{proof of nonadjacent distance larger than one II-(ii)-111}
and $a'\geq 1$, we see that
\beqnarray{proof of nonadjacent distance larger than one II-(ii)-case-4-aaa}
1\leq b'-a'-1\leq b'-2.
\eeqnarray
It then follows from \reqnarray{proof of nonadjacent distance larger than one II-(ii)-case-4-aaa}
and $b'-a'-1\leq r_{h-1}-b'$ that
\beqnarray{proof of nonadjacent distance larger than one II-(ii)-case-4-bbb}
1\leq b'-a'-1\leq \min\{b'-2, r_{h-1}-b'\}=\min\{(b'-1)-1, r_{h-1}-(b'-1)-1\}.
\eeqnarray
Thus, \reqnarray{proof of nonadjacent distance larger than one II-(ii)-case-4-777}
follows from \reqnarray{proof of nonadjacent distance larger than one II-(ii)-case-4-bbb}.

To prove \reqnarray{proof of nonadjacent distance larger than one II-(ii)-case-4-888}
and \reqnarray{proof of nonadjacent distance larger than one II-(ii)-case-4-999},
note that from \reqnarray{proof of nonadjacent distance larger than one II-(ii)-222}
and $a'=(b'-1)-(b'-a'-1)<b'$ we have
\beqnarray{}
\alignspace
n_{(b'-1)-j'}=p+1, \textrm{ for } j'=1,2,\ldots,b'-a'-2,
\label{eqn:proof of nonadjacent distance larger than one II-(ii)-case-4-ccc}\\
\alignspace
n_{(b'-1)-(b'-a'-1)}=n_{a'}=p+2.
\label{eqn:proof of nonadjacent distance larger than one II-(ii)-case-4-ddd}
\eeqnarray
Furthermore, in this subcase we have from $b'-a'-1\leq r_{h-1}-b'$ that
\beqnarray{proof of nonadjacent distance larger than one II-(ii)-case-4-eee}
n_{b'+j'}=p+1, \textrm{ for } j'=1,2,\ldots,\min\{b'-a'-1,r_{h-1}-b'\}=b'-a'-1.
\eeqnarray
By combining \reqnarray{proof of nonadjacent distance larger than one II-(ii)-case-4-ccc},
\reqnarray{proof of nonadjacent distance larger than one II-(ii)-case-4-ddd},
and \reqnarray{proof of nonadjacent distance larger than one II-(ii)-case-4-eee},
we obtain \reqnarray{proof of nonadjacent distance larger than one II-(ii)-case-4-888}
and \reqnarray{proof of nonadjacent distance larger than one II-(ii)-case-4-999}.

\bappendix{Proof of \rlemma{main lemma II}} {proof of main lemma II}

In this appendix,
we use \rcorollary{adjacent distance larger than one II}(i) (corollary to \rlemma{adjacent distance larger than one II}),
\rcorollary{nonadjacent distance larger than one II}(i) (corollary to \rlemma{nonadjacent distance larger than one II}),
and Comparison rule B in \rlemma{comparison rule B} to prove \rlemma{main lemma II}.

Let $\nbf_1^{r_{h-1}}(h)$ be an optimal sequence over $\Ncal_{M,k}(h)$.
As commented before the statement of \rlemma{main lemma II},
we can use \rcorollary{adjacent distance larger than one II}(i)
and \rcorollary{nonadjacent distance larger than one II}(i) to show that
\beqnarray{proof of main lemma II-111}
n_i(h)=
\bselection
q_h+1, &\textrm{if } i=i_1,i_2,\ldots,i_{r_h}, \\
q_h, &\textrm{otherwise},
\eselection
\eeqnarray
for some $1\leq i_1<i_2<\cdots <i_{r_h}\leq r_{h-1}$.

In the following, we show that $i_{r_h}=r_{h-1}$ by contradiction.
Assume on the contrary that $i_{r_h}\leq r_{h-1}-1$.
We will use Comparison rule B in \rlemma{comparison rule B} to show that
there exists a sequence ${\nbf'}_1^{r_{h-1}}(h)\in \Ncal_{M,k}(h)$
such that ${\nbf'}_1^{r_{h-1}}(h)\succ\nbf_1^{r_{h-1}}(h)$,
contradicting to the optimality of $\nbf_1^{r_{h-1}}(h)$.
For simplicity, let $\nbf_1^{r_{h-1}}=\nbf_1^{r_{h-1}}(h)$.
As $i_{r_h}\leq r_{h-1}-1$, we have $i_{r_h}<i_{r_h}+1\leq r_{h-1}$.
Let ${\nbf'}_1^{r_{h-1}}$ be a sequence of positive integers such that
\beqnarray{proof of main lemma II-222}
n'_{i_{r_h}}=n_{i_{r_h}}-1,\ n'_{i_{r_h}+1}=n_{i_{r_h}+1}+1,
\textrm{ and } n'_i=n_i \textrm{ for } i\neq i_{r_h}, i_{r_h}+1.
\eeqnarray
It is easy to see from \reqnarray{proof of main lemma II-222},
\reqnarray{proof of main lemma II-111},
$\nbf_1^{r_{h-1}}\in \Ncal_{M,k}(h)$, and \reqnarray{N-M-k-h} that
\beqnarray{proof of main lemma II-333}
\sum_{i=1}^{r_{h-1}}n'_i=\sum_{i=1}^{r_{h-1}}n_i=r_{h-2}.
\eeqnarray
As such, it follows from $\nbf_1^{r_{h-1}}\in \Ncal_{M,k}(h)$,
\reqnarray{proof of main lemma II-111}, \reqnarray{proof of main lemma II-222},
and \reqnarray{proof of main lemma II-333} that ${\nbf'}_1^{r_{h-1}}\in \Ncal_{M,k}(h)$.
Note that from \reqnarray{proof of main lemma II-111} we have
\beqnarray{proof of main lemma II-444}
n_{i_{r_h}}-n_{i_{r_h}+1}=(q_h+1)-q_h=1.
\eeqnarray

Now we have $\nbf_1^{r_{h-1}}\in \Ncal_{M,k}(h)$,
$n_{i_{r_h}}-n_{i_{r_h}+1}=1$ in \reqnarray{proof of main lemma II-444},
and $n'_{i_{r_h}}=n_{i_{r_h}}-1$, $n'_{i_{r_h}+1}=n_{i_{r_h}+1}+1$, and $n'_i=n_i$ for $i\neq i_{r_h}, i_{r_h}+1$
in \reqnarray{proof of main lemma II-222}.
As such, we are in a position to use Comparison rule B in \rlemma{comparison rule B} (with $a=i_{r_h}$)
to show that ${\nbf'}_1^{r_{h-1}}\succ\nbf_1^{r_{h-1}}$.
We need to consider the two cases $r_h=1$ and $r_h\geq 2$ separately.

\emph{Case 1: $r_h=1$}.
In this case, we have from \reqnarray{proof of main lemma II-111} and $i_{r_h}\leq r_{h-1}-1$ that
\beqnarray{proof of main lemma II-case-1-111}
n_i=
\bselection
q_h+1, &\textrm{if } i=i_{r_h}, \\
q_h, &\textrm{otherwise}.
\eselection
\eeqnarray
If $i_{r_h}=1$ or $i_{r_h}=r_{h-1}-1$,
then it follows from \reqnarray{comparison rule B-1} in \rlemma{comparison rule B}(i)
that $\nbf_1^{r_{h-1}}\prec{\nbf'}_1^{r_{h-1}}$.
On the other hand, if $2\leq i_{r_h}\leq r_{h-1}-2$,
then it is easy to see from \reqnarray{proof of main lemma II-case-1-111} that
\beqnarray{proof of main lemma II-case-1-222}
n_{i_{r_h}-j}=n_{(i_{r_h}+1)+j}=q_h, \textrm{ for } j=1,2,\ldots,\min\{i_{r_h}-1,r_{h-1}-i_{r_h}-1\}.
\eeqnarray
Therefore, it follows from $2\leq i_{r_h}\leq r_{h-1}-2$,
\reqnarray{proof of main lemma II-case-1-222},
and \reqnarray{comparison rule B-4} in \rlemma{comparison rule B}(iii)
that $\nbf_1^{r_{h-1}}\prec{\nbf'}_1^{r_{h-1}}$.

\emph{Case 2: $r_h\geq 2$}.
As $1\leq i_1<i_2<\cdots <i_{r_h}\leq r_{h-1}-1$ and $r_h\geq 2$,
we have $2\leq i_1+1\leq i_{r_h}\leq r_{h-1}-1$ in this case.
If $i_{r_h}=r_{h-1}-1$, then it follows from \reqnarray{comparison rule B-1}
in \rlemma{comparison rule B}(i) that $\nbf_1^{r_{h-1}}\prec{\nbf'}_1^{r_{h-1}}$.
On the other hand, if $2\leq i_{r_h}\leq r_{h-1}-2$,
then we consider the following two subcases.

\emph{Subcase 2(a): $i_{r_h}-i_{r_h-1}>r_{h-1}-i_{r_h}-1$.}
In this subcase, we have $i_{r_h}-1\geq i_{r_h}-i_{r_h-1}>r_{h-1}-i_{r_h}-1$,
and it follows that
\beqnarray{proof of main lemma II-case-2-111}
\min\{i_{r_h}-1,r_{h-1}-i_{r_h}-1\}=r_{h-1}-i_{r_h}-1.
\eeqnarray
From $i_{r_h}-i_{r_h-1}>r_{h-1}-i_{r_h}-1$ in this subcase, we have
\beqnarray{}
\alignspace
i_{r_h}-(r_{h-1}-i_{r_h}-1)>i_{r_h}-(i_{r_h}-i_{r_h-1})=i_{r_h-1},
\label{eqn:proof of main lemma II-case-2-222}\\
\alignspace
(i_{r_h}+1)+(r_{h-1}-i_{r_h}-1)=r_{h-1}.
\label{eqn:proof of main lemma II-case-2-333}
\eeqnarray
It is easy to see from \reqnarray{proof of main lemma II-111}
and \reqnarray{proof of main lemma II-case-2-111}--\reqnarray{proof of main lemma II-case-2-333} that
\beqnarray{proof of main lemma II-case-2-444}
n_{i_{r_h}-j}=n_{(i_{r_h}+1)+j}=q_h,\ j=1,2,\ldots,r_{h-1}-i_{r_h}-1=\min\{i_{r_h}-1,r_{h-1}-i_{r_h}-1\}.
\eeqnarray
Therefore, it follows from $2\leq i_{r_h}\leq r_{h-1}-2$,
\reqnarray{proof of main lemma II-case-2-444},
and \reqnarray{comparison rule B-4} in \rlemma{comparison rule B}(iii)
that $\nbf_1^{r_{h-1}}\prec{\nbf'}_1^{r_{h-1}}$.

\emph{Subcase 2(b): $i_{r_h}-i_{r_h-1}\leq r_{h-1}-i_{r_h}-1$.}
In this subcase, we see from $i_{r_h-1}<i_{r_h}$, $i_{r_h}-1\geq i_{r_h}-i_{r_h-1}$,
and $i_{r_h}-i_{r_h-1}\leq r_{h-1}-i_{r_h}-1$ that
\beqnarray{proof of main lemma II-case-2-555}
1\leq i_{r_h}-i_{r_h-1}\leq \min\{i_{r_h}-1,r_{h-1}-i_{r_h}-1\}.
\eeqnarray
As it is clear that
\beqnarray{}
\alignspace
i_{r_h}-(i_{r_h}-i_{r_h-1})=i_{r_h-1},
\label{eqn:proof of main lemma II-case-2-666}\\
\alignspace
i_{r_h}<(i_{r_h}+1)+(i_{r_h}-i_{r_h-1})\leq (i_{r_h}+1)+(r_{h-1}-i_{r_h}-1)=r_{h-1},
\label{eqn:proof of main lemma II-case-2-777}
\eeqnarray
we see from \reqnarray{proof of main lemma II-111},
\reqnarray{proof of main lemma II-case-2-666},
and \reqnarray{proof of main lemma II-case-2-777} that
\beqnarray{}
\alignspace
n_{i_{r_h}-j}=n_{(i_{r_h}+1)+j}=q_h, \textrm{ for } j=1,2,\ldots,i_{r_h}-i_{r_h-1}-1,
\label{eqn:proof of main lemma II-case-2-888}\\
\alignspace
n_{i_{r_h}-(i_{r_h}-i_{r_h-1})}=n_{i_{r_h-1}}=q_h+1>n_{(i_{r_h}+1)+(i_{r_h}-i_{r_h-1})}=q_h.
\label{eqn:proof of main lemma II-case-2-999}
\eeqnarray
Therefore, it follows from
$2\leq i_{r_h}\leq r_{h-1}-2$,
\reqnarray{proof of main lemma II-case-2-555},
\reqnarray{proof of main lemma II-case-2-888},
\reqnarray{proof of main lemma II-case-2-999},
and \reqnarray{comparison rule B-2} in \rlemma{comparison rule B}(ii)
that $\nbf_1^{r_{h-1}}\prec{\nbf'}_1^{r_{h-1}}$.

\end{document}